\documentclass[12pt,oneside,a4paper,openright,english]{book}
\usepackage{times}                                 
\usepackage{amsmath,amssymb,amsfonts}
\usepackage{ifpdf}
\usepackage{graphicx}
\usepackage{epstopdf}
\usepackage{epsfig}
\usepackage{stackrel}
\usepackage{graphics}   
\usepackage{dcolumn,enumerate}    
\usepackage{bm,bbm}         
\usepackage{mathrsfs}
\usepackage{mathtools}
\usepackage{color}
\usepackage[T1]{fontenc}
\usepackage{rotating}
\usepackage{upgreek}
\usepackage{float}
\usepackage{babel}
\usepackage{longtable}
\usepackage{cite}
\usepackage[utf8]{inputenc}
\usepackage{titlesec}
\usepackage{txfonts}
\usepackage{url}
\usepackage{lineno}
\usepackage{geometry}                   
\usepackage{fancyhdr}                   
\usepackage{titlesec}\titleformat{\chapter}[display]
  {\normalfont\LARGE\bfseries\centering}{\MakeUppercase\chaptertitlename\ \thechapter}{20pt}{\LARGE}
\usepackage{setspace}
\DeclareMathVersion{normal2}
\renewcommand{\headrulewidth}{0.5pt}

\fancypagestyle{plain}{
  \fancyhead{}
  \renewcommand{\headrulewidth}{0pt}
}
\definecolor{lightgray}{rgb}{0.98, 0.81, 0.69}
\makeatletter
\newcommand\longleftrightarrowfill@{%
  \arrowfill@\leftarrow\relbar\rightarrow}
\makeatother

\newcommand{\blankpage}{\newpage\thispagestyle{empty}\mbox{}\newpage}

\usepackage{hyperref} 

\begin{document}

\renewcommand{\chaptername}{CHAPTER}\rhead{}
\pagenumbering{roman}

\addcontentsline{toc}{chapter}{\normalsize
\textbf{INNER FIRST
PAGE}}

\thispagestyle{plain}\thispagestyle{empty}

\begin{center}
\textbf{\uppercase{\LARGE{}A theoretical study of nonclassical effects in optical and spin systems and their applications}}
\par\end{center}{\LARGE \par}

~

\begin{center}
\emph{\large{}Thesis submitted in fulfillment of the requirements
for the Degree of}
\par\end{center}{\large \par}

~

\begin{center}
\textbf{\Large{}DOCTOR OF PHILOSOPHY}
\par\end{center}{\Large \par}

~

\begin{center}
By
\par\end{center}

~

\begin{center}
\textbf{\uppercase{\large{}Kishore Thapliyal}}
\par\end{center}{\large \par}

\begin{center}
\textbf{\uppercase{\large{}Enrollment No.: 14410002}}
\par\end{center}{\large \par}

~

~
\begin{quotation}
\begin{figure}[H]
\centering{}\includegraphics{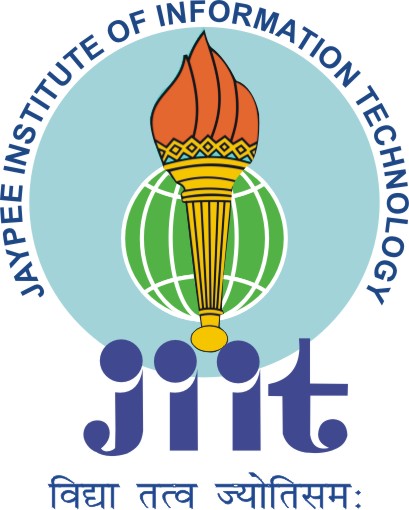}
\end{figure}

\end{quotation}
\begin{center}
Department of Physics and Materials Science and Engineering 
\par\end{center}

\begin{center}
JAYPEE INSTITUTE OF INFORMATION TECHNOLOGY 
\par\end{center}

\begin{center}
(Declared Deemed to be University U/S 3 of UGC Act) 
\par\end{center}

\begin{center}
A-10, SECTOR-62, NOIDA, INDIA
\par\end{center}

\begin{center}
April, 2018 
\par\end{center}

\thispagestyle{plain}\thispagestyle{empty}

\pagebreak{}

\text{}

\thispagestyle{plain}\thispagestyle{empty}

\newpage



\thispagestyle{plain}\thispagestyle{empty}
\begin{center}
\topskip0pt

``When you possess light within, you see it externally.''\\
-- Ana{\"\i}s Nin

\vspace*{\fill}
\Large{\bfseries\LARGE\textit{{Dedicated to the light within us}}}
\vspace*{\fill}
\end{center}
\cleardoublepage
\thispagestyle{plain}
\blankpage

\setlength{\headheight}{0.4in}
\pagestyle{fancy}\rhead{\textbf{CONTENTS}}\lhead{}
\renewcommand{\contentsname}{CONTENTS}
\tableofcontents{\normalsize}
\cleardoublepage

\phantomsection
\addcontentsline{toc}{chapter}{\normalsize
\textbf{PREFACE AND ACKNOWLEDGMENT}}\rhead{\textbf{PREFACE AND ACKNOWLEDGMENT}} 
\chapter*{Preface and acknowledgment \label{acknolwedgement}}

It gives me an immense pleasure to thank the person who had the most arduous task during the last four years (Prof. Anirban Pathak) not only for shaping this piece of work, my life too. Thankfully, as hard it turned out to summarize in this dissertation what I could comprehend from him, so does his contributions in changing my life in more than one way. I can only say ``I may be \textit{Vyasa} (the orator) of the story of light written in the next few pages, I can see him as \textit{Krishna} on each page and every word of it.''
I feel sorry that I am petty enough to even offer my \textit{Guru Dakshina} as whatever I accomplish in my life after this will be the outcome of his devotion. 

I am obliged to numerous people for the research work I have reported in this period of time.
Especially, my seniors during this doctoral thesis work, Dr. Amit Verma, Dr. Anindita Banerjee, and Dr. Chitra Shukla, who had always given their positive feedback to boost my morale. All my co-authors in different research papers to give me an opportunity to learn from/with them. With special reference to Dr. Biswajit Sen, the coolest of my coauthors; Dr. Subhashish Banerjee for further widening my horizon by introducing me to open quantum systems; Prof. Jan Pe{\v{r}}ina for his inputs and encouraging remarks at times. Also, Dr. Subhashish Banerjee, Dr. Gautum Pal, and Prof. Swapan Mandal for hosting me during my visits to IITJ, Rajasthan, ISI, Kolkata, and Visva-Bharati, Santiniketan, respectively. My DPMC members, Prof. S. P. Purohit, Prof. B. P. Chamola, and Prof. D. K. Rai, for their pedantic approach during the end semester seminars.

My group members, Dr. Abhishek Shukla (Professor X, a perfectionist), Dr. Nasir Alam (Hulk, a tired one), Dr. Rishi Dutt Sharma (Chacha Chowdhury, a friend), Mitali Sisodia (Supergirl, a perfect junior), Ashwin Saxena (Ant-man, a witty fellow), Kathakali Mandal (Catwoman, an artist), Meenakshi Rana (Wonder Woman, who knows everything), Suhail Khan (the newborn), and other research scholars, teaching and non-teaching staff in PMSE department, who had become part of my life during this period of time. In the words of Schr\"{o}dinger, ``When two systems enter into temporary physical interaction due to known forces between them, and when after a time of mutual influence the systems separate again, then they can no longer be described in the same way as before.''

This may/may not be the most important feat in my life, but is certainly so far. Therefore, I would like to take this opportunity to thank everyone who had influenced my life and had played a significant role in it. To begin with, my grandfather, whose words, ``I know you will make me proud'', reverberate through my mind in my lamest moments to elate me.

Today I feel that all my teachers, who lit my path allowing me to tread firmly through darkness, have their contributions in this work. I will thank especially my very first teacher, my sister Jyoti Di, for her patience during my initial inscriptions.
I am also indebted to Dr. B. V. Tripathi and Dr. Hemwati Nandan for introducing me to the research field, which I feel has given me both peace of mind and reason of life.

I am also grateful to all my bosom buddies Tarun, Pankaj, Shashi Bhushan, Pawan, and the one who is undergoing the same evolution as mine, Ravi, which helped both of us to vent out our frustrations. 
My roommates Rachit and Neeraj to endure a researcher roomie with an abnormal life and behavior for more than four years and still not to complain.

Needless to say, my parents and family for the love and affection showered upon me, believing in me, and always understanding when I was not there for them.

At the end, I will extend my gratitude to each and every person who shared his experiences to enlighten me in different phases of my life. Due to which I consider this my responsibility to dedicate the present piece of work to the light within all of us, part of which had emancipated me. 
\textit{May this work on light impart some light to light within us}.

I would also like to acknowledge the financial assistances I received in different phases of my research work from DST, JIIT, DRDO, and CSIR without which this work would not have been possible. TPSC for commending me as category A speaker, which provided me an opportunity to visit TPSC centers and present my work.

\vspace{1cm}

{\flushright{\textit{Kishore Thapliyal}}}

\thispagestyle{plain}   
\cleardoublepage


\phantomsection
\addcontentsline{toc}{chapter}{\normalsize
\textbf{ABSTRACT}}
\thispagestyle{plain}
\chapter*{Abstract \label{abstract}}

Nonclassical states, having no classical analogue, promise advantage in the performance in several fields of technology, such as computation, communication, sensing. This led to an escalated interest in the recent years for the generation and characterization of nonclassical states in various physical systems of interest. Keeping this in mind, we examined generation of various lower- and higher-order nonclassical states in both codirectional and contradirectional asymmetric nonlinear optical couplers operating under second harmonic generation with the help of moments-based criteria. Using another such system (a symmetric nonlinear optical coupler), we have also established the possibility of observing quantum Zeno and anti-Zeno effects and further reduced the obtained results to yield the corresponding expressions for the asymmetric coupler. These studies have been accomplished using a complete quantum description of the system considering all the field modes as weak. Further, characterization of nonclassicality in spin systems using quasiprobability distributions has been performed, which has also provided us the dynamics of the amount of nonclassicality (using nonclassical volume). As the  reconstruction of quasiprobability distributions is possible with the help of tomograms accessible in the experiments, tomograms are computed here for both finite and infinite dimensional systems. Finally, a set of controlled quantum communication (both insecure and secure) schemes with minimum number of entangled qubits has been proposed, which is also analyzed over Markovian channels. The optimized controlled secure quantum communication scheme is observed to be reducible to several cryptographic tasks. Using this interesting observation, we obtained the performance of a set of cryptographic schemes over non-Markovian channels, too.

\cleardoublepage

%
%
%
%

%
%
%
\blankpage

\lhead{} 
\fancyhead[LE,RO]{\bfseries  \leftmark}
\fancypagestyle{plain}{\fancyhead{}
\renewcommand{\headrulewidth}{0pt}}
\clearpage{\pagestyle{empty}\cleardoublepage} 
\thispagestyle{plain}
\pagenumbering{arabic}

\titlespacing*{\chapter}{0pt}{-50pt}{20pt}
\chapter{Introduction \label{Introduction}}

\section{History of quantum mechanics and quantum optics \label{Qop}}

The journey of quantum physics started with Planck's revolutionary idea of quantizing energy transfered from the oscillators in the cavity wall to the radiation field to resolve the \emph{ultraviolet catastrophe} in black-body spectrum \cite{planck1901law}. His thesis was later extended by Einstein in an effort to  explain the photoelectric effect \cite{einstein1905erzeugung}. Specifically, Einstein's proposal established that energy is not only transfered to the radiation field in a quantized manner, it also traverses in the quantized fashion (see \cite{pathak2017classical} for a detailed review). 
Chronologically, this was followed by some remarkable feats, such as Bohr model, stimulated emission and Einstein coefficients, de Broglie's matter waves, Heisenberg's uncertainty principle, Dirac's quantum theory for radiation, Heisenberg's matrix mechanics, Schr\"{o}dinger equation, which changed our understanding of microscopic world completely.

Today in the view of quantum mechanics, which has grown to become the most sophisticated model of nature, the interaction between light and matter can be modeled using three approaches \cite{fox2006quantum}. Firstly, the classical optics, as the name suggests, treats the atoms as dipoles and light as electromagnetic wave (solution of Maxwell's equations); the second one is the semiclassical optics, where atom is considered quantized, but light is still considered as the electromagnetic wave; and finally, quantum optics which considers an interaction between quantized atom and photons. The semiclassical theory provides a successful explanation to various phenomena, when the classical theory fails to do so. In some cases, applicability of semiclassical optics is found to be enhanced by taking vacuum fluctuation into consideration with the usual semiclassical theory. 
Here, also note that although the photoelectric effect can be described semiclassically, due to unavailability of the concept of quantization of atomic energy levels at that time Einstein preferred the notion of light quanta (now known as photon) \cite{pathak2017classical}. Therefore, it poses a serious question whether it is possible to characterize the quantum states and phenomena, which cannot be explained in the domain of classical physics. Before we address this question in Section \ref{Noncl}, it would be apt to briefly introduce quantum theory of radiation in the following section.

We will conclude the present section by introducing a few basic mathematical tools of quantum mechanics which will be useful in understanding the present work. Quantum mechanics allows one to write a mathematical function $\psi$ describing whole  information of a quantum system in a Hilbert space $\mathcal{H}$ of suitable dimension $N$. This function is called wavefunction or wavevector and can be represented by an $N$ element column vector. In the Dirac notation \cite{dirac1939new}, the state of a quantum system is represented by a ket as $|\psi\rangle$, and its adjoint as a bra $\langle\psi|$ (which becomes a row matrix). Also, a quantum state defined in two ($N$) dimensional Hilbert space is known as qubit (qunit). Further, every quantum state in an $N$-dimensional Hilbert space can be expanded in $N$ linearly independent state vectors known as basis vectors, and the set of such orthogonal vectors is called a basis set. 

Further, all physical observables are characterized by Hermitian operators ($\hat{O}$), which give one of the real eigenvalues (i.e., $\hat{O}|\psi\rangle=e|\psi\rangle$) of the operator on measurement. The average value of the physical quantity can be calculated as $\langle\hat{O}\rangle=\langle \psi|\hat{O}|\psi\rangle$. In principle, information of a physical observable can be extracted from the quantum system by its interaction with an external measuring device. The quantum state after one such quantum measurement resulting in $m$th outcome can be written as \cite{von1955mathematical}
\begin{equation}
|\psi_m\rangle=\frac{M_m|\psi\rangle}{\sqrt{\langle \psi|M_m^{\dagger}M_m|\psi\rangle}},\label{eq:meas}
\end{equation}
where the quantity in the denominator corresponds to the probability $p_m$ of the $m$th outcome. Instead of this selective measurement, one can also write the non-selective measurement leading to a mixture of quantum states corresponding to different measurement outcomes. Such mixture can be denoted with the help of density matrix $\rho=\underset{m}{\sum}p_{m}|\psi_m\rangle\langle \psi_m|$. One can reduce it to obtain the density matrix of a pure state as $\rho_m=|\psi_m\rangle\langle \psi_m|$. Independently, one can also obtain a mixed state after tracing over a subsystem of an entangled quantum state\footnote{Entanglement will be discussed in detail in Section \ref{Ent}.}. A detailed discussion of the mathematical techniques required in the fields of quantum mechanics and optics may be safely circumvented  here, restricting our review to the topics that would be required for the understanding of the present thesis work. For more detail, the interested readers may refer to interesting books \cite{louisell1973quantum,perina1991quantum,kim1991phase,plum1996density,scully1997quantum,puri2001mathematical,fox2006quantum,walls2007quantum,weiss2008quantum,luks2009quantum,schleich2011quantum,agarwal2013quantum} on this topic.

\section{Dirac's quantum theory of radiation \label{Sec-Q}}

A complete classical description of electromagnetic field can be given by Maxwell's equations \cite{maxwell1865dynamical}. Our focus in the present work is the description of light and its properties beyond Maxwell's equations and applications of such properties. Therefore, it would be apt to begin with Maxwell's equations in free space 
\begin{subequations}
\begin{equation}
\nabla\cdot E=0,\label{eq:Max-1}
\end{equation}
\begin{equation}
\nabla\cdot B=0,\label{eq:Max-2}
\end{equation}
\begin{equation}
\nabla\times E=-\frac{\partial B}{\partial t},\label{eq:Max-3}
\end{equation}
and
\begin{equation}
\nabla\times B=\frac{1}{c^2}\frac{\partial E}{\partial t},\label{eq:Max-4}
\end{equation}
\end{subequations}
where $c=\frac{1}{\sqrt{\mu_0 \epsilon_0}}$ is the speed of  light in vacuum defined in terms of permittivity ($\epsilon_0$) and permeability ($\mu_0$) of the free space. The set of equations (\ref{eq:Max-1})-(\ref{eq:Max-4}) enables one to describe both electric and magnetic fields as soultion of wave equations, such as 
\begin{equation}
\nabla^2 E-\frac{1}{c^2}\frac{\partial^2 E}{\partial t^2}=0.\label{eq:waveEq}
\end{equation}

To quantize the radiation field \cite{dirac1927quantum,fermi1932quantum,scully1997quantum}, we begin with a linearly polarized electric field in a cavity of length $L$ propagating in the $z$ direction. Due to linearity of the wave equation (\ref{eq:waveEq}), we can write the electric field as a linear combination of all the normal modes as
\begin{equation}
E_x \left(z,t\right)=\underset{n}{\sum}A_n q_n \left(t\right) \sin\left(k_n z\right),\label{eq:nor-mod-E}
\end{equation}
where $q_n$ is the normal mode amplitude for the $n$th mode (for integer $n$) with $k_n=\frac{n \pi}{L}.$ Further, in the introduced notation 
$A_n= \left(\frac{2 m_n \nu_n^2}{\epsilon_0 V}  \right)^{1/2}$
with $\nu_n=c\, k_n$, $V$ is the volume of the resonator; and $m_n$ is a constant in the units of mass. Using this, we intend  to establish an equivalence of radiation with mechanical oscillator. Further, with the help of Eq. (\ref{eq:nor-mod-E}), we can obtain corresponding magnetic field in the cavity to be
\begin{equation}
B_y \left(z,t\right)=\underset{n}{\sum}A_n \left(\frac{\dot{q}_n}{c^2 k_n} \right) \cos\left(k_n z\right).\label{eq:nor-mod-B}
\end{equation}
Thus, total energy for the field can be expressed as a classical Hamiltonian 
\begin{equation}
H=\frac{1}{2}\int_V d\tau \left(\epsilon_0 E_x^2+\frac{1}{\mu_0}B_y^2\right),\nonumber
\end{equation}
which can  be rewritten as (after using Eqs. (\ref{eq:nor-mod-E}) and (\ref{eq:nor-mod-B}) and integrating over total volume taking into account the orthogonality of normal modes) 
\begin{equation}
H=\frac{1}{2}\underset{n}{\sum} \left(m_n \nu_n^2 q_n^2+m_n \dot{q}_n^2\right)=\frac{1}{2}\underset{n}{\sum} \left(m_n \nu_n^2 q_n^2+\frac{p_n^2}{m_n} \right),\label{eq:Ham1}
\end{equation}
with $p_n=m_n \dot{q}_n$ in the dimensions of momentum. 

The position and momentum operators may be shown to obey the commutation relations 
\begin{equation}
\left[q_n,p_m\right]=i\hbar \delta_{nm},\quad\quad \left[q_n,q_m\right]=\left[p_n,p_m\right]=0, \nonumber
\end{equation}
 where $\hbar$ is the reduced Planck's constant.
Exploiting which a new set of operators can be introduced as 
\begin{subequations}
\begin{equation}
a_n e^{-i\nu_n t}=\frac{1}{\sqrt{2\hbar m_n \nu_n}}\left(m_n \nu_n q_n +i p_n\right)\label{eq:ann}
\end{equation}
and
\begin{equation}
a_n^\dagger e^{i\nu_n t}=\frac{1}{\sqrt{2\hbar m_n \nu_n}}\left(m_n \nu_n q_n -i p_n\right)\label{eq:crea}
\end{equation}
\end{subequations}
giving much simpler form to the Hamiltonian (\ref{eq:Ham1}), 
\begin{equation}
H=\underset{n}{\sum} \hbar \nu_n \left(a_n^\dagger a_n+\frac{1}{2}\right),\label{eq:Ham-fin}
\end{equation}
also respecting the commutation relation 
\begin{equation}
\left[a_n,a_m^\dagger\right]=\delta_{nm},\quad\quad \left[a_n,a_m\right]=\left[a_n^\dagger,a_m^\dagger\right]=0. \nonumber
\end{equation}

Further, one can write the electric and magnetic fields in terms of the operators, $a_n$ and $a_n^\dagger$ referred to as the annihilation and creation operators, respectively. Here, we rather report the electric and magnetic fields quantized  in the free space as \cite{dirac1927quantum,scully1997quantum}
\begin{subequations}
\begin{equation}
E\left(\vec{r},t\right)=\underset{\bm{k}}{\sum}\hat{\epsilon}_{\bm{k}} \mathcal{E}_{\bm{k}} a_{\bm{k}} e^{-i\nu_{\bm{k}} t+ i {\bm{k}}\cdot\vec{r}}+ \mathrm{H.c.},\label{eq:E-in-a}
\end{equation}
and
\begin{equation}
B \left(\vec{r},t\right)=\underset{\bm{k}}{\sum}\frac{{\bm{k}}\times \hat{\epsilon}_{\bm{k}}}{\nu_{\bm{k}}} \mathcal{E}_{\bm{k}} a_{\bm{k}} e^{-i\nu_{\bm{k}} t+ i {\bm{k}}\cdot\vec{r}}+ \mathrm{H.c.},\label{eq:B-in-a}
\end{equation}
\end{subequations}
where $\mathcal{E}_{\bm{k}}=\left(\frac{\hbar \nu_{\bm{k}}}{\epsilon_0 V}\right)^{1/2}$ is a constant of the same dimensions as that of the electric field; and  $\hat{\epsilon}_{\bm{k}}$ corresponds to a unit polarization vector with the wavevector ${\bm{k}}.$

We have already shown that a single-mode field is equivalent to harmonic oscillator (\ref{eq:Ham-fin}). Therefore, exactly solvable harmonic oscillator system plays a crucial role in quantum optics. As the eigenfunctions of this system form a complete basis, known as Fock basis, this allows us to expand an arbitrary quantum state in the Fock basis.

We are now in a position to introduce another important quantum state known as  the \textit{coherent state} $\left(|\alpha\rangle\right)$ represented as an infinite superposition of Fock states which is an eigenket of the annihilation operator $a$, such that $a|\alpha\rangle=\alpha|\alpha\rangle$. The coherent state is one of the widely used quantum states due to its several interesting and useful properties. Specifically, Schr\"{o}dinger introduced this kind of states as minimum uncertainty states \cite{schrodinger1926stetige}. Mathematically, the coherent state can be written in the Fock basis as 
\begin{equation}
|\alpha\rangle=\sum_{n=0}^{\infty}c_{n}|n\rangle=\exp\left(-\frac{|\alpha|^{2}}{2}\right)\sum_{n=0}^{\infty}\frac{\alpha^{n}}{\sqrt{n!}}|n\rangle,\label{eq:CS}
\end{equation}
where $\{|n\rangle\}$ is the Fock basis, and $\left|c_n\right|^2$ is the probability of obtaining the $n$th Fock state $\left(|n\rangle\right)$ or an $n$ photon state in a measurement performed in the Fock basis. Note that, $\alpha$ can take any complex value. Using this simplified expression (\ref{eq:CS}), one can easily obtain the photon number distribution as $\mathcal{P}\left( n \right)=|c_n|^2$, which follows the well-known Poisson distribution. It can be easily verified that the same expression of coherent state can be obtained by application of the displacement operator $D(\alpha)$ on the vacuum state $|0\rangle$, i.e.,
$D(\alpha)|0\rangle = e^{\alpha a^{\dagger} - \alpha^* a}|0\rangle=|\alpha\rangle.$  
Further, two arbitrary coherent states are not orthogonal to each other due to which the set of all coherent states form an overcomplete basis \cite{klauder1960action}.

\section{Nonclassical states and their characterization \label{Noncl}}

As mentioned previously, the true essence of quantum theory of light, which is briefly introduced in the previous section, can be captured only through the nonclassical states. Specifically, the quantum states not having a classical counterpart show nonclassical behavior and consequently called the nonclassical states. In the present section, we will describe some tools that can be used to characterize such states.

\subsection{Quasiprobability distributions \label{quasi}}
 
A complete theoretical characterization of quantum states able to show nonclassical features was made  possible only due to Glauber's \cite{glauber1963coherent,glauber1963photon} and Sudarshan's \cite{sudarshan1963equivalence} research papers published independently in 1963.  Specifically, Klauder established the overcompleteness of the coherent state basis \cite{klauder1960action} using which Sudarshan had obtained a diagonal representation of an arbitrary quantum state in the coherent state basis, such as 
\begin{equation}
\rho=\int dP\left(\alpha\right)\left|\alpha\rangle\langle\alpha\right|,\label{eq:P-fun}
\end{equation}
where the function $P\left(\alpha\right)$ is now known as  Glauber-Sudarshan $P$-function, which is normalized to unity as  $\int dP\left(\alpha\right)=1.$ Initially, Sudarshan's optical equivalence principle \cite{sudarshan1963equivalence} established that the expectation value of an operator written in the normal ordered form (i.e., all the creation operators are placed in the left side of the annihilation operators) can be calculated as a complex classical distribution function. Mathematically, it can be represented as
\begin{equation}
Tr\left[\rho\left(a^{\dagger m}a^n\right)\right]=\int dP\left(\alpha\right)\left(\alpha^{* m}\alpha^n\right).\label{eq:P-fun-exp}
\end{equation}
The Glauber-Sudarshan $P$-function is real due to Hermiticity of the density matrix $\rho$, but it fails to behave like a classical distribution function as it may take negative values. In analogy of the Wigner's attempt to give the phase space description of quantum mechanics \cite{wigner1932quantum}, the quantum state for which $P$-function can no longer be interpreted as classical probability distribution function is known as nonclassical state. Due to this property of the $P$-function and other distribution functions used in quantum optics with similar properties are called quasidistribution functions or quasiprobability distributions. Before discussing the relevance of the $P$-function as a necessary and sufficient criterion of nonclassicality, it becomes imperative to discuss other quasidistribution functions. 

The Wigner function, 
associated with a state $\rho$, is the symplectic Fourier transform of the mean value of the displacement operator $D(\alpha)$ in the state $\rho$ as
\begin{equation}
 W(\alpha) = \frac{1}{\pi^2}\int d^2 \beta\, Tr\left[\rho D(\beta) \right]e^{\beta \alpha^{*} - \beta^* \alpha} \label{WCoh}
\end{equation}
leading to the
standard representation of the Wigner function as the Fourier transform of the skewed matrix representation of $\rho$, i.e., 
\begin{equation}
 W(\eta, \overline{\eta}) = \frac{1}{2 \pi}\int_{-\infty}^{\infty} d\xi \langle a - \frac{\xi}{2}|\rho | a + 
 \frac{\xi}{2} \rangle e^{i \xi b} \label{Wusual}
\end{equation}
with $a$ and $b$ being real, and $\eta = \frac{1}{\sqrt{2}} (a + i b)$. Similarly, the $Q$-function \cite{husimi1940some} is related to the 
expectation value of $\rho$ with respect to the coherent states as  \begin{equation}
 Q(\beta,\beta^{*}) = \frac{1}{\pi}\langle \beta|\rho | \beta\rangle, \label{Qusual}
\end{equation}
where $| \beta\rangle$ is a coherent state. Note that the $P$ and $Q$ functions are associated with the normal and anti-normal (all the creation operators are placed in the right side of the annihilation operator) ordered form of operators; while the Wigner function is related to the symmetric ordering (mean value of both normal and anti-normal ordered operators). The negative values of the Wigner and $P$ functions correspond to nonclassicality of the quantum state \cite{glauber1963coherent,glauber1963photon,sudarshan1963equivalence,wigner1932quantum}, whereas zeros of the $Q$-function provide signatures of nonclassical behavior \cite{lutkenhaus1995nonclassical}. Detailed discussion of all these quasidistribution functions and their applications in quantum optics can be found in \cite{schleich2011quantum}.

Let's get back to the $P$-function, which despite being  experimentally non-reconstructable for an arbitrary state, has been of prime interest as it provides us an important indicator of nonclassicality. The negative (non-negative) values of the $P$-function necessarily correspond to the nonclassical (classical) behavior of the state under consideration. One interesting result using this fact establishes that, from the set of pure quantum states only coherent states have a positive $P$-function \cite{mandel1986non}, therefore the set of nonclassical pure states may be larger than that of classical states. This fact and the experimental impracticality of the $P$-function in its reconstruction led to numerous feasible substitutes as nonclassicality witness. One such witness was proposed by Richter and Vogel as an infinite set of moments-based criteria, which can be shown to be equivalent to the $P$-function \cite{richter2002nonclassicality}. Specifically, a hierarchy of the lower- and higher-order criteria involving experimentally measurable quantities, such as the phase quadrature, intensity, is obtained, where the failure of any criterion from this set establishes the nonclassical nature of the state. Here, it is also worth noting that in different contexts, it is shown that at least fourth-order moments in position and momentum (consequently second-order in intensity) are required to establish the nonclassical properties of a system \cite{bednorz2011fourth,bednorz2010quasiprobabilistic}.  Therefore, all the nonclassicality criteria that
include terms that are third- or higher-order in intensity (or sixth- or higher-order moments in position)
are defined as higher-order nonclassicality. These higher-order criteria become relevant due to their certain advantages in the detection of weak nonclassicality \cite{richter2002nonclassicality,allevi2012measuring,allevi2012high,avenhaus2010accessing}. In this section, we will restrict our discussion to quantized radiation field. The quasidistribution functions for the spin systems will be discussed in Chapter \ref{QDs}.

Once it is known that certain quantum state or system possesses nonclassical features, the next task in hand is to identify the type of nonclassicality present, which can further be exploited to accomplish certain quantum enhancement or speed-up over corresponding classical technology. It would not be feasible to discuss all types of nonclassical properties here, therefore we restrict ourselves to some of the nonclassical features that are widely discussed throughout the present work. 

\subsection{Squeezed states and their characterization \label{Sq}}
The famous Heisenberg's uncertainty principle quantifies our limits for precise measurement of two non-commuting observables. In mathematical terms, the variances in the measured values of two non-commuting observables $A$ and $B$ are known to satisfy $\Delta A \Delta B\geq \frac{1}{2}|\langle[A,B]\rangle|$. Therefore, a quantum state can be termed as squeezed state iff  $\left(\Delta A\right)^2\, \mathrm{or} \left(\Delta B\right)^2  < \frac{1}{2}|\langle[A,B]\rangle|$.

To characterize squeezing in a single-mode radiation field one can introduce two Hermitian quadrature operators
\begin{equation}
X=\frac{1}{2}\left(a+a^\dagger\right)\, \mathrm{and}\, Y=\frac{1}{2i}\left(a-a^\dagger\right),\label{eq:quad}
\end{equation}
which are dimensionless position and momentum operators with $[X,Y]=\frac{i}{2}\neq0$. Therefore, a quantum state is squeezed iff
\begin{equation}
\left(\Delta X\right)^{2}<\frac{1}{4}\,\mathrm{or}\,\left(\Delta Y\right)^{2}<\frac{1}{4}.\label{eq:condition-for-squeezing}\end{equation}
It may be noted that squeezed state was introduced in the early days of quantum mechanics by Kennard \cite{kennard1927quantenmechanik}.
Further, it would also be appropriate to mention here that one of the properties of coherent states is being a minimum uncertainty state, i.e., $\left(\Delta X\right)^2=\left(\Delta Y\right)^2=\frac{1}{4}$.

\begin{figure}
\centering{}\includegraphics[angle=0,scale=0.5]{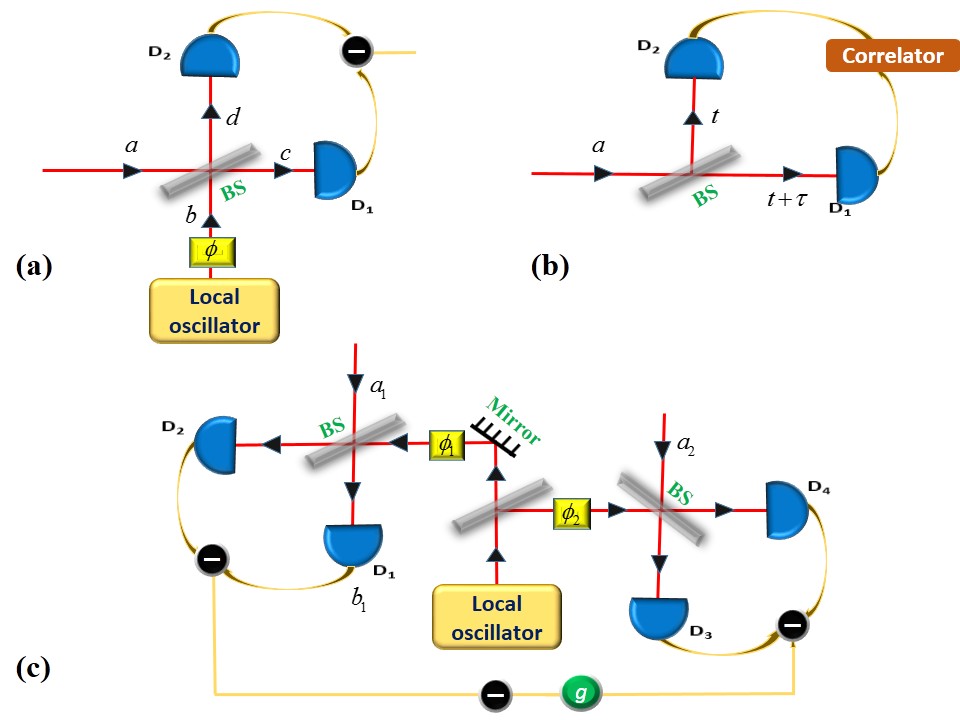}\caption[Schematic diagrams for the detection of quadrature squeezing, antibunching, and entanglement]{\label{fig:Non-Det} Schematic diagrams for the detection of (a) quadrature squeezing, (b) antibunching, and (c) entanglement in a quantum input $a$ ($a_1$ and $a_2$ in (c)).}
\end{figure}

For now, we will focus on the experimental detection of quadrature squeezing, which is carried out using homodyne and heterodyne detection schemes as shown schematically in Figure \ref{fig:Non-Det} (a). The homodyne (heterodyne) detection technique uses the local oscillator mode of the same (different) frequency as that of the input mode.

Specifically, in a balanced homodyne detection, the single-mode quantum input (in mode $a$) is mixed at a symmetric beamsplitter with a strong classical field (in mode $b$) from a local oscillator. The output modes of the beamsplitter can be easily obtained as $c=\frac{1}{\sqrt{2}}\left(a+ib\right)$ and $d=\frac{1}{\sqrt{2}}\left(ia+b\right),$ respectively. Subsequently, the difference of the photocurrents of the two detectors is obtained as output $\langle c^\dagger c-d^\dagger d\rangle=i \langle a^\dagger b-b^\dagger a\rangle.$
Without any loss of generality, we can assume the strong local oscillator ($b$) mode as a coherent state $|\alpha|\exp\left(i\phi\right)$ and obtain the output as $i |\alpha| \langle a^\dagger \exp\left(i\phi\right)-a \exp\left(-i\phi\right)\rangle.$ Using the quadrature operators defined in Eq. (\ref{eq:quad}), one can write $a=X+i Y$ and $a^\dagger=X-i Y.$ Therefore, the output can be written in terms of quadrature operators as $2|\alpha| \left\langle X\cos\left(\phi+\frac{\pi}{2}\right)+Y \sin\left(\phi+\frac{\pi}{2}\right)\right\rangle.$

One can check that the output depends on the phase of the local oscillator, and by controlling the phase angle $\phi$ each quadrature can be addressed, and corresponding quadrature squeezing can be detected. Interestingly, the applications of homodyne detection are not restricted to detection of squeezing. On top of that, the quasidistribution functions of the input quantum state can also be reconstructed by obtaining the probability distribution of the rotated quadrature phase for all values of the phase of the local oscillator known as quantum optical tomography \cite{vogel1989determination}. Tomograms of finite and infinite dimensional states will be discussed in detail in Chapter \ref{Tomogram}. Further, we would like to note that there exists a notion of higher-order squeezing, and it is popularly defined in two ways--amplitude powered squeezing \cite{hillery1987amplitude} and Hong-Mandel squeezing 
\cite{hong1985higher,hong1985generation}.  In fact, in what follows, we will report lower- and higher-order squeezing in codirectional and contradirectional optical couplers in Chapter \ref{Coupler}. For now, we will proceed to other types of nonclassicality and their detection mechanism.

\subsection{Antibunched states and their characterization \label{Ant}}

Antibunching is another type of nonclassicality, which is invariably discussed in the field of quantum optics and quantum information due to its application in a long list of single-photon-based communication schemes and their experiments \cite{bennett1984quantum,bennett1992B92,sharma2016comparative,hu2016experimental}. Traditionally, a thermal source is known to radiate multiphoton pulses.  However, to implement BB84 \footnote{BB84 is the first quantum key distribution scheme proposed by Bennett and Brassard in 1984, thus known as BB84 protocol.} quantum key distribution (QKD) \cite{bennett1984quantum} and  similar protocols \cite{bennett1992B92,sharma2016comparative,hu2016experimental} for secure quantum communication, we would require on demand single photon source \cite{pathak2010recent,pathak2013elements}. 

Characterization of a single photon source  can be performed using the famous HBT experiment. Originally, the HBT experiment \cite{brown1956correlation} designed by Hanbury Brown and Twiss manifested the relevance of the second-order intensity correlation in stellar interferometer. The idea was later employed in quantum optics, where the photons generated from a light source to be characterized falls on a beam splitter, with vacuum in the other input port, followed by a detector on each output port (cf. Figure \ref{fig:Non-Det} (b)). It is not difficult to control the distance of each detector from the beam splitter, and varying the difference between the distances ($\tau$), the second-order correlation function, $g^{(2)}(\tau)=\frac{\langle a^{\dagger}(t)a^{\dagger}(t+\tau)a(t+\tau)a(t)\rangle}{\langle a^{\dagger}(t)a(t)\rangle\langle a^{\dagger}(t+\tau)a(t+\tau)\rangle},$ can be obtained. Equivalently, one can fix the distance of the detectors from the beam splitter and let the electronics deal with this. This reveals the probability of detecting two photons together normalized to the photon numbers. A perfectly coherent source is expected to give unit value for $g^{(2)}(\tau)$ for all possible values of $\tau$. Additionally, a thermal source will have $g^{(2)}(\tau)>1$, while $g^{(2)}(\tau)<1$ would illustrate a truly quantum feature, i.e., the photons are not bunched together, hence termed as antibunching \cite{kimble1976theory,carmichael1976proposal,fox2006quantum}.

The schematic diagram in Figure \ref{fig:Non-Det} (b) demonstrates the optical elements required for detection of antibunched light. Specifically, a photon detected at one of the single photon detectors starts a current, which stops only after a click at the second detector. One can record the time interval for which the current was flowing with a large sample size, and this can be used to subsequently reveal the quantum nature of the source if $g^{(2)}(0)<g^{(2)}(\tau)$ \cite{fox2006quantum}. It would be apt to mention here a closely related quantum behavior of light, i.e., sub-Poissonian statistics of light characterized as $g^{(2)}(0)<1.$ Specifically, fluctuations in the photon number can be calculated as $\left(\Delta N\right)^2=\langle N^2\rangle-\langle N\rangle^2$ with the number operator $N=a^{\dagger}a$. This can also be written after normal ordering as $\left(\Delta N\right)^2=\langle a^{\dagger 2}a^2\rangle+\langle a^{\dagger}a\rangle-\langle a^{\dagger}a\rangle^2.$ Note that in the case of coherent state, which shows Poisson distribution, we obtain $\left(\Delta N\right)^2=\langle N\rangle$. Therefore, all the cases leading to $\left(\Delta N\right)^2>\langle N\rangle$ and $\left(\Delta N\right)^2<\langle N\rangle$ are termed as super-Poissonian and sub-Poissonian statistics, respectively. We are mainly interested in sub-Poissonian statistics due to unavailability of its  classical counterpart \cite{mandel1959fluctuations,mandel1986non,zou1990photon}, and one can verify that mathematically $g^{(2)}(0)<1$ signifies the same nonclassicality. Also, the sub-Poissonian character of light ensures antibunching in the light at least for some timescale. Interestingly, though a relation between the lower-order sub-Poissonian statistics and antibunching criteria can be established by reducing them to the same form \cite{teich1983antibunching,zou1990photon,verma2010generalized,thapliyal2017comparison}, this is not the case with the corresponding higher-order criteria \cite{verma2010generalized,thapliyal2017comparison}. A detailed discussion will follow during discussion of higher-order antibunching criterion in Chapter \ref{Coupler}. In fact, in what follows, we will report lower- and higher-order antibunching in codirectional and contradirectional optical couplers in Chapter \ref{Coupler}. However, right now, we will rather shift our attention to multi-mode nonclassicality.

 \subsection{Entangled states and their characterization \label{Ent}}
One of the most promising quantum resources nowadays is entanglement. In principle, entanglement can be viewed as a superposition in the tensor product space which fails to satisfy separability condition. Suppose that a system can be defined by a quantum state $|\psi\rangle$ in the Hilbert space $\mathcal{H}=\mathcal{H}_1\otimes \mathcal{H}_2,$ and we fail to write two independent quantum states of the individual sub-systems in the Hilbert spaces $\mathcal{H}_1$ and $\mathcal{H}_2$ as $|\psi\rangle\neq|\psi_1\rangle\otimes|\psi_2\rangle$, we term the state as inseparable or entangled  state. Without any complication, one can write a similar expression for the signature of entanglement for the mixed  states as $\rho\neq\underset{i}{\sum}p_i\rho_1^i\otimes\rho_2^i$. 

In the next chapter, we will enlist some criteria of inseparability of two- and multi-mode quantum states and use them in our theoretical studies in that chapter. Here, we rather focus on the experimental detection of entanglement. To accomplish this task, we require two sets of apparatus used for the detection of squeezing using balanced homodyne or heterodyne technique (discussed in Section \ref{Sq}). For the homodyne detection, we have to use a single local oscillator in one of the input ports of both the beam splitters where two modes under consideration (of the same frequency as that of the local oscillator) are mixed with them (see Figure \ref{fig:Non-Det} (c)). Quite similar to the homodyne detection for squeezing, the difference of current generated from both the detectors in each  set of detection apparatus is further collected, and the difference of this current gives the signature of entanglement in the quadratures addressed by controlling the phases of the two homodyne detection setups independently (see \cite{bachor2004guide} for detail). Another scheme of entanglement detection of polarized photons involves coincidence counts of two detectors on which two modes generated due to a nonlinear process shine (\cite{brida2009characterization} and references therein). We will discuss some applications of entangled states in Section \ref{QIP} and the next chapters. 

Before concluding the section, we would like to mention that quantum feature of  various types of nonclassicality can be established using the fact that the criteria mentioned here (and also in the next chapter) show nonclassical behavior only when the corresponding $P$-function becomes non-positive. For example, the criterion for sub-Poissonian photon statistics (also revealing antibunching) $\langle a^{\dagger 2}a^2\rangle-\langle a^{\dagger}a\rangle^2<0$ can be written using optical equivalence principle as  
\begin{equation}
\left[\int P\left(\alpha\right)|\alpha|^4 d^2\alpha-\left(\int P\left(\alpha\right)|\alpha|^2 d^2\alpha\right)^2\right]=\int P\left(\alpha\right)\left[|\alpha|^2-\left(\int P\left(\alpha^\prime\right)|\alpha^\prime|^2 d^2\alpha^\prime\right)\right]^2 d^2\alpha<0.\nonumber
\end{equation}
From which it can be easily concluded that the quantity in the left hand side can only become negative for non-positive values of $P\left(\alpha\right).$ Similarly, one can check the negativity of Glauber-Sudarshan $P$-function for the presence of the remaining nonclassical features.
The interested readers may find similar proofs and detailed discussion on the experimental detection of various types of nonclassical states in Refs. \cite{fox2006quantum,bachor2004guide,agarwal2013quantum}.
Note that this thesis presents a theoretical work, but in this chapter, we have briefly discussed experimental techniques that may be used to validate the theoretical predictions of the present work. This is done to establish that the theoretical research performed here can be realized and tested.

\section{Role of nonlinear optics in the generation of nonclassical  states \label{Non-Op}}

In Refs. \cite{ge2015conservation,vogel2014unified}, it is claimed that linear optics conserves both quantumness and mixedness of a quantum state. Therefore, generation of nonclassical states is closely associated with nonlinear optics (\cite{boyd2003nonlinear} and references therein). In brief, when a very intense light falls on a material, the polarization induced in it no longer remains a linear function of the applied electric field ($E$) and involves higher powers of $E$ as well. The induced polarization for these optical materials can be written as \cite{boyd2003nonlinear,schleich2011quantum}
\begin{equation}
P=\varepsilon_0 \left(\chi^{\left(1\right)} E + \chi^{\left(2\right)} E^2 +\chi^{\left(3\right)} E^3 +\ldots \right),\nonumber
\end{equation}
where $\chi^{\left(n\right)}$ is $n$th-order susceptibility of the medium, which is characterized by an $\left(n+1\right)$th rank tensor.

Usually, photons interact weakly, due to which they decohere less, and this is why photon (optical qubit) is one of the widely used quantum resources. Various quantum information applications exploiting optical qubits involve linear optics, i.e., with a small photon number and weak interaction. The remaining tasks can be accomplished in two ways \cite{chang2014quantum}. Specifically, as we were discussing, increasing the photon number one enters into the regime of classical nonlinear optics with the weak interaction strength \cite{chang2014quantum}. However, numerous tasks in quantum optics and quantum information processing require nonlinear optics with higher interaction strength and low intensity, which is studied under the domain quantum nonlinear optics. For instance, certain operations (such as a two-qubit CNOT gate) require strong interaction between two photons, and in the absence of which ancilla photons are used, and the success of the operation depends on the measurement conditioned on specific outcomes for each ancilla \cite{knill2000efficient}. In this case, nonlinearity is introduced due to measurement of ancillary photons.

There are several phenomena associated with second- and higher-order nonlinearity, such as parametric down conversion, second harmonic generation, Kerr effect, having applications in various disciplines of physics and engineering. Spontaneous parametric down conversion has so far served as the backbone of experimental quantum optics and information processing by providing entanglement \cite{hardy1992source,kwiat1999ultrabright} and heralded single photon \cite{pittman2002single,migdall2002tailoring} sources. As it is widely discussed nonlinear process, we will rather give an example of another second-order nonlinear phenomenon known as second harmonic generation \cite{franken1961generation}. As mentioned previously, the variation of a single-mode electric field can be written as $E\left(t\right) \propto \left[E_0\exp\left(i\omega t\right)+\textrm{c.c.}\right],$ therefore second-order terms contribute as $E\left(t\right)^2 \propto 2|E_0|^2+\left[E_0^2\exp\left(2i\omega t\right)+\textrm{c.c.}\right],$ where the presence of $2\omega$ terms gives rise to the generation of second harmonic beams in such an interaction \cite{kockum2017deterministic}. In Chapter \ref{Coupler}, we will discuss a nonlinear waveguide undergoing second harmonic generation  and will exploit this nonlinearity to generate nonclassical states in a linear waveguide coupled with it. To perform such study we will first equip ourselves with some mathematical techniques, which will be used in the present work.

\section{Dynamics of a quantum system \label{Math}}

A closer look at the above discussed nonclassical features would reveal the role of nonlinear optics in the generation of nonclassical states. To perform this study, one needs to take into account the time evolution of the system initially prepared in a known quantum state $|\psi\rangle$ (or $\rho$ in case of mixed state). In this section, we will briefly describe different approaches which may be adopted to obtain time evolution or dynamics of a quantum system under a given Hamiltonian.

\subsection{Different approaches for obtaining the equation of motion \label{Eq-of-m}}

Dynamics of a quantum system can be studied using different approaches. Most general picture for studying the time evolution of a quantum system is the interaction picture, which can be reduced to  two other approaches--Schr{\"o}dinger and Heisenberg pictures in the limiting cases. We will begin with the Schr{\"o}dinger picture.

\subsubsection{Equation of motion in the Schr{\"o}dinger picture \label{Sch}}

The time evolution of a quantum state $|\psi\left(t\right)\rangle$ is determined by the time-dependent Schr{\"o}dinger equation, 
\begin{equation}
i\hbar\frac{\partial}{\partial t}|\psi\left(t\right)\rangle={H} |\psi\left(t\right)\rangle,\label{eq:SchEq}
\end{equation}
where ${H}$ is the Hamiltonian of the system, and $i\hbar\frac{\partial}{\partial t}$ is the operator of energy. In general, the Hamiltonian itself can be time-dependent and the same can be represented as $H\left(t\right)$. 

One can obtain the solution of the Schr{\"o}dinger equation (\ref{eq:SchEq}) as 
\begin{equation}
|\psi\left(t\right)\rangle={U}\left(t\right) |\psi\left(0\right)\rangle.\label{eq:sol-SchEq}
\end{equation}
Here, ${U}\left(t\right)$ is a unitary operation responsible for the evolution of the quantum state from the initial time $t=0$ to time $t$. In case of an isolated system (system with time independent Hamiltonian), the unitary evolution takes a simpler form ${U}\left(t\right)=\exp\left(-\frac{i}{\hbar}{H} t\right),$ whereas for the time dependent Hamiltonian, it involves a chronological time-ordering as well, 
\begin{equation}
{U}\left(t\right)=T_\leftarrow\exp\left(-\frac{i}{\hbar}\int_{0}^{t}{H}\left(t^\prime\right) dt^\prime\right).\nonumber
\end{equation}

It is straightforward to extend the solution of the Schr{\"o}dinger equation to the mixed state and to obtain 
\begin{equation}
\rho\left(t\right) ={U}\left(t\right) \rho\left(0\right){U}^\dagger\left(t\right),\label{eq:sol-SchEq2}
\end{equation}
using which we can obtain the equation of motion for the mixed states as 
 \begin{equation}
\frac{\partial}{\partial t}\rho\left(t\right)=-\frac{i}{\hbar}\left[{H}\left(t\right), \rho\left(t\right)\right].\label{eq:LioEq}
\end{equation}
This famous equation describing the time evolution in the Schr{\"o}dinger picture is known as von Neumann or Liouville-von Neumann equation. One can also write this using the Liouville operator ${\mathcal{L}}\left(t\right)$ as 
 \begin{equation}
\frac{\partial}{\partial t}\rho\left(t\right)={\mathcal{L}}\left(t\right)\rho\left(t\right).\label{eq:LioEq-2}
\end{equation}

It is imperative to discuss here some properties of the evolution operator and its role in the quantum information processing tasks. As previously mentioned the evolution operator is unitary in nature (i.e., ${U}^\dagger\left(t\right){U}\left(t\right)=\mathbb{I}$ with $\mathbb{I}$ as an identity matrix). It means, unlike the measurement operator, there exists an operation which can invert the unitary map from $|\psi\rangle\xrightarrow{\hspace{0.1cm}{U} \hspace{0.1cm}}|\phi\rangle$ such that to obtain $|\phi\rangle\xrightarrow{\hspace{0.1cm}{U}^{\dagger} \hspace{0.1cm}}|\psi\rangle.$ This fact allows reversibility to quantum operations (except measurement), which has been exploited to obtain enhancement in the performance during the computation tasks using powers of quantum superposition.

One interesting problem in the domain of unitary and non-unitary (measurement) operators is quantum Zeno effect, and with the recent advent of its set of applications in both theoretical and experimental counterfactual quantum computation \cite{hosten2006counterfactual,kong2015experimental} and communication \cite{salih2013protocol,cao2017direct} it deserves a brief introduction here. To understand the quantum Zeno effect, using Eq. (\ref{eq:sol-SchEq}), one can easily obtain the probability that the state remains unchanged even after the free evolution for time $t$ as
\begin{equation}
\begin{array}{lcl}
P\left(t\right)&=&\left|\langle\psi\left(0\right)|\psi\left(t\right)\rangle\right|^2=\left|\langle\psi\left(0\right)|{U}\left(t\right) |\psi\left(0\right)\rangle\right|^2\\
&=&\left|\langle\psi\left(0\right)\left|\left(1-\frac{i}{\hbar}{H}t-\frac{1}{2\hbar^2}{H}^2 t^2+\ldots\right) \right|\psi\left(0\right)\rangle\right|^2\\
&=&1-\frac{1}{\hbar^2}\left(\Delta E\right)^2 t^2+\ldots,
\end{array}\nonumber
\end{equation}
where we have substituted $\left(\Delta E\right)^2=\left(\langle\psi\left(0\right)|{H}^2|\psi\left(0\right)\rangle-\langle\psi\left(0\right)|{H}|\psi\left(0\right)\rangle^2\right),$ and the dots represent terms in higher-orders of time. Interestingly, for the small values of time, we can easily neglect these higher power terms and observe that a quantum system decays quadratically instead of  exponential in time. Suppose we measure the quantum state at time $t$, and let it evolve for another time interval $t$ before measuring it. Repeating this for $n$ $\left(=\frac{T}{t}\right)$ number of times, we can calculate the probability of finding the quantum system still in the same state after time $T$ is 
\begin{equation}
P_n\left(T\right)=\left[P\left(t\right)\right]^n\approx\left[1-\frac{1}{\hbar^2}\left(\Delta E\right)^2 \frac{T^2}{n^2}\right].\nonumber
\end{equation}
In the range of $n\rightarrow\infty$, we tend to the limits of continuous measurement, and $\underset{n\rightarrow\infty}{\lim}P_n\left(T\right)=1.$ It leads to an interesting conclusion that the decay of a quantum state can be suppressed with the help of a continuous measurement. In the analogy of the Zeno's paradoxes, this phenomena was termed as Zeno's paradox in quantum theory by Misra and Sudarshan \cite{misra1977zeno}, now known as quantum Zeno effect. It was only later recognized that a continuous measurement can also possibly enhance the decay termed as quantum anti-Zeno or inverse Zeno effect (see Refs. \cite{venugopalan2007quantum,facchi2001quantum,pascazio2014all}
for the reviews). Both these effects will be discussed in detail in Chapter \ref{Zeno}. For now, we may proceed to the discussion of  another picture that also leads to the time evolution of a quantum system.

\subsubsection{Equation of motion in the Heisenberg picture \label{Heis}}

In contrast of the Schr{\"o}dinger picture, here the operators are assumed to evolve with time, and the expectation values of these operators are calculated with respect to the initial state. Using this one can easily write the time dependence of an operator ${O}$ as 
\begin{equation}
{O}\left(t\right)={U}^\dagger\left(t\right){O}\left(0\right){U}\left(t\right).\label{eq:Hei-pic}
\end{equation}
Therefore, average value of the operator with respect to an initial state $\rho_{0}=|\psi\left(0\right)\rangle\langle\psi\left(0\right)|$ can be calculated to be
\begin{equation}
\begin{array}{lcl}
\langle{O}\left(t\right)\rangle&=&Tr\left[{O}\left(t\right)\rho_{0}\right]\\
&=&\langle\psi\left(0\right)\left|{U}^\dagger\left(t\right){O}\left(0\right){U}\left(t\right)\right|\psi\left(0\right)\rangle,\nonumber
\end{array}
\end{equation}
which is same as that in the Schr{\"o}dinger picture. Therefore, one can  write the equation of motion describing dynamics of the quantum system in the Heisenberg picture as
 \begin{equation}
\frac{d}{dt}{O}\left(t\right)=\frac{i}{\hbar}\left[{H}\left(t\right), {O}\left(t\right)\right]+\frac{\partial {O}\left(t\right)}{\partial t}.\label{eq:HeisEq}
\end{equation}
This is known as the Heisenberg's equation of motion. Note that the second term in the right-hand side vanishes when the initial operator ${O}$ does not have an explicit time dependence. 

It would be apt to mention here an equivalent of Heisenberg's equation of motion (\ref{eq:HeisEq}) in a transparent (dielectric) medium. In the view of a brief discussion in \cite{luks2009quantum}, the matter-field interaction can also be studied without considering all degrees of freedom of the matter. Such nonlinear interaction is usually studied either considering a cavity problem or a steady-state propagation \cite{shen1967quantum,luks2009quantum}. In the present work, we have considered the latter. Without going into much detail, one can obtain the equation of motion for the steady-state propagation by replacing $\frac{-z}{c}$  in place of $t$ in Eq. (\ref{eq:HeisEq}). As this substitution will make the operators localized in space by omitting their time dependence. Therefore, translation of the localized field operators is governed by a localized momentum operator $G\left(z\right)$, obtained as $\frac{H\left(z,t\right)}{c}$ \cite{shen1967quantum,luks2009quantum}, 
 \begin{equation}
\frac{d}{dz}{O}\left(z\right)=-\frac{i}{\hbar}\left[{G}\left(z\right), {O}\left(z\right)\right].\label{eq:HeisEq-coupler}
\end{equation}
Importantly, the localized field operators satisfy equal space commutation relation 
\begin{equation}
\left[{a}_k\left(z\right),\, {a}_{k^{\prime}}^\dagger\left(z\right)\right] = \delta_{kk^{\prime}}\nonumber
\end{equation}
for discrete values of the wavevector $k$, in analogy of equal time commutation relation discussed in Section \ref{Sec-Q}. 

We will now skip further discussion on this topic and get back to it in Chapters \ref{Coupler} and \ref{Zeno}, where equation of motion described in Eq. (\ref{eq:HeisEq-coupler}) will be used for studying the spatial evolution of radiation in the optical waveguides.

\subsubsection{Equation of motion in the interaction picture \label{Int}}

So far, we have discussed the two limiting cases of a more general picture to study the time evolution of a quantum system. To discuss the interaction picture, we will start with the general form of Hamiltonian operator ($\hat{H}\left(t\right)$), which can be written as a sum of a time-independent ($\hat{H}_0$) and a time-dependent ($\hat{H}_I \left(t\right)$) terms known as free field and interaction terms, respectively. Thus, we have $\hat{H}\left(t\right)=\hat{H}_0+\hat{H}_I \left(t\right).$ In what follows, we have used hat $\left(\,\hat{}\,\right)$ with all the terms of the total Hamiltonian to avoid ambiguity amongst the interaction term of the total Hamiltonian and the Hamiltonian written in the interaction picture. The free field term corresponds to the energy of the system in the absence of any interaction, which is the sum of energies of all the subsystems. 

One can write the Hamiltonian in the interaction picture as $H_I \left(t\right)={U}_0^\dagger\left(t\right)$ $ {H}_I \left(t\right) {U}_0\left(t\right),$ where ${U}_0\left(t\right)$ is the free field evolution operator. Thus, one can use this Hamiltonian operator in the interaction picture to write the usual von Neumann equation (\ref{eq:LioEq}) as in the Schr{\"o}dinger picture   
 \begin{equation}
\frac{\partial}{\partial t}\rho_I\left(t\right)=-\frac{i}{\hbar}\left[H_I\left(t\right), \rho_I\left(t\right)\right].\label{eq:LioEq-Int}
\end{equation}
Note that the interaction picture density matrix $\rho_I\left(t\right)$ can be viewed as a Schr{\"o}dinger picture evolution of the density matrix, but only due to the interaction part of the Hamiltonian, i.e., $\rho_I\left(t\right)={U}_I\left(t\right)\rho \left(0\right){U}_I^\dagger\left(t\right).$ 
Similarly, one can calculate the average value of an arbitrary operator as 
\begin{equation}
\langle{O}\left(t\right)\rangle=Tr\left[{O}_I \left(t\right)\rho_{I}\left(t\right)\right],\nonumber
\end{equation}
where the interaction picture operator is defined as ${O}_I \left(t\right)={U}_0^\dagger\left(t\right){O}\left(0\right){U}_0\left(t\right).$

In what follows, we will use these basic quantum mechanical tools to describe the methods adopted in our present work and their significance over some other existing mathematical techniques.

\subsection{Mathematical techniques used in the present work to study the dynamics of a system \label{Sol-of-eq}}

To accomplish the task we have set ourselves, i.e., to analyze the nonclassical behavior of various physical systems, we would require the time evolution of the system described by a given Hamiltonian (or momentum operator). While working in the Schr\"{o}dinger picture, we have obtained the dynamics of the system numerically by solving the time dependent Schr\"{o}dinger equation using matrix forms of the involved states and operators (see Section \ref{sec:num-Coupler} for detail). 

Also, the Heisenberg's equations of motion obtained in these cases give coupled differential equations only solvable using some perturbative technique. Here, we opted to obtain the closed form analytic expressions for time evolution (spatial evolution in case of localized operators in the momentum operator) of all the modes involved in the interaction. For which we have used the Sen-Mandal perturbative technique \cite{sen2005squeezed,mandal2004approximate} for the systems possessing weak nonlinearity or interaction constants.

To elaborate the Sen-Mandal perturbative technique \cite{sen2005squeezed,mandal2004approximate}, we consider here an example of the codirectional asymmetric nonlinear optical coupler \cite{perina2000review,mandal2004approximate} which is described by the following momentum operator in the interaction picture
 
\begin{equation}
G\left(z\right)=-\hbar kab_{1}^{\dagger}-\hbar\Gamma b_{1}^{2}b_{2}^{\dagger}\exp(i\Delta kz)\,+{\rm H.c.}\,,\label{eq:Mom-asymmetric}
\end{equation}
with ${\rm H.c.}$ referring to the Hermitian conjugate. 
Here, we refrain from the description of the physical process and the physics involved with the momentum operator (\ref{eq:Mom-asymmetric}), which will be discussed in detail in Chapters \ref{Coupler} and \ref{Zeno}. We only need to note that under the weak pump condition,  
$\Gamma\ll k$, as $k$ and $\Gamma$ are proportional to the
linear $(\chi^{(1)})$ and nonlinear $(\chi^{(2)})$ susceptibilities,
respectively, and usually $\chi^{(2)}/\chi^{(1)}\,\simeq10^{-6}$.

The Heisenberg's equations of motion for different modes of the momentum operator given in Eq. (\ref{eq:Mom-asymmetric}) can be obtained as
three coupled differential equations 
\begin{equation}\frac{da}{dz}=-ik^{*}b_{1},\,\frac{db_{1}}{dz}=-ika-2i\Gamma^{*}b_{1}^{\dagger}b_{2}\exp\left(-i\Delta kz\right),\,\frac{db_{2}}{dz}=-i\Gamma b_{1}^{2}\exp\left(i\Delta kz\right).\label{eq:aaa}\end{equation}
Note that the equation of motion used here is the one obtained for the localized operators in Eq. (\ref{eq:HeisEq-coupler}).
Using the Sen-Mandal approach a closed form perturbative analytic solution of (\ref{eq:aaa}) is obtained as \cite{mandal2004approximate}
\begin{equation}
\begin{array}{lcl}
a(z) & = & f_{1}a(0)+f_{2}b_{1}(0)+f_{3}b_{1}^{\dagger}(0)b_{2}(0)+f_{4}a^{\dagger}(0)b_{2}(0),\\
b_{1}(z) & = & g_{1}a(0)+g_{2}b_{1}(0)+g_{3}b_{1}^{\dagger}(0)b_{2}(0)+g_{4}a^{\dagger}(0)b_{2}(0),\\
b_{2}(z) & = & h_{1}b_{2}(0)+h_{2}b_{1}^{2}(0)+h_{3}b_{1}(0)a(0)+h_{4}a^{2}(0).\end{array}\label{eq:ass-sol}\end{equation}
Here, the perturbative solution is obtained restricting us up to the linear
power of the nonlinear coupling constant $\Gamma$. To derive the solution, the spatial evolution of the operator $a(z)$ is used, 
\begin{equation}
\begin{array}{lcl}
a\left(z\right) & = & \exp\left(iGz/\hbar\right)a\left(0\right)\exp\left(-iGz/\hbar\right),\end{array}\label{evolution}
\end{equation}
which on expansion yields
\begin{equation}
\begin{array}{lcl}
a\left(z\right) & = & a\left(0\right)+\frac{iz}{\hbar}\left[G,a\left(0\right)\right]+\frac{\left(iz\right)^{2}}{2!\hbar^2}\left[G,\left[G,a\left(0\right)\right]\right]
+\frac{\left(iz\right)^{3}}{3!\hbar^2}\left[G,\left[G,\left[G,a\left(0\right)\right]\right]\right]+\cdots.\end{array}\label{eq:taylorseries}
\end{equation}
All the commutators present in the right-hand side of Eq. (\ref{eq:taylorseries}) are computed, neglecting the terms beyond linear power in the nonlinear coupling constant ($\Gamma$). For example,
\begin{equation}\begin{array}{lcl}
\left[G,a\right] & = & \hbar k^{*}b_{1},\\
\left[G,\left[G,a\right]\right] & = & \hbar^{2}\left|k\right|^{2}a+2\hbar^{2}\Gamma^{*}k^{*}b_{1}^{\dagger}b_{2}\exp\left(-i\Delta kz\right),\\
\left[G,\left[G,\left[G,a\right]\right]\right] & = & \hbar^{3}k^{*}\left|k\right|^{2}b_{1}+2\hbar^{3}\Gamma^{*}k^{*2}a^{\dagger}b_{2}\exp\left(i\Delta kz\right).
\end{array}\label{eq:comm}
\end{equation} 
We stop here as all the higher-order commutators (higher than $\left[G,\left[G,\left[G,a\right]\right]\right]$) would only have terms proportional to $a$, $b_{1}$, $b_{1}^{\dagger}b_{2},$ and $a^{\dagger}b_{2}$. Therefore, a first-order solution for the spatial evolution of $a\left(z\right)$ (i.e., a solution linear in $\Gamma$) will only possess terms proportional to $a$, $b_{1}$, $b_{1}^{\dagger}b_{2},$ and $a^{\dagger}b_{2}$. As a consequence, the spatial evolution of $a\left(z\right)$ can be assumed  in the form shown in Eq. (\ref{eq:ass-sol}).
Similarly, the spatial evolution of the annihilation operators of the rest of the modes can also be assumed. 

Here, it would be apt to note that a conventional short-length solution \cite{perina1991quantum} is usually obtained from the Taylor series, $a(z)=a(0)+z\frac{da}{dz}|_{z=0}+\frac{z^{2}}{2!}\frac{d^{2}a}{dz^{2}}|_{z=0}+\cdots,$ with $\hbar=1$. Using  Eq. (\ref{eq:comm}) we can obtain a first-order short-length  solution as $a\left(z\right)=a(0)-ik^{*}zb_{1}(0)-\frac{1}{2}\left|k\right|^{2}z^{2}a(0)-\Gamma^{*}k^{*}z^{2}b_{1}^{\dagger}(0)b_{2}(0)$. We can easily observe that the solution does not contain any term proportional to $a^{\dagger}b_{2}$.

Subsequently, to obtain the analytic expressions of the coefficients $f_{i}$, $g_{i},$ and $h_{i}$ in Eq. (\ref{eq:ass-sol}), we substitute the assumed solution in Eq. (\ref{eq:aaa}) and by comparing the coefficients of all the terms obtained in both the sides provide coupled differential equations for different coefficients $f_{i}$, $g_{i},$ and $h_{i}$, as follows 
\begin{equation}
\begin{array}{lcllcllcl}
\dot{f}_{1} & = & -ik^{*}g_{1}, & 
\dot{g}_{1} & = & -ikf_{1}, & 
\dot{h}_{1} & = & 0,\\
\dot{f}_{2} & = & -ik^{*}g_{2}, & 
\dot{g}_{2} & = & -ikf_{2}, & 
\dot{h}_{2} & = & -i\Gamma g_{2}^{2}\exp\left(i\Delta kz\right),\\
\dot{f}_{3} & = & -ik^{*}g_{3}, & 
\dot{g}_{3} & = & -ikf_{3}-2i\Gamma^{*}g_{2}^{*}\exp\left(-i\Delta kz\right), & 
\dot{h}_{3} & = & -2i\Gamma g_{1}g_{2}\exp\left(i\Delta kz\right),\\
\dot{f}_{4} & = & -ik^{*}g_{4}, &
\dot{g}_{4} & = & -ikf_{4}-2i\Gamma^{*}g_{1}^{*}\exp\left(-i\Delta kz\right), &
\dot{h}_{4} & = & -i\Gamma g_{1}^{2}\exp\left(i\Delta kz\right).\end{array}\label{eq:cou-diff}
\end{equation}
A simultaneous solution of these differential equations, with the boundary conditions $F_{i}\left(z=0\right)=\delta_{i,1},$ where
$F\in\left\{ f,g,h\right\}$, provides us a closed form analytic expression of all the terms as
 \begin{equation}
\begin{array}{lcl}
f_{1} & = & g_{2}=\cos|k|z,\\
f_{2} & = & -g_{1}*=-\frac{ik^{*}}{|k|}\sin|k|z,\\
f_{3} & = & \frac{2k^{*}\Gamma^{*}}{4|k|^{2}-(\Delta k)^{2}}\left[G_{-}f_{1}+\frac{f_{2}}{k^{*}}\left\{ \Delta k-\frac{2|k|^{2}}{\Delta k}G_{-}\right\} \right],\\
f_{4} & = & \frac{4k^{*2}\Gamma^{*}}{\Delta k\left[4|k|^{2}-(\Delta k)^{2}\right]}G_{-}f_{1}+\frac{2k^{*}\Gamma^{*}}{\left[4|k|^{2}-(\Delta k)^{2}\right]}G_{+}f_{2},\\
g_{3} & = & \frac{2\Gamma^{*}k}{\left[4|k|^{2}-(\Delta k)^{2}\right]}G_{+}f_{2}-\frac{2\Gamma^{*}\left(2|k|^{2}-(\Delta k)^{2}\right)f_{1}}{\Delta k\left[4|k|^{2}-(\Delta k)^{2}\right]}G_{-},\\
g_{4} & = & \frac{4\Gamma^{*}|k|^{2}}{\Delta k\left[4|k|^{2}-(\Delta k)^{2}\right]}f_{2}-\frac{2\Gamma^{*}\left(2|k|^{2}-(\Delta k)^{2}\right)}{\Delta k\left[4|k|^{2}-(\Delta k)^{2}\right]} \left(G_{+}-1\right)f_{2}+\frac{2k^{*}\Gamma^{*}}{\left[4|k|^{2}-(\Delta k)^{2}\right]}G_{-}f_{1},\\
h_{1} & = & 1,\\
h_{2} & = & \frac{\Gamma G_{-}^{*}}{2\Delta k}-\frac{i\Gamma}{2\left[4|k|^{2}-(\Delta k)^{2}\right]}\left[2|k|\left(G_{+}^{*}-1\right)\sin2|k|z- i\Delta k\left(1-\left(G_{+}^{*}-1\right)\cos2|k|z\right)\right],\\
h_{3} & = & \frac{-\Gamma|k|}{k^{*}\left[4|k|^{2}-(\Delta k)^{2}\right]}\left[i\Delta k\left(G_{+}^{*}-1\right)\sin2|k|z+ 2|k|\left(1-\left(G_{+}^{*}-1\right)\cos2|k|z\right)\right],\\
h_{4} & = & -\frac{\Gamma|k|^{2}G_{-}^{*}}{2k^{*^{2}}\Delta k}-\frac{i\Gamma|k|^{2}}{2k^{*^{2}}\left[4|k|^{2}-(\Delta k)^{2}\right]}\left[2|k|\left(G_{+}^{*}-1\right)\sin2|k|z-i\Delta k\left(1-\left(G_{+}^{*}-1\right)\cos2|k|z\right)\right].\end{array}\label{eq:terms}\end{equation}
Here, $G_{\pm}=\left[1\pm\exp(-i\Delta kz)\right]$ and $C=\frac{\Gamma}{\Delta k\left[4|k|^{2}-\Delta k^{2}\right]}$.
The obtained perturbative solution is verified by checking the equal space commutation relations, i.e., $\left[a\left(z\right),a^{\dagger}\left(z\right)\right]=\left[b_{1}\left(z\right),b_{1}^{\dagger}\left(z\right)\right]=\left[b_{2}\left(z\right),b_{2}^{\dagger}\left(z\right)\right]=1,$
and the rest of all the equal space commutations are zero. Independently, the solution is checked to satisfy the constant of motion (which in this case is $N_a\left(z\right)+N_{b_1}\left(z\right)+2N_{b_2}\left(z\right)=\textrm{constant}$) as well. Further, the second-order short-length solution can be deduced from the present solution by neglecting the terms beyond quadratic in length ($z^{2}$), such as 
\begin{equation}
\begin{array}{c}
f_{1}=g_{2}=(1-\frac{1}{2}\left|k\right|^{2}z^{2}),\, f_{2}=-g_{1}^{*}=-ik^{*}z,\, f_{3}=-g_{4}=-\Gamma^{*}k^{*}z^{2},\,
g_{3}=-2i\Gamma^{*}z-\Gamma^{*}\Delta k z^{2},\\ l_{1}=1,\, l_{2}=-i\Gamma z+\frac{1}{2}\Gamma \Delta k z^{2},\, l_{3}=-\Gamma k z^{2}, \, {\rm \, and\,}\, f_{4}=l_{4}=0.\end{array}\label{eq:shortlength}\end{equation}
Here, it would be apt to add that the Sen-Mandal perturbative solution used here has terms beyond quadratic in length as the solution contains terms of sine and cosine functions. Further, it contains some additional terms (terms not present in the short-length solution), like a term proportional to $a^{\dagger}b_{2}$ in the expression of $a\left(z\right)$. The presence of these terms provides some advantages over the conventional short-length solution. Additionally, the solutions obtained by the Sen-Mandal approach are found to match with the exact numerical solution (\cite{sen2013intermodal,giri2014single,thapliyal2014higher,giri2017nonclassicality,alam2017lower} and references therein), and it is also found to succeed in detecting various nonclassicalities present in the physical systems which were not detected by the corresponding short-length solution \cite{mandal2004approximate,thapliyal2014nonclassical}.

Additionally, there is a solution technique for the Heisenberg's equation of motion, claimed to be non-perturbative in nature \cite{mista2001non}. However, the solution obtained using this technique fails to satisfy the equal time (or space) commutation relation, in contrast to the Sen-Mandal peturbative technique. Therefore, we opted the Sen-Mandal technique to obtain the analytic solution for the dynamics of the system in the present work. So far we have not considered the effects of the surroundings on the system dynamics. Therefore, we will briefly introduce the open quantum system formalism that will be required in the present work (especially, Chapters \ref{QDs}-\ref{NonMarkovian}).

\section{Open quantum system \label{OQS}}

\begin{figure}
\centering{}\includegraphics[scale=1.0]{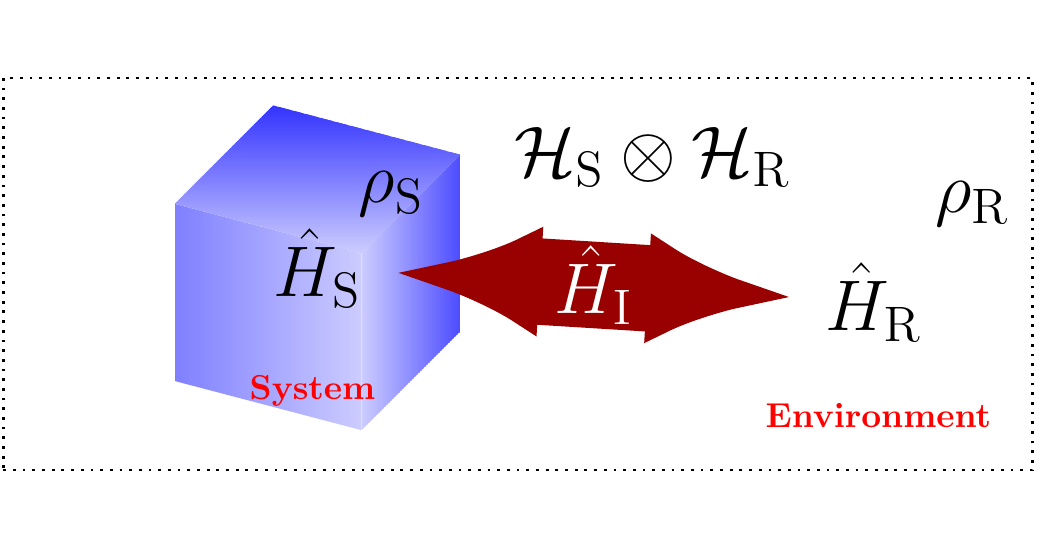}\caption[Visualization of open quantum system dynamics]{\label{fig:OQS} Dynamics of the open quantum system can be visualized as a quantum system and its environment initially in the quantum states $\rho_{\rm S}$ and $\rho_{\rm R}$ evolving independently induced by their respective Hamiltonians $\hat{H}_{\rm S}$ and $\hat{H}_{\rm R}$ and also interacting with each other due to interaction Hamiltonian $\hat{H}_{\rm I}.$ The whole dynamical process can be represented in the tensor product space $\mathcal{H}_{\rm S}\otimes \mathcal{H}_{\rm R}.$}
\end{figure}

We have described the unitary evolution of a quantum system in the previous section. In contrast, it is not always possible to neglect interaction between the system of interest and its close vicinity. As a quantum system (as shown in Figure \ref{fig:OQS}) cannot be isolated completely from its environment, it is rather important to consider an enlarged system consisting of both the system and its environment in the quantum states $\rho_{\rm S}$ and $\rho_{\rm R}$, respectively. The composite state of the system $\rho=\rho_{\rm S}\otimes\rho_{\rm R}$ evolves unitarily in the combined Hilbert space $\mathcal{H}_{\rm S}\otimes \mathcal{H}_{\rm R}.$ Note that the system and the environment can evolve independently due to system Hamiltonian $\hat{H}_{\rm S}$ and the reservoir Hamiltonian $\hat{H}_{\rm R},$ whereas their interaction is controlled by the  interaction Hamiltonian $\hat{H}_{\rm I}.$ The reservoir corresponds to a mathematical model of the environment with infinitely many degrees of freedom forming a frequency continuum. Additionally, a reservoir in thermal equilibrium is known as a bath \cite{breuer2002theory}.

The dynamics of the reduced system (after tracing out environment degrees of freedom) influenced by the total Hamiltonian   
\begin{equation}
\hat{H}=\hat{H}_{\rm S}\otimes I_{\rm R}+I_{\rm S}\otimes\hat{H}_{\rm R}+\hat{H}_{\rm I}\label{eq:OQS-H}
\end{equation} cannot necessarily be represented by a unitary map. This usually non-unitary map is known as quantum channel, and the reduced system dynamics is studied as open quantum system formalism. 
This can be summarized in the Liouville equation of the system as 
 \begin{equation}
\frac{\partial}{\partial t}\rho_{\rm S}\left(t\right)=-\frac{i}{\hbar}Tr_{\rm R}\left[\hat{H}, \rho\left(t\right)\right],\label{eq:Lio-OQS}
\end{equation}
where partial trace over the environment has been performed. 

We understand that Eq. (\ref{eq:Lio-OQS}) gives a realistic picture of the dynamics of a quantum system, but the question that arises is ``whether it solves our problem?'' To answer this question, we need to consider the challenges in solving this dynamical equation. Often mathematical modeling of the system-reservoir dynamics may be complex thus leading to infinitely many coupled equations of motion corresponding to numerous degrees of freedoms of the environment \cite{breuer2002theory}. Therefore, we can entirely focus on the system of our interest by employing some approximations which would allow us to dedicate all our resources to obtain the dynamical behavior of the quantum process or system under consideration \cite{breuer2002theory}.

\subsection{Dynamical maps and operator-sum representation \label{Kraus}}

The reduced system dynamics described by Eq. (\ref{eq:Lio-OQS}) can also be visualized as a map $\Phi\left(t\right)$ from the Hilbert space $\mathcal{H}_{\rm S}$ to the same space at some later time $t$. This set of completely positive trace preserving (CPTP) dynamical maps for all values of time $t$ forms a semigroup (as only positive values of time are allowed) and satisfies $\Phi\left(t_1\right)\Phi\left(t_2\right)=\Phi\left(t_1+t_2\right)$, which can describe the complete dynamics of the system after neglecting the memory effect \cite{breuer2002theory}. 

Before we get back to Eq. (\ref{eq:Lio-OQS}) to establish its CPTP nature, let us briefly discuss its requirement. A linear map $\Phi$ describing the effect of a physical process on a quantum system in a particular quantum state is expected to preserve its normalization and positivity (i.e., trace and positivity of the density operator). Complete positivity is a stronger notion and becomes relevant when $\Phi$ is acting only on a subsystem of correlated (entangled) system. In such cases, the map must ensure positivity of the density operators describing the combined state \cite{caruso2014quantum}. An arbitrary CPTP map is a quantum channel which is useful in quantum information processing (discussed in Section \ref{QIP}). 

The brief introduction of open quantum system in the previous section and Eq. (\ref{eq:Lio-OQS}) allow  us to write the dynamical map $\Phi$ as 
 \begin{equation}
\Phi\left(\rho_{\rm S}\right)=Tr_{\rm R}\left[{U}_{\rm I}\left(\rho_{\rm S}\otimes\rho_{\rm R}\right){U}_{\rm I}^\dagger\right],\label{eq:for-Kraus}
\end{equation}
where ${U}_{\rm I}$ is the evolution operator for the composite state of the quantum system and its environment. 
Without loss of generality, we can assume the reservoir initial state $\rho_{\rm R}=|f_0\rangle\langle f_0|$ in the pure state and an orthonormal basis $\left\{|e_k\rangle\right\}$ for the environment. This will help us to write 
 \begin{equation}
\Phi\left(\rho_{\rm S}\right)=\underset{k}{\sum}
\langle e_k\left|{U}_{\rm I}\left(\rho_{\rm S}\otimes|f_0\rangle\langle f_0|\right){U}_{\rm I}^\dagger\right|e_k\rangle=\underset{k}{\sum}E_k \rho_{\rm S}E_k^\dagger.\label{eq:Kraus}
\end{equation}
The second summation is known as operator-sum representation \cite{stinespring1955positive,sudarshan1961stochastic,kraus1971general}, where the Kraus opeators, $E_k=\langle e_k\left|{U}_{\rm I}\right|f_0\rangle,$ follow $\underset{k}{\sum}E_k^\dagger E_k =I_{\rm S}$ to ensure the CPTP nature of the map $\Phi$.

This operator-sum representation provides a very useful tool for the study of open quantum system dynamics. Here, we can aptly mention that the number of non-zero elements $E_k$ (which cannot be more than $d^2$ for Hilbert space of dimension $d$) is known as Kraus rank \cite{caruso2014quantum}. Also, this technique is not restricted to a reservoir which is initially in the pure state as one can purify this mixed state by using an ancillary system \cite{nielsen2010quantum}. 

In the next section, we will use the Kraus operators \cite{kraus1971general} obtained in the present section to obtain the well-known quantum master equation in the Markovian regime as in Refs. \cite{preskill1998lecture,rajagopal2010kraus}.

\subsection{Markovian noise and quantum master equation \label{Mark}}

Let us consider that a quantum state $\rho_{\rm S}\left(0\right)$ evolves in a small time interval $t$ to $\rho_{\rm S}\left(t\right)\approx \rho_{\rm S}\left(0\right)+O\left(t\right).$ For this dynamics, one of the Kraus operators can be assumed in the form $E_0\approx I_{\rm S}+O\left(t\right)$, and others to be in the order of $\sqrt{t}.$ Therefore, the Kraus operators can be expressed, in the short-time limit, in terms of another operator $L_k$ as $E_k\approx \sqrt{t}L_k\,\forall k\neq0,$ and as $E_0\approx I_{\rm S}+t L_0$. Here, $L_0$ can be written in terms of two Hermitian matrices $M=\frac{L_0+L_0^\dagger}{2}$ and $H=-\frac{\left(L_0-L_0^\dagger\right)}{2i}$ as $L_0=\left(M-i H\right).$ The $L_k\,\forall k\neq0,$ can be interpreted as quantum jumps of the system taking place with probability in the order of $t.$ Using this mathematical form of the Kraus operators $E_k\, \forall k\neq0$ and the CPTP nature, we can obtain $M=-\frac{1}{2}\underset{k\geq1}{\sum}L_k^\dagger L_k.$

Meanwhile, Eq. (\ref{eq:Kraus}) can lead to $\rho_{\rm S}\left(t\right)-\rho_{\rm S}\left(0\right)\approx t \mathcal{L}\rho_{\rm S}\left(0\right),$ and the standard Lindblad equation (using Eq. (\ref{eq:Lio-OQS})) as 
\begin{equation}
\frac{\partial}{\partial t}\rho_{\rm S}\left(t\right)=\mathcal{L}\rho_{\rm S}=-i \left[\hat{H}, \rho_{\rm S}\right]+\underset{k\geq1}{\sum}\left\{ L_k \rho_{\rm S}L_k^\dagger-\frac{1}{2}\left( L_k^\dagger L_k \rho_{\rm S}+\rho_{\rm S}L_k^\dagger L_k  \right) \right\},\label{eq:Lindblad}\end{equation}
where we have assumed $\hbar=1.$ The second term in Eq.  (\ref{eq:Lindblad}) is usually called dissipator, and $L_k$s are known as Lindblad operators. Also, the superoperator $\mathcal{L}$ represents a generator of the semigroup. For the contributions of Lindblad \cite{lindblad1976generators}, and Gorini, Kossakowski, and Sudarshan \cite{gorini1976completely}, the Markovian quantum master equation (\ref{eq:Lindblad}) is also known as LGKS equation. For instance, Eq. (\ref{eq:master_eq-SGAD}) is a Markovian master equation for a single spin-qubit evolving in a squeezed thermal bath. 

While writing such Markovian master equation some approximations are involved. Specifically, to omit the reservoir degrees of freedom we have to perform first approximation, i.e., to consider that the system and reservoir are weakly coupled. This is referred to as \textit{Born approximation}, and it assumes the reservoir state $\rho_{\rm R}$ is negligibly affected due to interaction. 
This means that the characteristic time scale over which reservoir correlation functions decay ($\tau_{\rm R}$) is much smaller in comparison to that over which the system changes appreciably ($\tau_{\rm rel}$). This is further simplified using \textit{Markov approximation}. In which, we assume that the reduced system density matrix depends only on the present state and not on the previous states of the system. The obtained equation sometimes requires averaging over the fast oscillating terms known as \textit{rotating wave approximation} which is applicable when the characteristic time of the intrinsic time evolution of the system ($\tau_{\rm S}$) is much smaller than the relaxation time ($\tau_{\rm rel}$)  of the system. 
Interestingly, most of the quantum optical phenomena can be well characterized using the Lindblad equation (\ref{eq:Lindblad}) as the involved approximation are well satisfied in them.

So far, we have only discussed  quantum master equation in the weak system-environment coupling case. In contrast to that, in the strong system-environment coupling, we enter into the domain of quantum Brownian motion \cite{breuer2002theory} at high temperature (as here system evolution is slow when compared to the bath correlation time, i.e., $\tau_{\rm R}\ll \tau_{\rm S}$), while non-Markovian character \cite{breuer2002theory} comes into picture at low temperature. The master equation for the quantum Brownian motion cannot be written in the Lindblad form as it includes one more term corresponding to the fluctuations depending on temperature  \cite{breuer2002theory}. We will skip the discussion regarding quantum Brownian motion and will return to non-Markovian noise in Section \ref{Mark}. Here, we may further note that Heisenberg-Langevin equation can  also be used to accomplish the same task in the Heisenberg picture (see \cite{scully1997quantum} for more discussion). This approach is not discussed here as it is not used in the present work. In the next section, we will briefly describe the effect of the environment on the system dynamics with the help of some specific examples. 

\subsection{Decoherence and examples of Markovian noise channels \label{ex}}

The operator-sum representation (discussed in Section \ref{Kraus}) of the system dynamics possess the same form as the generalized quantum measurements. Specifically, the measurement introduced in Eq. (\ref{eq:meas}) can be generalized as an operation for $i$th measurement outcome as
\begin{equation}
\Phi_i\left(\rho\right)=\underset{k}{\sum}A_{ik}\rho A_{ik}^{\dagger},\label{eq:gen-meas}
\end{equation} 
where $A_{ik}$ forms a set of system operators satisfying the completeness relation. From which one can calculate the probability, normalized to unity (i.e., $\underset{i}{\sum}\mathrm{Pr}\left(i\right)=1$), for the $i$th outcome as $\mathrm{Pr}\left(i\right)=Tr\left\{\Phi_i\left(\rho\right)\right\}=Tr\left\{\rho P_{i}\right\}$ using the notation $P_{i}=\underset{k}{\sum}A_{ik}^{\dagger} A_{ik}$ for the positive and complete set of operators, known as positive-operator-valued-measure.

This analogy allows us to visualize the interaction of a quantum system with its surroundings as a measurement performed on the system by the reservoir. Thus, after tracing over environment degrees of freedom, few states in certain basis set are  preferred over others while their superposition decay due to open system dynamics. This decay in quantum coherence is termed as decoherence. In the present section, we will discuss different types of Markovian channels and obtain corresponding Kraus operators to discuss the decoherence caused due to them.

Nature of the open quantum system dynamics can be characterized by the system-reservoir interaction Hamiltonian. The first type of dynamics involves decoherence without dissipation of energy from the system. For example, the system is described by a harmonic oscillator and the reservoir, as mentioned previously, is a collection of such harmonic oscillators \cite{nielsen2010quantum}. Without loss of generality, we can assume the interaction Hamiltonian as 
$\hat{H}_{\rm I}=g a_{\rm S}^\dagger a_{\rm S} \left(b_{\rm R}+b_{\rm R}^\dagger\right).$ The whole system-reservoir dynamics is controlled by the unitary $U_{\rm I}$ induced by this Hamiltonian. Note that the system Hamiltonian $\hat{H}_{\rm S}$ (number operator in case of harmonic oscillator) commutes with the interaction Hamiltonian $\hat{H}_{\rm I}$, thus no energy loss is anticipated due to decoherence. 
Suppose the system (reservoir) is composed of a two dimensional Hilbert space with the basis elements $|0\rangle$ and $|1\rangle$ ($|0_{\rm R}\rangle$ and $|1_{\rm R}\rangle$), and the reservoir is initially prepared in the vacuum state $|0\rangle,$ we are equipped to obtain the Kraus operators of the system dynamics using Eq. (\ref{eq:Kraus}) as $E_k=\langle k_{\rm R}\left|{U}_{\rm I}\right|0_{\rm R}\rangle.$ This is how we can obtain the Kraus operators of the phase damping (PD) noise as \cite{nielsen2010quantum}
\begin{equation}
\begin{array}{c}
K_{0}=|0\rangle\langle0|+p|1\rangle\langle1|,\quad\quad K_{1}=\sqrt{1-p^{2}}|1\rangle\langle1|,\end{array}\label{eq:Kraus-dephasing}
\end{equation}
where $\left(1-p^{2}\right)$ is the probability of a photon  getting scattered from the system. Here, $p$ is a function of the system-reservoir coupling. Specifically, a physically relevant condition often encountered in the quantum communication, which may be described by this noise model, is a photon scattering randomly while traveling through a waveguide.

An alternative representation of the PD noise can be presented using three Kraus operators assuming finite probability of transition of the reservoir vacuum state to the first (second) exited state if the system state is in the ground (first excited) state \cite{preskill1998lecture} (assuming the environment state in a 3-dimensional Hilbert space). In Chapter \ref{CryptSwitch}, this representation of PD Kraus operators will be used. 

We can proceed now to another noise model which consists of energy loss, too. In the analogy of the PD noise model, we can define an interaction Hamiltonian as $\hat{H}_{\rm I}=g \left(a_{\rm S}^\dagger b_{\rm R}-a_{\rm S} b_{\rm R}^\dagger\right)$ for the system and reservoir defined by harmonic oscillators. In case of the vacuum reservoir, using the same strategy, one can obtain Kraus operators defining amplitude damping (AD) channel as 
\cite{nielsen2010quantum}
\begin{equation}
\begin{array}{c}
K_{0}=|0\rangle\langle0|+\sqrt{p}|1\rangle\langle1|,\quad\quad K_{1}=\sqrt{1-p}|0\rangle\langle1|,\end{array}\label{eq:Kraus-damping}
\end{equation}
where $\left(1-p\right)$ is the probability for loosing a photon by the system. Here, we would like to bring to your attention that Hamiltonian $\hat{H}_{\rm I}$ do not commute with the system Hamiltonian in case of the AD noise. We can visualize the difference between these two noise models with the help of an arbitrary single-qubit state as $\rho=\left[\begin{array}{cc}
a & b\\
b^* & 1-a\end{array}\right]$
evolving under an AD and a PD channels independently. Transformed quantum state in the AD channel is $\rho_{\rm AD}=\left[\begin{array}{cc}
1-p\left(1-a\right) & \sqrt{p}b\\
\sqrt{p}b^* & p\left(1-a\right)\end{array}\right];$ whereas the same state evolves under the PD noise to $\rho_{\rm PD}=\left[\begin{array}{cc}
a & pb\\
pb^* & 1-a\end{array}\right].$ Note that the off-diagonal terms in both these cases show environment induced decoherence, while only the state evolved in the AD channel shows a change in populations, i.e., in the diagonal terms. 

Generally, two types of open quantum systems can be studied on the basis of system-reservoir interaction. In all cases when the system-reservoir interaction Hamiltonian commutes with the system Hamiltonian, i.e., $\left[\hat{H}_{\rm S},\hat{H}_{\rm I} \right]=0,$ we call the open quantum system quantum non-demolition (QND). In this type of open quantum system, only dephasing is observed. In contrast, when the Hamiltonians fail to commute then dephasing along with damping in the system is induced and are known as dissipative open systems. Both these aspects of open
system evolution have been realized in a series of beautiful experiments
\cite{brune1996observing,turchette2000decoherence,myatt2000decoherence}.  In the following chapters, we will come across a few other noise models defined by their Kraus operators, such as generalized amplitude damping (GAD), squeezed generalized amplitude damping (SGAD). Finally, we are in a position to go beyond the Markovian approximation, which will be introduced in the next section.

\subsection{Non-Markovian noise and quantum master equation \label{N-Mark}}

Generally, reduced system dynamics obtained after tracing over the environment degrees of freedom is non-Markovian in nature. However, in most of the cases, the Markovian approximation provides us a qualitative idea of the long-time dynamics of the system \cite{breuer2002theory}. The total Hamiltonian of the open quantum system can also be written as   
$\hat{H}=\hat{H}_{0}+\alpha \hat{H}_{\rm I},$ where $\hat{H}_{0}$ ($\hat{H}_{\rm I}$) is the uncoupled system and environment (the system-environment interaction) Hamiltonian, and $\alpha$ is a dimensionless coupling constant. For which the equation of motion in the interaction picture, in analogy of Eqs. (\ref{eq:LioEq}) and (\ref{eq:LioEq-2}), is  $\frac{\partial}{\partial t}\rho\left(t\right)=-i \alpha\left[ H_{\rm I}\left(t\right),\rho\left(t\right)\right]= \alpha\mathcal{L}\left(t\right)\rho\left(t\right)$ with $\hbar=1$ for $H_{\rm I}\left(t\right)$, the interaction picture Hamiltonian.

The non-Markovian master equations, viz., Nakajima-Zwanzig equation, provide exact dynamics of the system \cite{nakajima1958quantum,zwanzig1960ensemble}, which depend on both the initial conditions at $t^\prime$ and history of the system. In other words, the equation of motion is not local in time. This manifests the non-Markovian memory effects of the obtained equation of motion for the reduced system dynamics.
Interestingly, note that there is no restriction on the initial condition of the environment's initial state (i.e., not even the initial factorizing condition), which is beyond the assumption of Markovian quantum master equation.

Here, we will consider a specific example of the exact master equation of a two level system strongly coupled to its environment beyond Born-Markov approximation. In case of the system under consideration, as discussed previously in the present section, we can write the Hamiltonian as $\hat{H}=\hat{H}_0+\hat{H}_{\rm I}$ with $\hat{H}_0=\omega_0 \sigma_{+}\sigma_{-}+\underset{k}{\sum}\omega_k b_k^\dagger b_k$ and $\hat{H}_{\rm I}=\sigma_{+}\otimes \underset{k}{\sum} g_k b_k+\sigma_{-}\otimes \underset{k}{\sum} g_k b_k^\dagger.$ Here, $\omega_0$ is the transition frequency of the two-level system with $\sigma_{+}$ and $\sigma_{-}$ as the raising and lowering operators. The reservoir is considered as a collection of the harmonic oscillators with the creation and annihilation operators $b_k$ and $b_k^\dagger$, respectively, and different values of frequency $\omega_k$ and the coupling constant with the system $g_k$ for each mode.

We can introduce three quantum states to describe the system and environment \cite{garraway1997nonperturbative} as the ground and first excited states of the system 
\begin{equation}
\begin{array}{lcl}
\varPsi_0&=&|0\rangle_{\rm S}\otimes|0\rangle_{\rm R},\\
\varPsi_1&=&|1\rangle_{\rm S}\otimes|0\rangle_{\rm R},
\end{array}\nonumber
\end{equation}
and the first excited state of the $k$th mode of the environment as \begin{equation}
\begin{array}{lcl}
\varPsi_k&=&|0\rangle_{\rm S}\otimes|k\rangle_{\rm R}.
\end{array}\nonumber
\end{equation}
As the total number of particles is a constant of motion, i.e., $\left[ H,\left(\sigma_{+}\sigma_{-}+\underset{k}{\sum} b_k^\dagger b_k\right)\right]=0,$ any initial state $|\varPhi\left(0\right)\rangle=c_0 \varPsi_0+ c_1\left(0\right) \varPsi_1+\underset{k}{\sum} c_k\left(0\right) \varPsi_k$ would
evolve to 
\begin{equation}
|\varPhi\left(t\right)\rangle=c_0 \varPsi_0+ c_1\left(t\right) \varPsi_1+\underset{k}{\sum} c_k\left(t\right) \varPsi_k.\label{eq:NM-ev-st} 
\end{equation}
Assuming the vacuum reservoir, and noting that the amplitude $c_0$ is constant, we obtained 
\begin{equation}
\frac{\partial}{\partial t}c_1\left(t\right)=-\int_{0}^{t}dt^\prime K\left(t-t^\prime\right) c_1\left(t^\prime\right).\nonumber
\end{equation}
The kernel $K\left(t-t^\prime\right)$ can be written in terms of the correlation function and spectral density $J\left(\omega\right)$ of the reservoir as 
\begin{equation}
K\left(t-t^\prime\right)=Tr_{\rm R} \left\{
B\left(t\right)B^\dagger\left(t^\prime\right)\rho_{\rm R}\right\} \exp\left[i \omega_0\left(t-t^\prime \right) \right]=\int d\omega J\left(\omega\right) \exp\left[i \left(\omega_0-\omega\right)\left(t-t^\prime \right) \right].\nonumber
\end{equation}
Thus, using the evolved quantum state (\ref{eq:NM-ev-st}), one can write the reduced state of the two-level system as 
\begin{equation}
\rho_{\rm S}\left(t\right)=Tr_{\rm R} \left\{|\varPhi\left(t\right)\rangle\langle\varPhi\left(t\right)|\right\}=\left[\begin{array}{cc} 1-\left|c_1\left(t\right)\right|^2 & c_0 c_1^*\left(t\right)\\
c_0^* c_1\left(t\right) & \left|c_1\left(t\right)\right|^2\end{array}\right].\nonumber
\end{equation}
Here, we have used quantum information theoretic notation for basis elements as $|0\rangle_{\rm S}=\left[\begin{array}{c}1\\0\end{array}\right]$ and $|1\rangle_{\rm S}=\left[\begin{array}{c}0\\1\end{array}\right].$
One can also obtain an exact master equation by differentiating this reduced density matrix to obtain
\begin{equation}
\frac{\partial}{\partial t}\rho_{\rm S}\left(t\right)=
-\frac{i}{2}S\left(t\right)\left[\sigma_{+}\sigma_{-},\rho_{\rm S}\left(t\right)\right]+\gamma\left(t\right)\left\{\sigma_{-}\rho_{\rm S}\left(t\right)\sigma_{+}-\frac{1}{2}\sigma_{+}\sigma_{-}\rho_{\rm S}\left(t\right)-\frac{1}{2}\rho_{\rm S}\left(t\right)\sigma_{+}\sigma_{-}\right\},\nonumber
\end{equation}
where due to the presence of time dependent terms $S\left(t\right)$ and $\gamma\left(t\right)$ the dynamical semigroup property is not always satisfied. Also the decay rate $\gamma\left(t\right)$ may take negative values costing complete positivity of the generator.
Note that this solution of the exact master equation can be written in the operator-sum representation, and the Kraus operators (same as in Eq. (\ref{eq:Kraus-damping})) can be obtained with a time dependent decoherence parameter, where we need to restrict ourselves to the values of various independent parameters such that the dynamical map remains CPTP. This point will be further discussed in Chapter \ref{NonMarkovian}. In our example, we have only discussed a dissipative interaction with a vacuum reservoir. In Chapter \ref{NonMarkovian}, we will discuss dephasing and depolarizing non-Markovian channels as well. Specifically, Chapters \ref{QDs}-\ref{Tomogram} are dedicated to the effect of Markovian noise on the nonclassicality present in the spin systems; the effects of Markovian noise on some schemes of quantum communication is discussed in Chapter \ref{CryptSwitch}; and Chapter \ref{NonMarkovian} is devoted to the study of quantum cryptographic schemes analyzed over the non-Markovian channels.  For now, we will focus on the applications of nonclassical states in quantum information processing tasks.

\section{Applications of nonclassical states \label{QIP}}

We studied the basic mathematical tools for quantum mechanics that are required to understand the dynamical behavior of a quantum system, quantum measurement theory, characterization of quantum states not having a classical analogue, and their vulnerability to decoherence. Now, we are in a position to discuss the advantages of nonclassical states in the information processing tasks. 

First of all, we would like to note that "information'' is the amount of our ignorance before the event, which is also the knowledge gained after the observation. The unit of information is "bit", which corresponds to a two dimensional classical system. Processing certain input information to obtain a useful output is studied under information processing. Analogously, a two dimensional quantum system allows superposition of two classical bit values, known as qubit. Use of qubits intrinsically allows us to utilize the advantage of quantum parallelism in performing various tasks. This is why, utilizing quantum resources, information processing tasks have been studied under a wider subject \textit{quantum information processing}. It is broadly studied in two domains, quantum communication and computation. 

As introduced previously, an arbitrary physical process represented by a CPTP map can be used as a quantum channel, which can be exploited with nonclassical states to get an enhancement in numerous information processing tasks. Specifically, several new applications of nonclassical states have been reported
in the recent past \cite{hillery2000quantum,furusawa1998unconditional,yuan2002electrically,ekert1991quantum,bennett1993teleporting,bennett1992communication}.
For example, applications of squeezed states are reported in the implementation
of continuous variable quantum cryptography \cite{hillery2000quantum},
teleportation of coherent states \cite{furusawa1998unconditional}, gravitational wave detection \cite{aasi2013enhanced,grote2013first}  (as squeezed vacuum state was used at the Laser Interferometer Gravitational-Wave Observatory (LIGO) \cite{abbott2016observation,abbott2016gw151226});
antibunching is shown to be useful in building single photon
sources \cite{yuan2002electrically,pathak2010recent}; entangled states have appeared as
one of the main resources of quantum information processing as it
is shown to be essential for the implementation of a set of protocols
of discrete \cite{ekert1991quantum} and continuous variable quantum
cryptography \cite{hillery2000quantum}, quantum teleportation \cite{bennett1993teleporting},
dense coding \cite{bennett1992communication}; states
violating Bell's inequality are reported to be useful for the implementation of protocols
of device independent quantum key distribution \cite{acin2006bell}. As a consequence of these recently
reported applications, generation of nonclassical states in various
quantum systems emerged as one of the most important areas of interest
in quantum information theory and quantum optics. Further, nonclassical states are established to be a powerful resource for quantum supremacy in computing \cite{o2007optical,ladd2010quantum,proctor2017ancilla}. In general, continuous variable nonclassical states and atomic nonclassical states are the main ingredients of the recently growing quantum technology \cite{andersen2015hybrid,pu2017experimental,browne2017quantum,biamonte2017quantum}. 

The works reported in the present thesis  can be broadly divided into three parts. The first part, composed of Chapters \ref{Coupler}-\ref{Zeno}, deals with the generation of different types of lower- and higher-order nonclassical states in optical systems characterized with the help of a set of moments-based criteria. This is followed by the characterization and quantification of nonclassicality in spin systems using different quasidistribution functions and tomograms in the presence of dephasing and dissipative noise in the second part (Chapters \ref{QDs} and \ref{Tomogram}). Thus, the first two parts report the existence of various types of nonclassical features in optical (first part) and spin (second part)  systems, and also report the effect of noise on various physical systems of relevance (second part). The necessity of the present study is justified in the last part of the thesis (Chapters \ref{CryptSwitch} and \ref{NonMarkovian}) where we report a set of applications of the nonclassical states in quantum communication schemes. In the present thesis work, we restrict the applications of quantum channel discussed in the previous sections for transmitting quantum information to distant parties. A detailed discussion of quantum communication schemes will be skipped for now (upto Chapter \ref{CryptSwitch}), as we are going to discuss the generation and dynamics of nonclassical states in various physical systems in the next few chapters.

The remaining part of the thesis is organized as follows.
In Chapter \ref{Coupler}, we investigate the nonclassical properties of the output fields propagating
through (i) a codirectional and (ii) a contradirectional asymmetric nonlinear optical couplers consisting
of a linear waveguide and a nonlinear (quadratic) waveguide operated
by second harmonic generation. In contrast to the earlier results (\cite{perina2000review,perina1991quantum} and references therein),
all the initial fields are considered weak, and a completely quantum
mechanical model is used here to describe the system. The perturbative
solutions of the Heisenberg's equations of motion for various field modes
are obtained using the Sen-Mandal technique described in Section \ref{Sol-of-eq}. The obtained solutions are subsequently
used to show the existence of single-mode and intermodal
squeezing, single-mode and intermodal antibunching, two-mode 
entanglement in the output of the contradirectional asymmetric nonlinear
optical coupler. Further, the existence of higher-order nonclassicality
is also established in both the codirectional and contradirectional asymmetric nonlinear optical couplers by showing the existence of higher-order antibunching,
squeezing and entanglement. 
It is also shown that the
nonclassical properties of light can transfer from a nonlinear waveguide to a linear waveguide. The obtained results have been published as two international journal  articles \cite{thapliyal2014higher,thapliyal2014nonclassical}.

This is followed by the study of quantum Zeno and anti-Zeno effects in a symmetric nonlinear
optical coupler, which is composed of two nonlinear ($\chi^{\left(2\right)}$)
waveguides that are interacting with each other via the evanescent
waves, in Chapter \ref{Zeno}. Both waveguides operate under second harmonic generation in this case as well.
However, to study the quantum Zeno and anti-Zeno effects one of them is
considered as the system and the other one as the probe.
Considering all the fields involved as weak, a completely quantum
mechanical description is provided, and the analytic solutions of
the Heisenberg's equations of motion for all the field modes are obtained
using the Sen-Mandal perturbative technique. Photon number statistics of the second
harmonic mode of the system is shown to depend on the presence of
the probe, and this dependence is considered as quantum Zeno and anti-Zeno
effects. Further, it is established that as a special case of the
momentum operator for $\chi^{\left(2\right)}-\chi^{\left(2\right)}$
symmetric coupler we can obtain momentum operator of $\chi^{\left(2\right)}-\chi^{\left(1\right)}$
asymmetric coupler (discussed in Chapter \ref{Coupler})
with the linear ($\chi^{\left(1\right)}$) waveguide as the probe; and in such a particular case, the expressions obtained
for the quantum Zeno and anti-Zeno effects with the nonlinear probe (which we referred
to as nonlinear quantum Zeno and anti-Zeno effects) may be reduced
to the corresponding expressions with the linear probe (which we referred
to as the linear quantum Zeno and anti-Zeno effects). Linear and nonlinear
quantum Zeno and anti-Zeno effects are rigorously investigated, and
it is established that in the stimulated case, we may switch between the
quantum Zeno and anti-Zeno effects just by controlling the phase of
the second harmonic mode of the system or probe. This piece of work was published as a journal \cite{thapliyal2016linear} and a conference \cite{thapliyal2015quantum} papers. 

In Chapter \ref{QDs}, we study nonclassical features in a number of spin-qubit systems including
single, two- and three-qubit states, of importance in the fields of quantum
optics and information. This is done by analyzing the behavior of
the well-known Wigner, $P$, and $Q$ quasiprobability distributions
for them. We also discuss the not so well-known $F$-function and specify
its relation to the Wigner function. Here, we provide a comprehensive
analysis of the quasiprobability distributions for the spin-qubit systems
under general open system effects, including both pure dephasing as
well as dissipation. Finally, the amount of nonclassicality quantified as the volume of the negative part of the Wigner function is obtained and found to decay as a signature of quantum to classical transition. This makes it relevant from the perspective of
experimental implementation. The work reported in this chapter has been published as Ref. \cite{thapliyal2015quasiprobability}.

In Chapter \ref{Tomogram}, tomograms for the spin-qubit systems discussed in Chapter \ref{QDs} 
are obtained as probability distributions, which are often used to reconstruct a quantum
state from experimentally measured values. 
We also study the evolution of tomograms for different quantum systems, both finite and 
infinite dimensional. 
In the realistic experimental conditions, quantum states are exposed to the ambient environments
and hence subject to effects like decoherence and dissipation, which are dealt
with here, consistently, using the formalism of open quantum systems. This is extremely relevant from the perspective of 
experimental implementation and  issues related to the state reconstruction in quantum computation and communication. These
considerations are also expected to affect the quasiprobability distribution obtained from the experimentally generated tomograms 
and nonclassicality observed from them. The work reported in this chapter led to a publication \cite{thapliyal2016tomograms}.

Recently, several aspects of controlled quantum communication (e.g.,
bidirectional controlled state teleportation (BCST), controlled deterministic secure quantum communication (CDSQC), controlled quantum dialogue (CQD)) have been
studied using $n$-qubit ($n\geq3$) entanglement. Specially, a large
number of schemes for BCST
are proposed using $m$-qubit entanglement ($m\in\{5,6,7\}$). In Chapter \ref{CryptSwitch},
we propose a set of protocols to illustrate that it is possible to
realize all these tasks related to controlled quantum communication
using only Bell states and permutation of particles (PoP). As the
generation and maintenance of a Bell state is much easier than a multi-partite
entanglement (also shown in Chapter \ref{QDs}), the proposed strategy has a clear advantage over the
existing proposals. Further, it is shown that all the schemes proposed
here may be viewed as applications of the concept of quantum cryptographic
switch. The performances
of the proposed protocols subjected to the AD and
PD channels are also discussed. The work reported in this chapter has been published as \cite{thapliyal2015applications,thapliyal2015general}.

In Chapter \ref{NonMarkovian}, a three party scheme for secure quantum communication, namely CQD, is analyzed
under the influence of non-Markovian channels. By comparing
with the corresponding Markovian cases, it is seen that the average
fidelity can be maintained for relatively longer
periods of time. Interestingly, a number of facets of quantum
cryptography, such as  quantum secure direct communication (QSDC),
deterministic secure quantum communication (DSQC) and their controlled counterparts, quantum dialogue (QD), 
QKD, quantum key agreement (QKA), can be reduced from the CQD scheme. Therefore, the CQD scheme is analyzed under the influence of
damping, dephasing and depolarizing non-Markovian channels, and subsequently, the effect of these non-Markovian channels on the other schemes of secure quantum communication is deduced from the results obtained for CQD. The damped non-Markovian
channel causes a periodic revival in the fidelity, while fidelity
is observed to be sustained under the influence of the dephasing non-Markovian
channel. This work resulted in a journal article as \cite{thapliyal2017quantum}.

Finally, the thesis work is concluded in Chapter \ref{conclusions-and-scope}, where the future scope of the present work is also discussed briefly. We would also like to mention that detailed discussion at a few occasions is avoided to remain within the prescribed limit of the thesis. Interested readers may refer the corresponding published versions for details.

\thispagestyle{empty}
\cleardoublepage
\blankpage

\titlespacing*{\chapter}{0pt}{-50pt}{20pt}
\chapter{Nonclassicalities in a codirectional and contradirectional asymmetric nonlinear
optical couplers \label{Coupler}}

\section{Introduction}

In Chapter \ref{Introduction}, we have noted that several systems
are already investigated and have been shown to produce entanglement
and other nonclassical states (see \cite{pathak2013nonclassicality,sen2013intermodal} 
and references therein). However, experimentally realizable simple
systems that can be used to generate and manipulate nonclassical states
are still of much interest. One such experimentally realizable and
relatively simple system is the nonlinear optical coupler. The motivation for studying nonclassical properties of light propagating through an optical coupler is  manyfold. To begin with, we may note that the nonlinear optical
couplers are of specific interest because they can be easily realized
using optical fibers or photonic crystals, and the amount of nonclassicality
present in the output field can be controlled by monitoring the interaction
length and the coupling constant. Further, recently Matthews et al.,
have experimentally demonstrated manipulation of multi-photon entanglement
in quantum circuits constructed using waveguides \cite{matthews2009manipulation}.
Quantum circuits implemented by them can also be viewed as optical
coupler-based quantum circuits as in their circuits, waveguides are
essentially combined to form couplers. Using similar
arrangements of optical couplers, the same group has also successfully
implemented reconfigurable controlled two-qubit operation \cite{li2011reconfigurable}
and Shor's algorithm \cite{politi2009shor} on a photonic chip.   In another interesting application,
Mandal and Midda have shown that a universal irreversible gate library
(NAND gate) can be built using nonlinear optical couplers \cite{mandal2011all}.
Mandal and Midda's work essentially showed that, in principle, a classical
computer can be built using optical couplers. Further, a directional optical coupler
is known to be one of the most important integrated guided wave components \cite{lugani2013studies}.
Thus, if we can establish the possibility of generation of intermodal
entanglement or any other nonclassicality in a directional optical
coupler, that would imply the existence of another source of entanglement
or other required nonclassical fields in a complex on-chip photonic
circuit that can be used to perform a specific task related to quantum
computation or quantum communication. In addition to the fact that
waveguide-based directional couplers can be realized easily, its potential
adoptability in the integrated waveguide-based photonic circuits provides
it an edge over many other systems, where nonclassical characters have
already been studied. This is so because  most of the atomic and optomechanical systems
cannot be used in the integrated quantum optic devices, such as in on-chip
photonic circuits. Further, it is already established by several groups
that integrated waveguide-based  structures are a better source of entanglement
compared to those based on bulk crystal (\cite{tanzilli2012genesis}
and references therein). These facts motivated
us to systematically investigate the possibility of observation of
nonclassicality in nonlinear optical couplers. Among different possible
nonlinear optical couplers two of the simplest systems are codirectional and contradirectional 
asymmetric nonlinear optical couplers which are prepared by combining
a linear waveguide and a nonlinear (quadratic) waveguide operated
by second harmonic generation. Waveguides interact with each other
through evanescent waves, and we may say that transfer of nonclassical
effect from the nonlinear waveguide to the linear one happens through
evanescent waves. In the present chapter, we aim to study various higher-order
nonclassical properties of light propagating through the codirectional and contradirectional  optical couplers with specific
attention to entanglement.

It is interesting to note that several nonclassical properties of the
optical couplers have been studied in the past (see \cite{perina2000review} for
a review). For example, photon statistics, phase properties, and squeezing
in codirectional and contradirectional Kerr nonlinear couplers are
studied with fixed and varying linear coupling constants \cite{el2005quantum,korolkova1997quantumKerr,fiuravsek1999quantum,ariunbold2000quantum,korolkova1997kerr};
photon statistics of Raman and Brillouin couplers \cite{perina1997statistics}
and parametric couplers \cite{korolkova1997quantum} is studied in detail; and photon
statistics and other nonclassical properties of asymmetric \cite{perina1995asy,perina1995quantum,mandal2004approximate,perina1995photon,perina1995non}
and symmetric \cite{perina1996quantum,perina1995non,mista1997nonclassical} directional
nonlinear couplers are investigated for various conditions, such as strong
pump \cite{perina1995asy}, weak pump \cite{mandal2004approximate},
phase mismatching \cite{perina1995quantum863} for codirectional
\cite{mandal2004approximate,perina1996quantum,perina1995quantum863}
and contradirectional \cite{perina1996quantum,perina1995photon,perina1995quantum863}
propagation of coherent and nonclassical \cite{perina1995non,mista1997nonclassical}
input modes. However, almost all the earlier studies were limited
to the investigation of lower-order nonclassical effects (e.g., squeezing
and antibunching) either under the conventional short-length approximation
\cite{perina1995photon} or under the parametric approximation, where a pump
mode is assumed to be strong and treated classically as a $c$-number
\cite{perina1995quantum863}. Only a few discrete efforts
have recently been made to study higher-order nonclassical effects
and entanglement in optical couplers \cite{kowalewska2009sudden,abbasi2013thermal,leonski2004kerr,miranowicz2006two,kowalewska2012generalized,el2004higher},
but even these efforts are limited to Kerr nonlinear couplers. For
example, in 2004, Leo{\'n}ski and Miranowicz reported entanglement in
Kerr nonlinear couplers \cite{leonski2004kerr}; subsequently, entanglement in pumped Kerr nonlinear optical couplers \cite{miranowicz2006two},
entanglement sudden death \cite{kowalewska2009sudden} and thermally induced
entanglement \cite{abbasi2013thermal} are reported in Kerr nonlinear
couplers. Amplitude squared (higher-order) squeezing is also reported
in Kerr nonlinear couplers \cite{el2004higher}. 

In the earlier
studies on the codirectional and contradirectional asymmetric nonlinear optical couplers \cite{perina2000review,perina1995photon}, intermodal entanglement and
some of the higher-order nonclassical properties studied here were
not studied. In fact, in those early studies on the contradirectional asymmetric nonlinear optical coupler, the  second harmonic mode
was assumed to be pumped with a strong coherent beam. In
other words, the  second harmonic mode
was assumed to be classical, and thus it
was beyond the scope of the previous studies to investigate single-mode
and intermodal nonclassicalities involving this mode. Interestingly,
a completely quantum mechanical treatment adopted in this thesis
is found to show intermodal squeezing in a compound mode involving
the  second harmonic mode. Importantly, neither has any
effort yet been made to rigorously study the higher-order nonclassical
effects in the nonlinear optical couplers in general, nor has a serious effort
been made to study entanglement in the nonlinear optical couplers
other than Kerr nonlinear couplers. 
Keeping these facts in mind, the present chapter aims to study nonclassical
properties associated with the codirectional and contradirectional asymmetric nonlinear optical
couplers (with specific attention to entanglement) using the  perturbative
solution obtained by the Sen-Mandal technique (described in Section \ref{Sol-of-eq}).

The remaining part of this chapter is organized as follows. In Section \ref{sec:The-model-and},
the model momentum operator that represents the asymmetric nonlinear
optical coupler is described and perturbative solution of the equations
of motion corresponding to different field modes present in the momentum
operator of the contradirectional optical coupler are reported. The
perturbative
solution of the equations of motion corresponding to different field
modes present in the momentum operator of the codirectional asymmetric nonlinear optical coupler is already discussed in Section \ref{Sol-of-eq} of the previous chapter. In Section \ref{sec:Criteria-of-nonclassicality},
we briefly describe a set of criteria for lower- and higher-order nonclassicality. Thereafter, in Section
\ref{sec:Nonclassicality-in-contradirectional}, the criteria described
in Section \ref{sec:Criteria-of-nonclassicality} are used to investigate the existence of different
nonclassical characters (e.g., lower- and higher-order squeezing,
antibunching, and entanglement) in various field modes present in
the contradirectional asymmetric nonlinear optical coupler. In Section \ref{sec:Nonclassicality-in-codirectional},
we use only the higher-order criteria to illustrate
the nonclassical characters of various field modes present in the codirectional
asymmetric nonlinear optical coupler. Specifically, we have reported
higher-order squeezing, antibunching, and entanglement. Finally,
Section \ref{sec:Conclusions-coupler} is dedicated for conclusions.

\section{The model and the solution\label{sec:The-model-and}}

\begin{figure}
\centering{}\includegraphics[scale=0.8]{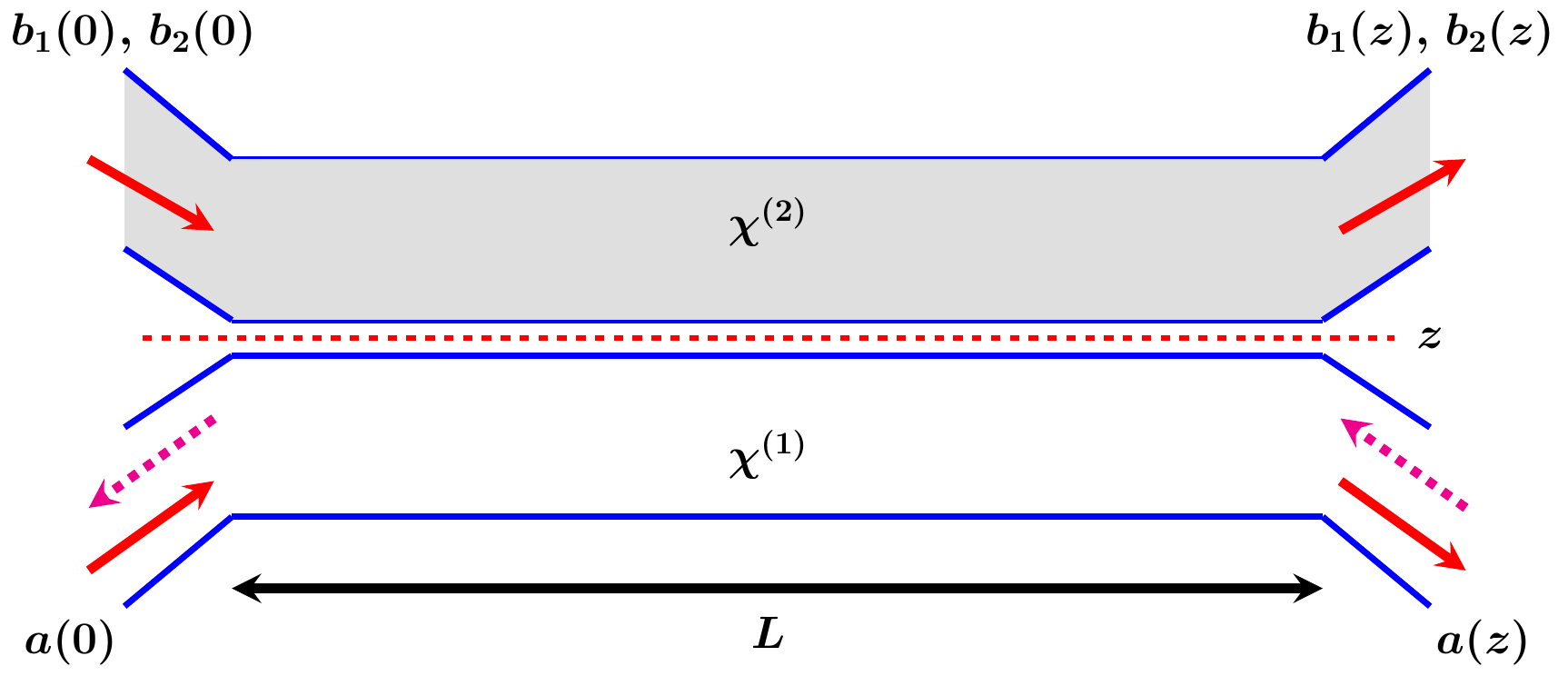}\caption[Schematic
diagram of a codirectional and a contradirectional asymmetric nonlinear optical couplers]{\label{fig:schematic-diagram} Schematic
diagram of an asymmetric nonlinear optical coupler
prepared by combining a linear waveguide ($\chi^{(1)}$) with a nonlinear
($\chi^{(2)}$) waveguide operated by second harmonic generation.
The fields involved are described by the corresponding annihilation
operators, as shown; $L$ is the interaction length. The contradirectional (codirectional) propagation of radiation field in the linear waveguide is shown by magenta-dashed (red-smooth) arrows.}
\end{figure}

A schematic diagram of a codirectional and contradirectional asymmetric nonlinear optical
couplers is shown in Figure \ref{fig:schematic-diagram}. From the figure,
one can easily observe that a linear waveguide is combined with a
nonlinear $\left(\chi^{(2)}\right)$ waveguide to constitute the asymmetric
couplers of our interest. Further, 
we can also observe that in the linear waveguide, field propagates in the same (opposite) 
direction as that of the propagation of the fields in the
nonlinear waveguide for the codirectional (contradirectional) nonlinear optical coupler.

Electromagnetic field characterized by the bosonic
field annihilation (creation) operator $a\,(a^{\dagger})$ propagates
through the linear waveguide. Similarly, the field operators $b_{i}\,(b_{i}^{\dagger})$
corresponds to the nonlinear medium. Specifically, $b_{1}\,(k_{1})$
and $b_{2}\,(k_{2})$ denote annihilation operators (wavevectors) for the
fundamental and second harmonic modes, respectively. The momentum
operator for the codirectional and contradirectional optical couplers \cite{perina2000review,mandal2004approximate} is given by Eq. (\ref{eq:Mom-asymmetric}),
where $\Delta k=|2k_{1}-k_{2}|$
represents the phase mismatch between the fundamental and second harmonic
beams. The linear (nonlinear) coupling constant, proportional to susceptibility
$\chi^{(1)}$ $(\chi^{(2)})$, is denoted by the parameter $k\,\left(\Gamma\right)$. Precisely, the coupling constant $\Gamma$ is usually
expressed in terms of the quadratic susceptibility $\chi^{\left(2\right)}$
as $\Gamma=\frac{\omega_{2}^{2}\chi^{(2)}}{2k_{2}c^{2}}$, where $k_{2}=\frac{\omega_{2}}{c}\sqrt{1+\chi^{(1)}}$,
and $k_{2}$, $c$, and $\omega_{2}$ are the wavevector, speed
of light in vacuum, and frequency of $b_{2}$ mode, respectively
(cf. Ref. \cite[p.~88]{boyd2003nonlinear}). It is reasonable to assume $\chi^{(2)}\ll\chi^{(1)}$
as in a real physical system we usually obtain $\chi^{(2)}/\chi^{(1)}\,\simeq10^{-6}$.
As a consequence, in the absence of a highly strong pump $\Gamma\ll k$.
In fact, below we write the equations
of motion for the dimensionless annihilation and creation operators,
the coupling constants present in those equations of motion should
be multiplied by the quantum factor $\left(\frac{\hbar\omega_{2}}{2\epsilon_{2}V}\right)^{1/2}$,
$\epsilon_{2}$ being dielectric susceptibility and $V$ is the interaction
volume. This factor further reduces the value of the coupling constant.

Earlier this model of codirectional \cite{perina2000review,mandal2004approximate} and contradirectional (\cite{perina2000review} and references
therein) optical couplers was investigated. Heisenberg's equations of motion for all the field modes are already reported in case of the codirectional optical coupler in Eq. (\ref{eq:aaa}) of this thesis. For the contradirectional optical coupler, 
we can obtain the coupled differential equations for three different
modes, using the procedure described in \cite{perina2000review}, as follows \begin{equation}
\frac{da}{dz}=ik^{*}b_{1},\,\frac{db_{1}}{dz}=-ika-2i\Gamma^{*}b_{1}^{\dagger}b_{2}\exp\left(-i\Delta kz\right),\,\frac{db_{2}}{dz}=-i\Gamma b_{1}^{2}\exp\left(i\Delta kz\right).\label{eq:aaa-c}\end{equation}
Here, it would be apt to mention that 
for the contradirectional couplers the sign of derivative in the Heisenberg's
equation of motion of the contra-propagating mode is changed (i.e.,
in the present case $\frac{da}{dz}$ is replaced by $-\frac{da}{dz}$
as mode $a$ is considered here as the contra-propagating mode). The
method used here to obtain (\ref{eq:aaa-c}) is described in Refs. \cite{perina2000review,perina1995photon}.
Further, this particular description of contradirectional coupler
is valid only for the situation when the forward propagating waves
reach $z=L$ and the counter (backward) propagating waves reach $z=0$.
Thus, the coupled differential equations described by (\ref{eq:aaa-c}) and their
solutions obtained below are not valid for $0<z<L$ \cite{perina1995photon}.

Earlier these coupled differential equations (\ref{eq:aaa-c}) in case of the contradirectional optical coupler were solved under the
short-length approximation \cite{perina2000review,perina1995photon}. However,
the use of the short-length solution for the investigation of the nonclassical
properties of this system is not appropriate. To be precise, here
we have considered that the fields are not strong enough to be treated
classically and the coupling constant $\Gamma$ is not large. In such
a situation, we need long interaction length for the effective appearance
of quantum effects, but a short-length solution is not good for describing
a system having long interaction length. In fact, the asymmetric nonlinear optical coupler
described by Eq. (\ref{eq:Mom-asymmetric}) allows long interaction length in the real physical situation, as long interaction length can be achieved easily with
the help of optical fibers. In the present work, we assume the interaction
length of the order of hundred meters, which is many orders higher
compared to the traditional arrangements. This allows us to observe nonclassical
effects using several order lower nonlinearity. For instance, having
several hundred meters of fibers, we can have about 4 orders lower quadratic
susceptibility, which just corresponds to the value $\Gamma\thickapprox0.001$
(which is used in the present work). Further, in this work, we have
considered the power of the initial beams in about tenths of microwatts (i.e., about $10^{12}$ photons/s). This choice of initial fields
is consistent with the fact that the initial fields must be very weak
to enable us to observe quantum effects. A small value of the coupling
constant suggests us to use a perturbative solution, but we need the
solution to be valid for long interaction length. The requirement of long interaction length explicitly
shows the advantage of the Sen-Mandal approach over the conventional short-length
approximation, as the former is not restricted by $L$ (i.e., valid
for long interaction length) whereas the later is valid only for short-length (i.e., restricted by the value of L). Keeping these facts in mind, we have used the Sen-Mandal approach.

Using the Sen-Mandal approach and restricting ourselves up to the linear
power of the nonlinear coupling constant $\Gamma$, we have obtained
closed form perturbative analytic solution of (\ref{eq:aaa-c}) as
\begin{equation}
\begin{array}{lcl}
a(0) & = & f_{1}a(L)+f_{2}b_{1}(0)+f_{3}b_{1}^{\dagger}(0)b_{2}(0)+f_{4}a^{\dagger}(L)b_{2}(0),\\
b_{1}(L) & = & g_{1}a(L)+g_{2}b_{1}(0)+g_{3}b_{1}^{\dagger}(0)b_{2}(0)+g_{4}a^{\dagger}(L)b_{2}(0),\\
b_{2}(L) & = & h_{1}b_{2}(0)+h_{2}b_{1}^{2}(0)+h_{3}b_{1}(0)a(L)+h_{4}a^{2}(L),\end{array}\label{eq:ass-sol-c}\end{equation}
 with\begin{equation}
\begin{array}{lcl}
f_{1} & = & g_{2}={\rm sech}|k|L,\\
f_{2} & = & -g_{1}^{*}=-\frac{ik^{*}\tanh|k|L}{|k|},\\
f_{3} & = & Ck^{*}\Delta kf_{1}^{2}\left\{ i\Delta k\sinh2|k|L+2|k|\left(G_{+}-1-\cosh2|k|L\right)\right\} ,\\
f_{4} & = & 2Ck^{*2}f_{1}^{2}\left\{ i\Delta k\sinh|k|LG_{+}-2|k|\cosh|k|LG_{-}\right\} ,\\
g_{3} & = & -2C|k|f_{1}^{2}\left\{ \left(\Delta k^{2}+2|k|^{2}\right)\cosh|k|LG_{-}+i\Delta k|k|\sinh|k|LG_{+}\right\} ,\\
g_{4} & = & Ck^{*}\Delta kf_{1}^{2}\left\{ i\Delta k\sinh2|k|L\left(G_{+}-1\right)-2|k|\left(1-\cosh2|k|L\left(G_{+}-1\right)\right)\right\} ,\\
h_{1} & = & 1,\\
h_{2} & = & \frac{C^{*}|k|}{2}f_{1}^{2}\left\{ 4|k|^{2}G_{-}^{*}+\Delta k^{2}\left(1-2\left(G_{+}^{*}-1\right)+\cosh2|k|L\right)-2i\Delta k|k|\sinh2|k|L\right\} ,\\
h_{3} & = & 2C^{*}kf_{1}^{2}\left\{ \Delta k|k|G_{-}^{*}\cosh|k|L+\left[i\Delta k^{2}-2i|k|^{2}G_{-}^{*}\right]\sinh|k|L\right\} ,\\
h_{4} & = & \frac{C^{*}|k|k}{k^{*}}f_{1}\left\{ 2|k|^{2}f_{1}G_{-}^{*}+\Delta k\sinh|k|L\left(G_{+}^{*}-1\right)\left[2i|k|+\Delta k\tanh|k|L\right]\right\} ,\end{array}\label{eq:soln}\end{equation}
where $C=\frac{\Gamma^{*}}{|k|\Delta k\left(\Delta k^{2}+4|k|^{2}\right)}$
and $G_{\pm}=\left(1\pm\exp(-i\Delta kL)\right)$. The solution (\ref{eq:soln}) is obtained using the initial conditions $f_{1}(L)=1$,
$g_{2}(0)=1$, and $h_{1}(0)=1$, and all the remaining $f_{i}(L)$,
$g_{j}(0)$, and $h_{i}(0)$ are zero, where $i\in\{2,3,4\}$, and
$j\in\{1,3,4\}$. 

The first-order perturbative solution obtained above is verified by computing the
equal space commutation relations, which implies, in case of contradirectional coupler, $\left[a\left(0\right),a^{\dagger}\left(0\right)\right]=\left[b_{1}\left(L\right),b_{1}^{\dagger}\left(L\right)\right]=\left[b_{2}\left(L\right),b_{2}^{\dagger}\left(L\right)\right]=1$,
while all other equal space commutations are zero. Further, we have
verified that the solution reported here also satisfies the constant of motion.
To be precise, in Ref. \cite{perina1995photon}, it was shown that the constant
of motion for the present system can be obtained as \begin{equation}
a^{\dagger}\left(0\right)a\left(0\right)+b_{1}^{\dagger}\left(L\right)b_{1}\left(L\right)+2b_{2}^{\dagger}\left(L\right)b_{2}\left(L\right)=a^{\dagger}\left(L\right)a\left(L\right)+b_{1}^{\dagger}\left(0\right)b_{1}\left(0\right)+2b_{2}^{\dagger}\left(0\right)b_{2}\left(0\right).\label{eq:const-of-motion}\end{equation}
In analogy with the reducibility of the short-length solution of the codirectional optical coupler from the corresponding Sen-Mandal perturbative solution in Section \ref{Sol-of-eq}, here we have also verified that the solution of the contradirectional coupler
using the short-length solution method reported in \cite{perina2000review}
can be obtained as a special case of the present solution. Specifically,
to obtain the short-length solution we need to expand the trigonometric
functions present in the above solution and neglect all the terms
beyond quadratic powers of $L$ and consider phase mismatch $\Delta k=0$.
After doing so, we obtain\begin{equation}
\begin{array}{c}
f_{1}=g_{2}=(1-\frac{1}{2}\left|k\right|^{2}L^{2}),\, f_{2}=-g_{1}^{*}=-ik^{*}L,\, f_{3}=-g_{4}=-\Gamma^{*}k^{*}L^{2},\\
g_{3}=-2i\Gamma^{*}L,\, h_{1}=1,\, h_{2}=-i\Gamma L,\, h_{3}=-\Gamma kL^{2}{\rm \, and\,}\, f_{4}=h_{4}=0,\end{array}\label{eq:shortlength-c}\end{equation}
which coincides with the short-length solution reported earlier  \cite{perina2000review,perina1995photon}. Clearly,
the solution obtained here is valid and more general than the conventional
short-length solution as the solution reported here is a fully quantum
solution and is valid for any length, restricting the coupling constant
only. 

The model is elaborately discussed for the codirectional propagation of all radiation fields in some of the recent 
publications \cite{perina2000review,perina1995quantum,mandal2004approximate}. Specifically,
in Ref. \cite{perina1995quantum}, single-mode and intermodal squeezing,
antibunching, and subshot noise were studied using analytic expressions
of the spatial evolution of the field operators obtained by the short-length solution
of the Heisenberg's equations of motion corresponding to (\ref{eq:aaa}).
Subsequently, the Sen-Mandal technique was used in Ref. \cite{mandal2004approximate}
to obtain the spatial evolution of the field operators corresponding to (\ref{eq:aaa})
and to study the single-mode and intermodal squeezing and antibunching.
Interestingly, in \cite{mandal2004approximate}, some nonclassical characters
of the codirectional asymmetric nonlinear optical coupler were observed which were not
observed in the earlier investigations \cite{perina2000review,perina1995quantum}
performed using the short-length solution. Keeping these facts in
mind, we have used the solution reported in Ref. \cite{mandal2004approximate} (which is also summarized in Chapter \ref{Introduction} while introducing the Sen-Mandal perturbative technique)
to study the higher-order nonclassicalities in the codirectional nonlinear optical coupler. Before doing so, we investigate both lower- and higher-order nonclassical characters of the fields that
have propagated through the contradirectional asymmetric nonlinear optical
coupler.

\section{Moments-based criteria of nonclassicality \label{sec:Criteria-of-nonclassicality}}

A large number
of nonclassicality criteria (as introduced in Section \ref{Noncl}) are expressed as inequalities involving expectation
values of the functions of annihilation and creation operators. This implies
that Eqs. (\ref{eq:ass-sol-c}) and (\ref{eq:soln}) provide us with the sufficient mathematical
framework required to study the nonclassical properties of the contradirectional
asymmetric nonlinear optical coupler. Similarly, Eqs. (\ref{eq:ass-sol}) and (\ref{eq:terms}) allow us to study the nonclassical behavior in the codirectional nonlinear optical coupler. 

In quantum
optics and quantum information, higher-order nonclassical properties
of bosons (e.g., higher-order Hong-Mandel squeezing, 
antibunching, sub-Poissonian statistics, 
entanglement) are often studied (\cite{verma2010generalized}
and references therein). Until recently, studies on higher-order
nonclassicalities were predominantly restricted to theoretical investigations.
However, a bunch of exciting experimental demonstrations of higher-order nonclassicalities have been recently reported \cite{allevi2012measuring,allevi2012high,avenhaus2010accessing,perina2017higher}.
Specifically, the existence of higher-order nonclassicality in bipartite
multi-mode states produced in a twin-beam experiment has been recently demonstrated
by Allevi, Olivares and Bondani \cite{allevi2012measuring} using a new criterion
for higher-order nonclassicality introduced by them. They also showed
that detection of weak nonclassicalities is easier with their higher-order criterion of nonclassicality as compared to the existing lower-order criteria \cite{allevi2012measuring}. This observation was consistent
with the earlier theoretical observation of Pathak and Garcia \cite{pathak2006control}
that established that the depth of nonclassicality witness in higher-order antibunching
increases with the order. Lately, experimental groups were even able to detect ninth-order nonclassicality \cite{perina2017higher}. The possibility that higher-order nonclassicality
may be more useful in identifying the weak nonclassicalities has
considerably increased the interest of the quantum optics community
on the higher-order nonclassical characters of bosonic fields (\cite{lee2009demonstrating,rundquist2014nonclassical,giri2014single,bohmann2017higher} and references therein). In
the remaining part of this section, we list a set of criteria of lower- and higher-order nonclassicality, and in the following sections, we study the
possibility of satisfying those criteria in the codirectional and contradirectional asymmetric
nonlinear optical couplers.

\subsection{Squeezing, intermodal squeezing, and higher-order squeezing}

The criterion for single-mode quadrature squeezing is already introduced in Section \ref{Sq}. Therefore, we rather discuss here
the criterion for quadrature squeezing
in the compound mode $(jl)$ as \cite{loudon1987squeezed}
\begin{equation}
\left(\Delta X_{jl:j\neq l}\right)^{2}<\frac{1}{4}\,\mathrm{or}\,\left(\Delta Y_{jl:j\neq l}\right)^{2}<\frac{1}{4},\label{eq:condition for squeezing-1}\end{equation}
where $j,l\in\{a,\, b_{1},\, b_{2}\}$, and the quadrature operators
are defined as 
\begin{equation}
\begin{array}{lcl}
X_{ab} & = & \frac{1}{2\sqrt{2}}\left(a+a^{\dagger}+b+b^{\dagger}\right),\\
Y_{ab} & = & -\frac{i}{2\sqrt{2}}\left(a-a^{\dagger}+b-b^{\dagger}\right).\end{array}\label{eq:two-mode-quadrature}\end{equation}

As discussed in the previous chapter, higher-order squeezing is usually studied using two different approaches 
\cite{hillery1987amplitude,hong1985higher,hong1985generation}. In the first approach
introduced by Hillery in 1987 \cite{hillery1987amplitude}, reduction of variance
of an amplitude powered quadrature variable for a quantum state with
respect to its coherent state counterpart reflects nonclassicality.
In contrast, in the second type of higher-order squeezing introduced
by Hong and Mandel in 1985 \cite{hong1985higher,hong1985generation}, 
higher-order squeezing is reflected through the reduction of higher-order
moments of usual quadrature operators with respect to their coherent
state counterparts. In the present chapter, we have studied higher-order
squeezing using Hillery's criterion of amplitude powered squeezing.
Specifically, Hillery introduced amplitude powered quadrature variables
as
\begin{subequations}
\begin{equation}
Y_{1,a}=\frac{a^{k}+\left(a^{\dagger}\right)^{k}}{2}\label{eq:quadrature-power1}\end{equation}
 and \begin{equation}
Y_{2,a}=i\left(\frac{\left(a^{\dagger}\right)^{k}-a^{k}}{2}\right).\label{eq:quadrature-power2}\end{equation}
\end{subequations}
As $Y_{1,a}$ and $Y_{2,a}$ do not commute we can obtain an uncertainty
relation and a condition of the $n$th-order amplitude powered squeezing as \begin{equation}
A_{i,a}=\left(\Delta Y_{i,a}\right)^{2}-\frac{1}{2}\left|\left\langle \left[Y_{1,a},Y_{2,a}\right]\right\rangle \right|<0.\end{equation} For example, for $k=2$, Hillery's
criterion for amplitude squared squeezing is described as \begin{equation}
A_{i,a}=\left(\Delta Y_{i,a}\right)^{2}-\left\langle N_{a}+\frac{1}{2}\right\rangle <0,\label{eq:criterion-amplitude squared}\end{equation}
where $i\in\{1,2\}.$

\subsection{Antibunching, intermodal antibunching, and higher-order antibunching}

Similarly, the existence of single- and multi-mode nonclassical (sub-Poissonian)
photon statistics can be obtained through the following inequalities\begin{equation}
D_{a}=\left(\Delta N_{a}\right)^{2}-\left\langle N_{a}\right\rangle <0,\label{eq:antib-1}\end{equation}
 and \begin{equation}
D_{ab}=(\Delta N_{ab})^{2}=\left\langle a^{\dagger}b^{\dagger}ba\right\rangle -\left\langle a^{\dagger}a\right\rangle \left\langle b^{\dagger}b\right\rangle <0,\label{eq:antib2}\end{equation}
where (\ref{eq:antib-1}) provides us the condition for single-mode
antibunching
\footnote{To be precise, this zero-shift correlation is more connected to sub-Poissonian
behavior. However, it is often referred to as antibunching \cite{miranowicz2010testing}. A relation between them is discussed in Section \ref{Ant}.
} and (\ref{eq:antib2}) provides us the condition for intermodal antibunching.

Signatures of higher-order nonclassical photon statistics 
in different optical systems of interest have been investigated since 1977 using criterion based on higher-order
moments of number operators (cf. Ref. \cite{perina2000review}, Chapter
10 of \cite{perina1991quantum} and references therein). However, 
higher-order antibunching was not specifically discussed, but it was
demonstrated there for degenerate and nondegenerate parametric processes
in single and compound signal-idler modes, respectively, and for Raman
scattering in compound Stokes--anti-Stokes mode up to $n=5$. Further,
it was shown that the value of the parameter that describes 
 higher-order nonclassical photon statistics decreases with increasing $n$ occurring
on a shorter time interval in parametric processes, whereas different
order  higher-order nonclassical photon statistics occur on the same time interval in Raman scattering. A
specific criterion for higher-order antibunching was first introduced by Lee \cite{lee1990higher}
in 1990 using higher-order moments of number operators. Initially,
higher-order antibunching was considered to be a phenomenon that appears rarely in optical
systems, but in 2006, it was established that
it is not really a rare phenomenon \cite{gupta2006higher}. 
Higher-order antibunching has been reported since then in several quantum optical (\cite{verma2010generalized}
and references therein) and atomic systems \cite{giri2014single}.
However, no effort has yet been made to study higher-order antibunching in optical couplers.
Thus, the present study of higher-order antibunching in the asymmetric nonlinear optical coupler
is the first of its kind and is expected to lead to similar observations
in other types of optical couplers. Before we proceed further, we would
like to note that the signature of higher-order antibunching can be observed through a bunch
of equivalent but different criteria, all of which can be interpreted
as modified Lee criterion. In what follows, we will use the following simple
criterion of $n$th-order single-mode antibunching introduced
by Pathak and Garcia \cite{pathak2006control}: \begin{equation}
\begin{array}{lcl}
D_{a}(n)=\left\langle a^{\dagger n}a^{n}\right\rangle -\left\langle a^{\dagger}a\right\rangle ^{n} & < & 0.\end{array}\label{eq:higher-order-antibunching}\end{equation}
For $n=2$, it corresponds to the usual antibunching (\ref{eq:antib-1}), and $n\geq3$ refers
to the higher-order antibunching. Here, it would be
apt to note that the term ``higher-order antibunching'' was coined
by Lee in his pioneering work \cite{lee1990higher} in 1990. In Ref. \cite{lee1990higher},
Lee used $D_{a}(2)<0$
as the condition of antibunching and supported his choice by citing
a 1959 work of Mandel \cite{mandel1959fluctuations}, and Lee stated, {}``The correspondence
between antibunching and sub-Poisson distribution has been established
by Mandel through the so-called Poisson transform. Therefore, we consider
antibunching and the sub-Poissonian distribution as equivalent.'' This is how the term higher-order antibunching originated. The same notion
of higher-order antibunching was used in all the future works on this topic \cite{verma2010generalized,pathak2006control,gupta2006higher,giri2014single,an2002multimode,verma2008higher,perinova2013dynamics},
and in the present work, we have also followed the same convention.

\subsection{Entanglement and higher-order entanglement}

There exist several inseparability criteria (\cite{agarwal2005inseparability}
and references therein) that are expressed in terms of expectation
values of the field operators and thus suitable for the study of entanglement
dynamics within the framework of the present approach. Among these
criteria, the criterion of Duan et al. \cite{duan2000inseparability}, which is usually referred
to as Duan's criterion, and Hillery-Zubairy criterion I and II (HZ-I and
HZ-II) \cite{hillery2006entanglement,hillery2006entanglementapplications,hillery2010conditions} have received more attention for various reasons, such as computational simplicity, experimental
realizability, and their recent success in detecting entanglement in
various optical, atomic, and optomechanical systems (\cite{sen2013intermodal,giri2014single}
and references therein). To begin with, we may note that the first
inseparability criterion of Hillery and Zubairy, i.e., the HZ-I criterion
of inseparability, is described as \begin{equation}
\begin{array}{lcl}
\left\langle N_{a}N_{b}\right\rangle  & -\left|\left\langle ab^{\dagger}\right\rangle \right|^{2}< & 0,\end{array}\label{hz1}\end{equation}
whereas the second criterion of Hillery and Zubairy, i.e., the HZ-II criterion,
is given by \begin{equation}
\begin{array}{lcl}
\left\langle N_{a}\right\rangle \left\langle N_{b}\right\rangle  & -\left|\left\langle ab\right\rangle \right|^{2}< & 0.\end{array}\label{hz2}\end{equation}
 The other criterion of inseparability to be used in the present chapter
is the Duan's criterion which is described as follows \cite{duan2000inseparability}:
\begin{equation}
\begin{array}{lcl}
d_{ab}=\left\langle \left(\Delta u_{ab}\right)^{2}\right\rangle +\left\langle \left(\Delta v_{ab}\right)^{2}\right\rangle -2 & < & 0,\end{array}\label{duan}\end{equation}
 where \begin{equation}
\begin{array}{lcl}
u_{ab} & = & \frac{1}{\sqrt{2}}\left\{ \left(a+a^{\dagger}\right)+\left(b+b^{\dagger}\right)\right\}, \\
v_{ab} & = & -\frac{i}{\sqrt{2}}\left\{ \left(a-a^{\dagger}\right)+\left(b-b^{\dagger}\right)\right\} .\end{array}\label{eq:duan-2}\end{equation}
Clearly, our analytic solution (\ref{eq:ass-sol-c})-(\ref{eq:soln}) enables 
us to investigate intermodal entanglement in the asymmetric nonlinear
optical coupler using all three inseparability criteria described
above and
a set of other criteria listed in \cite{miranowicz2010testing}. It is interesting to note that
 all the inseparability criteria described
above and in the rest of the chapter are special cases
of the Shchukin-Vogel inseparbility criterion \cite{shchukin2005inseparability}.
 Miranowicz et
al., have clearly established this point in Refs. \cite{miranowicz2010testing,miranowicz2009inseparability}. As all these  inseparability criteria that are explicitly described here are only sufficient
(not necessary), a particular criterion may fail to identify entanglement
detected by another criterion. Keeping this fact in mind, we use all
three criteria to study the intermodal entanglement in the asymmetric
nonlinear optical coupler. The criteria described above can only detect
bipartite entanglement of lowest order. As the possibility of generation
of entanglement in both the codirectional and contradirectional asymmetric nonlinear optical couplers has not been discussed
earlier, we have studied the spatial evolution of intermodal entanglement
using these lower-order inseparability criteria. However, to be consistent
with the focus of the present chapter, we need to investigate the possibility
of observing higher-order entanglement, too. For that purpose we require
another set of criteria for the detection of higher-order entanglement.
All criteria for the detection of multi-partite entanglement are essentially
higher-order criteria \cite{zeilinger1992higher,pan2001experimental,mair2001entanglement}
as they reveal some higher-order correlation. Interestingly, there
exist higher-order inseparability criteria for the detection of higher-order 
entanglement in the bipartite case, too. Specifically, Hillery and Zubairy
introduced two criteria for intermodal higher-order entanglement \cite{hillery2006entanglement}
as follows: \begin{equation}
E_{ab}^{m,n}=\left\langle \left(a^{\dagger}\right)^{m}a^{m}\left(b^{\dagger}\right)^{n}b^{n}\right\rangle -\left\vert \left\langle a^{m}\left(b^{\dagger}\right)^{n}\right\rangle \right\vert ^{2}<0,\label{hoe-criteria}\end{equation}
and\begin{equation}
E_{ab}^{\prime m,n}=\left\langle \left(a^{\dagger}\right)^{m}a^{m}\right\rangle\left\langle\left(b^{\dagger}\right)^{n}b^{n}\right\rangle -\left\vert \left\langle a^{m}b^{n}\right\rangle \right\vert ^{2}<0.\label{hoe-criteria-1}\end{equation}
Here, $m$ and $n$ are nonzero positive integers, and the lowest possible
values of $m$ and $n$ are $m=n=1$, which reduces (\ref{hoe-criteria})
and (\ref{hoe-criteria-1}) to usual HZ-I criterion (i.e., (\ref{hz1}))
and HZ-II criterion (i.e., (\ref{hz2})), respectively. Thus, these
two criteria are generalized versions of well-known lower-order criteria
of Hillery and Zubairy, and we may refer to (\ref{hoe-criteria}) and
(\ref{hoe-criteria-1}) as the HZ-I criterion and HZ-II criterion, respectively,
in analogy to the lowest-order cases. A quantum state will be referred
to as a (bipartite) higher-order entangled state if it is found to satisfy
(\ref{hoe-criteria}) and/or (\ref{hoe-criteria-1}) for any choice
of integers $m$ and $n$ satisfying $m+n\geq3.$ The other type of
higher-order entanglement, i.e., multi-partite entanglement can be
detected in various ways. In the present chapter, we have used a set
of multi-mode inseparability criteria introduced by Li et al. \cite{li2007entanglement}.
Specifically, Li et al., have shown that a three-mode quantum state
is not bi-separable in the form $ab_{1}|b_{2}$ (i.e., compound mode
$ab_{1}$ is entangled with mode $b_{2}$) if the following inequality
holds for the three-mode system:
\begin{equation}
E_{ab_{1}|b_{2}}^{m,n,l}=\left\langle\left(a^{\dagger}\right)^{m}a^{m}\left(b_{1}^{\dagger}\right)^{n}b_{1}^{n}\left(b_{2}^{\dagger}\right)^{l}b_{2}^{l}\right\rangle-\left|\left\langle a^{m}b_{1}^{n}(b_{2}^{\dagger})^{l}\right\rangle\right|^{2}<0,\label{eq:tripartite ent1}\end{equation}
where $m,\, n,$ and $l$ are positive integers, and annihilation operators
$a,b_{1},$ and $b_{2}$ correspond to the three modes. A quantum state satisfying
the above inequality is referred to as the $ab_{1}|b_{2}$ entangled state.
The three-mode inseparability criterion can be written in various alternative
forms. For example, an alternative criterion for the detection of the $ab_{1}|b_{2}$
entangled state is \cite{li2007entanglement} \begin{equation}
E_{ab_{1}|b_{2}}^{^{\prime}m,n,l}=\left\langle\left(a^{\dagger}\right)^{m}a^{m}\left(b_{1}^{\dagger}\right)^{n}b_{1}^{n}\rangle\langle\left(b_{2}^{\dagger}\right)^{l}b_{2}^{l}\right\rangle-\left|\left\langle a^{m}b_{1}^{n}b_{2}^{l}\right\rangle\right|^{2}<0.\label{eq:tripartite ent2}\end{equation}
Similarly, one can define criteria for the detection of $a|b_{1}b_{2}$
and $b_{1}|ab_{2}$ entangled states and use them to obtain the criterion
for detection of fully entangled tripartite states. For example, using
(\ref{eq:tripartite ent1}) and (\ref{eq:tripartite ent2}), respectively,
we can write that the three modes of our interest are not bi-separable
in any form if any one of the following two sets of inequalities is
satisfied simultaneously: \begin{equation}
E_{ab_{1}|b_{2}}^{1,1,1}<0,\, E_{a|b_{1}b_{2}}^{1,1,1}<0,\, E_{b_{1}|b_{2}a}^{1,1,1}<0,\label{eq:fully enta 0}\end{equation}
and/or
\begin{equation}
E_{ab_{1}|b_{2}}^{^{\prime}1,1,1}<0,\, E_{a|b_{1}b_{2}}^{^{\prime}1,1,1}<0,\, E_{b_{1}|b_{2}a}^{^{\prime}1,1,1}<0.\label{eq:fully enta 1}\end{equation}
 Further, for a fully separable pure state, we always have \begin{equation}
|\langle ab_{1}b_{2}\rangle|=|\langle a\rangle\langle b_{1}\rangle\langle b_{2}\rangle|\leq\left[\langle N_{a}\rangle\langle N_{b_{1}}\rangle\langle N_{b_{2}}\rangle\right]^{\frac{1}{2}}.\label{eq:fully ent2}\end{equation}
Thus, a three-mode pure state that violates (\ref{eq:fully ent2}) (i.e.,
satisfies $\langle N_{a}\rangle\langle N_{b_{1}}\rangle\langle N_{b_{2}}\rangle-|\langle ab_{1}b_{2}\rangle|^{2}<0)$
and simultaneously satisfies either (\ref{eq:fully enta 0}) or (\ref{eq:fully enta 1})
is a fully entangled state as it is neither fully separable nor bi-separable
in any form.

\section{Nonclassicality in the contradirectional optical coupler\label{sec:Nonclassicality-in-contradirectional}}

The spatial evolution of different operators that are relevant for witnessing
nonclassicality in contradirectional optical coupler can be obtained using the perturbative solution (\ref{eq:ass-sol-c})-(\ref{eq:soln})
reported here. For example, using (\ref{eq:ass-sol-c})-(\ref{eq:soln}), we can
obtain the following closed form expressions for the number operators
of various field modes \begin{equation}
\begin{array}{lcl}
N_{a} & = & a^{\dagger}a=|f_{1}|^{2}a^{\dagger}(L)a(L)+|f_{2}|^{2}b_{1}^{\dagger}(0)b_{1}(0)+\left[f_{1}^{*}f_{2}a^{\dagger}(L)b_{1}(0)+f_{1}^{*}f_{3}a^{\dagger}(L)b_{1}^{\dagger}(0)b_{2}(0)\right.\\
 & + & \left.f_{1}^{*}f_{4}a^{\dagger2}(L)b_{2}(0)+f_{2}^{*}f_{3}b_{1}^{\dagger2}(0)b_{2}(0)+f_{2}^{*}f_{4}b_{1}^{\dagger}(0)a^{\dagger}(L)b_{2}(0)+{\rm H.c.}\right],\end{array}\label{eq:na}\end{equation}
 \begin{equation}
\begin{array}{lcl}
N_{b_{1}} & = & b_{1}^{\dagger}b_{1}=|g_{1}|^{2}a^{\dagger}(L)a(L)+|g_{2}|^{2}b_{1}^{\dagger}(0)b_{1}(0)+\left[g_{1}^{*}g_{2}a^{\dagger}(L)b_{1}(0)+g_{1}^{*}g_{3}a^{\dagger}(L)b_{1}^{\dagger}(0)b_{2}(0)\right.\\
 & + & \left.g_{1}^{*}g_{4}a^{\dagger2}(L)b_{2}(0)+g_{2}^{*}g_{3}b_{1}^{\dagger2}(0)b_{2}(0)+g_{2}^{*}g_{4}b_{1}^{\dagger}(0)a^{\dagger}(L)b_{2}(0)+{\rm H.c.}\right],\end{array}\label{eq:nb}\end{equation}
 \begin{equation}
N_{b_{2}}=b_{2}^{\dagger}b_{2}=b_{2}^{\dagger}(0)b_{2}(0)+\left[h_{2}b_{2}^{\dagger}(0)b_{1}^{2}(0)+h_{3}b_{2}^{\dagger}(0)b_{1}(0)a(L)+h_{4}b_{2}^{\dagger}(0)a^{2}(L)+{\rm H.c.}\right].\label{eq:nb2}\end{equation}
It is now straightforward to compute the average values of the number
of photons in different modes with respect to a given initial state.
In the present work, we consider that the initial state is a product
of three coherent states: $|\alpha\rangle|\beta\rangle|\gamma\rangle,$
where $|\alpha\rangle,\,|\beta\rangle$, and $|\gamma\rangle$ are the
eigen kets of the annihilation operators $a,\, b_{1}$, and $b_{2}$, respectively.
Thus, \begin{equation}
a(L)|\alpha\rangle|\beta\rangle|\gamma\rangle=\alpha|\alpha\rangle|\beta\rangle|\gamma\rangle,\label{2bcd}\end{equation}
and $|\alpha|^{2},\,|\beta|^{2},$ and $|\gamma|^{2}$ are the number of
input photons in the field modes $a,\, b_{1}$, and $b_{2}$, respectively.
For a spontaneous process, $\beta=\gamma=0$ and $\alpha\ne0,$ whereas
for a stimulated process, the complex amplitudes are not necessarily
zero, and it seems reasonable to consider $\left|\alpha\right|>\left|\beta\right|>\left|\gamma\right|.$ 

\subsection{Squeezing, intermodal squeezing, and higher-order squeezing}

\begin{figure}[t]
\begin{centering}
\includegraphics[angle=-90,scale=0.5]{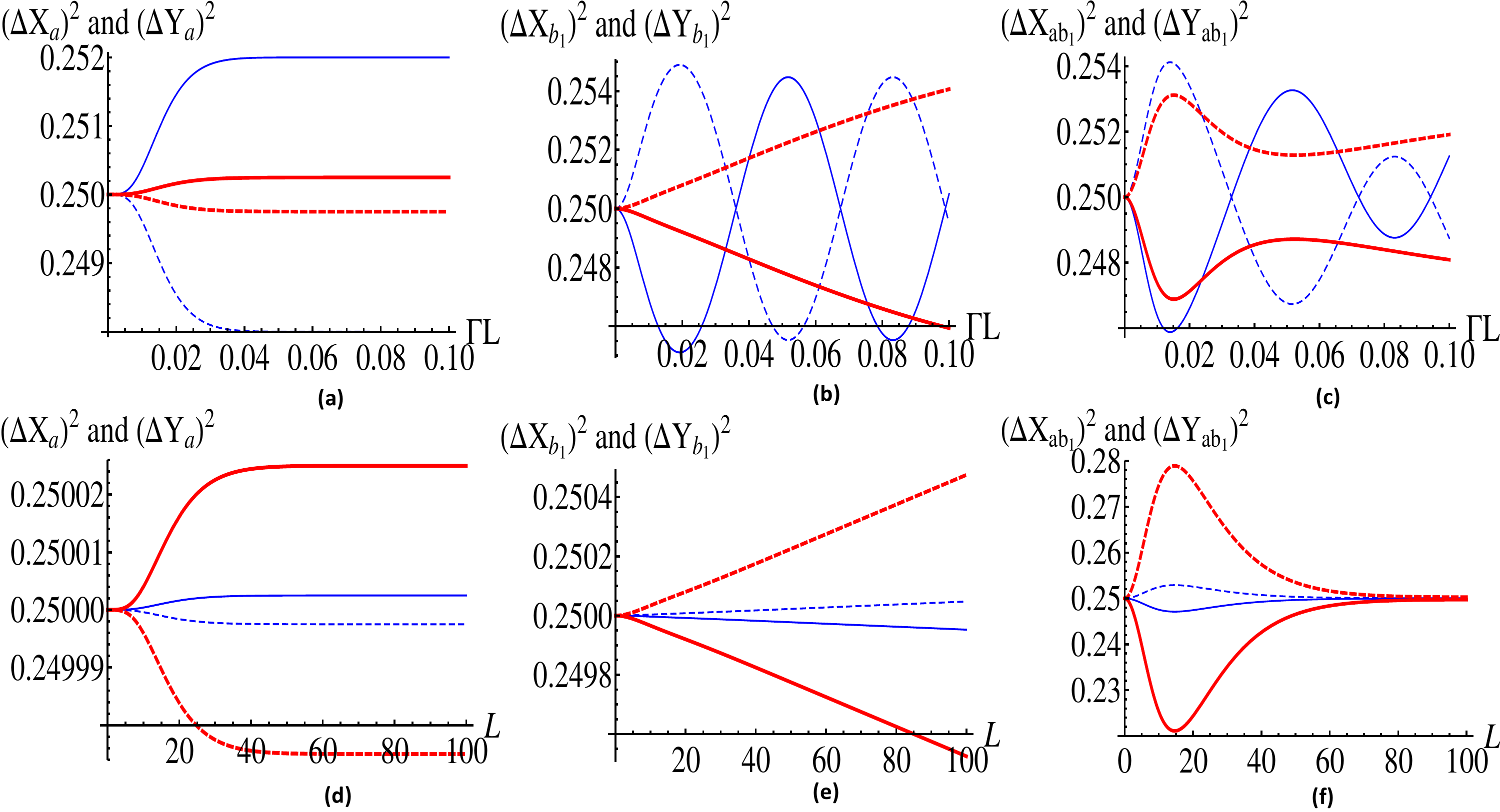}
\par\end{centering}

\caption[Quadrature
squeezing and intermodal squeezing in the contradirectional nonlinear optical coupler]{\label{fig:Squeezing-del-k-G}  Existence of quadrature
squeezing in modes $a$ and $b_{1}$ for the contradirectional setting, and intermodal squeezing in mode
$ab_{1}$ is illustrated with $k=0.1,\,\alpha=5,\,\beta=2,\,\gamma=1$
for different values of phase mismatching $\Delta k$ and nonlinear
coupling constant $\Gamma$. In (a)-(c), squeezing and intermodal
squeezing is plotted with rescaled interaction length $\Gamma L$
with $\Gamma=0.001$ for $\Delta k=10^{-1}$ (thin blue lines) and
$\Delta k=10^{-2}$ (thick red lines). In (d)-(f), squeezing and
intermodal squeezing is plotted with interaction length $L$ with
$\Delta k=10^{-4}$ for $\Gamma=0.001$ (thin blue lines) and $\Gamma=0.01$
(thick red lines). In all the sub-figures, a solid (dashed) line represents
$X_{i}$ $(Y_{i})$, where $i\in\left\{ a,b_{1}\right\} $, or $X_{ab_{1}}$
$(Y_{ab_{1}})$ quadrature. The parts of the plots that depict values
of variance <$\frac{1}{4}$ in (a) and (d) show squeezing in quadrature
variable $Y_{a}$, that in (b) and (c) show squeezing in quadrature
variable $X_{b_{1}}$, $Y_{b_{1}}$ and intermodal squeezing in quadrature
variable $X_{ab_{1}}$, $Y_{ab_{1}},$ respectively. Similarly, (e)
and (f) show squeezing and intermodal squeezing in quadrature variables
$X_{b_{1}}$ and $X_{ab_{1}}$, respectively. Squeezing in the other
quadrature variables (say $X_{a}$) can be obtained by a suitable choice
of phase of the input coherent states.}
\end{figure}

Using (\ref{eq:ass-sol-c})-(\ref{eq:soln}) with (\ref{eq:condition-for-squeezing}) and (\ref{eq:two-mode-quadrature}),
we obtain analytic expressions for variance in the single-mode and compound
mode quadratures as

\begin{equation}
\begin{array}{lcl}
\left[\begin{array}{c}
\left(\Delta X_{a}\right)^{2}\\
\left(\Delta Y_{a}\right)^{2}\end{array}\right] & = & \frac{1}{4}\left[1\pm\left\{ \left(f_{1}f_{4}+f_{2}f_{3}\right)\gamma+{\rm c.c}.\right\} \right],\\
\left[\begin{array}{c}
\left(\Delta X_{b_{1}}\right)^{2}\\
\left(\Delta Y_{b_{1}}\right)^{2}\end{array}\right] & = & \frac{1}{4}\left[1\pm\left\{ \left(g_{1}g_{4}+g_{2}g_{3}\right)\gamma+{\rm c.c}.\right\} \right],\\
\left[\begin{array}{c}
\left(\Delta X_{b_{2}}\right)^{2}\\
\left(\Delta Y_{b_{2}}\right)^{2}\end{array}\right] & = & \frac{1}{4},\end{array}\label{eq:squeezing}\end{equation}
 and 
\begin{equation}
\begin{array}{lcl}
\left[\begin{array}{c}
\left(\Delta X_{ab_{1}}\right)^{2}\\
\left(\Delta Y_{ab_{1}}\right)^{2}\end{array}\right] & = & \frac{1}{4}\left[1\pm\frac{1}{2}\left\{ \left(\left(f_{1}+g_{1}\right)\left(f_{4}+g_{4}\right)+\left(f_{2}+g_{2}\right)\left(f_{3}+g_{3}\right)\right)\gamma+{\rm c.c}.\right\} \right],\\
\left[\begin{array}{c}
\left(\Delta X_{ab_{2}}\right)^{2}\\
\left(\Delta Y_{ab_{2}}\right)^{2}\end{array}\right] & = & \frac{1}{4}\left[1\pm\frac{1}{2}\left\{ \left(f_{1}f_{4}+f_{2}f_{3}\right)\gamma+{\rm c.c}.\right\} \right]=\frac{1}{2}\left[\begin{array}{c}
\left(\Delta X_{a}\right)^{2}\\
\left(\Delta Y_{a}\right)^{2}\end{array}\right]+\frac{1}{8},\end{array}\nonumber\end{equation}
\begin{equation}
\begin{array}{lcl}
\left[\begin{array}{c}
\left(\Delta X_{b_{1}b_{2}}\right)^{2}\\
\left(\Delta Y_{b_{1}b_{2}}\right)^{2}\end{array}\right] & = & \frac{1}{4}\left[1\pm\frac{1}{2}\left\{ \left(g_{1}g_{4}+g_{2}g_{3}\right)\gamma+{\rm c.c}.\right\} \right]=\frac{1}{2}\left[\begin{array}{c}
\left(\Delta X_{b_{1}}\right)^{2}\\
\left(\Delta Y_{b_{1}}\right)^{2}\end{array}\right]+\frac{1}{8},\end{array}\label{eq:intermodal squeezing}\end{equation}
respectively. From Eq. (\ref{eq:squeezing}), it is clear that no squeezing
is observed in $b_{2}$ mode. However, squeezing is possible in $a$
and $b_{1}$ modes as illustrated in Figures \ref{fig:Squeezing-del-k-G}
(a)-(b) and (d)-(e). Further, we observed intermodal squeezing in quadratures
$X_{ab_{1}}$ and $Y_{ab_{1}}$ by plotting the right-hand sides of Eq.
(\ref{eq:intermodal squeezing}) in Figures \ref{fig:Squeezing-del-k-G}
(c) and (f). Variation of the amount of
squeezing in different modes with the phase mismatch $\Delta k$ and the nonlinear
coupling constant $\Gamma$ is shown in Figures \ref{fig:Squeezing-del-k-G}
(a)-(c) and Figures \ref{fig:Squeezing-del-k-G} (d)-(f), respectively. We have
also studied the effect of the linear coupling constant $k$ on the amount
of squeezing, but its effect is negligible in all other quadratures
except the quadratures of compound mode $ab_{1}.$ In the compound mode
$ab_{1}$, after a short distance the depth of witness of squeezing is observed to
increase with the decrease in the linear coupling constant $k$ (this
is not illustrated through figure). Intermodal squeezing in the compound
mode quadratures $Y_{ab_{2}}$ and $X_{b_{1}b_{2}}$ can be visualized
from the last two rows of Eq. (\ref{eq:intermodal squeezing}). Specifically,
 we can see that the variance in the compound mode quadrature $X_{jb_{2}}$
and $Y_{jb_{2}}$ have \emph{bijective }(both one-to-one and onto)
correspondence with the variance in $X_{j}$ and $Y_{j}$, respectively,
where $j\in\left\{ a,b_{1}\right\} .$ To be precise, quadrature squeezing
in the single-mode $X_{j}$ ($Y_{j}$) implies quadrature squeezing in
$X_{j,b_{2}}\,(Y_{j,b_{2}})$ and vice versa. For example, $\left(\Delta X_{jb_{2}}\right)^{2}<\frac{1}{4}\Rightarrow\frac{1}{2}\left(\Delta X_{j}\right)^{2}+\frac{1}{8}<\frac{1}{4}$
or $\left(\Delta X_{j}\right)^{2}<2\left(\frac{1}{4}-\frac{1}{8}\right)=\frac{1}{4}.$
This is why we have not explicitly shown the variance of the compound
mode quadratures $X_{j,b_{2}}$ and $Y_{j,b_{2}}$ with different
parameters as we have done for the other cases. As we have shown squeezing
in quadratures $Y_{a},\, X_{b_{1}},$ and $Y_{b_{1}}$ through Figures \ref{fig:Squeezing-del-k-G}
(a)-(b), and (d)-(e), this implies the existence of squeezing in quadratures
$Y_{ab_{2}},\, X_{b_{1}b_{2}},$ and $Y_{b_{1}b_{2}}.$ Thus, we have
observed intermodal squeezing in the compound modes involving $b_{2}$.
This nonclassical feature was not observed in the earlier studies \cite{perina2000review}
as in those studies $b_{2}$ mode was considered classical. The plots
do not show quadrature squeezing in $X_{a},$ $X_{ab_{2}}$, and $X_{ab_{1}}$
(for some specific values of $\Gamma$). However, a suitable choice
of phase of the input coherent state would lead to squeezing in these
quadratures. For example, if we replace $\gamma$ by $-\gamma$ (i.e.,
if we choose $\gamma=\exp(i\pi)$ instead of the present choice of $\gamma=1$)
then we would observe squeezing in all these quadratures, but the
squeezing that is observed now with the original choice of $\gamma$
would vanish. This is so as all the expressions of variance of the quadrature
variables that are $\neq\frac{1}{4}$ have a common functional form:
$\frac{1}{4}\pm\gamma F(f_{i},g_{i})$ (cf. Eqs. (\ref{eq:squeezing}) 
and (\ref{eq:intermodal squeezing})). 

\begin{figure}
\begin{centering}
\includegraphics[angle=-180,scale=0.5]{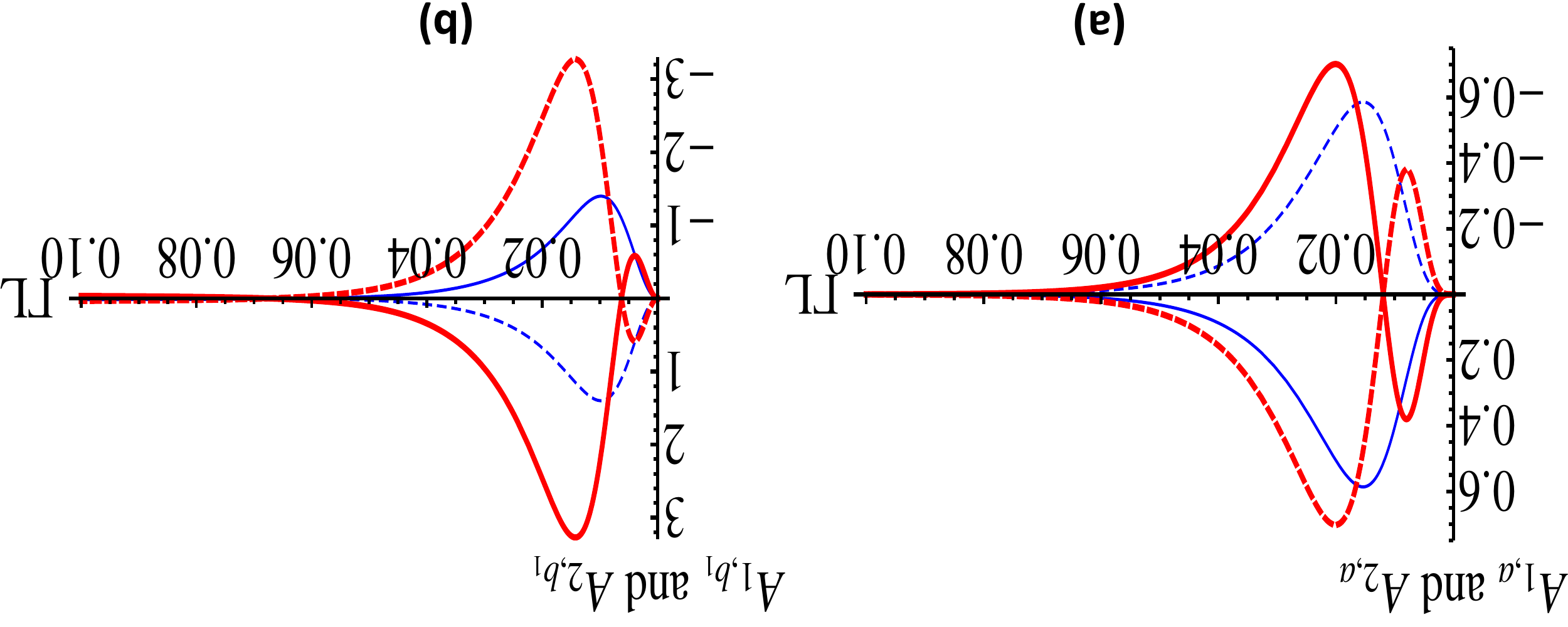}
\par\end{centering}

\centering{}\caption[Amplitude
powered squeezing in the contradirectional nonlinear optical coupler]{\label{fig:Amplitude-squared-squeezing-c}  Amplitude
powered squeezing is observed in (a) counter propagating $a$ mode and (b) $b_{1}$ mode
for $k=0.1,\,\Gamma=0.001,\,\Delta k=10^{-4},\,\alpha=3,\,\beta=2,\,\gamma=1.$
The negative parts of the solid line represent amplitude powered squeezing
in quadrature variable $Y_{1,a}$ ($Y_{1,b_{1}})$ and that in the
dashed line represents squeezing in quadrature variable $Y_{2,a}$
($Y_{2,b_{1}})$ for $n=2$ (thin blue lines) and $n=3$ (thick red
lines). To display the plots in the same scale $A_{i,a}$ and $A_{i,b_{1}}$
for $n=2$ are multiplied by 10, where $i\in\left\{ 1,2\right\} $.}
\end{figure}

After establishing the existence of squeezing in the single and
compound modes, we now examine the possibility of higher-order squeezing
using Eqs. (\ref{eq:ass-sol-c})-(\ref{eq:soln}), (\ref{eq:na})-(\ref{eq:nb2}),
and (\ref{eq:criterion-amplitude squared}). In fact, we obtain \begin{equation}
\begin{array}{lcl}
\left[\begin{array}{c}
A_{1,a}\\
A_{2,a}\end{array}\right] & = & \pm\frac{n^{2}}{4}\left[\gamma\left(f_{1}f_{4}+f_{2}f_{3}\right)\left(f_{1}\alpha+f_{2}\beta\right)^{2n-2}+{\rm c.c.}\right],\end{array}\label{eq:asq-1}\end{equation}
\begin{equation}
\begin{array}{lcl}
\left[\begin{array}{c}
A_{1,b_{1}}\\
A_{2,b_{1}}\end{array}\right] & = & \pm\frac{n^{2}}{4}\left[\gamma\left(g_{1}g_{4}+g_{2}g_{3}\right)\left(g_{1}\alpha+g_{2}\beta\right)^{2n-2}+{\rm c.c.}\right],\end{array}\label{eq:as-q-2}\end{equation}
 and \begin{equation}
\begin{array}{lcl}
\left[\begin{array}{c}
A_{1,b_{2}}\\
A_{2,b_{2}}\end{array}\right] & = & 0.\end{array}\label{eq:as-q-3}\end{equation}
Thus, we do not get any signature of amplitude powered squeezing in
$b_{2}$ mode using the present solution. In contrast, mode $a$ $(b_{1})$
is found to show amplitude powered squeezing in one of the quadrature
variables for any value of interaction length as both $A_{1,a}$ and
$A_{2,a}$ ($A_{1,b_{1}}$ and $A_{2,b_{1}}$) cannot be positive
simultaneously. To study the possibilities of amplitude powered squeezing
in further detail, we have plotted the spatial variation of $A_{i,a}$
and $A_{i,b_{1}}$ in Figure \ref{fig:Amplitude-squared-squeezing-c}.
The negative regions of these two plots clearly illustrate the existence
of amplitude powered squeezing in both $a$ and $b_{1}$ modes for
$n=2$ and $n=3$. Extending our observations in context of the single-mode
and intermodal squeezing, we can state that the appearance
of the amplitude powered squeezing in a particular quadrature can be controlled
by a suitable choice of phase of the input coherent state $\gamma$ as the
expressions for the amplitude powered squeezing reported in Eqs. (\ref{eq:asq-1})
and (\ref{eq:as-q-2}) have a common functional form $\pm\gamma F(f_{i},g_{i})$. 

\subsection{Lower- and higher-order antibunching}

\begin{figure}
\centering{}\includegraphics[angle=-90, scale=0.5]{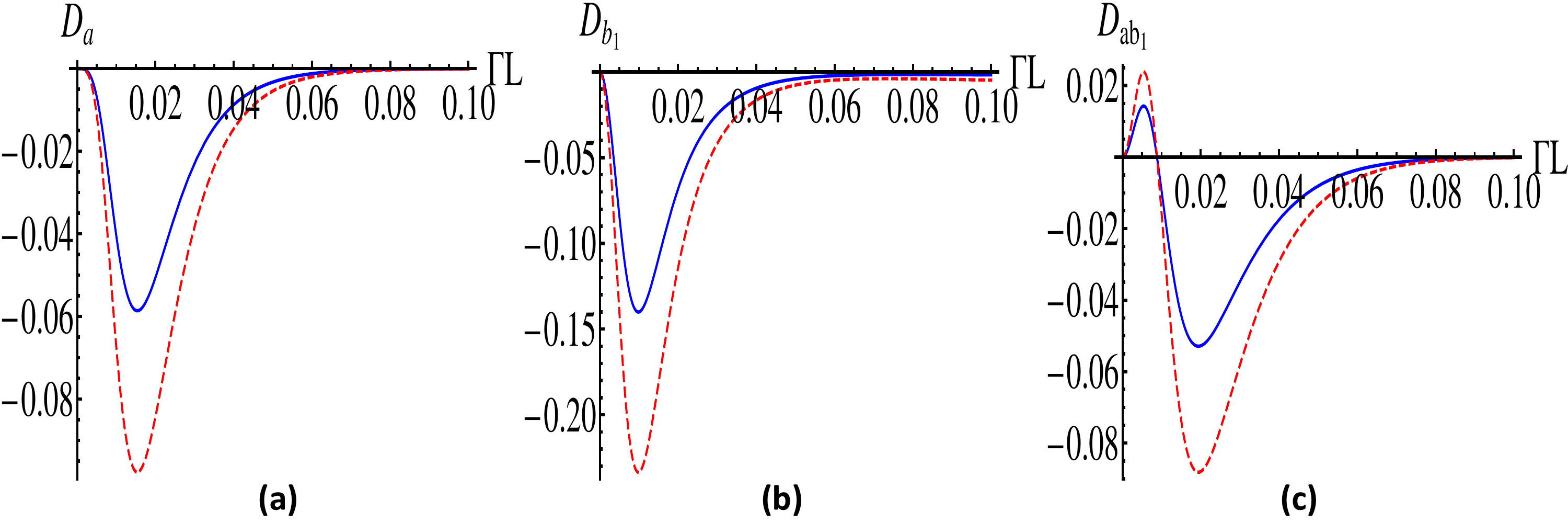}\caption[Antibunching and intermodal antibunching in the contradirectional nonlinear optical coupler]{\label{fig:Antibunching}   Variation of $D_{i}$ and
$D_{ij}$ with rescaled interaction length $\Gamma L$
for $\alpha=3$ (smooth line) and $\alpha=5$ (dashed line) in (a)
single-mode $a$, (b) single-mode $b_{1}$, and (c) compound mode $ab_{1}$ in case of the contradirectional optical coupler
with $k=0.1,\,\Gamma=0.001,\,\Delta k=10^{-4},\,\beta=2,$ and $\gamma=1$
for (a) and $\gamma=-1$ for (b)-(c). The negative parts of the plots illustrate the existence of nonclassical photon statistics (antibunching).}
\end{figure}

The condition of higher-order antibunching is already provided through the inequality (\ref{eq:higher-order-antibunching}).
Using this inequality along with Eqs. (\ref{eq:ass-sol-c})-(\ref{eq:soln})
and (\ref{eq:na})-(\ref{eq:nb2}), we can obtain closed form analytic
expressions for $D_{i}(n)$ for various modes as follows \begin{equation}
\begin{array}{lcl}
D_{a}(n) & = & ^{n}C_{2}\gamma|\left(f_{1}\alpha+f_{2}\beta\right)|^{2n-4}\left\{ \left(f_{1}\alpha+f_{2}\beta\right)^{2}\left(f_{2}^{*}f_{3}^{*}+f_{1}^{*}f_{4}^{*}\right)+{\rm c.c.}\right\} ,\end{array}\label{eq:dan}\end{equation}
\begin{equation}
\begin{array}{lcl}
D_{b_{1}}(n) & = & ^{n}C_{2}\gamma|\left(g_{1}\alpha+g_{2}\beta\right)|^{2n-4}\left\{ \left(g_{1}\alpha+g_{2}\beta\right)^{2}\left(g_{2}^{*}g_{3}^{*}+g_{1}^{*}g_{4}^{*}\right)+{\rm c.c.}\right\} ,\end{array}\label{eq:dbn}\end{equation}
and
\begin{equation}
D_{b_{2}}(n)=0.\label{eq:db2n}\end{equation}
Further, using the condition of intermodal antibunching described
in (\ref{eq:antib2}) and Eqs. (\ref{eq:ass-sol-c})-(\ref{eq:soln}), we obtain the
following closed form expressions of $D_{ij}$ \begin{equation}
\begin{array}{lcl}
D_{ab_{1}} & = & \left\{ \left(|g_{1}|^{2}f_{1}^{*}f_{4}+f_{1}^{*}f_{3}g_{1}^{*}g_{2}\right)\alpha^{*2}\gamma+\left(|g_{2}|^{2}f_{2}^{*}f_{3}+f_{2}^{*}f_{4}g_{2}^{*}g_{1}\right)\beta^{*2}\gamma\right.\\
&+&\left.\left(|g_{1}|^{2}-|g_{2}|^{2}\right)\left(f_{2}^{*}f_{4}-f_{1}^{*}f_{3}\right)\alpha\beta\gamma^{*}+{\rm c.c.}\right\} ,\end{array}\label{eq:dab1}\end{equation}
\begin{equation}
\begin{array}{lcl}
D_{ab_{2}} & = & 0,\end{array}\label{eq:dab2}\end{equation}
and
\begin{equation}
D_{b_{1}b_{2}}=0.\label{eq:db1b2}\end{equation}
From the above expressions, it is clear that neither the single-mode
antibunching nor the intermodal antibunching is obtained involving
$b_{2}$ mode. As the expressions obtained in the right-hand sides
of Eqs. (\ref{eq:dan}), (\ref{eq:dbn}), and (\ref{eq:dab1}) are not simple,
we plot them to investigate the existence of single-mode and compound
mode antibunching. The plots for usual antibunching (obtained as $n=2$ in Eqs. (\ref{eq:dan}) and (\ref{eq:dbn})) and intermodal
antibunching are shown in Figure \ref{fig:Antibunching}. Existence
of antibunching is obtained in single-mode $a$ for $\gamma=1$, and
the same is illustrated through Figure \ref{fig:Antibunching} (a). However,
in an effort to obtain antibunching in $b_{1}$ mode, we do not observe
any antibunching in $b_{1}$ mode for $\gamma=1$. Interestingly,
from (\ref{eq:dbn}) it is clear that if we replace $\gamma=1$ by
$\gamma=\exp(i\pi)=-1$ as before, and keep $\alpha$ and $\beta$ unchanged,
then we would observe antibunching for all values of the rescaled interaction
length $\Gamma L.$ This is true in general for all values of $\gamma$.
To be precise, if we observe antibunching (bunching) in a mode for
$\gamma=c$ we will always observe bunching (antibunching) for $\gamma=-c,$
if we keep $\alpha$ and $\beta$ unchanged. This fact is illustrated through
Figure \ref{fig:Antibunching} (b), where we plot the variation of $D_{b_{1}}$
with $\Gamma L$ and have observed the existence of antibunching.
In compound mode $ab_{1},$ we can observe existence of antibunching
for both $\gamma=1$ and $\gamma=-1$. However, in Figure \ref{fig:Antibunching}
(c) we have illustrated the existence of intermodal antibunching in
compound mode $ab_{1}$ by plotting the variation of $D_{ab_{1}}$ with rescaled
interaction length $\Gamma L$ for $\gamma=-1$ as the region of nonclassicality
is relatively larger (compared to the case where $\gamma=1$, and $\alpha$ and $\beta$
are same) in this case.

\begin{figure}
\begin{centering}
\includegraphics[angle=-90,scale=0.5]{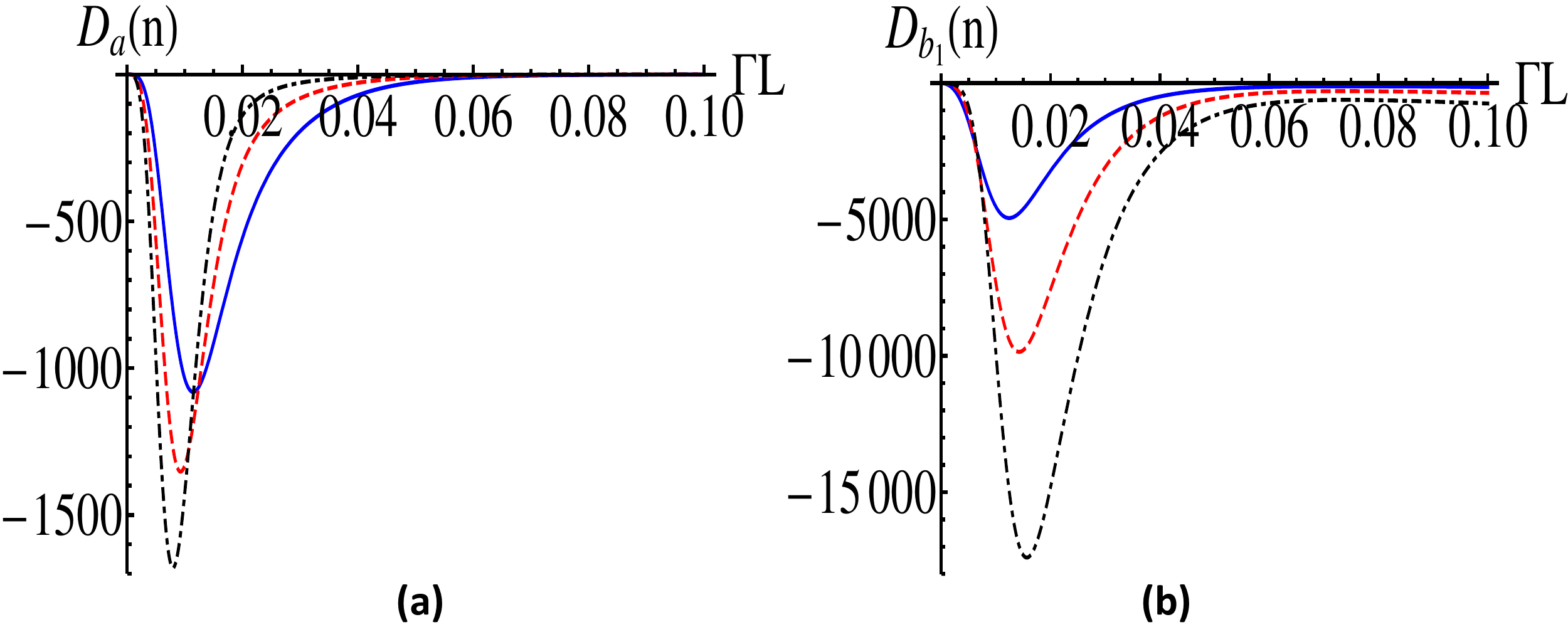}
\par\end{centering}

\centering{}\caption[Higher-order antibunching in the contradirectional nonlinear optical coupler]{\label{fig:higher-order antibunching-c}   Variation of $D_{i}(n)$ with rescaled
interaction length $\Gamma L$ for $k=0.1,\,\Gamma=0.001,\,\Delta k=10^{-4},\,\alpha=5,\,\beta=2$ in (a) contradirectional mode $a$ with $\gamma=1$
and (b) mode $b_{1}$ with $\gamma=-1$
for $n=3$ (smooth lines), $n=4$ (dashed lines), and $n=5$ (dot-dashed
lines). To display the plots in the same scale $D_{i}(3)$ and $D_{i}(4)$
are multiplied by 400 and 20, respectively, where $i\in\left\{ a,b_{1}\right\} $. The negative parts of the plots show higher-order antibunching.}
\end{figure}

We may now extend the discussion to higher-order antibunching and plot the right-hand sides
of (\ref{eq:dan}) and (\ref{eq:dbn}) for various values of $n$.
The plots are shown in Figure \ref{fig:higher-order antibunching-c}, which clearly illustrate
the existence of higher-order antibunching and also demonstrate that the depth of the witness of nonclassicality
increases with $n.$ This is consistent with the earlier observations
on higher-order antibunching in other systems \cite{pathak2006control}.

\subsection{Lower- and higher-order intermodal entanglement}

\begin{figure}
\begin{centering}
\includegraphics[scale=0.5]{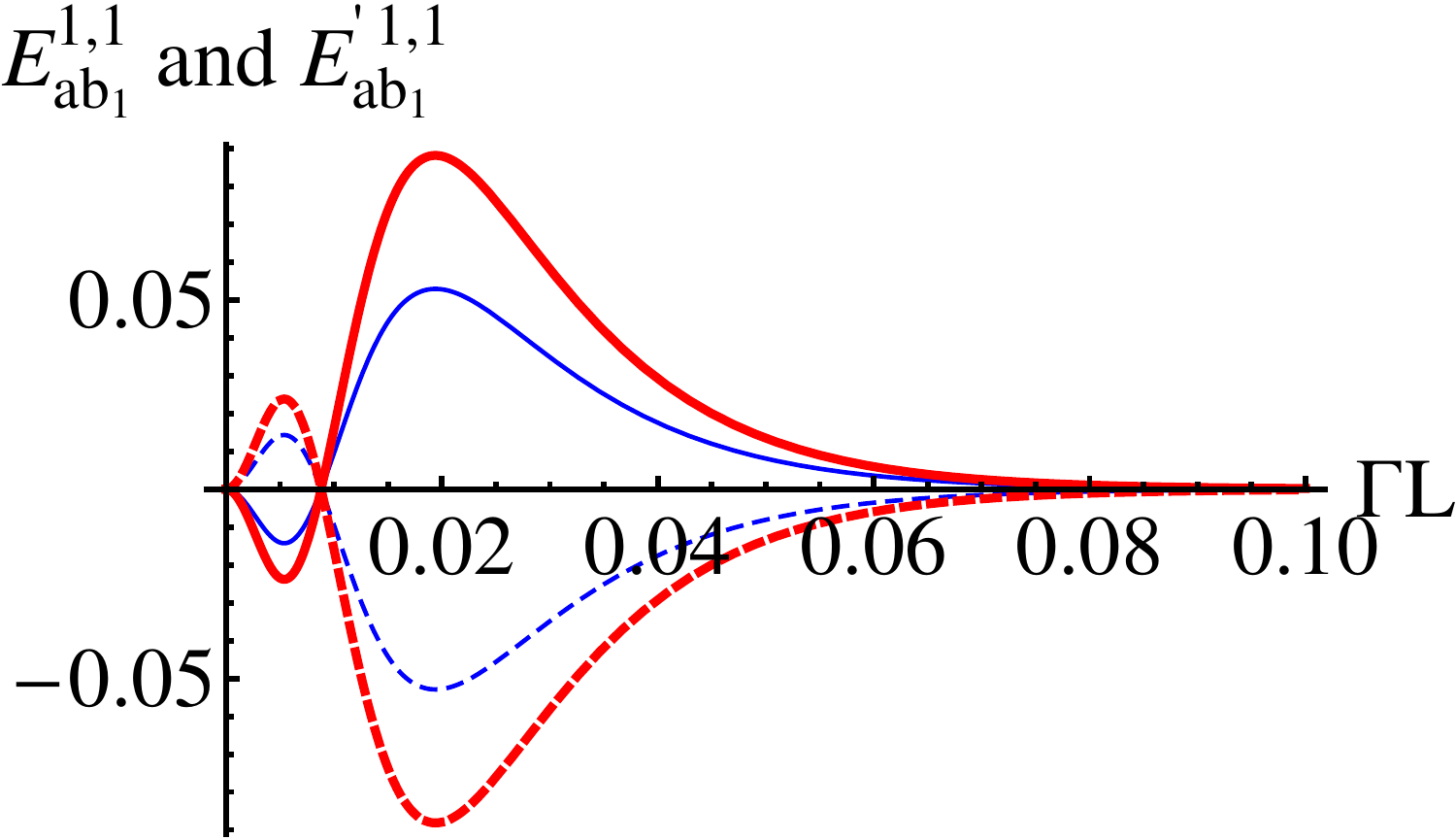}
\par\end{centering}

\caption[Lower-order entanglement for the contradirectional nonlinear optical coupler]{\label{fig:ent-hz1-II-c}  Hillery-Zubairy criterion
I (solid line) and criterion II (dashed line) for entanglement are
showing the presence of intermodal entanglement between modes $a$ and $b_{1}$ in case of the contradirectional propagation of $a$ mode. Here
$E_{ab_{1}}^{1,1}$ (solid line) and $E_{ab_{1}}^{\prime1,1}$ (dashed
line) are plotted with rescaled interaction length $\Gamma L$
for mode $ab_{1}$ with $k=0.1,\,\Gamma=0.001,\,\Delta k=10^{-4},\,\beta=2,\gamma=1$
for $\alpha=3$ (thin blue lines) and $\alpha=5$ (thick red lines).}
\end{figure}

We first examine the existence of intermodal entanglement in the compound
mode $ab_{1}$ using HZ-I criterion (\ref{hz1}) with
Eqs. (\ref{eq:ass-sol-c})-(\ref{eq:soln}) and (\ref{eq:na})-(\ref{eq:nb2}),
and obtain\begin{equation}
\begin{array}{lcl}
E_{ab_{1}}^{1,1} & = & \langle N_{a}N_{b_{1}}\rangle-|\langle ab_{1}^{\dagger}\rangle|^{2}\\
 & = & \left(|g_{1}|^{2}f_{4}^{*}f_{1}+f_{3}^{*}f_{1}g_{2}^{*}g_{1}\right)\alpha^{2}\gamma^{*}+\left(|f_{1}|^{2}g_{1}^{*}g_{4}+f_{1}^{*}f_{2}g_{1}^{*}g_{3}\right)\alpha^{*2}\gamma\\
 & + & \left(|g_{2}|^{2}f_{3}^{*}f_{2}+f_{4}^{*}f_{2}g_{1}^{*}g_{2}\right)\beta^{2}\gamma^{*}+\left(|f_{2}|^{2}g_{2}^{*}g_{3}+f_{2}^{*}f_{1}g_{2}^{*}g_{4}\right)\beta^{*2}\gamma\\
 & + & \left(|g_{1}|^{2}-|g_{2}|^{2}\right)\left\{ \left(f_{4}^{*}f_{2}-f_{3}^{*}f_{1}\right)\alpha\beta\gamma^{*}-\left(g_{2}^{*}g_{4}-g_{1}^{*}g_{3}\right)\alpha^{*}\beta^{*}\gamma\right\} .\end{array}\label{eq:HZ-1-ab1}\end{equation}
Similarly, using HZ-II criterion (\ref{hz2}) we obtain

\begin{equation}
\begin{array}{lcl}
E_{ab_{1}}^{\prime1,1} & = & \langle N_{a}\rangle\langle N_{b_{1}}\rangle-|\langle ab_{1}\rangle|^{2}\\
 & = & -\left[\left(|g_{1}|^{2}f_{4}^{*}f_{1}+f_{3}^{*}f_{1}g_{2}^{*}g_{1}\right)\alpha^{2}\gamma^{*}+\left(|f_{1}|^{2}g_{1}^{*}g_{4}+f_{1}^{*}f_{2}g_{1}^{*}g_{3}\right)\alpha^{*2}\gamma\right.\\
 & + & \left.\left(|g_{2}|^{2}f_{3}^{*}f_{2}+f_{4}^{*}f_{2}g_{1}^{*}g_{2}\right)\beta^{2}\gamma^{*}+\left(|f_{2}|^{2}g_{2}^{*}g_{3}+f_{2}^{*}f_{1}g_{2}^{*}g_{4}\right)\beta^{*2}\gamma\right.\\
 & + & \left.\left(|g_{1}|^{2}-|g_{2}|^{2}\right)\left\{ \left(f_{4}^{*}f_{2}-f_{3}^{*}f_{1}\right)\alpha\beta\gamma^{*}-\left(g_{2}^{*}g_{4}-g_{1}^{*}g_{3}\right)\alpha^{*}\beta^{*}\gamma\right\} \right].\end{array}\label{eq:hz2-ab1}\end{equation}
It is easy to observe that Eqs. (\ref{eq:HZ-1-ab1}) and (\ref{eq:hz2-ab1})
provide us the following simple relation that is valid for the present
case: $E_{ab_{1}}^{1,1}=-E_{ab_{1}}^{\prime1,1}$, which implies that
for any particular choice of the rescaled interaction length $\Gamma L$
either HZ-I criterion or HZ-II criterion would show the existence
of entanglement in the contradirectional asymmetric nonlinear optical
coupler as both of them cannot be simultaneously positive. Thus, the
compound mode $ab_{1}$ is always entangled. The same is explicitly
illustrated through Figure \ref{fig:ent-hz1-II-c}. Similar investigations
using HZ-I and HZ-II criteria in the other two compound modes (i.e.,
$ab_{2}$ and $b_{1}b_{2})$ failed to show any signature of entanglement
in these cases. Further, signature of intermodal entanglement was
not witnessed using Duan's criterion as using the present solution
and (\ref{duan}), we obtain \begin{equation}
d_{ab_{1}}=d_{ab_{2}}=d_{b_{1}b_{2}}=0.\label{eq:variance-duan}\end{equation}
 However, it does not ensure separability of these modes as HZ-I,
HZ-II, and Duan's inseparability criteria are only sufficient
and not necessary. 

\begin{figure}
\begin{centering}
\includegraphics[scale=0.5]{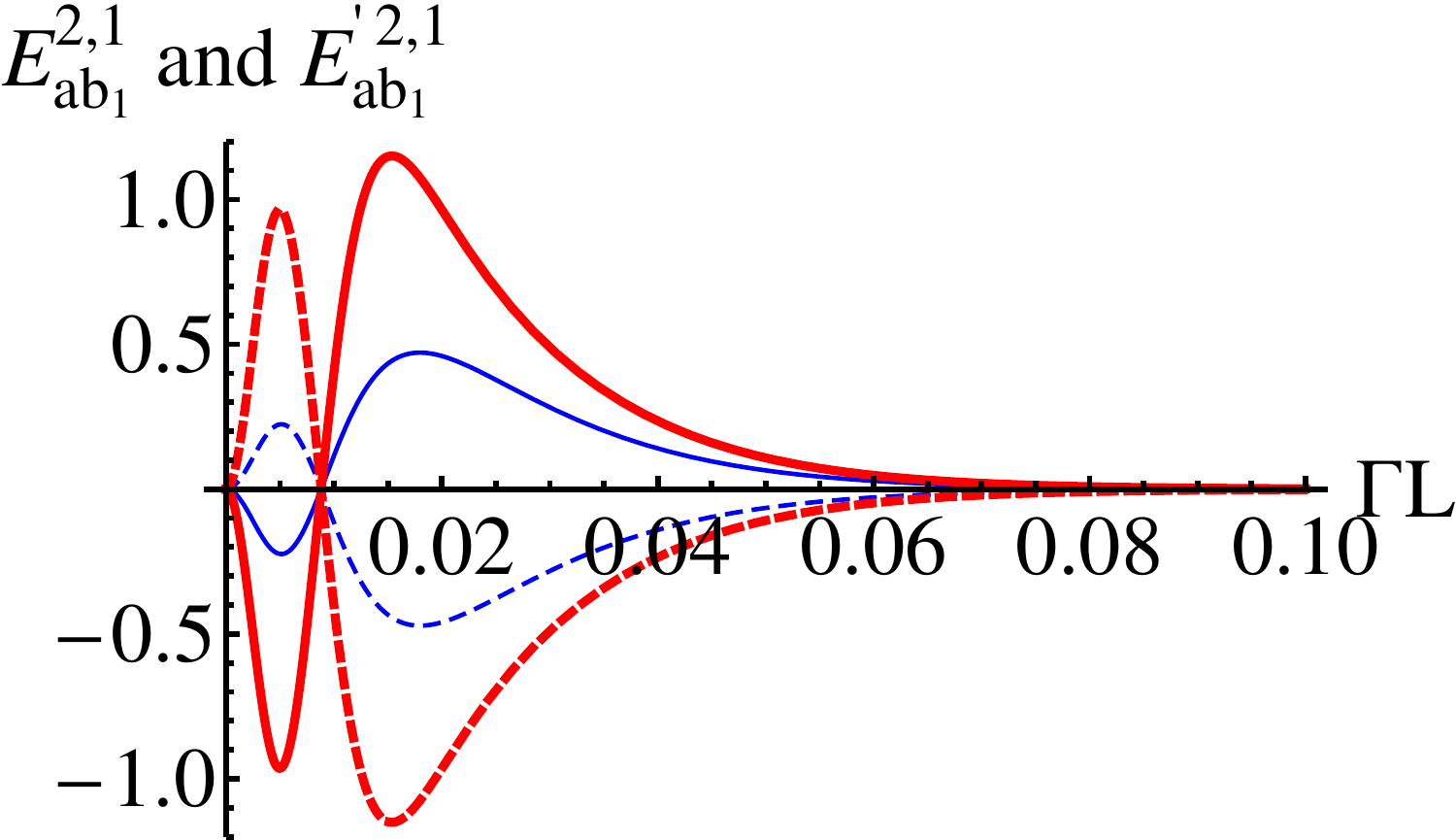}
\par\end{centering}

\caption[Higher-order entanglement in the contradirectional nonlinear optical coupler]{\label{fig:hoe-c}Higher-order entanglement
is observed using Hillery-Zubairy criteria for the contradirectional nonlinear optical coupler. Solid lines show spatial
variation of $E_{ab_{1}}^{2,1}$ and dashed lines show spatial variation
of $E_{ab_{1}}^{\prime2,1}$ with $k=0.1,\,\Gamma=0.001,\,\Delta k=10^{-4},\,\beta=2,\gamma=1$
for $\alpha=3$ (thin blue lines) and $\alpha=5$ (thick red lines).
It is observed that the depth of nonclassicality witness increases with increase
in $\alpha$.}
\end{figure}

We may now study the possibilities of existence of higher-order entanglement
using Eqs. (\ref{hoe-criteria})-(\ref{eq:fully ent2}). To begin
with, we use (\ref{eq:ass-sol-c})-(\ref{eq:soln}) and (\ref{hoe-criteria})
to yield \begin{equation}
\begin{array}{lcl}
E_{ab_{1}}^{m,n} & = & \langle a^{\dagger m}a^{m}b_{1}^{\dagger n}b_{1}^{n}\rangle-|\langle a^{m}b_{1}^{\dagger n}\rangle|^{2}\\
 & = & mn\left|\left(f_{1}\alpha+f_{2}\beta\right)\right|^{2m-2}\left|\left(g_{1}\alpha+g_{2}\beta\right)\right|^{2n-2}E_{ab_{1}}^{1,1}.\end{array}\label{eq:emnab1}\end{equation}
Similarly, using (\ref{eq:ass-sol-c})-(\ref{eq:soln}) and (\ref{hoe-criteria-1}),
we can produce an analytic expression for $E_{ab_{1}}^{\prime m,n}$
and observe that \begin{equation}
E_{ab_{1}}^{\prime m,n}=-E_{ab_{1}}^{m,n}.\label{eq:emnprimeab1}\end{equation}
From relation (\ref{eq:emnprimeab1}), it is clear that the higher-order entanglement between $a$ and $b_{1}$ modes would always
exist for any choice of $\Gamma L,$ $m$, and $n$. This is so because
$E_{ab_{1}}^{m,n}$ and $E_{ab_{1}}^{\prime m,n}$ cannot be simultaneously
positive. Using (\ref{eq:emnab1}) and (\ref{eq:emnprimeab1}), it
is a straightforward exercise to obtain the analytic expressions of $E_{ab_{1}}^{m,n}$
and $E_{ab_{1}}^{\prime m,n}$ for specific values of $m$ and $n.$
Such analytic expressions are not reported here as the existence of
higher-order entanglement is clearly observed through (\ref{eq:emnprimeab1}).
However, in Figure \ref{fig:hoe-c}, we have illustrated the variation
of $E_{ab_{1}}^{2,1}$ and $E_{ab_{1}}^{\prime2,1}$ with the rescaled
interaction length $\Gamma L$. The negative parts of the plot shown
in Figure \ref{fig:hoe-c} illustrate the existence of higher-order intermodal
entanglement in the compound mode $ab_{1}.$ As expected from (\ref{eq:emnprimeab1}),
we observe that for any value of $\Gamma L$ the compound mode $ab_{1}$
is higher-order entangled. Further, it is observed that Hillery-Zubairy's
higher-order entanglement criteria (\ref{hoe-criteria})-(\ref{hoe-criteria-1})
fail to detect any signature of higher-order entanglement in the compound
modes $ab_{2}$ and $b_{1}b_{2}.$

One can also investigate higher-order entanglement using criterion
of multi-partite (multi-mode) entanglement as all the multi-mode entangled
states are essentially higher-order entangled. Here, we have only three
modes in the coupler, and thus, we can study higher-order entanglement
by investigating the existence of three-mode entanglement. A three-mode
pure state that violates (\ref{eq:fully ent2}) (i.e., satisfies $\langle N_{a}\rangle\langle N_{b_{1}}\rangle\langle N_{b_{2}}\rangle-|\langle ab_{1}b_{2}\rangle|^{2}<0)$
and simultaneously satisfies either (\ref{eq:fully enta 0}) or (\ref{eq:fully enta 1})
is a fully entangled state. Using (\ref{eq:ass-sol-c})-(\ref{eq:soln}) and
(\ref{eq:tripartite ent1})-(\ref{eq:fully ent2}), we obtain the following
set of interesting relations for $m=n=l=1$: 

\begin{equation}
E_{a|b_{1}b_{2}}^{1,1,1}=-E_{a|b_{1}b_{2}}^{\prime1,1,1}=E_{ab_{2}|b_{1}}^{1,1,1}=-E_{ab_{2}|b_{1}}^{\prime1,1,1}=|\gamma|^{2}E_{ab_{1}}^{1,1},\label{eq:relation1}\end{equation}
 \begin{equation}
E_{ab_{1}|b_{2}}^{1,1,1}=E_{ab_{1}|b_{2}}^{\prime1,1,1}=0,\label{eq:relation2}\end{equation}
and \begin{equation}
\langle N_{a}\rangle\langle N_{b_{1}}\rangle\langle N_{b_{2}}\rangle-|\langle ab_{1}b_{2}\rangle|^{2}=-|\gamma|^{2}E_{ab_{1}}^{1,1}.\label{eq:relation3}\end{equation}
From (\ref{eq:relation1}), it is easy to observe that three modes
of the coupler are not bi-separable in the form $a|b_{1}b_{2}$ and
$ab_{2}|b_{1}$ for any choice of the rescaled interaction length $\Gamma L>0.$
Further, Eq. (\ref{eq:relation3}) shows that the three modes of
the coupler are not fully separable for $E_{ab_{1}}^{1,1}>0$ (cf.
positive regions of the plot of $E_{ab_{1}}^{1,1}$ shown in Figure \ref{fig:ent-hz1-II-c}).
However, Eq. (\ref{eq:relation2}) illustrates that the present solution
does not show entanglement between coupled mode $ab_{1}$ and single-mode
$b_{2}.$ Thus, the three modes present here are not found to be fully
entangled. Specifically, three-mode (higher-order) entanglement is
observed here, but signature of fully entangled three-mode state is
not observed. Further, we have observed that in all the figures, depth
of nonclassicality witness increases with $\alpha.$

So far, various lower- and higher-order nonclassical
phenomena have been observed in the contradirectional asymmetric nonlinear
optical coupler. Now we will discuss the nonclassical behavior of the codirectional optical coupler.

\section{ Nonclassicality in the codirectional optical coupler \label{sec:Nonclassicality-in-codirectional}}

In this section, we aim to investigate the possibilities of generating nonclassical states in the codirectional optical coupler. To do so, we would require the analytic expressions for the spatial evolution of the field operators involved in describing a codirectional optical coupler. Here, these expressions will be obtained from the rigorous  results reported for the contradirectional nonlinear optical coupler. The results reported here are actually obtained rigorously and reported in \cite{thapliyal2014higher}, but to avoid repetition, here we exploit the symmetry of the codirectional and contradirectional optical couplers to provide the spatial evolution of the operators. Analogous to the contradirectional optical coupler discussed previously, using the perturbative solution (\ref{eq:ass-sol}) and (\ref{eq:terms}), we can
obtain the spatial evolution of various operators that are relevant for the
detection of nonclassical characters in case of the codirectional coupler. For example, we may use (\ref{eq:ass-sol}) and (\ref{eq:terms})
to obtain the number operators for various field modes as in Eqs. (\ref{eq:na})-(\ref{eq:nb2}), which will possess the same form with the only change in the initial value of the operators. Specifically, the number operator in the codirectional optical coupler can be obtained by replacing the initial value of the contra-propagating mode $a(L)$ by $a(0)$. Further, the obtained solution in case of the codirectional optical coupler would be valid for all the values of $z$, thus we have replaced $L$ by $z$ in all the results obtained for the contradirectional coupler to give corresponding expressions for the codirectional coupler. For instance, the number operator for mode $a$ can be obtained as
\begin{equation}
\begin{array}{lcl}
N_{a} & = & a^{\dagger}a=|f_{1}|^{2}a^{\dagger}(0)a(0)+|f_{2}|^{2}b_{1}^{\dagger}(0)b_{1}(0)+\left[f_{1}^{*}f_{2}a^{\dagger}(0)b_{1}(0)+f_{1}^{*}f_{3}a^{\dagger}(0)b_{1}^{\dagger}(0)b_{2}(0)\right.\\
 & + & \left.f_{1}^{*}f_{4}a^{\dagger2}(0)b_{2}(0)+f_{2}^{*}f_{3}b_{1}^{\dagger2}(0)b_{2}(0)+f_{2}^{*}f_{4}b_{1}^{\dagger}(0)a^{\dagger}(0)b_{2}(0)+{\rm H.c.}\right].\end{array}\label{eq:na-co}\end{equation}
Similarly, all other quantities calculated for the contradirectional nonlinear optical coupler can be obtained for the codirectional optical coupler.

Due to change in the initial value, the average value of the number of photons in the modes $a,$ $b_{1}$,
and $b_{2}$ may now be calculated with respect to a given initial
state, i.e.,
\begin{equation}
a(0)|\alpha\rangle|\beta\rangle|\gamma\rangle=\alpha|\alpha\rangle|\beta\rangle|\gamma\rangle,\label{2bcd}\end{equation}
where the rest is same as in Eqs. (\ref{eq:na})-(\ref{eq:nb2}).
In what follows, in all the figures
(except Figure \ref{fig:higher-order antibunching}) that illustrate the existence of nonclassical
character in the codirectional asymmetric nonlinear optical coupler, we have chosen $\alpha=5,\,\beta=2,$ and $\gamma=1.$ It is important to note that all the quantities obtained in Eqs. (\ref{eq:squeezing})-(\ref{eq:relation3}) for the contradirectional optical coupler possess the same functional form as that of the codirectional coupler. However, due to different values of functional coefficients $F_i$s, in Eqs. (\ref{eq:terms}) and (\ref{eq:soln}), we expect the nonclassical characters to behave independently in both cases, and thus will be discussed in what follows.

\subsection{Higher-order squeezing}

\begin{figure}
\centering{}\includegraphics[angle=-90,scale=0.5]{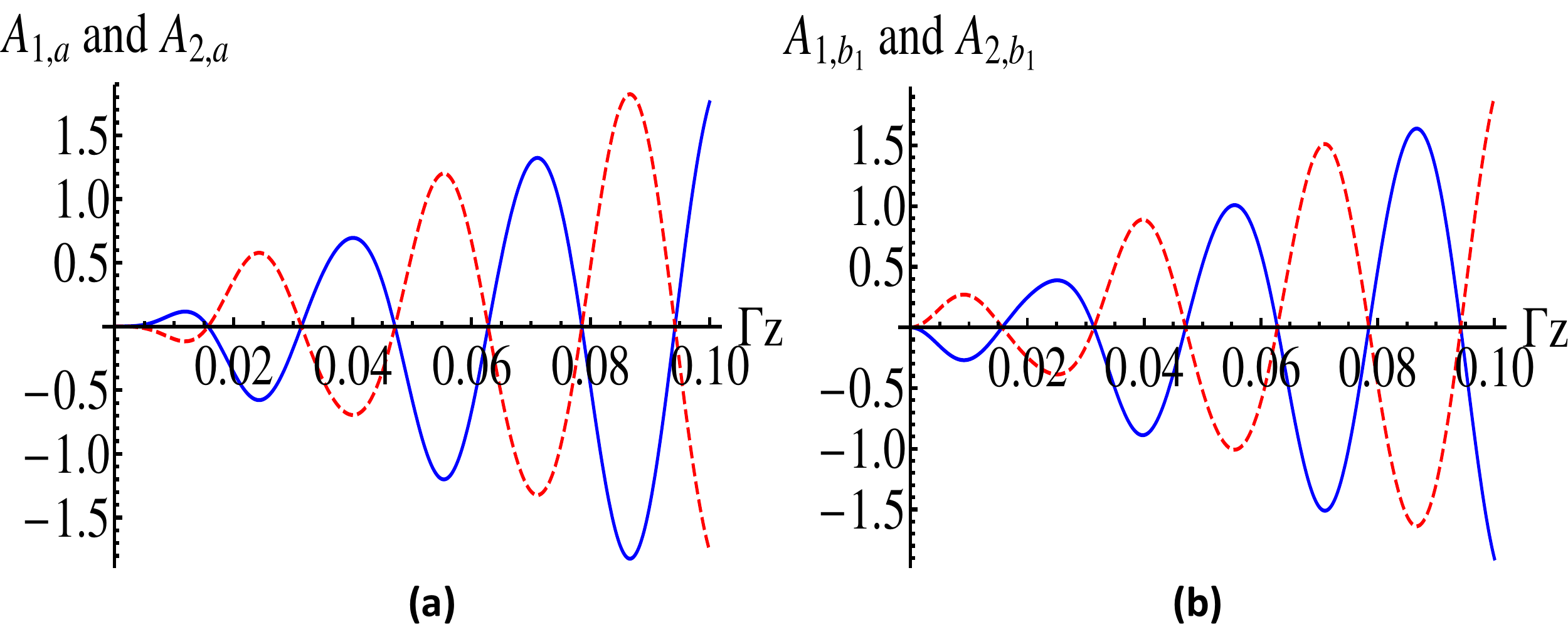}
\caption[Amplitude
squared squeezing in the codirectional nonlinear optical coupler]{\label{fig:Amplitude-squared-squeezing}  Amplitude
squared squeezing is observed in modes (a) $a$  and (b) $b_{1}$ while codirectional propagation
for the initial state $|\alpha\rangle|\beta\rangle|\gamma\rangle$
with $k=0.1,\,\Gamma=0.001,\,\Delta k=10^{-4},\,\alpha=5,\,\beta=2,$ and $\gamma=1.$
The negative parts of the solid line represent squeezing in the quadrature
variable $Y_{1,a}$ ($Y_{1,b_{1}})$ and that of the dashed line represent
squeezing in the quadrature variable $Y_{2,a}$ ($Y_{2,b_{1}})$. }
\end{figure}

To analyze the higher-order squeezing in the codirectional optical coupler, we use Eq. (\ref{eq:terms}) in Eqs. (\ref{eq:asq-1})-(\ref{eq:as-q-3}).
Clearly, we do not obtain any signature of amplitude powered squeezing
in $b_{2}$ mode using the present solution, and mode $a$ $(b_{1})$
should always show amplitude powered squeezing in one of the quadrature
variables as both $A_{1,a}$ and $A_{2,a}$ ($A_{1,b_{1}}$ and $A_{2,b_{1}}$)
cannot be positive simultaneously. To investigate the possibility
of amplitude powered squeezing in further detail in modes $a$ and
$b_{1}$, we have plotted the spatial variation of $A_{i,a}$ and $A_{i,b_{1}}$
in Figure \ref{fig:Amplitude-squared-squeezing}. The negative regions of
these two plots clearly illustrate the existence of amplitude squared
squeezing in both $a$ and $b_{1}$ modes.

\subsection{Higher-order antibunching\label{sec:num-Coupler}}

We have already described the condition of higher-order antibunching as (\ref{eq:higher-order-antibunching}). Now
using Eq. (\ref{eq:terms}) in Eq. (\ref{eq:higher-order-antibunching}),
we can obtain closed form analytic expressions for $D_{i}(n)$ for
various modes as Eqs. (\ref{eq:dan})-(\ref{eq:db2n}).

\begin{figure}
\centering{}\includegraphics[angle=-90,scale=0.58]{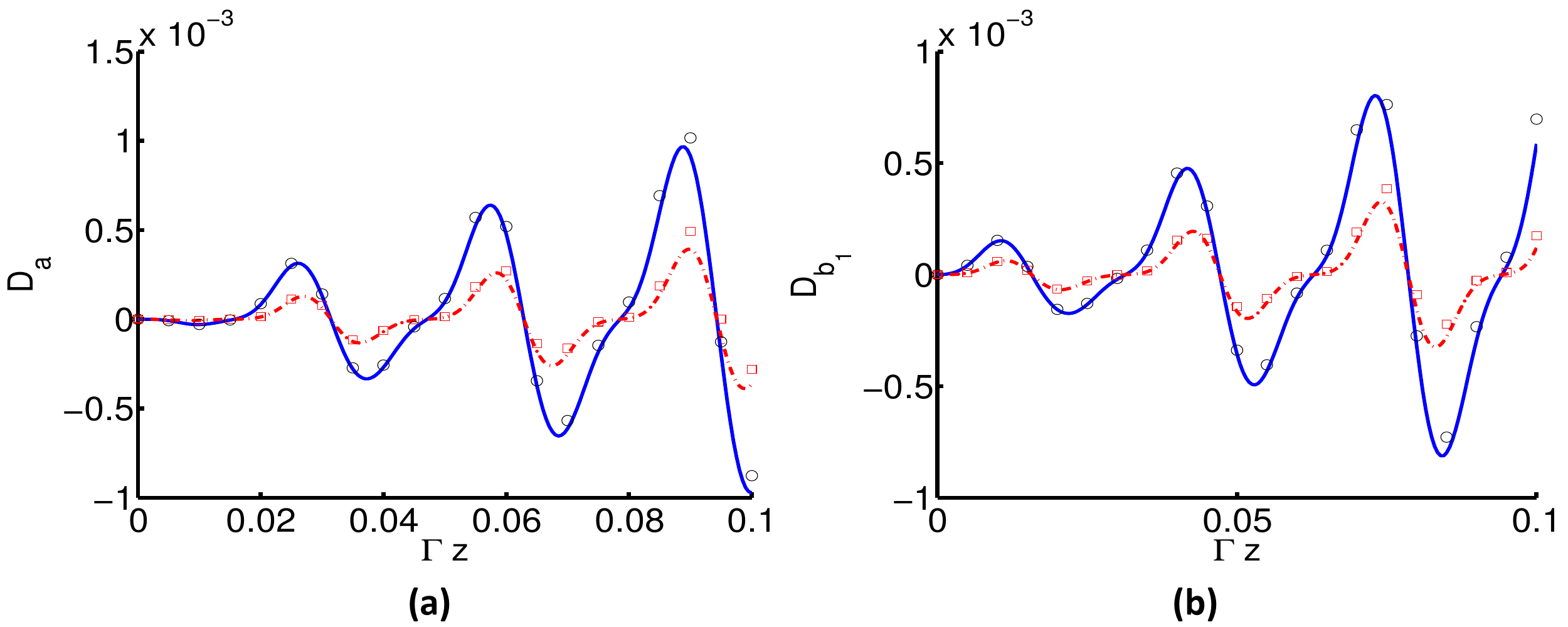}
\caption[Higher-order antibunching in the codirectional nonlinear optical coupler]{\label{fig:higher-order antibunching}  Higher-order antibunching with the rescaled interaction length
$\Gamma z$ in mode (a) $a$ and (b) $b_{1}$ in the codirectional coupler for $n=3$ (smooth line) and $n=4$ (dashed line), and the square
and circle denote the corresponding numerical results with the initial
state $|\alpha\rangle|\beta\rangle|\gamma\rangle$, and $k=0.1,\,\Gamma=0.001,\,\Delta k=10^{-4},\,\alpha=0.5,\,\beta=0.2,$ and $\gamma=0.1$}
\end{figure}

Clearly, the perturbative solution used here cannot detect any signature
of higher-order antibunching for the $b_{2}$ mode. However, in the other
two modes, we observe higher-order antibunching for different values of $n$ as illustrated
in Figure \ref{fig:higher-order antibunching}. In Figure \ref{fig:higher-order antibunching}, we have plotted the right-hand sides of Eqs. (\ref{eq:dan}) and (\ref{eq:dbn}) along with the exact
numerical results obtained by integrating the $z$-dependent Schr\"{o}dinger
equation corresponding to the given momentum operator (\ref{eq:Mom-asymmetric}) by using the matrix form
of the operators. A close resemblance of the exact numerical result
with the perturbative result even for the higher-order case clearly validates
the perturbative solution used here.

\subsection{Lower- and higher-order intermodal entanglement}

To apply the HZ-I criterion to investigate the existence of intermodal
entanglement between modes $a$ and $b_{1}$, i.e., compound mode $ab_{1}$ for the codirectional optical coupler,
we use Eq. (\ref{eq:terms})
in Eq. (\ref{eq:HZ-1-ab1}).
Similarly, to study the HZ-II criterion for the compound mode $ab_{1}$
we use the expression reported in Eq. (\ref{eq:terms}) in Eq. (\ref{eq:hz2-ab1}).

\begin{figure}
\begin{centering}
\includegraphics[scale=0.5]{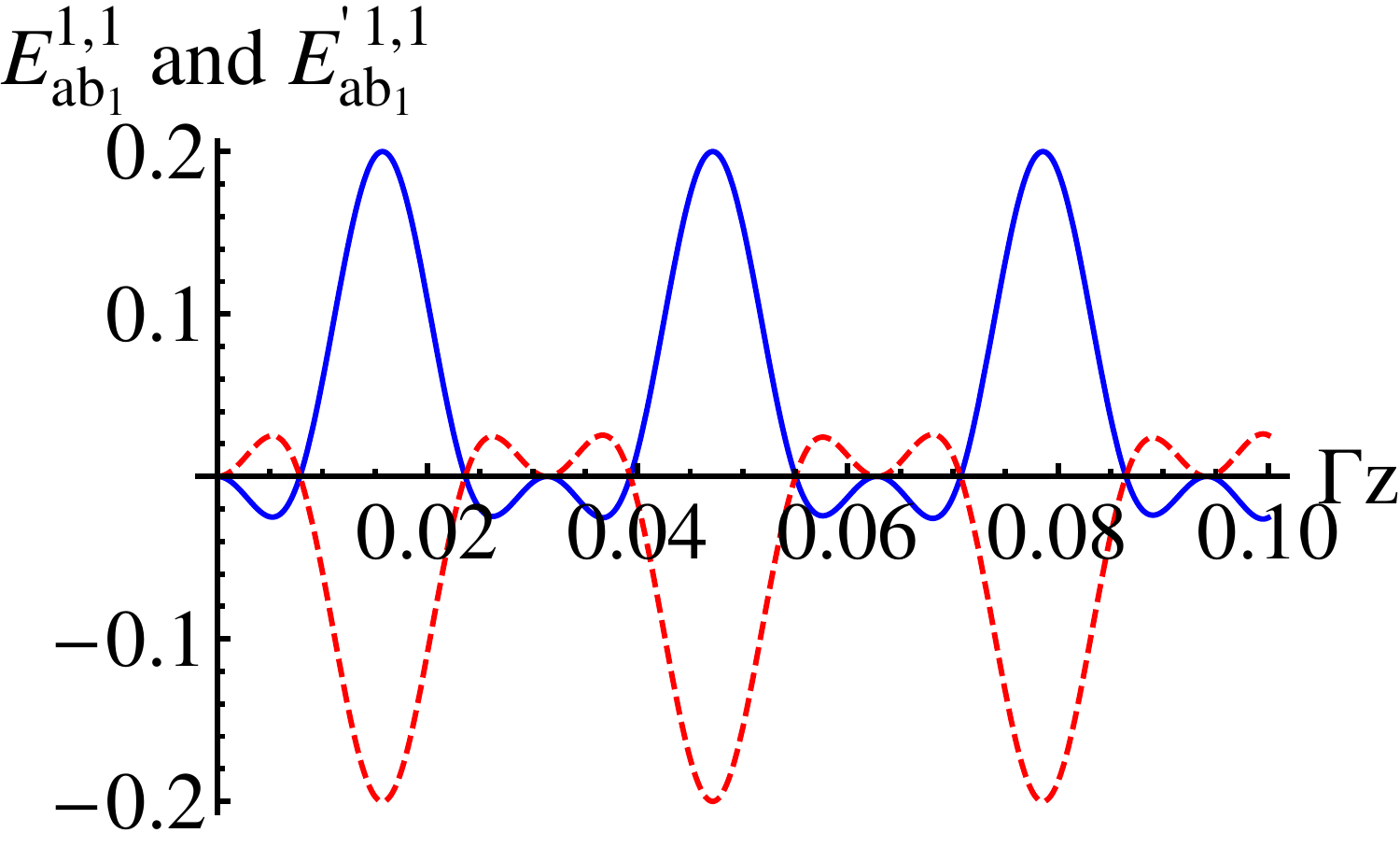}
\par\end{centering}

\caption[Lower-order entanglement for the codirectional nonlinear optical coupler]{\label{fig:ent-hz1-II}  Hillery-Zubairy criterion
I (solid line) and criterion II (dashed line) for entanglement are
showing intermodal entanglement between modes $a$ and $b_{1}$ in the codirectional optical coupler. Here,
$E_{ab_{1}}^{1,1}$ (solid line) and $E_{ab_{1}}^{\prime1,1}$ (dashed
line) are plotted with rescaled interaction length $\Gamma z$
for mode $ab_{1}$ with the initial state $|\alpha\rangle|\beta\rangle|\gamma\rangle$
and $k=0.1,\,\Gamma=0.001,\,\Delta k=10^{-4},\,\alpha=5,\,\beta=2,$ and $\gamma=1.$}
\end{figure}

From Eqs. (\ref{eq:HZ-1-ab1}) and (\ref{eq:hz2-ab1}), we can easily
observe that in the present case as well, $E_{ab_{1}}^{1,1}=-E_{ab_{1}}^{\prime1,1}$,
which implies that at any point inside the coupler either the HZ-I criterion
or the HZ-II criterion would show the existence of entanglement as both
of them cannot be simultaneously positive. Thus, the compound mode $ab_{1}$
is always entangled inside the codirectional asymmetric nonlinear optical coupler.
The same is explicitly illustrated through Figure \ref{fig:ent-hz1-II}.
Following the same approach, we investigated the existence of entanglement
in other compound modes (e.g., $ab_{2}$ and $b_{1}b_{2}),$ but both the
HZ-I and HZ-II criteria failed to detect any entanglement in these
cases for the codirectional coupler, too. However, it does not indicate that the modes are separable
as both the HZ-I and HZ-II inseparability criteria are only sufficient
and not essential. Further, the perturbative analytic solution used
here is an approximate solution, and in the recent past, there are several
examples where the existence of entanglement not detected by HZ criteria
is detected by Duan's criterion or vice versa \cite{sen2013intermodal,giri2014single}. 
Keeping these facts in mind, we studied the possibilities of observing
intermodal entanglement using Duan's criterion, too, but it
failed to detect any entanglement in the codirectional coupler. 

\begin{figure}
\begin{centering}
\includegraphics[scale=0.5]{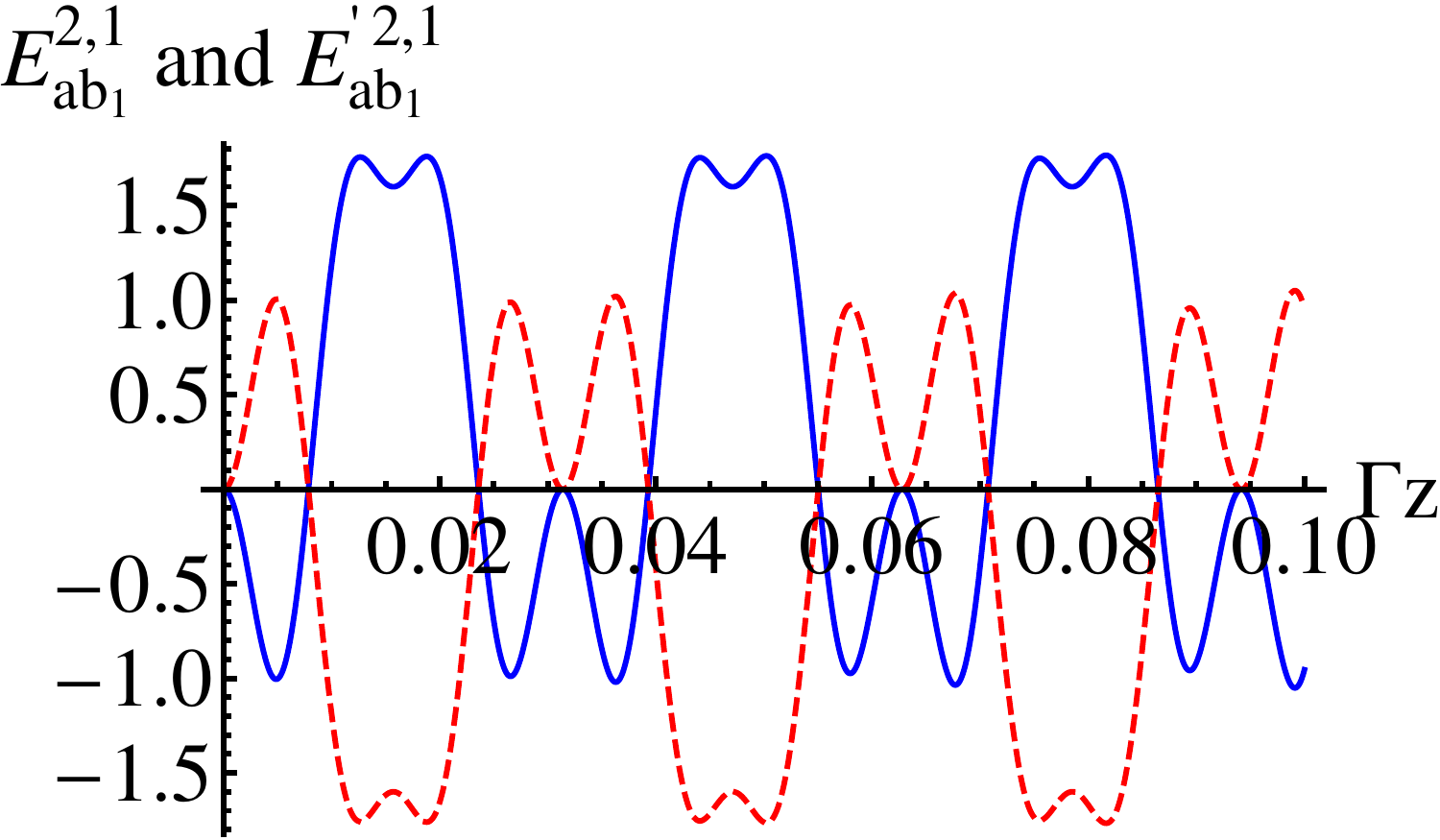}
\par\end{centering}

\caption[Higher-order entanglement for the codirectional nonlinear optical coupler]{\label{fig:hoe}Higher-order
entanglement in the codirectional nonlinear optical coupler is observed using the Hillery-Zubairy criteria. The solid line
shows the spatial variation of $E_{ab_{1}}^{2,1}$, and the dashed line illustrates the
spatial variation of $E_{ab_{1}}^{\prime2,1}$ with the initial state
$|\alpha\rangle|\beta\rangle|\gamma\rangle$ and $k=0.1,\,\Gamma=0.001,\,\Delta k=10^{-4},\,\alpha=5,\,\beta=2,$ and $\gamma=1.$ }
\end{figure}

We may now investigate the existence of higher-order entanglement in the codirectional optical coupler
using Eqs. (\ref{hoe-criteria})-(\ref{eq:fully ent2}). 
Similar to the contradirectional coupler, $E_{ab_{1}}^{\prime m,n}=-E_{ab_{1}}^{m,n}$
clearly shows that higher-order entanglement
between $a$ and $b_{1}$ modes would always be observed for any
choice of $m$ and $n$ as $E_{ab_{1}}^{m,n}$ and $E_{ab_{1}}^{\prime m,n}$
cannot be simultaneously positive. In Figure \ref{fig:hoe}, we have illustrated the spatial
evolution of $E_{ab_{1}}^{2,1}$ and $E_{ab_{1}}^{\prime2,1}$. The negative
regions of this figure clearly show the existence of higher-order
intermodal entanglement in the compound mode $ab_{1}.$ As expected from
(\ref{eq:emnprimeab1}), we observe that for any value of $\Gamma z$ the
compound mode $ab_{1}$ is higher-order entangled. However, for the remaining two compound modes, the same result as that for the contradirectional coupler can be inferred due to similar form of expressions. Specifically, the obtained results
could not show any signature of higher-order entanglement in the compound
modes $ab_{2}$ and $b_{1}b_{2}.$ This is not surprising as the Hillery-Zubairy's
criteria are only sufficient, not necessary, and we have already seen
that these criteria fail to detect lower-order entanglement present
in the compound modes $ab_{2}$ and $b_{1}b_{2}.$

We can further extend our discussion to multi-partite entanglement.
From (\ref{eq:relation1}), we can see that three modes of the coupler
are not bi-separable in the form $a|b_{1}b_{2}$ and $ab_{2}|b_{1}$
for any value of $\Gamma z>0.$ Further, Eq. (\ref{eq:relation3})
and the positive regions of $E_{ab_{1}}^{1,1}$ shown in Figure \ref{fig:ent-hz1-II}
show that the three modes of the coupler are not fully separable.
However, the present solution does not show signature of a fully entangled
three-mode state as (\ref{eq:relation2}) does not show entanglement between the
coupled mode $ab_{1}$ and mode $b_{2}.$ To be specific, we observed
three-mode (higher-order) entanglement, but could not observe signature
of a fully entangled three-mode state. However, here we cannot conclude
whether the three modes of the coupler are fully entangled or not
as the criteria used here are only sufficient. 

We have already observed different signatures of nonclassicality in the codirectional and contradirectional
asymmetric nonlinear optical couplers of our interest. If we now closely
look into all the analytic expressions of the signatures of nonclassicality
provided here through Eqs. (\ref{eq:squeezing})-(\ref{eq:relation3})
we can find an interesting symmetry: All the non-vanishing expressions
of signatures of nonclassicality are proportional to $|\gamma|.$
Thus, we may conclude that within the domain of the validity of the present
solution, in the spontaneous process, we would not observe any of the
nonclassical characters that are observed in the stimulated case for the codirectional and contradirectional nonlinear optical couplers. 

\begin{table}[t]
\begin{centering}
\begin{tabular}{|c|>{\centering}p{1.4in}|c|>{\centering}p{1.4in}|c|}
\hline 
S.No. & Nonclassical phenomenon & Modes & Short-length approximation \cite{perina1995photon,perina1995quantum863} & Present work\tabularnewline
\hline 
1. & Squeezing  & $b_{2}$ & Not investigated & Not observed\tabularnewline
\hline 
2. & Intermodal squeezing  & $b_{1}b_{2},\, ab_{2}$ & Not observed & Observed\tabularnewline
\hline 
3. & Amplitude powered squeezing  & $a,\, b_{1}$ & Not investigated & Observed\tabularnewline
\hline 
4. & Amplitude powered squeezing  & $b_{2}$ & Not investigated & Not observed\tabularnewline
\hline 
5. & Lower- and higher-order intermodal entanglement & $ab_{1}$ & Not investigated & Observed\tabularnewline
\hline 
6. & Lower- and higher-order intermodal entanglement  & $b_{1}b_{2},\, ab_{2}$ & Not investigated & Not observed\tabularnewline
\hline 
7. & Three-mode (higher-order) bi-separable entanglement & $a|b_{1}b_{2}$, $ab_{2}|b_{1}$ & Not investigated & Observed\tabularnewline
\hline 
8. & Three-mode (higher-order) bi-separable entanglement & $a|b_{1}b_{2}$ & Not investigated & Not observed\tabularnewline
\hline
\end{tabular}
\par\end{centering}

\caption[Nonclassical properties that were observed in the present study but not in the earlier studies
on the contradirectional asymmetric nonlinear optical coupler]{\label{tab:Nonclassicalities-observed-in}Nonclassical properties that were observed in the present study but not in the earlier studies
on the contradirectional asymmetric nonlinear optical coupler.}
\end{table}

\section{Conclusions \label{sec:Conclusions-coupler}}

In the present study, we report lower- and higher-order nonclassicalities
in a contradirectional asymmetric nonlinear optical coupler using
a set of criteria of entanglement, single-mode squeezing, intermodal
squeezing, antibunching, intermodal antibunching, etc. Further, possibility of generation of higher-order nonclassical states in the codirectional asymmetric nonlinear optical coupler  is also investigated. In brief, we have observed higher-order (amplitude powered) squeezing,
higher-order antibunching and lower- and higher-order entanglement in case of the codirectional nonlinear optical coupler.  None of these
higher-order nonclassical phenomena were reported in the earlier studies
on the codirectional asymmetric nonlinear optical coupler (\cite{mandal2004approximate}
and references therein). In fact, prior to this work, neither entanglement nor
higher-order nonclassicalities are systematically studied in optical
couplers other than the Kerr coupler. The method followed in the present
chapter is quite general and it can be extended easily to other
types of couplers, such as codirectional and contradirectional Raman and Brillouin couplers \cite{perina1997statistics},
and parametric couplers \cite{korolkova1997quantum}.

In the present work, variation of
nonclassicality with various parameters, such as the number of input photons
in the linear mode, linear coupling constant, nonlinear coupling constant, phase mismatch, is also studied, and it is observed that the amount
of nonclassicality can be controlled by controlling these parameters.
The contradirectional asymmetric nonlinear optical coupler studied
in the present work was studied earlier using a short-length solution \cite{perina1995photon} and by using a closed form exact
solution obtained by considering $b_{2}$ mode as classical \cite{perina1995quantum863}.
In contrast, a completely quantum mechanical solution of the equations
of motion is obtained here using the Sen-Mandal approach, which is not
restricted by length. The use of better solution and completely quantum
mechanical treatment led to the identification of several nonclassical
characters of the contradirectional asymmetric nonlinear optical coupler
that were not reported in the earlier studies. All such nonclassical phenomena
that are observed here and were not observed in earlier studies are
listed in Table \ref{tab:Nonclassicalities-observed-in}. However, the results reported here for the lower- and higher-order nonclassicalities in the codirectional coupler cannot be compared along the same line as this is the first attempt to study the concerned nonclassical features. Further,
all the results obtained here in context of the contradirectional
coupler show that in the contradirectional propagation, the beams
involved are more effectively matched compared to the 
case of codirectional propagation, and consequently
the nonlinear interaction is more effective and nonclassical effects
are stronger.

There exist a large number of nonclassicality criteria that are not
studied here and are based on the expectation values of moments of annihilation
and creation operators (cf. Table I and II of Ref. \cite{miranowicz2010testing}).
As we already have compact expressions for the field operators it
is a straightforward exercise to extend the present work to investigate
other signatures of nonclassicality, such as photon hyperbunching
\cite{jakob2001comparative}, sum and difference squeezing of An-Tinh \cite{an1999general,an2000general}
and Hillery \cite{hillery1989sum}, inseparability criterion
of Manicini et al. \cite{mancini2002entangling}, Simon \cite{simon2000peres} and Miranowicz
et al. \cite{miranowicz2009inseparability}. It is even possible
to investigate the existence of Hong-Mandel \cite{hong1985higher,hong1985generation}
type higher-order squeezing and Agarwal-Tara parameter $A_{n}$ \cite{agarwal1992nonclassical}
for higher-order nonclassicality using the present approach. Possibly the study of lower- and higher-order steering using
the present approach and the strategy adopted in Ref. \cite{he2012einstein} is also feasible. 
However, we have not investigated steering as recently it is shown
that every pure entangled state is maximally steerable \cite{skrzypczyk2014quantifying}.
Since the combined states of three modes of the asymmetric codirectional and contradirectional
nonlinear optical couplers are pure states, the findings of Ref. \cite{skrzypczyk2014quantifying}
and the intermodal entanglement observed in the present chapter imply
that the compound mode $ab_{1}$ in both these cases are maximally steerable. The importance
of entanglement and steering in various applications of quantum computing
and quantum communication, and the easily implementable structure of
the coupler studied here indicate the possibility that the entangled
states generated through the optical couplers of the present form would be
useful in various practical purposes.

Note that the experimental feasibility of characterizing the lower-order nonclassicalities reported here is already discussed in Section \ref{Noncl}. It is also possible to experimentally verify the existence
of higher-order nonclassicalities reported here as the higher-order nonclassical
effects can generally be detected using higher numbers of detectors
correlating their outcomes. Alternatively, higher-order quantities
can be calculated from the measured distributions. Specifically, all the
criteria of higher-order nonclassicalities reported here are expressed
as the expectation values of the moments of annihilation and creation
operators, and these expectation values can be measured using different
variants of homodyne or heterodyne measurements and time multiplexing. For example,
Shchukin and Vogel clearly showed that amplitude squared squeezing
\cite{shchukin2005nonclassical} and amplitude $n$th
powered squeezing \cite{shchukin2005nonclassical}
can be detected using a technique based on the balanced homodyne correlation
measurement \cite{shchukin2005nonclassical,shchukin2006universal}.
Using Shchukin and Vogel's approach one can measure $\langle a^{\dagger k}a^{l}\rangle$
for any values of $k$ and $l$ (cf., Figure 1 of Ref \cite{shchukin2005nonclassical}).
Thus, if we replace the source $S$ in Figure 1 of Ref. \cite{shchukin2005nonclassical}
by a field of specific frequency (i.e., a field representing a particular
mode) obtained at the output of one of the waveguides that constitute
the coupler studied here, it would be possible to measure all the
single-mode correlations (including higher-order antibunching and
higher-order squeezing) reported in the present work. However, with
an increase in $k$ and $l$, the requirement of the number of beamsplitters,
photodetectors, and measurements increases considerably. This technical
limitation of Shchukin and Vogel's approach is considerably circumvented
in the later works of Prakash and Yadav \cite{prakash2012proposal} for $n$th-order amplitude powered squeezing, and Prakash, Kumar, and Mishra's work
on amplitude squared squeezing \cite{prakash2010detection}, where
only one beamsplitter and one photodetector were used, and the number
of measurements required was also reduced. In these interesting works
\cite{prakash2012proposal,prakash2010detection}, higher-order moments
of the number operators were obtained by using standard homodyne technique.
Following an independent approach Prakash and Mishra \cite{prakash2006higher}
showed that higher-order sub-Poissonian statistics can also be used
for detection of amplitude squared squeezing. The proposals of detection
of higher-order moments of the form $\langle a^{\dagger k}a^{l}\rangle$
is relatively new, but schemes for the measurement of higher-order moments
of the number operator (and thus higher-order antibunching) was long in existence
\cite{mandel1986coherence}. Beyond these new and old schemes
of experimental detection of the results reported here what is more
exciting is the fact that our control over the field, quality of source,
detector, and other devices required for the experimental detection
have been considerably improved in the recent past, and as a consequence,
a set of very interesting experimental works demonstrating higher-order nonclassicality have been  recently reported \cite{allevi2012measuring,allevi2012high,avenhaus2010accessing,perina2017higher}.
Specifically, using a hybrid photodetector Allevi et al. \cite{allevi2012measuring,allevi2012high}
experimentally measured $\langle a_{1}^{\dagger j}a_{1}^{j}a_{2}^{\dagger k}a_{2}^{k}\rangle$
which can be used to fully characterize bipartite multi-mode states.
Clearly, their method can be directly used to detect higher-order entanglement
using HZ-I and HZ-II criteria described by Eqs. (\ref{hoe-criteria})-(\ref{eq:fully ent2}),
higher-order antibunching and higher-order squeezing reported in the
present work. Further, Avenhaus et al. \cite{avenhaus2010accessing} have
also experimentally measured $\langle a_{1}^{\dagger j}a_{1}^{j}a_{2}^{\dagger k}a_{2}^{k}\rangle$
using time multiplexing. Thus, in brief, there exist a large number
of alternative paths that may be used to experimentally verify the
existence of nonclassical states reported in the present work. It is true that most of the nonclassicalities reported
here can be observed in some other bosonic systems, too. For instance, we have shown recently that two bosonic modes in an atom-molecule Bose--Einstein condensate \cite{giri2017nonclassicality} and two arbitrary pump modes in the most general form of non-degenerate hyper-Raman processes \cite{thapliyal2017nonclassicality} are always entangled. However,
the present system has some intrinsic advantages over most of the
other systems as it can be used as a component in the integrated waveguide-based structures in general and photonic circuits in particular (\cite{lugani2013studies,tanzilli2012genesis} and references therein).
Thus, the nonclassicalities reported in this easily implementable waveguide-based system is expected to be observed experimentally. Further, the
system (in both codirectional and contradirectional propagation of fields) studied here is expected to play an important role as a source
of nonclassical fields in the integrated waveguide-based structures. 

The results reported in this chapter for the codirectional and contradirectional nonlinear optical coupler have been published as two independent international journal articles  \cite{thapliyal2014higher,thapliyal2014nonclassical}. 
Along the same line, possibilities of generation of nonclassicalities in other optical \cite{thapliyal2017nonclassicality}, atomic \cite{giri2017nonclassicality,alam2017lower} and optomechanical \cite{alam2017lower} systems have also been performed, but the results of such studies are not included in the thesis. Similarly, one can study the feasibility of generating nonclassical states in new systems considering the open quantum system effects as we have performed in \cite{naikoo2017probing}.

\thispagestyle{empty}
\thispagestyle{plain}   
\cleardoublepage
\blankpage

\mathversion{normal2}
\titlespacing*{\chapter}{0pt}{-50pt}{20pt}
\chapter{Linear and nonlinear quantum Zeno and anti-Zeno effects in a nonlinear optical coupler \label{Zeno}}

\section{Introduction\label{sec:Introduction}}

Quantum Zeno and anti-Zeno effects were introduced in the first chapter while discussing the difference between unitary and measurement operators (see Section \ref{Eq-of-m}). In the present chapter, we aim to discus it in more detail. To begin with, we would like to note that the Greek philosopher
Zeno of Elea introduced a set of paradoxes of motion in the 5th century BC. These paradoxes,
which are now known as Zeno's paradoxes, were unsolved for long, and
they fascinated mathematicians, logicians, physicists and other creative
minds since their introduction. In the recent past, a quantum analog
of Zeno's paradox has been studied intensively. Specifically, in
the late 50s and early 60s of the 20th century, Khalfin studied nonexponential
decay of unstable atoms  \cite{khalfin1957theory,khalfin1958contribution,khalfin1961quantum}. Later on, in 1977, Misra
and Sudarshan \cite{misra1977zeno} showed that under continuous measurement,
an unstable particle will never be found to decay, and in analogy
with classical Zeno's paradox they named this phenomenon as \emph{Zeno's paradox in quantum theory}. Quantum Zeno effect in the original formulation
refers to the inhibition of the temporal evolution of a system on
continuous measurement \cite{misra1977zeno}, while quantum anti-Zeno
effect or inverse Zeno effect refers to the enhancement of the
evolution instead of the inhibition (see Refs. \cite{venugopalan2007quantum,facchi2001quantum,pascazio2014all}
for the reviews). Here, it is important to note that usually quantum
Zeno effect is viewed as a process, which is associated with the repeated
projective measurement. This is only a specific manifestation of quantum
Zeno effect. In fact, it can be manifested in a few equivalent ways
\cite{facchi2003three}. One such manifestation of quantum
Zeno effect is a process in which continuous interaction between the
system and probe leads to the quantum Zeno effect. Here, we aim to study
the continuous interaction type manifestation of the quantum Zeno effect
in a symmetric nonlinear optical coupler, which is made of two nonlinear
waveguides with $\chi^{(2)}$ nonlinearity, and each of the waveguides
is operating under second harmonic generation. As we are describing
an optical coupler, the waveguides are coupled with each other. More
precisely, these two waveguides interact with each other through the
evanescent waves. We consider one of the waveguides as the probe and
the other one as the system. In what follows, we will show that the
beauty of the symmetric nonlinear optical coupler ($\chi^{(2)}-\chi^{(2)}$
coupler) studied here is that the results obtained for this coupler
can be directly reduced to the corresponding results for the asymmetric
nonlinear optical coupler discussed in Chapter \ref{Coupler}, where the probe is linear ($\chi^{(1)})$
and the system is nonlinear ($\chi^{(2)}).$ We have reported quantum
Zeno and anti-Zeno effects in both the symmetric ($\chi^{(2)}-\chi^{(2)}$)
nonlinear optical coupler and asymmetric ($\chi^{(2)}-\chi^{(1)})$
nonlinear optical coupler. Following an earlier work \cite{abdullaev2011linear},
we refer to the quantum Zeno (anti-Zeno) effect observed due to the continuous interaction
of a nonlinear $\left(\chi^{(2)}\right)$ probe as the nonlinear quantum Zeno
(anti-Zeno) effect. Similarly, the quantum Zeno (anti-Zeno) effect observed due to the continuous
interaction of a linear $\left(\chi^{(1)}\right)$ probe is referred
to as the linear quantum Zeno (anti-Zeno) effect. In general, quantum Zeno and anti-Zeno effects are considered as observation of a physical process. The system here is a nonlinear waveguide operating under second harmonic generation. The probe mode (in both linear and nonlinear quantum Zeno effects) interacts with the fundamental mode of the system waveguide via evanescent field, which can be thought of as a continuous measurement on the system. Consequently,  the inhibition or enhancement of the second harmonic generation due to the presence of a probe is characterized here as quantum Zeno and anti-Zeno effects, respectively. The number of photons can be measured on a quantum level, which will be a signature of the presence of these effects. Say, in the regime of quantum Zeno effect, the second harmonic generation of the pump mode in the system waveguide will be inhibited, resulting in a lesser number of photons in the second harmonic mode. 

Optical couplers can be prepared easily and several exciting applications
of the optical couplers have been mentioned in the previous chapter (also see
\cite{perina2000review} and references therein).
Consequently, it is no wonder that quantum Zeno and anti-Zeno effects
have been investigated in various types of optical couplers \cite{rehacek2001quantum,thun2002quantum,thapliyal2015quantum,mista2000quantum,rehacek2000quantum}. Specifically, quantum Zeno and anti-Zeno effects were shown in Raman
and Brillouin scattering using an asymmetric ($\chi^{(3)}-\chi^{(1)}$) 
nonlinear optical coupler \cite{thun2002quantum}; their existence
was also shown in the $\chi^{(2)}-\chi^{(1)}$ optical couplers \cite{thapliyal2015quantum},
$\chi^{(2)}-\chi^{(2)}$ optical couplers \cite{mista2000quantum}, etc.
In all these studies, it was always considered that one of the modes
in the nonlinear waveguide (this waveguide is considered as the system)
is coupled with the auxiliary mode in a (non)linear waveguide (this
waveguide is considered as the probe). Actually, the auxiliary mode
acts as the probe since its coupling with the system implements continuous
observation on the evolution of the system (nonlinear waveguide) and
changes the photon statistics of the other modes (which are not directly coupled
to the probe mode) of the nonlinear waveguide. Quantum Zeno and anti-Zeno
effects have also been investigated in optical systems other than
couplers, such as in parametric down-conversion \cite{luis1996zeno,luis1998anti,rehacek2000quantum},
parametric down-conversion with losses \cite{perina2004quantum},
an arrangement of beam splitters \cite{agarwal1994all}. In these studies on quantum Zeno effect in optical systems, often
the pump mode has been considered strong, and thus the complexity
of a completely quantum mechanical treatment has been circumvented.
Keeping this in mind, in this thesis, we plan to use a completely quantum mechanical
description of the coupler.

Initially, interest on the quantum Zeno effect was theoretical and purely
academic in nature, but with time quantum Zeno effect has been experimentally
realized by several groups using different techniques \cite{kwiat1999high,itano1990quantum,fischer2001observation}.
Not only that, several interesting applications of quantum Zeno effect
have also been proposed \cite{kwiat1999high,hosten2006counterfactual,salih2013protocol,facchi2002quantum,pascazio2001quantum,cao2017direct}.
Specifically, in Refs. \cite{facchi2002quantum,pascazio2001quantum},
it was established that the quantum Zeno effect may be used to increase
the resolution of absorption tomography. A few of the proposed applications
have also been experimentally realized. For example, Kwait et al.,
implemented high-efficiency quantum interrogation measurement using
quantum Zeno effect \cite{kwiat1999high}. Until
recently, all the investigations related to the quantum Zeno effect
were restricted to the microscopic world. Recently, in a very interesting
work, it has been extended to the macroscopic world by showing the
evidence for the existence of quantum Zeno effect for large black
holes \cite{nikolic2014suppressing}. The possibility of observing the macroscopic
Zeno effect was also studied in the context of stationary flows in
the nonlinear waveguides with localized dissipation \cite{zezyulin2012macroscopic}.
The interest in quantum Zeno effect has recently been amplified with
the advent of various protocols of quantum communication that are
based on quantum Zeno effect. Specifically, in Ref. \cite{salih2013protocol},
a counterfactual protocol for direct quantum communication was proposed
using chained quantum Zeno effect which has been  experimentally  realized in the recent past \cite{cao2017direct}; and in Ref. \cite{hosten2006counterfactual},
the same effect has also  been used to propose a scheme for counterfactual quantum
computation. In the past, a proposal for quantum computing was made
using an environment induced quantum Zeno effect to confine the dynamics
in a decoherence-free subspace \cite{beige2000quantum}. Recently,
quantum Zeno effect has also been used to reduce communication complexity
\cite{tavakoli2015quantum}. These applications of quantum Zeno effect and
easy production of optical couplers motivated us to systematically
investigate the possibility of observing quantum Zeno and anti-Zeno
effects in a symmetric nonlinear optical coupler which is not studied earlier
using a completely quantum description.

To investigate the existence of quantum Zeno and anti-Zeno effects
in the optical coupler of our interest, we have obtained closed form
analytic expressions for the spatial evolution of the different field
operators using the Sen-Mandal perturbative approach (discussed in Section \ref{Sol-of-eq}). In the past, photon statistics
and dynamics of the symmetric coupler of our interest was studied
with an assumption that both the second
harmonic modes are strong \cite{perina1996quantum}. This assumption circumvented
the use of a completely quantum mechanical description. The present investigation, which uses a completely quantum mechanical description, would not only reveal
the existence of the nonlinear quantum Zeno and anti-Zeno effects, it would also
establish the existence of the linear quantum Zeno and anti-Zeno effects. The
study would also show that switching between quantum Zeno and anti-Zeno
effects is possible by varying phase mismatches.

The rest of the chapter is organized as follows. In Section \ref{sec:System-and-solution},
we briefly describe the momentum operator for the symmetric nonlinear
optical coupler and the method used here to obtain the analytic expressions
of the spatial evolution of the field operators of various modes.
In Section \ref{sec:Linear-and-nonlinear}, the existence of quantum
Zeno and anti-Zeno effects is systematically investigated. Finally,
this chapter is concluded in Section \ref{sec:Conclusions-Zeno}.

\section{The system and the solution\label{sec:System-and-solution}}

\begin{figure}[t]
\centering{}\includegraphics[angle=-180,scale=0.44]{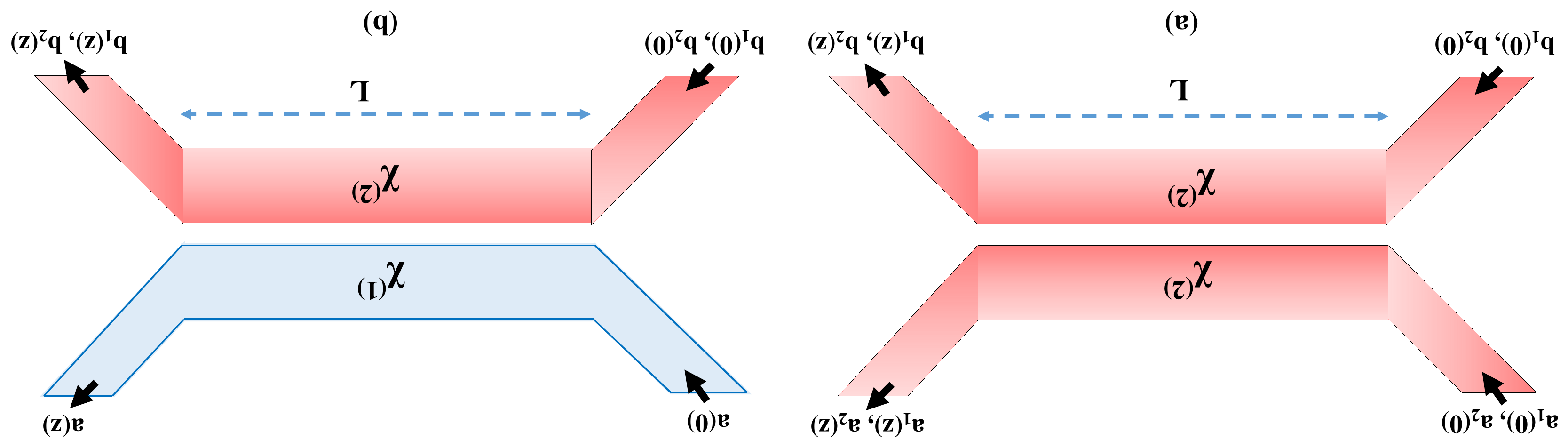}
\protect\caption[Schematic diagrams of
a symmetric and an asymmetric nonlinear optical couplers]{\label{fig:Schematic-diagram}  Schematic diagrams of
(a) a symmetric and (b) an asymmetric nonlinear optical couplers of
interaction length $L$ in a codirectional propagation of different
field modes involved. The symmetric coupler is prepared by combining
two nonlinear (quadratic) waveguides operating under second harmonic
generation; and in the asymmetric coupler, one nonlinear waveguide
of the symmetric coupler is replaced by a linear waveguide.}
\end{figure}

Momentum operator of a symmetric nonlinear optical coupler, prepared by combining two nonlinear (quadratic) waveguides operated by the second
harmonic generation (as shown in Figure \ref{fig:Schematic-diagram}
(a)), in the interaction picture is given by \cite[p.~114]{luks2009quantum}
\begin{equation}
\begin{array}{lcl}
G_{{\rm sym}} & = & \hbar ka_{1}b_{1}^{\dagger}+\hbar\Gamma_{a}a_{1}^{2}a_{2}^{\dagger}\exp(i\Delta k_{a}z) +  \hbar\Gamma_{b}b_{1}^{2}b_{2}^{\dagger}\exp(i\Delta k_{b}z)\,+{\rm H.c}.\,,
\end{array}\label{eq:Ham-symmetric}
\end{equation}
 where the annihilation (creation) operators $a_{i}\,(a_{i}^{\dagger})$
and $b_{i}\,(b_{i}^{\dagger})$ correspond to the field operators
in the two nonlinear waveguides. Here, $a_{1}\,(k_{a_{1}})$ and $a_{2}\,(k_{a_{2}})$
denote the annihilation operators (wavevectors) for the fundamental and second
harmonic modes, respectively, in the first waveguide. Similarly, $b_{1}\,(k_{b_{1}})$
and $b_{2}\,(k_{b_{2}})$ represent the annihilation operators (wavevectors)
for the fundamental and second harmonic modes, respectively, in the second
waveguide. Further, ${\rm H.c.}$ stands for the Hermitian conjugate;
$\Delta k_{j}=|2k_{j_{1}}-k_{j_{2}}|$ refers to the phase mismatch
between the fundamental and second harmonic beams; the parameters
$k$ and $\Gamma_{j}$ denote the linear and nonlinear coupling constants,
respectively, where $j\in\left\{ a,b\right\} $. The momentum operator
described above (\ref{eq:Ham-symmetric}) is completely quantum mechanical in the sense that
all the modes involved in the process are considered weak and treated
quantum mechanically. Thus, we consider the pump as weak and note that
$\Gamma_{j}\ll k$ as $k$ and $\Gamma_{j}$ are proportional to the
linear $(\chi^{(1)})$ and nonlinear $(\chi^{(2)})$ susceptibilities,
respectively, and usually $\chi^{(2)}/\chi^{(1)}\,\simeq10^{-6}$ as stated in the previous chapter.

For the study of quantum Zeno and anti-Zeno effects in a system, we
need a system momentum operator under a continuous interaction with
a probe. In this particular system--the symmetric nonlinear optical
coupler--we can consider that the waveguide which is described by $G_{{\rm sys}}=\hbar\Gamma_{b}b_{1}^{2}b_{2}^{\dagger}\exp(i\Delta k_{b}z)\,+{\rm H.c}.\,$,  can be considered  as the system as it 
is in a continuous interaction with the probe, which is described by
$G_{{\rm probe}}=\hbar ka_{1}b_{1}^{\dagger}+\hbar\Gamma_{a}a_{1}^{2}a_{2}^{\dagger}\exp(i\Delta k_{a}z)\,+{\rm H.c.}$
Here, the probe itself is considered to be nonlinear. Further, if
we take $\Gamma_{a}=0$ in Eq. (\ref{eq:Ham-symmetric}), i.e., if
we consider the probe to be linear then we obtain \cite{mandal2004approximate}
\begin{equation}
G_{{\rm asym}}=\hbar ka_{1}b_{1}^{\dagger}+\hbar\Gamma_{b} b_{1}^{2}b_{2}^{\dagger}\exp(i\Delta k_{b}z)\,+{\rm H.c}.\,,\label{eq:Ham-asymmetric}
\end{equation}
which is the momentum operator of an asymmetric nonlinear optical
coupler in the interaction picture studied in Chapter \ref{Coupler} (which is same as Eq. (\ref{eq:Mom-asymmetric}) after writing $a_{1}=a,\, \Gamma_{b}=\Gamma,$ and $\Delta k_{b}=\Delta k$). 

The spatial evolution of various modes involved in the momentum operators
(\ref{eq:Ham-symmetric}) and (\ref{eq:Ham-asymmetric}) can be obtained
as the simultaneous solution of the Heisenberg's equations of motion
corresponding to each mode. However, for the complex systems, such
as considered here, the closed form analytic solution can be obtained
only by using some perturbative methods. Here, we use the Sen-Mandal perturbative
method, already discussed in Section \ref{Sol-of-eq},
to obtain the spatial evolution
of all the field operators involved in (\ref{eq:Ham-symmetric}).

The solution of the momentum operator given in Eq. (\ref{eq:Ham-symmetric})
using the Sen-Mandal perturbative approach can be obtained once we write
the Heisenberg's equations of motion for all the field modes involved,
which are obtained as
\begin{equation}
\begin{array}{lcl}
\frac{da_{1}}{dz} & = & ik^{*}b_{1}+2i\Gamma_{a}^{*}a_{1}^{\dagger}a_{2}\exp\left(-i\Delta k_{a}z\right),\\
\frac{db_{1}}{dz} & = & ika_{1}+2i\Gamma_{b}^{*}b_{1}^{\dagger}b_{2}\exp\left(-i\Delta k_{b}z\right),\\
\frac{da_{2}}{dz} & = & i\Gamma_{a}a_{1}^{2}\exp\left(i\Delta k_{a}z\right),\\
\frac{db_{2}}{dz} & = & i\Gamma_{b}b_{1}^{2}\exp\left(i\Delta k_{b}z\right).
\end{array}\label{eq:Heisenberg's-eqs}
\end{equation}
In the Sen-Mandal  approach, the evolution of the field modes can be assumed (up to quadratic
terms in the nonlinear coupling constants $\Gamma_{i}$)  in the following forms 
\begin{equation}
\begin{array}{lcl}
a_{1}(z) & = & f_{1}a_{1}(0)+f_{2}b_{1}(0)+f_{3}a_{1}^{\dagger}(0)a_{2}(0)+f_{4}a_{2}(0)b_{1}^{\dagger}(0)+f_{5}b_{1}^{\dagger}(0)b_{2}(0)+f_{6}a_{1}^{\dagger}(0)b_{2}(0)\\
 & + & f_{7}a_{1}(0)a_{2}^{\dagger}(0)a_{2}(0)+f_{8}a_{1}^{\dagger}(0)a_{1}^{2}(0)+f_{9}a_{2}^{\dagger}(0)a_{2}(0)b_{1}(0)+f_{10}a_{1}^{\dagger}(0)a_{1}(0)b_{1}(0)\\
 & + & f_{11}a_{1}(0)b_{1}^{\dagger}(0)b_{1}(0)+f_{12}a_{1}^{\dagger}(0)b_{1}^{2}(0)+f_{13}a_{1}^{2}(0)b_{1}^{\dagger}(0)+f_{14}b_{1}^{\dagger}(0)b_{1}^{2}(0)\\
 & + & f_{15}a_{2}(0)b_{1}(0)b_{2}^{\dagger}(0)+f_{16}a_{1}(0)a_{2}(0)b_{2}^{\dagger}(0)+f_{17}a_{1}(0)a_{2}^{\dagger}(0)b_{2}(0)\\
 & + & f_{18}a_{2}^{\dagger}(0)b_{1}(0)b_{2}(0)+f_{19}b_{1}(0)b_{2}^{\dagger}(0)b_{2}(0)+f_{20}b_{1}^{\dagger}(0)b_{1}^{2}(0)+f_{21}a_{1}^{\dagger}(0)b_{1}^{2}(0)\\
 & + & f_{22}a_{1}(0)b_{2}^{\dagger}(0)b_{2}(0)+f_{23}a_{1}(0)b_{1}^{\dagger}(0)b_{1}(0)+f_{24}a_{1}^{\dagger}(0)a_{1}(0)b_{1}(0)+f_{25}a_{1}^{2}(0)b_{1}^{\dagger}(0)\\
 & + & f_{26}a_{1}^{\dagger}(0)a_{1}^{2}(0),\\
b_{1}(z) & = & g_{1}a_{1}(0)+g_{2}b_{1}(0)+g_{3}a_{1}^{\dagger}(0)a_{2}(0)+g_{4}a_{2}(0)b_{1}^{\dagger}(0)+g_{5}b_{1}^{\dagger}(0)b_{2}(0)+g_{6}a_{1}^{\dagger}(0)b_{2}(0)\\
 & + & g_{7}a_{1}(0)a_{2}^{\dagger}(0)a_{2}(0)+g_{8}a_{1}^{\dagger}(0)a_{1}^{2}(0)+g_{9}a_{2}^{\dagger}(0)a_{2}(0)b_{1}(0)+g_{10}a_{1}^{\dagger}(0)a_{1}(0)b_{1}(0)\\
 & + & g_{11}a_{1}(0)b_{1}^{\dagger}(0)b_{1}(0)+g_{12}a_{1}^{\dagger}(0)b_{1}^{2}(0)+g_{13}a_{1}^{2}(0)b_{1}^{\dagger}(0)+g_{14}b_{1}^{\dagger}(0)b_{1}^{2}(0)\\
 & + & g_{15}a_{2}(0)b_{1}(0)b_{2}^{\dagger}(0)+g_{16}a_{1}(0)a_{2}(0)b_{2}^{\dagger}(0)+g_{17}a_{1}(0)a_{2}^{\dagger}(0)b_{2}(0)\\
 & + & g_{18}a_{2}^{\dagger}(0)b_{1}(0)b_{2}(0)+g_{19}b_{1}(0)b_{2}^{\dagger}(0)b_{2}(0)+g_{20}b_{1}^{\dagger}(0)b_{1}^{2}(0)+g_{21}a_{1}^{\dagger}(0)b_{1}^{2}(0)\\
 & + & g_{22}a_{1}(0)b_{2}^{\dagger}(0)b_{2}(0)+g_{23}a_{1}(0)b_{1}^{\dagger}(0)b_{1}(0)+g_{24}a_{1}^{\dagger}(0)a_{1}(0)b_{1}(0)+g_{25}a_{1}^{2}(0)b_{1}^{\dagger}(0)\\
 & + & g_{26}a_{1}^{\dagger}(0)a_{1}^{2}(0),\\
 a_{2}(z) & = & h_{1}a_{2}(0)+h_{2}a_{1}^{2}(0)+h_{3}b_{1}(0)a_{1}(0)+h_{4}b_{1}^{2}(0)+h_{5}a_{1}^{\dagger}(0)a_{1}(0)a_{2}(0)+h_{6}a_{2}(0)\\
 & + & h_{7}a_{1}^{\dagger}(0)a_{2}(0)b_{1}(0)+h_{8}a_{1}(0)a_{2}(0)b_{1}^{\dagger}(0)+h_{9}a_{2}(0)b_{1}^{\dagger}(0)b_{1}(0)+h_{10}a_{1}(0)b_{1}^{\dagger}(0)b_{2}(0)\\
 & + & h_{11}b_{1}^{\dagger}(0)b_{1}(0)b_{2}(0)+h_{12}b_{2}(0)+h_{13}a_{1}^{\dagger}(0)a_{1}(0)b_{2}(0)+h_{14}a_{1}^{\dagger}(0)b_{1}(0)b_{2}(0),
 \end{array}\nonumber
\end{equation}
 \begin{equation}
\begin{array}{lcl}
b_{2}(z) & = & l_{1}b_{2}(0)+l_{2}b_{1}^{2}(0)+l_{3}b_{1}(0)a_{1}(0)+l_{4}a_{1}^{2}(0)+l_{5}b_{1}^{\dagger}(0)b_{1}(0)b_{2}(0)+l_{6}b_{2}(0)\\
 & + & l_{7}a_{1}(0)b_{1}^{\dagger}(0)b_{2}(0)+l_{8}a_{1}^{\dagger}(0)b_{1}(0)b_{2}(0)+l_{9}a_{1}^{\dagger}(0)a_{1}(0)b_{2}(0)+l_{10}a_{1}^{\dagger}(0)a_{2}(0)b_{1}(0)\\
 & + & l_{11}a_{1}^{\dagger}(0)a_{1}(0)a_{2}(0)+l_{12}a_{2}(0)+l_{13}a_{2}(0)b_{1}^{\dagger}(0)b_{1}(0)+l_{14}a_{1}(0)a_{2}(0)b_{1}^{\dagger}(0).
\end{array}\label{eq:assumed-sol}
\end{equation}

All the functions $f_{i},\,g_{i},\,h_{i},$ and $l_{i}$ can be obtained using
the assumed solution (\ref{eq:assumed-sol}) for different field modes
in the coupled differential equations given in Eq. (\ref{eq:Heisenberg's-eqs})
with the boundary conditions for all $F_{1}\left(z=0\right)=1,$ where
$F\in\left\{ f,g,h,l\right\} $. The closed form analytic solution
given in Eq. (\ref{eq:assumed-sol}) contains various coefficients,
for example, 
\[
\begin{array}{lcl}
l_{1} & = & 1,\\
l_{2} & = & -\frac{\Gamma_{b}G_{b-}^{*}}{2\Delta k_{b}}+\frac{iC_{b}}{2}\left[2|k|\left(G_{b+}^{*}-1\right)\sin2|k|z-i\Delta k_{b}\left(1-\left(G_{b+}^{*}-1\right)\cos2|k|z\right)\right],\\
l_{3} & = & \frac{-C_{b}|k|\left[i\Delta k_{b}\left(G_{b+}^{*}-1\right)\sin2|k|z+2|k|\left(1-\left(G_{b+}^{*}-1\right)\cos2|k|z\right)\right]}{k^{*}},\\
l_{4} & = & \frac{\Gamma_{b}|k|^{2}G_{b-}^{*}}{2k^{*^{2}}\Delta k_{b}}+\frac{iC_{b}|k|^{2}}{2k^{*^{2}}}\left[2|k|\left(G_{b+}^{*}-1\right)\sin2|k|z-i\Delta k_{b}\left(1-\left(G_{b+}^{*}-1\right)\cos2|k|z\right)\right],\\
l_{5} & = & \frac{\left|C_{b}\right|^{2}}{|k|\Delta k_{b}^{2}}\left[-16|k|^{5}\left(G_{b-}^{*}+i\Delta k_{b}z\right)-8i|k|^{4}\Delta k_{b}G_{b-}^{*}\sin2|k|z\right.+6i|k|^{2}\Delta k_{b}^{3}G_{b-}^{*}\sin2|k|z\\
 & - & i\Delta k_{b}^{5}\sin2|k|z+4|k|^{3}\Delta k_{b}^{2}\left(\cos2|k|z-1+3G_{b-}^{*}+3i\Delta k_{b}z\right)\\
 & + & \left.\Delta k_{b}^{4}|k|\left(\left(1-2G_{b-}^{*}\right)\cos2|k|z-1-2G_{b-}^{*}-2i\Delta k_{b}z\right)\right],\\
l_{6} & = & \frac{\left|C_{b}\right|^{2}}{\Delta k_{b}^{2}}\left[-16|k|^{4}\left(G_{b-}^{*}+i\Delta k_{b}z\right)-4i|k|\Delta k_{b}^{3}\right.\left(G_{b+}^{*}-1\right)\sin2|k|z+4\left(\cos2|k|z\left(G_{b+}^{*}-1\right) \right.\\
 & + & \left. 2G_{b-}^{*}-1+3i\Delta k_{b}z\right)|k|^{2}\Delta k_{b}^{2}+\left.\Delta k_{b}^{4}\left(\cos2|k|z\left(G_{b+}^{*}-1\right)-1-G_{b-}^{*}-2i\Delta k_{b}z\right)\right],\\
l_{7} & = & \frac{\left|C_{b}\right|^{2}}{k^{*}\Delta k_{b}}\left[2\Delta k_{b}^{4}\sin^{2}|k|z+4i|k|^{3}\Delta k_{b}\sin2|k|z\right.- i|k|\Delta k_{b}^{3}\left(2G_{b-}^{*}-1\right)\sin2|k|z\\
 & + & 8|k|^{4}\left(-G_{b-}^{*}\cos2|k|z+G_{b-}^{*}-i\Delta k_{b}z\right)+\left.2|k|^{2}\Delta k_{b}^{2}\left(3G_{b-}^{*}\cos2|k|z-G_{b-}^{*}+i\Delta k_{b}z\right)\right],\\
l_{8} & = & -\frac{\left|C_{b}\right|^{2}}{k\Delta k_{b}}\left[2\Delta k_{b}^{4}\sin^{2}|k|z-i|k|\Delta k_{b}^{3}\sin2|k|z\right.+4i|k|^{3}\Delta k_{b}\left(2G_{b-}^{*}-1\right)\sin2|k|z+2|k|^{2}
\Delta k_{b}^{2}\\
 & \times & \left.\left(\left(G_{b+}^{*}+2\right)\cos2|k|z-G_{b-}^{*}-4-i\Delta k_{b}z\right)+ 8|k|^{4}\left(-G_{b-}^{*}\cos2|k|z+G_{b-}^{*}+i\Delta k_{b}z\right)\right],\\
l_{9} & = & \frac{\left|C_{b}\right|^{2}}{|k|\Delta k_{b}^{2}}\left[-16|k|^{5}\left(G_{b-}^{*}+i\Delta k_{b}z\right)+\left\{ 8i|k|^{4}\Delta k_{b}G_{b-}^{*}\right.\right.-\left.2i|k|^{2}\Delta k_{b}^{3}\left(G_{b+}^{*}+2\right)+i\Delta k_{b}^{5}\right\}\\
 & \times &  \sin2|k|z+4|k|^{3}\left(\left(1-2G_{b-}^{*}\right)\cos2|k|z-1+G_{b-}^{*}+3i\Delta k_{b}z\right)\\
 & \times & \left.\Delta k_{b}^{2}+|k|\Delta k_{b}^{4}\left(\cos2|k|z-1-2i\Delta k_{b}z\right)\right],
 \end{array}
\]
\begin{equation}
 \begin{array}{lcl}
l_{10} & = & \frac{2C_{ab}|k|\left(G_{ab+}-1\right)}{k^{*}}\left[2i|k|^{2}\Delta k_{b}\Delta k_{ab}\sin2|k|z\left\{ 4|k|^{2}\right.\right.\left(\Delta k_{b}\left(G_{a-}^{*}-1\right)-\Delta k_{a}\left(G_{a-}^{*}+1\right)+\Delta k_{b}\right)\\
 & - & \left(\Delta k_{b}^{2}\left(\Delta k_{b}-2\Delta k_{a}\right)+\left(\Delta k_{ab}^{3}-2\Delta k_{a}^{2}\Delta k_{b}\right)\left.\left(G_{a-}^{*}-1\right)\right)\right\} -i\Delta k_{a}^{2}\Delta k_{b}^{2}\Delta k_{ab}^{3}\left(G_{a+}^{*}-1\right)\\
 & \times & \sin2|k|z+\left(-\Delta k_{b}\Delta k_{ab}G_{a-}^{*}\cos2|k|z\right.+\left(\Delta k_{a}\left(G_{ab-}^{*}-1\right)+\Delta k_{ab}\left(G_{a+}^{*}-1\right)+\Delta k_{b}\right)\\
 & \times & \left.\Delta k_{a}\right)16|k|^{5}-4|k|^{3}\left\{ \Delta k_{b}\Delta k_{ab}\cos2|k|z\right.\left(\Delta k_{a}^{2}\left(3G_{a+}^{*}-2\right)-\Delta k_{a}\Delta k_{b}G_{a+}^{*}-\Delta k_{b}^{2}G_{a-}^{*}\right)\\
 & + & \Delta k_{a}\left(\left(\Delta k_{b}^{2}\Delta k_{ab}+\Delta k_{ab}^{3}\right)\left(G_{a+}^{*}-1\right)\right.+\Delta k_{b}^{3}+\Delta k_{b}\Delta k_{ab}^{2}-\left(G_{ab+}^{*}-1\right)\\
 & \times & \left.\left.\left(2\Delta k_{a}^{2}\Delta k_{b}-4\Delta k_{a}\Delta k_{b}^{2}+\Delta k_{a}^{3}+2\Delta k_{b}^{3}\right)\right)\right\} -|k|\Delta k_{a}\Delta k_{b}\Delta k_{ab}^{2}\left(\Delta k_{b}\Delta k_{ab}\left(G_{a-}^{*}-1\right)\right.\\
 & + & 2\Delta k_{a}^{2}\left(G_{ab+}^{*}-1\right)-\Delta k_{b}^{2}+\cos2|k|z\left(-\Delta k_{b}^{2}\right.+ \left.\left.\left.\left(\Delta k_{a}\Delta k_{b}-2\Delta k_{a}^{2}+\Delta k_{b}^{2}\right)\left(G_{a+}^{*}-1\right)\right)\right)\right],\\
l_{11} & = & -\frac{2C_{ab}k\left(G_{ab+}-1\right)}{k^{*}}\left[32|k|^{6}\left(\Delta k_{a}\left(G_{ab-}^{*}-1\right)+\Delta k_{b}\right.\right.\left.\Delta k_{ab}\left(G_{a+}^{*}-1\right)\right)+16i\Delta k_{b}\Delta k_{ab}|k|^{5}G_{a-}^{*}\\
 & \times & \sin2|k|z-2\Delta k_{a}^{2}\Delta k_{b}^{2}\Delta k_{ab}^{3}\left(G_{a+}^{*}-1\right)\sin^{2}|k|z+\left\{\left(-\Delta k_{a}\Delta k_{b}G_{a+}^{*}-\Delta k_{b}^{2}G_{a-}^{*}+\Delta k_{a}^{2} \right.\right.\\
 & \times & \left.\left(3G_{a+}^{*}-2\right)\right)  4i\Delta k_{b}\Delta k_{ab}|k|^{3} - i\Delta k_{a}\left(\Delta k_{b}^{2}+\left(2\Delta k_{a}^{2}-\Delta k_{a}\Delta k_{b}-\Delta k_{b}^{2}\right)\left(G_{a+}^{*}-1\right)\right)\\
 & \times & \left.\Delta k_{b}\Delta k_{ab}^{2}|k|\right\} \sin2|k|z-8|k|^{4}\left\{ \Delta k_{b}\Delta k_{ab}\cos2|k|z\right.\left(\Delta k_{b}G_{a-}^{*}-\Delta k_{a}\left(G_{a-}^{*}+1\right)\right)-\Delta k_{a}\\
 & \times & \left(G_{ab+}^{*}-1\right)\left(\Delta k_{a}\Delta k_{b}+3\Delta k_{ab}^{2}\right)+\Delta k_{b}\left(\Delta k_{ab}^{2}+\Delta k_{b}^{2}\right)+\left(G_{a+}^{*}\right.-\left.\left.1\right)\left(-5\Delta k_{a}^{2}\Delta k_{b}\right.\right.\\
 & + &\left.\left. 4\Delta k_{a}\Delta k_{b}^{2}+3\Delta k_{a}^{3}-2\Delta k_{b}^{3}\right)\right\} +2\Delta k_{ab}|k|^{2}\left\{ \Delta k_{b}\left(\Delta k_{b}^{2}\left(\Delta k_{b}-2\Delta k_{a}\right)+\left(G_{a+}^{*}-1\right)\right.\right.\\
 & \times & \left.\left(\Delta k_{ab}^{3}-2\Delta k_{a}^{2}\Delta k_{b}\right)\right)\cos2|k|z+\Delta k_{a}^{3}\left(G_{ab-}^{*}+1\right)\left(2\Delta k_{ab}-\Delta k_{b}\right)+\Delta k_{b}^{3}\Delta k_{ab}\\
 & + & \left(G_{a+}^{*}-1\right)\left.\left.\left(2\Delta k_{a}^{2}\Delta k_{ab}^{2}-2\Delta k_{a}\Delta k_{b}^{3}+3\Delta k_{a}^{2}\Delta k_{b}^{2}+\Delta k_{b}^{4}\right)\right\} \right],\\
l_{12} & = & \frac{-2C_{ab}k\left(G_{ab+}-1\right)\left(4|k|^{2}-\Delta k_{ab}^{2}\right)}{k^{*}}\left[8|k|^{4}\left(\Delta k_{a}\left(G_{ab-}^{*}-1\right)\right.\right.+\left.\Delta k_{ab}\left(G_{a+}^{*}-1\right)+\Delta k_{b}\right)+\Delta k_{a}^{2}\Delta k_{b}^{2}\\
 & \times & \Delta k_{ab}\sin^{2}|k|z\left(G_{a+}^{*}-1\right)-2|k|^{2}\left\{ \Delta k_{a}\Delta k_{ab}\left(G_{a+}^{*}-1\right)\cos2|k|z\right.\Delta k_{b}+\Delta k_{ab}\left(\Delta k_{a}^{2}+\Delta k_{b}^{2}\right)\\
 & \times & \left(G_{a+}^{*}-1\right)+\Delta k_{a}^{3} \left.\left(G_{a-}^{*}-1\right)\Delta k_{b}^{3}\right\} +i\Delta k_{a}\Delta k_{b}|k|\left(\Delta k_{a}^{2}-\Delta k_{b}^{2}\right)\left.\left(G_{a+}^{*}-1\right)\sin2|k|z\right],\\
l_{13} & = & \frac{-2C_{ab}k|k|\left(G_{ab+}-1\right)}{k^{*}}\left[32|k|^{5}\left(\Delta k_{a}\left(G_{ab-}^{*}-1\right)+\Delta k_{b}\right.\right.+\left.\Delta k_{ab}\left(G_{a+}^{*}-1\right)\right)-4i\Delta k_{b}\Delta k_{ab}|k|^{2}\\
 & \times & \sin2|k|z\left\{ 4|k|^{2}G_{a-}^{*}-\Delta k_{a}\Delta k_{b}\left(3G_{a+}^{*}-2\right)+\Delta k_{a}^{2}G_{a+}^{*}\right.-\left.\Delta k_{b}^{2}G_{a-}^{*}\right\} -i\Delta k_{a}\Delta k_{b}^{2}\Delta k_{ab}^{2}\\
 & \times & \sin2|k|z\left(-\Delta k_{b}+\Delta k_{ab}\left(G_{a+}^{*}-1\right)\right)-8|k|^{3}\left\{ \Delta k_{b}\Delta k_{ab}\right.\left(-\Delta k_{b}G_{a-}^{*}+\Delta k_{a}\left(G_{a+}^{*}+1\right)\right)\\
 & \times & \cos2|k|z-\Delta k_{ab}\left(G_{ab+}^{*}-1\right)\left(\Delta k_{a}^{2}+\Delta k_{b}\Delta k_{ab}\right)+\left.\left(\Delta k_{b}+\Delta k_{ab}\left(G_{a+}^{*}-1\right)\right)\left(\Delta k_{b}^{2}+\Delta k_{ab}^{2}\right)\right\} \\
 & - & 2\Delta k_{b}\Delta k_{ab}|k|\left\{ \cos2|k|z\left(-\Delta k_{b}^{2}\left(\Delta k_{a}+\Delta k_{ab}\right)\right.\right.-\left.\Delta k_{ab}^{2}\left(\Delta k_{a}+\Delta k_{b}\right)\left(G_{a+}^{*}-1\right)\right)-\Delta k_{b}\Delta k_{ab}^{2}\\
 & \times & \left.\left.\left(G_{a+}^{*}-1\right)+\Delta k_{a}^{3}\left(G_{ab+}^{*}-1\right)-\Delta k_{b}^{2}\Delta k_{ab}\right\} \right],\\
l_{14} & = & \frac{-2C_{ab}k^{2}\left(G_{ab+}-1\right)}{k^{*}}\left[2\Delta k_{a}\Delta k_{b}^{2}\Delta k_{ab}^{2}\left(-\Delta k_{b}+\Delta k_{ab}\right.\right.\left.\left(G_{a+}^{*}-1\right)\right)+2i\Delta k_{b}\Delta k_{ab}|k|\sin2|k|z\left\{ -4\right.\\
 & \times & |k|^{2}\left(-\Delta k_{b}G_{a-}^{*}+\Delta k_{a}\left(G_{a+}^{*}+1\right)\right)+\left(\Delta k_{a}+\Delta k_{ab}\right)\left.\Delta k_{b}^{2}+\Delta k_{ab}^{2}\left(\Delta k_{a}+\Delta k_{b}\right)\left(G_{a+}^{*}-1\right)\right\} \\
 & + & 16|k|^{4}\left\{ -\Delta k_{b}\Delta k_{ab}G_{a-}^{*}\cos2|k|z+\Delta k_{a}\left(\left(G_{a+}^{*}-1\right)\right.\right.\left.\Delta k_{ab}+\left(\left(\Delta k_{b}-\Delta k_{ab}\right)\left(G_{ab+}^{*}-1\right)\right.\right.\\
 & - & \left.\left.\Delta k_{b}\right)\right\} -4|k|^{2}\left\{ \Delta k_{b}\Delta k_{ab}\left(-\Delta k_{a}\Delta k_{b}\left(3G_{a+}^{*}-2\right)-\Delta k_{b}^{2}\right.\right.\left.G_{a-}^{*}+\Delta k_{a}^{2}G_{a+}^{*}\right)\cos2|k|z\\
 & + & \Delta k_{a}\left(\left(\Delta k_{b}-\Delta k_{ab}\right)\right.\Delta k_{a}^{2}\left(G_{ab+}^{*}-1\right)+\left(\Delta k_{b}^{2}+\Delta k_{ab}^{2}\right)\left(-\Delta k_{b}\right.+\left.\left.\left.\left.\Delta k_{ab}\left(G_{a+}^{*}-1\right)\right)\right)\right\} \right]
\end{array}\label{eq:coefficient-l}
\end{equation}
with $G_{i\pm}=\left(1\pm\exp(-i\Delta k_{i}z)\right)$ for $i\in\left\{ a,b,ab\right\}$,
and $\Delta k_{ab}=\Delta k_{a}-\Delta k_{b}$. Also, $C_{a}=\frac{\Gamma_{a}}{\left[4|k|^{2}-\Delta k_{a}^{2}\right]}$, $C_{b}=\frac{\Gamma_{b}}{\left[4|k|^{2}-\Delta k_{b}^{2}\right]}$, and $C_{ab}=\frac{\Gamma_{a}^{*}\Gamma_{b}}{\Delta k_{a}\Delta k_{b}\Delta k_{ab}\left[4|k|^{2}-\Delta k_{a}^{2}\right]\left[4|k|^{2}-\Delta k_{b}^{2}\right]\left[4|k|^{2}-\Delta k_{ab}^{2}\right]}$. 

Here, we report the closed form analytic
expression for the spatial evolution of $b_{2}$ mode only, i.e., $b_{2}(z)$. We restrict our description
to $b_{2}(z)$ as only the coefficients present in the analytic expression
of $b_{2}(z)$ appear in the expressions of the linear and nonlinear Zeno
parameters. To maintain the flow of the chapter, the detailed solution is not shown here. The same may be found at the supplementary material of Ref. \cite{thapliyal2016linear}, which reports the finding of this Chapter of the thesis, at \cite{supplZeno}. In
what follows, we use the expression of $b_{2}(z)$ to investigate the linear and nonlinear quantum Zeno and anti-Zeno effects
in the optical couplers.

\section{Linear and nonlinear quantum Zeno and anti-Zeno effects \label{sec:Linear-and-nonlinear}}

Being consistent with the theme of the present work, the presence
of the quantum Zeno and anti-Zeno effects with a nonlinear probe corresponds
to the nonlinear quantum Zeno and anti-Zeno effects. Similarly, a
linear probe will give the linear quantum Zeno and anti-Zeno effects.
Further, it has already been mentioned in Section \ref{sec:Introduction}
that analytical expressions for the Zeno parameter for a linear probe
can be obtained as the limiting cases of the expressions obtained
for the nonlinear probe by neglecting the nonlinearity present in the
probe \cite{li2006continuous}. A quite similar analog of the linear and nonlinear
quantum Zeno and anti-Zeno effects were also discussed in the recent
past \cite{abdullaev2011linear,shchesnovich2010control} in other
physical systems.

\subsection{Number operator and Zeno parameter}

The analytic expression of the number operator for the second harmonic
field mode in the system waveguide, i.e., $b_{2}$ mode using Eq.
(\ref{eq:assumed-sol}), is given by
\begin{equation}
\begin{array}{lcl}
N_{b_{2}}\left(z\right) & = & b_{2}^{\dagger}\left(z\right)b_{2}\left(z\right)\\
 & = & b_{2}^{\dagger}\left(0\right)b_{2}\left(0\right)+\left|l_{2}\right|^{2}b_{1}^{\dagger2}\left(0\right)b_{1}^{2}\left(0\right)+\left|l_{3}\right|^{2}a_{1}^{\dagger}\left(0\right)a_{1}\left(0\right)b_{1}^{\dagger}\left(0\right)b_{1}\left(0\right)\\
 & + & \left|l_{4}\right|^{2}a_{1}^{\dagger2}\left(0\right)a_{1}^{2}\left(0\right)+ \left[l_{2}b_{2}^{\dagger}\left(0\right)b_{1}^{2}\left(0\right)\right.+l_{3}b_{2}^{\dagger}\left(0\right)b_{1}\left(0\right)a_{1}\left(0\right)+l_{4}b_{2}^{\dagger}\left(0\right)a_{1}^{2}\left(0\right)\\
 & + & l_{2}^{*}l_{3}a_{1}\left(0\right)b_{1}^{\dagger2}\left(0\right)b_{1}\left(0\right)+l_{2}^{*}l_{4}a_{1}^{2}\left(0\right)b_{1}^{\dagger2}\left(0\right)+l_{3}^{*}l_{4}a_{1}^{\dagger}\left(0\right)a_{1}^{2}\left(0\right)b_{1}^{\dagger}\left(0\right)\\
 & + & l_{5}b_{1}^{\dagger}\left(0\right)b_{1}\left(0\right)b_{2}^{\dagger}\left(0\right)b_{2}\left(0\right)+l_{6}b_{2}^{\dagger}\left(0\right)b_{2}\left(0\right)+l_{7}a_{1}\left(0\right)b_{1}^{\dagger}\left(0\right)b_{2}^{\dagger}\left(0\right)b_{2}\left(0\right)\\
 & + & l_{8}a_{1}^{\dagger}\left(0\right)b_{1}\left(0\right)b_{2}^{\dagger}\left(0\right)b_{2}\left(0\right)+l_{9}a_{1}^{\dagger}\left(0\right)a_{1}\left(0\right)b_{2}^{\dagger}\left(0\right)b_{2}\left(0\right)\\
 & + & l_{10}a_{1}^{\dagger}\left(0\right)a_{2}\left(0\right)b_{1}\left(0\right)b_{2}^{\dagger}\left(0\right)+l_{11}a_{1}^{\dagger}\left(0\right)a_{1}\left(0\right)a_{2}\left(0\right)b_{2}^{\dagger}\left(0\right)+l_{12}a_{2}\left(0\right)b_{2}^{\dagger}\left(0\right)\\
 & + & \left.l_{13}a_{2}\left(0\right)b_{1}^{\dagger}\left(0\right)b_{1}\left(0\right)b_{2}^{\dagger}\left(0\right)+l_{14}a_{1}\left(0\right)a_{2}\left(0\right)b_{1}^{\dagger}\left(0\right)b_{2}^{\dagger}\left(0\right)+{\rm H.c.}\right],
\end{array}\label{eq:nb2z}
\end{equation}
where the functional form of coefficients $l_{i}$ is given in Eq.
(\ref{eq:coefficient-l}). 

Here, we have considered the initial state to be
a multi-mode coherent state $|\alpha\rangle|\beta\rangle|\gamma\rangle|\delta\rangle$,
which is the product of four single-mode coherent states $|\alpha\rangle,\,|\beta\rangle$,
$|\gamma\rangle$, and $|\delta\rangle$, where $|\alpha\rangle,\,|\beta\rangle$,
$|\gamma\rangle$, and $|\delta\rangle$ are the eigenkets of the
annihilation operators for the corresponding field modes, i.e., $a_{1}$,
$b_{1}$, $a_{2}$, and $b_{2}$, respectively. For example, $b_{1}(0)|\alpha\rangle|\beta\rangle|\gamma\rangle|\delta\rangle=\beta|\alpha\rangle|\beta\rangle|\gamma\rangle|\delta\rangle$
and $a_{1}(0)|\alpha\rangle|\beta\rangle|\gamma\rangle|\delta\rangle=\alpha|\alpha\rangle|\beta\rangle|\gamma\rangle|\delta\rangle,$
where $|\alpha|^{2},\,|\beta|^{2},\,|\gamma|^{2},$ and $|\delta|^{2}$
are the initial number of photons in the field modes $a_{1}$, $b_{1}$ 
$a_{2}$, and $b_{2}$, respectively. Further, the symmetric nonlinear optical
coupler and its approximated special case of the asymmetric nonlinear
optical coupler can operate under two conditions: spontaneous and
stimulated. In the spontaneous (stimulated) case, initially, i.e.,
at $t=0,$ there is no photon (a nonzero number of photons) in the
second harmonic modes of the system, whereas average photon numbers
in the other modes are nonzero. 

Following some of the earlier works \cite{rehacek2001quantum,thun2002quantum,mista2000quantum,rehacek2000quantum},
the effect of the presence of the probe mode on the photon statistics
of the second harmonic mode of the system is investigated using Zeno
parameter ($\Delta N_{Z}$), introduced by us in \cite{thapliyal2015quantum}, which is defined as 
\begin{equation}
\Delta N_{Z}=\left\langle N_{X}\left(z\right)\right\rangle -\left\langle N_{X}\left(z\right)\right\rangle _{k=0}.\label{eq:zeno-parameter}
\end{equation}

The Zeno parameter is a measure of the effect caused on the evolution
of the photon statistics of the system (obtained for mode $X$) due
to its interaction with the probe. It can be inferred from Eq. (\ref{eq:zeno-parameter})
that the negative values of the Zeno parameter signify that the continuous
measurement via the probe inhibited the evolution of mode $X$ by
decreasing the photon generation in that particular mode, which demonstrates the
presence of the quantum Zeno effect. On the other hand, the positive
values of the Zeno parameter correspond to the enhancement of the photon
generation due to coupling with the auxiliary mode in the probe. This
is a signature of the presence of the quantum anti-Zeno effect.

For the system of our interest, the symmetric nonlinear optical coupler,
using the analytic expression of the photon number operator in Eq.
(\ref{eq:nb2z}), the Zeno parameter can be calculated for the second harmonic
mode of the system waveguide as 
\begin{equation}
\begin{array}{lcl}
\Delta N_{NZ} & = & \left(\left|l_{2}\right|^{2}-\left|p_{2}\right|^{2}\right)\left|\beta\right|^{4}+\left|l_{3}\right|^{2}\left|\alpha\right|^{2}\left|\beta\right|^{2}+ \left|l_{4}\right|^{2}\left|\alpha\right|^{4}+\left[\left(l_{2}-p_{2}\right)\beta^{2}\delta^{*}+l_{3}\alpha\beta\delta^{*}+l_{4}\alpha^{2}\delta^{*}\right.\\
 & + & l_{2}^{*}l_{3}\left|\beta\right|^{2}\alpha\beta^{*}+l_{2}^{*}l_{4}\alpha^{2}\beta^{*2}+l_{3}^{*}l_{4}\left|\alpha\right|^{2}\alpha\beta^{*}+\left(l_{5}-p_{5}\right)\left|\beta\right|^{2}\left|\delta\right|^{2}+\left(l_{6}-p_{6}\right)\left|\delta\right|^{2}
 \\
 & + & 
 l_{7}\left|\delta\right|^{2}\alpha\beta^{*}+l_{8}\left|\delta\right|^{2}\alpha^{*}\beta+l_{9}\left|\alpha\right|^{2}\left|\delta\right|^{2}+l_{10}\alpha^{*}\beta\gamma\delta^{*}+l_{11}\left|\alpha\right|^{2}\gamma\delta^{*}+l_{12}\gamma\delta^{*}\\
 & + & \left.l_{13}\left|\beta\right|^{2}\gamma\delta^{*}+l_{14}\alpha\beta^{*}\gamma\delta^{*}+{\rm c.c.}\right],
\end{array}\label{eq:NL-zeno-parameter}
\end{equation}
where 
\begin{equation}
\begin{array}{lcl}
p_{2} & = & -\frac{\Gamma_{b}G_{-}^{*}}{\Delta k_{b}},\\
p_{5} & = & 2p_{6}=-\frac{4\left|\Gamma_{b}\right|^{2}\left(G_{-}^{*}+i\Delta k_{b}z\right)}{\left(\Delta k_{b}\right)^{2}}.
\end{array}\label{eq:atk0}
\end{equation}
Here, $p_{i}$s are obtained by taking $k=0$ in corresponding $l_{i}$s
in Eq. (\ref{eq:coefficient-l}). All the remaining
$p_{i}$s vanish in the absence of the probe. The subscript $NZ$
in the Zeno parameter corresponds to the physical situation where
a nonlinear probe is used, i.e., when ``nonlinear Zeno'' effect is investigated.
Thus, $\Delta N_{NZ}$ can be referred to as the nonlinear Zeno parameter.
Similarly, $\Delta N_{LZ}$ will denote linear Zeno parameter, i.e.,
Zeno parameter for a physical situation where the linear probe is used.

It is easy to obtain the Zeno parameter for the spontaneous case. Specifically,
in the spontaneous case, i.e., in the absence of any photon in the second
harmonic mode of the system at $t=0$ (or considering $\delta=0$
at $t=0)$, the analytic expression of the nonlinear Zeno parameter
can be obtained from Eq. (\ref{eq:NL-zeno-parameter}) by keeping
the $\delta$ independent terms as 
\begin{equation}
\begin{array}{lcl}
\left(\Delta N_{NZ}\right)_{\delta=0} & = & \left(\left|l_{2}\right|^{2}-\left|p_{2}\right|^{2}\right)\left|\beta\right|^{4}+\left|l_{3}\right|^{2}\left|\alpha\right|^{2}\left|\beta\right|^{2}+\left|l_{4}\right|^{2}\left|\alpha\right|^{4}+\left[l_{2}^{*}l_{3}\left|\beta\right|^{2}\alpha\beta^{*}+l_{2}^{*}l_{4}\alpha^{2}\beta^{*2}\right.\\
 & + & \left.l_{3}^{*}l_{4}\left|\alpha\right|^{2}\alpha\beta^{*}+{\rm c.c.}\right].
\end{array}\label{eq:spon-nonlin-Z-par}
\end{equation}

In Section \ref{sec:System-and-solution}, we have already mentioned
that the momentum operator for an asymmetric nonlinear optical coupler
($\chi^{\left(2\right)}-\chi^{\left(1\right)})$ can be obtained by
just neglecting the nonlinear coupling terms in one of the nonlinear
waveguides present in the symmetric nonlinear coupler ($\chi^{\left(2\right)}-\chi^{\left(2\right)}$)
studied here. Thus, we may consider the probe in the nonlinear Zeno
parameter obtained in Eq. (\ref{eq:NL-zeno-parameter}) to be linear
by taking $\Gamma_{a}=0$. This is how we can obtain the expression
for the linear Zeno parameter. Thus, the Zeno parameter of the asymmetric
nonlinear optical coupler characterized by Eq. (\ref{eq:Ham-asymmetric})
can be obtained as 
\begin{equation}
\begin{array}{lcl}
\Delta N_{LZ} & = & \left(\left|l_{2}\right|^{2}-\left|p_{2}\right|^{2}\right)\left|\beta\right|^{4}+\left|l_{3}\right|^{2}\left|\alpha\right|^{2}\left|\beta\right|^{2}+\left|l_{4}\right|^{2}\left|\alpha\right|^{4}+\left[\left(l_{2}-p_{2}\right)\beta^{2}\delta^{*}+l_{3}\alpha\beta\delta^{*}+l_{4}\alpha^{2}\delta^{*}\right.\\
 & + & l_{2}^{*}l_{3}\left|\beta\right|^{2}\alpha\beta^{*}+l_{2}^{*}l_{4}\alpha^{2}\beta^{*2}+l_{3}^{*}l_{4}\left|\alpha\right|^{2}\alpha\beta^{*}+\left(l_{5}-p_{5}\right)\left|\beta\right|^{2}\left|\delta\right|^{2}+\left(l_{6}-p_{6}\right)\left|\delta\right|^{2}\\
 & + & \left.l_{7}\left|\delta\right|^{2}\alpha\beta^{*}+l_{8}\left|\delta\right|^{2}\alpha^{*}\beta+l_{9}\left|\alpha\right|^{2}\left|\delta\right|^{2}+{\rm c.c.}\right].
\end{array}\label{eq:L-zeno-parameter}
\end{equation}
Further, if we neglect all the terms beyond linear power in nonlinear
coupling constant $\Gamma_{b}$ in Eq. (\ref{eq:L-zeno-parameter}),
we find that the result obtained here matches exactly with the result reported by us in
Ref. \cite{thapliyal2015quantum}. Interestingly, the analytic expressions
of the nonlinear and linear Zeno parameters have the same form
in the spontaneous case. It can also be checked that the expression
obtained in Eq. (\ref{eq:spon-nonlin-Z-par}) vanishes if we neglect
all the terms beyond linear powers in the nonlinear coupling constant.
This is also consistent with the earlier findings of ours \cite{thapliyal2015quantum}, which was obtained using the Sen-Mandal perturbative solution of asymmetric codirectional coupler reported in Section \ref{Sol-of-eq}.

\subsection{Variation of Zeno parameter with different variables}

\begin{figure}
\centering{}\includegraphics[angle=-90,scale=0.6]{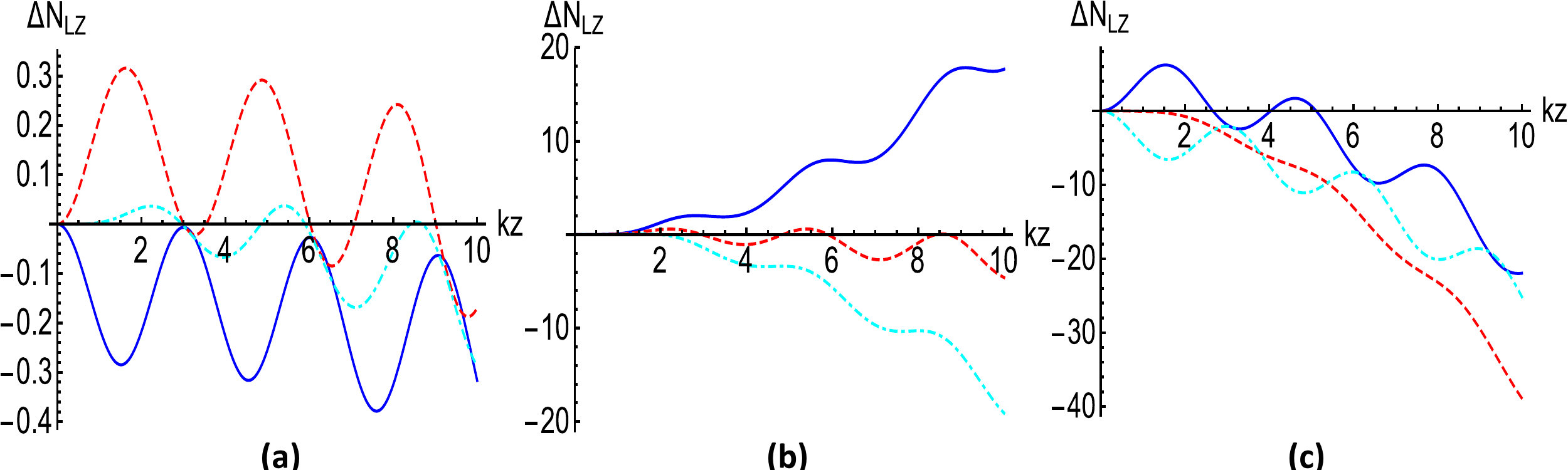}
\protect\caption[Linear
Zeno parameter with different intensity and phase of input fields for the spontaneous and stimulated cases]{\label{fig:Linear-Zeno-z}  (a) Variation in the linear
Zeno parameter with rescaled interaction length for $\frac{\Gamma_{b}}{k}=10^{-2},\,\frac{\Delta k_{b}}{k}=10^{-3}$
with $\alpha=5,\,\beta=3,$ and $\delta=1$ and -1 in the stimulated case
(smooth and dashed lines) and $\delta=0$ in the spontaneous case (dot-dashed
line). (b) In the spontaneous case, $\alpha=10$ with $\beta=3,6,$ and
7 for the smooth, dashed, and dot-dashed lines, respectively. (c) $\alpha=10,\,\beta=8,$
and $\delta=-4,\,0,$ and 4 for the smooth, dashed, and dot-dashed lines,
respectively. All the quantities shown in the present chapter are dimensionless unless stated otherwise.}
\end{figure}

The analytic expressions obtained for both nonlinear and linear Zeno
parameters depend on various parameters, such as photon numbers and
phases of different field modes, linear and nonlinear coupling, interaction
length, phase mismatch between the fundamental and second harmonic
modes in the nonlinear waveguides. In the spontaneous case,
the system shows quantum anti-Zeno effect initially which eventually goes
towards quantum Zeno effect with increase in the rescaled interaction
length (cf. Figure \ref{fig:Linear-Zeno-z} (a)). This behavior of the linear Zeno parameter in the spontaneous
case is further elaborated in Figure \ref{fig:Linear-Zeno-z} (b), where
it can be observed that as the number of photons in the linear mode
of the system waveguide become comparable to the photon numbers in
the probe mode, quantum Zeno effect is prominent. A similar effect
on the photon numbers is observed even in the stimulated case in Figure
\ref{fig:Linear-Zeno-z} (c), where the transition to quantum Zeno effect
with increasing rescaled interaction length is more dominating than
in Figure \ref{fig:Linear-Zeno-z} (a). Further, in the stimulated case,
a transition between the linear quantum Zeno and anti-Zeno effects can
be obtained by controlling the phase of the second harmonic mode of the
system waveguide as illustrated in Figure \ref{fig:Linear-Zeno-z} (a).
However, with an increase in the number of photons in the fundamental mode
of the system waveguide shown in Figure \ref{fig:Linear-Zeno-z} (c),
this nature disappears gradually  as it  tends
toward the quantum Zeno effect.

\begin{figure}
\centering{}\includegraphics[angle=-90,scale=0.6]{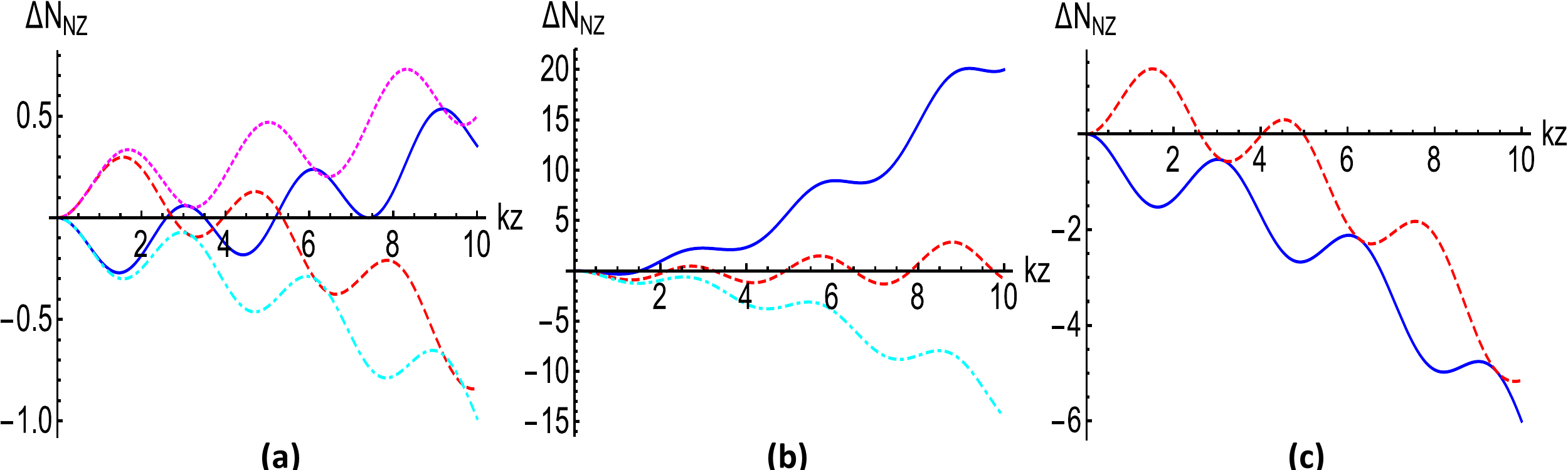}
\protect\caption[Nonlinear
Zeno parameter with different intensity and phase of inputs fields for the stimulated case]{\label{fig:Nonlinear-z}  (a) Nonlinear Zeno parameter
with $\frac{\Gamma_{a}}{k}=\frac{\Gamma_{b}}{k}=10^{-2},\,\frac{\Delta k_{a}}{k}=1.1\times10^{-3},\,\frac{\Delta k_{b}}{k}=10^{-3}$
with $\alpha=5,\,\beta=3,$ and $\gamma=2,\,\delta=1$ and -1 in the stimulated
case (smooth and dashed lines) and $\gamma=-2,\,\delta=1$ and -1
in the dot-dashed and dotted lines. (b) $\alpha=10,\,\gamma=3,\,\delta=1$
with $\beta=3,6,$ and 7 for the smooth, dashed and dot-dashed lines,
respectively. (c) shows variation in the nonlinear Zeno parameter due to 
change in the phase of $\alpha$ or $\beta$ by an amount of $\pi$ for $\alpha=\beta=6,$ and $\gamma=3,\,\delta=2$. It is also observed
that the change in phase of $\alpha$ is equivalent to that 
of $\beta$.}
\end{figure}

\begin{figure}[t]
\begin{centering}
\includegraphics[angle=-90,scale=0.9]{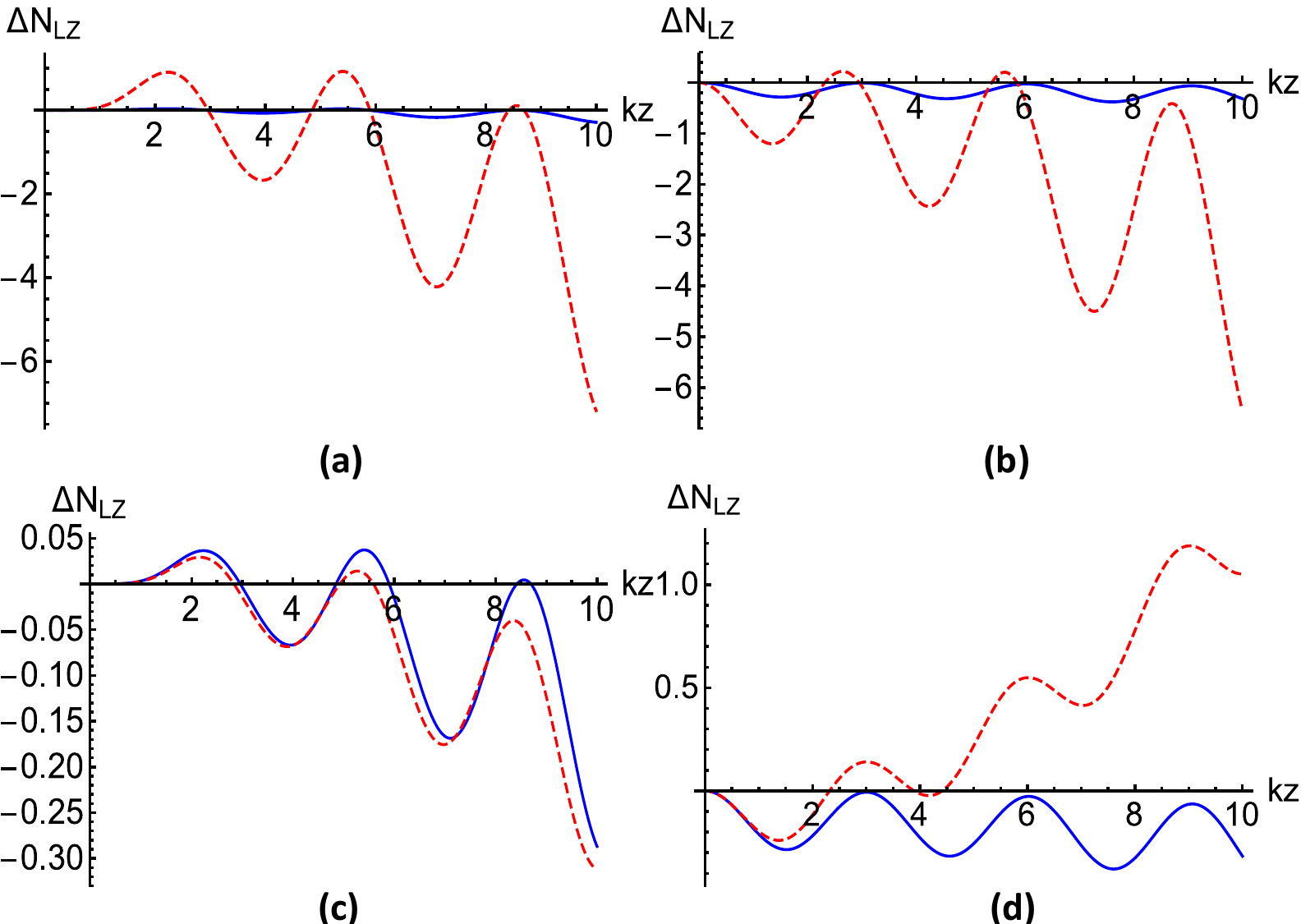}
\par\end{centering}
\protect\caption[The effect of change in the linear coupling and phase mismatch on the linear Zeno parameters for the spontaneous and stimulated cases]{\label{fig:Linear-Zeno-kz}  (a) Linear Zeno parameter
in the spontaneous case for $\frac{\Delta k_{b}}{k}=10^{-3}$ with
$\alpha=5,\,\beta=3$ for $\frac{\Gamma_{b}}{k}=10^{-2}\,\left(5\times10^{-2}\right)$
in the smooth blue (dashed red) line. (b) A similar observation in the
stimulated case with $\delta=1$ and all the remaining values same
as (a). In (c) and (d), the effect of phase mismatch in the spontaneous
and stimulated (with $\delta=1$) cases of linear Zeno parameter is
shown, respectively. The remaining parameters are $\frac{\Gamma_{b}}{k}=10^{-2}$
with $\frac{\Delta k_{b}}{k}=10^{-3}\,\left(10^{-1}\right)$ in the
smooth blue (dashed red) line.}
\end{figure}

To illustrate the variation of the nonlinear Zeno parameter with the
rescaled interaction length of the coupler in Figure \ref{fig:Nonlinear-z},
we have considered specific values of all the remaining parameters.
As in the case of linear Zeno parameter (cf. Figure \ref{fig:Linear-Zeno-z}), the
nonlinear Zeno parameter also shows dependence on the phases of both
second harmonic modes involved in the symmetric coupler. Specifically,
Figure \ref{fig:Nonlinear-z} (a) illustrates that a change in the phase
of the second harmonic mode of the system creates some changes in
the photon statistics which causes a transition between quantum Zeno
and anti-Zeno effects. This becomes more dominant with the change of
phase	 of the second harmonic mode of the probe as well. Similarly, Figure
\ref{fig:Nonlinear-z} (b) establishes an analogous fact for the nonlinear
Zeno parameter as in Figure \ref{fig:Linear-Zeno-z} (b) for the linear Zeno
parameter, i.e., when the photon numbers in the linear modes of both
the waveguides are comparable, then the quantum Zeno effect prevails. Figure
\ref{fig:Nonlinear-z} (c) shows that by changing the phase of $\alpha$
by $\pi$ (i.e., transforming $\alpha$ to $-\alpha$) we observe a similar
effect as observed by changing the phase of $\beta$ by the same amount. Interestingly,
this kind of nature can be attributed to the symmetry present in the
symmetric coupler studied here.

\begin{figure}[t]
\centering{}\includegraphics[angle=-90,scale=0.85]{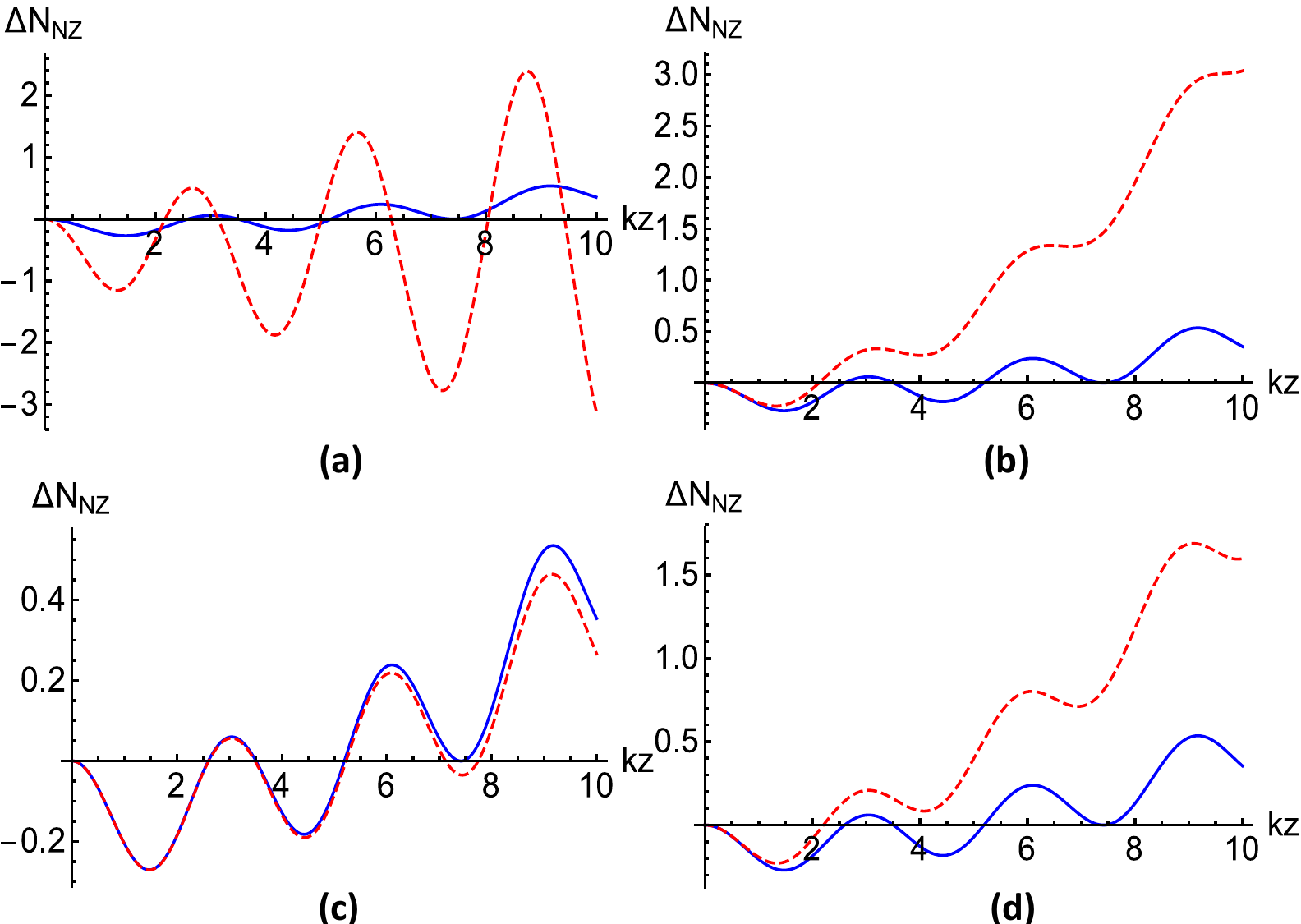}

\protect\caption[The effect of change in the nonlinear coupling and phase mismatch on the nonlinear Zeno parameters for the spontaneous and stimulated cases]{\label{fig:Nonlinear-Zeno-kz}  Dependence of the nonlinear
Zeno parameter on the nonlinear coupling constant is depicted in (a)
and (b). In (a), the nonlinear Zeno parameter is shown with rescaled
interaction length for $\frac{\Gamma_{a}}{k}=10^{-2},\,\frac{\Delta k_{a}}{k}=1.1\times10^{-3},\,\frac{\Delta k_{b}}{k}=10^{-3}$
with $\alpha=5,\,\beta=3,\gamma=2,$ and $\delta=1.$ The smooth (blue)
and dashed (red) lines correspond to $\frac{\Gamma_{b}}{k}=10^{-2}$
and $\frac{\Gamma_{b}}{k}=5\times10^{-2},$ respectively. Similarly,
in (b), the smooth (blue) and dashed (red) lines correspond to $\frac{\Gamma_{a}}{k}=10^{-2}$
and $\frac{\Gamma_{a}}{k}=5\times10^{-2},$ respectively, with $\frac{\Gamma_{b}}{k}=10^{-2}$
and all the remaining values as in (a). In (c), the nonlinear Zeno
parameter is shown in the smooth blue (dashed red) line with $\frac{\Gamma_{a}}{k}=\frac{\Gamma_{b}}{k}=10^{-2},\,\frac{\Delta k_{b}}{k}=10^{-3}$
for $\frac{\Delta k_{a}}{k}=1.1\times10^{-3}\,\left(1.1\times10^{-1}\right).$
Similarly, in (d), the nonlinear Zeno parameter is shown in the smooth
blue (dashed red) line with $\frac{\Delta k_{a}}{k}=1.1\times10^{-3}$
for $\frac{\Delta k_{b}}{k}=10^{-3}\,\left(10^{-1}\right).$}
\end{figure}

The explicit dependence of the linear and nonlinear Zeno parameters
on the remaining parameters, such as nonlinear coupling constants,
phase mismatches, of both system and probe waveguides is illustrated
in Figures \ref{fig:Linear-Zeno-kz} and \ref{fig:Nonlinear-Zeno-kz},
respectively. Specifically, Figures \ref{fig:Linear-Zeno-kz} (a) and (b) show
the variation in the linear Zeno parameter for two values of the nonlinear
coupling constant of the system in the spontaneous and stimulated cases,
respectively. A similar study is shown in Figures \ref{fig:Nonlinear-Zeno-kz}
(a) and (b) for the nonlinear Zeno parameter with two values of the nonlinear
coupling constants of the system and probe waveguides, respectively.
All the cases demonstrate that with increase in the nonlinear coupling
of the system a dominant oscillatory nature is observed, while increase
in the nonlinear coupling of the probe waveguide shows preference
for the quantum anti-Zeno effect.

\begin{figure}[t]
\centering{}\includegraphics[angle=0,scale=0.7]{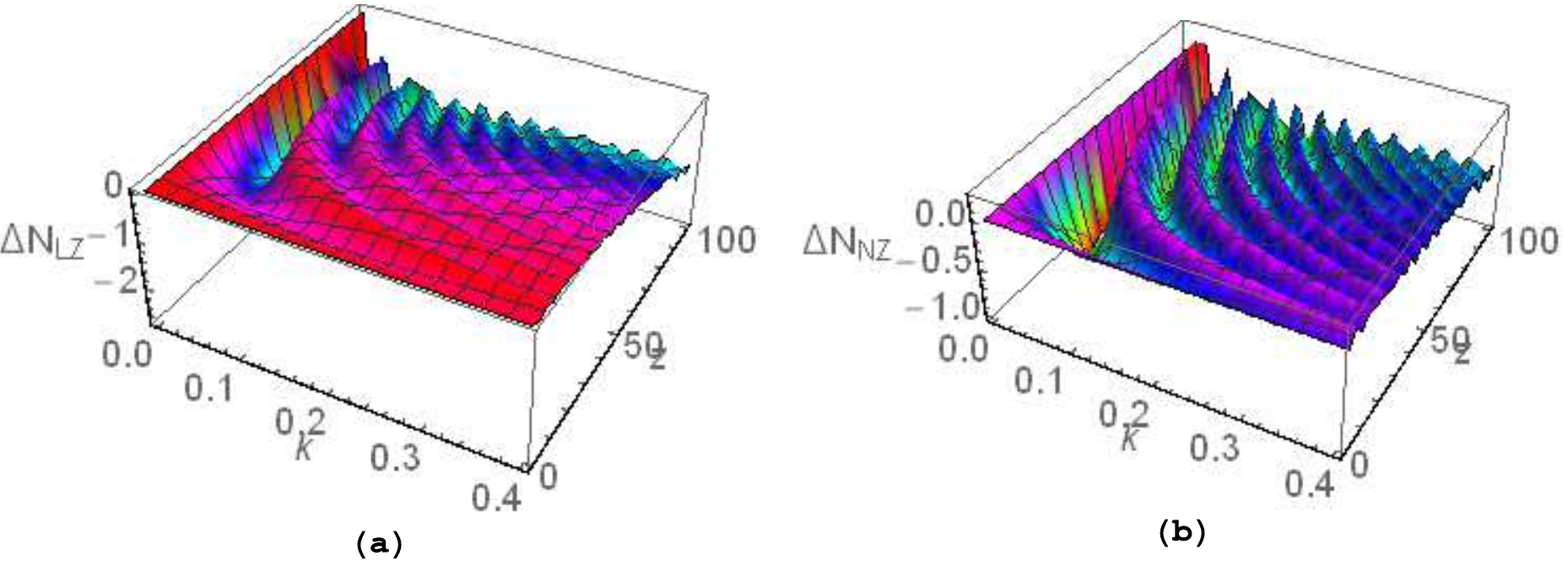}

\protect\caption[Variation in the linear and nonlinear Zeno parameters with the linear coupling constant]{\label{fig:with-k}  Variation in the (a) linear and
(b) nonlinear Zeno parameters with the linear coupling constant (in the units of $\rm{meter}^{-1}$)  and the interaction
length (in meter) is shown for $\frac{\Gamma_{a}}{k}=\frac{\Gamma_{b}}{k}=10^{-2},\,\frac{\Delta k_{a}}{k}=1.1\times10^{-3},\,\frac{\Delta k_{b}}{k}=10^{-3}$
with $\alpha=6,\,\beta=4,\,\gamma=2,$ and $\delta=1$. }
\end{figure}

\begin{figure}[t]
\begin{centering}
\includegraphics[angle=0,scale=0.75]{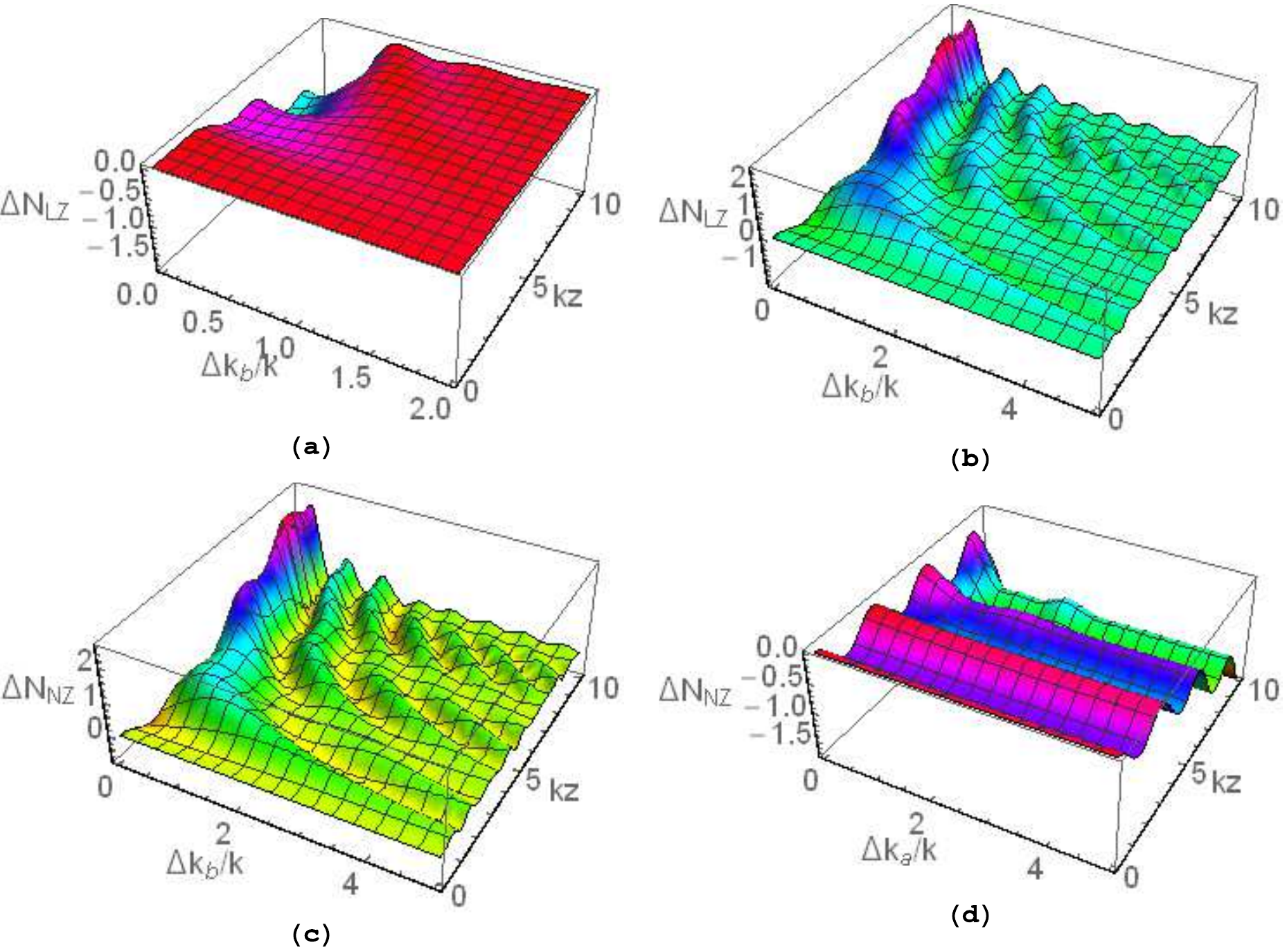}
\par\end{centering}

\protect\caption[Variation in the linear and nonlinear Zeno
parameters with phase mismatch illustrated through three dimensional surface plots]{\label{fig:with-mismatch}  Variation in the linear Zeno
parameter with the phase mismatch between the fundamental and second harmonic
modes in the system waveguide and rescaled interaction length in the (a) spontaneous
and (b) stimulated cases are shown for $\frac{\Gamma_{b}}{k}=10^{-2}$
with $\alpha=6,\,\beta=4,$ and $\delta=0$ and 1 in (a) and (b),
respectively. (c) shows the dependence of the nonlinear Zeno parameter
on the phase mismatch between the fundamental and second harmonic modes
in the system waveguide and rescaled interaction length for $\frac{\Gamma_{a}}{k}=10^{-2},\,\frac{\Delta k_{a}}{k}=1.1\times10^{-3}$
and $\gamma=2,\,\delta=1$ with all the remaining values same as (a) and
(b). In (d), the effect of the phase mismatch between the fundamental and
second harmonic modes in the probe waveguide and rescaled interaction
length on the nonlinear Zeno parameter is shown for $\frac{\Delta k_{b}}{k}=10^{-3}$
with all the remaining values as (c).}
\end{figure}

\begin{figure}[t]
\begin{centering}
\includegraphics[angle=-180,scale=0.75]{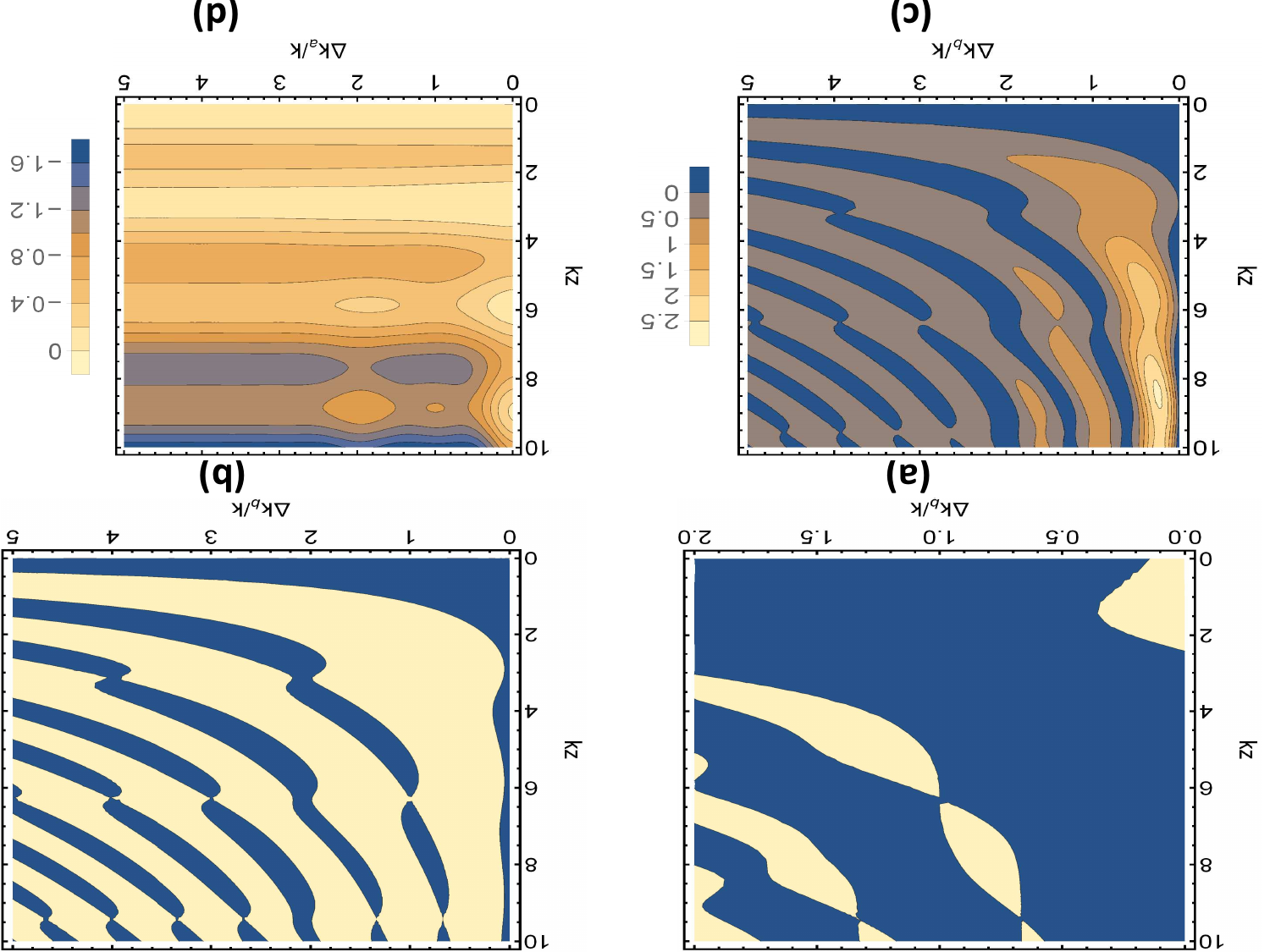} 
\par\end{centering}

\protect\caption[Variation in the linear and nonlinear Zeno
parameters with phase mismatch illustrated through contour plots]{\label{fig:with-mismatch-cont}  Variation in the linear and nonlinear Zeno parameters shown in three dimensional plots in Figures \ref{fig:with-mismatch} (a)-(d) is illustrated via equivalent contour plots. Here, all four contour plots corresponding to Figures \ref{fig:with-mismatch} (a)-(d) are obtained using the same parameters. In (a) and (b), the yellow regions illustrate the regions for quantum anti-Zeno effect, while the blue regions correspond to quantum Zeno effect. In (c) and (d), along with the regions of quantum Zeno and anti-Zeno effects, variation of the magnitude of the Zeno parameters is also shown with different colors (see the color bars in the right side of the figures).}
\end{figure}

The phase mismatch between the fundamental and second harmonic modes of
the system (probe) waveguide has a negligible effect on the linear (nonlinear)
Zeno parameter in the spontaneous (stimulated) case as depicted in
Figure \ref{fig:Linear-Zeno-kz} (c) (\ref{fig:Nonlinear-Zeno-kz} (c)),
while a similar observation for the linear (nonlinear) Zeno parameter
shown in Figure \ref{fig:Linear-Zeno-kz} (d)  (\ref{fig:Nonlinear-Zeno-kz}
(d)) for the stimulated case with the phase mismatch in the system waveguide
exhibits a transition from the quantum Zeno effect to quantum anti-Zeno
effect.

The dependence of both linear and nonlinear Zeno parameters on the
linear coupling can be observed with the interaction length in Figures \ref{fig:with-k}
(a) and (b), respectively. With a particular choice of values for other
parameters, both Zeno parameters show quantum Zeno effect. However,
as observed in Figure \ref{fig:Linear-Zeno-z}, the quantum anti-Zeno effect
can be illustrated here by just controlling the phase of the second
harmonic mode in the system, though an increase in the effect of the
presence of the probe in the photon statistics of the second harmonic
mode with increasing interaction length and linear coupling can be
observed from the figure. This dominance of the effect of the probe
is oscillatory in nature and gives a ripplelike structure in Figure
\ref{fig:with-k}.

The effect of change in the phase mismatch between the fundamental and second
harmonic modes explored in Figures \ref{fig:Linear-Zeno-kz} (c)-(d) and
\ref{fig:Nonlinear-Zeno-kz} (c)-(d) is further illustrated in Figure \ref{fig:with-mismatch}. The phase mismatch between the fundamental and second harmonic modes
in the system waveguide has an evident effect on the linear Zeno parameter
only in the small mismatch region for the spontaneous case (cf. Figure
\ref{fig:with-mismatch} (a)), whereas an increase in the initial number
of photons in the second harmonic mode of the system, i.e., in the
stimulated case, changes the photon statistics drastically and both
quantum Zeno and anti-Zeno effects, with continuous switching between
them, have been observed in Figure \ref{fig:with-mismatch} (b). The corresponding
plot for the nonlinear Zeno parameter shows a quite similar behavior in
Figure \ref{fig:with-mismatch} (c) with slight changes in the photon
statistics due to the presence of the second harmonic mode in the
probe. A similar study for the effect of the phase mismatch between the
fundamental and second harmonic modes of the probe on the nonlinear Zeno
parameter shows an ample amount of variation only for small mismatch
and becomes almost constant for larger values of the phase mismatch. These features of quantum Zeno and anti-Zeno effects can be further illustrated using contour plots as shown in Figure \ref{fig:with-mismatch-cont}, where the values of different parameters are same as those used in Figure \ref{fig:with-mismatch}. The contour plots are drawn here to clearly show the regions of quantum Zeno and anti-Zeno effects (without referring to the magnitude of the Zeno parameter) as shown in Figures \ref{fig:with-mismatch-cont} (a) and (b), where the blue regions correspond to the quantum Zeno effect while the yellow regions correspond to the quantum anti-Zeno effect in the linear Zeno case. The contour plots can also be drawn to illustrate the depth of the Zeno parameter for both the effects as illustrated in Figures \ref{fig:with-mismatch-cont} (c) and (d) for the nonlinear Zeno parameter.

\begin{figure}
\centering{}\includegraphics[angle=-90,scale=0.7]{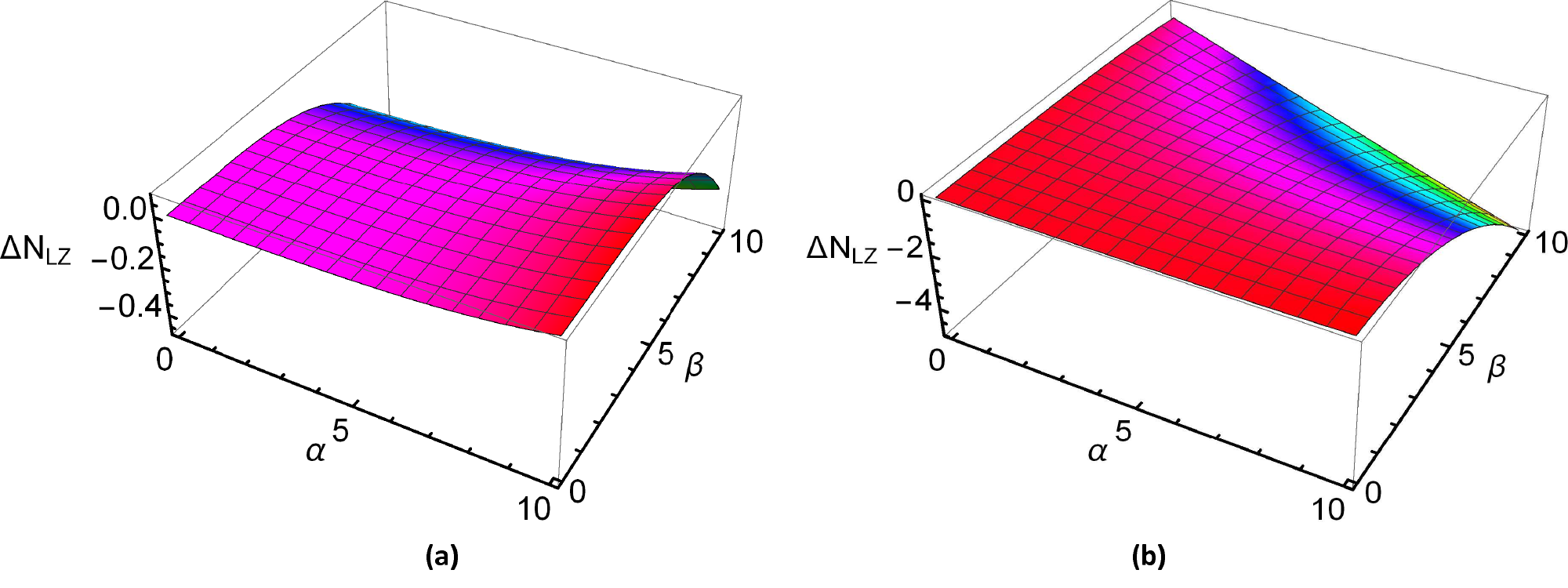}

\protect\caption[Variation in the linear Zeno
parameter with the photon
numbers of the linear modes in the spontaneous and stimulated cases]{\label{fig:Linear-ab}  Variation in the linear Zeno
parameter in the (a) spontaneous and (b) stimulated cases with  the photon
numbers of the linear modes in both the waveguides ($\alpha$ and $\beta$)
are shown for $\frac{\Gamma_{b}}{k}=10^{-2},\,\frac{\Delta k_{b}}{k}=10^{-3}$
with $\gamma=\frac{\alpha}{2}$ and $\delta=\frac{\beta}{3}$ after
rescaled interaction length $kz=1$. A similar behavior to the linear
Zeno parameter is observed for the nonlinear Zeno parameter in the stimulated
case.}
\end{figure}

Figure \ref{fig:Linear-ab} demonstrates the nature of linear Zeno parameter
with changes in the number of photons in the linear modes of both
the waveguides. Here, it can be seen that with an increase in photon
numbers in the probe mode quantum anti-Zeno effect is preferred, while
with an increase in the intensity in the linear mode of system waveguide
it tends towards the quantum Zeno effect.

\section{Conclusions \label{sec:Conclusions-Zeno}}

Linear and nonlinear quantum Zeno and anti-Zeno effects in a symmetric
and an asymmetric nonlinear optical couplers are rigorously investigated
in the present chapter. The investigation is performed using the linear and
nonlinear Zeno parameters, which are introduced in this chapter in analogy
with that of the Zeno parameter introduced by us in Ref. \cite{thapliyal2015quantum}.
Closed form analytic expressions for both linear and nonlinear Zeno
parameters are obtained here using the Sen-Mandal perturbative method (introduced in Section \ref{Sol-of-eq}).
Subsequently, variation of the Zeno parameters with
respect to various quantities is investigated and the same is illustrated
in Figures \ref{fig:Linear-Zeno-z}-\ref{fig:Linear-ab}. The investigation
led to several interesting observations. For example, we have observed
that the analytic expressions obtained for both linear and nonlinear
Zeno parameters are same for the spontaneous case. Further, in
the spontaneous case, it is observed that the transition from the
quantum anti-Zeno effect to quantum Zeno effect can be achieved by
increasing the intensity of the radiation field in the linear mode
of the system waveguide (cf. Figure \ref{fig:Linear-Zeno-z} (b)). Similarly,
a switching between the linear (nonlinear) quantum Zeno and anti-Zeno
effects is also observed in the stimulated case. However, it is observed
that this switching can be obtained just by controlling the phase
of the second harmonic mode in the system waveguide in the linear
case (cf. Figures \ref{fig:Linear-Zeno-z} (a) and (c)) and by controlling
the phase of the nonlinear modes of both the waveguides (cf. Figures
\ref{fig:Nonlinear-z} (a) and (c)). Here, we may note that a change
in the phase of the linear mode of the probe is equivalent to the change in the
phase of the linear mode of the nonlinear system waveguide. This kind
of nature can be attributed to the symmetry present in the system,
which is evident even in the momentum operator of the system  (cf. Figure \ref{fig:Nonlinear-z}
(c)). In fact, in general, we have observed that an increase in the intensity
of the probe increases quantum anti-Zeno effect, while
with an increase in the intensity of the linear mode of the system
waveguide the quantum Zeno effect becomes more prominent (cf. Figures \ref{fig:Linear-Zeno-z}
(b) and (c) and Figure \ref{fig:Linear-ab}).

For the smaller values of the linear coupling constant, a considerable
amount of variation in the photon number statistics is observed through
the linear Zeno parameter. This variation is observed to fade away
as ripples with increasing interaction length for higher values of
linear coupling constant (cf. Figure \ref{fig:with-k} (a)). Similar but
more prominent nature is observed in the nonlinear case (cf. Figure
\ref{fig:with-k} (b)). We have also observed that with an increase
in the phase mismatch between the fundamental and second harmonic modes in
the system waveguide, a transition from quantum Zeno effect to quantum
anti-Zeno effect occurs. It can be concluded that the transition between quantum Zeno and anti-Zeno effects is determined by the phase relations in the system. Specifically, for the stimulated processes, the quantum Zeno effect is related to the phase matching and quantum anti-Zeno effect to the phase mismatching. Further, the occurrence of quantum Zeno and anti-Zeno effects and their transition can also be attributed to the intensity of the pump mode in the system waveguide (as observed in Figures \ref{fig:Linear-Zeno-z} (b) and \ref{fig:Nonlinear-z} (b)). In addition, Figure \ref{fig:Linear-ab} establishes the point that for relatively stronger fields in both the probe and the linear mode of the system waveguide a transition to quantum Zeno effect is dominant. The change in the photon number statistics of
the nonlinear waveguide is more prominent in the stimulated case compared
to that in the spontaneous case of the linear and nonlinear Zeno parameters,
respectively (cf. Figure \ref{fig:with-mismatch}).

In brief, in this chapter, the possibility of observing quantum Zeno and anti-Zeno effects is rigorously
investigated in the symmetric and asymmetric nonlinear optical couplers,
which are experimentally realizable at ease. A completely quantum
description of the primary physical system (i.e., symmetric nonlinear
optical coupler), an appropriate use of a perturbative technique which
is known to perform better than the short-length method, reducibility
of the results obtained for the symmetric nonlinear optical coupler to
that of the asymmetric nonlinear optical coupler, an easy experimental realizability
of the physical systems, etc., provide an edge to this work over the
existing works on quantum Zeno effect in optical couplers, where usually the
use of a complete quantum description is circumvented by considering
one or more modes as strong and/or short-length method is used to
reduce computational difficulty. The approach adopted here is also
very general and can be easily extended to the study of other optical
couplers and other quantum optical systems having a similar structure
of momentum operators or Hamiltonian as is used here. We conclude
the chapter with an expectation that the experimentalists will find this
work interesting for an experimental verification, and it will be
possible to find its applicability in some of the recently proposed
Zeno-effect-based schemes for quantum computation and communication. The work reported in this chapter is published in a journal \cite{thapliyal2016linear} and a conference \cite{thapliyal2015quantum}.

\mathversion{normal}
\thispagestyle{empty}
\thispagestyle{plain}   

\blankpage

\mathversion{normal2}
\titlespacing*{\chapter}{0pt}{-50pt}{20pt}

\chapter{Quasiprobability distributions in open quantum systems: spin 
systems \label{QDs}}

\section{Introduction}

A very useful concept in the analysis of the dynamics of classical
systems is the notion of phase space. A straightforward extension
of this to the realm of quantum mechanics is however foiled due to
the uncertainty principle. Despite this, it is possible to construct
different quasiprobability distributions for quantum mechanical systems
in analogy with their classical counterparts \cite{klauder1968fundamentals,scully1997quantum,schleich2011quantum,agarwal2013quantum,puri2001mathematical,klimov2009group}. Some of those quasiprobability distributions are already introduced in Section \ref{quasi}.
These quasiprobability distributions are very useful as they provide a quantum classical
correspondence and facilitate the calculation of quantum mechanical
averages in close analogy to classical phase space averages. Nevertheless,
the quasiprobability distributions are not true probability distributions as they can take negative
values as well, a feature that could be used for the identification
of quantumness (or nonclassciality) in a system.

The first such quasiprobability distribution was developed by Wigner resulting in the epithet
Wigner function  \cite{wigner1932quantum,moyal1949quantum,hillery1984distribution,kim1991phase,miranowicz2001quantum,opatrny1996coherent} in the form of Eq. (\ref{Wusual}).
Another, very well-known, quasiprobability distribution is the $P$-function (\ref{eq:P-fun-exp}) whose development
was a precursor to the evolution of the field of quantum optics (more precisely, nonclassical optics)
\cite{glauber1963coherent,sudarshan1963equivalence}. The $P$-function can become singular for quantum
states, a feature that promoted the development of other quasiprobability distributions such
as the $Q$-function \cite{mehta1965relation,kano1964probability,husimi1940some} as well as further
highlighted the applicability of the Wigner function which does not have this
feature. As introduced previously, these quasiprobability distributions are intimately related to the operator
orderings. It is quite clear that
there can be other quasiprobability distributions, apart from the above three, depending upon
the operator ordering. However, among all the possible quasiprobability distributions the above
three quasiprobability distributions are the most widely studied. There exist several reasons
behind the intense interest in these quasiprobability distributions. A particularly important reason is that they can be used to identify
the nonclassical (quantum) nature of a state \cite{ryu2013operational}.

In the field of quantum information \cite{plum1996density,scully1997quantum,puri2001mathematical,walls2007quantum,schleich2011quantum,agarwal2013quantum}, the atoms, in their
simplest forms, are modeled as qubits (two-level quantum systems) while studying atom-field interactions. These two-level systems are 
also of immense practical importance as they can be the effective
realizations of Rydberg atoms \cite{saffman2010quantum,gaetan2009observation}. Atomic systems are also
studied in the context of the Dicke model \cite{dicke1954coherence,dowling1994wigner},
a collection of two-level atoms; in atomic traps \cite{wodkiewicz1985coherent}, atomic
interferometers \cite{kitagawa1991nonlinear}, polarization optics \cite{karasev1999polarization}. Recently, interesting applications of atomic systems have been reported in the field of quantum computation (\cite{zu2014experimental,monroe2014large,jones1998implementation,ladd2002all,harneit2002fullerene,gershenfeld1997bulk}
and references therein) as well as in the generation of long-distance
entanglement \cite{sahling2015experimental}. All these  evoke a question:
Whether one could have quasiprobability distributions for such atomic systems as well? Such a question is of relevance to the present work as it is closely tied
to the problem of development of quasiprobability distributions for $SU(2)$, spin-like (spin-$j$),
systems. An effort to answer the above question was made in \cite{stratonovich1957distributions}, where a quasiprobability distribution
on the sphere, naturally related to the $SU(2)$ dynamical group \cite{klimov2002SU,chumakov2000connection},
was obtained. There are by now a number of constructions of spin quasiprobability distributions
\cite{wootters1987wigner,vourdas2003factorization,chaturvedi2006wigner,leonhardt1996discrete,paz2002discrete}, among others.

However, another possible approach, the one adapted here, is to make use of
the connection of $SU(2)$ geometry to that of a sphere. The spherical
harmonics provide a natural basis for functions on the sphere. This,
along with the general theory of multipole operators \cite{plum1996density,zare2013angular},
can be made use for constructing quasiprobability distributions of spin systems as the functions
of polar and azimuthal angles \cite{agarwal1993perspective,agarwal1981relation}. Other constructions,
in the literature, of Wigner functions for spin-$\frac{1}{2}$ systems can be
found in \cite{cohen1986joint,varilly1989moyal}, among others. Another concept that played
an important role in the above developments, was the atomic coherent
state \cite{arecchi1972atomic}, which lead to the definition of the atomic $P$-function in a close analogy to its radiation field counterpart (in Eq. (\ref{eq:P-fun-exp})). Another
related development, following \cite{margenau1961correlation} where joint probability
distributions were obtained for spin-1 systems exposed to quadrupole
fields, was a quasiprobability distribution obtained from the Fourier inversion of the characteristic
function of the corresponding probability mass function, using the
Wigner-Weyl correspondence. This can be referred to as the characteristic
function or $F$-function approach \cite{ramachandran1996quasi,usha2002spin}.

In the present chapter, we investigate the presence of nonclassicality in a number of
spin-qubit systems including single, two, and three-qubit states of importance
in the fields of quantum optics and information. This would be done by analyzing
the behavior of the well-known Wigner, $P$, and $Q$ quasiprobability distributions for the above mentioned spin-qubit systems. The significance
of this is rooted to the phenomena of quantum state engineering, which
involves the generation and manipulation of nonclassical states \cite{ozdemir2001quantum,makhlin2001quantum,verstraete2009quantum}.
In this context, it is imperative to have an understanding over quantum
to classical transitions, under ambient conditions using open quantum system formalism \cite{louisell1973quantum,breuer2002theory,weiss2008quantum,banerjee2003general}  introduced in Section \ref{OQS}. Such an understanding
is made possible by the present work, where investigations are done
in the presence of open system effects, both purely dephasing (decoherence)
\cite{banerjee2008geometric,banerjee2007dynamics}, also known as QND,
as well as dissipation \cite{banerjee2008geometric,srikanth2008squeezed}. We also discuss the not so well-known
$F$-function and specify its relation to the Wigner function. Further,
in the next chapter, we will show that the works reported in this chapter have  an impact on the tomography related issues.
Similar attempt in the past was carried out in Ref. \cite{agarwal1998state}, where a method for quantum state
reconstruction of a system of spins or qubits was proposed using the
$Q$-function. Also, the $Q$-function, studied here, can be turned
to address fundamental issues such as complementarity between the number
and phase distributions \cite{shapiro1991quantum,hall1991quantum,agarwal1992classical,agarwal1996complementarity}, under
the influence of QND and dissipative interactions with their
environment, as well as for phase dispersion in atomic systems \cite{banerjee2007phase,banerjee2007phaseQND,srikanth2009complementarity,banerjee2010complementarity,srikanth2010complementarity}.
Here, to the best of our knowledge, we provide, for the first time,
a comprehensive analysis of quasiprobability distributions for spin-qubit systems under general
open system effects.

The plan of this chapter is as follows. In the next section, we will
briefly discuss the quasiprobability distributions for spin states that will be subsequently used in the rest
of the chapter, i.e., the Wigner, $P$, $Q$, and $F$ functions. This
will be followed by a study of open system quasiprobability distributions for single-qubit states.
Subsequently, we will take up the case of some interesting two- and three-qubit
states. These examples will provide an understanding
of quantum to classical transitions as indicated by the various quasiprobability distributions,
under general open system evolutions. Although quasiprobability distributions have been frequently
used to identify the existence of nonclassical states \cite{perina1991quantum},
they do not directly provide any quantitative measure of the amount
of nonclassicality. Keeping this in mind, several measures of nonclassicality
have been proposed, but all of them are seen to suffer from some limitations
\cite{miranowicz2015statistical}. A specific measure of nonclassicality is
the nonclassical volume, which considers the doubled volume of the integrated negative part of
the Wigner function as a measure of nonclassicality \cite{kenfack2004negativity}. 
 In the penultimate section, we make a study of quantumness, in some
of the systems considered in this chapter, by using nonclassical volume
\cite{kenfack2004negativity}. We would then make our conclusions.

\section{Quasiprobability distributions for spin systems}

Here, we briefly discuss the different quasiprobability distributions, i.e., the Wigner, $P$,
$Q$, and $F$ functions, which are subsequently used in this chapter.

\subsection{Wigner function}

Exploiting the connection between spin-like, $SU(2)$, systems and
the sphere, a quasiprobability distribution can be expressed as a function of the polar and azimuthal
angles. Thus, expanded over a complete basis set, a convenient one
being the spherical harmonics, the Wigner function for a single spin-$j$
state can be expressed as \cite{agarwal1981relation,agarwal1993perspective} 
\begin{equation}
\begin{array}{lcl}
W\left(\theta,\phi\right) & = & \left(\frac{2j+1}{4\pi}\right)^{1/2}\underset{K,Q}{\sum}\rho_{KQ}Y_{KQ}\left(\theta,\phi\right),\end{array}\label{eq:wigner-singlequbit}
\end{equation}
where $K=0,1,\ldots,2j$, and $Q=-K,-K+1,\ldots,0,\ldots,K-1,K$,
and 
\begin{equation}
\begin{array}{lcl}
\rho_{KQ} & = & Tr\left\{ T_{KQ}^{\dagger}\rho\right\} .\end{array}\label{eq:rho_kq}
\end{equation}
Here, $Y_{KQ}$ are the spherical harmonics and $T_{KQ}$ are the multipole
operators given by 
\begin{equation}
\begin{array}{lcl}
T_{KQ} & = & \underset{m,m^{\prime}}{\sum}\left(-1\right)^{j-m}\left(2K+1\right)^{1/2}\left(\begin{array}{ccc}
j & K & j\\
-m & Q & m^{\prime}
\end{array}\right)|j,m\rangle\langle j,m^{\prime}|,\end{array}\label{eq:multipole-operator}
\end{equation}
where $\left(\begin{array}{ccc}
j_{1} & j_{2} & j\\
m_{1} & m_{2} & m
\end{array}\right)=\frac{\left(-1\right)^{j_{1}-j_{2}-m}}{\sqrt{2j+1}}\langle j_{1}m_{1}j_{2}m_{2}|j-m\rangle$ is the Wigner $3j$ symbol \cite{varshalovich1988quantum}, and $\langle j_{1}m_{1}j_{2}m_{2}|j-m\rangle$
is the Clebsh-Gordon coefficient. The multipole operators $T_{KQ}$
are orthogonal to each other, and they form a complete set with property
$T_{KQ}^{\dagger}=\left(-1\right)^{Q}T_{K,-Q}$. The Wigner function introduced in Eq. (\ref{eq:wigner-singlequbit})
is normalized 
\[
\int W\left(\theta,\phi\right)\sin\theta d\theta d\phi=1,
\]
and real $W^{*}\left(\theta,\phi\right)=W\left(\theta,\phi\right)$. Similarly,
the Wigner function of a two-particle system, each with spin-$j$ is \cite{agarwal1981relation,agarwal1993perspective,ramachandran1996quasi,usha2002spin}
\begin{equation}
\begin{array}{lcl}
W\left(\theta_{1},\phi_{1},\theta_{2},\phi_{2}\right) & = & \left(\frac{2j+1}{4\pi}\right)\underset{K_{1},Q_{1}}{\sum}\underset{K_{2},Q_{2}}{\sum}\rho_{K_{1}Q_{1}K_{2}Q_{2}}Y_{K_{1}Q_{1}}\left(\theta_{1},\phi_{1}\right)Y_{K_{2}Q_{2}}\left(\theta_{2},\phi_{2}\right),\end{array}\label{eq:wigner2qubit}
\end{equation}
where $\begin{array}{lcl}
\rho_{K_{1}Q_{1}K_{2}Q_{2}} & = & Tr\left\{ \rho T_{K_{1}Q_{1}}^{\dagger}T_{K_{2}Q_{2}}^{\dagger}\right\} .\end{array}$ The function $W\left(\theta_{1},\phi_{1},\theta_{2},\phi_{2}\right)$ is also
normalized as 
\[
\int W\left(\theta_{1},\phi_{1},\theta_{2},\phi_{2}\right)\sin\theta_{1}\sin\theta_{2}d\theta_{1}d\phi_{1}d\theta_{2}d\phi_{2}=1.
\]
Further, it is known that an arbitrary operator can be mapped into
the Wigner function or any other quasiprobability distribution discussed here, which enables us to obtain its expectation value. In what follows,
using the same notations we describe $P$, $Q$, and $F$ functions
for single spin-$j$ state and for two spin-$j$ particles. It may
be noted that all the analytic expressions for the quasiprobability distributions given below
are normalized.

\subsection{$P$-function}

In analogy with the $P$-function for continuous variable systems (cf. Eq. (\ref{eq:P-fun})), the $P$-function 
for a single spin-$j$ state is defined as \cite{agarwal1981relation,agarwal1993perspective}
\begin{equation}
\rho = \int d\theta  d\phi P\left(\theta, \phi\right) |\theta, \phi \rangle \langle \theta, \phi|
\end{equation}
and can be shown to be 
\begin{equation}
\begin{array}{lcl}
P\left(\theta,\phi\right) & = & 
\underset{K,Q}{\sum}\rho_{KQ}Y_{KQ}\left(\theta,\phi\right)\left(\frac{1}{4\pi}\right)^{1/2}
\left(-1\right)^{K-Q}\left(\frac{\left(2j-K\right)!\left(2j+K+1\right)!}{\left(2j\right)!\left(2j\right)!}
\right)^{1/2}.\end{array}\label{eq:Pfunction-singlequbit}
\end{equation}
The $P$-function for two spin-$j$ particles is \cite{agarwal1981relation,agarwal1993perspective,ramachandran1996quasi,usha2002spin}
\begin{equation}
\begin{array}{lcl}
P\left(\theta_{1},\phi_{1},\theta_{2},\phi_{2}\right) & = & \underset{K_{1},Q_{1}}{\sum}\underset{K_{2},Q_{2}}{\sum}\rho_{K_{1}Q_{1}K_{2}Q_{2}}Y_{K_{1}Q_{1}}\left(\theta_{1},\phi_{1}\right)Y_{K_{2}Q_{2}}\left(\theta_{2},\phi_{2}\right)\\
 & \times & \left(-1\right)^{K_{1}-Q_{1}+K_{2}-Q_{2}}\left(\frac{1}{4\pi}\right)\left(\frac{\sqrt{\left(2j-K_{1}\right)!\left(2j-K_{2}\right)!\left(2j+K_{1}+1\right)!\left(2j+K_{2}+1\right)!}}{\left(2j\right)!\left(2j\right)!}\right).
\end{array}\label{eq:Pfunction-2qubit}
\end{equation}
Here, $|\theta, \phi \rangle$ is the atomic coherent state \cite{arecchi1972atomic} and can be expressed in terms of
the Wigner-Dicke states $|j, m \rangle$ as
\begin{equation}
|\theta, \phi \rangle = \underset{m=-j}{\sum^{j}}  \left(\begin{array}{c}
2j \\
m+j 
\end{array}\right)^{1/2}\sin^{j+m}\left(\frac{\theta}{2}\right) \cos^{j-m}\left(\frac{\theta}{2}\right) e^{-i(j+m)\phi} |j,m \rangle. \label{eq:atomic-coherent-state}
\end{equation}

\subsection{$Q$-function}

Similarly, the $Q$-function for a single spin-$j$ state is  
\begin{equation}
Q\left(\theta,\phi\right) = \frac{2j+1}{4\pi}  \langle \theta, \phi| \rho | \theta, \phi \rangle
\end{equation}
and can be expressed as \cite{agarwal1981relation,agarwal1993perspective}
\begin{equation}
\begin{array}{lcl}
Q\left(\theta,\phi\right) & = & \underset{K,Q}{\sum}\rho_{KQ}Y_{KQ}\left(\theta,\phi\right)\left(\frac{1}{4\pi}\right)^{1/2}\left(-1\right)^{K-Q}\left(2j+1\right)\left(\frac{\left(2j\right)!\left(2j\right)!}{\left(2j-K\right)!\left(2j+K+1\right)!}\right)^{1/2}.\end{array}\label{eq:Qfunction-singlequbit}
\end{equation}

Further, the normalized $Q$-function for two-particle system of spin-$j$ \cite{agarwal1981relation,agarwal1993perspective,ramachandran1996quasi,usha2002spin}
particles is 
\begin{equation}
\begin{array}{lcl}
Q\left(\theta_{1},\phi_{1},\theta_{2},\phi_{2}\right) & = & \underset{K_{1},Q_{1}}{\sum}\underset{K_{2},Q_{2}}{\sum}\rho_{K_{1}Q_{1}K_{2}Q_{2}}Y_{K_{1}Q_{1}}\left(\theta_{1},\phi_{1}\right)Y_{K_{2}Q_{2}}\left(\theta_{2},\phi_{2}\right)\left(\frac{\left(2j+1\right)^{2}}{4\pi}\right)\\
 & \times & \left(-1\right)^{K_{1}-Q_{1}+K_{2}-Q_{2}}\left(\frac{\left(2j\right)!\left(2j\right)!}{\sqrt{\left(2j-K_{1}\right)!\left(2j-K_{2}\right)!\left(2j+K_{1}+1\right)!\left(2j+K_{2}+1\right)!}}\right).
\end{array}\label{eq:Qfunction-2qubit}
\end{equation}

\subsection{$F$-function}

The $F$-function \cite{ramachandran1996quasi,usha2002spin} is defined using the
relation between Fano statistical tensors and the state multipole operators.
Specifically, for a single spin-$j$ state, it is defined as \cite{ramachandran1996quasi,usha2002spin}
\begin{equation}
\begin{array}{lcl}
F\left(\theta,\phi\right) & = & \underset{K,Q}{\sum}\rho_{KQ}Y_{KQ}\left(\theta,\phi\right)\left(\frac{1}{4\pi}\right)^{1/2}\frac{1}{2^{K}}\left(\frac{\left(2j+K+1\right)!}{\left(2j-K\right)!\left\{ j\left(j+1\right)\right\} ^{K}}\right)^{1/2}.\end{array}\label{eq:Ffunction-singlequbit}
\end{equation}
Similarly, the normalized $F$-function for a two-particle spin-$j$ \cite{ramachandran1996quasi,usha2002spin}
system is 
\begin{equation}
\begin{array}{lcl}
F\left(\theta_{1},\phi_{1},\theta_{2},\phi_{2}\right) & = & \underset{K_{1},Q_{1}}{\sum}\underset{K_{2},Q_{2}}{\sum}\rho_{K_{1}Q_{1}K_{2}Q_{2}}Y_{K_{1}Q_{1}}\left(\theta_{1},\phi_{1}\right)Y_{K_{2}Q_{2}}\left(\theta_{2},\phi_{2}\right)\\
 & \times & \left(\frac{1}{4\pi\left(2^{K_{1}+K_{2}}\right)}\right)\left(\frac{\left(2j+K_{1}+1\right)!\left(2j+K_{2}+1\right)!}{\left(2j-K_{1}\right)!\left(2j-K_{2}\right)!\left\{ j\left(j+1\right)\right\} ^{K_{1}+K_{2}}}\right)^{1/2}.
\end{array}\label{eq:Ffunction-2qubit}
\end{equation}

In brief, all the quasiprobability distributions discussed in this work are normalized to
unity. They are also real functions as they correspond to probability
density functions for classical states. The density matrix of a quantum
state can be reconstructed from these quasiprobability distributions \cite{klimov2009group}. One can
also calculate the expectation value of an operator from them \cite{agarwal1981relation,agarwal1993perspective}.

It would be appropriate here to make a brief comparison of the quasiprobability distributions, discussed above, with their continuous
variable counterparts discussed in Section \ref{quasi}. The coherent states and thereby the displacement operator $D\left(\alpha\right)$, which generates the coherent states 
from vacuum, plays a pivotal role in those (continuous variable) considerations. 
Similarly, the multipole operators $T_{KQ}$ (\ref{eq:multipole-operator})
play a pivotal role in the construction of quasiprobability distributions of spin systems, discussed here. These operators are extensively used 
in the study of atomic and nuclear radiation and can be shown to have properties analogous to those of the coherent
state displacement operator $D\left(\alpha\right)$ for usual continuous variable bosonic systems \cite{agarwal1981relation,agarwal1993perspective}. In this sense, the properties
of spin quasiprobability distributions are analogous to those of their continuous variable counterparts, with the atomic coherent state playing the
role of the usual coherent state.  

Before proceeding further, it would be worth noting here that for all 
spin-$\frac{1}{2}$ states (qubits), single- or multi-qubit, the Wigner
and $F$ quasiprobability distributions are identical. Specifically, for the single-qubit case,
the Wigner function is 
\[
W_{\frac{1}{2}}\left(\theta,\phi\right)=\frac{1}{\sqrt{2\pi}}\underset{K,Q}{\sum}M_{K,Q}\left(\theta,\phi\right),
\]
where $M_{K,Q}\left(\theta,\phi\right)=\rho_{KQ}Y_{KQ}\left(\theta,\phi\right)$,
while, the $F$ function is 
\[
F_{\frac{1}{2}}\left(\theta,\phi\right)=\frac{1}{\sqrt{4\pi}}\underset{K,Q}{\sum}M_{K,Q}\left(\theta,\phi\right)\left(\frac{\left(2+K\right)!}{3^{K}\left(1-K\right)!}\right)^{1/2},
\]
where the term inside the brackets with square root is $2$ for both
the values of $K$ (i.e., $0$ or $1$). Similarly, for two spin-$\frac{1}{2}$
states, the Wigner and $F$ functions are 
\[
\begin{array}{lcl}
F_{\frac{1}{2},\frac{1}{2}}\left(\theta_{1},\phi_{1},\theta_{2},\phi_{2}\right) & = & \frac{1}{4\pi}\underset{K_{1},Q_{1}}{\sum}\underset{K_{2},Q_{2}}{\sum}M_{K_{1}Q_{1}K_{2}Q_{2}}\left(\theta_{1},\phi_{1},\theta_{2},\phi_{2}\right)\left(\frac{\left(2+K_{1}\right)!\left(2+K_{2}\right)!}{3^{K_{1}+K_{2}}\left(1-K_{1}\right)!\left(1-K_{2}\right)!}\right)^{1/2}\\
 & = & \frac{1}{2\pi}\underset{K_{1},Q_{1}}{\sum}\underset{K_{2},Q_{2}}{\sum}M_{K_{1}Q_{1}K_{2}Q_{2}}\left(\theta_{1},\phi_{1},\theta_{2},\phi_{2}\right)\\
 & = & W_{\frac{1}{2},\frac{1}{2}}\left(\theta_{1},\phi_{1},\theta_{2},\phi_{2}\right),
\end{array}
\]
where $M_{K_{1}Q_{1}K_{2}Q_{2}}\left(\theta_{1},\phi_{1},\theta_{2},\phi_{2}\right)=\rho_{K_{1}Q_{1}K_{2}Q_{2}}Y_{K_{1}Q_{1}}\left(\theta_{1},\phi_{1}\right)Y_{K_{2}Q_{2}}\left(\theta_{2},\phi_{2}\right),$
and the term in the brackets is $4$ for all possible values of
$K_{1}$ and $K_{2}$. This can further be extended for higher number
of spin-$\frac{1}{2}$ states.

Since the Wigner and $F$ functions are same for spin-$\frac{1}{2}$
systems, we will not discuss the evolution of the $F$-function further.

\section{Quasiprobability distributions for single spin-$\frac{1}{2}$ (qubit) states}

Here, we consider single spin-$\frac{1}{2}$ states, initially in
an atomic coherent state, in the presence of two different noises,
i.e., QND \cite{banerjee2008geometric,banerjee2007dynamics}, which are purely dephasing, and the
dissipative SGAD  \cite{banerjee2008geometric,srikanth2008squeezed}
noises. For calculating the quasiprobability distributions, we will require multipole operators
for $j=\frac{1}{2}$ and $m,\, m^{\prime}=\pm\frac{1}{2}$, giving
$K=0$ and $1$. For $K=0$, $Q=0$, and for $K=1$, $Q=1,\,0,$ and $-1$.
Using these, the multipole operators $T_{KQ}$ can be obtained as
$\begin{array}{lcl}
T_{00} & = & \frac{1}{\sqrt{2}}\left[\begin{array}{cc}
1 & 0\\
0 & 1
\end{array}\right],\end{array}$ $\begin{array}{lcl}
T_{11} & = & \left[\begin{array}{cc}
0 & 0\\
-1 & 0
\end{array}\right],\end{array}$ $\begin{array}{lcl}
T_{10} & = & \frac{1}{\sqrt{2}}\left[\begin{array}{cc}
1 & 0\\
0 & -1
\end{array}\right],\end{array}$and $\begin{array}{lcl}
T_{1-1} & = & \left[\begin{array}{cc}
0 & 1\\
0 & 0
\end{array}\right].\end{array}$

\subsection{Atomic coherent state in the QND noise \label{subsec:QND}}

The master equation of a system interacting with a squeezed thermal
bath and undergoing a QND evolution \cite{banerjee2007dynamics} is 
\begin{equation}
\begin{array}{lcl}
\dot{\rho}{}_{nm}\left(t\right) & = & \left[-\frac{i}{\hbar}\left(E_{n}-E_{m}\right)+i\dot{\eta}\left(t\right)\left(E_{n}^{2}-E_{m}^{2}\right)-\left(E_{n}-E_{m}\right)^{2}\dot{\gamma}\left(t\right)\right]\rho_{nm}\left(t\right),\end{array}\label{eq:master-eq-QND}
\end{equation}
where $E_{n}$s are the eigenvalues of the system Hamiltonian in the
system eigenbasis $\left\{|n\rangle\right\}$, which here would correspond to the
Wigner-Dicke states \cite{agarwal2013quantum}; 
\[
\eta\left(t\right)=-\underset{k}{\sum}\frac{g_{k}^{2}}{\hbar^{2}\omega_{k}^{2}}\sin\left(\omega_{k}t\right)
\]
and 
\[
\begin{array}{lcl}
\gamma\left(t\right) & = & \frac{1}{2}\underset{k}{\sum}\frac{g_{k}^{2}}{\hbar^{2}\omega_{k}^{2}}\coth\left(\frac{\beta\hbar\omega_{k}}{2}\right)\left|\left(e^{i\omega_{k}t}-1\right)\cosh\left(r_{k}\right)+\left(e^{-i\omega_{k}t}-1\right)\sinh\left(r_{k}\right)e^{2i\Phi_{k}}\right|^{2}.\end{array}
\]
Here, $\beta=\frac{1}{k_{B}T}$, and $k_{B}$ is the Boltzmann constant,
while $r_{k}$ and $\Phi_{k}$ are the squeezing parameters. The initial
density matrix for the atomic coherent state is 
\begin{equation}
\rho\left(0\right)=|\alpha,\beta\rangle\langle\alpha,\beta|,\label{eq:initial-state-QND}
\end{equation}
where $|\alpha,\beta\rangle$ is given by Eq. (\ref{eq:atomic-coherent-state}).
Different terms in the density matrix (\ref{eq:initial-state-QND}) in the presence of
QND noise at time $t$ become
\begin{equation}
\begin{array}{lcl}
\rho_{jm,jn}\left(t\right) & = & e^{-i\omega\left(m-n\right)t}e^{i\left(\hbar\omega\right)^{2}\left(m^{2}-n^{2}\right)\eta\left(t\right)}e^{-\left(\hbar\omega\right)^{2}\left(m-n\right)^{2}\gamma\left(t\right)}\rho_{jm,jn}\left(0\right),\end{array}\label{eq:QND-densitymatirix}
\end{equation}
where 
\begin{equation}
\begin{array}{lcl}
\rho_{jm,jn}\left(0\right) & = & \langle j,m|\rho\left(0\right)|j,n\rangle =  \langle j,m|\alpha,\beta\rangle\langle\alpha,\beta|j,n\rangle.
\end{array}\label{eq:at_t0}
\end{equation}
For $j=\frac{1}{2}$, the initial density matrix, 
\begin{equation}
\rho\left(0\right)=\left[\begin{array}{cc}
\sin^{2}\left(\frac{\alpha}{2}\right) & \frac{1}{2}e^{-i\beta}\sin\alpha\\
\frac{1}{2}e^{i\beta}\sin\alpha & \cos^{2}\left(\frac{\alpha}{2}\right)
\end{array}\right],\label{eq:density-matrix}
\end{equation}
evolves in the presence of QND noise to become 
\begin{equation}
\rho\left(t\right)=\left[\begin{array}{cc}
\sin^{2}\left(\frac{\alpha}{2}\right) & \frac{1}{2}e^{-i\omega t}e^{-\left(\hbar\omega\right)^{2}\gamma\left(t\right)}e^{-i\beta}\sin\alpha\\
\frac{1}{2}e^{i\omega t}e^{-\left(\hbar\omega\right)^{2}\gamma\left(t\right)}e^{i\beta}\sin\alpha & \cos^{2}\left(\frac{\alpha}{2}\right)
\end{array}\right].\label{eq:density-matrix-QND}
\end{equation}

\begin{figure}
\begin{centering}
\includegraphics[angle=-180,scale=0.59]{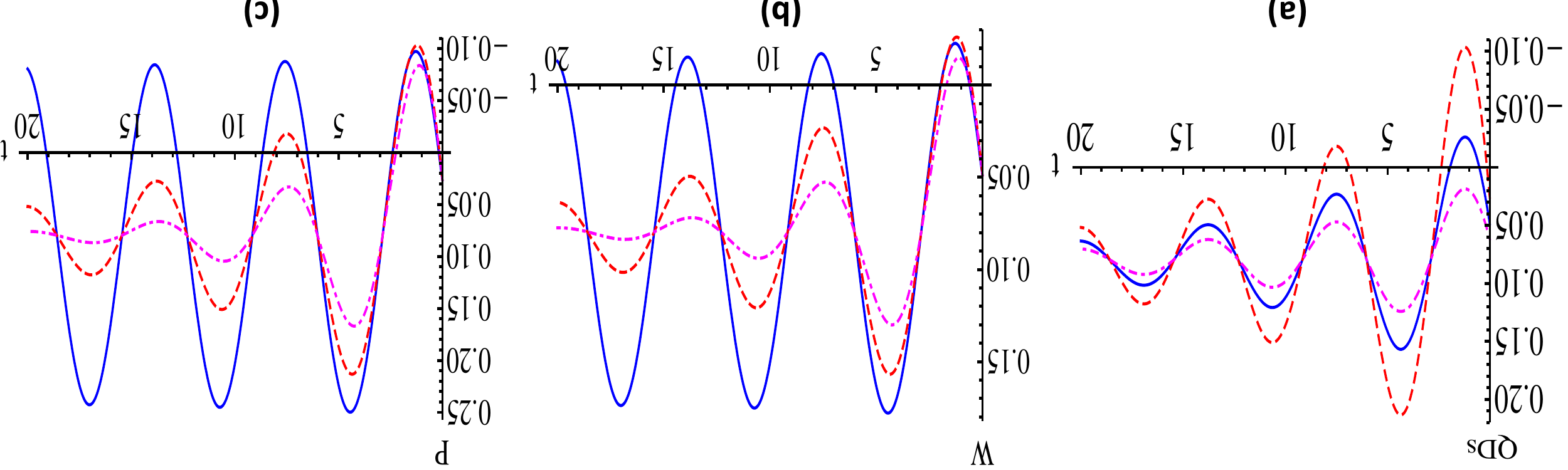} 
\par\end{centering}

\caption[Quasiprobability distributions for single spin-$\frac{1}{2}$ atomic
coherent state in the presence of QND noise]{\label{fig:QND-single} (a) Variation
of all quasiprobability distributions with time ($t$) for single spin-$\frac{1}{2}$ atomic
coherent state in the presence of QND noise with bath parameters $\gamma_{0}=0.1,\,\omega_{c}=100,$
squeezing parameters $r=0,\, a=0,$ and $\omega=1.0$ at temperature
$T=1$, and $\alpha=\frac{\pi}{2},\,\beta=\frac{\pi}{3},\,\theta=\frac{\pi}{3},\,\phi=\frac{\pi}{4},$
in the units of $\hbar=k_{B}=1$. The smooth (blue), dashed (red),
and dot-dashed (magenta) lines correspond to the Wigner, $P,$ and
$Q$ functions, respectively; (b) and (c) show the variation of the Wigner
and $P$ functions with time for different temperatures $T=0,\,1,$
and $2$ by the smooth (blue), dashed (red), and dot-dashed
(magenta) lines, respectively.}
\end{figure}

Here, we consider the case of an Ohmic bath for which analytic expressions
for $\gamma\left(t\right)$, both for zero and high temperatures,
can be obtained \cite{banerjee2007dynamics}. These are functions of the bath parameters
$\gamma_{0}$ and $\omega_{c}$ as well as squeezing parameters $r$
and $\phi$, with $\phi=a\omega$ and $a$ is a constant dependent
on the squeezed bath. Now, using multipole operators, mentioned above,
analytic expressions of the different quasiprobability distributions can be obtained. For example,
we have obtained the Wigner function for a qubit, starting from an atomic
coherent state, in the presence of QND noise as 
\begin{equation}
\begin{array}{lcl}
W\left(\theta,\phi\right) & = & \frac{1}{4\pi}\left(1-\sqrt{3}\cos\alpha\cos\theta+\sqrt{3}e^{-\left(\hbar\omega\right)^{2}\gamma\left(t\right)}\cos\left(\beta+\omega t+\phi\right)\sin\alpha\sin\theta\right),\end{array}\label{eq:wigner-QND}
\end{equation}
while the corresponding $P$ and $Q$ functions are obtained as 
\begin{equation}
\begin{array}{lcl}
P\left(\theta,\phi\right) & = & \frac{1}{4\pi}\left(1+3\cos\alpha\cos\theta+3e^{-\left(\hbar\omega\right)^{2}\gamma\left(t\right)}\cos\left(\beta+\omega t+\phi\right)\sin\alpha\sin\theta\right)\end{array}\label{eq:P-QND}
\end{equation}
and 
\begin{equation}
\begin{array}{lcl}
Q\left(\theta,\phi\right) & = & \frac{1}{4\pi}\left(1+\cos\alpha\cos\theta+e^{-\left(\hbar\omega\right)^{2}\gamma\left(t\right)}\cos\left(\beta+\omega t+\phi\right)\sin\alpha\sin\theta\right),\end{array}\label{eq:Q-QND}
\end{equation}
respectively. All the quasiprobability distributions calculated in Eqs. (\ref{eq:wigner-QND})-(\ref{eq:Q-QND})
can be used to get the corresponding noiseless quasiprobability distributions for the same system,
and this also serves as a nice consistency check of the calculations.
Variation of the quasiprobability distributions, Eqs. (\ref{eq:wigner-QND})-(\ref{eq:Q-QND}),
for some specific values of the parameters is shown in Figures \ref{fig:QND-single}
(a)-(c), where the effect of the presence of noise on the quasiprobability distributions can be easily
observed. Both the $P$ and Wigner functions are found to exhibit negative
values--indicative of the presence of nonclassicality in the system. Also, in Figures \ref{fig:QND-single}
(b)-(c), we can see that with an increase in temperature $T$, the quasiprobability distributions
tend to become less negative, which is an indicator of a move towards
classicality, as expected. Interestingly, in Figure \ref{fig:QND-single}
(a), we do not observe any zero of the $Q$-function which implies that
the $Q$-function does not show any signature of nonclassicality in
this particular case. The oscillatory nature of the quasiprobability distributions for the atomic coherent state when subjected to 
QND noise can be attributed to the purely dephasing effect of the QND interaction. Thus, this process involves
decoherence without any dissipation. At temperature $T=0$, decoherence is minimal, and hence an oscillatory pattern is
observed in the depicted time scale. With increase in $T$, resulting in increase in the influence of decoherence,
these oscillations gradually decrease.

\subsection{Atomic coherent state in the SGAD noise}

\begin{figure}
\begin{centering}
\includegraphics[angle=-180,scale=0.59]{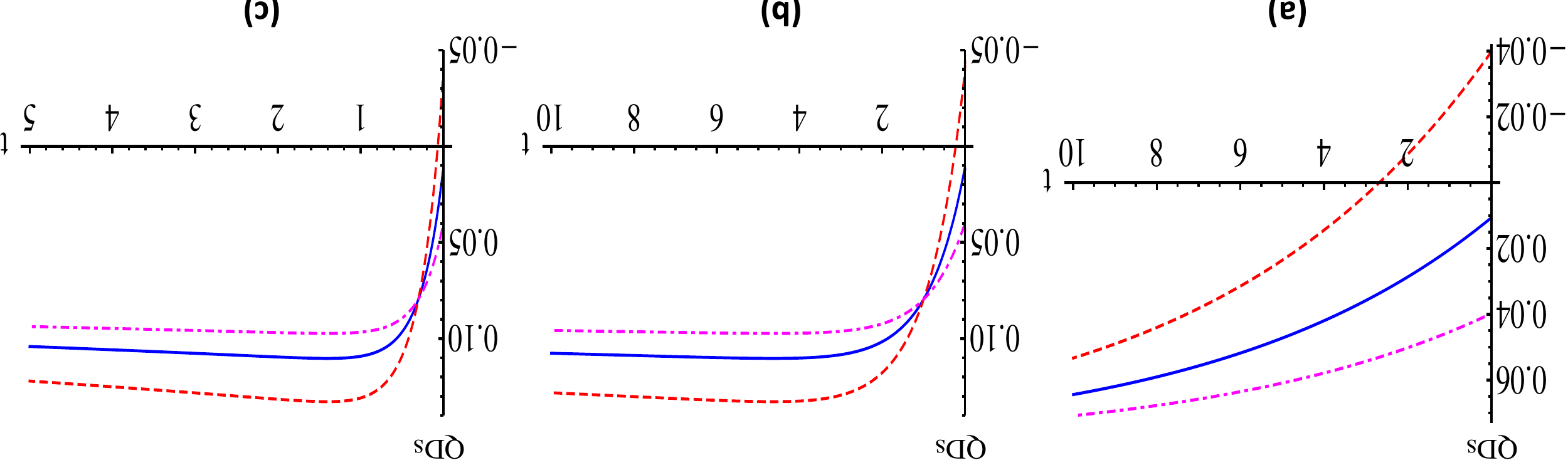} 
\par\end{centering}

\caption[Quasiprobability distributions for single spin-$\frac{1}{2}$ atomic
coherent state in the presence of SGAD noise]{\label{fig:SGAD-ACS-qd} Variation of all quasidistribution
functions with time ($t$) is shown together for a single spin-$\frac{1}{2}$
atomic coherent state in the presence of the SGAD noise for zero bath
squeezing angle in the units of $\hbar=k_{B}=1$, with $\omega=1.0,\,\gamma_{0}=0.05,$
and $\alpha=\frac{\pi}{2},\,\beta=\frac{\pi}{3},\,\theta=\frac{\pi}{2},\,\phi=\frac{\pi}{3}.$
In (a), the variation with time is shown for temperature $T=3.0$ in
the absence of bath squeezing, i.e., $r=0$. In (b), the effect
of change in the squeezing parameter for the same temperature, i.e., $T=3.0,$
is shown by using the squeezing parameter $r=1.0$, keeping all
other values as same as that used in (a). Further, in (c), keeping
$r=1.0$ as in (b), the temperature is increased to $T=10$ to show
the effect of variation in T. In (c), time is varied only up to $t=5$
to emphasize the effect of temperature. In all three plots, the smooth
(blue), dashed (red), and dot-dashed (magenta) lines correspond
to the Wigner, $P$, and $Q$ functions, respectively.}
\end{figure}

Now, we take up a spin $j=\frac{1}{2}$, starting from an atomic coherent
state, given by Eq. (\ref{eq:density-matrix}), evolving under a SGAD channel, incorporating the effects
of dissipation and bath squeezing, and which includes the well-known
AD and GAD channels
as special cases. The Kraus operators of the SGAD channel are \cite{srikanth2008squeezed}
\[
\begin{array}{lcl}
E_{0} & = & \sqrt{p}\left[\begin{array}{cc}
\sqrt{1-\lambda\left(t\right)} & 0\\
0 & 1
\end{array}\right],\end{array}
\]
\[
\begin{array}{lcl}
E_{1} & = & \sqrt{p}\left[\begin{array}{cc}
0 & 0\\
\sqrt{\lambda\left(t\right)} & 0
\end{array}\right],\end{array}
\]
\[
\begin{array}{lcl}
E_{2} & = & \sqrt{1-p}\left[\begin{array}{cc}
\sqrt{1-\mu\left(t\right)} & 0\\
0 & \sqrt{1-\nu\left(t\right)}
\end{array}\right],\end{array}
\]
and
\begin{equation}
\begin{array}{lcl}
E_{3} & = & \sqrt{1-p}\left[\begin{array}{cc}
0 & \sqrt{\nu\left(t\right)}\\
\sqrt{\mu\left(t\right)}e^{-i\xi\left(t\right)} & 0
\end{array}\right],\end{array}\label{eq:SGAD-kraussoperators2-1}
\end{equation}
where $\lambda=\frac{1}{p}\left\{ 1-\left(1-p\right)\left[\mu+\nu\right]-\exp\left(-\gamma_{0}\left(2N+1\right)t\right)\right\} ,$
$\mu=\frac{2N+1}{2N\left(1-p\right)}\frac{\sinh^{2}\left(\gamma_{0}at/2\right)e^{\left(-\frac{\gamma_{0}}{2}\left(2N+1\right)t\right)}}{\sinh^{2}\left(\gamma_{0}\left(2N+1\right)t/2\right)},$
and $\nu=\frac{N}{\left(1-p\right)\left(2N+1\right)}\left\{ 1-\exp\left(-\gamma_{0}\left(2N+1\right)t\right)\right\} .$
Note that, for convenience in writing the expressions, we have omitted the time dependence in the argument
of different time dependent parameters (e.g., $\lambda(t)$, $\mu(t)$,
$\nu(t)$) in the Kraus operators of SGAD noise. Here, $\gamma_{0}$
is the spontaneous emission rate, $a=\sinh\left(2r\right)\left(2N_{th}+1\right),$
and $N=N_{th}\left\{ \cosh^{2}\left(r\right)+\sinh^{2}\left(r\right)\right\} +\sinh^{2}\left(r\right),$
with $N_{th}=1/\left\{ \exp\left(\hbar\omega/k_{B}T\right)-1\right\} $
being the Planck distribution. Here, $r$ and the bath squeezing angle
($\xi\left(t\right)$) are the bath squeezing parameters. The analytic expression of $p$ is
$p=1-\frac{1}{\left(A+B-C-1\right)^2-4D}\left\{A^2\right. B+C^2+A\left[B^2-C-B\left(1+C\right)-D\right]\pm 2 \sqrt{D\left[ B-AB+\left(A-1\right) C+D\right]\left[A-AB+\left(B-1 \right)C+D \right]} -\left(B+1\right)D-C\left(B+D-1\right)\left. \right\},$ where
$A=\frac{(2 N+1) \sinh ^2\left(\frac{\gamma_0 a t}{2}\right) \exp \left(-\frac{\gamma_0 (2 N+1) t}{2} \right)}{2 N \sinh \left(\frac{\gamma_0 (2 N+1) t}{2}\right)},$
$B=\frac{N \left[1-\exp \left(-\gamma_0 (2 N+1) t \right)\right]}{2 N+1},$
$C=A+B+\exp \left(-\gamma_0 (2 N+1) t \right),$
and
$D=\cosh ^2\left(\frac{\gamma_0 a t}{2}\right) \exp \left(-\gamma_0 (2 N+1) t \right)$ (see \cite{srikanth2008squeezed} for detail). Note that, quantum optical notations  are used here with $|+\frac{1}{2}\rangle=\left[\begin{array}{c}1\\0\end{array}\right]$ and $|-\frac{1}{2}\rangle=\left[\begin{array}{c}0\\1\end{array}\right].$
Application of the above Kraus operators to the initial state (as introduced in Eq. (\ref{eq:Kraus})) results
in 
\begin{equation}
\rho\left(t\right)=\left[\begin{array}{cc}
\rho_{11} & \rho_{21}^{*}\\
\rho_{21} & 1-\rho_{11}
\end{array}\right],\label{eq:density-matrix-SGAD}
\end{equation}
where 
\[
\begin{array}{lcl}
\rho_{11} & = & \frac{1}{2}\left\{ 1-\mu+\nu-p\left(\lambda-\mu+\nu\right)+\left(-1+\mu+\nu+p\left(\lambda-\mu-\nu\right)\right)\cos\alpha\right\} \end{array}
\]
and 
\[
\begin{array}{lcl}
\rho_{21} & = & \frac{1}{2}\sin\alpha\left\{ \left(1-p\right)\sqrt{\mu\nu}e^{-i\left(\beta+\xi\right)}+p\sqrt{1-\lambda}e^{i\beta}+\left(1-p\right)\sqrt{\left(1-\mu\right)\left(1-\nu\right)}e^{i\beta}\right\} .\end{array}
\]
Using this density matrix, we can calculate the evolution of different
quasiprobability distributions, in a manner similar to the previous example of evolution under
QND channel, leading to 
\begin{equation}
\begin{array}{lcl}
W\left(\theta,\phi\right) & = & \frac{1}{4\pi}\left[1+\sqrt{3}\left\{ -\mu+\nu-p\left(\lambda-\mu+\nu\right)+\left(-1+\mu+\nu+p\left(\lambda-\mu-\nu\right)\right)\cos\alpha\right\} \cos\theta\right.\\
 & + & \sqrt{3}\left(\left\{ p\sqrt{1-\lambda}+\left(1-p\right)\sqrt{\left(1-\mu\right)\left(1-\nu\right)}\right\} \cos\left(\beta+\phi\right)\right.\\
 & + & \left.\left.\left(1-p\right)\sqrt{\mu\nu}\cos\left(\beta+\xi-\phi\right)\right)\sin\alpha\sin\theta\right];
\end{array}\label{eq:Wigner-SGAD}
\end{equation}
\begin{equation}
\begin{array}{lcl}
P\left(\theta,\phi\right) & = & \frac{1}{4\pi}\left[1-3\left\{ -\mu+\nu-p\left(\lambda-\mu+\nu\right)+\left(-1+\mu+\nu+p\left(\lambda-\mu-\nu\right)\right)\cos\alpha\right\} \cos\theta\right.\\
 & + & 3\left(\left\{ p\sqrt{1-\lambda}+\left(1-p\right)\sqrt{\left(1-\mu\right)\left(1-\nu\right)}\right\} \cos\left(\beta+\phi\right)\right.\\
 & + & \left.\left.\left(1-p\right)\sqrt{\mu\nu}\cos\left(\beta+\xi-\phi\right)\right)\sin\alpha\sin\theta\right];
\end{array}\label{eq:P-SGAD}
\end{equation}
and 
\begin{equation}
\begin{array}{lcl}
Q\left(\theta,\phi\right) & = & \frac{1}{4\pi}\left[1-\left\{ -\mu+\nu-p\left(\lambda-\mu+\nu\right)+\left(-1+\mu+\nu+p\left(\lambda-\mu-\nu\right)\right)\cos\alpha\right\} \cos\theta\right.\\
 & + & \left(\left\{ p\sqrt{1-\lambda}+\left(1-p\right)\sqrt{\left(1-\mu\right)\left(1-\nu\right)}\right\} \cos\left(\beta+\phi\right)\right.\\
 & + & \left.\left.\left(1-p\right)\sqrt{\mu\nu}\cos\left(\beta+\xi-\phi\right)\right)\sin\alpha\sin\theta\right].
\end{array}\label{eq:Q-SGAD}
\end{equation}
Variation of all the quasiprobability distributions with time for some specific values
of the parameters is depicted in Figure \ref{fig:SGAD-ACS-qd}, which incorporates
the effect of both temperature and squeezing. A comparison of Figures \ref{fig:SGAD-ACS-qd}
(a) and (b) brings out the effect of squeezing on
the evolution of quasiprobability distributions. Further, it is easily observed that with an
increase in $T$, the quantumness reduces. An important point to notice
here, is that if we make the noise parameters zero, i.e., in the absence
of noise, the different quasiprobability distributions given by Eqs. (\ref{eq:Wigner-SGAD})-(\ref{eq:Q-SGAD}),
reduce to a form exactly equal to the corresponding noiseless quasiprobability distributions
obtained for QND evolutions (Eqs. (\ref{eq:wigner-QND})-(\ref{eq:Q-QND})).
Also, results for the GAD channel can be obtained
in the limit of vanishing squeezing, i.e., for $\mu\left(t\right)=0$
and $\lambda\left(t\right)=\nu\left(t\right)$, while corresponding
results for quasiprobability distributions under evolution of an AD channel can
be obtained by further setting $T=0$ and $p=1$. Further, it would be apt to mention here that the 
oscillatory nature of the quasiprobability distributions for an atomic coherent state evolving under QND noise is not seen here. 
This is consistent with the fact that the SGAD noise is dissipative in nature, involving decoherence along with
dissipation.

\section{Quasiprobability distributions for multi-qubit systems undergoing dissipative
evolutions}

We further wish to study the evolution of quasiprobability distributions for some interesting two-
and three-qubit systems under general open system dynamics. In each case,
we study the nonclassicality exhibited by the system under consideration.

\subsection{Two qubits under dissipative evolution}

In this section, we study the evolution of quasiprobability distributions, for two-qubit systems, undergoing
dissipative evolution with a vacuum bath ($T=0$
and zero bath squeezing). Here, we will make use
of the results worked out in Ref. \cite{banerjee2010dynamics}.

\subsubsection{Vacuum bath \label{sub:Vacuum-bath}}

The density matrix, in the dressed state basis, can be used for calculating
different quasiprobability distributions. We consider the initial state with one qubit in the
excited state $\left|e_{1}\right\rangle $ and the other in the ground
state $\left|g_{2}\right\rangle $, i.e., $\left|e_{1}\right\rangle \left|g_{2}\right\rangle $.
The two-qubit reduced density matrix is given by 
\begin{equation}
\begin{array}{lcl}
\rho\left(t\right) & = & \left[\begin{array}{cccc}
\rho_{ee}\left(t\right) & \rho_{es}\left(t\right) & \rho_{ea}\left(t\right) & \rho_{eg}\left(t\right)\\
\rho_{es}^{*}\left(t\right) & \rho_{ss}\left(t\right) & \rho_{sa}\left(t\right) & \rho_{sg}\left(t\right)\\
\rho_{ea}^{*}\left(t\right) & \rho_{sa}^{*}\left(t\right) & \rho_{aa}\left(t\right) & \rho_{ag}\left(t\right)\\
\rho_{eg}^{*}\left(t\right) & \rho_{sg}^{*}\left(t\right) & \rho_{ag}^{*}\left(t\right) & \rho_{gg}\left(t\right)
\end{array}\right],\end{array}\label{eq:densitymatrix-vaccumbath}
\end{equation}
where the analytic expressions of all the elements of the density matrix are 
\begin{equation}
\begin{array}{lcl}
\rho_{ee}\left(t\right) & = & e^{-2\Gamma t}\rho_{ee}\left(0\right),\\
\rho_{ss}\left(t\right) & = & e^{-\left(\Gamma+\Gamma_{12}\right)t}\rho_{ss}\left(0\right) +  \frac{\left(\Gamma+\Gamma_{12}\right)}{\left(\Gamma-\Gamma_{12}\right)}\left(1-e^{-\left(\Gamma-\Gamma_{12}\right)t}\right)e^{-\left(\Gamma+\Gamma_{12}\right)t}\rho_{ee}\left(0\right),\\
\rho_{aa}\left(t\right) & = & e^{-\left(\Gamma-\Gamma_{12}\right)t}\rho_{aa}\left(0\right) +  \frac{\left(\Gamma-\Gamma_{12}\right)}{\left(\Gamma+\Gamma_{12}\right)}\left(1-e^{-\left(\Gamma+\Gamma_{12}\right)t}\right)e^{-\left(\Gamma-\Gamma_{12}\right)t}\rho_{ee}\left(0\right),\\
\rho_{gg}\left(t\right) & = & \rho_{gg}\left(0\right)+\left(1-e^{-\left(\Gamma+\Gamma_{12}\right)t}\right)\rho_{ss}\left(0\right) +  \left(1-e^{-\left(\Gamma-\Gamma_{12}\right)t}\right)\rho_{aa}\left(0\right)\\
 & + & \left[\frac{\left(\Gamma+\Gamma_{12}\right)}{2\Gamma}\left\{ 1-\frac{2}{\left(\Gamma-\Gamma_{12}\right)}e^{-\left(\Gamma+\Gamma_{12}\right)t}\right.
\left[\frac{\left(\Gamma+\Gamma_{12}\right)}{2}\left(1-e^{-\left(\Gamma-\Gamma_{12}\right)t}\right)+\frac{\left(\Gamma-\Gamma_{12}\right)}{2}\right]\right\} \\
 & + & \frac{\left(\Gamma-\Gamma_{12}\right)}{\left(\Gamma+\Gamma_{12}\right)}\left\{ \left(1-e^{-\left(\Gamma-\Gamma_{12}\right)t}\right) - \left.\frac{\left(\Gamma-\Gamma_{12}\right)}{2\Gamma}\left(1-e^{-2\Gamma t}\right)\right\} \right]\rho_{ee}\left(0\right),\\
\rho_{es}\left(t\right) & = & e^{-i\left(\omega_{0}-\Omega_{12}\right)t}e^{-\frac{1}{2}\left(3\Gamma+\Gamma_{12}\right)t}\rho_{es}\left(0\right),\\
\rho_{ea}\left(t\right) & = & e^{-i\left(\omega_{0}+\Omega_{12}\right)t}e^{-\frac{1}{2}\left(3\Gamma-\Gamma_{12}\right)t}\rho_{ea}\left(0\right),
\end{array}\nonumber
\end{equation} 
\begin{equation}
\begin{array}{lcl}
\rho_{eg}\left(t\right) & = & e^{-2i\omega_{0}t}e^{-\Gamma t}\rho_{eg}\left(0\right),\\
\rho_{sa}\left(t\right) & = & e^{-2i\Omega_{12}t}e^{-\Gamma t}\rho_{sa}\left(0\right),\\
\rho_{sg}\left(t\right) & = & e^{-i\left(\omega_{0}+\Omega_{12}\right)t}e^{-\frac{1}{2}\left(\Gamma+\Gamma_{12}\right)t} \left[\rho_{sg}\left(0\right)+\frac{\left(\Gamma+\Gamma_{12}\right)}{\left(\Gamma^{2}+4\Omega_{12}^{2}\right)}\left(\left\{ 2\Omega_{12}e^{-\Gamma t}\sin\left(2\Omega_{12}t\right)\right.\right.\right.+\Gamma\left(1-e^{-\Gamma t}\right.\\
 & \times & \left.\left.\cos\left(2\Omega_{12}t\right)\right)\right\} + i\left\{ 2\Omega_{12}\left(1-e^{-\Gamma t}\cos\left(2\Omega_{12}t\right)\right) - \left.\left.\Gamma e^{-\Gamma t}\sin\left(2\Omega_{12}t\right)\right\}\right)\rho_{es}\left(0\right)\right],\\
\rho_{ag}\left(t\right) & = & e^{-i\left(\omega_{0}-\Omega_{12}\right)t}e^{-\frac{1}{2}\left(\Gamma-\Gamma_{12}\right)t}\left[\rho_{ag}\left(0\right)-\frac{\left(\Gamma-\Gamma_{12}\right)}{\left(\Gamma^{2}+4\Omega_{12}^{2}\right)}\right. \left(\left\{ 2\Omega_{12}e^{-\Gamma t}\sin\left(2\Omega_{12}t\right)\right.\right.+\Gamma\left(1-e^{-\Gamma t}\right.\\
 & \times & \left.\left.\cos\left(2\Omega_{12}t\right)\right)\right\} - i\left\{ 2\Omega_{12}\left(1-e^{-\Gamma t}\cos\left(2\Omega_{12}t\right)\right) - \left.\left.\Gamma e^{-\Gamma t}\sin\left(2\Omega_{12}t\right)\right\} \right)\rho_{ea}\left(0\right)\right].
\end{array}\label{eq:matrix-elements}
\end{equation}
Here, all the matrix elements are written in the dressed state basis, which is connected with the bare state basis by
\[
\begin{array}{lcl}
\left|g\right\rangle  & = & \left|g_{1}\right\rangle \left|g_{2}\right\rangle ,\\
\left|s\right\rangle  & = & \frac{1}{\sqrt{2}}\left(\left|e_{1}\right\rangle \left|g_{2}\right\rangle +\left|g_{1}\right\rangle \left|e_{2}\right\rangle \right),\\
\left|a\right\rangle  & = & \frac{1}{\sqrt{2}}\left(\left|e_{1}\right\rangle \left|g_{2}\right\rangle -\left|g_{1}\right\rangle \left|e_{2}\right\rangle \right),\\
\left|e\right\rangle  & = & \left|e_{1}\right\rangle \left|e_{2}\right\rangle .
\end{array}
\]
\begin{figure}[t]
\centering{}\includegraphics[angle=-180,scale=0.8]{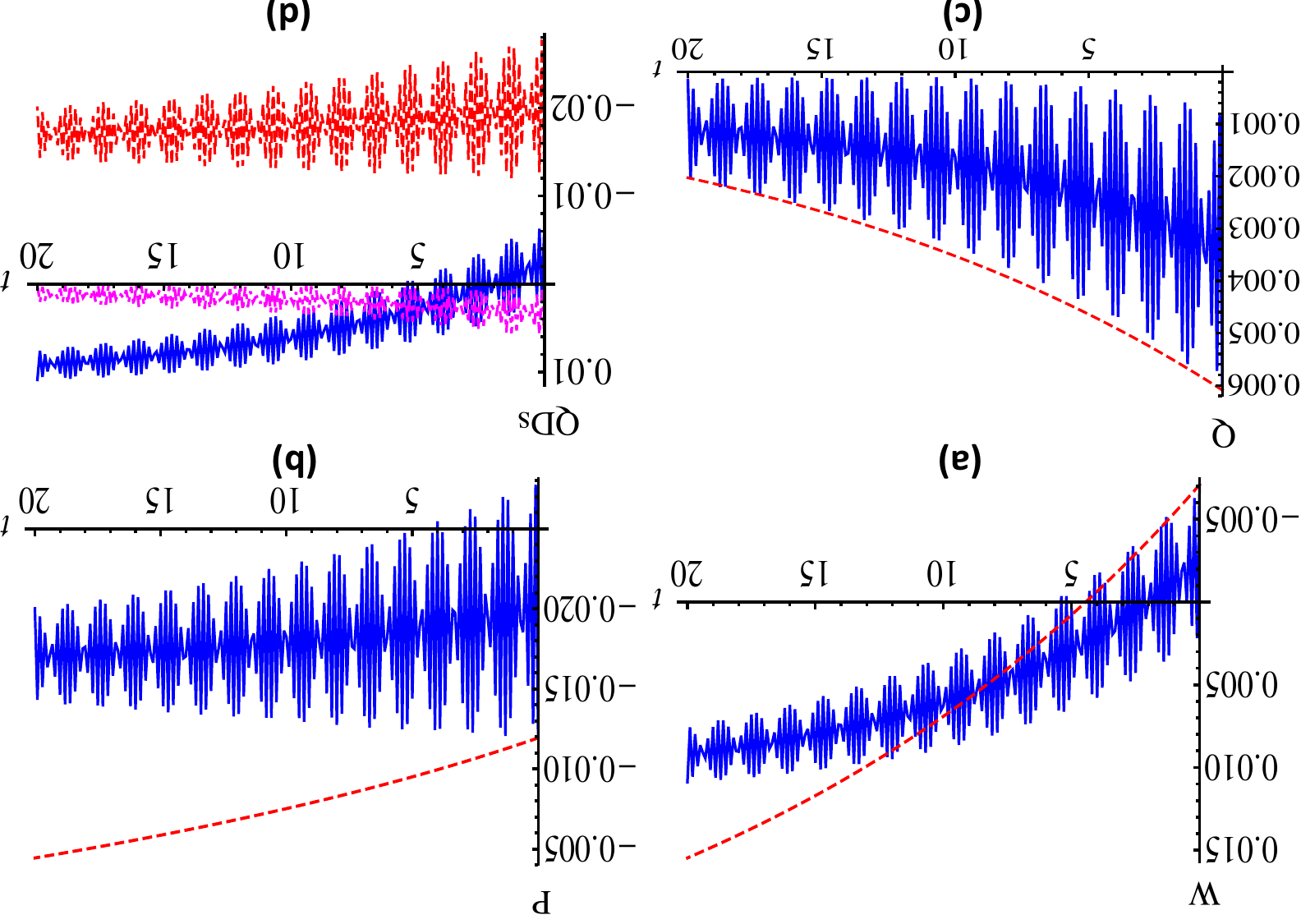}\caption[Quasiprobability distributions for two qubits undergoing dissipative evolution in the vacuum bath]{\label{fig:Vacuum-bath-qd} Variation of the Wigner,
$P$, and $Q$ functions with time is shown in (a)-(c) for the two-qubit
state, in the presence of a vacuum bath, with $\theta_{1}=\frac{\pi}{8},\,\theta_{2}=\frac{\pi}{3},\,\phi_{1}=\frac{\pi}{4},\,\phi_{2}=\frac{\pi}{4}$
with the inter-qubit spacing $r_{12}=0.05$ (smooth blue line), and
$r_{12}=2.0$ (dashed red line). In (d), all the quasiprobability distributions, varying with
time, are plotted together with inter-qubit spacing $r_{12}=0.05$.
Here, the $P$-function is seen to be negative for all times shown,
while the Wigner function is negative only for sometime, and the $Q$-function is always positive.}
\end{figure}
\begin{figure}[t]
\centering{}\includegraphics[angle=0,scale=0.8]{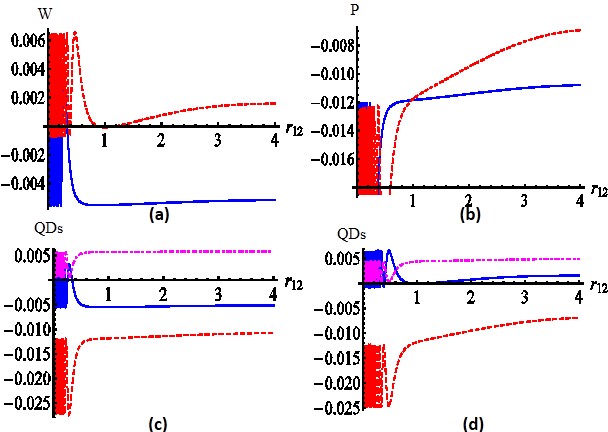}\caption[Effect of inter-qubit spacing on quasiprobability distributions for two qubits undergoing dissipative evolution in the vacuum bath]{\label{fig:Vacuumbath-2} (a) and (b) depict the Wigner
and $P$ functions for the two-qubit state, interacting with a vacuum
bath, as a function of the inter-qubit spacing at $t=1$ (smooth blue
line) and $t=5$ (dashed red line). In (c) and (d), the Wigner (smooth
blue line), $P$ (dashed red line), and $Q$ (dot-dashed magenta line)
quasiprobability distributions are plotted together, depicting their variation with inter-qubit
spacing at time $t=1$ and $t=5$, respectively. For all the plots
$\theta_{1}=\frac{\pi}{8},\,\theta_{2}=\frac{\pi}{3},\,\phi_{1}=\frac{\pi}{4},\,{\rm and}\,\phi_{2}=\frac{\pi}{4}$.}
\end{figure}
Further,
\[
\begin{array}{lcl}
\Omega_{ij} & = & \frac{3}{4}\sqrt{\Gamma_{i}\Gamma_{j}}\left[-\left\{1-\left(\hat{\mu}\mathord{\cdot}\hat{r}_{ij}\right)^{2}\right\}\frac{\cos\left(k_{0}r_{ij}\right)}{k_{0}r_{ij}} + \left\{1-3\left(\hat{\mu}\mathord{\cdot}\hat{r}_{ij}\right)^{2}\right\}\left(\frac{\sin\left(k_{0}r_{ij}\right)}{\left(k_{0}r_{ij}\right)^{2}}+\frac{\cos\left(k_{0}r_{ij}\right)}{\left(k_{0}r_{ij}\right)^{3}}\right)\right],
\end{array}
\]
where $\hat{\mu}=\hat{\mu}_{1}=\hat{\mu}_{2}$ are the unit vectors
along the atomic transition dipole moments, $\hat{r}_{ij}=\hat{r}_{i}-\hat{r}_{j}$,
and $k_{0}=\frac{\omega_{0}}{c}$ with $\omega_{0}=\frac{\omega_{1}+\omega_{2}}{2}$;
the spontaneous emission rate is 
\[
\begin{array}{lcl}
\Gamma_{i} & = & \frac{\omega_{i}^{3}\mu_{i}^{2}}{3\pi\epsilon\hbar c^{3}},\end{array}
\]
and the collective incoherent effect due to the dissipative multi-qubit
interaction with the bath is 
\[
\begin{array}{lcl}
\Gamma_{ij} & = & \Gamma_{ji}=\sqrt{\Gamma_{i}\Gamma_{j}}\,F\left(k_{0}r_{ij}\right),\end{array}
\]
for $i\neq j$ with 
\[
\begin{array}{lcl}
F\left(k_{0}r_{ij}\right) & = & \frac{3}{2}\left[\left\{1-\left(\hat{\mu}\mathord{\cdot}\hat{r}_{ij}\right)^{2}\right\}\frac{\sin\left(k_{0}r_{ij}\right)}{k_{0}r_{ij}} + \left\{1-3\left(\hat{\mu}\mathord{\cdot}\hat{r}_{ij}\right)^{2}\right\}\left(\frac{\cos\left(k_{0}r_{ij}\right)}{\left(k_{0}r_{ij}\right)^{2}}-\frac{\sin\left(k_{0}r_{ij}\right)}{\left(k_{0}r_{ij}\right)^{3}}\right)\right].
\end{array}
\]
Further, for the case of identical qubits, as  considered here,
$\Omega_{12}=\Omega_{21}$, $\Gamma_{12}=\Gamma_{21}$, and $\Gamma_{1}=\Gamma_{2}=\Gamma$. 
A detailed discussion can be found in Ref. \cite{banerjee2010dynamics}.

Here, the dynamics involve collective
coherent effects due to the multi-qubit interaction, as well collective
incoherent effects due to dissipative multi-qubit interaction with
the bath, and spontaneous emission. Analytic expressions of the corresponding
quasiprobability distributions are very cumbersome, hence we resort to numerically studying the
quasiprobability distributions for some parameters. Values of different parameters are as follows: the
wavevector and mean frequency $k_{0}=\omega_{0}=1$, the spontaneous emission
rate $\Gamma_{j}=0.05$, and $\hat{\mu}\mathord{\cdot}\hat{r}_{ij}=0$, where
$\hat{\mu}$ is equal to the unit vector along the atomic transition
dipole moment, and $\hat{r}_{ij}$ is the interatomic distance. Considering
the initial state with $\rho_{ee}\left(0\right)=\rho_{gg}\left(0\right)=\rho_{es}\left(0\right)=\rho_{ea}\left(0\right)=\rho_{eg}\left(0\right)=\rho_{sg}\left(0\right)=\rho_{ag}\left(0\right)=0$,
and $\rho_{ss}\left(0\right)=\rho_{aa}\left(0\right)=\rho_{sa}\left(0\right)=0.5$,
the Wigner, $P$, and $Q$ functions are calculated.

The variation of the different quasiprobability distributions is depicted in Figures \ref{fig:Vacuum-bath-qd}
and \ref{fig:Vacuumbath-2}. Figures \ref{fig:Vacuum-bath-qd} (a) and 
(b) show the negative values of the Wigner and $P$ functions for
some and for all times, respectively. The $Q$-function exhibits a
decaying pattern. These features are reinforced in the last plot of
the figure, where all the quasiprobability distributions are plotted together. In Figure \ref{fig:Vacuumbath-2},
various quasiprobability distributions are plotted with respect to the inter-qubit distance.
In the collective regime, $r_{12}\ll1$, the quasiprobability distributions exhibit an oscillatory
behavior, in consonance with the general behavior in this regime \cite{banerjee2010dynamics}.
Also, for the chosen parameters, the $P$-function is always negative,
while the Wigner function is negative for $t=1$, but becomes positive
for a longer time $t=5$, due to the dissipative influence of the
bath.

Thereafter, we have discussed quasiprobability distributions of the same two-qubit state, studied earlier in a dissipative interaction with a vacuum bath, evolving under the dissipative influence of a squeezed thermal bath. In this particular case,
we have shown that with an increase in $T$, the quasiprobability distributions, both $P$ and Wigner, which were
earlier exhibiting negative values start becoming positive, a clear
indicator of a quantum to classical transition (cf. Figure 7 of Ref. \cite{thapliyal2015quasiprobability}). Due to increase in the temperature, the oscillations observed in the quasiprobability distributions for the small inter-qubit 
spacing, corresponding to the case of interaction with the vacuum bath in Figure \ref{fig:Vacuumbath-2} (d), also decrease in the 
scenario of interaction with a squeezed thermal bath. Due to scarcity of space we have not included whole discussion here, which   is available in \cite{thapliyal2015quasiprobability}.

\subsubsection{EPR singlet state in an amplitude damping channel}

Now, we take an initially entangled two-qubit, EPR singlet, state
\cite{einstein1935can}. The evolution of this state is studied assuming independent
action of an AD channel on each qubit. Such a
scenario could be envisaged in a quantum memory net with the qubits
being its remote components, subjected locally to the AD noise \cite{yu2010entanglement}. 
Using Kraus operators of an AD channel obtained from Eq. (\ref{eq:SGAD-kraussoperators2-1}) by considering $p=1$, where $
\lambda\left(t\right) = 1-e^{-\gamma_{0}t}$ with $\gamma_{0}$ as the spontaneous emission rate. 
Further, assuming
that the two qubits, of the singlet, are independent and do not have
any interaction, the Kraus operators for the action of the AD channels,
one on each spin, can be modeled as 
\[
\begin{array}{lcl}
K_{1} & = & E_{0}\left(A\right)\otimes E_{0}\left(B\right),\\
K_{2} & = & E_{0}\left(A\right)\otimes E_{1}\left(B\right),\\
K_{3} & = & E_{1}\left(A\right)\otimes E_{0}\left(B\right),\\
K_{4} & = & E_{1}\left(A\right)\otimes E_{1}\left(B\right),
\end{array}
\]
where $A$ and $B$ stand for the first and second qubits (spins)
comprising the singlet, respectively. From the form of the Kraus operators
(\ref{eq:SGAD-kraussoperators2-1}) and assuming $\lambda_{A}=\lambda_{B}=\lambda$,
we have 
\[
\begin{array}{lcl}
K_{1} & = & \left[\begin{array}{cccc}
1-\lambda & 0 & 0 & 0\\
0 & \sqrt{1-\lambda} & 0 & 0\\
0 & 0 & \sqrt{1-\lambda} & 0\\
0 & 0 & 0 & 1
\end{array}\right],\end{array}
\]
\[
\begin{array}{lcl}
K_{2} & = & \left[\begin{array}{cccc}
0 & 0 & 0 & 0\\
\sqrt{\lambda\left(1-\lambda\right)} & 0 & 0 & 0\\
0 & 0 & 0 & 0\\
0 & 0 & \sqrt{\lambda} & 0
\end{array}\right],\end{array}
\]
\[
\begin{array}{lcl}
K_{3} & = & \left[\begin{array}{cccc}
0 & 0 & 0 & 0\\
0 & 0 & 0 & 0\\
\sqrt{\lambda\left(1-\lambda\right)} & 0 & 0 & 0\\
0 & \sqrt{\lambda} & 0 & 0
\end{array}\right],\end{array}
\]
and
\begin{equation}
\begin{array}{lcl}
K_{4} & = & \left[\begin{array}{cccc}
0 & 0 & 0 & 0\\
0 & 0 & 0 & 0\\
0 & 0 & 0 & 0\\
\lambda & 0 & 0 & 0
\end{array}\right].\end{array}\label{eq:AD-Krauss-2qubit}
\end{equation}
The density matrix of the singlet state at time $t$, under the action
of the above channel is 
\[
\begin{array}{lcl}
\rho\left(t\right) & = & \stackrel[i=1]{4}{\sum}K_{i}\left(t\right)\rho\left(0\right)K_{i}^{\dagger}\left(t\right),\end{array}
\]
where $\rho\left(0\right)=\left|\phi\right\rangle \left\langle \phi\right|$,
and $\begin{array}{lcl}
\left|\phi\right\rangle  & = & \frac{1}{\sqrt{2}}\left(\left|\frac{1}{2},-\frac{1}{2}\right\rangle -\left|-\frac{1}{2},\frac{1}{2}\right\rangle \right),\end{array}$ is the initial state at time $t=0$. Therefore, at time $t$ the evolved
density matrix is 
\begin{equation}
\begin{array}{lcl}
\rho\left(t\right) & = & \left[\begin{array}{cccc}
0 & 0 & 0 & 0\\
0 & \frac{1}{2}\left(1-\lambda\right) & -\frac{1}{2}\left(1-\lambda\right) & 0\\
0 & -\frac{1}{2}\left(1-\lambda\right) & \frac{1}{2}\left(1-\lambda\right) & 0\\
0 & 0 & 0 & \lambda
\end{array}\right].\end{array}\label{eq:rho-ADchannel}
\end{equation}

\begin{figure}[t]
\centering{}\includegraphics[angle=-180,scale=0.5]{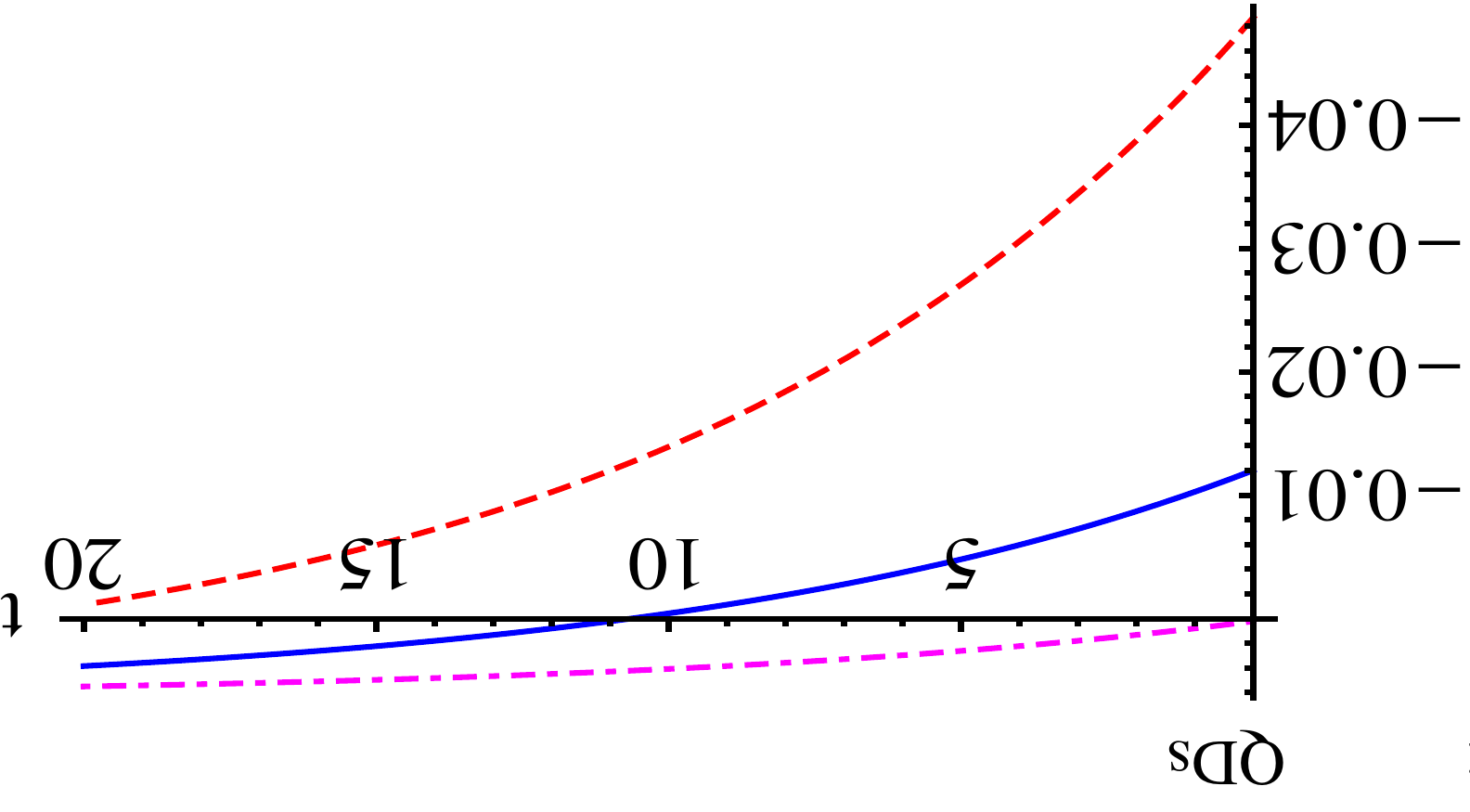}\caption[Quasiprobability distributions for EPR state evolving under AD channel]{\label{fig:EPR-AD} The variation of all the quasiprobability distributions with
time $t$ for EPR singlet state in the presence of AD channel with
$\gamma_{0}=0.1$ and $\theta_{1}=\frac{\pi}{2},\theta_{2}=\frac{\pi}{2},\phi_{1}=\frac{\pi}{4},\phi_{2}=\frac{\pi}{3}$.
The smooth (blue), dashed (red), and dot-dashed (magenta) lines are for the
Wigner, $P$, and $Q$ functions, respectively.}
\end{figure}

On the evolved state, represented by the above density matrix (\ref{eq:rho-ADchannel}), we
may now apply the prescription for obtaining the quasiprobability distributions to yield compact
analytical expressions of the quasiprobability distributions. Specifically, the Wigner
function is obtained as 
\begin{equation}
\begin{array}{lcl}
W\left(\theta_{1},\phi_{1},\theta_{2},\phi_{2}\right) & = & \frac{1}{16\pi^{2}}\left[\lambda\left\{ 1+3\cos\theta_{1}\cos\theta_{2}-\sqrt{3}\left(\cos\theta_{1}+\cos\theta_{2}\right)\right\} \right.\\
 & + & \left.\left(1-\lambda\right)\left\{ 1-3\cos\theta_{1}\cos\theta_{2}-3\sin\theta_{1}\sin\theta_{2}\cos\left(\phi_{1}-\phi_{2}\right)\right\} \right],
\end{array}\label{eq:Wigner-AD-EPR}
\end{equation}
while the $P$-function is 
\begin{equation}
\begin{array}{lcl}
P\left(\theta_{1},\phi_{1},\theta_{2},\phi_{2}\right) & = & \frac{1}{16\pi^{2}}\left[\lambda\left\{ 1+9\cos\theta_{1}\cos\theta_{2}+3\left(\cos\theta_{1}+\cos\theta_{2}\right)\right\} \right.\\
 & + & \left.\left(1-\lambda\right)\left\{ 1-9\cos\theta_{1}\cos\theta_{2}-9\sin\theta_{1}\sin\theta_{2}\cos\left(\phi_{1}-\phi_{2}\right)\right\} \right],
\end{array}\label{eq:P-AD-EPR}
\end{equation}
and the $Q$-function is 
\begin{equation}
\begin{array}{lcl}
Q\left(\theta_{1},\phi_{1},\theta_{2},\phi_{2}\right) & = & \frac{1}{16\pi^{2}}\left[\lambda\left\{ 1+\cos\theta_{1}\cos\theta_{2}+\left(\cos\theta_{1}+\cos\theta_{2}\right)\right\} \right.\\
 & + & \left.\left(1-\lambda\right)\left\{ 1-\cos\theta_{1}\cos\theta_{2}-\sin\theta_{1}\sin\theta_{2}\cos\left(\phi_{1}-\phi_{2}\right)\right\} \right].
\end{array}\label{eq:Q-AD-EPR}
\end{equation}
The quasiprobability distributions, reported here for an EPR pair (singlet state) evolving under
AD channel exactly match with the corresponding noiseless results
\cite{agarwal1981relation,agarwal1993perspective,ramachandran1996quasi,usha2002spin}, by setting $\lambda=0$ in the above expressions.
The variation of different quasiprobability distributions with time is shown in Figure \ref{fig:EPR-AD}.
The $P$ and Wigner functions are found to show negative values for
a long time, indicative of the perfect initial entanglement in the
system, and finally become positive due to exposure to noise.

As all the quasiprobability distributions, in this case, are symmetric pairwise in $\left(\theta_{1}\leftrightarrow\theta_{2}\right)$
and $\left(\phi_{1}\leftrightarrow\phi_{2}\right)$, hence either
or both of these exchanges would leave the expressions unchanged.
For $\theta_{1}=-\theta_{2}=\frac{\pi}{2}$, and $\phi_{1}=\phi_{2}=0$,
we observe the classically perfect anti-correlation of spins. We can
also observe that for certain angles, viz., $\theta_{1}=\theta_{2}=\frac{\pi}{2}$,
and $\phi_{1}-\phi_{2}=\frac{n\pi}{2},$ where $n$ is an odd integer,
all the quasiprobability distributions become equal to $\frac{1}{\left(4\pi\right)^{2}}$, a
result which remains unaffected by the presence or absence of noise.
Thus, for these settings, the evolution of the quasiprobability distributions becomes noise
independent.

We have also studied nonclassicality in a two-qubit system under QND noise using different quasiprobability distributions, which is reported in \cite{thapliyal2015quasiprobability}, where the findings of the present chapter are published. In that case, both the $P$ and Wigner functions exhibited
negative values, indicative of the quantumness in the system. After initiation of the evolution the nonclassicality decays 
due to dephasing caused by the bath (cf. Figure 3 of Ref. \cite{thapliyal2015quasiprobability}). Here, due to lack of space we have not discussed this particular case in detail.

\subsection{Three-qubit quasiprobability distributions evolving in an amplitude damping channel}

Three-qubit entangled states can be classified into two classes (GHZ
and W classes) of quantum states, such that a state of W (GHZ) class
cannot be transformed to a state of GHZ (W) class by using LOCC (local
operation and classical communication). Here, we study both GHZ \cite{greenberger1989going,bouwmeester1999observation}
and W \cite{dur2000three} class of states. To simulate the effect of noise,
we consider the scenario wherein the first qubit is affected by the
AD channel. An arbitrary effect of noise on each subsystem could be thought of as more natural. 
The assumption of only one subsystem affected by the AD noise is consistent with the 
effect of noise considered in, for example, various cryptrographic protocols  (\cite{sharma2015controlled} and references therein), 
where it is commonly assumed that the qubits which travel through the channel are affected by noise 
while the channel noise does not affect the qubits to be teleported or not traveling through it.

\subsubsection{GHZ state in an amplitude damping channel}

\begin{figure}
\centering{}\includegraphics[angle=-180,scale=0.6]{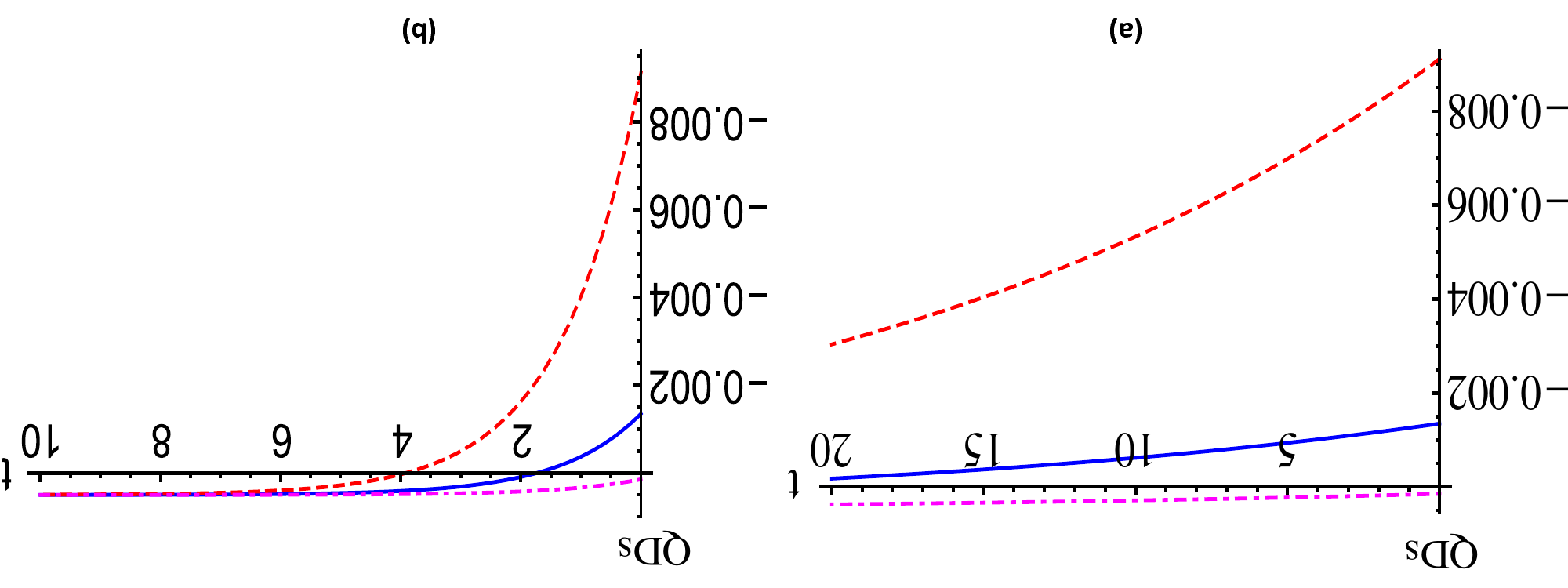}\caption[Quasiprobability distributions for GHZ state over AD and GAD channels]{\label{fig:GHZ-AD} 
Variation of all quasiprobability distributions with
time for the GHZ state when acted upon by (a) an AD noise on the first
qubit and (b) a GAD noise on each qubit, with $\gamma_{0}=0.1$ and 
$\theta_{1}=\frac{\pi}{2},\,\theta_{2}=\frac{\pi}{2},\,\theta_{3}=\frac{\pi}{2},\,\phi_{1}=\frac{\pi}{4},\,\phi_{2}=
\frac{\pi}{3},\,\phi_{3}=\frac{\pi}{6}$. In (b), different quasiprobability distributions are shown for $\omega=1.0$ 
with the first, second, and third qubits subjected to GAD noise at $T=0$, 1, and 2, respectively. 
In both the plots, the smooth (blue), dashed (red), and dot-dashed (magenta) lines are for
the Wigner, $P$, and $Q$ functions, respectively.}
\end{figure}

The GHZ (Greenberger\textendash{}Horne\textendash{}Zeilinger) state
is a three-qubit quantum state $
\left|GHZ\right\rangle   =  \frac{1}{\sqrt{2}}\left(\right.\left|000\right\rangle +\left|111\right\rangle \left.\right).$ The first qubit of the state is acted upon by an AD channel while the remaining two qubits remain unaffected. Here, assuming that the three
qubits are independent of each other and do not have any interactions,
the Kraus operators for the action of the AD channel only on the first
qubit, can be modeled as 
\begin{equation}
\begin{array}{lcl}
K_{1} & = & E_{0}\left(A\right)\otimes\mathbb{I}\left(B\right)\otimes\mathbb{I}\left(C\right),\\
K_{2} & = & E_{1}\left(A\right)\otimes\mathbb{I}\left(B\right)\otimes\mathbb{I}\left(C\right),
\end{array}\label{eq:KrausOp-GHZ_W}
\end{equation}
where $E_{0}$ and $E_{1}$ are the
Kraus operators for the AD channel, which can be obtained from 
Eq. (\ref{eq:SGAD-kraussoperators2-1}) as in the previous section,
and $\mathbb{I}$ is a $2\times2$ identity matrix. Also, $A$, $B,$
and $C$ stand for the first, second, and third qubits of the GHZ state,
respectively. Thus, the density matrix of the GHZ state at time $t$
in the AD channel is 
\begin{equation}
\begin{array}{lcl}
\rho\left(t\right) & = & \overset{2}{\underset{i=1}{\Sigma}}K_{i}\left(t\right)\rho\left(0\right)K_{i}^{\dagger}\left(t\right),\end{array}\label{eq:density-mat-GHZ_W}
\end{equation}
where $\rho\left(0\right)=\left|GHZ\right\rangle \left\langle GHZ\right|,$
is the initial state at time $t=0$. Thus, at time $t$ the density
matrix for the GHZ state evolving in the presence of the AD channel
is 
\begin{equation}
\begin{array}{lcl}
\rho\left(t\right) & = & \frac{1}{2}\left(\left(1-\lambda\right)\left|000\right\rangle \left\langle000\right| + \lambda\left|100\right\rangle \left\langle100\right| + \left|111\right\rangle \left\langle111\right| +\sqrt{\left(1-\lambda\right)} \left\{\left|000\right\rangle \left\langle111\right|+{\rm{H.c.}} \right\}\right). \end{array}\label{eq:rho-AD-GHZ}
\end{equation}
Analytical expressions can be obtained for the different quasiprobability distributions for the
time evolved GHZ state described by Eq. (\ref{eq:rho-AD-GHZ}). Specifically, we obtain the Wigner function as 
\begin{equation}
\begin{array}{lcl}
W\left(\theta_{1},\phi_{1},\theta_{2},\phi_{2},\theta_{3},\phi_{3}\right) & = & \frac{1}{64\pi^{3}}\left[1-\sqrt{3}\lambda\cos\theta_{1}+3\cos\theta_{2}\cos\theta_{3}\right.\\
&+& 3\left(1-\lambda\right)\cos\theta_{1}\left(\cos\theta_{2}+\cos\theta_{3}\right)-3\sqrt{3}\left\{ \lambda\cos\theta_{1}\cos\theta_{2}\cos\theta_{3}\right.\\
 & - & \left.\left.\sqrt{\left(1-\lambda\right)}\sin\theta_{1}\sin\theta_{2}\sin\theta_{3}\cos\left(\phi_{1}+\phi_{2}+\phi_{3}\right)\right\} \right],
\end{array}\label{eq:Wigner-AD-GHZ}
\end{equation}
while the $P$-function is 
\begin{equation}
\begin{array}{lcl}
P\left(\theta_{1},\phi_{1},\theta_{2},\phi_{2},\theta_{3},\phi_{3}\right) & = & \frac{1}{64\pi^{3}}\left[1+3\lambda\cos\theta_{1}+9\cos\theta_{2}\cos\theta_{3}\right.\\
&+& 9\left(1-\lambda\right)\cos\theta_{1}\left(\cos\theta_{2}+\cos\theta_{3}\right)+27\left\{ \lambda\cos\theta_{1}\cos\theta_{2}\cos\theta_{3}\right.\\
 & - & \left.\left.\sqrt{\left(1-\lambda\right)}\sin\theta_{1}\sin\theta_{2}\sin\theta_{3}\cos\left(\phi_{1}+\phi_{2}+\phi_{3}\right)\right\} \right],
\end{array}\label{eq:P-AD-GHZ}
\end{equation}
and the $Q$-function is 
\begin{equation}
\begin{array}{lcl}
Q\left(\theta_{1},\phi_{1},\theta_{2},\phi_{2},\theta_{3},\phi_{3}\right) & = & \frac{1}{64\pi^{3}}\left[1+\lambda\cos\theta_{1}+\cos\theta_{2}\cos\theta_{3}\right.\\
&+& \left(1-\lambda\right)\cos\theta_{1}\left(\cos\theta_{2}+\cos\theta_{3}\right)+\lambda\cos\theta_{1}\cos\theta_{2}\cos\theta_{3}\\
 & - & \left.\sqrt{\left(1-\lambda\right)}\sin\theta_{1}\sin\theta_{2}\sin\theta_{3}\cos\left(\phi_{1}+\phi_{2}+\phi_{3}\right)\right].
\end{array}\label{eq:Q-AD-GHZ}
\end{equation}
The variation of different quasiprobability distributions, as in Eqs. (\ref{eq:Wigner-AD-GHZ})-(\ref{eq:Q-AD-GHZ}), with 
time is illustrated, for a particular
choice of the parameters, in Figure \ref{fig:GHZ-AD} (a). The
Wigner and $P$ functions exhibit negative values (nonclassical character)
for the times shown, which could be attributed to the initial entanglement
in the state. Further, for the sake of generality, we depict in Figure \ref{fig:GHZ-AD} (b) the scenario wherein
all three qubits are affected by the GAD noise \cite{srikanth2008squeezed}, which can be obtained from 
Eq. (\ref{eq:SGAD-kraussoperators2-1}) by setting the bath squeezing parameters to zero. 
All three qubits are subjected to different 
temperatures corresponding to independent environment for each qubit and simulates the scenario where 
each qubit travels through an independent channel. It can be observed that the nonclassicality indicated by the 
negativity of the Wigner and $P$ functions at $t=0$ decays more rapidly when the last two qubits are subjected to 
finite temperature noises.

\subsubsection{W state in an amplitude damping channel}

\begin{figure}
\centering{}\includegraphics[angle=-180,scale=0.6]{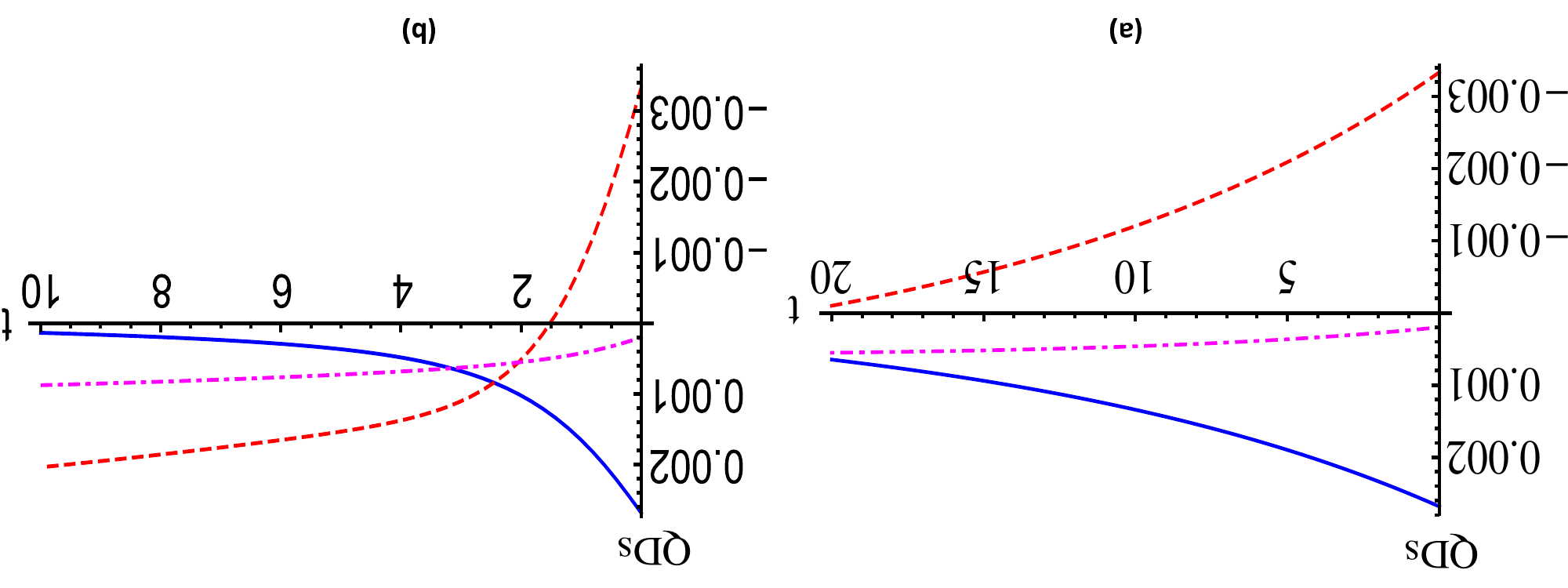}\caption[Quasiprobability distributions for W state evolving under AD and GAD channels]{\label{fig:W-AD} 
(a) Variation of the quasiprobability distributions with time for the
W state in the presence of an AD channel, acting only upon the first
qubit, with $\gamma_{0}=0.1$ and $\theta_{1}=\frac{\pi}{4},
\,\theta_{2}=\frac{\pi}{6},\,\theta_{3}=\frac{\pi}{3},\,\phi_{1}=\frac{\pi}{8},\,\phi_{2}=\frac{\pi}{4},
\,\phi_{3}=\frac{\pi}{6}$. (b) The quasiprobability distributions for the W state, when all the qubits are subjected to 
GAD noise, with $\omega=1.0$ and $T=0$, 1, and 2 for the first, second, and third qubits, 
respectively. The remaining values of the parameters in (b) are same as in (a).
The smooth (blue), dashed (red), and dot-dashed (magenta) lines correspond to the
Wigner, $P$, and $Q$ functions, respectively.}
\end{figure}

For our second example of quasiprobability distributions of three-qubit states, we take up the
W state, $
\left|W\right\rangle =  \frac{1}{\sqrt{3}}\left(\left|001\right\rangle +\left|010\right\rangle +\left|100\right\rangle \right).$ As before, we consider the evolution where only the first qubit of
the state is acted upon by an AD channel. The Kraus operators, describing
the evolution are given by Eq. (\ref{eq:KrausOp-GHZ_W}). Here, as
in the case of GHZ or EPR states, assuming that the three qubits are
independent of each other and do not have any interactions, the density
matrix of the evolved state is, as in the last case, given by Eq.
(\ref{eq:density-mat-GHZ_W}), where $\rho\left(0\right)=\left|W\right\rangle \left\langle W\right|,$
is the initial state at time $t=0$. Therefore, at time $t$, the W state
evolves, in the presence of the AD channel, to 
\begin{equation}
\begin{array}{lcl}
\rho\left(t\right) & = & \frac{1}{3}
\left(\left(1-\lambda\right)\left\{\left|001\right\rangle \left\langle001\right| +\left|010\right\rangle \left\langle010\right|  \right\}+ \lambda\left\{\left|101\right\rangle \left\langle101\right| +\left|110\right\rangle \left\langle110\right| \right\}+ \left|100\right\rangle \left\langle100\right| \right.\\
&+&\left.\left\{\sqrt{\left(1-\lambda\right)} \left(\left|001\right\rangle \left\langle100\right|+\left|010\right\rangle \left\langle100\right|\right)+\left(1-\lambda\right) \left|001\right\rangle \left\langle010\right|+\lambda\left|101\right\rangle \left\langle110\right|+{\rm{H.c.}} \right\}\right).\end{array}\label{eq:rho-AD-W}
\end{equation}
In this case, again, making use of Eq. (\ref{eq:rho-AD-W}), analytical
forms of the different quasiprobability distributions can be obtained as follows 
\begin{equation}
\begin{array}{lcl}
W\left(\theta_{1},\phi_{1},\theta_{2},\phi_{2},\theta_{3},\phi_{3}\right) & = & \frac{1}{64\pi^{3}}\left[1-\left(\cos\theta_{1}\cos\theta_{2}+\cos\theta_{2}\cos\theta_{3}+\cos\theta_{1}\cos\theta_{3}\right)\right.\\
&+& \frac{\sqrt{3}}{3}\left(\cos\theta_{1}+\cos\theta_{2}+\cos\theta_{3}\right)-
3\sqrt{3}\cos\theta_{1}\cos\theta_{2}\cos\theta_{3}\\
 & + & 2\left(1+\sqrt{3}\cos\theta_{1}\right)\sin\theta_{2}\sin\theta_{3}\cos\left(\phi_{2}-\phi_{3}\right)\\
 & + & 2\sqrt{\left(1-\lambda\right)}\left\{ \left(1+\sqrt{3}\cos\theta_{2}\right)\sin\theta_{1}\sin\theta_{3}\cos\left(\phi_{1}-\phi_{3}\right)\right.\\
 & + & \left.\left(1+\sqrt{3}\cos\theta_{3}\right)\sin\theta_{1}\sin\theta_{2}\cos\left(\phi_{1}-\phi_{2}\right)\right\} \\
 & + & \left.4\sqrt{3}\lambda\cos\theta_{1}\left\{ -\frac{1}{3}+\cos\theta_{2}\cos\theta_{3}-\sin\theta_{2}\sin\theta_{3}\cos\left(\phi_{2}-\phi_{3}\right)\right\} \right],
\end{array}\label{eq:Wigner-AD-W}
\end{equation}

\begin{equation}
\begin{array}{lcl}
P\left(\theta_{1},\phi_{1},\theta_{2},\phi_{2},\theta_{3},\phi_{3}\right) & = & \frac{1}{64\pi^{3}}\left[1-3\left(\cos\theta_{1}\cos\theta_{2}+\cos\theta_{2}\cos\theta_{3}+\cos\theta_{1}\cos\theta_{3}\right)\right.\\
&-& \left(\cos\theta_{1}+\cos\theta_{2}+\cos\theta_{3}\right)+27\cos\theta_{1}\cos\theta_{2}\cos\theta_{3}\\
 & + & 6\left(1-3\cos\theta_{1}\right)\sin\theta_{2}\sin\theta_{3}\cos\left(\phi_{2}-\phi_{3}\right)\\
 & + & 6\sqrt{\left(1-\lambda\right)}\left\{ \left(1-3\cos\theta_{2}\right)\sin\theta_{1}\sin\theta_{3}\cos\left(\phi_{1}-\phi_{3}\right)\right.\\
 & + & \left.\left(1-3\cos\theta_{3}\right)\sin\theta_{1}\sin\theta_{2}\cos\left(\phi_{1}-\phi_{2}\right)\right\} \\
 & + & \left.4\lambda\cos\theta_{1}\left\{ 1-9\cos\theta_{2}\cos\theta_{3}+9\sin\theta_{2}\sin\theta_{3}\cos\left(\phi_{2}-\phi_{3}\right)\right\} \right],
\end{array}\label{eq:P-AD-W}
\end{equation}
and 
\begin{equation}
\begin{array}{lcl}
Q\left(\theta_{1},\phi_{1},\theta_{2},\phi_{2},\theta_{3},\phi_{3}\right) & = & \frac{1}{192\pi^{3}}\left[3-\left(\cos\theta_{1}\cos\theta_{2}+\cos\theta_{2}\cos\theta_{3}+\cos\theta_{1}\cos\theta_{3}\right)\right.\\
&-& \left(\cos\theta_{1}+\cos\theta_{2}+\cos\theta_{3}\right)+3\cos\theta_{1}\cos\theta_{2}\cos\theta_{3}\\
 & + & 4\sin\frac{\theta_{1}^{2}}{2}\sin\theta_{2}\sin\theta_{3}\cos\left(\phi_{2}-\phi_{3}\right)+4\sqrt{\left(1-\lambda\right)}\sin\theta_{1}\\
 & \times & \left\{ \sin\frac{\theta_{2}^{2}}{2}\sin\theta_{3}\cos\left(\phi_{1}-\phi_{3}\right)+\sin\theta_{2}\sin\frac{\theta_{3}^{2}}{2}\cos\left(\phi_{1}-\phi_{2}\right)\right\} \\
 & + & \left.4\lambda\cos\theta_{1}\left\{ 1-\cos\theta_{2}\cos\theta_{3}+\sin\theta_{2}\sin\theta_{3}\cos\left(\phi_{2}-\phi_{3}\right)\right\} \right].
\end{array}\label{eq:Q-AD-W}
\end{equation}
The variation of quasiprobability distributions with time is shown in Figure \ref{fig:W-AD} (a)
for a particular choice of the parameters. Here, the $P$-function
exhibits negative values for the times shown, but in contrast to the
GHZ case, the Wigner function is found to be positive. Thus, the signature
of quantumness (nonclassicality) of the state is identified by the $P$-function, but the Wigner function fails to detect the same. Similar to the quasiprobability distributions of the GHZ state in the presence
of GAD noise, a decrease in the nonclassicality of the W state with time can be observed due to the 
effect of noise on the last two qubits at non-zero temperatures (cf. Figure \ref{fig:W-AD} (b)).

\section{Nonclassical volume}

Till now, we have studied nonclassicality using the negative values of
the Wigner or $P$ function. The negative values of the quasiprobability distributions only provide
a signature of nonclassicality, but they do not provide a quantitative
measure of nonclassicality. There do exist some quantitative measures
of nonclassicality, see for example, \cite{miranowicz2015statistical} for a
review. One such measure is nonclassical volume introduced in \cite{kenfack2004negativity}.
In this approach, the doubled volume of the integrated negative part of
the Wigner function of a given quantum state is used as a quantitative measure of the quantumness \cite{kenfack2004negativity}. Using our knowledge of the Wigner
functions for various systems, studied here, the nonclassical volume
$\delta$, which is defined as 
\begin{equation}
\delta=\int\left|W\left(\theta,\phi\right)\right|\sin\theta d\theta d\phi-1,\label{eq:ncv}
\end{equation}
can be computed. It can be easily observed that a nonzero value of
$\delta$ would imply the existence of nonclassicality, but this measure
is not useful in measuring inherent nonclassicality in all quantum
states. This is so because, the Wigner function is only a witness of
nonclassicality (it does not provide a necessary condition). However,
this measure of nonclassicality has been used in a number of optical
systems, see for example, \cite{miranowicz2014phase,pathak2014wigner} and references
therein.

\begin{figure}[t]
\begin{centering}
\includegraphics[angle=-180,scale=0.59]{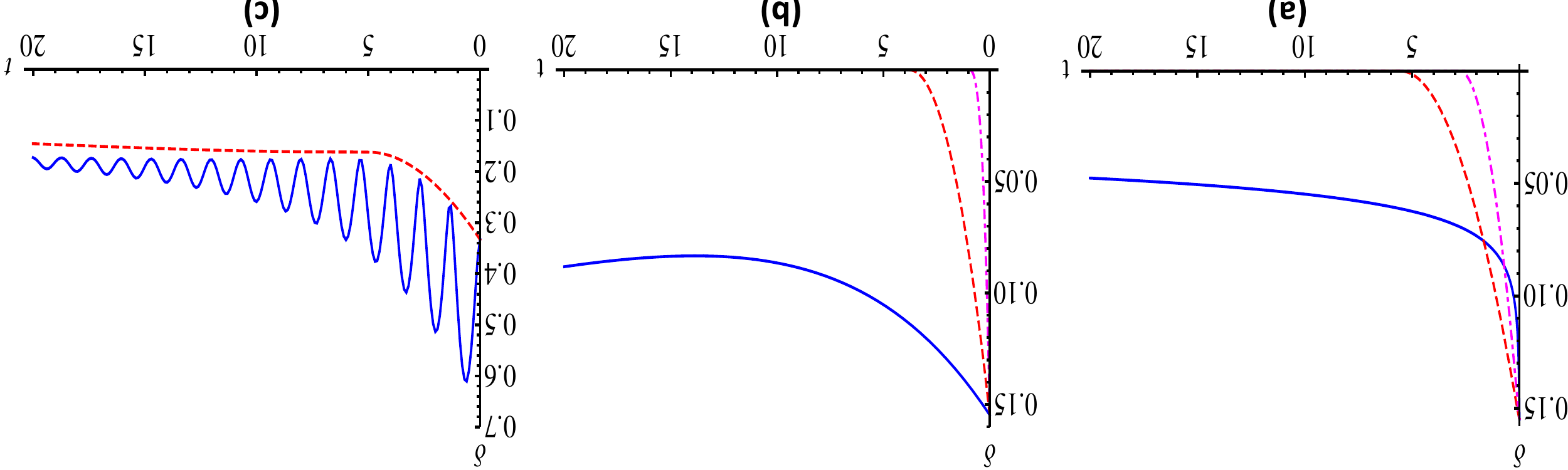} 
\par\end{centering}

\caption[Variation
of the nonclassical volume in the presence of different noises]{\label{fig:W_Vol} Variation
of the nonclassical volume in the presence of different noises. (a)
The variation of the nonclassical volume with time is shown for the
single spin-$\frac{1}{2}$ atomic coherent state in the presence of
the QND noise with $\gamma_{0}=0.1,\, r=0,\, a=0,\,\omega_{c}=100,\,\omega=1.0,$
and $\alpha=\frac{\pi}{2},\,\beta=\frac{\pi}{3},$ in the units of
$\hbar=k_{B}=1$, where the smooth (blue), dashed (red), and dot-dashed
(magenta) lines correspond to different temperatures $T=0,\,1,$ and
$2$, respectively. (b) The variation of the nonclassical volume with
time is shown for a single spin-$\frac{1}{2}$ atomic coherent state
in a SGAD channel, where the smooth (blue) line corresponds to the
variation in the nonclassical volume in a vacuum bath, i.e., at $T=0$ and
squeezing parameters $r=\xi=0$ (AD channel); the dashed
(red) line corresponds to the variation in a channel with zero squeezing
at $T=3$, i.e., GAD channel, and the dot-dashed
(magenta) line corresponds to the variation with squeezing $r=1,$
squeezing angle $\xi=0,$ and $T=3$. In all these cases, $\alpha$
and $\beta$ have the same values as in (a). (c) The behavior of the nonclassical
volume with time is depicted in a vacuum bath for the state discussed
in Section \ref{sub:Vacuum-bath} with the inter-qubit spacing $r_{12}=0.05$
(smooth blue line) and $r_{12}=2.0$ (red dashed line).}
\end{figure}

Here, we will illustrate the time evolution of $\delta$ for some
of the spin-qubit systems, studied above. Specifically, Figures \ref{fig:W_Vol}
(a) and (b) show the variation of $\delta$ for two spin-$\frac{1}{2}$
systems, initially in an atomic coherent state under the influence
of QND and SGAD channels, respectively. Figure \ref{fig:W_Vol} (c) shows
two spin-$\frac{1}{2}$ states in a two-qubit vacuum bath. The dashed
and dot-dashed lines in Figures \ref{fig:W_Vol} (a) and (b), i.e., for the atomic
coherent state in QND noise with a finite temperature and in SGAD channels with a finite
temperature and squeezing, exhibit an exponential reduction of the nonclassical
volume with time implying a quick transition from nonclassical to
classical states, whereas the smooth lines in Figures \ref{fig:W_Vol} (a) and (b) (when temperature
and squeezing parameters are taken to be zero) show that after an
initial reduction, the nonclassical volume stabilizes over a reasonably
large duration. Thus, nonclassicality does not get completely destroyed
with time. A similar nature of the time evolution of $\delta$ is also
observed for the dashed line in Figure \ref{fig:W_Vol} (c) (a two-qubit
state in a vacuum bath with a relatively large inter-qubit spacing),
whereas an oscillatory nature is observed for a small inter-qubit spacing,
depicted here by a smooth line. It should be noted that the smooth blue line in Figure \ref{fig:W_Vol} (b) 
corresponds to the nonclassical volume for an
atomic coherent state dissipatively interacting with a vacuum bath, i.e., at zero temperature and squeezing, 
while the Wigner function of the atomic coherent state, 
illustrated in Figure \ref{fig:SGAD-ACS-qd}, is for non-zero temperature and squeezing. At zero temperature, 
the nonclassicality present in the system is expected to 
survive for a relatively longer period of time. Further, the nonclassical volume is the overall contribution in nonclassicality
from all values of $\theta$ and $\phi$. It is possible that the Wigner function 
shown for a particular value of $\theta$ and $\phi$, in the previous sections, may not exhibit nonclassical behavior at 
time $t$ whereas the other possible values provide a finite contribution to the nonclassical volume, 
resulting in a nonvanishing $\delta$.

\section{Conclusions \label{sec:Conclusions-QDs}}

The nonclassical nature of all the systems studied here, of relevance
to the fields of quantum optics and information, is illustrated via
their quasiprobability distributions as a function of the time of
evolution as well as various state or bath parameters. We also provide
a quantitative idea of the amount of nonclassicality observed in some
of the systems studied using a measure which essentially makes use
of the Wigner function. These issues assume significance in questions
related to quantum state engineering, where the central point is to
have a clear understanding of coherences in the quantum mechanical
system being used. Thus, it is essential to have an understanding
over quantum to classical transitions, under ambient conditions. This
is made possible by the present work, where a comprehensive analysis
of quasiprobability distributions for spin-qubit systems is made under general open system effects,
including both pure dephasing as well as dissipation, making it relevant
from the perspective of experimental implementation. Along with the
well-known Wigner, $P$, and $Q$ quasiprobability distributions, we
also discuss the so called $F$-function and specify its relation
to the Wigner function. We expect this work to have an impact on issues
related to state reconstruction, in the presence of decoherence and
dissipation. These quasiprobability distributions also play an important
role in the fundamental issues such as complementarity between number
and phase distributions as well as for phase dispersion in atomic
systems. It is interesting to note that in \cite{veitch2012negative}, a connection
was established between the negative values of a particular quasiprobability distribution
and the potential for quantum speed-up. The present study can be of
use to probe this connection deeper. In this chapter, we have not included the quasiprobability distributions obtained for a spin-1 and $N$-qubit Dicke model, which were part of Ref. \cite{thapliyal2015quasiprobability}, where the findings of the present chapter are published. The next chapter deals with the probability distribution, which is directly measurable in experiment, and is often used to reconstruct the quasiprobability distributions discussed in the present chapter.

\mathversion{normal}
\thispagestyle{empty}

\mathversion{normal2}
\titlespacing*{\chapter}{0pt}{-50pt}{20pt}
\chapter{Tomograms for open quantum systems: Optical
and spin systems \label{Tomogram}}

\section{Introduction}

A quantum state can be characterized by a number of probability and
quasiprobability distribution functions \cite{paris2004quantum}. The quasiprobability
distributions, as discussed previously in Section \ref{quasi} and Chapter \ref{QDs}, are not true probability distributions as most of them
can have negative values. Interestingly, this nonpositivity can be viewed as a signature
of nonclassicality or quantumness. As there does not exist any 
straightforward prescription for direct measurement of these quasiprobability distributions, several efforts have been made to construct measurable probability 
distributions that can be used to uniquely construct either all or some of these quasiprobability distributions. Such measurable probability distributions 
are referred to as  tomograms  \cite{ibort2009introduction,miranowicz2014phase,filippov2011optical,man1997spin}. 
In other words, a tomogram is a scheme for measuring a quantum state by using a representation in one-to-one correspondence with the probability
distribution \cite{man1997damped}. 
 A relationship between a tomogram and a quasidistribution function, such as the
Wigner function, can be established for both continuous and discrete variable systems \cite{vogel1989determination,leonhardt1996discrete,leonhardt1995quantum}. 
Specifically, in Ref. \cite{vogel1989determination}, it was shown that quasiprobability distributions ($P$, $Q$, and Wigner functions) can be uniquely determined in terms of the
probability distributions for the rotated quadrature phase which can be viewed as an optical tomogram of the state. Similarly, in Refs.  \cite{leonhardt1996discrete,leonhardt1995quantum}, it was shown that for finite dimensional phase states, discrete Wigner functions and
tomograms are connected by a discretization of the continuous variable
Radon transformation and was referred to as the \textit{Plato transformation}.

In the recent past, a few successful attempts have been made to measure the
Wigner function directly in experiments \cite{miquel2002interpretation,smithey1993measurement},
but the methods adopted were state specific. The same limitation is also present in the theoretical
proposals  \cite{smithey1993measurement} for the measurement of Wigner function. Further, optical homodyne tomography has been employed for the experimental measurement of the 
Wigner functions of vacuum and squeezed states in \cite{smithey1993measurement,beck1993experimental},  while distributions corresponding to Pegg-Barnett and 
Susskind-Glogower phase operators were also obtained in \cite{beck1993experimental}. An experimental measurement of 
the $P$, $Q$, and Wigner quantum phase distributions for the
squeezed vacuum state has been reported in \cite{smithey1993complete}. Precision of homodyne tomography technique was compared with the conventional detection techniques in 
\cite{d1994precision}. A number of alternative methods of tomography have also been proposed \cite{d1994detection,d1996quantum,mancini1996symplectic} and 
exploited to obtain the phase distributions, like the Wigner and $Q$ functions \cite{mancini1995wigner}. Further, in \cite{lvovsky2009continuous}, continuous variable quantum state tomography was
reviewed from the perspective of quantum information. 
In brief, various facets of quasiprobability distributions have been studied in the recent past, but there does not exist any general prescription for the direct
experimental measurement of the  Wigner function and other quasidistribution functions.  In practice, to detect
the nonclassicality in a system the Wigner function is obtained either
by photon counting or from experimentally measured tomograms \cite{smithey1993measurement}. Thus, tomograms are very important for the
identification of nonclassical character(s) of a physical system.
In another line of studies, simulation of quantum systems have been performed using tomography. For example, tomograms have been used 
for the simulation of tunneling \cite{arkhipov2003new,arkhipov2005center,lozovik2004simulation} and multi-mode quantum states \cite{arkhipov2003tomography}. Attempts have also been made 
to understand the tomogram via path integrals \cite{man1999classical,fedorov2013feynman}. 

Furthermore, how to reconstruct a quantum state from experimentally measured
values is of prime interest for both quantum computation \cite{miquel2002interpretation}
and communication \cite{ma2012quantum}. Specifically, in Ref. \cite{miquel2002interpretation}, it is strongly established that 
tomography and spectroscopy can be interpreted as dual forms of quantum computation; and in Ref. \cite{ma2012quantum}, quantum teleportation was 
experimentally performed over a distance of 143 km and the quality of teleportation was verified with the help of quantum process tomography of 
quantum teleportation without feed-forward. Here, it would be apt to note that quantum process tomography employs quantum state tomography in which a quantum process is obtained
as a CPTP map  \cite{poyatos1997complete,chuang1997prescription,nielsen1998complete}. In the recent
past, quantum process tomography has been discussed from the perspective of open
quantum system effects \cite{kuah2007state,bellomo2009reconstruction,bellomo2010tomographic,bellomo2010reconstruction}. 
A novel method of complete experimental characterization of quantum optical processes was introduced in \cite{lobino2008complete}. It was further developed 
in \cite{rahimi2011quantum,anis2012maximum} and extended to the characterization of $N$-modes in \cite{fedorov2015tomography}. In \cite{lobino2009memory}, quantum process tomography was applied to 
the characterization of optical memory based on electromagnetically induced transparency, while \cite{kumar2013experimental} and \cite{cooper2015characterization} were devoted to the quantum process tomography
of the electromagnetic field and conditional state engineering, respectively.
Quantum state tomography has its applications in quantum cryptography as well
\cite{liang2003tomographic}. Specifically, in Ref. \cite{liang2003tomographic}, an interesting  protocol of quantum cryptography was proposed  in which eavesdropping in the quantum channel was checked by requiring consistency of the outcome of the tomography  with the unbiased noise situation. Keeping these facts in mind, we aim to construct tomograms
for a number of physical systems of practical relevance (mostly having applications in quantum computation and communication)
and investigate the effects of various types of noise on them.

From the experimental perspective, a quantum state always interacts
with its surroundings. Thus, the evolution of the corresponding tomogram should be analyzed
after considering the interaction of the quantum state with
its environment. This can be achieved with the
open quantum system formalism \cite{louisell1973quantum,breuer2002theory,weiss2008quantum,banerjee2003general}, introduced in Section \ref{OQS} and used already in Chapter \ref{QDs}. Specifically,
both purely dephasing (QND) \cite{banerjee2007dynamics} and dissipative \cite{srikanth2008squeezed}
open quantum system effects have been studied here. 

Here, we set ourselves the task of obtaining the tomograms for various
finite and infinite dimensional quantum systems in different open
quantum system scenarios. For the finite dimensional spin states,
a tomogram is the distribution function of the projections of the
spin on an arbitrary axis, characterized by the Euler angles, and can
be obtained from the diagonal elements of the rotated density matrix, while for the continuous variable systems, such as the radiation field, the 
analog would be the homodyne probability. It follows from the general group theoretical arguments that, 
making use of the unitary irreducible square integrable representation of the tomographic group under consideration, a unified tomographic
prescription can be developed for both finite dimensional and continuous variable systems \cite{d2003spin}.  
Tomograms for spin states have been developed as projections on an arbitrary axis \cite{man1998describing} as 
well as by using a discrete variable analog of symplectic 
tomography \cite{fedorov2013quaternion}.
Tomograms of optical systems have
been well studied in the past \cite{vogel1989determination,dodonov1997positive,ibort2009introduction,lvovsky2009continuous,d2004quantum}.
In Ref. \cite{liu2011experimental}, quantum state tomography was used to determine the degree
of non-Markovianity in an open system. Further,  thermal noise is used in tomography (for reconstruction of the photon number 
distributions) as a probe \cite{harder2014tomography}. Interesting results reported in these studies have motivated us to systematically investigate the tomograms for both finite (spin) and infinite dimensional (harmonic oscillator) quantum systems. In fact, in what follows, we will report the evolution of tomograms for the single- and two-qubit states and the harmonic oscillator affected due to interaction with the surroundings.

The rest of this chapter is organized as follows. In Section \ref{sec:Tomograms-of-single-half},
tomograms of single spin-$\frac{1}{2}$ (qubit) atomic coherent states
under purely dephasing (QND) and dissipative evolution are obtained.
Further, the tomogram of two spin-$\frac{1}{2}$ (qubit) quantum state
is studied in Section \ref{sec:Tomogram-of-two-spin} under the influence of a vacuum bath. In  Section \ref{sec:Optical-tomogram}, 
we discuss the tomogram of an infinite dimensional system, the ubiquitous dissipative harmonic oscillator.
We conclude the chapter in Section \ref{sec:Conclusion-Tomogram}.

\section{Tomograms of single spin-$\frac{1}{2}$ (qubit) states \label{sec:Tomograms-of-single-half}}

In this section, we study the tomograms for single spin-$\frac{1}{2}$
(qubit) atomic coherent states evolving under two general noise models, i.e., pure dephasing (QND)
and dissipative SGAD evolution, incorporating the effects of dissipation, decoherence, and bath squeezing.

\subsection{QND Evolution}

The master equation of a quantum state under QND evolution \cite{banerjee2007dynamics}
is introduced as Eq. (\ref{eq:master-eq-QND}) in Section \ref{subsec:QND} while obtaining the quasiprobability distributions  for a single-qubit state in the QND channel. Considering the initial state of the system as atomic coherent state (\ref{eq:atomic-coherent-state}), we obtain the evolved quantum state in QND noise as Eq. (\ref{eq:density-matrix-QND}).

\begin{figure}
\centering{}
\includegraphics[scale=0.6]{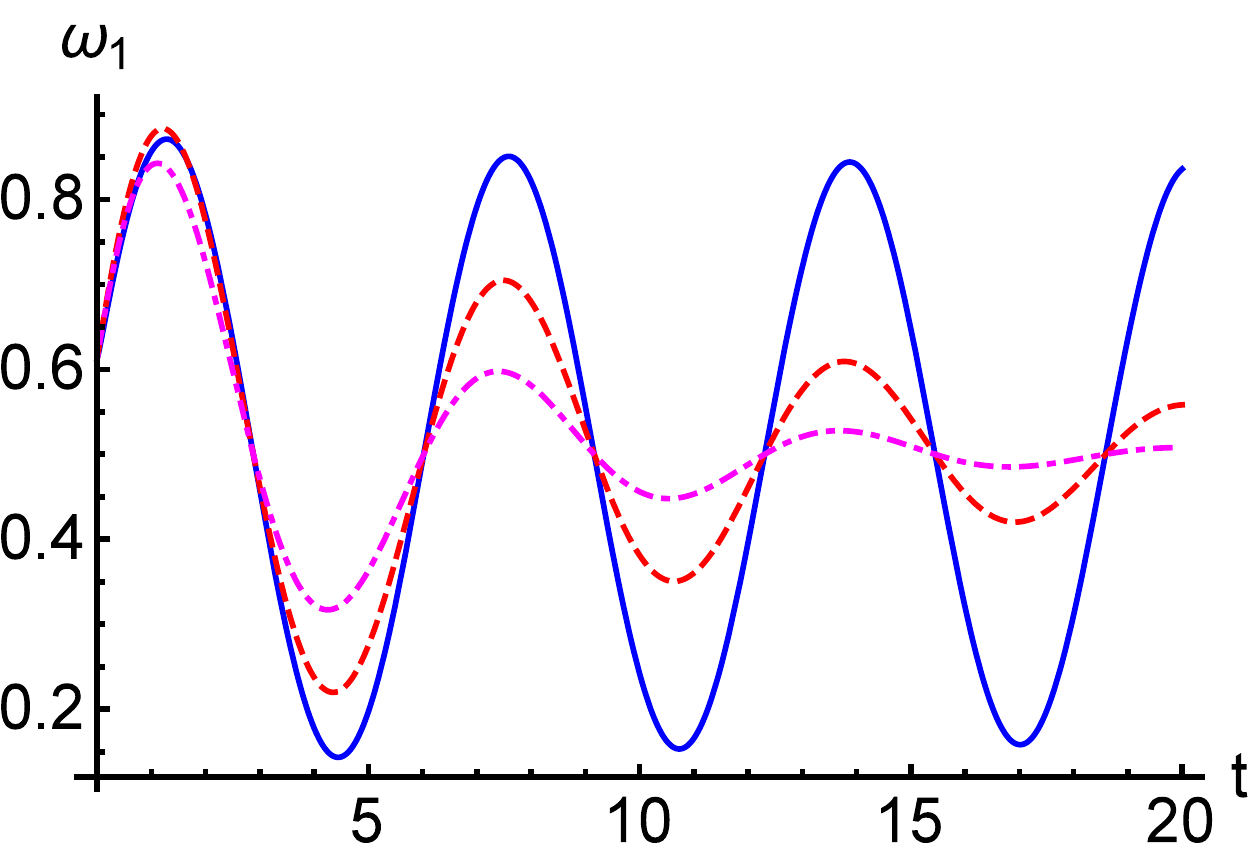}

\protect\caption[Tomogram
for single spin-$\frac{1}{2}$ atomic coherent state
in QND noise]{\label{fig:qnd-single}Variation of the tomogram
with time ($t$) for single spin-$\frac{1}{2}$ atomic coherent state
in the presence of QND noise with bath parameters $\gamma_{0}=0.1,\,\omega_{c}=100,$
squeezing parameters $r=0,\,a=0,$ and $\omega=1.0$, and $\alpha=\frac{\pi}{2},\,\beta=\frac{\pi}{3},\,\widetilde{\beta}=\frac{\pi}{3},\,\widetilde{\gamma}=\frac{\pi}{4},$
in the units of $\hbar=k_{B}=1$. The smooth (blue), dashed (red), and dot-dashed (magenta) lines correspond to the evolution of the tomogram with
time for different values of temperature $T=0,\,1,$ and $2$, respectively.} 
\end{figure}

Further,
the tomogram of a spin-$j$ state can be expressed as \cite{man1997spin}
\begin{equation}
\begin{array}{lcl}
\omega\left(m_{1},\widetilde{\alpha},\widetilde{\beta},\widetilde{\gamma}\right) & = & \stackrel[m=-j]{j}{\sum}\,\,\stackrel[m^{\prime}=-j]{j}{\sum}D_{m_{1},m}^{\left(j\right)}\left(\widetilde{\alpha},\widetilde{\beta},\widetilde{\gamma}\right) \rho_{m,m^{\prime}}^{\left(j\right)}D_{m_{1},m^{\prime}}^{\left(j\right)*}\left(\widetilde{\alpha},\widetilde{\beta},\widetilde{\gamma}\right),
\end{array}\label{eq:tomogram}
\end{equation}
where $D_{m,m^{\prime}}^{\left(j\right)}\left(\widetilde{\alpha},\widetilde{\beta},\widetilde{\gamma}\right)$
is the Wigner $D$-function  
\begin{equation}
\begin{array}{lcl}
D_{m,m^{\prime}}^{\left(j\right)}\left(\widetilde{\alpha},\widetilde{\beta},\widetilde{\gamma}\right) & = & e^{-im\widetilde{\alpha}}d_{m,m^{\prime}}^{\left(j\right)}\left(\widetilde{\beta}\right)e^{-im^{\prime}\widetilde{\gamma}}\end{array}, \label{eq:wigner_D-function}
\end{equation}  and the notation used here is consistent with
that in Ref. \cite{varshalovich1988quantum}.
Here, $\widetilde{\alpha}$, $\widetilde{\beta}$, and $\widetilde{\gamma}$
are the Euler angles $\equiv\phi$ , $\theta$, and $\psi$, with $\phi,\psi\in\left[0,2\pi\right]$
and $\theta\in\left[0,\pi\right]$, and
\begin{equation}
\begin{array}{lcl}
d_{m,m^{\prime}}^{\left(j\right)}\left(\widetilde{\beta}\right) & = & \left[\frac{\left(j+m\right)!\left(j-m\right)!}{\left(j+m^{\prime}\right)!\left(j-m^{\prime}\right)!}\right]^{1/2}\left(\cos\frac{\widetilde{\beta}}{2}\right)^{m+m^{\prime}} \left(\sin\frac{\widetilde{\beta}}{2}\right)^{m-m^{\prime}}P_{j-m}^{\left(m-m^{\prime},m+m^{\prime}\right)}\left(\cos\widetilde{\beta}\right),
\end{array}\label{eq:D-function-d}
\end{equation}
where $P_{n}^{\left(a,b\right)}\left(x\right)$ are Jacobi polynomials.
A tomogram is the spin projection onto an arbitrary, rotated, axis.
The physical significance of the $D$-function can be visualized through its connection to
the process of rotation and can be illustrated by
\begin{equation}
\begin{array}{lcl}
\langle j,m_{1}|R\left(\widetilde{\alpha},\widetilde{\beta},\widetilde{\gamma}\right)|j,m_{1}^{\prime}\rangle & = & D_{m_{1},m_{1}^{\prime}}^{\left(j\right)}\left(\widetilde{\alpha},\widetilde{\beta},\widetilde{\gamma}\right),\\
\langle j,m_{2}^{\prime}|R^{\dagger}\left(\widetilde{\alpha},\widetilde{\beta},\widetilde{\gamma}\right)|j,m_{1}\rangle & = & D_{m_{1},m_{2}^{\prime}}^{*\left(j\right)}\left(\widetilde{\alpha},\widetilde{\beta},\widetilde{\gamma}\right).
\end{array}\label{eq:Meaning-of-D}
\end{equation}
Here, $R\left(\widetilde{\alpha},\widetilde{\beta},\widetilde{\gamma}\right)$
stands for the operation of rotation about an axis whose orientation
is specified by $\widetilde{\alpha},\,\widetilde{\beta},$ and $\widetilde{\gamma}.$
Using the different values of $m$ and $m^{\prime}$, we can obtain
various Wigner $D$-functions as
\begin{subequations}
\begin{eqnarray}
D_{\frac{1}{2},-\frac{1}{2}}^{\left(1/2\right)}\left(\widetilde{\alpha},\widetilde{\beta},\widetilde{\gamma}\right) & = & -\sin\left(\frac{\widetilde{\beta}}{2}\right)e^{-\frac{i}{2}\left(\widetilde{\alpha}-\widetilde{\gamma}\right)},\label{eq:values-of-Ds-1}\\
D_{\frac{1}{2},\frac{1}{2}}^{\left(1/2\right)}\left(\widetilde{\alpha},\widetilde{\beta},\widetilde{\gamma}\right) & = & \cos\left(\frac{\widetilde{\beta}}{2}\right)e^{-\frac{i}{2}\left(\widetilde{\alpha}+\widetilde{\gamma}\right)},\label{eq:values-of-Ds-2}\\
D_{-\frac{1}{2},-\frac{1}{2}}^{\left(1/2\right)}\left(\widetilde{\alpha},\widetilde{\beta},\widetilde{\gamma}\right) & = & \cos\left(\frac{\widetilde{\beta}}{2}\right)e^{\frac{i}{2}\left(\widetilde{\alpha}+\widetilde{\gamma}\right)},\label{eq:values-of-Ds-3}\\
D_{-\frac{1}{2},\frac{1}{2}}^{\left(1/2\right)}\left(\widetilde{\alpha},\widetilde{\beta},\widetilde{\gamma}\right) & = & \sin\left(\frac{\widetilde{\beta}}{2}\right)e^{\frac{i}{2}\left(\widetilde{\alpha}-\widetilde{\gamma}\right)}.\label{eq:values-of-Ds-4}
\end{eqnarray}
\end{subequations}
Using Eqs. (\ref{eq:values-of-Ds-1})-(\ref{eq:values-of-Ds-2})
and Eq. (\ref{eq:density-matrix-QND}), the first component of the
tomogram can be obtained from Eq. (\ref{eq:tomogram}) as 
\begin{equation}
\begin{array}{lcl}
\omega\left(\frac{1}{2},\widetilde{\alpha},\widetilde{\beta},\widetilde{\gamma}\right)\equiv\omega_{1} & = & \cos^{2}\left(\frac{\widetilde{\beta}}{2}\right)-\cos\widetilde{\beta}\cos^{2}\left(\frac{\alpha}{2}\right) - \frac{1}{2}\sin\widetilde{\beta}\sin\alpha\cos\left(\omega t+\beta+\widetilde{\gamma}\right) e^{-\left(\hbar\omega\right)^{2}\gamma\left(t\right)}.
\end{array}\label{eq:tomogram1-QND}
\end{equation}
From Eq. (\ref{eq:tomogram1-QND}), it can be inferred that the tomogram
is free from the Euler angle $\widetilde{\alpha}$, and consequently 
is a function of  $\widetilde{\beta}$ and $\widetilde{\gamma}$ only,
or $f(\widetilde{\beta},\widetilde{\gamma})$. It is
worth mentioning here that $\widetilde{\gamma}$ and $\gamma\left(t\right)$
are two different parameters, the former being an Euler angle while the latter is responsible for decoherence. Variation of the tomogram with time is
shown in Figure \ref{fig:qnd-single}, for the different values of temperature.
For the second component of the tomogram with $m_{1}=-\frac{1}{2}$,
using Eqs. (\ref{eq:values-of-Ds-3})-(\ref{eq:values-of-Ds-4}) and substituting Eq.
(\ref{eq:density-matrix-QND}) in Eq. (\ref{eq:tomogram}), we obtain
\begin{equation}
\begin{array}{lcl}
\omega\left(-\frac{1}{2},\widetilde{\alpha},\widetilde{\beta},\widetilde{\gamma}\right)\equiv\omega_{2} & = & \cos^{2}\left(\frac{\widetilde{\beta}}{2}\right)-\cos\widetilde{\beta}\sin^{2}\left(\frac{\alpha}{2}\right) + \frac{1}{2}\sin\widetilde{\beta}\sin\alpha\cos\left(\omega t+\beta+\widetilde{\gamma}\right) e^{-\left(\hbar\omega\right)^{2}\gamma\left(t\right)}.
\end{array}\label{eq:tomogram-2-QND}
\end{equation}
We can check the validity of the tomogram obtained by
verifying that $\sum\omega_{i}=
\omega_{1}+\omega_{2} =  1.$ Interestingly, we can see  that the knowledge of one of the components
of the tomogram is enough to reconstruct the whole state.
Keeping this in mind, we have only shown the variation of $\omega_{1}$ in Figure \ref{fig:qnd-single}
as $\omega_{2}=1-\omega_{1}$. 

\begin{figure}
\includegraphics[angle=0,scale=0.57]{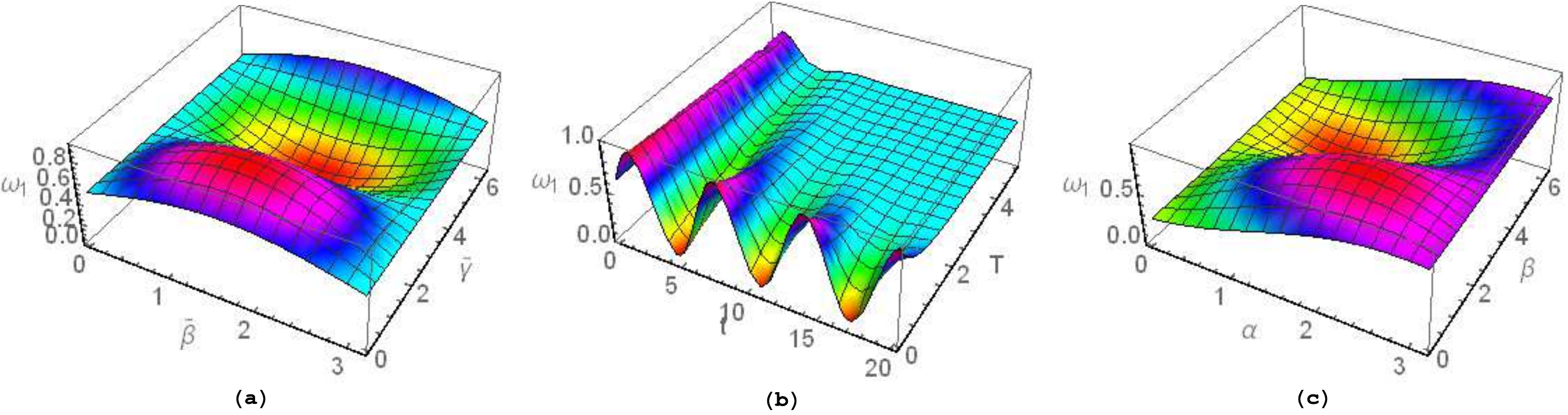}
\protect\caption[Variation of tomogram with different parameters
for single spin-$\frac{1}{2}$ atomic coherent state
in QND noise]{\label{fig:qnd-single-3D}The dependence of the tomogram
on various parameters is depicted for single spin-$\frac{1}{2}$ atomic
coherent state in the presence of QND noise with the bath parameters $\gamma_{0}=0.1,\,\omega_{c}=100,$
squeezing parameter $a=0$ and $\omega=1.0$ in the units of $\hbar=k_{B}=1$.
In (a), the tomogram is shown as a function of $\widetilde{\beta}$
and $\widetilde{\gamma}$ with $\alpha=\frac{\pi}{2},\,\beta=\frac{\pi}{3},$
and $r=t=T=1$; while (b) exhibits the variation of the tomogram with
time and temperature for $r=0$ and $\alpha=\frac{\pi}{2},\,\beta=\frac{\pi}{3},\,\widetilde{\beta}=\frac{\pi}{3},\,\widetilde{\gamma}=\frac{\pi}{4}$.
(c) shows the dependence of the tomogram on
the atomic coherent state parameters $\alpha$ and $\beta$
with $\widetilde{\beta}=\frac{\pi}{3},\,\widetilde{\gamma}=\frac{\pi}{4}$ at time $t=1$
for bath squeezing parameter $r=1$ at $T=1$.}
\end{figure}

In Figure \ref{fig:qnd-single}, we can easily see the expected behavior
of tomogram with increase in temperature for zero bath squeezing. 
Specifically, with increase in temperature, the tomogram tends to
randomize more quickly towards probability $1/2$. Figure \ref{fig:qnd-single-3D}
further establishes the effect of the environment on the tomogram.
Particularly, Figure \ref{fig:qnd-single-3D} (b) brings out the oscillatory nature
of the tomogram with time, while temperature tends to randomize it. Similarly,
Figures \ref{fig:qnd-single-3D} (a) and (c) show the dependence of the tomogram
on the Euler angles and the atomic coherent state parameters, respectively.

\subsection{Dissipative SGAD channel}

Master equation for the dissipative evolution of a given state in
the SGAD channel is given
by \cite{srikanth2008squeezed} 
\begin{equation}
\begin{array}{lcl}
\frac{d}{dt}\rho^{s}\left(t\right) & = & -\frac{i\omega}{2}\left[\sigma_{z},\rho^{s}\left(t\right)\right]+\gamma_{0}\left(N+1\right) \left\{ \sigma_{-}\rho^{s}\left(t\right)\sigma_{+}-\frac{1}{2}\sigma_{+}\sigma_{-}\rho^{s}\left(t\right)-\frac{1}{2}\rho^{s}\left(t\right)\sigma_{+}\sigma_{-}\right\} \\
 & + & \gamma_{0}N\left\{ \sigma_{+}\rho^{s}\left(t\right)\sigma_{-}-\frac{1}{2}\sigma_{-}\sigma_{+}\rho^{s}\left(t\right)-\frac{1}{2}\rho^{s}\left(t\right)\sigma_{-}\sigma_{+}\right\} \\
 &-& \gamma_{0}M\sigma_{+}\rho^{s}\left(t\right)\sigma_{+}-\gamma_{0}M^{*}\sigma_{-}\rho^{s}\left(0\right)\sigma_{-}.
\end{array}\label{eq:master_eq-SGAD}
\end{equation}
The density matrix for a quantum state under a dissipative SGAD channel
at time $t$ can be obtained, from the above equation, as 
\begin{equation}
\begin{array}{lcl}
\rho^{s}\left(t\right) & = & \frac{1}{4}\rho^{s}\left(0\right)f_{+}+\frac{1}{4}\sigma_{z}\rho^{s}\left(0\right)\sigma_{z}f_{-}-\frac{1}{4}\rho^{s}\left(0\right)\sigma_{z}g_{-} -  \frac{1}{4}\sigma_{z}\rho^{s}\left(0\right)g_{+}-\gamma_{0}\frac{\sinh\left(\alpha^{\prime}t\right)}{\alpha^{\prime}}e^{-\frac{\gamma^{\beta}t}{2}}\\
& \times & \left\{ M\sigma_{+}\rho^{s}\left(0\right)\sigma_{+}+M^{*}\sigma_{-}\rho^{s}\left(0\right)\sigma_{-}\right\} + \left(1-e^{-\gamma^{\beta}t}\right)\left\{ \frac{\gamma_{+}}{\gamma^{\beta}}\sigma_{-}\rho^{s}\left(0\right)\sigma_{+}+\frac{\gamma_{-}}{\gamma^{\beta}}\sigma_{+}\rho^{s}\left(0\right)\sigma_{-}\right\},
\end{array}\label{eq:density-matrix-SGAD}
\end{equation}
where $f_{\pm}=\left\{ 1+e^{-\gamma^{\beta}t} \pm 2\cosh\left(\alpha^{\prime}t\right)e^{-\frac{\gamma^{\beta}t}{2}} \right\},$ $g_{\pm}=\left\{ \frac{\gamma}{\gamma^{\beta}}\left(1-e^{-\gamma^{\beta}t} \right) \pm \frac{2i\omega}{\alpha^{\prime}}\sinh\left(\alpha^{\prime}t\right)e^{-\frac{\gamma^{\beta}t}{2}} \right\},$ $\gamma_{+}=\gamma_{0}\left(N+1\right)$, $\gamma_{-}=\gamma_{0}N$,
$\gamma^{\beta}=\gamma_{+}+\gamma_{-}$, $\gamma=\gamma_{+}-\gamma_{-}=\gamma_{0}$,
$\alpha^{\prime}=\sqrt{\gamma_{0}^{2}\left|M\right|^{2}-\omega^{2}}$;
and
\[
\begin{array}{lcl}
\sigma_{+} & = & |1\rangle\langle0|,\,\,\sigma_{-}=|0\rangle\langle1|,\\
\sigma_{z} & = & \sigma_{+}\sigma_{-}-\sigma_{-}\sigma_{+}\\
 & = & |1\rangle\langle1|-|0\rangle\langle0|\\
 & = & |e\rangle\langle e|-|g\rangle\langle g|.
\end{array}
\]
Also,
\[
\begin{array}{lclccc}
\sigma_{z}|g\rangle & = & -|g\rangle, & \sigma_{z}|e\rangle & = & |e\rangle;\\
\sigma_{+}|g\rangle & = & |e\rangle, & \sigma_{+}|e\rangle & = & 0;\\
\sigma_{-}|g\rangle & = & 0, & \sigma_{-}|e\rangle & = & |g\rangle.
\end{array}
\]
Here, $\gamma_{0}$ is the spontaneous emission rate, $M=-\frac{1}{2}\left\{ 2N_{th}+1\right\} \exp\left(i\phi\right)\sinh\left(2r\right)$,
and $N=N_{th}\left\{ \cosh^{2}\left(r\right)+\sinh^{2}\left(r\right)\right\} +\sinh^{2}\left(r\right),$
where $N_{th}=1/\left\{ \exp\left(\hbar\omega/k_{B}T\right)-1\right\} $
being the Planck distribution, and $r$ and bath squeezing angle
$\phi$ are the bath squeezing parameters. The initial state, as
for the tomogram of a quantum state under QND evolution, is the atomic
coherent state given in Eq. (\ref{eq:atomic-coherent-state}). Using Eq.
(\ref{eq:density-matrix-SGAD}), the density matrix can be written
as
\begin{equation}
\rho^{s}\left(t\right)=\left[\begin{array}{cc}
\langle\frac{1}{2}|\rho^{s}\left(t\right)|\frac{1}{2}\rangle & \langle\frac{1}{2}|\rho^{s}\left(t\right)|-\frac{1}{2}\rangle\\
\langle-\frac{1}{2}|\rho^{s}\left(t\right)|\frac{1}{2}\rangle & \langle-\frac{1}{2}|\rho^{s}\left(t\right)|-\frac{1}{2}\rangle
\end{array}\right],\label{eq:final_density-matrix-SGAD}
\end{equation}
where the various terms are 
\[
\begin{array}{lcl}
\langle\frac{1}{2}|\rho^{s}\left(t\right)|\frac{1}{2}\rangle & = & \sin^{2}\left(\frac{\alpha}{2}\right)e^{-\gamma^{\beta}t}+\frac{\gamma_{-}}{\gamma^{\beta}}\left(1-e^{-\gamma^{\beta}t}\right),\\
\langle\frac{1}{2}|\rho^{s}\left(t\right)|-\frac{1}{2}\rangle & = & \frac{1}{2}\sin\alpha\left[\left\{ \cosh\left(\alpha^{\prime}t\right)-\frac{i\omega}{\alpha^{\prime}}\sinh\left(\alpha^{\prime}t\right)\right\} e^{-i\beta}-\frac{\gamma_{0}M}{\alpha^{\prime}}\sinh\left(\alpha^{\prime}t\right)e^{i\beta}\right]e^{-\frac{\gamma^{\beta}t}{2}},\\
\langle-\frac{1}{2}|\rho^{s}\left(t\right)|\frac{1}{2}\rangle & = & \frac{1}{2}\sin\alpha\left[\left\{ \cosh\left(\alpha^{\prime}t\right)+\frac{i\omega}{\alpha^{\prime}}\sinh\left(\alpha^{\prime}t\right)\right\} e^{i\beta}-\frac{\gamma_{0}M^{*}}{\alpha^{\prime}}\sinh\left(\alpha^{\prime}t\right)e^{-i\beta}\right]e^{-\frac{\gamma^{\beta}t}{2}},\\
\langle-\frac{1}{2}|\rho^{s}\left(t\right)|-\frac{1}{2}\rangle & = & \cos^{2}\left(\frac{\alpha}{2}\right)e^{-\gamma^{\beta}t}+\frac{\gamma_{+}}{\gamma^{\beta}}\left(1-e^{-\gamma^{\beta}t}\right),
\end{array}
\]
 and the density matrix can be seen to be normalized as
$\stackrel[m=-1/2]{1/2}{\sum}\langle m|\rho^{s}\left(t\right)|m\rangle=1.$

The tomogram of a state evolving in a dissipative SGAD channel, in analogy to the QND case, can be obtained 
in the basis set formed by the Wigner-Dicke states. Using Eq. (\ref{eq:tomogram}), Eqs. (\ref{eq:values-of-Ds-1})-(\ref{eq:values-of-Ds-2}), and Eq. (\ref{eq:final_density-matrix-SGAD}), the first component of the tomogram is obtained as

\begin{equation}
\begin{array}{lcl}
\omega\left(\frac{1}{2},\widetilde{\alpha},\widetilde{\beta},\widetilde{\gamma}\right)\equiv\omega_{1} & = & \sin^{2}\left(\frac{\widetilde{\beta}}{2}\right)\left\{ \cos^{2}\left(\frac{\alpha}{2}\right)e^{-\gamma^{\beta}t}+\frac{\gamma_{+}}{\gamma^{\beta}}\left(1-e^{-\gamma^{\beta}t}\right)\right\} +\cos^{2}\left(\frac{\widetilde{\beta}}{2}\right)\left\{ \sin^{2}\left(\frac{\alpha}{2}\right)e^{-\gamma^{\beta}t}\right.\\
&+& \left.\frac{\gamma_{-}}{\gamma^{\beta}}\left(1-e^{-\gamma^{\beta}t}\right)\right\} -\frac{1}{2}\sin\widetilde{\beta}\left\{ e^{-i\widetilde{\gamma}}\left[\frac{1}{2}\sin\alpha e^{-i\beta}e^{-\frac{\gamma^{\beta}t}{2}}\left\{ \cosh\left(\alpha^{\prime}t\right)\right.\right.\right.\\
&-&\left.\left.\left. \frac{i\omega}{\alpha^{\prime}}\sinh\left(\alpha^{\prime}t\right)\right\} -\frac{\gamma_{0}M}{2\alpha^{\prime}}\sin\alpha\sinh\left(\alpha^{\prime}t\right)e^{i\beta}e^{-\frac{\gamma^{\beta}t}{2}}\right]+ {\rm c.c.}\right\} .
\end{array}\label{eq:tomogram1-SGAD}
\end{equation}

Again, we can check the validity of the analytic expression of the tomogram in the absence of the
open system effects, i.e., by considering $\gamma_{0}=\gamma=0$, $\gamma_{+}=\gamma_{-}=0=\gamma^{\beta}$, which
leads to $\alpha^{\prime}=i\omega$, we have
\begin{equation}
\begin{array}{lcl}
\omega\left(\frac{1}{2},\widetilde{\alpha},\widetilde{\beta},\widetilde{\gamma}\right) & = & \cos^{2}\left(\frac{\widetilde{\beta}}{2}\right)-\cos\widetilde{\beta}\cos^{2}\left(\frac{\alpha}{2}\right) - \frac{1}{2}\sin\widetilde{\beta}\sin\alpha\cos\left(\omega t+\beta+\widetilde{\gamma}\right),
\end{array}\label{eq:check-tomogram1-SGAD}
\end{equation}
which is identical to the QND case, i.e., Eq. (\ref{eq:tomogram1-QND}),
with $\gamma\left(t\right)=0$. Similarly, using Eq. (\ref{eq:tomogram}),
Eqs. (\ref{eq:values-of-Ds-3})-(\ref{eq:values-of-Ds-4}), and Eq. (\ref{eq:final_density-matrix-SGAD}),
we obtain the second component as

\begin{equation}
\begin{array}{lcl}
\omega\left(-\frac{1}{2},\widetilde{\alpha},\widetilde{\beta},\widetilde{\gamma}\right)\equiv\omega_{2} & = & \cos^{2}\left(\frac{\widetilde{\beta}}{2}\right)\left\{ \cos^{2}\left(\frac{\alpha}{2}\right)e^{-\gamma^{\beta}t}+\frac{\gamma_{+}}{\gamma^{\beta}}\left(1-e^{-\gamma^{\beta}t}\right)\right\} +\sin^{2}\left(\frac{\widetilde{\beta}}{2}\right)\left\{ \sin^{2}\left(\frac{\alpha}{2}\right)e^{-\gamma^{\beta}t}\right.\\
 & + & \left.\frac{\gamma_{-}}{\gamma^{\beta}}\left(1-e^{-\gamma^{\beta}t}\right)\right\} +\frac{1}{2}\sin\widetilde{\beta}\left\{ e^{-i\widetilde{\gamma}}\left[\frac{1}{2}\sin\alpha e^{-i\beta}e^{-\frac{\gamma^{\beta}t}{2}}\left\{ \cosh\left(\alpha^{\prime}t\right)\right.\right.\right.\\
 &-&\left.\left.\left.\frac{i\omega}{\alpha^{\prime}}\sinh\left(\alpha^{\prime}t\right)\right\} -\frac{\gamma_{0}M}{2\alpha^{\prime}}\sin\alpha\sinh\left(\alpha^{\prime}t\right)e^{i\beta}e^{-\frac{\gamma^{\beta}t}{2}}\right]+{\rm c.c.}\right\} .
\end{array}\label{eq:tomogram2-SGAD}
\end{equation}

Similar to the first tomogram over the dissipative SGAD
channel, we can check the solution in the absence of the open system
effects which leads to $\alpha^{\prime}=i\omega$. This can be seen to be the same as the corresponding
QND case, i.e., Eq. (\ref{eq:tomogram-2-QND}), with $\gamma\left(t\right)=0$, as
\begin{equation}
\begin{array}{lcl}
\omega\left(-\frac{1}{2},\widetilde{\alpha},\widetilde{\beta},\widetilde{\gamma}\right) & = & \cos^{2}\left(\frac{\widetilde{\beta}}{2}\right)-\cos\widetilde{\beta}\sin^{2}\left(\frac{\alpha}{2}\right) + \frac{1}{2}\sin\widetilde{\beta}\sin\alpha\cos\left(\omega t+\beta+\widetilde{\gamma}\right).
\end{array}\label{eq:check-tomogram-2-SGAD}
\end{equation}
We can also check the validity of the tomogram as in the QND case
by verifying that $\sum\omega_{i}=\omega_{1}+\omega_{2}=1.$ Therefore, as before, one component of the tomogram would be sufficient to recover
all the information. This is why, in the plots, we only show
the first component of the tomogram. The other component can be easily obtained from
that.

\begin{figure}
\centering{}
\includegraphics[scale=0.6]{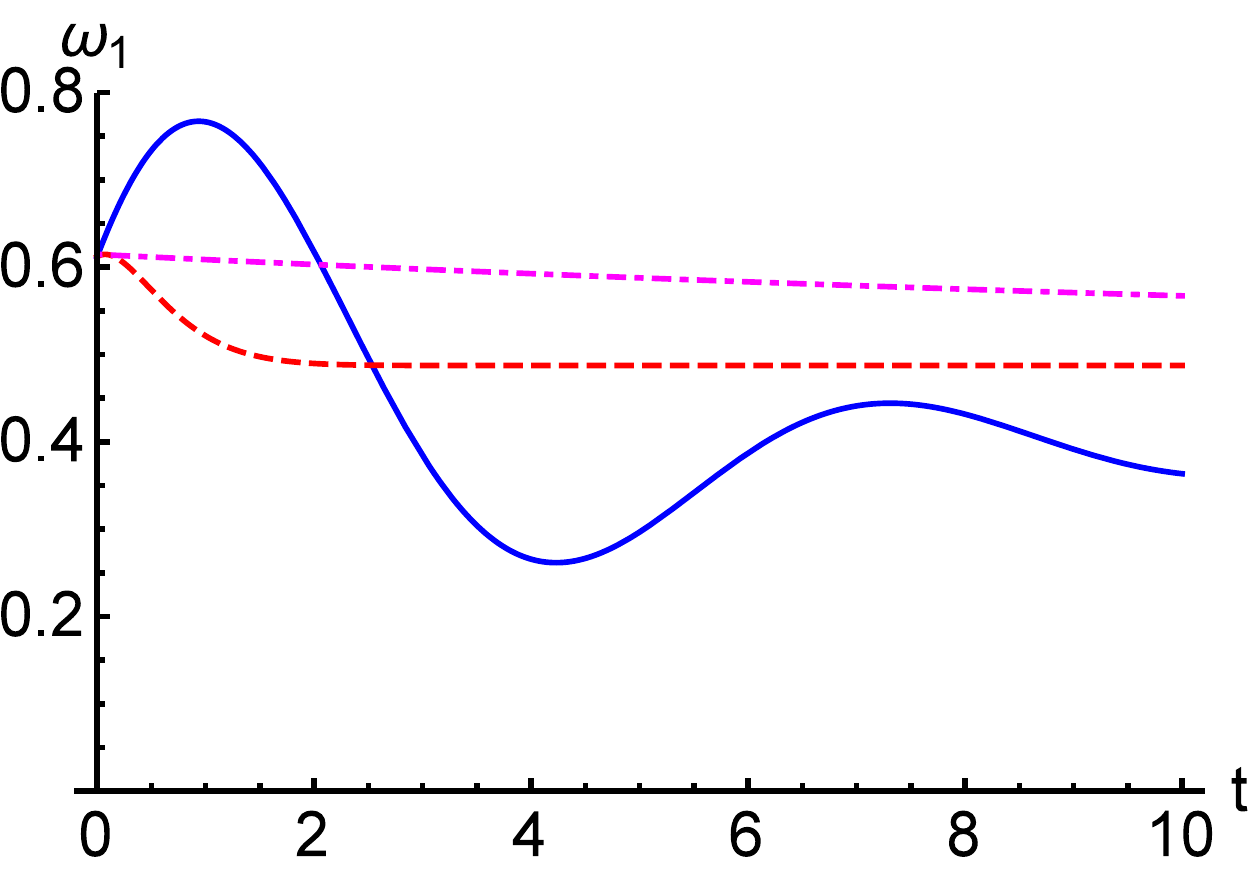}

\protect\caption[Tomogram
for single spin-$\frac{1}{2}$ atomic coherent state
in SGAD noise]{\label{fig:SGAD-ACS} The tomogram varying
with time ($t$) is shown for a single spin-$\frac{1}{2}$
atomic coherent state in the presence of the SGAD noise for bath squeezing
angle $\phi=\pi,$ in the units of $\hbar=k_{B}=1$, with
$\omega=1.0,\,\gamma_{0}=0.25,$ and $\alpha=\frac{\pi}{2},\,\beta=\frac{\pi}{3},\,\widetilde{\beta}=\frac{\pi}{3},\,\widetilde{\gamma}=\frac{\pi}{4}$.
The smooth (blue), dashed (red), and dot-dashed (magenta)
lines correspond to the tomogram evolving with time for different values of temperature
and squeezing parameters $T=1,\,10,$ and $10$, and $r=0,\,0,$ and
$1$, respectively.}
\end{figure}

\begin{figure}
\centering{}
\includegraphics[scale=0.55]{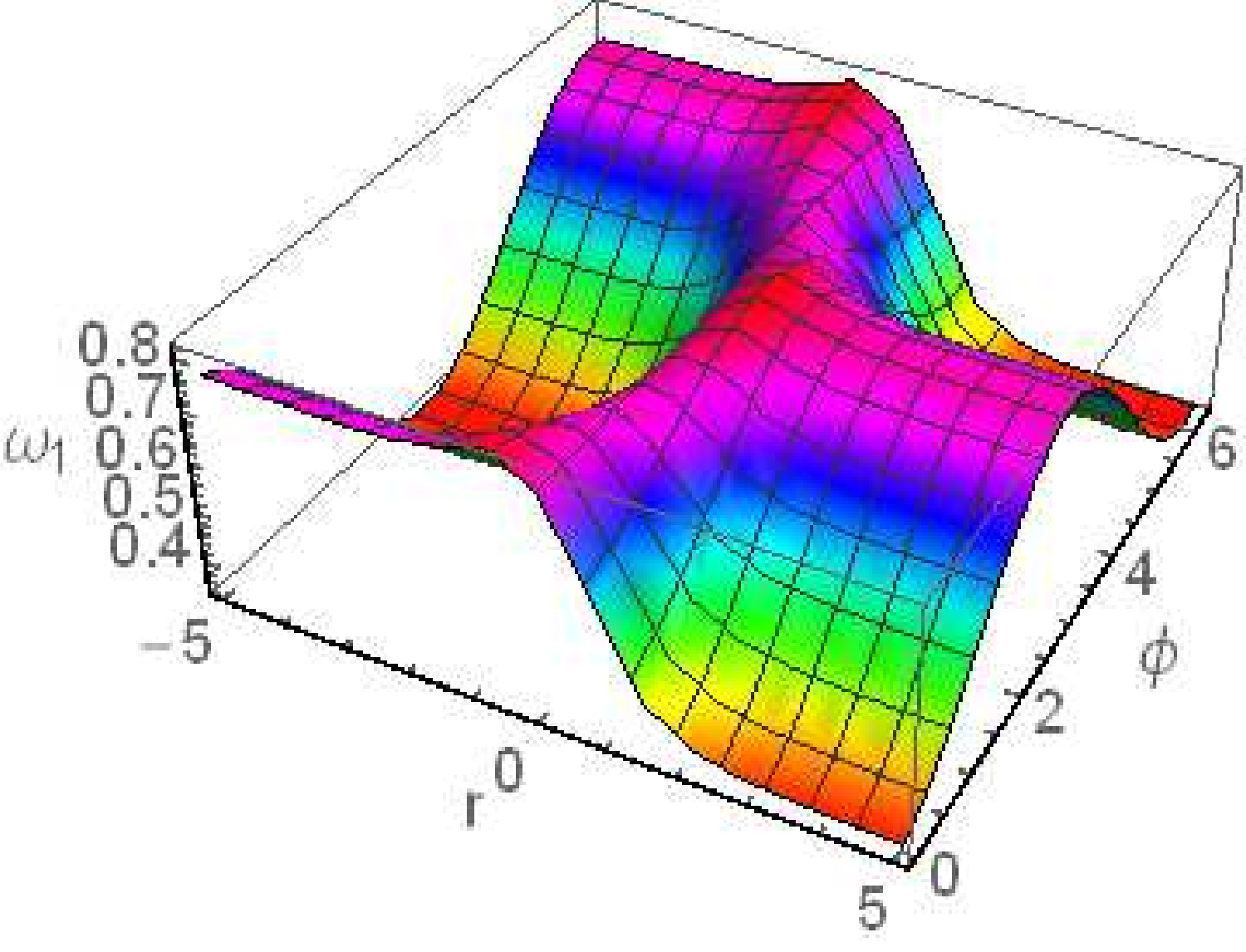}

\protect\caption[Variation of tomogram with squeezing parameters
for single spin-$\frac{1}{2}$ atomic coherent state
in SGAD noise]{\label{fig:SGAD-3D-ACS} The tomogram for single spin-$\frac{1}{2}$ atomic coherent
state in the presence of SGAD noise is
shown as a function of the squeezing parameters $r$ and $\phi$ with
$\alpha=\frac{\pi}{2},\,\beta=\frac{\pi}{3},\,\widetilde{\beta}=\frac{\pi}{3},\,\widetilde{\gamma}=\frac{\pi}{4},$ and $\omega=1.0,\,\gamma_{0}=0.25,$
in the units of $\hbar=k_{B}=1,$
for $T=1$ at time $t=1$.}
\end{figure}

The variation of tomogram with different parameters is shown in Figures
\ref{fig:SGAD-ACS} and \ref{fig:SGAD-3D-ACS}. Figure
\ref{fig:SGAD-ACS} exhibits the randomization of the tomogram with increase in
temperature. This fact can be
observed in the smooth (blue) and dashed (red) lines. However, an interesting behavior is
observed here with respect to the bath squeezing. It can be seen that
it takes relatively longer to randomize the tomogram
in the presence of squeezing than in its absence, temperature remaining the same, as illustrated by a comparison between the dot-dashed
(magenta) and dashed (red) lines. 
This fact, in turn, establishes that
squeezing is a useful quantum resource. This behavior is
further elaborated in Figure \ref{fig:SGAD-3D-ACS}, where the effect
of the bath squeezing can be observed and is consistent with the
quadrature behavior of squeezing. This beneficial effect of squeezing is not observed for the evolution under QND channel. From the present
analysis, it could be envisaged that a tomographic connection could be established between the state under consideration and the generic open system interaction 
evolving it. 

\section{Tomogram of two spin-$\frac{1}{2}$ (qubit) states \label{sec:Tomogram-of-two-spin}}

Various two-qubit tomography schemes have been proposed in the recent past \cite{adam2014wigner,adam2014finite,miranowicz2014optimal,kiktenko2014tomographic,fedorov2015tomographic}. Specifically, the tomogram for two spin-$\frac{1}{2}$ (qubit) states can be obtained
using the star product scheme \cite{adam2014wigner,adam2014finite}. In \cite{kiktenko2014tomographic}, two-qubit states were analyzed from the perspective of tomographic causal 
analysis, while in \cite{fedorov2015tomographic}, an interesting connection between the tomographic construction of two-qubit states to aspects of quantum correlations, such as
discord and measurement induced disturbance, was developed. 

For a two-qubit state $\rho$ one can obtain the tomogram as 
\begin{equation}
\omega\left(m_{1},m_{2}\right)={\rm Tr}\left[\rho\left\{ Q_{1}\left(m_{1}\right)\otimes Q_{2}\left(m_{2}\right)\right\} \right],\label{eq:2qubitTomogram}
\end{equation}
where $Q_{i}\left(m_{i}\right)=U_{i}^{\dagger}\left|m_{i}\right\rangle \left\langle m_{i}\right|U_{i},$
and $m_{i}=\pm\frac{1}{2},$ while the unitary matrices $U_{i}$ are
\[
U_{i}=\left[\begin{array}{cc}
\cos\frac{\widetilde{\beta}_{i}}{2}\exp\left\{ \frac{i\left(\widetilde{\alpha}_{i}+\widetilde{\gamma}_{i}\right)}{2}\right\}  & \sin\frac{\widetilde{\beta}_{i}}{2}\exp\left\{ \frac{i\left(\widetilde{\alpha}_{i}-\widetilde{\gamma}_{i}\right)}{2}\right\} \\
-\sin\frac{\widetilde{\beta}_{i}}{2}\exp\left\{ -\frac{i\left(\widetilde{\alpha}_{i}-\widetilde{\gamma}_{i}\right)}{2}\right\}  & \cos\frac{\widetilde{\beta}_{i}}{2}\exp\left\{ -\frac{i\left(\widetilde{\alpha}_{i}+\widetilde{\gamma}_{i}\right)}{2}\right\} 
\end{array}\right]
\]
for $i\in\left\{ 1,2\right\}.$ Thus, the tomogram of the two-qubit
state can be written as the diagonal elements of $\widetilde{\rho},$
where $\widetilde{\rho}=\left(U_{1}\otimes U_{2}\right)\rho\left(U_{1}\otimes U_{2}\right)^{\dagger}.$

\subsection{Tomogram of a two-qubit state under dissipative evolution in a vacuum bath}

Here, we construct the tomogram of a two-qubit state in a vacuum
bath under dissipative evolution, as discussed in Ref. \cite{banerjee2010dynamics}. The same is already discussed in reference of the quasiprobability distributions for the two-qubit state under dissipative evolution in a vacuum bath in Section \ref{sub:Vacuum-bath}. Similar to Section \ref{sub:Vacuum-bath}, the initial state of the system
is considered with one qubit in the excited state $\left|e_{1}\right\rangle $
and the other in the ground state $\left|g_{2}\right\rangle $, i.e.,
$\left|e_{1}\right\rangle \left|g_{2}\right\rangle $. The reduced
density matrix of the system of interest, here the two qubits, is 
already discussed previously in Eq. (\ref{eq:densitymatrix-vaccumbath}) with various terms defined in Section \ref{sub:Vacuum-bath}.

The tomogram can be thought of as a tomographic-probability vector
$\omega=\left[\omega_{1},\omega_{2},\omega_{3},\omega_{4}\right]^{T}$
(here $T$ corresponds to the transpose of the vector), where each component
can be expressed analytically as 
\begin{equation}
\begin{array}{lcl}
\omega_{1}\left(t\right) & = & \frac{1}{4}\left[4\rho_{ee}\cos^{2}\frac{\widetilde{\beta}_{1}}{2}\cos^{2}\frac{\widetilde{\beta}_{2}}{2}+4\rho_{gg}\sin^{2}\frac{\widetilde{\beta}_{1}}{2}\sin^{2}\frac{\widetilde{\beta}_{2}}{2}\right. + \left(\rho_{aa}+\rho_{ss}\right)\left(1-\cos\widetilde{\beta}_{1}\cos\widetilde{\beta}_{2}\right)-\left(\rho_{aa}\right.\\
 & - & \left.\rho_{ss}\right)\sin\widetilde{\beta}_{1}\sin\widetilde{\beta}_{2}\cos\left(\widetilde{\gamma}_{1}-\widetilde{\gamma}_{2}\right)
+ \left\{ \rho_{sa}\left(\cos\widetilde{\beta}_{1}-\cos\widetilde{\beta}_{2}-i\sin\widetilde{\beta}_{1}\sin\widetilde{\beta}_{2}\right.\sin\left(\widetilde{\gamma}_{1}-\widetilde{\gamma}_{2}\right)\right)\\
 & + & \sqrt{2}\left[\left(\left(-\rho_{ea}+\rho_{es}\right)\cos^{2}\frac{\widetilde{\beta}_{2}}{2}+\left(\rho_{ag}+\rho_{sg}\right)\sin^{2}\frac{\widetilde{\beta}_{2}}{2}\right)\right.\sin\widetilde{\beta}_{1}\exp\left(i\widetilde{\gamma}_{1}\right)+\sin\widetilde{\beta}_{2}\exp\left(i\widetilde{\gamma}_{2}\right)\\
 & \times & \left.\left(\left(\rho_{ea}+\rho_{es}\right)\cos^{2}\frac{\widetilde{\beta}_{1}}{2}-\left(\rho_{ag}-\rho_{sg}\right)\sin^{2}\frac{\widetilde{\beta}_{1}}{2}\right)\right]+ \left.\left.\exp\left(i\widetilde{\gamma}_{1}+i\widetilde{\gamma}_{2}\right)\rho_{eg}\sin\widetilde{\beta}_{1}\sin\widetilde{\beta}_{2}+{\rm c.c.}\right\} \right],
\end{array}\label{eq:2qubit-tomo1}
\end{equation}
\begin{equation}
\begin{array}{lcl}
\omega_{2}\left(t\right) & = & \frac{1}{4}\left[4\rho_{ee}\cos^{2}\frac{\widetilde{\beta}_{1}}{2}\sin^{2}\frac{\widetilde{\beta}_{2}}{2}+4\rho_{gg}\sin^{2}\frac{\widetilde{\beta}_{1}}{2}\cos^{2}\frac{\widetilde{\beta}_{2}}{2}\right. + \left(\rho_{aa}+\rho_{ss}\right)\left(1+\cos\widetilde{\beta}_{1}\cos\widetilde{\beta}_{2}\right)+\left(\rho_{aa}\right.\\
 & - & \left.\rho_{ss}\right)\sin\widetilde{\beta}_{1}\sin\widetilde{\beta}_{2}\cos\left(\widetilde{\gamma}_{1}-\widetilde{\gamma}_{2}\right) +\left\{ \rho_{sa}\left(\cos\widetilde{\beta}_{1}+\cos\widetilde{\beta}_{2}+i\sin\widetilde{\beta}_{1}\sin\widetilde{\beta}_{2}\right.\sin\left(\widetilde{\gamma}_{1}-\widetilde{\gamma}_{2}\right)\right)\\
 & + & \sqrt{2}\left[\left(\left(\rho_{ag}+\rho_{sg}\right)\cos^{2}\frac{\widetilde{\beta}_{2}}{2}-\left(\rho_{ea}-\rho_{es}\right)\sin^{2}\frac{\widetilde{\beta}_{2}}{2}\right)\sin\widetilde{\beta}_{1}\exp\left(i\widetilde{\gamma}_{1}\right)\right.+\sin\widetilde{\beta}_{2}\exp\left(i\widetilde{\gamma}_{2}\right)\\
 & \times & \left.\left(-\left(\rho_{ea}+\rho_{es}\right)\cos^{2}\frac{\widetilde{\beta}_{1}}{2}+\left(\rho_{ag}-\rho_{sg}\right)\sin^{2}\frac{\widetilde{\beta}_{1}}{2}\right)\right] - \left.\left.\exp\left(i\widetilde{\gamma}_{1}+i\widetilde{\gamma}_{2}\right)\rho_{eg}\sin\widetilde{\beta}_{1}\sin\widetilde{\beta}_{2}+{\rm c.c.}\right\} \right],
\end{array}\label{eq:2qubit-tomo2}
\end{equation}
\begin{equation}
\begin{array}{lcl}
\omega_{3}\left(t\right) & = & \frac{1}{4}\left[4\rho_{ee}\sin^{2}\frac{\widetilde{\beta}_{1}}{2}\cos^{2}\frac{\widetilde{\beta}_{2}}{2}+4\rho_{gg}\cos^{2}\frac{\widetilde{\beta}_{1}}{2}\sin^{2}\frac{\widetilde{\beta}_{2}}{2}\right.
 + \left(\rho_{aa}+\rho_{ss}\right)\left(1+\cos\widetilde{\beta}_{1}\cos\widetilde{\beta}_{2}\right)+\left(\rho_{aa}\right.\\
 & - & \left.\rho_{ss}\right)\sin\widetilde{\beta}_{1}\sin\widetilde{\beta}_{2}\cos\left(\widetilde{\gamma}_{1}-\widetilde{\gamma}_{2}\right)
+ \left\{ -\rho_{sa}\left(\cos\widetilde{\beta}_{1}+\cos\widetilde{\beta}_{2}-i\sin\widetilde{\beta}_{1}\sin\widetilde{\beta}_{2}\right.\sin\left(\widetilde{\gamma}_{1}-\widetilde{\gamma}_{2}\right)\right)\\
 & + & \sqrt{2}\left[\left(-\left(\rho_{ag}+\rho_{sg}\right)\sin^{2}\frac{\widetilde{\beta}_{2}}{2}+\left(\rho_{ea}-\rho_{es}\right)\cos^{2}\frac{\widetilde{\beta}_{2}}{2}\right)\sin\widetilde{\beta}_{1}\exp\left(i\widetilde{\gamma}_{1}\right)\right.+\sin\widetilde{\beta}_{2}\exp\left(i\widetilde{\gamma}_{2}\right) \\
 & \times & \left.\left(\left(\rho_{ea}+\rho_{es}\right)\sin^{2}\frac{\widetilde{\beta}_{1}}{2}-\left(\rho_{ag}-\rho_{sg}\right)\cos^{2}\frac{\widetilde{\beta}_{1}}{2}\right)\right]
 - \left.\left.\exp\left(i\widetilde{\gamma}_{1}+i\widetilde{\gamma}_{2}\right)\rho_{eg}\sin\widetilde{\beta}_{1}\sin\widetilde{\beta}_{2}+{\rm c.c.}\right\} \right],
\end{array}\label{eq:2qubit-tomo3}
\end{equation}

\begin{figure}
\begin{centering}
\includegraphics[angle=-180,scale=0.9]{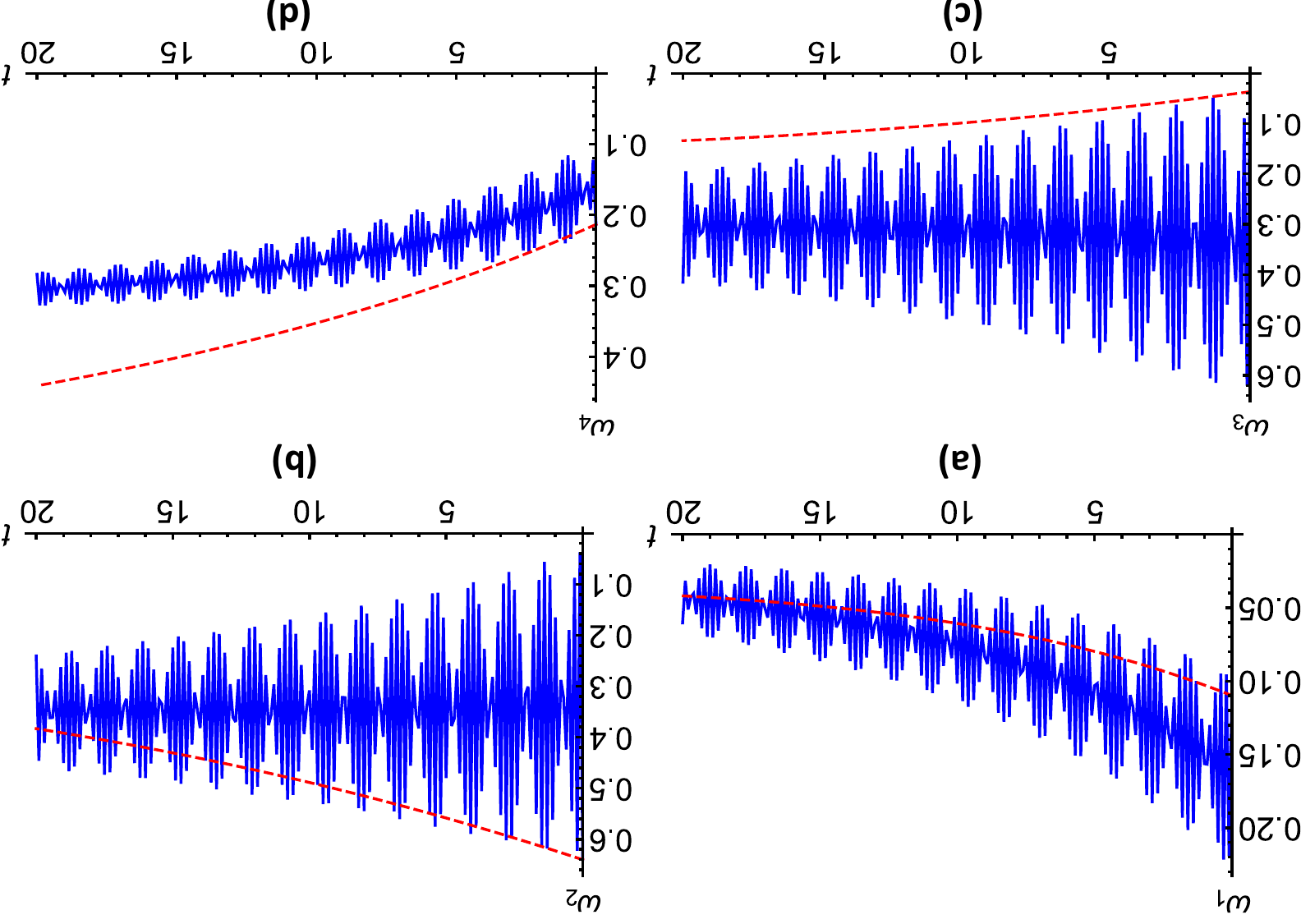}

\protect\caption[Various components of tomogram
for two qubits under dissipative evolution in a 
vacuum bath]{\label{fig:Vacuum-bath} Various components
of the tomogram changing with time are shown in (a)-(d) for the two-qubit state,
in the presence of the vacuum bath, with $\widetilde{\beta}_{1}=\frac{\pi}{3},\,\widetilde{\beta}_{2}=\frac{\pi}{4},\,\widetilde{\gamma}_{1}=\frac{\pi}{3},\,\widetilde{\gamma}_{2}=\frac{\pi}{4},$
and the inter-qubit spacing $r_{12}=0.05$ (2.0) corresponding to the smooth blue (red dashed) line.}
\end{centering}
\end{figure}

\noindent and 
\begin{equation}
\begin{array}{lcl}
\omega_{4}\left(t\right) & = & \frac{1}{4}\left[4\rho_{ee}\sin^{2}\frac{\widetilde{\beta}_{1}}{2}\sin^{2}\frac{\widetilde{\beta}_{2}}{2}+4\rho_{gg}\cos^{2}\frac{\widetilde{\beta}_{1}}{2}\cos^{2}\frac{\widetilde{\beta}_{2}}{2}\right. +  \left(\rho_{aa}+\rho_{ss}\right)\left(1-\cos\widetilde{\beta}_{1}\cos\widetilde{\beta}_{2}\right)-\left(\rho_{aa}\right.\\
 & - & \left.\rho_{ss}\right)\sin\widetilde{\beta}_{1}\sin\widetilde{\beta}_{2}\cos\left(\widetilde{\gamma}_{1}-\widetilde{\gamma}_{2}\right)
+ \left\{ -\rho_{sa}\left(\cos\widetilde{\beta}_{1}-\cos\widetilde{\beta}_{2}+i\sin\widetilde{\beta}_{1}\sin\widetilde{\beta}_{2}\right.\sin\left(\widetilde{\gamma}_{1}-\widetilde{\gamma}_{2}\right)\right)\\
 & + & \sqrt{2}\left[\left(-\left(\rho_{ag}+\rho_{sg}\right)\cos^{2}\frac{\widetilde{\beta}_{2}}{2}+\left(\rho_{ea}-\rho_{es}\right)\sin^{2}\frac{\widetilde{\beta}_{2}}{2}\right)\sin\widetilde{\beta}_{1}\exp\left(i\widetilde{\gamma}_{1}\right)\right.+\sin\widetilde{\beta}_{2}\exp\left(i\widetilde{\gamma}_{2}\right)\\
 & \times & \left.\left(-\left(\rho_{ea}+\rho_{es}\right)\sin^{2}\frac{\widetilde{\beta}_{1}}{2}+\left(\rho_{ag}-\rho_{sg}\right)\cos^{2}\frac{\widetilde{\beta}_{1}}{2}\right)\right] - \left.\left.\exp\left(i\widetilde{\gamma}_{1}+i\widetilde{\gamma}_{2}\right)\rho_{eg}\sin\widetilde{\beta}_{1}\sin\widetilde{\beta}_{2}+{\rm c.c.}\right\} \right].
\end{array}\label{eq:2qubit-tomo4}
\end{equation}
Here, $\rho_{ij}$ are the elements of the matrix in Eq. (\ref{eq:densitymatrix-vaccumbath}).
For simplicity of notations in writing, the time dependence in the arguments
of the matrix elements is omitted. Similar to the tomograms for single
spin-$\frac{1}{2}$ states, the tomogram obtained here is also free
from $\widetilde{\alpha}.$

As in the cases of single-qubit tomograms, we can again verify that
the tomogram obtained here satisfies the condition $\sum\omega_{i}=\rho_{ee}+\rho_{gg}+\rho_{aa}+\rho_{ss},$
which is the trace of the density matrix given in Eq. (\ref{eq:densitymatrix-vaccumbath})
and hence equal to one.

For the case of identical qubits considered here, we take the wavevector
and mean frequency to be $k_{0}=\omega_{0}=1$, the spontaneous emission rate
$\Gamma_{j}=0.05,$ and $\hat{\mu}\mathord{\cdot}\hat{r}_{ij}=0$. 
Here, $\hat{\mu}$ is the unit vector along the atomic transition
dipole moment, and $\hat{r}_{ij}$ is the inter-atomic distance. Further,
the initial state of the system is taken to be 
$\rho_{ee}\left(0\right)=\rho_{gg}\left(0\right)=\rho_{es}\left(0\right)=\rho_{ea}\left(0\right)=\rho_{eg}\left(0\right)=\rho_{sg}\left(0\right)=
\rho_{ag}\left(0\right)=0$,
and $\rho_{ss}\left(0\right)=\rho_{aa}\left(0\right)=\rho_{sa}\left(0\right)=0.5$.

\begin{figure}[t]
\begin{centering}
\includegraphics[angle=0,scale=0.9]{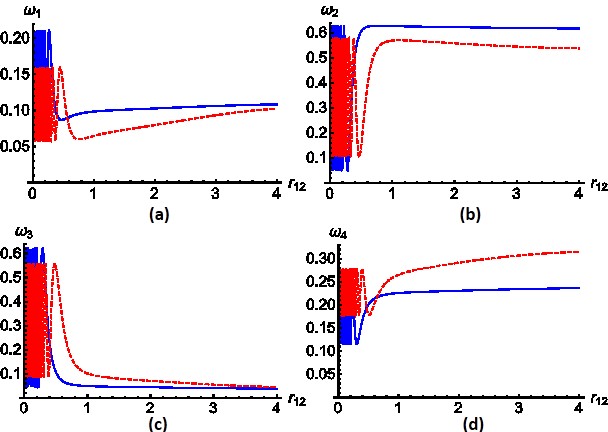}

\protect\caption[Various components of tomogram
for two-qubit state interacting with a vacuum bath as a function of inter-qubit spacing]{\label{fig:Vacuum-bath-r} (a)-(d) depict the tomogram
for the two-qubit state, interacting with a vacuum bath, as a function
of the inter-qubit spacing at $t=1$ (smooth blue line) and $t=5$
(red dashed line). For all the plots, $\widetilde{\beta}_{1}=\frac{\pi}{3},\,\widetilde{\beta}_{2}=\frac{\pi}{4},\,\widetilde{\gamma}_{1}=\frac{\pi}{3},\,\widetilde{\gamma}_{2}=\frac{\pi}{4}$. }
\end{centering}
\end{figure}

Variation of all four components of the tomogram is shown with
different parameters in Figures \ref{fig:Vacuum-bath} and \ref{fig:Vacuum-bath-r}.
In Figure \ref{fig:Vacuum-bath}, large oscillations can be observed
for small inter-qubit spacing, which is consistent with the earlier
observations in a plethora of scenario \cite{banerjee2010dynamics,banerjee2008geometric,srinatha2014quantum}, and also with our observations illustrated in Figures \ref{fig:Vacuum-bath-qd} and \ref{fig:Vacuumbath-2} in Chapter \ref{QDs}.
Figure \ref{fig:Vacuum-bath-r} further demonstrates similar behavior for small inter-qubit spacing. For small
inter-qubit spacing, the ambient environment opens up a channel between the qubits resulting in enhancement of oscillations.

\section{Optical tomogram for a dissipative harmonic oscillator \label{sec:Optical-tomogram}}

At the end, we come to the tomogram of an infinite dimensional system, the harmonic oscillator. This is typical of a plethora 
of oscillatory and optical systems \cite{louisell1973quantum,perina1991quantum}. In Ref. \cite{man1997damped}, the quantum mechanics of the damped harmonic oscillator was examined,
from the perspective of a classical description of quantum mechanics \cite{mancini1996symplectic}. Use was made of the generating 
function method, resulting in the avoidance of the need to evaluate the Wigner function as an intermediary step for obtaining the tomogram. Further, in
\cite{chernega2008wave}, the density matrix, state tomogram, and Wigner function of a parametric oscillator were studied. 

Here, we construct the tomogram of the dissipative harmonic 
oscillator evolving under a Lindbladian evolution, in a phase sensitive reservoir \cite{hu2012kraus}. 
It would be pertinent to mention that tomographic reconstruction of Gaussian states evolving under
a Markovian evolution has also been considered in \cite{bellomo2009reconstruction}.
The dissipative harmonic oscillator can be described by the Hamiltonian described in Section \ref{OQS} in Eq. (\ref{eq:OQS-H}),
where the system Hamiltonian $H_{\rm S}$ of a harmonic oscillator is
described as 
\[
H_{\rm S}=\frac{p^{2}}{2m}+\frac{1}{2}m\omega^{2}x^{2},
\]
while the reservoir Hamiltonian $H_{\rm R}$ is given by 
\[
H_{\rm R}=\sum_{j}\frac{p_{j}^{2}}{2m_{j}}+\frac{1}{2}m_{j}\omega_{j}^{2}x_{j}^{2},
\]
with the system-reservoir interaction Hamiltonian $H_{\rm I}$ as 
\[
H_{\rm I}=\sum_{j}c_{j}xx_{j}.
\]
Here, the reservoir is modeled as a bath of harmonic oscillators with
$c_{j}$ as the coupling constant. The dynamics of the system harmonic oscillator is obtained by tracing over the 
reservoir degrees of freedom. The optical tomogram from the Wigner
function can be obtained using \cite{vogel1989determination} 
\begin{equation}
\begin{array}{l}
\omega\left(X,\theta\right)=\int W\left(X\cos\theta-p\sin\theta,X\sin\theta+p\cos\theta\right)dp,\end{array}\label{eq:tomogram_from_wigner}
\end{equation}
where $W\left(x,y\right)$ is the Wigner function (\ref{Wusual}). Similarly, the corresponding
Wigner function can also be reconstructed from the tomogram by inverse
Radon transformation.
The analytic expression of the tomogram for the system,
initially in the coherent state $\left|\beta\right\rangle,$ is 
\begin{equation}
\begin{array}{lcl}
\omega(X,\theta,t) & = & \sqrt{\frac{2}{\pi}}\frac{1}{\sqrt{\left(2N_{th}M+1\right)-\left(rMe^{-2i\theta}+{\rm c.c.}\right)}} \exp\left(-\frac{2\left(Re[\beta e^{i\theta}]e^{-kt}-X\right)^{2}}{\left(2N_{th}M+1\right)-\left(rMe^{-2i\theta}+{\rm c.c.}\right)}\right).
\end{array}\label{eq:optical_tomogram}
\end{equation}
Here, $N_{th}=\frac{1}{\exp(\hbar \omega_{k}/k_{B}T) - 1}$ is the 
average thermal photon number of the environment at temperature $T$. 
Also, $r$ is the bath squeezing parameter, $Re[u]$ denotes the real part of $u$,
and $M=1-\exp\left(-2kt\right),$ where $k$ is the dissipation coefficient, analogous to the spontaneous emission term. 

\begin{figure}[t]
\includegraphics[angle=0,scale=0.56]{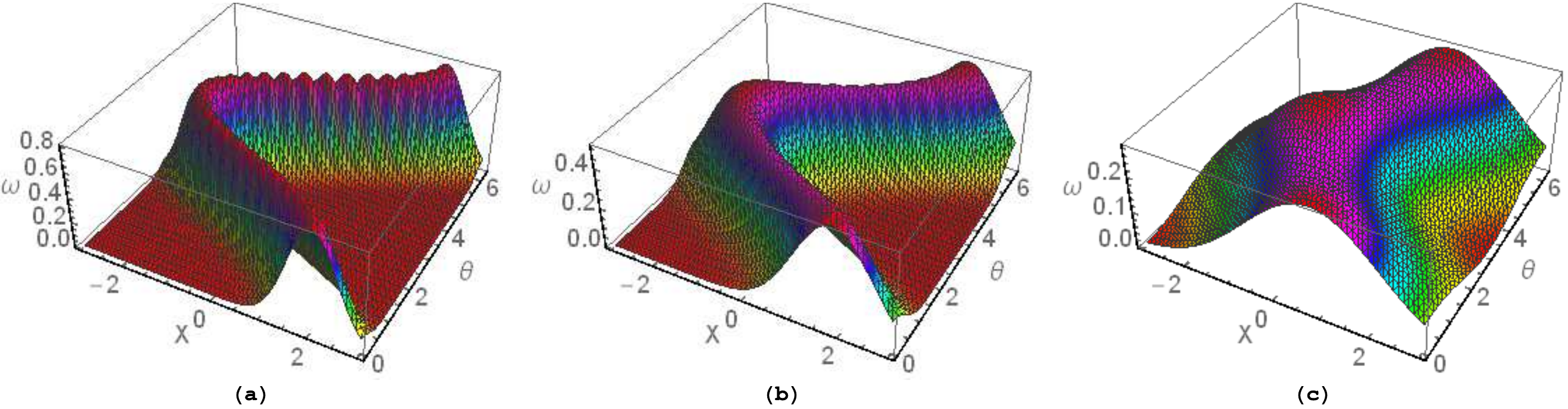}
\protect\caption[Optical tomogram of a dissipative harmonic oscillator at different times]{\label{fig:op-tom-3D} The effect of interaction of the optical
tomogram with its environment is shown as a function of $X$ and $\theta$ in (a)-(c) for $N_{th}=5,\,r=1$ for the initial coherent state parameter $\beta=2$
at time $t=0,\,1,$ and $10$, respectively.}
\end{figure}
\begin{figure}[t]
\centering{}
\includegraphics[scale=0.6]{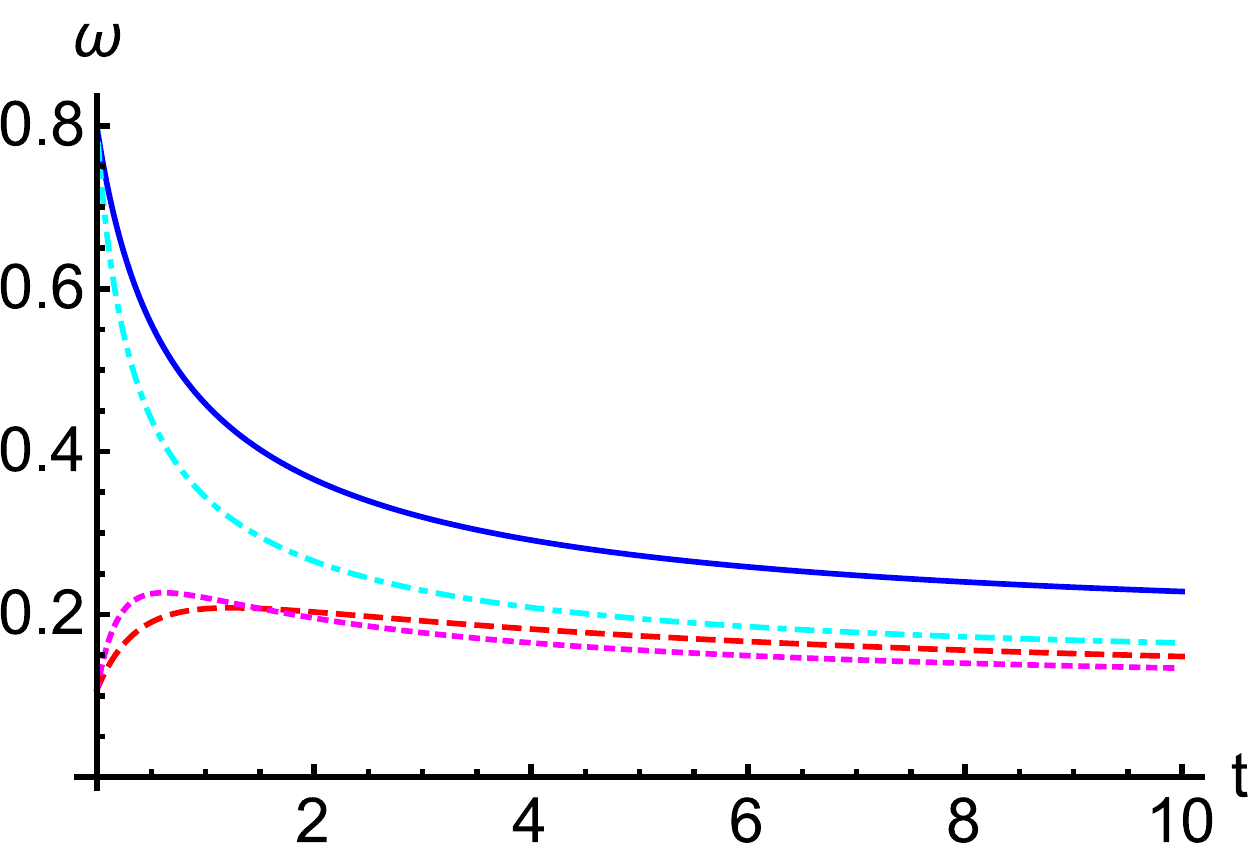}

\protect\caption[Optical tomogram of a dissipative harmonic oscillator evolving with time]{\label{fig:op-tom}  The tomogram of a dissipative harmonic oscillator varying with time is shown for the initial coherent state parameter $\beta=2$ and $\theta=\frac{\pi}{3}$.
The smooth (blue) and dashed (red) lines correspond to the tomogram
for $N_{th}=5,\,r=1$ for $X=1$ and 2, respectively. Similarly, the dot-dashed
(cyan) and dotted (magenta) lines correspond to the tomogram for $N_{th}=10,\,r=4$
for $X=1$ and 2, respectively.}
\end{figure}

In the corresponding figures for the tomogram with the
specific values of different parameters, we can observe the decay
of the tomogram. Specifically, Figure \ref{fig:op-tom-3D} (a) shows the
tomogram of an initial coherent state, where we can see a beautiful valley
like shape surrounded by a mountain. Interestingly, a similar tomogram
has been observed for a binomial state of large dimension (cf. Figure 2 in \cite{bazrafkan2004tomography}). However,
in Figures \ref{fig:op-tom-3D} (b) and (c), we can see this sharp structure
gradually fades away due to interaction with its surrounding. 
 Thus, with an increase in temperature, the effect of decoherence and dissipation,
due to the ambient environment, deteriorates the obtained tomogram. 
Further, Figure \ref{fig:op-tom} illustrates the effect of change of the average
thermal photon number and squeezing parameter, where in the smooth (blue)
and dot-dashed (cyan) lines, we can observe the enhancement of decay.
Similarly, the dashed (red) and dotted (magenta) lines also show the effect
due to changes in the bath parameters for another set of parameters.

\section{Conclusions \label{sec:Conclusion-Tomogram}}

Tomography is a powerful quantum state reconstruction tool. Its wide
applicability in obtaining quasiprobability distributions, quantum process tomography, and density matrix reconstruction
in quantum computation and communication is already established. For example, the quasiprobability distributions discussed in Chapter \ref{QDs} can be obtained from the tomograms directly accessible from the experimental data as computed in the present chapter.
However, these properties can get affected by the influence of the ambient environment.
Here, an effort has been made to study the evolution of tomograms
for different quantum systems, both finite and infinite dimensional, under
general system-reservoir interactions, 
using the formalism of open quantum systems. The effect of the environment
on the finite dimensional quantum systems (spin states)
is to randomize the tomogram. For the spin quantum states, single and two spin-$\frac{1}{2}$ states are considered with 
open quantum system effects. The increase in temperature tends to decohere 
the tomograms, while squeezing is shown to be a  useful quantum resource. Further, the tomogram for an infinite dimensional system, the
ubiquitous dissipative harmonic oscillator, is also studied.
The results obtained here are expected to have an impact on issues
related to quantum state reconstruction in the quantum computation, communication, and information processing.

Besides this, a tomogram for a spin-1 pure 
quantum state is also obtained. For the number-phase states, a general expression is obtained and is
illustrated through the example of a three-level quantum (qutrit) system 
in a spontaneous emission channel. However, these results are not included in the present chapter. Interested readers can find corresponding discussion in Ref. \cite{thapliyal2016tomograms}, where the findings of the present chapter are published.

We can now conclude our discussion on the dynamics of the nonclassical properties of the spin systems and their decoherence due to the effect of the environment. This would allow us to summarize that nonclassical resources can be generated in the optical and spin states, which are subjected to inevitable interactions with the surroundings causing a quantum to classical transition. Whole motivation of the pursuit for  generation of nonclassical states in newer and simpler systems, studying the evolution of the nonclassical effects, and suppressing the effect of decoherence on them, is justified by the next two chapters, where we report applications of the nonclassical states in quantum communication and also study the effect of various types of noise on those schemes.

\mathversion{normal}
\thispagestyle{empty}
\thispagestyle{plain}   
\cleardoublepage
\blankpage

\titlespacing*{\chapter}{0pt}{-50pt}{20pt}
\chapter{Controlled quantum communication} \label{CryptSwitch}

\section{Introduction \label{CryptSwitch-Intro}}

In Chapter \ref{Introduction}, we have already mentioned various applications of nonclassical states. In what follows, in the present chapter and the next chapter, we will discuss some specific applications of nonclassical states in the field of quantum communication. Therefore, here we are going to briefly discuss historical development of quantum communication science, which will be useful in the next chapter as well. The first quantum communication scheme, proposed by Bennett and Brassard in 1984 \cite{bennett1984quantum}, was based on the Wiesner's idea of \emph{conjugate coding} \cite{wiesner1983conjugate}. This protocol for QKD is now known as the BB84 protocol. Interestingly, BB84 protocol was the first scheme for unconditionally secure communication. As unconditional security is a desirable feature that cannot be achieved in the classical world, BB84 protocol drew considerable attention of the scientific community.  Over the past few decades, BB84 protocol has been followed by several schemes for secure quantum communication; and commercial quantum cryptography products based on the BB84 protocol and the other protocols of secure quantum communication have also been introduced (e.g., \cite{IDQ}).  Here, we may note that BB84 protocol is also relevant for the fact that it provided us the first useful example of a task, where enhancement in the performance happens due to involvement of quantum resources \cite{bennett1984quantum}. Thus, it clearly established quantum advantage in the context of secure communication. Subsequently, quantum supremacy has been established in  the context of insecure communication (e.g., teleportaion \cite{bennett1993teleporting} and densecoding \cite{bennett1992communication}) and computation (viz., Deutsch's  \cite{deutsch1985quantum}, Deutsch-Jozsa  \cite{deutsch1992rapid}, Grover's  \cite{grover1997quantum}, Shor's \cite{shor1999polynomial} algorithms).

On the basis of requirement for the security in the task in hand, protocols for quantum communication can be divided into two subgroups--protocols for the insecure communication and protocols for the secure communication. 
Specifically, the idea of quantum teleportation which can be considered as the first scheme for insecure quantum communication was introduced by Bennett et al. \cite{bennett1993teleporting} in 1993. Subsequently, several schemes for quantum communication (e.g.,  the schemes for quantum information splitting or controlled teleportation
\cite{karlsson1998quantum,pathak2011efficient}, quantum secret sharing \cite{hillery1999quantum},
hierarchical quantum information splitting  \cite{wang2010hierarchical,shukla2013hierarchical},
remote state preparation \cite{pati2000minimum}), which can be viewed as modified
teleportation schemes, have been proposed.  The teleportation scheme also inspired joint remote state preparation \cite{nguyen2008joint} and its hierarchical version \cite{shukla2016hierarchical}. Further studies on these schemes drew considerable attention of the researchers because of the facts that there does not exist any classical analogue of these schemes, and these schemes can be used to develop protocols for  the secure quantum communication and remote
quantum operations \cite{huelga2001quantum}.  Additionally, quantum
teleportation has also been linked to other ideas of quantum information in various interesting works
\cite{brassard1998teleportation,knill2000efficient,yan2004scheme}.

Several experimental realizations of quantum teleportation (\cite{bouwmeester1997experimental,sisodia2017design} and references therein) and remote state preparation (i.e., teleportation of a known state) \cite{peters2005remote,liu2007experimental,xiang2005remote,raadmark2013experimental} schemes have also been reported. Various groups have also reported proof of principle experimental realization of teleportation \cite{nielsen1998complete,furusawa1998unconditional,zhao2004experimental,riebe2004deterministic,barrett2004deterministic,sisodia2017design}. It is noteworthy that the optical realizations of quantum teleportation involve non-deterministic Bell measurement \cite{bouwmeester1997experimental}, and also most of the recent experiments have assigned the Bell state preparation task to a third party so that the teleportation can be performed over twice the distance of conventional teleportation \cite{de2004long}. This idea was found applicable to measurement device independent quantum cryptography \cite{lo2012measurement}. Due to its wide applicability in different information processing tasks, bidirectional counterpart of teleportation, which is referred to as bidirectional state teleportation and its controlled variant BCST have also been proposed. A focused discussion on the BCST scheme will be performed in Section \ref{sec:BCST}.

We are now in a position to discuss secure quantum communication schemes in detail. Particularly, to contemplate the quantum enhancement attained by the BB84 scheme \cite{bennett1984quantum} in the field of secure communication, we need to understand the requirement of cryptography. Aptly, the Greek meaning of the name itself suggests, cryptography (secret writing) is the art of transmitting a message in a secure manner, and a critical analysis of a cryptographic protocol to extract the inaccessible secret information is known as cryptanalysis. In general, both these tasks are studied under a common subject called cryptology. At the core of the cryptography, Kerckhoffs's principle \cite{kerckhoffs1883cryptographic} plays an important role. Specifically, the principle states that the cryptogram (obtained by performing an operation defined as a cipher or cryptosystem between the message and the private key) is secure until the key used in preparing that is secure. In the domain of classical physics, this secure key is obtained using one-way computationally complex problems. For example,  RSA cryptosystem \cite{rivest1978method} and Diffie-Hellman \cite{diffie1976new} schemes utilize computational complexity of the prime factorization and discrete logarithms problems, respectively. As discussed previously, quantum enhancement in the information processing tasks also led to certain quantum algorithms \cite{shor1999polynomial,grover1997quantum} which ensured solutions to these complex problems, in turn, endangering corresponding classical cryptography scheme. Therefore, most of the classical cryptographic protocols will not be secure once a scalable quantum computer is built. In contrast, the security of the BB84 protocol \cite{bennett1984quantum} does not depend on certain computationally complex problem, rather it arises from the fact that any  eavesdropping attempt by an intruder (Eve) leads to a detectable disturbance. To be precise, an arbitrary attack by Eve reveals her some useful information, but leaves detectable traces at the receiver's end. Therefore, the security of a quantum cryptographic protocol is ensured due to information versus disturbance trade-off.

Thus, quantum cryptography has
been flourishing over the last three decades due to the possibility
of the unconditional security, a task unachievable in the domain of
classical physics. Specifically, the BB84 protocol \cite{bennett1984quantum} uses four states, which was later modified to obtain a scheme with only two states \cite{bennett1992B92}. Meanwhile, an entangled-state-based protocol was proposed in 1991 \cite{ekert1991quantum}, where Alice (sender) and Bob (receiver) check the nonlocal correlations of a quantum state prepared from a source kept between Alice and Bob and shared among them. This requires six states and forms the basis of device independent cryptography \cite{acin2006bell}. This was followed by another protocol using bipartite entanglement in 1992 \cite{bennett1992quantum}. Meanwhile, it was shown that unconditionally secure quantum cryptography can also be accomplished using only orthogonal state \cite{goldenberg1995quantum}.
The unconditional security of all these schemes was analyzed on different fronts over the time. 

At the same time, cryptanalysts also kept designing attacks to gain access to the secret information leaving minimum traces (\cite{lo2014secure,shenoy2017quantum} and references therein). The first such attack was photon number splitting attack \cite{brassard2000limitations} on the BB84 protocol \cite{bennett1984quantum}, which exploited Alice's inability to generate single photon source at will. Precisely, all the single-photon-based implementation either use weak coherent pulses or heralded single photon sources. When they use the former, there is a non-zero probability of obtaining more than one photons in a pulse. Exploiting which Eve may perform an operation that would stop all the single photon pulses and store one photon each from every multi-photon pulses, allowing the rest of the photons to travel to the receiver. This operation will allow Eve to know the whole key without being detected. This is feasible as, in cryptography, we assume Alice and Bob are restricted by the present technology, while Eve is only restricted by the physical laws, which are governed by quantum mechanics. The remedy for such an attack was soon discovered, which employs an intensity modulator to mix multi-photon pulses randomly with the single photon pulses, called decoy pulses, to detect the photon number splitting attack or any other such side-channel attacks. This ensured the unconditional security of the decoy-qubit-based BB84 protocol. 
Later this idea of using decoy qubits was extended to secure direct quantum communication, and it played a vital role in establishing security of these protocols (see \cite{sharma2016verification} for detailed discussion).

Interestingly, the security achieved in these schemes assumes the validity of quantum mechanics and trustworthy implementing devices \cite{acin2006bell}. To omit the second assumption and to perform quantum cryptography with untrusted devices, the quantum correlations among the legitimate users, which violate Bell nonlocality, i.e., cannot be reproduced by any hidden variable model, are used. Due to the limitations in realizing the resource required to accomplish this feat of device independent cryptography, researchers settle with one-side device independent (\cite{branciard2012one} and references therein), where either preparation or measurement devices can be made side-channel attack free. As characterization of a source is relatively easier and a measurement device is more prone to quantum hacking \cite{lo2014secure}, measurement device independent quantum  cryptography \cite{lo2012measurement} has gained much attention recently. The quest to understand the powers of quantum resources has also led to semi-quantum cryptography \cite{PhysRevLett.99.140501}, where unconditionally secure quantum communication is achieved between a classical and a quantum enabled user. Another important facet of quantum cryptography, which is closely associated with the foundational aspects of quantum mechanics, is counterfactual quantum cryptography \cite{noh2009counterfactual}, where the final key is prepared with the help of only qubits which had not traveled through the channel accessible to Eve. 

This feature of unconditional security (see \cite{gisin2002quantum} for a review) and already available marketable products
based on quantum cryptography have motivated further research in this
field. To name a few, apart from the initial works on QKD \cite{bennett1984quantum,bennett1992B92,ekert1991quantum,goldenberg1995quantum,bennett1992quantum},
various schemes concerning secure direct quantum communication (secure communication
circumventing the need of a prior shared key) \cite{bostrom2002deterministic,lucamarini2005secure,shukla2013improved,long2007quantum,banerjee2012maximally,pathak2015efficient},
QKA \cite{shukla2014protocols}, quantum secret sharing
\cite{hillery1999quantum} have been proposed (see
\cite{pathak2013elements} for details). Specifically, in the secure direct quantum communication, the
receiver may or may not require an additional classical information
to decode the message sent by the sender; depending upon this, the
protocol falls under the category of DSQC \cite{jun2006revisiting,li2006deterministic,yan2004scheme,zhong2005deterministic,hwang2011quantum,zhu2006secure,hai2006quantum,yuan2011high,banerjee2012maximally} and QSDC
\cite{long2002theoretically,bostrom2002deterministic,degiovanni2004quantum,lucamarini2005secure,shukla2013improved,long2007quantum}, respectively. There is another novel
technique of direct communication, QD \cite{nguyen2004quantum},
where both the users can send their information simultaneously, with
no need of a prior shared key. Further, a counterfactual direct communication scheme was not only proposed in the recent past \cite{salih2013protocol}, is also experimentally realized recently \cite{cao2017direct}. Only other reported experiment for a direct communication scheme are \cite{hu2016experimental,zhang2017quantum}.

All these schemes for secure direct quantum communication provide us a vast potential
for extension and modification to design the protocols for performing various cryptographic tasks that have relevance in the
real-life. 
We may consider an important scenario, 
where a controller supervises the communication among all the remaining users, and he can maintain
his control by making sure that the communication is not accomplished
without his consent (\cite{srinatha2014quantum,pathak2015efficient} and references therein). Further, an asymmetric counterpart of the quantum dialogue scheme \cite{nguyen2004quantum} was also proposed \cite{banerjee2017asymmetric}. Recently, a multiparty scheme for direct communication among arbitrary number of parties has been proposed and christened \emph{quantum conference} \cite{banerjee2017quantum} due to its analogy with conferences. 

As we are restricting our applications of quantum resources solely to the domain of quantum communication, in its vicinity lies a computation task which is worth mentioning here. Specifically, it is possible to define a class of functions with inputs from different parties, and the final output of the function is to be calculated in such a way that no input is revealed \cite{yao1982protocols}. There are numerous tasks which can be viewed as applications of these types of secure multiparty  computation tasks, such as voting, private comparison, socialist-millionaire problem, sealed-bid auction. In the domain of classical physics, it is known that a solution of a secure multiparty  computation task using a trusted party can also be performed without the trusted party, just by increasing the complexity \cite{canetti2000security}. Evidently, the security in the quantum domain does not rely on the computational powers of an eavesdropper, therefore when the secure two-party computation tasks were analyzed, it led to an impossibility of two-party secure computation without third party \cite{lo1997insecurity}. However, when analyzed closely quantum resources made it feasible to accomplish the task with a dishonest third party, and several direct communication schemes could be modified to obtain the solutions of the problems of socioeconomic relevance, e.g., protocols of quantum voting \cite{thapliyal2017protocols}, quantum private comparison \cite{thapliyal2016orthogonal,shukla2017semi}, quantum sealed-bid auction \cite{sharma2017quantumauction}, quantum e-commerce \cite{shukla2017semi}, have been proposed by modifying schemes of secure quantum communication which were originally designed for some other task(s).

It would be worth summarizing that the
security achieved in all the cryptographic schemes is based on the
principle of splitting the whole information into two or more pieces, and the
whole information can only be extracted if all the pieces are available
simultaneously. Usually, one of the parties prepares an entangled
state to be used as a quantum channel and shares it with all other parties
in a secure way. By secure, we mean that a proper eavesdropping checking
technique is employed, after inserting the decoy qubits with the entangled
qubits to ensure the absence of Eve (see \cite{sharma2016verification} for detailed discussion). Once this channel is shared,
the legitimate parties can securely share their secrets, either by teleportation
or by encoding their information using Pauli operations
and sending the qubits to the receiver again in a secure manner.

During transmission of information to distant parties employing quantum resources, they are expected to undergo decoherence due to interaction with the surroundings. Therefore, the performance of a certain quantum communication scheme (both secure and insecure in nature) in a realistic scenario, i.e., considering the effect of the environment, should be analyzed using the open quantum system formalism adopted in Chapters \ref{QDs}-\ref{Tomogram}. The relevance of this type of studies can be recognized with our recent observations, where we have shown that a deterministic hierarchical joint remote state preparation scheme becomes probabilistic one \cite{shukla2016hierarchical} in the noisy environments; the singlet-like non-orthogonal entangled (quasi-Bell) state may perform best (worst) among the quasi-Bell states in the ideal (noisy) conditions \cite{sisodia2017teleportation}; and a three-stage quantum cryptography scheme \cite{kak2006three} fails under noisy environment \cite{thapliyal2018kak}. This led us to analyze the performance of a set of schemes for the quantum communication over a noisy channel(s).

The remaining part of the present chapter is organized as follows. A detailed discussion of the BCST and other controlled quantum communication schemes is given in 
Section \ref{sec:BCST}. In
Section \ref{sec:The-Condition-of}, we provide a general mathematical structure of the quantum states suitable as quantum channels for BCST schemes. In
Section \ref{sec:Bidirectional-teleportation-using}, 
a Bell-state-based protocol of BCST is proposed, which is followed by a set of Bell-state-based protocols for other controlled quantum
communication tasks in Section \ref{sec:Other-protocols-of}.  The performance of the proposed BCST and CQD schemes is analyzed over AD and PD channels in Section \ref{sec:Effect-of-noise} before concluding the chapter in Section \ref{sec:Conclusions-CryptSwitch}.

\section{A brief introduction to controlled quantum communication \label{sec:BCST}}

In 2001, Huelga et al. \cite{huelga2001quantum,huelga2002remote} proposed a scheme for bidirectional state teleportation. This protocol was of extreme interest because of the fact that in contrast to the original teloportation scheme of  Bennett et
al. \cite{bennett1993teleporting} (which was
a one-way scheme in the sense that in the scheme in its original form only Alice was allowed to transmit an unknown
(single-qubit) quantum state to Bob) Huelga et al.'s scheme allowed both Alice and Bob to simultaneously transmit
unknown quantum states to each other. Interestingly, a relation between quantum nonlocal gates and bidirectional state teleportation was also established in the pioneering work of Huelga et al. To visualize this point, we let us assume that  Bob teleports
a quantum state $|\psi\rangle$ to Alice. Upon receiving it, Alice applies a unitary operator
$U$ on it and  teleports back the modified state $|\psi^{\prime}\rangle=U|\psi\rangle$
to Bob. Clearly, the existence of a scheme for bidirectional state teleportation is equivalent to the  ability to implement a nonlocal quantum gate or a quantum remote
control.  Further extending the concept 
of bidirectional state teleportation, a large number of schemes for BCST have been proposed \cite{zha2013bidirectional,xin2010bidirectional,li2013bidirectional,shukla2013bidirectional,li2013cqsdc,duan2014bidirectional6,fu2014general,chen2015bidirectional,yan2013bidirectional,duan2014bidirectional7}.

Interestingly, one can visualize the three party BCST scheme as bidirectional state teleportation provided
the supervisor/controller (Charlie) has allowed the other two users (Alice
and Bob) to execute a protocol of bidirectional state teleportation. From the study of all the
recently proposed schemes of BCST \cite{zha2013bidirectional,xin2010bidirectional,li2013bidirectional,li2013cqsdc,duan2014bidirectional6,fu2014general,chen2015bidirectional,yan2013bidirectional,duan2014bidirectional7},
one can easily conclude that at least five qubits are used in all
those schemes. The intrinsic symmetry
of the five-qubit quantum states used as quantum channels in
the BCST schemes was explored in \cite{shukla2013bidirectional} to provide
a general structure for the five-qubit quantum states suitable for performing a BCST protocol
as follows: 
\begin{equation}
|\psi\rangle=\frac{1}{\sqrt{2}}\left(|\psi_{1}\rangle_{A_{1}B_{1}}|\psi_{2}\rangle_{A_{2}B_{2}}|a\rangle_{C_{1}}\pm|\psi_{3}\rangle_{A_{1}B_{1}}|\psi_{4}\rangle_{A_{2}B_{2}}|b\rangle_{C_{1}}\right),\label{eq:the 5-qubit state}
\end{equation}
where the single-qubit states $|a\rangle$ and $|b\rangle$ satisfy $\langle a|b\rangle=\delta_{a,b}$,
$|\psi_{i}\rangle\in\left\{ |\psi^{+}\rangle,|\psi^{-}\rangle,|\phi^{+}\rangle,|\phi^{-}\rangle\right.:\left.|\psi_{1}\rangle\neq|\psi_{3}\rangle,|\psi_{2}\rangle\neq|\psi_{4}\rangle\right\} $,
$|\psi^{\pm}\rangle=\frac{|00\rangle\pm|11\rangle}{\sqrt{2}},$ $|\phi^{\pm}\rangle=\frac{|01\rangle\pm|10\rangle}{\sqrt{2}}$,
and the subscripts $A$, $B,$ and $C$ indicate the qubits of Alice,
Bob, and Charlie, respectively. After preparing the quantum state $|\psi\rangle$
Charlie sends qubits $A1$ and $A2$ ($B1$ and $B2$) to Alice (Bob), and the last qubit remains with himself. The condition 
\begin{equation}
|\psi_{1}\rangle\neq|\psi_{3}\rangle,\,\mathrm{and} \,|\psi_{2}\rangle\neq|\psi_{4}\rangle\label{eq:condition}
\end{equation}
ensures that Charlie holds a qubit appropriately entangled with the rest of the qubits, which makes sure that Alice and Bob remain ignorant
of the entangled (Bell) states shared between them until disclosure of the outcome of the measurement performed on Charlie's
qubit in $\{|a\rangle,|b\rangle\}$ basis.
On Charlie's announcement, Alice and Bob come to know
with certainty the Bell states they share, which
enables them to employ the conventional teleportation scheme to transmit unknown
quantum states to each other with the help of classical communication and
local operations summarized in Table \ref{tab:table1}. A successful protocol of BCST requires Charlie to 
control both directions of quantum communication. Incidentally, a few proposals for BCST in the past had Charlie's
control limited to only one direction \cite{li2013bidirectional,li2013cqsdc} (for details see \cite{thapliyal2015applications}).

\begin{table}
\begin{centering}
\begin{tabular}{|c|>{\centering}p{1.5cm}|>{\centering}p{1.5cm}|>{\centering}p{1.5cm}|>{\centering}p{1.5cm}|}
\hline 
 & \multicolumn{4}{c|}{Initial state shared by Alice and Bob}\tabularnewline
\cline{2-5} 
SMO  & $|\psi^{+}\rangle$  & $|\psi^{-}\rangle$  & $|\phi^{+}\rangle$  & $|\phi^{-}\rangle$\tabularnewline
\cline{2-5} 
 & \multicolumn{4}{c|}{Receiver's operation}\tabularnewline
\hline 
00  & $I$  & $Z$  & $X$  & $iY$\tabularnewline
\hline 
01  & $X$  & $iY$  & $I$  & $Z$\tabularnewline
\hline 
10  & $Z$  & $I$  & $iY$  & $X$\tabularnewline
\hline 
11  & $iY$  & $X$ & $Z$  & $I$\tabularnewline
\hline 
\end{tabular}
\par\end{centering}

\protect\caption[Unitary operations performed by the receiver in the
quantum teleportation]{\label{tab:table1} Unitary operations performed by the receiver in the
quantum teleportation. Here, SMO stands for the sender's measurement outcome.}
\end{table}

The BCST scheme 
can also be achieved using more than five-qubit states
if $|a\rangle$ and $|b\rangle$ are chosen as multi-qubit states \cite{duan2014bidirectional6,fu2014general,chen2015bidirectional,yan2013bidirectional,
duan2014bidirectional7}. 
Motivated by these schemes for BCST using different quantum channels, we have extended the general form (\ref{eq:the 5-qubit state}) to provide mathematical structure of quantum states suitable as quantum channels to accomplish the task of BCST.
Further, both experimental generation and maintenance of a six- or seven-qubit
state are challenging tasks at the moment. For instance, in Chapter \ref{Coupler}, we observed that using the nonlinear optical couplers it is possible to generate two-mode entanglement while the existence of three-mode entanglement could not be established. This sets our motivation 
to attempt to minimize the required quantum resources for a BCST scheme by performing it solely using Bell states. 
We would like to note that in addition to the schemes for BCST, in the recent times, a few protocols for a set of other controlled
quantum communication tasks, i.e., CQD \cite{dong2008controlled,xia2006quantum} and CDSQC \cite{li2013cqsdc}, using multi-qubit states have also been proposed. 
In what follows, we will also establish the feasibility of 
designing Bell-state-based schemes for these tasks.

Before we proceed further, it would be apt to note that here we employ the idea of quantum
cryptographic switch, introduced in \cite{srinatha2014quantum} in a different context, to design
various controlled quantum communication schemes with minimal quantum resources. Specifically, the schemes can be performed solely using Bell
states by implementing PoP technique
introduced in 2003 \cite{deng2003controlled}. This technique has been further used in
many protocols (\cite{shukla2012beyond,shukla2013improved,banerjee2012maximally,yadav2014two}
and references therein). Let us briefly introduce the concept of quantum cryptographic
switch, which describes a situation where the amount
of information accessible to a receiver 
can be controlled by a supervisor to a continuously varying degree. Note that the sender and receiver in all the controlled quantum communication schemes should be semi-honest
as otherwise, they can set up a quantum channel between them to ignore the supervisor. To circumvent such possibilities
the sender and receiver may be assumed semi-honest
(a user who follows the protocol but tries to cheat the controller).
Incidentally, the existing
controlled quantum communication protocols did not specify this requirement. In the following section, we report a general structure possessed by the quantum channels suitable for BCST scheme.

\section{A general method for selecting a quantum channel for bidirectional controlled state teleportation\label{sec:The-Condition-of}}

Further extending Eq.
(\ref{eq:the 5-qubit state})  considering $|a\rangle$ and $|b\rangle$ as multi-qubit states, 
the general structure of quantum channels for BCST can be obtained as
\begin{equation}
|\psi\rangle=\sum_{m=1}^{n}\frac{1}{\sqrt{n}}\left(\left(|\psi_{i}\rangle|\psi_{j}\rangle\right)_{m}|a_{m}\rangle\right),\label{eq:the multi-qubit state}
\end{equation}
where $|a_{m}\rangle$ represent $n$ mutually orthogonal $l$-qubit states such that 
$2^{l}\geq n\geq2$. Here,  both $|\psi_{i}\rangle$ and $|\psi_{j}\rangle$
are multi-qubit maximally
(nonmaximally) entangled states to be used for performing perfect (probabilistic)
teleportation. It is important that $\left(|\psi_{i}\rangle|\psi_{j}\rangle\right)_{m}=\left(|\psi_{i}\rangle|\psi_{j}\rangle\right)_{m^{\prime}}{\rm \, iff}\, m=m^{\prime}$.
One can also observe for $p$-qubit entangled
states $|\psi_{i}\rangle$ and $|\psi_{j}\rangle$, $n\leq\left(2^{p}\right)^{2}.$

In what follows, we will discuss a systematic method for obtaining
quantum channel for BCST that satisfies all the aforementioned requirements as:
\begin{description}
\item [{Step~1:}] The $p$-qubit entangled states
may be chosen from the basis set $\left\{ |\psi_{i}\rangle:i\in\{1,2,\ldots,2^{p}\}\right\} $
to construct a $2^{p}\times2^{p}$ matrix $S$ such that
its $i$th row--$j$th
column element is $s_{ij}=|\psi_{i}\rangle|\psi_{j}\rangle$. Thus, 
\begin{equation}
S\equiv\left[\begin{array}{cccc}
|\psi_{1}\rangle|\psi_{1}\rangle & |\psi_{1}\rangle|\psi_{2}\rangle & \cdots & |\psi_{1}\rangle|\psi_{2^{p}}\rangle\\
|\psi_{2}\rangle|\psi_{1}\rangle & |\psi_{2}\rangle|\psi_{2}\rangle & \cdots & |\psi_{2}\rangle|\psi_{2^{p}}\rangle\\
\vdots & \vdots & \ddots & \vdots\\
|\psi_{2^{p}}\rangle|\psi_{1}\rangle & |\psi_{2^{p}}\rangle|\psi_{2}\rangle & \cdots & |\psi_{2^{p}}\rangle|\psi_{2^{p}}\rangle
\end{array}\right].\label{eq:Matrix S}
\end{equation}

\item [{Step~2:}] A suitable quantum state of the form (\ref{eq:the multi-qubit state})
can be constructed by choosing $n\geq2$ elements of $S$ (i.e., $\left(|\psi_{i}\rangle|\psi_{j}\rangle\right)_{m}$)
taking into consideration the following restrictions:

\begin{description}
\item [{Rule~1:}] All the $n$ elements cannot be chosen from the same row
or column of the matrix%
\footnote{
If all the elements are chosen from the same row/column
of $S$, then the obtained quantum channel 
will become separable, and the supervisor's control will remain in one direction only. %
} (\ref{eq:Matrix S}). 
\item [{Rule~2:}] One element cannot be chosen more than once%
\footnote{This condition is necessary to ensure the bijective mapping between Charlie's
measurement result and the reduced quantum channel between Alice and Bob.
Note that, the task cannot be accomplished in the absence of such unique mapping as the unitary operations to be applied by the receivers could not be determined.%
} as $\left(|\psi_{i}\rangle|\psi_{j}\rangle\right)_{m}=\left(|\psi_{i}\rangle|\psi_{j}\rangle\right)_{m^{\prime}}$ iff $m=m^{\prime}$.
\end{description}
\end{description}

Here, we refrain ourselves from discussing particular examples to elaborate the method discussed here. Interested readers may refer to the published version of this work \cite{thapliyal2015general}, where the quantum channels used in the existing BCST schemes are also summarized as the special cases of the general form proposed here.
Further, a quantitative analysis of the total number of possible
quantum states $(N_{s})$ possessing the form (\ref{eq:the multi-qubit state}) that
can be constructed from $S$ satisfying the above mentioned rules reveals 
\begin{equation}
N_{s}=\left\{ \begin{array}{ll}
\frac{2^{2p}!}{\left(2^{2p}-n\right)!} & {\rm for}\, n>2^{p}\\
2^{pn}\left(2^{pn}-2^{p+1}+1\right) & {\rm for}\, n\leq2^{p}
\end{array}\right.\label{eq:NS}
\end{equation}
for an $n$-th order subset of 
a basis set $\{|a_{m}\rangle\}$ of $p$-qubit entangled states. 
The obtained result (\ref{eq:NS}) may be used to know the possible number of
quantum channels, say for
$p=2$ (i.e., Bell basis), if
$n\leq4$, then $N_{s}=2^{2n}\left(2^{2n}-7\right)$ and consequently, the desired quantum states can be constructed
in 144, 3648, and 63744 ways
for $n=2,\,3,$ and 4, respectively. 
While performing
this quantitative analysis of $N_{s}$, we have not considered the following:
(i) the relative phases of the superposition in (\ref{eq:the multi-qubit state}),
(ii) possible permutations of $\{|a_{m}\rangle\}$, and (iii) the number
of ways in which the subset of order $n$ can be constructed. Inclusion
of these factors will certainly enhance the number of ways a quantum channel for 
BCST schemes may be chosen which we are not intending to obtain here. We rather want to show that the BCST schemes proposed so far use only a very small subset of the quantum states able to perform the task. For instance, inclusion of the factors (ii) and (iii) will
further increase the obtained number of useful states by $\frac{2^{l}!}{\left(2^{l}-n\right)!}$
times for an arbitrary value of $n$, where $l$ qubits of the quantum channel are hold by Charlie. 

Thus, our discussion establishes that the systematic
way introduced here provides infinitely many useful quantum channels for BCST, but in view of the results obtained in Chapters \ref{Coupler}-\ref{Tomogram}, the requirement of a BCST scheme with a smaller number of entangled qubits remains unanswered. In what follows, we propose a BCST scheme that minimizes the number of entangled qubits as it can be performed with only two-qubit entanglement.

\section{Bidirectional controlled teleportation using Bell states\label{sec:Bidirectional-teleportation-using}}

Our BCST scheme with semi-honest Alice and Bob, and supervisor Charlie works as follows:
\begin{enumerate}
\item Charlie arranges all the qubits, initially prepared in $2n$ Bell states with  $n\geq2$, in four ordered sequences. The choice of each Bell state is purely random%
\footnote{Charlie can use a quantum random number generator to generate a large sequence of 0 and 1 to decide the choice of initial Bell state. For instance, he can prepare the first Bell state as $|\psi^{+}\rangle,\,|\psi^{-}\rangle,\,|\phi^{+}\rangle,$
and $|\phi^{-}\rangle$ if the first two bits in the generated string are 00, 01, 10,
and 11, respectively.%
}). The ordered sequences will be: 

\begin{enumerate}
\item A sequence consisting of only the first qubits of the first $n$ Bell states:
$P_{A_{1}}=\left[p_{1}\left(t_{A}\right),\right.p_{2}\left(t_{A}\right),$ $\ldots,\left.p_{n}\left(t_{A}\right)\right]$, 
\item A sequence consisting of only the first qubits of the last $n$ Bell states:
$P_{A_{2}}=\left[p_{n+1}\left(t_{A}\right)\right.,$ $p_{n+2}\left(t_{A}\right),\ldots,\left.p_{2n}\left(t_{A}\right)\right]$,
\item A sequence consisting of only the second qubits of the first $n$ Bell states:
$P_{B_{1}}=\left[p_{1}\left(t_{B}\right)\right.,$ $p_{2}\left(t_{B}\right),\ldots,\left.p_{n}\left(t_{B}\right)\right]$,
\item A sequence consisting of only the second qubits of the last $n$ Bell states:
$P_{B_{2}}=\left[p_{n+1}\left(t_{B}\right)\right.,$ $p_{n+2}\left(t_{B}\right),\ldots,\left.p_{2n}\left(t_{B}\right)\right]$.
\end{enumerate}

We have denoted the order of a particle
pair $p_{i}=\{t_{A}^{i},t_{B}^{i}\}$, prepared in the Bell state, in the subscript $1,2,\ldots,2n$. 

\item Charlie sends $P_{A_{1}}$
and $P_{A_{2}}$ to Alice, whereas applies
 $n$-qubit permutation operators $\Pi_{n_{1}}$ and $\Pi_{n_{2}}$
on the sequences $P_{B_{1}}$ and $P_{B_{2}}$ to generate two new sequences $P_{B_{i}}^{\prime}=\Pi_{n_{i}}P_{B_{i}}$
with $i\in\left\{ 1,2\right\} $ to send to Bob.
As Charlie withholds the information of permutation operators, only Charlie is aware of the actual order. \\
The first (last) $n$ Bell states are intended to be used for teleporting Alice's (Bob's) qubit. 
\item The receipt of the qubits from Charlie makes the senders able to teleport their unknown qubits $|\psi_{A_{j}}\rangle=\alpha_{A_{j}}|0\rangle+\beta_{A_{j}}|1\rangle$ and
$|\psi_{B_{j}}\rangle=\alpha_{B_{j}}|0\rangle+\beta_{B_{j}}|1\rangle,$ where $|\alpha_{x}|^{2}+|\beta_{x}|^{2}=1.$
For which Alice (Bob) can use the standard teleportation scheme to entangle
her (his) unknown qubit $|\psi_{A_{j}}\rangle$ ($|\psi_{B_{j}}\rangle$)
with $p_{j}(t_{A})$ ($p_{n+j}^{\prime}(t_{B})$) before measuring her (his) qubits in the computational basis and announcing
the measurement result. \\
It is imperative to discuss here that although Alice and Bob possess all the qubits of the quantum channel shared between them, they cannot circumvent Charlie's control due to their inability to find the choices of the initial Bell state and permutation operator even after their collusion. 
Precisely, Bob will not be able to reproduce the state teleported by Alice using her measurement outcome as he is unaware of the qubit entangled with Alice's qubit and the initial Bell state.
Incidentally, all
the existing BCST schemes using quantum channels possessing (\ref{eq:the 5-qubit state}) or (\ref{eq:the multi-qubit state}) form are vulnerable to such a collusion
between Alice and Bob. 
\item Charlie discloses the
initially prepared Bell state and the exact sequence $\Pi_{n_{1}}$
$\left(\Pi_{n_{2}}\right)$ if he wishes to allow Bob (Alice) to accomplish the teleportion of qubit sent by Alice (Bob). 
\item Using the information of the initial Bell states and the exact sequence, both Alice and Bob can accomplish the BCST task by applying appropriate unitary operations 
as summarized in Table \ref{tab:table1}. 
\end{enumerate}
The proposed protocol has numerous advantages which provide it an edge over the existing proposals of
BCST. Primary advantage is the requirement of minimal quantum resource, i.e., two-qubit entanglement, which increases the possibility of an experimental implementation of the task. Further, the scheme provides more control to Charlie, i.e., he may choose to disclose only one of the permutation operators, say $\Pi_{n_{1}}$, if he wishes only Alice to Bob teleportation to be accomplished.
Interestingly, one may observe that such directional control
was absent in the existing BCST protocols as their exists a bijective mapping between the Charlie's measurement outcome in $\left\{ |a\rangle,|b\rangle\right\} $
basis and the reduced state shared
by them. Additionally, Charlie possesses no control over the fidelity of the quantum state reconstructed by the receiver(s) in the ideal condition, whereas our proposal based on the idea of quantum cryptographic
switch allows Charlie to control the maximum average fidelity that can be obtained in the BCST scheme implemented by Alice and Bob. 
To understand the significance of quantum cryptographic switch here, let us consider Charlie informs Bob that the parity-0 Bell states are
twice more likely the candidate for the first Bell state than the parity-1 Bell states, and equal parity states are equally likely.
Bob obtains a probability
distribution  $\left(\frac{1}{3},\frac{1}{3},\frac{1}{6},\frac{1}{6}\right)$ with an entropy of 1.92 bits. Thus, Charlie has revealed only
${\rm c}=0.08$ bits of information, and this will lead to indecisiveness at the Bob's end to choose appropriate unitary operation. Our BCST scheme not only reduces the required
quantum resources and provides better control to the supervisor, it is possible to extend the argument to design
a multi-controlled bidirectional state teleportation scheme where one direction of teleportation is controlled by Charlie1 and the other by Charlie2. 

\section{Other protocols of controlled quantum communication using Bell states
\label{sec:Other-protocols-of}}

There are various quantum cryptographic tasks (discussed in Section \ref{CryptSwitch-Intro}) that require entangled state \cite{nguyen2004quantum,shukla2013group,bennett1984quantum,ekert1991quantum,shukla2013improved,banerjee2012maximally,yadav2014two,bostrom2002deterministic,cai2004improving,deng2003two}. The PoP technique can be implemented to design the controlled versions of all such two-party 
quantum communication schemes that require only Bell states. Here, we will establish this strategy to obtain three party secure quantum communication schemes from corresponding two-party quantum cryptographic protocol. 

\subsection{Controlled quantum dialogue protocol of Ba An type \label{subsec:CQD}}

Let us begin with summarizing the Ba An's original two-party scheme of QD, which allows
both Alice and Bob to communicate simultaneously using a Bell state. The
scheme works as follows \cite{nguyen2004quantum}: 
\begin{description}
\item [{Step~1}] Bob prepares Bell state $|\phi^{+}\rangle^{\otimes n}:\,|\phi^{+}\rangle=\frac{|01\rangle+|10\rangle}{\sqrt{2}}$ to be used as quantum channel with the first (second) qubit as home (travel) qubit. Subsequently, he applies Pauli operations $I,\, X,\, iY,$ and $Z$
to encode his secret $00,\,01,\,10,$ and $11$, respectively, on the travel qubit before sending it to Alice.
\item [{Step~2}] Bob sends the encoded sequence of the second qubits 
to Alice, and Alice informs him the receipt of the travel qubit. 
\item [{Step~3}] Alice returns the travel qubit to Bob only after encoding her message using the encoding rule discussed in Step 1.
\item [{Step~4}] Alice announces the run to be either the message mode
 or control mode chosen randomly. Bob measures the Bell states and can decode Alice's secret in the message mode. Bob's announcement of his measurement outcome helps Alice to decode his message. In contrast, Alice discloses 
her encoding operation in the control mode, using which Bob performs eavesdropping check. 
\end{description}
The two-party protocol can be converted into an equivalent
three-party protocol with Charlie supervising execution of QD scheme between semi-honest users Alice and
Bob. Our
modified protocol with proper security measures works as follows:
\begin{description}
\item [{Step~1}] Charlie arranges all the first (second) qubits of $n$ copies of Bell state $|\phi^{+}\rangle$
prepared by him in the first (second) sequence $P_{B_{1}}$ ($P_{B_{2}}$). 
\item [{Step~2}] Charlie prepares a new sequence $P_{B_{1}}^{\prime}=\Pi_{n}P_{B_{1}}$ by applying an $n$-qubit permutation operator $\Pi_{n}$ 
on $P_{B_{1}}$. Finally, he
sends both $P_{B_{1}}^{\prime}$ and $P_{B_{2}}$ to Bob.
\item [{Step~3}] Bob keeps the qubits of $P_{B_{1}}^{\prime}$ as home qubits, whereas the qubits in $P_{B_{2}}$ are used 
as the travel qubits. It is pre-decided to encode secret message $00,\,01,\,10,$
and $11$ by using Pauli operations $I,\, X,\, iY,$ and $Z$, respectively. Thus, Bob encodes his secret on $P_{B_{2}}$, which transforms it to $Q_{B_{2}}.$
\item [{Step~4}] Bob also prepares $n$ single-qubit decoy states as $\otimes_{j=1}^{n}|P_{j}\rangle\,:|P_{j}\rangle\in\left\{|0\rangle,|1\rangle,|+\rangle,|-\rangle\right\}$ 
to insert randomly in $Q_{B_{2}}$ to obtain
an enlarged sequence $R_{B_{2}}$ to send to Alice. Subsequently, he also 
confirms the receipt of all the qubits by Alice. 
\item [{Step~5}] Thereafter, Bob discloses the positions of randomly inserted decoy qubits in $R_{B_{2}}$ and applies
BB84 subroutine%
\footnote{BB84 subroutine \cite{pathak2013elements} refers to the eavesdropping checking similar to the original BB84
protocol. The receiver 
selects randomly half of the qubits received by him to measure them randomly in the computational $\left\{ |0\rangle,|1\rangle\right\} $
or diagonal $\left\{ |+\rangle,|-\rangle\right\} $ basis. Subsequently, he announces the position of each measured qubit with its
result and
the basis used for that. The sender checks his results with that he/she had prepared to detect any
eavesdropping attempt.%
} with the help of Alice. Specifically, if the computed error rate exceeds (is below) the tolerable limit, then they abort (proceed with)
the communication. \\
Note that Eve's attempt for an intercept-resend attack will be detected in the eavesdropping check, and such attack
will not reveal her encoding information of Bob. 
\item [{Step~6}] Alice also encodes her message and sends $R_{B_{3}}$, the encoded sequence obtained after 
randomly inserting $n$ decoy qubits. 
\item [{Step~7}] On the successful receipt of $R_{B_{3}},$ Alice and Bob
perform eavesdropping check as in Step 5 and proceed only after confirming no such attempt has been made. 
\item [{Step~8}] Charlie discloses the exact sequence of $P_{B_{1}}$ when he desires Alice and Bob to accomplish the communication.
\item [{Step~9}] With the help of Charlie's information, Bob measures $n$ Bell pairs and announces the outcomes. Bob is aware of the
initial and final Bell states, and his own encoding operation, and thus
he can extract Alice's secret. Alice uses Bob's measurement results, knowledge of the initial states from Charlie,
and her own encoding operations to obtain message encoded by Bob. 
\end{description}
If Charlie withholds the information of the exact sequence, the semi-honest Alice and Bob will not be able to perform the QD protocol.
Therefore, the proposed protocol is a CQD protocol. We
can restrict only Alice to send
her message then the proposed protocol will transform to a ping-pong-type CDSQC protocol. Note that the original ping-pong protocol \cite{bostrom2002deterministic} is a QSDC scheme, but its controlled counterpart (and similarly controlled versions of other QSDC protocols) falls under CDSQC category as the classical communication of single-bit regarding the quantum channel is involved. Interestingly, the sender in all the QSDC protocols may be assumed transmitting only random sequence to be used as key, which is the desired task of a QKD protocol. Thus, the CDSQC protocol
reduced from the proposed CQD protocol is also able to give a protocol of
controlled QKD. A protocol of controlled
QKA can also
be obtained from the above mentioned protocol by considering that
Alice and Bob send random raw keys instead of the meaningful messages to compute the final key. 
We would also
like to mention if instead of 
the BB84 subroutine for eavesdropping checking, one implements
Goldenberg-Vaidman (GV) type subroutine (see \cite{shukla2012beyond} for detail)
using entangled decoy qubits, then all the cryptographic tasks mentioned above can be implemented solely using orthogonal states. This shows that the proposed protocols can also be modified to design corresponding orthogonal-state-based protocols, for which not many schemes are proposed.
In what follows, we analyze the performance of some of the controlled quantum communication schemes proposed here.

\section{Effect of noise \label{sec:Effect-of-noise}}

We have already discussed and established in Chapters \ref{Coupler}-\ref{Tomogram} that generation and maintenance of multi-partite entanglement is much difficult in comparison to biparatite entanglement. In view of that, the Bell-state-based BCST or CQD schemes have obvious advantage over the existing schemes. Therefore, here we rather aim to obtain the suitable parameters for achieving best performance when the travel qubits are subjected to noisy channel.
For such study we can assume that the qubits to be teleported remain unaffected by noise. As Charlie's qubits in multi-qubit-based BCST scheme (discussed in Section \ref{sec:The-Condition-of}) 
also remains independent of the channel noise, the performance of the BCST schemes in which Alice and Bob share a product of two Bell states after Charlie's measurement can be deduced from the present results. 

In Section \ref{ex}, we introduced the Kraus operators of AD and PD noise models.
Specifically, the set of Kraus operators for the AD channel are given in Eq. (\ref{eq:Kraus-damping}). Here, we have used $p=\left(1-\eta_{A}\right)$ to write  \cite{nielsen2010quantum}: 
\begin{equation}
E_{0}^{A}=\left[\begin{array}{cc}
1 & 0\\
0 & \sqrt{1-\eta_{A}}
\end{array}\right],\,\,\,\,\,\,\,\,\,\,\,\,\,\,\, E_{1}^{A}=\left[\begin{array}{cc}
0 & \sqrt{\eta_{A}}\\
0 & 0
\end{array}\right].\label{eq:Krauss-amp-damping-M}
\end{equation}
Similarly, the PD channel is described by the following set
of Kraus operators \cite{nielsen2010quantum}:
\begin{equation}
E_{0}^{P}=\sqrt{1-\eta_{P}}\left[\begin{array}{cc}
1 & 0\\
0 & 1
\end{array}\right],\,\,\,\,\,\,\,\,\,\,\,\,\,\,\, E_{1}^{P}=\sqrt{\eta_{P}}\left[\begin{array}{cc}
1 & 0\\
0 & 0
\end{array}\right],\,\,\,\,\,\,\,\,\,\,\,\,\,\,\, E_{2}^{P}=\sqrt{\eta_{P}}\left[\begin{array}{cc}
0 & 0\\
0 & 1
\end{array}\right].\label{eq:Krauss-phase-damping-M}
\end{equation}
Although we study an independent effect of AD and PD channels, a similar study over collective effect of both these channels in analogy of Ref. \cite{an2009finite} can also be performed with the method adopted here.

Without any loss of generality, we can assume the qubits to be teleported by Alice
(Sender1) and Bob (Sender2) are $|\zeta_{1}\rangle{}_{S_{1}^{\prime}}\equiv \sin\theta_{1}|0\rangle+\cos\theta_{1}\exp(i\phi_{1})|1\rangle,$
and $|\zeta_{2}\rangle{}_{S_{2}^{\prime}}\equiv \sin\theta_{2}|0\rangle+\cos\theta_{2}\exp(i\phi_{2})|1\rangle$. Also consider that Charlie prepares and shares two Bell states between Alice and
Bob as $|\psi\rangle_{S_{1}R_{1}S_{2}R_{2}}=|\psi_{1}\rangle_{S_{1}R_{1}}\otimes$ $|\psi_{2}\rangle_{S_{2}R_{2}}$
with $|\psi_{i}\rangle\in\left\{ |\psi^{+}\rangle,|\psi^{-}\rangle,|\phi^{+}\rangle,|\phi^{-}\rangle\right\} $.
Therefore, the composite state of the quantum channel and qubits to be teleported is $|\psi^{\prime}\rangle_{S_{1}R_{1}S_{2}R_{2}S_{1}^{\prime}S_{2}^{\prime}}=|\psi_{1}\rangle_{S_{1}R_{1}}\otimes|\psi_{2}\rangle_{S_{2}R_{2}}\otimes|\zeta_{1}\rangle_{S_{1}^{\prime}}\otimes|\zeta_{2}\rangle_{S_{2}^{\prime}},$
which can also be written after rearrangement
of the qubits as 
\[
\rho=|\psi\rangle_{S_{1}S_{1}^{\prime}R_{1}S_{2}S_{2}^{\prime}R_{2}}{}_{S_{1}S_{1}^{\prime}R_{1}S_{2}S_{2}^{\prime}R_{2}}\langle\psi|.
\]
Application of the Kraus operators (\ref{eq:Krauss-amp-damping-M})
or (\ref{eq:Krauss-phase-damping-M}) describing the noisy environment  transforms the initial state
$\rho$ to
\begin{equation}
\rho_{k}=\sum_{i,j}E_{i,S_{1}}^{k}\otimes I_{2,S_{1}^{\prime}}\otimes E_{j,R_{1}}^{k}\otimes E_{j,S_{2}}^{k}\otimes I_{2,S_{2}^{\prime}}\otimes E_{i,R_{2}}^{k}\rho\left(E_{i,S_{1}}^{k}\otimes I_{2,S_{1}^{\prime}}\otimes E_{j,R_{1}}^{k}\otimes E_{j,S_{2}}^{k}\otimes I_{2,S_{2}^{\prime}}\otimes E_{i,R_{2}}^{k}\right)^{\dagger}.\label{eq:noise-effected-density-matrix-1}
\end{equation}
Here, we have assumed for simplicity that both the qubits traveling from Charlie to Alice (similarly both the qubits traveling from Charlie to Bob) are affected by the same Kraus operator. To perform the BCST scheme, Alice (Bob) measures $S_{1}$ and $S_{1}^{\prime}$ ($S_{2}$
and $S_{2}^{\prime}$) qubits
in the computational basis after applying a CNOT operation with the control (target) on $S_{i}^{\prime}\, \left(S_{i}\right)$ followed by a hadamard gate on $S_{i}^{\prime}$.
For our discussion, we have assumed that such a measurement performed by both Alice and Bob would result in 
$|00\rangle$. This outcome can be selectively chosen using
the measurement operator 
\[
U=\left(|00\rangle_{S_{1}S_{1}^{\prime}S_{1}S_{1}^{\prime}}\langle00|\right)\otimes I_{2,R_{1}}\otimes\left(|00\rangle_{S_{2}S_{2}^{\prime}S_{2}S_{2}^{\prime}}\langle00|\right)\otimes I_{2,R_{2}}
\]
to yield 
\[
\rho_{k_{2}}=\frac{\rho_{k_{1}}}{{\rm Tr}\left(\rho_{k_{1}}\right)}.
\]
From which, the quantum state of the receivers' qubits $R_{1}$ and $R_{2}$ can be obtained 
by tracing
over the remaining qubits, i.e., 
\[
\rho_{k_{3}}={\rm Tr}{}_{S_{1}S_{1}^{\prime}S_{2}S_{2}^{\prime}}\left(\rho_{k_{2}}\right).
\]
The receivers have to use a Pauli operation to obtain the teleported qubits, which depends upon the initial Bell state prepared by Charlie and the measurement results of the senders. As the senders measurement outcomes are assumed to be $|00\rangle$, suppose Charlie prepares $|\psi^{+}\rangle^{\otimes2},$
the receivers would need not apply any operation.
We can write the combined teleported state in the ideal condition as
$|T\rangle_{R_{1}R_{2}}=|\zeta_{1}\rangle{}_{S_{1}^{\prime}}\otimes|\zeta_{2}\rangle{}_{S_{2}^{\prime}}.$
The performance of the proposed BCST scheme can be quantified in terms of fidelity of the quantum state teleported in the ideal condition ($|T\rangle_{R_{1}R_{2}}$) with that in the noisy situation ($\rho_{k{\rm ,out}}$), which is computed as 
\begin{equation}
F=\langle T|\rho_{k,{\rm ou}t}|T\rangle.\label{eq:fidelity}
\end{equation}
Note that the definition of fidelity used here is the square of the conventional one.

Using this method, we obtained the teleported states in AD and PD channels as 
\begin{equation}
\begin{array}{lcl}
\rho_{A,{\rm out}} & = & N_{A}\left(\begin{array}{cccc}
\frac{\left(1+\eta_{A}\right)^{4}C_{\theta_{1}}^{2}C_{\theta_{2}}^{2}}{(1-\eta_{A})^{4}} & \frac{C_{\theta_{1}}^{2}S_{2\theta_{2}}{\rm e}^{-i\phi_{2}}}{2(1-\eta_{A})^{3}} & \frac{S_{2\theta_{1}}C_{\theta_{2}}^{2}{\rm e}^{-i\phi_{1}}}{2(1-\eta_{A})^{3}} & \frac{S_{2\theta_{1}}S_{2\theta_{2}}{\rm e}^{-i\phi_{12}}}{4(1-\eta_{A})^{2}}\\
\frac{C_{\theta_{1}}^{2}S_{2\theta_{2}}{\rm e}^{i\phi_{2}}}{2(1-\eta_{A})^{3}} & \frac{\left\{ C_{\theta_{1}}^{2}S_{\theta_{2}}^{2}+\eta_{A}^{2}S_{\theta_{1}}^{2}C_{\theta_{2}}^{2}\right\} }{(1-\eta_{A})^{2}} & \frac{S_{2\theta_{1}}S_{2\theta_{2}}{\rm e}^{-i\Delta\phi}}{4(1-\eta_{A})^{2}} & \frac{S_{2\theta_{1}}S_{\theta_{2}}^{2}{\rm e}^{-i\phi_{1}}}{2(1-\eta_{A})}\\
\frac{S_{2\theta_{1}}C_{\theta_{2}}^{2}{\rm e}^{i\phi_{1}}}{2(1-\eta_{A})^{3}} & \frac{S_{2\theta_{1}}S_{2\theta_{2}}{\rm e}^{i\Delta\phi}}{4(1-\eta_{A})^{2}} & \frac{\left\{ \eta_{A}^{2}C_{\theta_{1}}^{2}S_{\theta_{2}}^{2}+S_{\theta_{1}}^{2}C_{\theta_{2}}^{2}\right\} }{(1-\eta_{A})^{2}} & \frac{S_{\theta_{1}}^{2}S_{2\theta_{2}}{\rm e}^{-i\phi_{2}}}{2(1-\eta_{A})}\\
\frac{S_{2\theta_{1}}S_{2\theta_{2}}{\rm e}^{i\phi_{12}}}{4(1-\eta_{A})^{2}} & \frac{S_{2\theta_{1}}S_{\theta_{2}}^{2}{\rm e}^{i\phi_{1}}}{2(1-\eta_{A})} & \frac{S_{\theta_{1}}^{2}S_{2\theta_{2}}{\rm e}^{i\phi_{2}}}{2(1-\eta_{A})} & S_{\theta_{1}}^{2}S_{\theta_{2}}^{2}
\end{array}\right)\end{array}\label{eq:rhoAout}
\end{equation}
 and 
\begin{equation}
\begin{array}{lcl}
\rho_{P,{\rm out}} & = & N_{P}\left(\begin{array}{cccc}
P_{11}C_{\theta_{1}}^{2}C_{\theta_{2}}^{2} & 2C_{\theta_{1}}^{2}S_{2\theta_{2}}{\rm e}^{-i\phi_{2}} & 2S_{2\theta_{1}}C_{\theta_{2}}^{2}{\rm e}^{-i\phi_{1}} & S_{2\theta_{1}}S_{2\theta_{2}}{\rm e}^{-i\phi_{12}}\\
2C_{\theta_{1}}^{2}S_{2\theta_{2}}{\rm e}^{i\phi_{2}} & 4C_{\theta_{1}}^{2}S_{\theta_{2}}^{2} & S_{2\theta_{1}}S_{2\theta_{2}}{\rm e}^{-i\Delta\phi} & 2S_{2\theta_{1}}S_{\theta_{2}}^{2}{\rm e}^{-i\phi_{1}}\\
2S_{2\theta_{1}}C_{\theta_{2}}^{2}{\rm e}^{i\phi_{1}} & S_{2\theta_{1}}S_{2\theta_{2}}{\rm e}^{i\Delta\phi} & 4S_{\theta_{1}}^{2}C_{\theta_{2}}^{2} & 2S_{\theta_{1}}^{2}S_{2\theta_{2}}{\rm e}^{-i\phi_{2}}\\
S_{2\theta_{1}}S_{2\theta_{2}}{\rm e}^{i\phi_{12}} & 2S_{2\theta_{1}}S_{\theta_{2}}^{2}{\rm e}^{i\phi_{1}} & 2S_{\theta_{1}}^{2}S_{2\theta_{2}}{\rm e}^{i\phi_{2}} & P_{11}S_{\theta_{1}}^{2}S_{\theta_{2}}^{2}
\end{array}\right),\end{array}\label{eq:rhoPout}
\end{equation}
respectively. Here, $C_{x}=\cos x,$ $S_{x}=\sin x,$ $\phi_{12}=\phi_{1}+\phi_{2},$
$\Delta\phi=(\phi_{1}-\phi_{2})$, $P_{11}=\frac{4\left(1-2\eta_{P}+2\eta_{P}^{2}\right)^{2}}{(1-\eta_{P})^{4}},$
$N_{A}=\frac{4(1-\eta_{A})^{4}}{2\left[\left(2-4\eta_{A}+5\eta_{A}^{2}-4\eta_{A}^{3}+2\eta_{A}^{4}\right)+\eta_{A}\left(2-3\eta_{A}+2\eta_{A}^{2}\right)\left\{ \cos2\theta_{1}+\cos2\theta_{2}\right\} +\eta_{A}^{2}\cos2\theta_{1}\cos2\theta_{2}\right]},$
 and
\[
N_{P}=\frac{(1-\eta_{P})^{4}}{2\left[\left(2-8\eta_{P}+14\eta_{P}^{2}-12\eta_{P}^{3}+5\eta_{P}^{4}\right)+2\eta_{P}^{2}\left(2-4\eta_{P}+3\eta_{P}^{2}\right)\cos2\theta_{1}\cos\theta_{2}\right]}.\]
The fidelity
of the quantum states teleported in the BCST scheme over AD channel calculated with the help of Eqs. (\ref{eq:fidelity}) and (\ref{eq:rhoAout}) is
\begin{equation}
\begin{array}{lcl}
F_{{\rm AD}} & = & \frac{1}{16\left(2-4\eta_{A}+5\eta_{A}^{2}-4\eta_{A}^{3}+2\eta_{A}^{4}+\eta_{A}^{2}\cos2\theta_{1}\cos2\theta_{2}+\eta_{A}\left(2-3\eta_{A}+2\eta_{A}^{2}\right)\left(\cos2\theta_{1}+\cos2\theta_{2}\right)\right)}\left[32-164\eta_{A}+57\eta_{A}^{2}\right.\\
 & - & 26\eta_{A}^{3}+10\eta_{A}^{4}+\eta_{A}\left(34-51\eta_{A}+30\eta_{A}^{2}\right)\left(\cos2\theta_{1}+\cos2\theta_{2}\right)+\eta_{A}^{2}\left(3-2\eta_{A}+2\eta_{A}^{2}\right)\\
 & \times & \left(\cos4\theta_{1}+\cos4\theta_{2}\right)+4\eta_{A}^{3}\left(3-2\eta_{A}+2\eta_{A}^{2}\right)\left(\cos2\theta_{1}\cos4\theta_{2}+\cos4\theta_{1}\cos2\theta_{2}\right)\\
 & + & \left.16\eta_{A}^{2}\left(2-2\eta_{A}+\eta_{A}^{2}\right)\cos2\theta_{1}\cos2\theta_{2}+\eta_{A}^{2}\left(1-2\eta_{A}+2\eta_{A}^{2}\right)\cos4\theta_{1}\cos4\theta_{2}\right].
\end{array}\label{eq:fidelity-Amp-damp-probab}
\end{equation}
Similarly, the analytic expression of fidelity for the quantum state teleported using the proposed
BCST scheme calculated using Eqs. (\ref{eq:fidelity}) and (\ref{eq:rhoPout}) is
\begin{equation}
\begin{array}{lcl}
F_{{\rm PD}} & = & \frac{32-128\eta_{P}+210\eta_{P}^{2}-164\eta_{P}^{3}+59\eta_{P}^{4}+\eta_{P}^{2}\left\{ 2-4\eta_{P}+3\eta_{P}^{2}\right\} \left(16\cos2\theta_{1}\cos2\theta_{2}+\cos4\theta_{1}\cos4\theta_{2}+3\left(\cos4\theta_{1}+\cos4\theta_{2}\right)\right)}{16\left(2-8\eta_{P}+14\eta_{P}^{2}-12\eta_{P}^{3}+5\eta_{P}^{4}+\eta_{P}^{2}\left\{ 2-4\eta_{P}+3\eta_{P}^{2}\right\} \cos2\theta_{1}\cos2\theta_{2}\right)}.\end{array}\label{eq:fidelity-phase-damp-probab}
\end{equation}

A similar study is performed for the proposed CQD protocol in the AD and PD channels taking into consideration that the obatined fidelity is expected to depend upon the quantum channel prepared by Charlie and message encoded by
Alice and Bob. We have summarized the expressions
for fidelity in all possible cases in Table \ref{tab:AD-fidelity-CQD}. 

\begin{table}
\begin{centering}
\begin{tabular}{|>{\centering}p{1.cm}|>{\centering}p{1.7cm}|>{\centering}p{1.7cm}|>{\centering}p{4cm}|>{\centering}p{4.2cm}|}
\hline 
Initial state & Operations of Alice & Operations of Bob & Fidelity in AD channel $\left(F_{AD}^{\prime}\right)$ & Fidelity in PD channel $\left(F_{PD}^{\prime}\right)$\tabularnewline
\hline 
 & $X,\, iY$ & $X,\, iY$ & $F_{AD1}^{\prime}=\frac{4-8\eta_{A}+7\eta_{A}^{2}-2\eta_{A}^{3}+\eta_{A}^{4}}{4\left(1-\eta_{A}+\eta_{A}^{2}\right)}$ & \tabularnewline
\cline{2-4} 
$|\psi^{\pm}\rangle$  & $I,\, Z$ & $I,\, Z$ & $F_{AD2}^{\prime}=\frac{4-8\eta_{A}+9\eta_{A}^{2}-4\eta_{A}^{3}+\eta_{A}^{4}}{4\left(1-\eta_{A}+\eta_{A}^{2}\right)}$ & $F_{PD1}^{\prime}=\frac{2-6\eta_{P}+8\eta_{P}^{2}-4\eta_{P}^{3}+\eta_{P}^{4}}{2\left(1-2\eta_{P}+2\eta_{P}^{2}\right)}$\tabularnewline
\cline{2-4} 
 & $X,\, iY$ & $I,\, Z$ & $F_{AD3}^{\prime}=\frac{\left(1-\eta_{A}\right)^{2}\left(4+\eta_{A}^{2}\right)}{4\left(1-\eta_{A}+\eta_{A}^{2}\right)}$ & \tabularnewline
\cline{2-4} 
 & $I,\, Z$ & $X,\, iY$ & $F_{AD4}^{\prime}=\frac{4-8\eta_{A}+7\eta_{A}^{2}-4\eta_{A}^{3}+\eta_{A}^{4}}{4\left(1-\eta_{A}+\eta_{A}^{2}\right)}$ & \tabularnewline
\hline 
$|\phi^{\pm}\rangle$  & $I,\, X,\, iY,\, Z$ & $I,\, X,\, iY,\, Z$ & $F_{AD5}^{\prime}=\frac{1}{4}\left(2-\eta_{A}\right)^{2}$ & $F_{PD2}^{\prime}=\frac{1}{2}\left(2-2\eta_{P}+\eta_{P}^{2}\right)$\tabularnewline
\hline 
\end{tabular}
\par\end{centering}

\caption[Analytic expressions of the fidelity obtained for the
proposed CQD protocol over AD and PD channels]{\label{tab:AD-fidelity-CQD} The performance of the proposed CQD scheme is summarized in terms of the analytic expressions of fidelity over AD and PD channels. We have used $\prime$ in the superscript
to distinguish the present results from that of
the BCST protocols. Similar to the BCST scheme we have assumed that both the qubits of the Bell state are affected by same Kraus operator while traveling through the same channel.}
\end{table}

\begin{figure}
\begin{centering}
\includegraphics[angle=180,scale=0.5]{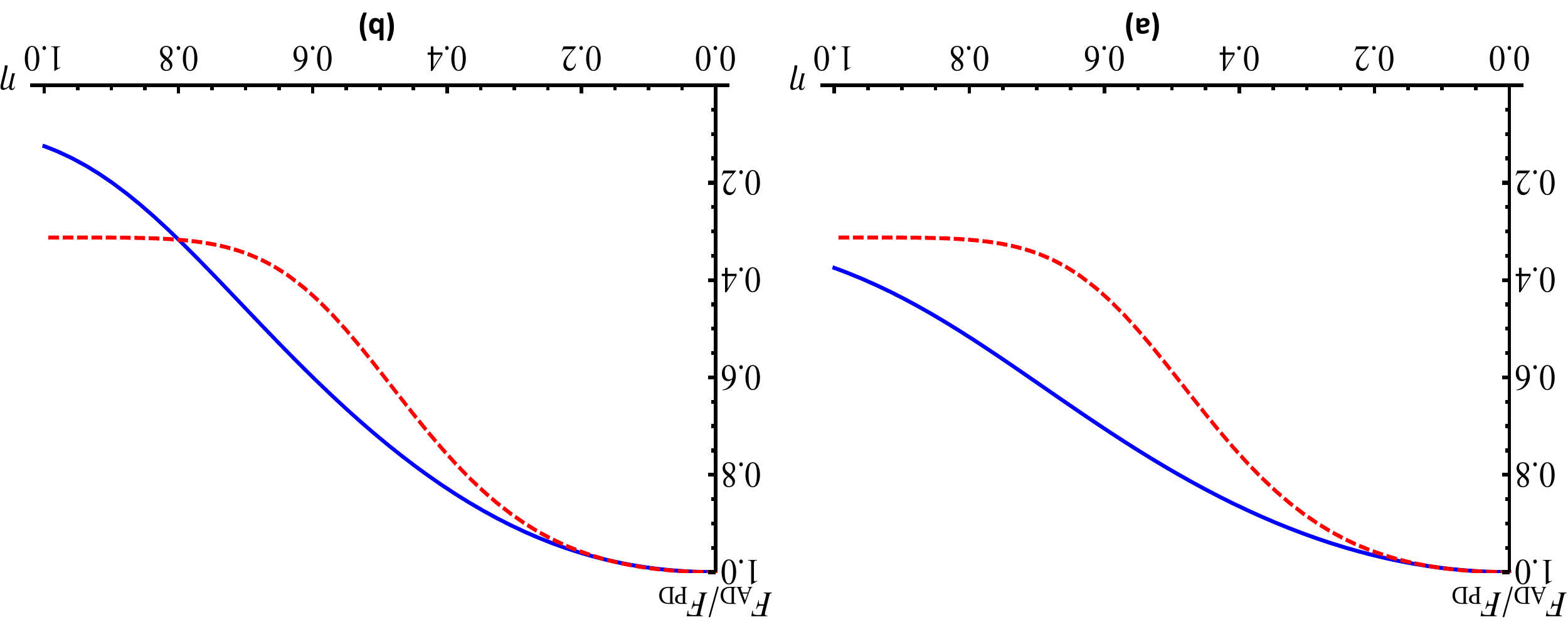}
\par\end{centering}

\caption[Effect of AD and PD noise for different values of decoherence rate on the
obtained fidelity in the BCST scheme]{\label{fig:Comparison-of-ampdampand-phasedamp} Variation of fidelity that could be attained in the Bell-state-based BCST scheme over AD and PD channels represented by the smooth (blue) and dashed
(red) lines, respectively. Two different values of the state parameters in
(a) and (b) are $\theta_{1}=\frac{\pi}{4},\,\theta_{2}=\frac{\pi}{6}$, and $\theta_{1}=\frac{\pi}{4},\,\theta_{2}=\frac{\pi}{3}$, respectively.}
\end{figure}

\begin{figure}[t]
\begin{centering}
\includegraphics[angle=0,scale=0.8]{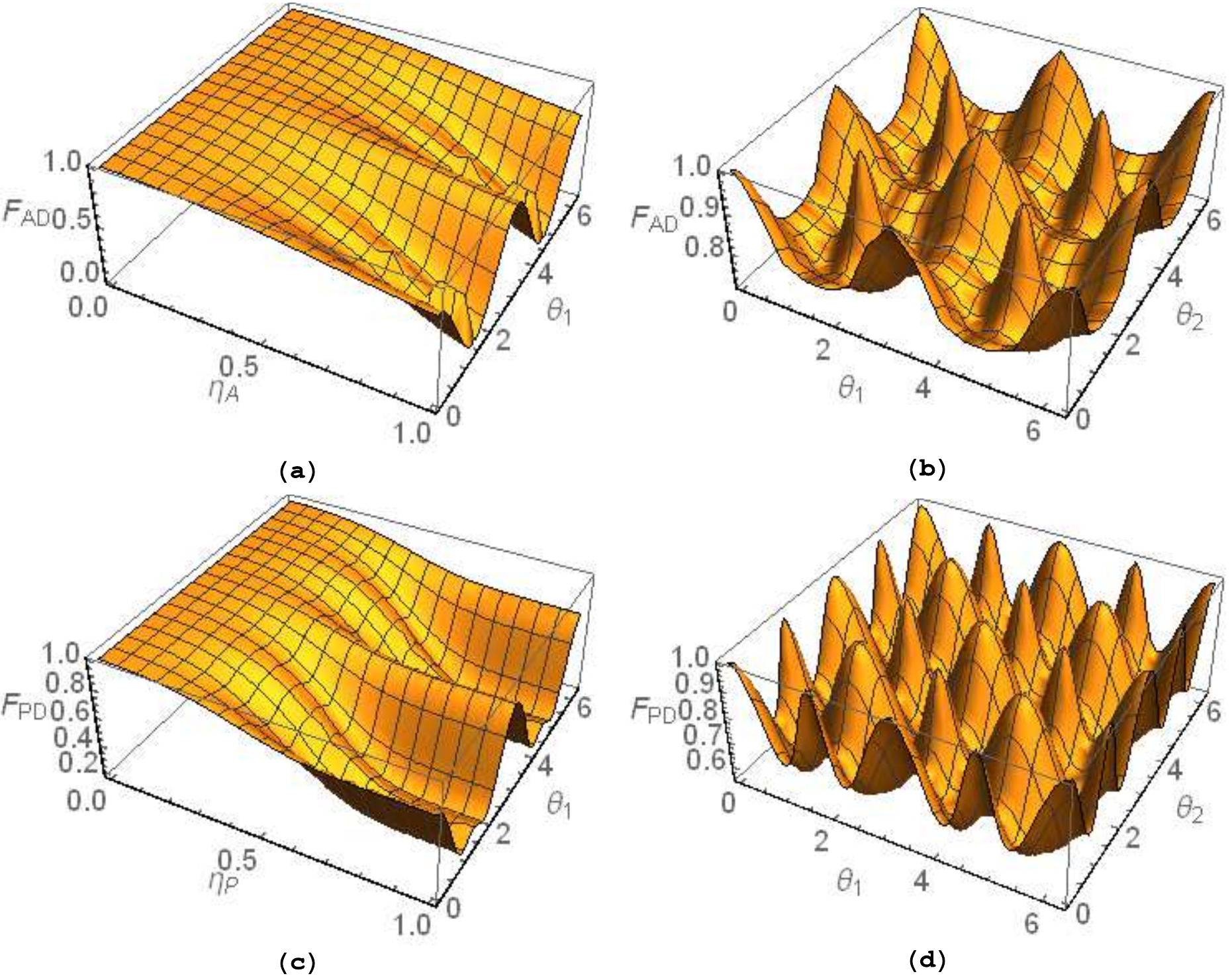} 
\par\end{centering}

\caption[Effect of AD and PD noise on the fidelity
of the Bell-state-based BCST protocol]{\label{fig:ampdampand-phasedamp} Performance of the Bell-state-based BCST scheme over AD (in (a) and (b)) and PD (in (c) and (d)) channels.
For (a) and
(c), we have chosen $\theta_{2}=\frac{\pi}{6}$; whereas
in (b) and (d), $\eta=0.5$.}
\end{figure}

The obtained fidelities $F_{{\rm AD}}$ and $F_{{\rm PD}}$ (in Eqs. (\ref{eq:fidelity-Amp-damp-probab})-(\ref{eq:fidelity-phase-damp-probab})) are functions of the decoherence rate $\eta_{k}$ and the amplitude
parameters $\theta_{i}$ and 
can be observed to be independent of the values of phase $\phi_{i}$ of the quantum states to be teleported.
This nature is consistent with similar studies performed on other teleportation based quantum communication schemes \cite{guan2014joint,sharma2015controlled}. Further, in case of the CQD protocol, all the expressions of the obtained fidelity, as reported in Table \ref{tab:AD-fidelity-CQD}, depend solely upon 
$\eta_{k}$. As the method adopted here to analyze the performance of quantum communication schemes is quite general, a similar study can be performed for new communication tasks/schemes or other noise models, like GAD or SGAD channels \cite{srinatha2014quantum,srikanth2008squeezed} as discussed in Chapter \ref{QDs} and Pauli-type
noise \cite{macchiavello2002entanglement}. 

\begin{figure}
\begin{centering}
\includegraphics[angle=180,scale=0.5]{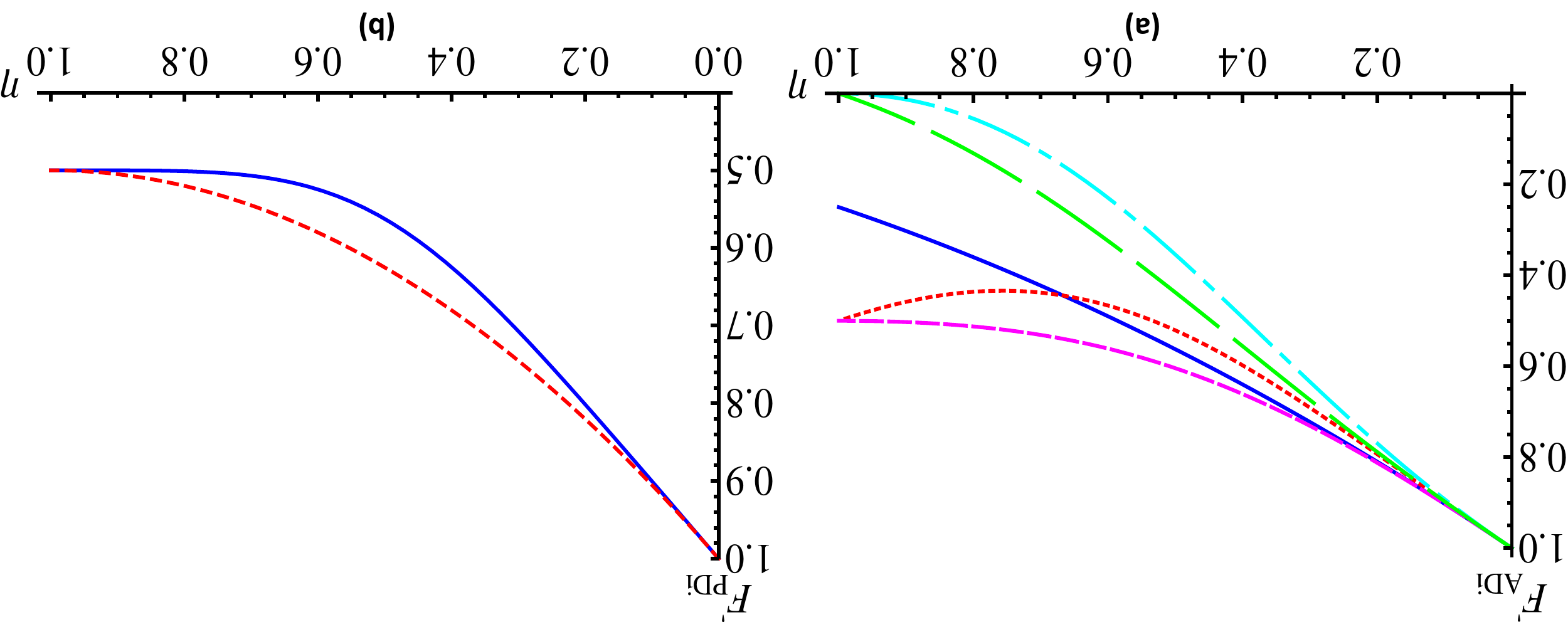}
\par\end{centering}

\caption[Effect of AD and PD noise
on the performance of the CQD protocol]{\label{fig:QD-Comparison-of-ampdampand-phasedamp-1} 
Variation of fidelity obtained in the proposed CQD protocol (summarized in Table \ref{tab:AD-fidelity-CQD}) under the effects of AD and PD noise
in (a) and (b), respectively. In (a), the smooth (blue)
line corresponds to the initial state $|\phi^{\pm}\rangle,$
whereas the dotted (red), small dashed (magenta), dot-dashed (cyan), and large
dashed (green) lines represent fidelity with the initial state $|\psi^{\pm}\rangle$, i.e., $F_{AD1}^{\prime},\, F_{AD2}^{\prime},\, F_{AD3}^{\prime},$
and $F_{AD4}^{\prime},$ respectively.
Similarly, the smooth (blue) and dashed (red) lines in (b) represent
the fidelity for the CQD protocol over PD channel implemented
with the initial state $|\psi^{\pm}\rangle$ and $|\phi^{\pm}\rangle,$
respectively.}
\end{figure}

The dependence of the obtained fidelity in the Bell-state-based BCST scheme on noise and state parameters is analyzed in Figures \ref{fig:Comparison-of-ampdampand-phasedamp}-\ref{fig:ampdampand-phasedamp}.
Specifically, Figure \ref{fig:Comparison-of-ampdampand-phasedamp} illustrates a comparative study of
the fidelity obtained over AD and PD channels for the BCST scheme with two different choices
of parameters of the quantum states to be teleported. In Figure \ref{fig:Comparison-of-ampdampand-phasedamp}
(a), 
higher fidelity for the teleported state can be observed for the AD channel when compared with corresponding PD channel with the same decoherence rate $\eta_{A}=\eta_{P}=\eta$. In contrast, a different choice of quantum states can be teleported with higher fidelity over the PD channel at higher values of the decoherence rate (cf.
Figure \ref{fig:Comparison-of-ampdampand-phasedamp} (b)). Therefore, the present study shows that the choice of the state to be teleported greatly affects the performance of the BCST scheme. We further analyzed this observation in Figure \ref{fig:ampdampand-phasedamp} in two cases: (1)
Variation of fidelity after fixing the state parameters of Bob's teleported state (i.e., all the quantum states with fixed $\theta_{2}$
and arbitrary values of the relative phase $\phi_{1}$ and $\phi_{2}$). 
(2) For independent choice of quantum states to be teleported by Alice and Bob while a fixed value of decoherence rate $\eta_{k}=\eta=0.5$ for both Charlie to Bob and Charlie to Alice quantum channels.
Interestingly, the symmetry of the task where both Alice and Bob are performing roles of the sender and receiver is clearly visible in the expressions of fidelity and Figure \ref{fig:ampdampand-phasedamp}. Also note that, we have used here fidelity as a quantitative measure of the performance of the proposed teleportation scheme. Additionally, average fidelity, minimum fidelity, and minimum assured fidelity are also used for the same purpose, which are considered in our recent proposal for teleporting an unknown single-qubit using non-orthogonal entangled states \cite{sisodia2017teleportation}.

A similar study for the performance of the CQD protocol over 
AD and PD channels
is summarized in Figure \ref{fig:QD-Comparison-of-ampdampand-phasedamp-1}.
Figure \ref{fig:QD-Comparison-of-ampdampand-phasedamp-1} (b) clearly establishes $|\phi^{\pm}\rangle$ as the preferred choice of the initial state for Charlie
in comparison to $|\psi^{\pm}\rangle$, when CQD is implemented 
over the PD channel. However, no such symmetry could be observed over the AD channel (cf. Figure \ref{fig:QD-Comparison-of-ampdampand-phasedamp-1}
(a)), where the obtained fidelity also depends on the encoding operations performed by Alice and Bob unlike PD noise as reported in Table \ref{tab:AD-fidelity-CQD}.

\section{Conclusions \label{sec:Conclusions-CryptSwitch}}

Several controlled quantum communication schemes (e.g., BCST, CDSQC, CQD) have been proposed in the recent past using
$n$-qubit ($n\geq3$) entanglement. 
Firstly, motivated by the 
various protocols for BCST proposed in the past using $m$-qubit entanglement
($m\in\{5,6,7\}$) we provided a systematic way to obtain new channels to implement the BCST scheme by giving 
the general structure for such quantum states. 
We have also shown that all the quantum channels used in the existing BCST schemes possess this form. However, this particular result is not included in the present chapter and can be found as Table 2 in Ref. \cite{thapliyal2015general}. This also provides infinitely many new possibilities of the quantum channels to be used for the implementation of BCST schemes depending upon their experimental feasibility. 

Thereafter, we have proposed a PoP-based protocol for BCST that uses solely Bell states and thus reduced the complexity of the required 
quantum resources. This is followed by a CQD scheme using PoP technique that can be reduced to other controlled quantum communication (e.g.,
controlled QKD, controlled
QKA, CDSQC) schemes. The performance of the proposed BCST and CQD schemes is also analyzed over AD and PD channels.
Also note that as remote state preparation is teleportation of a known qubit, the Bell-state-based BCST scheme proposed here can be reduced to 
a protocol of controlled bidirectional remote state preparation, which is performed using five-qubit quantum
states in the past \cite{sharma2015controlled,cao2013deterministic}.
A scheme for controlled bidirectional joint remote state preparation has also been proposed in the past using
the seven-qubit quantum states as \cite{sharma2015controlled}
\begin{equation}
|\psi\rangle=\frac{1}{\sqrt{2}}\left(|GHZ_{1}\rangle_{123}|GHZ_{2}\rangle_{456}|a\rangle_{7}\pm|GHZ_{3}\rangle_{123}|GHZ_{4}\rangle_{456}|b\rangle_{7}\right)\label{eq:7-qubit-3}
\end{equation}
with $GHZ_{1}\neq GHZ_{3}$ and $GHZ_{2}\neq GHZ_{4},$ and $GHZ_{i}$
with $i\in\left\{ 1,2,3,4\right\} $ is a GHZ state. Along the same line, a GHZ-based protocol for controlled bidirectional joint remote state preparation can be proposed, which will only require three-qubit entanglement. 

An efficient application of PoP technique allows us to reduce the number of entangled qubits required in the controlled quantum communication tasks, which enhances the possibilities of the experimental realization of the schemes for such tasks. Additionally, the proposed schemes have a few intrinsic advantages, such as 
(i) The controller can reveal the useful information to the receiver(s) in a continuously varying degree. (ii)
The controller possesses a directional control in the proposed BCST protocol by which he can allow only one out of two teleportation to be accomplished by revealing only corresponding information.
Such directional control was not present in the BCST schemes proposed in the past. In view of these
advantages, we expect the results reported in the present chapter (which are published in two articles \cite{thapliyal2015applications,thapliyal2015general}) 
will play an important role in the future developments in the theoretical and experimental research 
on the controlled quantum communication. To mention a few of them, we have recently shown controlled quantum communication to play an important role in providing solutions for the tasks having socioeconomic relevance, e.g., quantum solutions for voting \cite{thapliyal2017protocols},
private comparison \cite{thapliyal2016orthogonal}, e-commerce \cite{shukla2017semi} have been reported. In the next chapter, we will show that starting with the CQD scheme proposed in this chapter, schemes for almost all other quantum cryptographic tasks can be deduced.

\thispagestyle{empty}
\thispagestyle{plain}   
\cleardoublepage
\blankpage

\titlespacing*{\chapter}{0pt}{-50pt}{20pt}
\chapter{Quantum cryptography over non-Markovian channels} \label{NonMarkovian}

\section{Introduction}
In the previous chapter, we have already discussed a set of potential applications of nonclasscial (entangled) states in both insecure and secure quantum communication. Specifically, we have proposed an unconditionally secure CQD scheme and analyzed its performance over Markovian channels. We have also shown that various other controlled cryptographic schemes can be deduced from it.
In the present chapter, we would like to exploit yet another
interesting observation that if we
start with a scheme for CQD, we can
reduce it to almost all the schemes of secure quantum communication. This
point is discussed in detail in the forthcoming sections. Further, in the following sections of this chapter, we aim to study the effect of non-Markovian noise (introduced in Section \ref{N-Mark}) on the schemes for secure quantum communication.

Before we describe our findings, it would be apt to note that the feasibility of implementation of various quantum communication schemes,
when subjected to noisy environment, has been analyzed in the past.
In particular, the schemes of QKD \cite{sharma2016comparative}, QKA
\cite{sharma2016comparative}, CDSQC \cite{srinatha2014quantum}, QSDC \cite{sharma2016comparative},
CQD (in Chapter \ref{CryptSwitch}), QD \cite{sharma2016comparative}, asymmetric QD \cite{banerjee2017asymmetric},
among others, have been considered under 
the influence of both purely dephasing and damping
noises. Most of these investigations (cf. \cite{sharma2016comparative,srinatha2014quantum,banerjee2017asymmetric} and the works reported in the previous chapter) were restricted
to the domain of Markovian environments \cite{banerjee2007dynamics,srikanth2008squeezed}, though some attempts have been made to
study the effects of non-Markovian environments on quantum communication
schemes, such as teleportation \cite{yeo2010non,hao2012enhanced}, densecoding
\cite{yeo2010non}, and entanglement swapping \cite{jun2012non,jun2013non}.
The security of a QKD protocol has also been analyzed over
non-Markovian depolarizing channel \cite{huang2012study}. All these attempts
(except \cite{huang2012study}) to examine the usefulness of entangled
states under the influence of non-Markovian environments were restricted
to insecure quantum communication, where security is not required.
However, in the secure quantum communication protocols, the situation differs as it becomes relevant
to differentiate between the disturbance caused due to eavesdropping and the
effects of noise. 
This sets the motivation for this part of the present thesis work, where we
wish to analyze the exclusive effect of a non-Markovian noisy environment on the CQD scheme introduced in the previous chapter. We have set ourselves a task (i) to study the effect of non-Markovian noise on the fidelity of quantum states, done in detail for the CQD scheme, and (ii) the impact of the said noise on a range of cryptographic protocols, such as the CDSQC, QD, DSQC and QSDC protocols, that can be reduced from the CQD scheme. Here, it would be pertinent to mention that (ii) can be obtained from (i) as special cases and this provides the relevance to the present study.   

It is well-known that environment-induced decoherence sets a limit on the efficiency of any quantum-enhanced protocol \cite{bylicka2014non,wang2005fault,wang2004quantum,lo1999unconditional}. Here, we do this by studying the effect of the noisy channel on the protocols under consideration. Further, the channel is exposed to Eve, and in principle, she may try to hide herself as noise (i.e., try to hide her presence by attributing the error caused by her presence to the transmission noise considered here). Interestingly, such an effort can be detected as there exist methods to distinguish between noise and Eve. Specially, a method for distinction between transmission noise and Eve for single-photon-based QKD scheme was proposed \cite{brassard2000limitations} (discussed in detail in Section \ref{CryptSwitch-Intro}), and the same for entangled-state-based schemes of quantum cryptography was also introduced \cite{lo1999unconditional}. Thus, in brief, even when the channel is exposed to Eve, it is technically correct to concentrate on the effect of transmission noise on the efficiency of the scheme(s). However, we need to keep a specific possibility in mind: in what follows, we will see that exploiting the non-Markovianity, better performance (higher efficiency) can be attained compared with the corresponding Markovian environment. Exploiting this fact, Eve may replace a Markovian channel between Alice and Bob with a non-Markovian one and reduce the possibility of being detected. Thus, the tolerable error limit should be set by the legitimate parties considering this possibility into account. Further, to decide the tolerable error limit the users may require to characterize the channel. For example, the noise present in the AD channel may be characterized and corrected \cite{omkar2015characterization}. Further, characterization of quantum dynamics with reduced code length and an example of two-qubit noise have been investigated in the recent past in \cite{omkar2015quantum}.

In what follows, we would specifically consider  pure dephasing, damping and depolarizing interactions with a  non-Markovian 
reservoir. Though entanglement can be maintained for relatively
longer time due to dephasing non-Markovian interaction \cite{yu2010entanglement},
it can show revival under dissipative interactions \cite{bellomo2007non}. As we are essentially using entangled states and
entanglement revival could be an interesting feature to affect the
feasibility of quantum cryptographic schemes, we would like to address
the problem here.

Non-Markovian noise has been attracting a lot of interest from both quantum
optics and quantum information communities, theoretically as well as
experimentally. A paradigm for studying non-Markovian evolution is the quantum Brownian motion
\cite{grabert1988quantum,banerjee2003general,banerjee2000quantum}. Specifically, degradation of purity and nonclassicality
of Gaussian states have been studied under the effect of non-Markovian
channels \cite{ban2006decoherence}. Dynamics of entanglement has been discussed
in both discrete \cite{maniscalco2006non,bellomo2007non,paz2008dynamics,piilo2008non}
and continuous \cite{an2007non} variable channels. Recently, dynamics
of multipartite entanglement and its protection have been addressed
\cite{nourmandipour2016dynamics}. The additional problems due to non-Markovian noise
in quantum error correction \cite{novais2008hamiltonian} and dynamical decoupling
\cite{chen2007dynamical,shiokawa2007non} have also been discussed in the past.
The non-Markovianity was also characterized from an information theoretic
approach in terms of quantum Fisher information flow \cite{lu2010quantum}.
 A number of beautiful experiments depicting non-Markovian nature
of the system-reservoir interaction have also been performed \cite{liu2011experimental,xu2010experimental,orieux2014experimental}. 

The Kraus operators
of non-Markovian dissipative and dephasing noise models  are discussed in a concise manner (in Section \ref {sec:noise-models}).
In Section \ref{sec:effect-of-noise}, for the sake of introduction of the notation that has been used in the rest of this chapter we briefly report the CQD scheme (in Section \ref{sub:CQD}), which was already discussed in detail in Section \ref{subsec:CQD}.
Then, 
we study the effect of non-Markovian noise on the feasibility
of the CQD scheme. To quantify the effect of
noise, a distance-based measure known as fidelity (also used in the previous chapter) has been calculated. 
Next, we reduce the scheme of CQD to design a CDSQC protocol
(in Section \ref{sub:CDSQC}), a QD protocol (in Section \ref{sub:QD}),
a DSQC and a QSDC protocols (in Section \ref{sub:QSDC/DSQC}), a QKA
protocol (in Section \ref{sub:QKA}), and finally, two well-known QKD
protocols (in Section \ref{sub:QKD}). The QKD protocols discussed
here are well-known as BB84 \cite{bennett1984quantum} and BBM \cite{bennett1992quantum} protocols. The feasibility
of all these schemes under the action of non-Markovian channels is also analyzed.
Finally, we conclude the chapter in Section \ref{sec:Conclusion-NonMarkovian}.

\section{Non-Markovian noise models \label{sec:noise-models}}

 We briefly discuss below a few non-Markovian models that are subsequently used to study the performance of various quantum cryptographic schemes. 
The dynamics of a system interacting with its surroundings can be expressed
in the operator-sum representation (see \cite{caruso2014quantum} for a review) discussed in Section \ref{Kraus} as Eq. (\ref{eq:Kraus}). Here, we use this approach
to describe the dissipative and purely dephasing interactions with non-Markovian
environments. As mentioned in Section \ref{Mark}, the Kraus operators for the damping noise under non-Markovian
effects \cite{bellomo2007non} are given by Eq. (\ref{eq:Kraus-damping}),
where $p\equiv p\left(t\right)=\exp\left(-\Gamma t\right)\left\{ \cos\left(\frac{dt}{2}\right)+\frac{\Gamma}{d}\sin\left(\frac{dt}{2}\right)\right\} ^{2}$
with $d=\sqrt{2\gamma\Gamma-\Gamma^{2}}$. Here, $\Gamma$ is the
line width which depends on the reservoir correlation time $\tau_{r}\approx\Gamma^{-1}$;
and $\gamma$ is the coupling strength related to qubit relaxation time
$\tau_{s}\approx\gamma^{-1}$. In the domain of large reservoir correlation
time compared to the qubit relaxation time, memory effects come into
play. The memory effects are characteristic of non-Markovian nature
of dissipation. Interestingly, taking $p=\left(1-\eta\right)$, the results obtained
for AD noise under Markovian regime can be deduced,
with $\eta$ being the decoherence rate of the AD channel. 

Similarly, the Kraus operators for the purely dephasing non-Markovian noise \cite{yu2010entanglement} are given by Eq. (\ref{eq:Kraus-dephasing}),
where $p\equiv p\left(t\right)=\exp\left[-\frac{\gamma}{2}\left\{ t+\frac{1}{\Gamma}\left(\exp\left(-\Gamma t\right)-1\right)\right\} \right]$.
All the parameters have the same meaning as above. Similar to the case of dissipative noise, the effect
for the well-known PD channel in the Markovian regime can be obtained from the obtained result
by considering $p=\sqrt{1-\eta}$. In what
follows, we consider an independent environment for each qubit as
it travels through different channels; a similar assumption has been made in \cite{dajka2008non,cao2008non}.

Finally, a non-Markovian depolarizing channel can be described by
the Kraus  operators $K_{i}=\sqrt{\mathcal{P}_{i}}\sigma_{i}$,
where $\sigma_{4}\equiv I$ and $\sigma_{i}$s are the three Pauli
matrices. The $\mathcal{P}_{i}$s should remain positive to ensure
the complete positivity for all values of $\frac{\gamma_{j}}{\Gamma_{j}}$
and are given by \cite{daffer2004depolarizing} 
\[
\mathcal{P}_{1}=\frac{1}{4}\left[1+\Omega_{1}-\Omega_{2}-\Omega_{3}\right],
\]
\[
\mathcal{P}_{2}=\frac{1}{4}\left[1-\Omega_{1}+\Omega_{2}-\Omega_{3}\right],
\]
\[
\mathcal{P}_{3}=\frac{1}{4}\left[1-\Omega_{1}-\Omega_{2}+\Omega_{3}\right],
\]
and
\begin{equation}
\mathcal{P}_{4}=\frac{1}{4}\left[1+\Omega_{1}+\Omega_{2}+\Omega_{3}\right].\label{eq:depolar-par}
\end{equation}
Here, $\Omega_{i}=\exp\left(-\frac{\Gamma t}{2}\right)\left[\cos\left(\frac{\Gamma d_{i}t}{2}\right)+\frac{1}{d_{i}}\sin\left(\frac{\Gamma d_{i}t}{2}\right)\right]$
with $d_{i}=\sqrt{16\left(\frac{\gamma_{j}^{2}}{\Gamma_{j}^{2}}+\frac{\gamma_{k}^{2}}{\Gamma_{k}^{2}}\right)-1}$
for $i\neq j\neq k$ \cite{daffer2004depolarizing}. Further, $\gamma$ is the coupling
strength of the system, and $\Gamma$ is the noise bandwidth parameter.
It should be noted that the Markovian case can be deduced from the
above by taking $\Omega_{i}=\exp\left(-\frac{\gamma_{i}t}{2}\right)$
with $\gamma_{i}=\frac{4}{\Gamma}\left(\gamma_{j}^{2}+\gamma_{k}^{2}\right)$
for $i\neq j\neq k$ \cite{daffer2004depolarizing}. 

\section{Effect of non-Markovianity on the secure quantum communication schemes \label{sec:effect-of-noise}}
 
 In what follows, we consider the CQD scheme from the perspective of its implementation 
over the  above discussed non-Markovian channels. Further, the feasibility of a set of quantum cryptographic protocols, under the influence of non-Markovian channels, is deduced from the CQD scheme. For all the one-way schemes for quantum cryptography that are discussed here, we consider Alice as the sender and Bob as the receiver, unless stated otherwise, whereas Charlie is the third party supervising the protocol and referred to as the
controller. However, for two-way schemes (e.g., QD, CQD), both Alice and Bob are considered to play dual roles of the receiver and sender. 

In order to analyze the efficiency of quantum cryptographic protocols over non-Markovian channels, we have considered an independent environment for each separate channel between different parties. In the independent environment case, each qubit is coupled to its own independent environment, in contrast to the same reservoir in the collective decoherence case \cite{schlosshauer2007decoherence,banerjee2010dynamics}. This may also be viewed as a special case where no interaction has been assumed between the individual reservoirs of each qubit \cite{schlosshauer2007decoherence}. This assumption is valid for large spatial separation between the corresponding qubits when compared to the coherence length of the environment.  Naturally, independent noise model is adopted in a large number of recent studies on the effect of environment-induced decoherence on quantum-enhanced protocols (\cite{wang2004quantum,lo1999unconditional,schlosshauer2007decoherence,fortes2015fighting}, and references therein). Further, this assumption is frequently used in the context of quantum error correction \cite{schlosshauer2007decoherence} and quantum communication scenarios \cite{fortes2015fighting}. For instance, for teleportation of a single-qubit over Markovian noisy channels \cite{fortes2015fighting}; attempts have been made to improve the capacity of the quantum channel under the influence of non-Markovian environments \cite{bylicka2014non}. Additionally, from the choices of different $p_{i}$ (corresponding to different coupling strengths between the individual qubits and their reservoirs), as considered here (and in \cite{fortes2015fighting}), one can easily reduce to a special case, where individual environments are considered, but the coupling strengths between different qubits and their reservoirs are the same.

\subsection{Controlled quantum dialogue \label{sub:CQD}}

Let us start  with a three-party protocol for quantum cryptography, where
two parties (Alice and Bob) wish to communicate simultaneously under
the control of a third party (Charlie). In fact, as mentioned in the last chapter, all the controlled
communication protocols work under an assumption that all the parties
are semi-honest. In literature,  this is sometimes 
viewed as Alice and Bob lacking resources for state preparation, and consequently, they do not set up a quantum
channel between them; rather, they rely on Charlie to prepare it for them.

To begin with, we reintroduce the CQD scheme proposed in Chapter \ref{CryptSwitch} for the sake of notations used hereafter. Charlie prepares $n$ copies of a
Bell state and makes two strings $S_{A}$ and $S_{B}$ of all the
first and second qubits. Subsequently, he sends both the strings to
Bob, only after permuting $S_{B}$\footnote{Here, and in what follows, all the qubits traveling from one party
to another are sent in a secure manner, i.e., to send a sequence of $n$ travel qubits, an equal number of decoy
qubits  are inserted randomly in the original sequence of the travel qubits, and subsequently, these decoy qubits are measured to check the existence of 
eavesdropper(s). Various choices of decoy qubits and the corresponding
principles of security are discussed in \cite{sharma2016verification}.}. Both Bob and Alice encode their secrets on the qubits
in string $S_{A}$ and thus accomplish the task in hand with the help of Charlie (see Section \ref{subsec:CQD} for detail). Note that, if the choice of Bell state prepared by Charlie is made public, it leads to some leakage, which is often considered to be an inherent characteristic of the schemes for QD and its variants. However, such leakage can be circumvented if Charlie chooses the Bell state randomly and sends his choice to Alice and Bob by using a scheme of DSQC or QSDC \cite{banerjee2017asymmetric}.

Suppose Charlie started with the initial state $\rho=|\psi\rangle\langle\psi|$,
where $|\psi\rangle\in\left\{ |\psi^{\pm}\rangle,|\phi^{\pm}\rangle\right\} $.
The transformed density matrix over the noisy channel would become
\begin{equation}
\begin{array}{lcl}
\rho^{\prime} & = & \underset{i,j,k,l}{\sum}\left(K_{l}\left(p_{4}\right)\otimes I\right)U_{A_{n}}\left(K_{k}\left(p_{3}\right)\otimes I\right)U_{B_{n}}\left(K_{i}\left(p_{1}\right)\otimes K_{j}\left(p_{2}\right)\right)\rho\left(\left(K_{i}\left(p_{1}\right)\otimes K_{j}\left(p_{2}\right)\right)\right)^{\dagger}\\
 & \times & \left(\left(K_{l}\left(p_{4}\right)\otimes I\right)U_{A_{n}}\left(K_{k}\left(p_{3}\right)\otimes I\right)U_{B_{n}}\right)^{\dagger},
\end{array}\label{eq:transformed-rho}
\end{equation}
where $K_{i}$s are the Kraus operators for a specific kind of noise
discussed in the previous section, and $U_{j}$s are the Pauli operations
by Alice and Bob with $j\in\left\{ j_{00},j_{01},j_{10},j_{11}\right\} $
for $\left\{ I,X,iY,Z\right\} $. Here, we have used different values of $p_{i}$s corresponding to each operation of the Kraus operator 
(from Eq. (\ref{eq:Kraus-dephasing}), (\ref{eq:Kraus-damping}) or the depolarizing channel) on the initial quantum state as the coupling strength during 
various rounds of the quantum communication is assumed to be different.  It may be noted that the
summation in the right-hand side of Eq. (\ref{eq:transformed-rho})
ensures that the map is positive. Further, here, we have
assumed that the qubits not traveling through a quantum channel are
not affected by noise (as in Chapter \ref{CryptSwitch}). There are various distance-based measures
to quantify the effect of noise on the quantum state, such as trace
distance, fidelity, and the Bures distance \cite{miranowicz2015statistical}. The fidelity (given in Eq. (\ref{eq:fidelity})) of the transformed
density matrix with the quantum state in the ideal situation (i.e., in the absence of noise) would be \cite{caruso2014quantum} 
\begin{equation}
F=\langle\psi^{\prime}|\rho^{\prime}|\psi^{\prime}\rangle,\label{eq:fidelity-nm}
\end{equation}
where the expected quantum state $|\psi^{\prime}\rangle=U_{A_{n}}U_{B_{n}}|\psi\rangle$. Subsequently, average fidelity is obtained 
by averaging over all the possible encoding operations that Alice and
Bob are allowed to perform. Thus, the fidelity that we are discussing here, and
in the rest of the chapter, is the average fidelity. 

The fidelity of the quantum state transformed under the damping effect of the non-Markovian environment can be calculated using the standard definition of the fidelity (\ref{eq:fidelity-nm}), where the transformed density matrix (\ref{eq:transformed-rho}) has been obtained using the Kraus operators defined in Eq. (\ref{eq:Kraus-damping}). The analytic expression of the fidelity is 
\begin{equation}
F=\frac{1}{4}\left[1+2\sqrt{p_{1}p_{2}p_{3}p_{4}}+p_{1}p_{3}p_{4}\left(2p_{2}-1\right)+p_{3}p_{4}\left(1-p_{2}\right)\right],\label{eq:Damp-psi}
\end{equation}
when the initial quantum state prepared by Charlie is $|\psi^{\pm}\rangle$.
As the choice of initial state is solely a decision of Charlie, an independent
choice of the initial state, i.e., $|\phi^{\pm}\rangle$, would lead to the following expression of fidelity
\begin{equation}
F=\frac{1}{4}\left[1+2\sqrt{p_{1}p_{2}p_{3}p_{4}}+p_{1}p_{3}p_{4}+p_{3}p_{4}\left(p_{2}-1\right)\right].\label{eq:Damp-phi}
\end{equation}

 If the state prepared by Charlie were subjected to a non-Markovian dephasing noise, the fidelity would be 
\begin{equation}
F=\frac{1}{2}\left[1+p_{1}p_{2}p_{3}p_{4}\right],\label{eq:Deph}
\end{equation}
which has been calculated using the Kraus operators defined in Eq. (\ref{eq:Kraus-dephasing}) in the expression of the transformed density matrix (\ref{eq:transformed-rho})  and the fidelity (\ref{eq:fidelity-nm}). It is interesting to see that the obtained fidelity is independent
of the choice of the initial Bell state by Charlie. This  is also seen in analogous scenarios of the
Markovian dephasing noise in Chapter \ref{CryptSwitch} and Refs.
\cite{banerjee2017asymmetric,sharma2016comparative,shukla2016hierarchical,sharma2016verification} and references
therein. If we now consider that the system has evolved under the effect
of a depolarizing channel, characterized by the parameters of the Kraus operators defined in Eq. (\ref{eq:depolar-par}), then following the above prescription and using Eqs. (\ref{eq:depolar-par}) and (\ref{eq:transformed-rho}) in Eq. (\ref{eq:fidelity-nm}), the analytical expression 
for fidelity can be obtained as
\begin{equation}
F=\frac{1}{2}\left[1+\Omega_{1}^{4}+\Omega_{2}^{4}+\Omega_{3}^{4}\right].\label{eq:Depolarizing-CQD}
\end{equation}

It is interesting to observe the appearance of the fourth-order terms in all the fidelity expressions obtained over non-Markovian channels, a
signature of four noisy channels acting on the four different rounds of quantum
communication. It should be mentioned here that instead of  sending both
the strings to Bob, Charlie could have sent $S_{A}$ to Alice and
$S_{B}$ to Bob. Subsequently, Alice would have sent $S_{A}$ to Bob
after encoding her message and Bob would have encoded his message before
performing the measurement. The obtained fidelity expressions in this
case turn out to be the same as that of the CDSQC protocol, discussed in the
next subsection. The effect of noise in the case discussed here is
more than that in the case of CDSQC. Making use of this observation,  we analyze the scheme of CQD,
described in the previous chapter,  in detail  as the results obtained in the following subsections can be reduced from it.

Now, we will discuss the fidelities for different scenarios depicted in Eqs. (\ref{eq:Damp-psi})-(\ref{eq:Depolarizing-CQD}), for 
both Markovian and non-Markovian noises. The case of the non-Markovian damping/dephasing channel is also considered
for strong and weak coupling regimes. Specifically, we  obtain results in the strong and weak coupling regimes  over non-Markovian
damping channels $\Gamma=0.01\gamma$ and $\Gamma=0.1\gamma$, whereas
for very high values, such as $\Gamma=5\gamma$, it is found to reduce to the Markovian
case.  In the following figures, we have used the notation  NM, M, and $\rm{NM_{S}}$, which correspond to the non-Markovian, Markovian, and non-Markovian 
(under strong coupling strengths) regimes of interactions, respectively.

\begin{figure}[t]
\centering{}\includegraphics[angle=-90,scale=0.9]{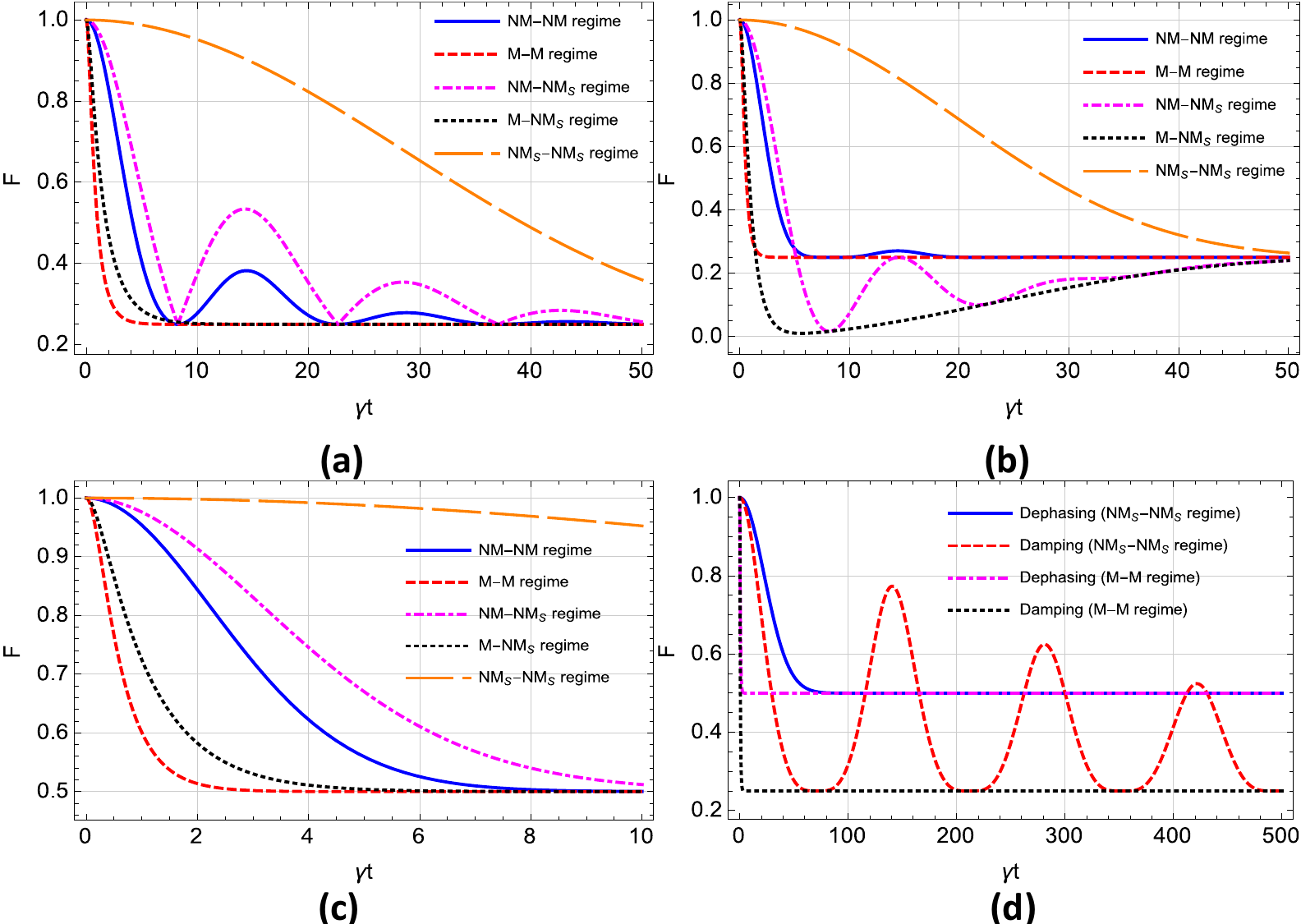}
\protect\caption[Average
fidelity obtained for the CQD protocol 
when the travel qubits undergo a non-Markovian damping or dephasing interaction]{\label{fig:CQD-Dam-Deph} Variation of the average
fidelity obtained for the CQD protocol with respect to the dimensionless quantity $\gamma t$
is depicted when the travel qubits undergo a damping or dephasing interaction
with its surroundings. In (a) and (b), both the travel qubits may have
different coupling strengths during their various rounds of travels
under damping effects, which are characterized by $p_{i}\,:i\in\left\{ 1,2,3,4\right\} $.
The values of the coupling strength for strong (weak) regime of non-Markovian
effect is chosen as $\Gamma=0.01\gamma$ ($\Gamma=0.1\gamma$), and
$\Gamma=5\gamma$ for Markovian regime. In (a) and (b), the choice
of the initial Bell states by Charlie is $|\psi^{\pm}\rangle$ and $|\phi^{\pm}\rangle$,
respectively. In (c), similar cases over the dephasing channels are shown.
In (d), both purely dephasing and damping effects are
shown together for strong coupling non-Markovian and Markovian regimes.}
\end{figure}

\begin{figure}[t]
\centering{}\includegraphics[angle=0,scale=0.76]{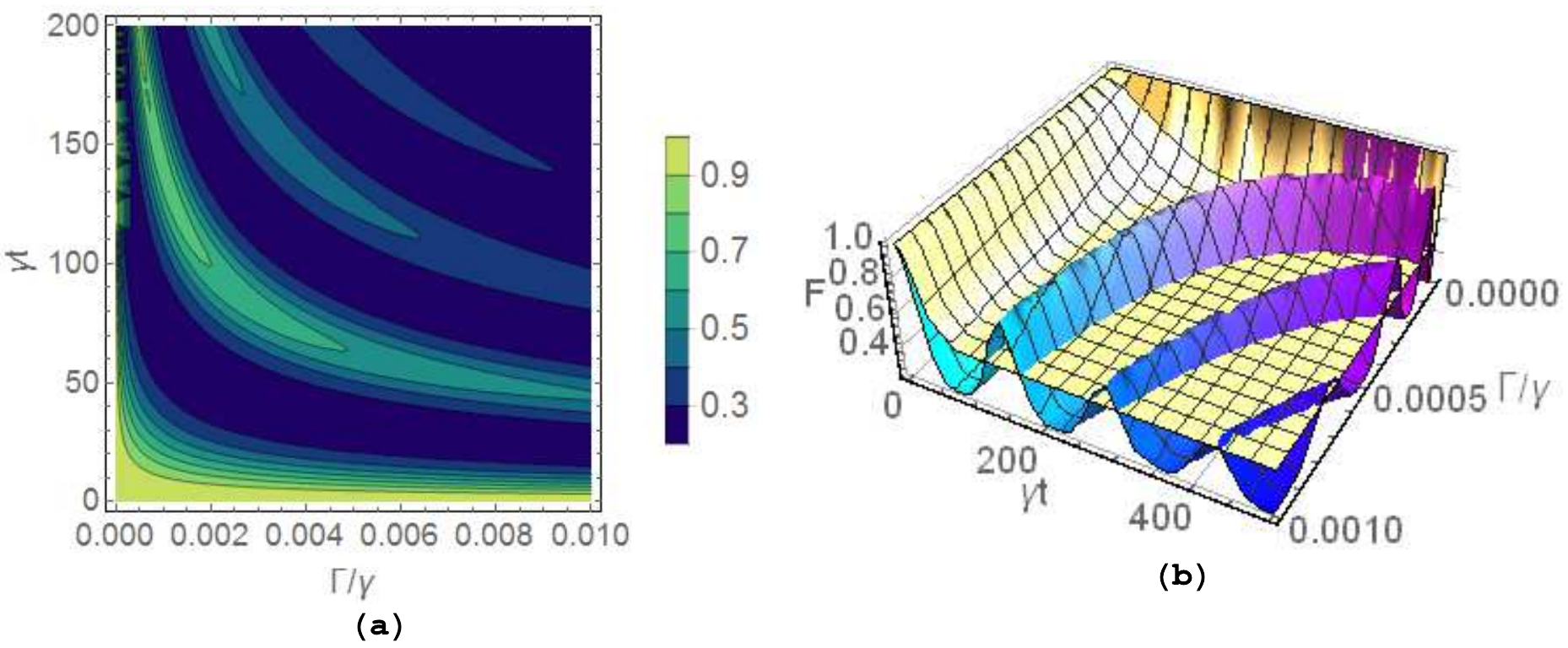}

\protect\caption[Dependence of average
fidelity obtained for the CQD protocol on the coupling strength and rescaled time over non-Markovian damping or dephasing channel]{\label{fig:CQD-3D} The dependence of the obtained fidelity
over the damping channel on the coupling strength and rescaled time is illustrated
through a contour plot in (a). (b) depicts
the variation of the fidelity for varying coupling strength
and time for both purely dephasing and damping non-Markovian channels in light (yellow) and dark (blue) colored surface plots, respectively.}
\end{figure}

\begin{figure}
\centering{}\includegraphics[angle=-90,scale=0.56]{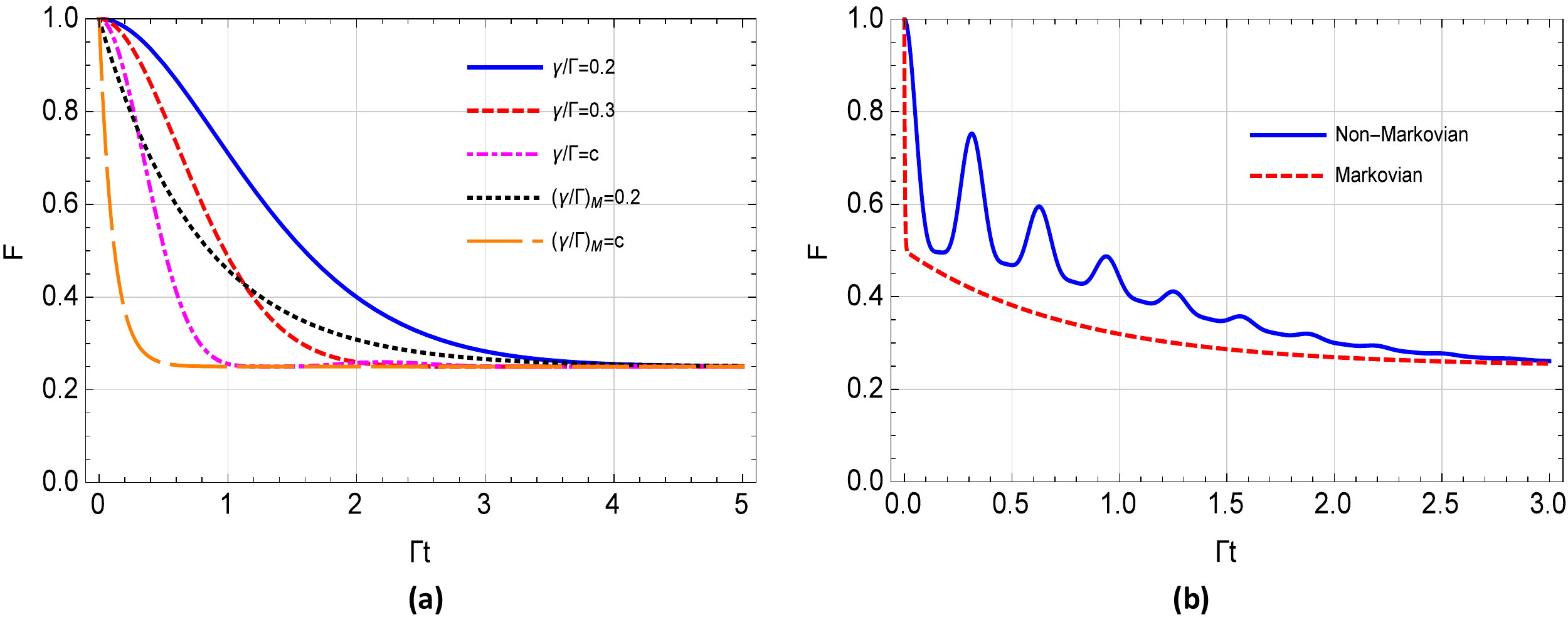}

\protect\caption[Effect of non-Markovian
depolarizing channel on the CQD scheme]{\label{fig:CQD-depol} (a) The effect of non-Markovian
depolarizing channel on the CQD scheme has been illustrated for different
values of the dimensionless quantity $\frac{\gamma}{\Gamma}$ indicated
in the plot. (a) shows the case of homogeneous non-Markovian depolarizing
channel (i.e., $\frac{\gamma_{i}}{\Gamma_{i}}=\frac{\gamma}{\Gamma}\forall i\in\left\{ 1,2,3\right\} $).
(b) illustrates a comparison between the inhomogeneous cases of non-Markovian
and Markovian depolarizing channels. In (a), the constant $c=\Gamma\left|\sqrt{\frac{1+\left(\pi/\log3\right)^{2}}{32}}\right|$,
which is the maximum value ensuring completely positive map for all
times for the homogeneous case; in (b), the noise parameters are $\frac{\gamma_{3}}{\Gamma_{3}}=5,\,\frac{\gamma_{i}}{\Gamma_{i}}=0.2$
for $i\in\left\{ 1,2\right\} $.}
\end{figure}

\begin{figure}
\centering{}\includegraphics[angle=0,scale=0.76]{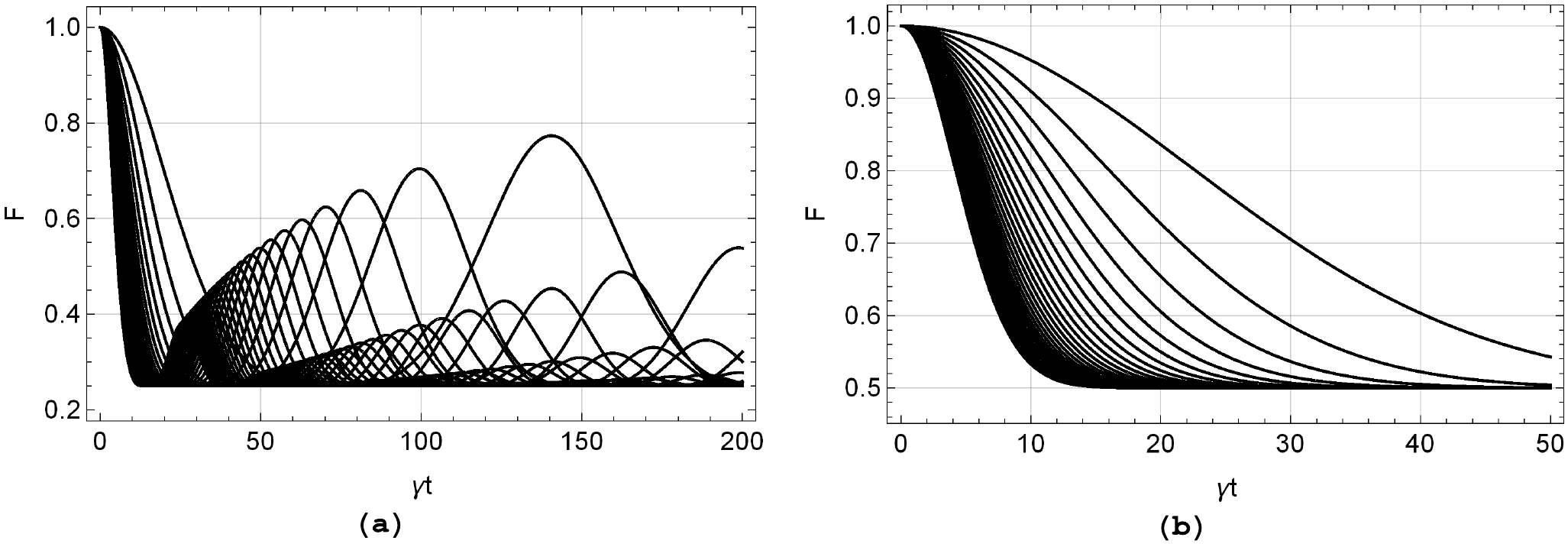}

\protect\caption[Effect of a change
in the coupling strength on the CQD scheme for damping and dephasing non-Markovian noise]{\label{fig:CQD-transition-dam-dep} The effect of a change
in the coupling strength on the fidelity is illustrated here
with a set of plots for damping and dephasing non-Markovian noise in
(a) and (b), respectively.  Specifically, the parameter of the coupling strength  $\Gamma/\gamma$ varies from 0.001
to 0.03 in the steps of 0.001 in both the plots.  }
\end{figure}

A comparative analysis of the effects of the non-Markovian (for both strong and weak
couplings) and Markovian noise is shown in Figures \ref{fig:CQD-Dam-Deph}-\ref{fig:CQD-transition-dam-dep}.
In this case, though four different coupling regimes are possible, one for each $p_{i}$s, we have restricted ourselves, for simplicity, to the scenario of 
Charlie to Bob quantum channel having the same coupling strength for both the travel qubits. Similarly, Bob to Alice quantum channel has the same coupling strength as 
that for the other way round. We explicitly mention the two choices of regimes in Figure \ref{fig:CQD-Dam-Deph}.  Specifically,
Figures \ref{fig:CQD-Dam-Deph} (a) and (b) show the effect of damping quantified by fidelity on the CQD scheme for different
choices of the initial Bell states, i.e., $|\psi^{\pm}\rangle$ and $|\phi^{\pm}\rangle$, respectively. It is interesting to observe that when both the qubits
undergo damping,  either in Markovian or in strong coupling non-Markovian regimes,  the choice of the initial Bell states becomes irrelevant (see
dashed (red) and large dashed (orange) curves in Figures \ref{fig:CQD-Dam-Deph} (a) and (b)). However, this initial choice becomes considerably important
for all the remaining cases, and $|\psi^{\pm}\rangle$ states are seen to be preferable as these states are less affected by the
non-Markovian damping noise than $|\phi^{\pm}\rangle$. Further, it is seen that, due to non-Markovian effects, the fidelity can be maintained for a 
relatively larger period of time (i.e., the quantum state decoheres slowly in non-Markovian environments in comparison with the corresponding Markovian environments), a feature that depends on the coupling strength (cf. Figures \ref{fig:CQD-Dam-Deph} (a) and (b)). Another interesting characteristic
of this kind of non-Markovian noise is periodicity \cite{bellomo2007non} and the kinks present in Figures \ref{fig:CQD-Dam-Deph} (a) and (b) are its signature. 
In Ref. \cite{sharma2016comparative} and Chapter \ref{Tomogram}, it was shown by us that the dilapidating influence of decoherence, due to Markovian damping, can be checked using squeezing. Here, it is seen that the same 
task can also be achieved by exploiting non-Markovianity. 

The effect of noisy environment is independent of the initial Bell
state over dephasing channel, and the fidelity is observed to improve gradually with
non-Markovian effects and coupling strength (cf. Figure \ref{fig:CQD-Dam-Deph}
(c)). Periodicity in the time variation of fidelity, when all interactions
are (strong) non-Markovian is not visible in the time scale of Figures
\ref{fig:CQD-Dam-Deph} (a) and (b). For larger time scales, this
can be observed in Figure \ref{fig:CQD-Dam-Deph} (d), where fidelity
over both the damping and dephasing non-Markovian channels is shown
together. It can be seen that the fidelity under the effect of the
damping noise decays faster than that over dephasing channel. At times,
the fidelity over damping channel is observed to be much larger than that over
dephasing channels, which remains constant at 1/2.

To analyze the effect of the coupling strength with varying time,
we depict, in Figure \ref{fig:CQD-3D}, a contour and a three dimensional plots. The
ripplelike plot (cf. the blue-colored surface plot in Figure \ref{fig:CQD-3D} (b)) shows that with decreasing coupling strength the
amplitudes of the revived fidelity gradually become smaller. The same
fact is also illustrated through a contour plot shown in Figure \ref{fig:CQD-3D}
(a), where we can see that the area of the light-colored region reduces
as we move from bottom to top. Physically, this corresponds to a transition
from the strong to weak coupling non-Markovian regime and finally into the
Markovian regime. 

A similar analysis of the fidelity expression for the depolarizing
channel is illustrated in Figure \ref{fig:CQD-depol}. In Figure \ref{fig:CQD-depol}
(a), homogeneous depolarizing noise is assumed $\gamma_{i}=\gamma\,\forall i\in\left\{ 1,2,3\right\} $,
for which $\gamma\le\left|\sqrt{\frac{1+\left(\pi/\log3\right)^{2}}{32}}\right|$
to ensure that the dynamical map is completely positive \cite{daffer2004depolarizing,huang2012study}.
Interestingly, Figure \ref{fig:CQD-depol} (a) shows
that the fidelity falls gradually with the parameter $\frac{\gamma}{\Gamma}$,
which determines the fluctuation due to the depolarizing channel. However,
it can be noted that for all the cases, the fidelity under non-Markovian
environment is always greater than that for the corresponding Markovian
case, till all the plots merge, with time, to a single value. Further, for the case
of inhomogeneous fluctuations  \cite{daffer2004depolarizing,huang2012study},
we observe a
revival in the fidelity in Figure \ref{fig:CQD-depol} (b).
Figure \ref{fig:CQD-depol} summarizes that the non-Markovian
depolarizing channel affects the system less than the corresponding Markovian channel.

The change in coupling strength controls the transition from non-Markovian
to Markovian regime for both damping and dephasing channels. This
dependence has been illustrated in Figure \ref{fig:CQD-transition-dam-dep}.
The initial small changes (decrease) in the value of the coupling strength parameter changes considerably the nature
of the obtained fidelity, i.e., the periodicity and the
maximum value of fidelity after revival show ample changes for even a
small change in the coupling strength parameter. However, for small values of the coupling
strength, this change becomes less sensitive as reflected in the dense black
lines corresponding to smaller values of the coupling strengths.

A similar comparison of the effect of the non-Markovian
and Markovian depolarizing channels shows that the fidelity sustains
for a longer period of time under the influence of a non-Markovian depolarizing channel and
is more sensitive to small changes in the noise parameter,
which controls the fluctuation. For higher values
of the noise parameter, the variation due to small changes in the noise parameters
becomes negligible in both Markovian and non-Markovian depolarizing
channels.

In the following subsections, we will deduce corresponding results for the remaining cryptographic tasks
from the results obtained in this section for the fidelity (for the CQD scheme) over various non-Markovian channels.

\subsection{Controlled deterministic secure quantum communication \label{sub:CDSQC}}

A protocol of CDSQC, based on quantum cryptographic switch,
can be obtained from the CQD scheme discussed in the previous subsection,
i.e., when only a single party encodes and sends his/her message in
a secure manner via the quantum channel, which is decoded by the other
party \cite{pathak2015efficient}. To be precise, Charlie initially follows the
same steps as discussed in Section \ref{sub:CQD}, but  instead of sending both the strings to Bob, he sends $S_{A}$ to Alice and $S_{B}$ to Bob. Subsequently, Alice  encodes
her message as usual and sends the encoded qubit to Bob, who decodes
the secret by performing Bell state measurement on the partner pairs with
the help of Charlie \cite{pathak2015efficient}. 

The CDSQC scheme and the effect of noise can be summarized as follows
\begin{equation}
\begin{array}{lcl}
\rho^{\prime} & = & \underset{i,j,l}{\sum}\left(K_{l}\left(p_{4}\right)\otimes I\right)U_{A_{n}}\left(K_{i}\left(p_{1}\right)\otimes K_{j}\left(p_{2}\right)\right)\rho\left(\left(K_{l}\left(p_{4}\right)\otimes I\right)U_{A_{n}}\left(K_{i}\left(p_{1}\right)\otimes K_{j}\left(p_{2}\right)\right)\right)^{\dagger},
\end{array}\label{eq:transformed-rho-CDSQC}
\end{equation}
where all the parameters have the same meaning as in Section \ref{sub:CQD}.
It is interesting to observe that the transformed density matrix in
Eq. (\ref{eq:transformed-rho-CDSQC}) can be obtained from Eq. (\ref{eq:transformed-rho})
just by considering $p_{3}=1$ and $U_{B_{n}}=I$. The fidelity can
be calculated with respect to the quantum state expected in the ideal situation,
i.e., $|\psi^{\prime}\rangle=U_{A_{n}}|\psi\rangle$.

Due to this observation, the fidelity of the quantum states affected by the non-Markovian
noise for the CDSQC scheme can be obtained
from the corresponding CQD expressions by taking $p_{3}=1$  in
Eqs. (\ref{eq:Damp-psi})-(\ref{eq:Deph}). Interestingly, for the case of the
depolarizing channel, the fidelity can be shown to be 
\begin{equation}
F=\frac{1}{2}\left[1+\Omega_{1}^{3}+\Omega_{2}^{3}+\Omega_{3}^{3}\right],\label{eq:Depolarizing-CDSQC}
\end{equation}
where the presence of cubic terms manifests the fact that the number of rounds of quantum communication involved in this scheme is 
less than that for the scheme discussed in the previous subsection. Specifically, the scheme for CDSQC  requires three rounds of quantum communication, 
while the scheme for CQD requires four rounds.

\begin{figure}[t]
\centering{}\includegraphics[angle=-90,scale=0.9]{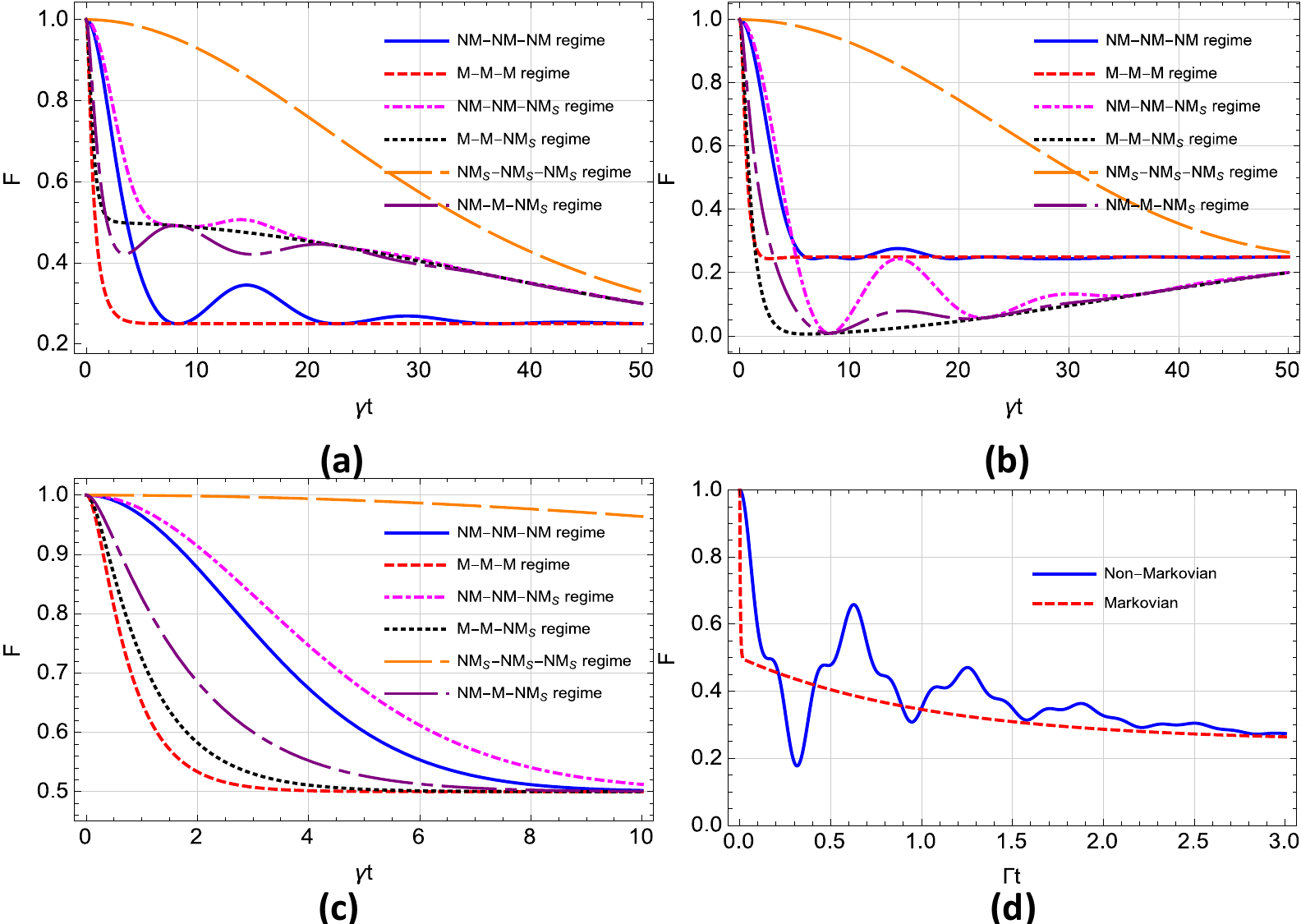}

\protect\caption[Effect of non-Markovian damping and dephasing channels on the CDSQC protocol]{\label{fig:CDSQC-Dam-Deph} The dependence of the average
fidelity obtained for the CDSQC protocol on the coupling strength is illustrated through its variation with the dimensionless quantity $\gamma t$, when the
travel qubits undergo damping (in (a) and (b)) or dephasing (in
(c)) interaction with their ambient surroundings. In (a) and (b), the initial
state chosen by Charlie was $|\psi^{\pm}\rangle$ and $|\phi^{\pm}\rangle$,
respectively. Here, we have chosen different values of all the coupling
constants in various regimes, i.e., non-Markovian with strong and weak
couplings as well as for the Markovian case. All the values of the coupling strengths
corresponding to various regimes are same as used in the previous
plots. In (d), time variation of the fidelity for the CDSQC
protocol over a depolarizing channel is shown corresponding to the values that
used in Figure \ref{fig:CQD-depol} (b).}
\end{figure}

The qubit traveling through the noisy channel may have different coupling
strengths during each round of travel. Here, we wish to emphasize this
point with the help of three possible coupling strengths for three
noisy channels acting on the travel qubits. The observations
made above for the extreme cases, i.e., the qubits traveling through either
non-Markovian channels with strong coupling or Markovian channels
all the time, remain valid here as well. Nevertheless, it cannot be conjectured
that the more the number of non-Markovian channels, the higher the fidelity. 
In particular, the large dot-dashed (purple) curve in Figures
\ref{fig:CDSQC-Dam-Deph} (a) and (b) establishes that  even lower
fidelity is observed with lesser number of Markovian channels acting
on the travel qubits. In fact, Figure \ref{fig:CDSQC-Dam-Deph} (b)
shows that the obtained fidelity for parity-1 Bell states (i.e., $|\phi^{\pm}\rangle$)
is less for all the cases when various noise channels had different
coupling strengths than that for the case of the travel qubits
subjected to  noisy channels with the same coupling strength.
However, no such nature is visible in Figure \ref{fig:CDSQC-Dam-Deph}
(c) for the dephasing channels.  It is worth stressing here that out of the three possible choices for different coupling regimes corresponding to each $p_{i}$, 
we have emphasized only on the interesting cases and mentioned them accordingly in Figure \ref{fig:CDSQC-Dam-Deph}.

Interestingly, the fidelity in the CDSQC protocol falls below the
corresponding Markovian value, under the influence of the non-Markovian depolarizing channel,
when all three noise parameters have different values (cf. Figure \ref{fig:CDSQC-Dam-Deph}
(d)). This nature can be attributed to the presence of cubic terms
in the fidelity, Eq. (\ref{eq:Depolarizing-CDSQC}).

\subsection{Quantum dialogue \label{sub:QD}}

A CQD scheme can be viewed as a QD scheme under the supervision of a controller.
Therefore, a QD scheme can be easily derived from the CQD scheme if
we consider the scenario that one of the two communicating  parties (i.e., either Alice or Bob) prepares and
measures the quantum state, while both the parties encode their secret on the same
qubits. This QD scheme, which is obtained as a result of reduction from the CQD scheme described above, can be easily recognized to be equivalent to the first QD protocol
proposed by Ba An \cite{nguyen2004quantum} (also discussed in Section \ref{subsec:CQD}). The effect of noise on this scheme for QD 
can be obtained  by considering $p_{1}=p_{2}=1$ in all the
expressions of Section \ref{sub:CQD}. This would imply that the initial state is
prepared by one of the communicating parties (say, Bob). Then, the transformed density matrix and the fidelity
expressions over the non-Markovian channels can be deduced from Eqs. (\ref{eq:transformed-rho})-(\ref{eq:Deph}). Here, it is important to note  that the effect of noise is independent
of the choice of initial Bell state by Charlie/Bob in all the schemes
other than CQD and CDSQC. Similarly, under the effect of depolarizing channels,
the expression of fidelity turns out to be

\begin{equation}
F=\frac{1}{2}\left[1+\Omega_{1}^{2}+\Omega_{2}^{2}+\Omega_{3}^{2}\right],\label{eq:Depolarizing-QD}
\end{equation}
due to two rounds of quantum communication of a travel qubit.

\subsection{Quantum secure direct communication/Deterministic secure quantum communication \label{sub:QSDC/DSQC}}

As mentioned beforehand in Section \ref{sub:CDSQC}, a CDSQC scheme
can be viewed as a CQD scheme, where only one party is allowed to encode.
In the same way, a QSDC scheme (say, a ping-pong protocol \cite{bostrom2002deterministic})
can be viewed as a scheme for QD \cite{nguyen2004quantum}, where one party (say Bob) is restricted to encode
identity only. Therefore, all the expressions of the fidelity for a QSDC scheme are exactly the same as those for the scheme of  QD. 

A DSQC scheme can be reduced  from the above-mentioned protocols if
Bob incorporates information splitting into two quantum pieces and sends them one after the other 
in two different rounds of Bob to Alice communication \cite{pathak2013elements}. Specifically, Bob prepares two strings ($S_{A}$ and $S_{B}$)
as in Section \ref{sub:QD} and sends the first string (say, $S_{B}$) to Alice. He
subsequently sends the second string (i.e., $S_{A}$) to Alice only if the first quantum
part is received by Alice undisturbed. The effect of non-Markovian
noisy environment on this DSQC scheme can be obtained from the corresponding expressions
for the CQD scheme obtained in Section \ref{sub:CQD},
 if we consider $p_{1}=p_{4}=1$
and $p_{2}=p_{3}^{\prime}$ in Eqs. (\ref{eq:Damp-psi})-(\ref{eq:Deph}).
Here, $p_{3}^{\prime}$ is used to show the effect of noise on the
second qubit traveling from Bob to Alice in the first round. In fact,
it turns out to be exactly similar to what is obtained for the QSDC
scheme. Interestingly, in case of the depolarizing noise, all the expressions for fidelity 
are found to be the same for QD, QSDC, and DSQC schemes. For the convenience
of discussion for the DSQC scheme, we have chosen Bob (Alice) as the sender
(receiver).

So far, we have discussed quantum communication schemes where
prior key generation is circumvented by proper use of quantum resources.
We may now proceed to key generation schemes and investigate the effect of non-Markovian
environment on them.

\subsection{Quantum key agreement \label{sub:QKA}}

A QKA scheme provides equal power to all the parties taking part
in the key generation process and does not allow members of a proper subset of the set of all users to solely decide the final
key. Here, we consider a completely orthogonal-state-based QKA scheme proposed
in \cite{shukla2014protocols}. In this QKA protocol, a party (say
Alice) sends her raw key to another party (say Bob) by using a QSDC protocol, while the other party
publicly announces his key. The security of the final key is achieved
by the unconditional security of Alice's transmission of raw key using quantum resources (i.e., from the security of the QSDC/DSQC scheme used by Alice and Bob for Alice to Bob communication). Specifically, Alice transmits a key $k_{A}$ to Bob in a secure manner, whereas Bob announces his key $k_{B}$, publicly, and for all future communication, they use a key $k_{AB}=k_{A}\oplus k_{B}$, where $\oplus$ denotes a bitwise XOR operation. Although Eve knows $k_{B}$, she cannot obtain any information about $k_{AB}$ as she knows nothing about $k_{A}$. Thus, the security of $k_{AB}$ depends on the security of $k_{A}$. In other words, unconditional security of the QSDC scheme involved here would ensure the security of the protocol for QKA.
Interestingly, in \cite{sharma2016comparative}, we have 
shown that the effect of noise on this scheme is identical to the
QSDC scheme discussed in the previous Section \ref{sub:QSDC/DSQC}. Since the observations made there remain valid here, we do not  discuss it in further detail.

\subsection{Quantum key distribution \label{sub:QKD}}

Any discussion on quantum cryptography remains
incomplete without discussing a protocol that changed the course of
cryptography by establishing the feasibility of unconditional security.
In this section, we  discuss two QKD protocols, which can be viewed as the
variants of the same scheme, differing only in the measurement
procedure. Specifically, the BB84 \cite{bennett1984quantum}
and BBM \cite{bennett1992quantum} QKD protocols are discussed here. Before we proceed further, it would be apt to note that in contrast to the fidelity expressions obtained in the earlier sections (which were average over all the encoding operations), for QKD protocols, the average fidelity is obtained over all possible equally probable measurement outcomes.

In the BBM protocol \cite{bennett1992quantum}, Alice prepares $n$ Bell states and sends
all the first qubits to Bob, and both of them measure the qubits of the shared
Bell states randomly in the computational ($\left\{ |0\rangle,|1\rangle\right\} $)
or diagonal ($\left\{ |+\rangle,|-\rangle\right\} $) basis. Using
the outcome of these measurements, they finally obtain an unconditionally secure quantum key for those cases where
both Alice and Bob perform measurement using the same basis.

The BB84 protocol can also be viewed along the same lines, where Alice
first measures her qubit (i.e., second qubit) of each Bell state randomly either in the computational
or diagonal basis and then sends the other qubit to Bob. Finally,
they can obtain a key by using the measurement outcomes of half of those cases, where they have chosen the same basis. The other half of the cases should be used 
for eavesdropping check. Specifically, when Alice and Bob have performed measurement in the same basis, in the absence of Eve, 
their measurement outcomes should match, and a mismatch would indicate the presence of Eve.

Interestingly, for the BBM protocol, the effect of noise can be considered
by taking $p_{1}=p_{2}=p_{4}=1$ in Eqs. (\ref{eq:Damp-psi})-(\ref{eq:Deph}).
Similarly, the effect of the depolarizing channel reduces the fidelity
to

\begin{equation}
F=\frac{1}{2}\left[1+\Omega_{1}+\Omega_{2}+\Omega_{3}\right].\label{eq:Depolarizing-QKD-BBM}
\end{equation}
A similar study for the BB84 protocol results in the following fidelity
over damping non-Markovian channels

\begin{equation}
F=\frac{1}{4}\left[2+\sqrt{p_{3}}+p_{3}\right],\label{eq:Damp-QKD}
\end{equation}
while for the dephasing channel, the fidelity is 

\begin{equation}
F=\frac{1}{4}\left[3+p_{3}\right].\label{eq:Deph-QKD}
\end{equation}
Further, the fidelity when the travel qubit in BB84 protocol is
subjected to a depolarizing channel is 

\begin{equation}
F=\frac{1}{2}\left[2+\Omega_{1}+\Omega_{3}\right].\label{eq:Depolarizing-QKD-BB84}
\end{equation}

Additionally, the present results can also be used to deduce the fidelity for a few other quantum cryptographic schemes, which will
reflect quantitatively the effect of non-Markovian channels on the
corresponding scheme. For example, the effect of noise on Ekert's
QKD protocol \cite{ekert1991quantum} can also be deduced from the results in
Section \ref{sub:CQD}, by taking $p_{3}=p_{4}=1$ as the source
of entanglement is between both the parties, and both the entangled
qubits travel to Alice and Bob from there. Similarly, the feasibility
of the B92 protocol \cite{bennett1992B92} can also be analyzed over the non-Markovian
channels in analogy with the study for the BB84 protocol by only considering
two of the four single-qubit states (one each chosen from the computational and diagonal basis).

\begin{figure}
\centering{}\includegraphics[angle=-90,scale=0.59]{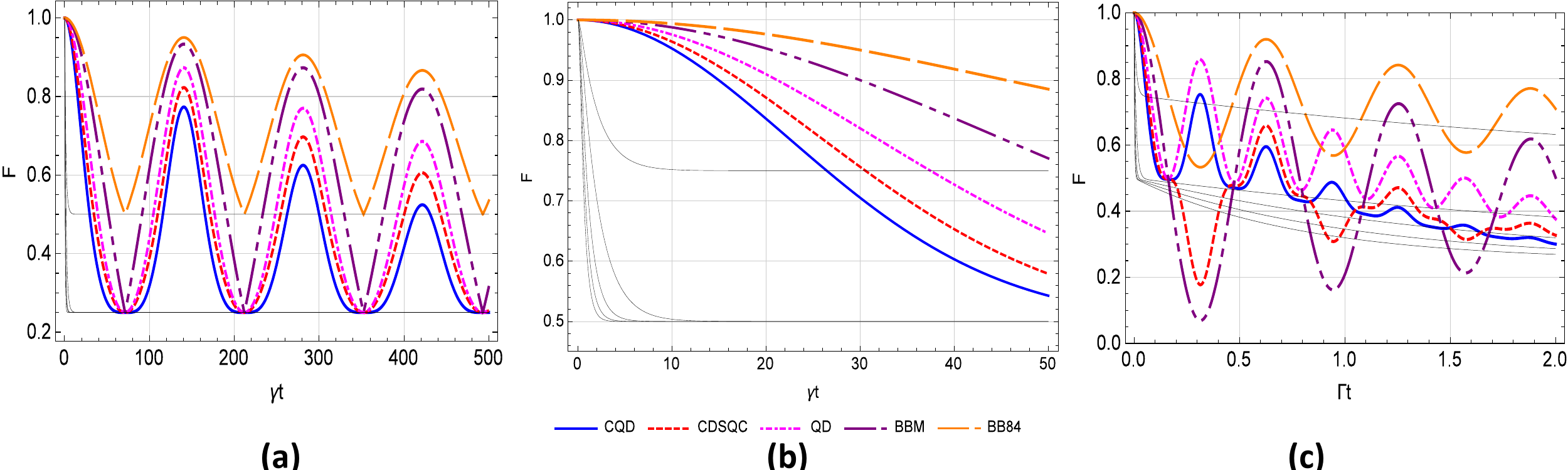}

\protect\caption[A comparative analysis of
different quantum cryptographic schemes over the non-Markovian
channels]{\label{fig:All-schemes} A comparative analysis of
all the quantum cryptographic schemes discussed so far over the non-Markovian
channels. Each line in all three plots corresponds to the different
cryptographic scheme mentioned in the plot legend at the bottom of
the figure. The light black lines in all three plots represent the corresponding
Markovian cases, and the black lines from bottom to top show the average
fidelity for CQD, CDSQC, QD, BBM (QKD), and BB84 (QKD) protocols. The
fidelity obtained for QSDC, DSQC, and QKA schemes is exactly the same
as that of the QD protocol. }

\end{figure}

Finally, we  perform a comparative study for the fidelity
obtained in each cryptographic scheme to reveal the general
nature of the effect of non-Markovian channels on all these schemes
(shown in Figure \ref{fig:All-schemes}). Interestingly, the effect
of noise depends on the number of rounds a qubit is required to travel through the noisy
channel, and thus on the complexity of the task in hand. This fact is consistent with the recent observations
on a set of Markovian channels \cite{sharma2016comparative}. Specifically, in the
CQD scheme, one qubit travels from Charlie to Bob, while another qubit
travels from Charlie-Bob-Alice-Bob. Therefore, the maximum number of rounds
of travels in the set of secure quantum communication schemes discussed
here is four for CQD scheme, which decreases to three for CDSQC. It
further reduces to two for QD, QSDC, DSQC, and QKA schemes. The same
fidelity for all these schemes further establishes
this point. Finally,  BBM and BB84 QKD protocols require only
one round of quantum communication. In fact, BBM and BB84 protocols use entangled and single-qubit states, respectively, to accomplish
the same task. Out of these two schemes, the BB84 QKD scheme is least affected by noise
as it uses single-qubit states, which were shown to be less affected
due to Markovian channels in \cite{sharma2016comparative}.

In Figures \ref{fig:All-schemes} (a) and (b), the fidelity
variation over non-Markovian channels due to the strong coupling of
the travel qubits with the environment is depicted. Similarly, the
effect of different noise parameters corresponding to depolarizing channel
is shown in Figure \ref{fig:All-schemes} (c). Also shown is the
effect of Markovian environment on the fidelity in all three
cases, depicted by  thin smooth (black) lines. For all cases of
Markovian dynamics, the observation that the effect of noise depends
on the rounds of quantum communication remains valid. 

In Figure \ref{fig:All-schemes} (a), the revival in the fidelity
over non-Markovian damping channel is seen to decrease with an increase in the
number of travel qubits. Similarly, the fidelity falls with
increasing rounds of quantum communication when subjected to dephasing
non-Markovian channel, as shown in Figure \ref{fig:All-schemes} (b).
Out of the set of fidelities, over the depolarizing channel, those having
odd power terms, such as for the CDSQC and QKD protocols, show
fidelity less than that for the corresponding Markovian case. Otherwise,
in all the remaining cases, the fidelity over non-Markovian
channels is more than that for the corresponding Markovian channels (cf. Figure \ref{fig:All-schemes} (c)).

\section{Conclusions \label{sec:Conclusion-NonMarkovian}}

The present study on the effect of a set of non-Markovian channels
on a three-party secure quantum communication task (CQD) led to a number of interesting
results. This serves to satisfy the twofold motivation we had set ourselves, that is, to quantify the effect of non-Markovian noise measured relative to Markovian noise and to obtain the impact of the noise on a range of cryptographic protocols. Specifically, we have considered here a damping, a purely
dephasing, and a depolarizing non-Markovian channels to analyze the feasibility
of some quantum cryptographic schemes evolving under the influence of the non-Markovian environments.
We have started with a CQD scheme, based on a quantum cryptographic switch that uses Bell
states. Later, this scheme is modified to deduce the results for other
quantum cryptographic tasks, such as CDSQC, QSDC, DSQC. Apart from
these direct communication schemes, the effect of non-Markovian noise on some protocols of QKD and QKA are also analyzed. 

It has been established that the effect of non-Markovian noise 
depends on the complexity of the desired task (the number of rounds of the travel qubits). We have observed that the BB84 QKD scheme is least affected due to
non-Markovian channels, while the CQD scheme shows a maximum fall in
the fidelity. In fact, from the results obtained here one
can also show that the asymmetric QD scheme \cite{banerjee2017asymmetric} will have the same effect
as that on the QD protocol if the number of travel qubits is kept unchanged.
This fact is consistent with the results obtained here, that the fidelity for QSDC, DSQC, and QKA
schemes are exactly the same as that for the QD protocol. In the recent past, we have established that squeezing is a useful quantum resource for 
quantum cryptography as it can help to stall the effect of decoherence \cite{sharma2016comparative} (also shown for  spin states in Chapter \ref{Tomogram}). Here, we have shown that non-Markovianity can also be used to accomplish a similar task. 

Interestingly, the effect of noise on the CQD and CDSQC schemes are found to depend
on Charlie's initial choice of the Bell state, while it is independent
of this in all the remaining schemes. Finally, our analysis has also
revealed that the fidelity obtained in the case of damping and dephasing
channels depends on the coupling strength. 
We hope these results would bring out the importance and utility of the non-Markovian behavior in the understanding of quantum cryptographic protocols from the perspective of their practical implementation. The findings of this chapter are published in Ref. \cite{thapliyal2017quantum}.

In this part of the thesis (which includes this chapter and the previous chapter), we have discussed the effect of Markovian and non-Markovian noise on a set of schemes for  quantum communication that can be implemented using minimum amount of nonclassical resources (in terms of the number of entangled qubits). Thus, in general, in Chapters \ref{Coupler}-\ref{Zeno}, we have reported various physical systems that may lead to nonclassical states, and in Chapters \ref{QDs}-\ref{Tomogram}, we have discussed how does the consideration of open quantum system affects the dynamics of the nonclassical states. Establishing the relevance of the studies reported in Chapters \ref{Coupler}-\ref{Tomogram}, in this  chapter and the previous chapter, we have reported some applications of the nonclassical states (analyzing their performance over Markovian and non-Markovian channels). This provides a kind of completeness to the work, and considering that we will conclude the present thesis work in the following chapter.

\thispagestyle{empty}

\titlespacing*{\chapter}{0pt}{-50pt}{20pt}
\chapter{Conclusions and scope for the future work \label{conclusions-and-scope}}

Recently, interest in the studies focused on nonclassical states is growing fast due to their promising applications in different domains of science and technology. Keeping this in mind, we have discussed the feasibility of generation of nonclassical states in an asymmetric nonlinear optical coupler operating under second harmonic generation in Chapter \ref{Coupler}. Possibilities of observing quantum Zeno and anti-Zeno effects, which have foundational and practical relevance in counterfactual quantum computation and communication, are also established in one such optical system (in Chapter \ref{Zeno}). While implementing quantum technologies, we need to take into consideration the decoherence of the nonclassical states to be used under realistic conditions. We have made attempts to obtain the dynamics of spin systems (in a single- and multi-qubit or spin-$\frac{1}{2}$ states) evolving under Markovian environments using quasiprobability distribution functions and tomograms in Chapters \ref{QDs}-\ref{Tomogram}. This study allows us to not only obtain the expectation value of an arbitrary operator at any instant of time, also to obtain the dynamics of nonclassicality present in the system and to quantify it. Finally, controlled communication (both insecure and secure) schemes with minimum number of entangled qubits have been proposed in Chapter \ref{CryptSwitch}, where the performance of the schemes is also analyzed over Markovian channels. The optimized controlled secure quantum communication scheme was observed to be reducible to several cryptographic tasks. Using this interesting observation, we have evaluated the performance of a set of cryptographic schemes over non-Markovian channels in Chapter \ref{NonMarkovian}. In view of this study, we list the major findings of the present thesis work in the next section.

\section{Conclusions of the thesis work \label{sec:Conclusion-Thesis}}

Specifically, the main  findings of the present thesis can be summarized as follows.
 
 \begin{enumerate}
 
 \item It is possible to generate lower- and higher-order nonclassical (squeezed, antibunched, and entangled) states in an experimentally realizable asymmetric nonlinear optical coupler composed of a linear waveguide and a nonlinear (quadratic) waveguide operated by second harmonic generation. The same has been established with the help of moments-based criteria in both codirectional and contradirectional propagation of fields.

\item It is observed that the optical couplers studied here have some intrinsic advantages over most of the
other systems studied in the past as the systems studied here can be used to build useful components (sources of nonclassical states, e.g., entanglement) in the integrated waveguide-based structures in general and photonic circuits in particular. 
   
 \item A complete quantum mechanical description of the system Hamiltonian considering all the fields as weak and the use of better solution (which is not restricted to short-length) allowed us to observe some nonclassical features that were missed in the earlier studies (specifically, nonclassicalities involving the second harmonic mode).
 
 \item In both codirectional and contradirectional couplers, it is observed that the pump modes always get (both lower- and higher-order) entangled. Further, three-mode state could not be found fully entangled. However, as the intermodal entanglement is observed in the pure state, it implies
that the two linear modes in the coupler are maximally steerable.
 
 \item It is found   that the amount
of nonclassicality can be controlled by various parameters, such as the number of input photons
in the linear mode, the linear and nonlinear coupling constants, 
phase mismatch. The results also established that the
nonclassical properties of light can transfer from a nonlinear waveguide to a linear waveguide. 
 
 \item All the results obtained here in context of the contradirectional
coupler show that in the contradirectional propagation, the beams
involved are more effectively matched compared to the 
case of the codirectional propagation, and consequently
the nonlinear interaction is more effective, and the corresponding  nonclassical effects
are found to be stronger.

 \item The existence of quantum Zeno and anti-Zeno effects has been observed in a symmetric and an asymmetric nonlinear optical couplers. 
 
 \item A completely quantum
description of the primary physical system (i.e., a symmetric nonlinear
optical coupler), an appropriate use of a perturbative technique, reducibility
of the results obtained for the symmetric nonlinear optical coupler to
that of the asymmetric nonlinear optical coupler, an easy experimental realizability
of the physical systems, etc., are found to provide an edge to this work over the
existing works on quantum Zeno effect in optical couplers.
 
 \item The investigation on quantum Zeno and anti-Zeno effects 
led to several interesting observations, such as the analytic expressions obtained for both linear and nonlinear
Zeno parameters are same for the spontaneous case. It is concluded that a transition between quantum Zeno and anti-Zeno effects is determined by the phase relations in the system. Specifically, for the stimulated processes, the quantum Zeno  (anti-Zeno) effect is found to be related to the phase matching  (phase mismatching).

\item In both
the spontaneous and stimulated cases, it is observed that a transition between the
quantum anti-Zeno effect and quantum Zeno effect can be achieved by
increasing the intensity of the radiation field in the linear mode
of the system waveguide, controlling the phase
of the fundamental and second harmonic modes in the system and probe waveguides, and the phase mismatch between the fundamental and second harmonic modes in
the system waveguide.

\item The nonclassical features have been observed in a set of  spin states using quasiprobability distributions and tomograms with the open quantum system formalism. Further, the obtained results were used to quantify the nonclassicality (using nonclassical volume) present in the spin states.  The tomogram for an infinite dimensional system, the
ubiquitous dissipative harmonic oscillator, is also obtained.

\item It is shown that the study would  assume significance in questions
related to quantum state engineering, where the central point is to
have a clear understanding of coherences in the quantum mechanical
system being used, and thus, it would be essential to have an understanding
of quantum to classical transitions, under ambient conditions. 

 \item Along with the
well-known Wigner, $P$, and $Q$ quasiprobability distributions,  so called $F$-function is also obtained and found to be equivalent to the Wigner function in case of spin-qubit systems. It is observed that an increase in temperature tends to decohere 
the tomograms, while squeezing is shown to be a  useful quantum resource. 

\item Applications of nonclassical (entangled) states have been proposed in the context of both insecure and secure quantum communication schemes. 

\item A general structure for the quantum states suitable as quantum channel for BCST schemes (using $n$-qubit entangled states with $n\geq5$) is obtained. It is shown that the general structure of the quantum states can be used to generate  infinitely many new quantum channels, any of which can be selected to perform BCST depending upon its experimental feasibility.

\item  Optimal (with respect to the use of multi-partite entanglement) PoP-based protocols that use only Bell states are designed  for insecure (BCST) and secure (CQD) controlled quantum communication.

\item It is found that the protocols for controlled quantum
communication that are designed using the concept of the quantum cryptographic switch have a set of advantages over the existing protocols
for the same task. For example, the controller can continuously vary the amount
of information revealed to the receiver(s) and the sender(s), and 
he also has directional control in the BCST and similar protocols. 

\item It is shown
that a set of other schemes for controlled quantum cryptographic tasks and various existing protocols of secure quantum communication can be reduced from the proposed CQD scheme. Finally, the proposed controlled secure quantum communication scheme is analyzed over Markovian and non-Markovian channels, and later the effect of noise on various cryptographic schemes is deduced from it.

\item It has been established that the effect of non-Markovian noise (also Markovian noise deduced in the limiting case)
depends on the complexity of the task in hand, thus in turn, on the number of rounds of the travel qubits. It is observed that the BB84 QKD scheme is least affected due to
non-Markovian channels, while the CQD scheme shows a maximum fall in
the fidelity. 

\item The performance of the quantum cryptographic schemes is found to depend upon the coupling strength between the qubits used and their surroundings. It is also observed that  squeezing can be helpful to stall the effect of decoherence in SGAD channels, and the non-Markovianity can also be used to accomplish a similar task. 

\end{enumerate}

Therefore, the present thesis discusses generation of nonclassical states (in optical systems), evolution of nonclassical behavior in the presence of decoherence due to ambient environment (in spin systems), and finally applications of nonclassical states to obtain quantum advantage in  communication. For instance, the performance of the proposed unconditionally secure CQD scheme (and other schemes) using an optimized number of entangled qubits has been analyzed over Markovian and non-Markovian channels. 

Our focus so far has been to optimize the amount of nonclassical resources in several communication schemes. In the future, we will attempt to exploit the maximum enhancement in the performance that is achievable using various quantum resources. For example, device independent and one-side device independent quantum cryptography with the help of Bell nonlocal and EPR steering correlations. 

\section{Scope for the future work \label{sec:Future-scope-Thesis}}

The works reported in the present thesis provide us a general framework for the investigation on the origin of nonclassicality, its applications, and the effect of noise on the nonclassicality witnesses, measures, and the applications of nonclasssical features. As the methodology developed and used in this work is quite general, it can be further extended in numerous ways. 
Some of the future scopes of the present work are listed below with a specific attention on those possibilities which may be realized in the near future. 
 
 \begin{enumerate}
\item The work reported here is theoretical in nature, but most of the physical systems studied (such as codirectional and contradirectional couplers) are experimentally realizable. Further, the schemes for various quantum information processing tasks designed and analyzed here can also be realized using the existing technology. These facts lead to the possibility that the theoretical results reported here may be verified experimentally in the near future.

\item Various types of nonclassicality (with a special focus on entanglement) observed in the nonlinear optical couplers can be quantified using the existing nonclassicality (entanglement) measures (as used in Chapter \ref{QDs}).

\item Entanglement (and other correlations, such as steering, Bell nonlocality, discord) can be studied for the system of interest considering open quantum system formalism.

\item Analogously, non-Gaussianity of the states generated in the systems studied here can be analyzed (using both witnesses and measures of non-Gaussianity).

\item Feasibility of observing linear and nonlinear quantum Zeno and anti-Zeno effects in the domains of Markovian and non-Markovian noise can be investigated.

\item Optical designs for the physical realization of the quantum communication schemes proposed here can be provided.

\item The effect of noise on other facets of quantum cryptography, such as counterfactual, device independent, and measurement-device independent quantum cryptography may also be addressed in the near future. Further, efforts can be made to design new schemes for these types of quantum cryptographic schemes.

\item In the present work, the effects of channel noise and eavesdropping are considered independently. In the future, attempts can be made to study their effects simultaneously.

\end{enumerate}

With the advent of experimental facilities and growing interest in the enhancement achievable in the performance of numerous computation, communication, and other technological aspects of our lives exploiting the true power of quantum (nonclassical) resources, we hope that the theoretical work performed in the present thesis will be experimentally realized and will lead to some devices/technologies having applications in daily life. Specifically, we expect that the present work will play a significant role in the near future via the impacts of quantum information processing on the society. 

\thispagestyle{empty}

\pagestyle{plain}

\phantomsection
\addcontentsline{toc}{chapter}{REFERENCES}\renewcommand{\bibname}{REFERENCES}
\bibliographystyle{jiit}
\bibliography{biblio}

\begin{thebibliography}{100}
\newcommand{\enquote}[1]{``#1''}
\providecommand{\url}[1]{\texttt{#1}}
\providecommand{\urlprefix}{URL }
\expandafter\ifx\csname urlstyle\endcsname\relax
  \providecommand{\doi}[1]{doi:\discretionary{}{}{}#1}\else
  \providecommand{\doi}{doi:\discretionary{}{}{}\begingroup
  \urlstyle{rm}\Url}\fi

\bibitem{planck1901law}
Planck M., \enquote{\emph{On the law of distribution of energy in the normal
  spectrum}}, Annalen der Physik, vol.~4, p.~1, 1901.

\bibitem{einstein1905erzeugung}
Einstein A., \enquote{\emph{{\"U}ber einen die {Erzeugung} und {Verwandlung}
  des {Lichtes} betreffenden heuristischen {Gesichtspunkt}}}, Annalen der
  Physik, vol. 322, pp. 132--148, 1905.

\bibitem{pathak2017classical}
Pathak A., Ghatak A., \enquote{\emph{Classical light vs. nonclassical light:
  characterizations and interesting applications}}, Journal of Electromagnetic
  Waves and Applications, vol.~32, pp. 229--264, 2018.

\bibitem{fox2006quantum}
Fox M., \emph{Quantum Optics: An Introduction}, Oxford University Press,
  Oxford, 2006.

\bibitem{dirac1939new}
Dirac P.A.M., \enquote{\emph{A new notation for quantum mechanics}},
  \emph{Mathematical Proceedings of the Cambridge Philosophical Society},
  vol.~35, pp. 416--418, Cambridge University Press, 1939.

\bibitem{von1955mathematical}
{v}on Neumann J., \emph{Mathematical Foundations of Quantum Mechanics},
  Princeton University Press, Princeton, New Jersey, 1955.

\bibitem{louisell1973quantum}
Louisell W.H., \emph{Quantum Statistical Properties of Radiation}, John Wiley
  \& Sons, Canada, 1973.

\bibitem{perina1991quantum}
Pe{\v{r}}ina J., \emph{Quantum Statistics of Linear and Nonlinear Optical
  Phenomena}, Kluwer Academic, Dordrecht-Boston, 1991.

\bibitem{kim1991phase}
Kim Y.S., Noz M.E., \emph{Phase Space Picture of Quantum Mechanics: Group
  Theoretical Approach}, World Scientific, 1991.

\bibitem{plum1996density}
Blum K., \emph{Density Matrix Theory and Applications}, Plenum Press, New York,
  1996.

\bibitem{scully1997quantum}
Scully M.O., Zubairy M.S., \emph{Quantum Optics}, Cambridge University Press,
  Cambridge, 1997.

\bibitem{puri2001mathematical}
Puri R.R., \emph{Mathematical Methods of Quantum Optics}, Springer Science \&
  Business Media, 2001.

\bibitem{walls2007quantum}
Walls D.F., Milburn G.J., \emph{Quantum Optics}, Springer Science \& Business
  Media, 2007.

\bibitem{weiss2008quantum}
Weiss U., \emph{Quantum Dissipative Systems}, World Scientific, 2008.

\bibitem{luks2009quantum}
Luk{\v{s}} A., Perinov{\'a} V., \emph{Quantum Aspects of Light Propagation},
  Springer, New York, 2009.

\bibitem{schleich2011quantum}
Schleich W.P., \emph{Quantum Optics in Phase Space}, Wiley-VCH, Berlin, 2001.

\bibitem{agarwal2013quantum}
Agarwal G.S., \emph{Quantum Optics}, Cambridge University Press, Cambridge,
  2013.

\bibitem{maxwell1865dynamical}
Maxwell J.C., \enquote{\emph{A dynamical theory of the electromagnetic field}},
  Philosophical transactions of the Royal Society of London, vol. 155, pp.
  459--512, 1865.

\bibitem{dirac1927quantum}
Dirac P.A.M., \enquote{\emph{The quantum theory of the emission and absorption
  of radiation}}, \emph{Proceedings of the Royal Society of London A:
  Mathematical, Physical and Engineering Sciences}, vol. 114, pp. 243--265,
  1927.

\bibitem{fermi1932quantum}
Fermi E., \enquote{\emph{Quantum theory of radiation}}, Reviews of Modern
  Physics, vol.~4, p.~87, 1932.

\bibitem{schrodinger1926stetige}
Schr{\"o}dinger E., \enquote{\emph{Der stetige {{\"U}}bergang von der
  {Mikro}-zur {Makromechanik}}}, Naturwissenschaften, vol.~14, pp. 664--666,
  1926.

\bibitem{klauder1960action}
Klauder J.R., \enquote{\emph{The action option and a feynman quantization of
  spinor fields in terms of ordinary c-numbers}}, Annals of Physics, vol.~11,
  pp. 123--168, 1960.

\bibitem{glauber1963coherent}
Glauber R.J., \enquote{\emph{Coherent and incoherent states of the radiation
  field}}, Physical Review, vol. 131, p. 2766, 1963.

\bibitem{glauber1963photon}
Glauber R.J., \enquote{\emph{Photon correlations}}, Physical Review Letters,
  vol.~10, p.~84, 1963.

\bibitem{sudarshan1963equivalence}
Sudarshan E.C.G., \enquote{\emph{Equivalence of semiclassical and quantum
  mechanical descriptions of statistical light beams}}, Physical Review
  Letters, vol.~10, p. 277, 1963.

\bibitem{wigner1932quantum}
Wigner E.P., \enquote{\emph{On the quantum correction for thermodynamic
  equilibrium}}, Physical Review, vol.~40, p. 749, 1932.

\bibitem{husimi1940some}
Husimi K., \enquote{\emph{Some formal properties of the density matrix}},
  Proceedings of the Physico-Mathematical Society of Japan [Nippon
  Sugaku-Buturigakkwai Kizi Dai 3 Ki], vol.~22, pp. 264--314, 1940.

\bibitem{lutkenhaus1995nonclassical}
L{\"u}tkenhaus N., Barnett S.M., \enquote{\emph{Nonclassical effects in phase
  space}}, Physical Review A, vol.~51, p. 3340, 1995.

\bibitem{mandel1986non}
Mandel L., \enquote{\emph{Non-classical states of the electromagnetic field}},
  Physica Scripta, vol. 1986, p.~34, 1986.

\bibitem{richter2002nonclassicality}
Richter T., Vogel W., \enquote{\emph{Nonclassicality of quantum states: a
  hierarchy of observable conditions}}, Physical Review Letters, vol.~89, p.
  283601, 2002.

\bibitem{bednorz2011fourth}
Bednorz A., Belzig W., \enquote{\emph{Fourth moments reveal the negativity of
  the {Wigner} function}}, Physical Review A, vol.~83, p. 052113, 2011.

\bibitem{bednorz2010quasiprobabilistic}
Bednorz A., Belzig W., \enquote{\emph{Quasiprobabilistic interpretation of weak
  measurements in mesoscopic junctions}}, Physical Review Letters, vol. 105, p.
  106803, 2010.

\bibitem{allevi2012measuring}
Allevi A., Olivares S., Bondani M., \enquote{\emph{Measuring high-order
  photon-number correlations in experiments with multimode pulsed quantum
  states}}, Physical Review A, vol.~85, p. 063835, 2012.

\bibitem{allevi2012high}
Allevi A., Olivares S., Bondani M., \enquote{\emph{High-order photon-number
  correlations: a resource for characterization and applications of quantum
  states}}, International Journal of Quantum Information, vol.~10, p. 1241003,
  2012.

\bibitem{avenhaus2010accessing}
Avenhaus M., Laiho K., Chekhova M.V., Silberhorn C., \enquote{\emph{Accessing
  higher order correlations in quantum optical states by time multiplexing}},
  Physical Review Letters, vol. 104, p. 063602, 2010.

\bibitem{kennard1927quantenmechanik}
Kennard E.H., \enquote{\emph{Zur {Quantenmechanik einfacher Bewegungstypen}}},
  Zeitschrift f{\"u}r Physik, vol.~44, pp. 326--352, 1927.

\bibitem{vogel1989determination}
Vogel K., Risken H., \enquote{\emph{Determination of quasiprobability
  distributions in terms of probability distributions for the rotated
  quadrature phase}}, Physical Review A, vol.~40, p. 2847, 1989.

\bibitem{hillery1987amplitude}
Hillery M., \enquote{\emph{Amplitude-squared squeezing of the electromagnetic
  field}}, Physical Review A, vol.~36, p. 3796, 1987.

\bibitem{hong1985higher}
Hong C.K., Mandel L., \enquote{\emph{Higher-order squeezing of a quantum
  field}}, Physical Review Letters, vol.~54, p. 323, 1985.

\bibitem{hong1985generation}
Hong C.K., Mandel L., \enquote{\emph{Generation of higher-order squeezing of
  quantum electromagnetic fields}}, Physical Review A, vol.~32, p. 974, 1985.

\bibitem{bennett1984quantum}
Bennett C.H., Brassard G., \enquote{\emph{Quantum cryptography: public key
  distribution and coin tossing}}, \emph{International Conference on Computer
  System and Signal Processing, IEEE, 1984}, pp. 175--179, 1984.

\bibitem{bennett1992B92}
Bennett C.H., \enquote{\emph{Quantum cryptography using any two non orthogonal
  states}}, Physical Review Letters, vol.~68, p. 3121, 1992.

\bibitem{sharma2016comparative}
Sharma V., Thapliyal K., Pathak A., Banerjee S., \enquote{\emph{A comparative
  study of protocols for secure quantum communication under noisy environment:
  single-qubit-based protocols versus entangled-state-based protocols}},
  Quantum Information Processing, vol.~15, pp. 4681--4710, 2016.

\bibitem{hu2016experimental}
Hu J.Y., Yu B., Jing M.Y., Xiao L.T., Jia S.T., Qin G.Q., Long G.L.,
  \enquote{\emph{Experimental quantum secure direct communication with single
  photons}}, Light: Science \& Applications, vol.~5, p. e16144, 2016.

\bibitem{pathak2010recent}
Pathak A., Verma A., \enquote{\emph{Recent developments in the study of higher
  order nonclassical states}}, Indian Journal of Physics, vol.~84, pp.
  1005--1019, 2010.

\bibitem{pathak2013elements}
Pathak A., \emph{Elements of Quantum Computation and Quantum Communication},
  Taylor \& Francis, New York, 2013.

\bibitem{brown1956correlation}
Brown R.H., Twiss R.Q., \enquote{\emph{Correlation between photons in two
  coherent beams of light}}, Nature, vol. 177, pp. 27--29, 1956.

\bibitem{kimble1976theory}
Kimble H.J., Mandel L., \enquote{\emph{Theory of resonance fluorescence}},
  Physical Review A, vol.~13, p. 2123, 1976.

\bibitem{carmichael1976proposal}
Carmichael H.J., Walls D.F., \enquote{\emph{Proposal for the measurement of the
  resonant {Stark} effect by photon correlation techniques}}, Journal of
  Physics B: Atomic and Molecular Physics, vol.~9, p. L43, 1976.

\bibitem{mandel1959fluctuations}
Mandel L., \enquote{\emph{Fluctuations of photon beams: the distribution of the
  photo-electrons}}, Proceedings of the Physical Society, vol.~74, p. 233,
  1959.

\bibitem{zou1990photon}
Zou X., Mandel L., \enquote{\emph{Photon-antibunching and sub-{Poissonian}
  photon statistics}}, Physical Review A, vol.~41, p. 475, 1990.

\bibitem{teich1983antibunching}
Teich M.C., Saleh B.E.A., Stoler D., \enquote{\emph{Antibunching in the
  {Franck-Hertz} experiment}}, Optics Communications, vol.~46, pp. 244--248,
  1983.

\bibitem{verma2010generalized}
Verma A., Pathak A., \enquote{\emph{Generalized structure of higher order
  nonclassicality}}, Physics Letters A, vol. 374, pp. 1009--1020, 2010.

\bibitem{thapliyal2017comparison}
Thapliyal K., Samantray N.L., Banerji J., Pathak A., \enquote{\emph{Comparison
  of lower- and higher-order nonclassicality in photon added and subtracted
  squeezed coherent states}}, Physics Letters A, vol. 381, pp. 3178 -- 3187,
  2017.

\bibitem{bachor2004guide}
Bachor H.A., Ralph T.C., \emph{A Guide to Experiments in Quantum Optics},
  Wiley-VCH, Weinheim, 2004.

\bibitem{brida2009characterization}
Brida G., Caricato V., Fedorov M.V., Genovese M., Gramegna M., Kulik S.P.,
  \enquote{\emph{Characterization of spectral entanglement of spontaneous
  parametric-down conversion biphotons in femtosecond pulsed regime}},
  Europhysics Letters, vol.~87, p. 64003, 2009.

\bibitem{ge2015conservation}
Ge W., Tasgin M.E., Zubairy M.S., \enquote{\emph{Conservation relation of
  nonclassicality and entanglement for {Gaussian} states in a beam splitter}},
  Physical Review A, vol.~92, p. 052328, 2015.

\bibitem{vogel2014unified}
Vogel W., Sperling J., \enquote{\emph{Unified quantification of nonclassicality
  and entanglement}}, Physical Review A, vol.~89, p. 052302, 2014.

\bibitem{boyd2003nonlinear}
Boyd R.W., \emph{Nonlinear Optics}, Elsevier, Delhi, 2003.

\bibitem{chang2014quantum}
Chang D.E., Vuleti{\'c} V., Lukin M.D., \enquote{\emph{Quantum nonlinear
  optics---photon by photon}}, Nature Photonics, vol.~8, pp. 685--694, 2014.

\bibitem{knill2000efficient}
Knill E., Laflamme R., Milburn G., \enquote{\emph{A scheme for efficient
  quantum computation with linear optics}}, Nature, vol. 409, pp. 46--52, 2001.

\bibitem{hardy1992source}
Hardy L., \enquote{\emph{Source of photons with correlated polarisations and
  correlated directions}}, Physics Letters A, vol. 161, pp. 326--328, 1992.

\bibitem{kwiat1999ultrabright}
Kwiat P.G., Waks E., White A.G., Appelbaum I., Eberhard P.H.,
  \enquote{\emph{Ultrabright source of polarization-entangled photons}},
  Physical Review A, vol.~60, p. R773, 1999.

\bibitem{pittman2002single}
Pittman T.B., Jacobs B.C., Franson J.D., \enquote{\emph{Single photons on
  pseudodemand from stored parametric down-conversion}}, Physical Review A,
  vol.~66, p. 042303, 2002.

\bibitem{migdall2002tailoring}
Migdall A.L., Branning D., Castelletto S., \enquote{\emph{Tailoring
  single-photon and multiphoton probabilities of a single-photon on-demand
  source}}, Physical Review A, vol.~66, p. 053805, 2002.

\bibitem{franken1961generation}
Franken P., Hill A.E., Peters C.e., Weinreich G., \enquote{\emph{Generation of
  optical harmonics}}, Physical Review Letters, vol.~7, p. 118, 1961.

\bibitem{kockum2017deterministic}
Kockum A.F., Miranowicz A., Macr\`{\i} V., Savasta S., Nori F.,
  \enquote{\emph{Deterministic quantum nonlinear optics with single atoms and
  virtual photons}}, Physical Review A, vol.~95, p. 063849, 2017.

\bibitem{hosten2006counterfactual}
Hosten O., Rakher M.T., Barreiro J.T., Peters N.A., Kwiat P.G.,
  \enquote{\emph{Counterfactual quantum computation through quantum
  interrogation}}, Nature, vol. 439, pp. 949--952, 2006.

\bibitem{kong2015experimental}
Kong F., Ju C., Huang P., Wang P., Kong X., Shi F., Jiang L., Du J.,
  \enquote{\emph{Experimental realization of high-efficiency counterfactual
  computation}}, Physical Review Letters, vol. 115, p. 080501, 2015.

\bibitem{salih2013protocol}
Salih H., Li Z.H., Al-Amri M., Zubairy M.S., \enquote{\emph{Protocol for direct
  counterfactual quantum communication}}, Physical Review Letters, vol. 110, p.
  170502, 2013.

\bibitem{cao2017direct}
Cao Y., Li Y.H., Cao Z., Yin J., Chen Y.A., Yin H.L., Chen T.Y., Ma X., Peng
  C.Z., Pan J.W., \enquote{\emph{Direct counterfactual communication via
  quantum {Zeno} effect}}, Proceedings of the National Academy of Sciences,
  vol. 114, pp. 4920--4924, 2017.

\bibitem{misra1977zeno}
Misra B., Sudarshan E.C.G., \enquote{\emph{The {Zeno's} paradox in quantum
  theory}}, Journal of Mathematical Physics, vol.~18, pp. 756--763, 1977.

\bibitem{venugopalan2007quantum}
Venugopalan A., \enquote{\emph{The quantum {Zeno} effect--watched pots in the
  quantum world}}, Resonance, vol.~12, pp. 52--68, 2007.

\bibitem{facchi2001quantum}
Facchi P., Pascazio S., \enquote{\emph{Quantum {Zeno} and inverse quantum
  {Zeno} effects}}, E.~Wolf (editor), \emph{Progress in Optics}, vol.~42,
  chap.~3, pp. 147--218, Elsevier, Amsterdam, 2001.

\bibitem{pascazio2014all}
Pascazio S., \enquote{\emph{All you ever wanted to know about the quantum
  {Zeno} effect in 70 minutes}}, Open Systems \& Information Dynamics, vol.~21,
  p. 1440007, 2014.

\bibitem{shen1967quantum}
Shen Y.R., \enquote{\emph{Quantum statistics of nonlinear optics}}, Physical
  Review, vol. 155, p. 921, 1967.

\bibitem{sen2005squeezed}
Sen B., Mandal S., \enquote{\emph{Squeezed states in spontaneous {Raman} and in
  stimulated {Raman} processes}}, Journal of Modern Optics, vol.~52, pp.
  1789--1807, 2005.

\bibitem{mandal2004approximate}
Mandal S., Pe{\v{r}}ina J., \enquote{\emph{Approximate quantum statistical
  properties of a nonlinear optical coupler}}, Physics Letters A, vol. 328, pp.
  144--156, 2004.

\bibitem{perina2000review}
Pe{\v{r}}ina~Jr. J., Pe{\v{r}}ina J., \enquote{\emph{Quantum statistics of
  nonlinear optical couplers}}, E.~Wolf (editor), \emph{Progress in Optics},
  vol.~41, pp. 361--419, Elsevier, Amsterdam, 2000.

\bibitem{sen2013intermodal}
Sen B., Giri S.K., Mandal S., Ooi C.H.R., Pathak A., \enquote{\emph{Intermodal
  entanglement in {Raman} processes}}, Physical Review A, vol.~87, p. 022325,
  2013.

\bibitem{giri2014single}
Giri S.K., Sen B., Ooi C.H.R., Pathak A., \enquote{\emph{Single-mode and
  intermodal higher-order nonclassicalities in two-mode {Bose-Einstein}
  condensates}}, Physical Review A, vol.~89, p. 033628, 2014.

\bibitem{thapliyal2014higher}
Thapliyal K., Pathak A., Sen B., Pe{\v{r}}ina J., \enquote{\emph{Higher-order
  nonclassicalities in a codirectional nonlinear optical coupler: quantum
  entanglement, squeezing, and antibunching}}, Physical Review A, vol.~90, p.
  013808, 2014.

\bibitem{giri2017nonclassicality}
Giri S.K., Thapliyal K., Sen B., Pathak A., \enquote{\emph{Nonclassicality in
  an atom--molecule {Bose--Einstein} condensate: Higher-order squeezing,
  antibunching and entanglement}}, Physica A: Statistical Mechanics and its
  Applications, vol. 466, pp. 140--152, 2017.

\bibitem{alam2017lower}
Alam N., Thapliyal K., Pathak A., Sen B., Verma A., Mandal S.,
  \enquote{\emph{Lower- and higher-order nonclassicality in a {Bose}-condensed
  optomechanical-like system and a {Fabry-Perot} cavity with one movable
  mirror: squeezing, antibunching and entanglement}}, arXiv preprint
  arXiv:1708.03967, 2017.

\bibitem{thapliyal2014nonclassical}
Thapliyal K., Pathak A., Sen B., Pe{\v{r}}ina J., \enquote{\emph{Nonclassical
  properties of a contradirectional nonlinear optical coupler}}, Physics
  Letters A, vol. 378, pp. 3431--3440, 2014.

\bibitem{mista2001non}
Mi{\v{s}}ta~Jr. L., Filip R., \enquote{\emph{Non-perturbative solution of
  nonlinear {Heisenberg} equations}}, Journal of Physics A: Mathematical and
  General, vol.~34, p. 5603, 2001.

\bibitem{breuer2002theory}
Breuer H.P., Petruccione F., \emph{The Theory of Open Quantum Systems}, Oxford
  University Press, New York, 2002.

\bibitem{caruso2014quantum}
Caruso F., Giovannetti V., Lupo C., Mancini S., \enquote{\emph{Quantum channels
  and memory effects}}, Reviews of Modern Physics, vol.~86, p. 1203, 2014.

\bibitem{stinespring1955positive}
Stinespring W.F., \enquote{\emph{Positive functions on {C*}-algebras}},
  Proceedings of the American Mathematical Society, vol.~6, pp. 211--216, 1955.

\bibitem{sudarshan1961stochastic}
Sudarshan E.C.G., Mathews P.M., Rau J., \enquote{\emph{Stochastic dynamics of
  quantum-mechanical systems}}, Physical Review, vol. 121, p. 920, 1961.

\bibitem{kraus1971general}
Kraus K., \enquote{\emph{General state changes in quantum theory}}, Annals of
  Physics, vol.~64, pp. 311--335, 1971.

\bibitem{nielsen2010quantum}
Nielsen M.A., Chuang I.L., \emph{Quantum Computation and Quantum Information},
  Cambridge University Press, New Delhi, 2010.

\bibitem{preskill1998lecture}
Preskill J., \enquote{\emph{Lecture notes for physics 229: quantum information
  and computation}}, California Institute of Technology, vol.~12, 1998.

\bibitem{rajagopal2010kraus}
Rajagopal A.K., Usha~Devi A.R., Rendell R.W., \enquote{\emph{Kraus
  representation of quantum evolution and fidelity as manifestations of
  {Markovian} and non-{Markovian} forms}}, Physical Review A, vol.~82, p.
  042107, 2010.

\bibitem{lindblad1976generators}
Lindblad G., \enquote{\emph{On the generators of quantum dynamical
  semigroups}}, Communications in Mathematical Physics, vol.~48, pp. 119--130,
  1976.

\bibitem{gorini1976completely}
Gorini V., Kossakowski A., Sudarshan E.C.G., \enquote{\emph{Completely positive
  dynamical semigroups of n-level systems}}, Journal of Mathematical Physics,
  vol.~17, pp. 821--825, 1976.

\bibitem{brune1996observing}
Brune M., Hagley E., Dreyer J., Maitre X., Maali A., Wunderlich C., Raimond J.,
  Haroche S., \enquote{\emph{Observing the progressive decoherence of the
  “meter” in a quantum measurement}}, Physical Review Letters, vol.~77, p.
  4887, 1996.

\bibitem{turchette2000decoherence}
Turchette Q.A., Myatt C.J., King B.E., Sackett C.A., Kielpinski D., Itano W.M.,
  Monroe C., Wineland D.J., \enquote{\emph{Decoherence and decay of motional
  quantum states of a trapped atom coupled to engineered reservoirs}}, Physical
  Review A, vol.~62, p. 053807, 2000.

\bibitem{myatt2000decoherence}
Myatt C.J., King B.E., Turchette Q.A., Sackett C.A., Kielpinski D., Itano W.M.,
  Monroe C., Wineland D.J., \enquote{\emph{Decoherence of quantum
  superpositions through coupling to engineered reservoirs}}, Nature, vol. 403,
  pp. 269--273, 2000.

\bibitem{nakajima1958quantum}
Nakajima S., \enquote{\emph{On quantum theory of transport phenomena: steady
  diffusion}}, Progress of Theoretical Physics, vol.~20, pp. 948--959, 1958.

\bibitem{zwanzig1960ensemble}
Zwanzig R., \enquote{\emph{Ensemble method in the theory of irreversibility}},
  The Journal of Chemical Physics, vol.~33, pp. 1338--1341, 1960.

\bibitem{garraway1997nonperturbative}
Garraway B.M., \enquote{\emph{Nonperturbative decay of an atomic system in a
  cavity}}, Physical Review A, vol.~55, p. 2290, 1997.

\bibitem{hillery2000quantum}
Hillery M., \enquote{\emph{Quantum cryptography with squeezed states}},
  Physical Review A, vol.~61, p. 022309, 2000.

\bibitem{furusawa1998unconditional}
Furusawa A., S{\o}rensen J.L., Braunstein S.L., Fuchs C.A., Kimble H.J., Polzik
  E.S., \enquote{\emph{Unconditional quantum teleportation}}, Science, vol.
  282, pp. 706--709, 1998.

\bibitem{yuan2002electrically}
Yuan Z., Kardynal B.E., Stevenson R.M., Shields A.J., Lobo C.J., Cooper K.,
  Beattie N.S., Ritchie D.A., Pepper M., \enquote{\emph{Electrically driven
  single-photon source}}, Science, vol. 295, pp. 102--105, 2002.

\bibitem{ekert1991quantum}
Ekert A.K., \enquote{\emph{Quantum cryptography based on {Bell's} theorem}},
  Physical Review Letters, vol.~67, p. 661, 1991.

\bibitem{bennett1993teleporting}
Bennett C.H., Brassard G., Cr{\'e}peau C., Jozsa R., Peres A., Wootters W.K.,
  \enquote{\emph{Teleporting an unknown quantum state via dual classical and
  {Einstein-Podolsky-Rosen} channels}}, Physical Review Letters, vol.~70, p.
  1895, 1993.

\bibitem{bennett1992communication}
Bennett C.H., Wiesner S.J., \enquote{\emph{Communication via one-and
  two-particle operators on {Einstein-Podolsky-Rosen} states}}, Physical Review
  Letters, vol.~69, p. 2881, 1992.

\bibitem{aasi2013enhanced}
Aasi J., Abadie J., Abbott B.P., Abbott R., Abbott T.D., Abernathy M.R., Adams
  C., Adams T., Addesso P., Adhikari R.X., et~al., \enquote{\emph{Enhanced
  sensitivity of the {LIGO} gravitational wave detector by using squeezed
  states of light}}, Nature Photonics, vol.~7, pp. 613--619, 2013.

\bibitem{grote2013first}
Grote H., Danzmann K., Dooley K.L., Schnabel R., Slutsky J., Vahlbruch H.,
  \enquote{\emph{First long-term application of squeezed states of light in a
  gravitational-wave observatory}}, Physical Review Letters, vol. 110, p.
  181101, 2013.

\bibitem{abbott2016observation}
Abbott B.P., Abbott R., Abbott T.D., Abernathy M.R., Acernese F., Ackley K.,
  Adams C., Adams T., Addesso P., Adhikari R., et~al.,
  \enquote{\emph{Observation of gravitational waves from a binary black hole
  merger}}, Physical Review Letters, vol. 116, p. 061102, 2016.

\bibitem{abbott2016gw151226}
Abbott B.P., Abbott R., Abbott T.D., Abernathy M.R., Acernese F., Ackley K.,
  Adams C., Adams T., Addesso P., Adhikari R.X., et~al.,
  \enquote{\emph{{GW151226:} observation of gravitational waves from a
  22-solar-mass binary black hole coalescence}}, Physical Review Letters, vol.
  116, p. 241103, 2016.

\bibitem{acin2006bell}
Acin A., Gisin N., Masanes L., \enquote{\emph{From {Bell's} theorem to secure
  quantum key distribution}}, Physical Review Letters, vol.~97, p. 120405,
  2006.

\bibitem{o2007optical}
O'brien J.L., \enquote{\emph{Optical quantum computing}}, Science, vol. 318,
  pp. 1567--1570, 2007.

\bibitem{ladd2010quantum}
Ladd T.D., Jelezko F., Laflamme R., Nakamura Y., Monroe C., O'Brien J.L.,
  \enquote{\emph{Quantum computing}}, Nature, vol. 464, pp. 45--53, 2010.

\bibitem{proctor2017ancilla}
Proctor T., Giulian M., Korolkova N., Andersson E., Kendon V.,
  \enquote{\emph{Ancilla-driven quantum computation for qudits and continuous
  variables}}, Physical Review A, vol.~95, p. 052317, 2017.

\bibitem{andersen2015hybrid}
Andersen U.L., Neergaard-Nielsen J.S., van Loock P., Furusawa A.,
  \enquote{\emph{Hybrid discrete- and continuous-variable quantum
  information}}, Nature Physics, vol.~11, pp. 713--719, 2015.

\bibitem{pu2017experimental}
Pu Y.F., Jiang N., Chang W., Yang H.X., Li C., Duan L.M.,
  \enquote{\emph{Experimental realization of a multiplexed quantum memory with
  225 individually accessible memory cells}}, Nature Communications, vol.~8, p.
  15359, 2017.

\bibitem{browne2017quantum}
Browne D., Bose S., Mintert F., Kim M.S., \enquote{\emph{From quantum optics to
  quantum technologies}}, Progress in Quantum Electronics, vol.~54, pp. 2--18,
  2017.

\bibitem{biamonte2017quantum}
Biamonte J., Wittek P., Pancotti N., Rebentrost P., Wiebe N., Lloyd S.,
  \enquote{\emph{Quantum machine learning}}, Nature, vol. 549, pp. 195--202,
  2017.

\bibitem{thapliyal2016linear}
Thapliyal K., Pathak A., Pe{\v{r}}ina J., \enquote{\emph{Linear and nonlinear
  quantum {Zeno and anti-Zeno} effects in a nonlinear optical coupler}},
  Physical Review A, vol.~93, p. 022107, 2016.

\bibitem{thapliyal2015quantum}
Thapliyal K., Pathak A., \enquote{\emph{Quantum {Zeno} and anti-{Zeno} effects
  in an asymmetric nonlinear optical coupler}}, \emph{International Conference
  on Optics and Photonics 2015}, pp. 96541F--96541F, International Society for
  Optics and Photonics, 2015.

\bibitem{thapliyal2015quasiprobability}
Thapliyal K., Banerjee S., Pathak A., Omkar S., Ravishankar V.,
  \enquote{\emph{Quasiprobability distributions in open quantum systems:
  spin-qubit systems}}, Annals of Physics, vol. 362, pp. 261--286, 2015.

\bibitem{thapliyal2016tomograms}
Thapliyal K., Banerjee S., Pathak A., \enquote{\emph{Tomograms for open quantum
  systems: in (finite) dimensional optical and spin systems}}, Annals of
  Physics, vol. 366, pp. 148--167, 2016.

\bibitem{thapliyal2015applications}
Thapliyal K., Pathak A., \enquote{\emph{Applications of quantum cryptographic
  switch: various tasks related to controlled quantum communication can be
  performed using {Bell} states and permutation of particles}}, Quantum
  Information Processing, vol.~14, pp. 2599--2616, 2015.

\bibitem{thapliyal2015general}
Thapliyal K., Verma A., Pathak A., \enquote{\emph{A general method for
  selecting quantum channel for bidirectional controlled state teleportation
  and other schemes of controlled quantum communication}}, Quantum Information
  Processing, vol.~14, pp. 4601--4614, 2015.

\bibitem{thapliyal2017quantum}
Thapliyal K., Pathak A., Banerjee S., \enquote{\emph{Quantum cryptography over
  non-{Markovian} channels}}, Quantum Information Processing, vol.~16, p. 115,
  2017.

\bibitem{pathak2013nonclassicality}
Pathak A., K{\v{r}}epelka J., Pe{\v{r}}ina J., \enquote{\emph{Nonclassicality
  in {Raman} scattering: quantum entanglement, squeezing of vacuum
  fluctuations, sub-shot noise and joint photon--phonon number and
  integrated-intensity distributions}}, Physics Letters A, vol. 377, pp.
  2692--2701, 2013.

\bibitem{matthews2009manipulation}
Matthews J.C.F., Politi A., Stefanov A., O'Brien J.L.,
  \enquote{\emph{Manipulation of multiphoton entanglement in waveguide quantum
  circuits}}, Nature Photonics, vol.~3, pp. 346--350, 2009.

\bibitem{li2011reconfigurable}
Li H.W., Przeslak S., Niskanen A.O., Matthews J.C.F., Politi A., Shadbolt P.,
  Laing A., Lobino M., Thompson M.G., O'Brien J.L.,
  \enquote{\emph{Reconfigurable controlled two-qubit operation on a quantum
  photonic chip}}, New Journal of Physics, vol.~13, p. 115009, 2011.

\bibitem{politi2009shor}
Politi A., Matthews J.C.F., O'Brien J.L., \enquote{\emph{Shor's quantum
  factoring algorithm on a photonic chip}}, Science, vol. 325, pp. 1221--1221,
  2009.

\bibitem{mandal2011all}
Mandal P., Midda S., \enquote{\emph{All optical method of developing {OR} and
  {NAND} logic system based on nonlinear optical fiber couplers}},
  Optik-International Journal for Light and Electron Optics, vol. 122, pp.
  1795--1798, 2011.

\bibitem{lugani2013studies}
Lugani J., \emph{Studies on guided wave devices and ultracold atoms in optical
  cavity for applications in quantum optics and quantum information}, Ph.D.
  thesis, Indian Institute of Technology Delhi, 2013.
  \url{http://web.iitd.ac.in/~sankalpa/Jasleen_corrected_thesis.pdf}.

\bibitem{tanzilli2012genesis}
Tanzilli S., Martin A., Kaiser F., De~Micheli M.P., Alibart O., Ostrowsky D.B.,
  \enquote{\emph{On the genesis and evolution of integrated quantum optics}},
  Laser \& Photonics Reviews, vol.~6, pp. 115--143, 2012.

\bibitem{el2005quantum}
El-Orany F.A.A., Sebawe~Abdalla M., Pe{\v{r}}ina J., \enquote{\emph{Quantum
  properties of the codirectional three-mode {Kerr} nonlinear coupler}}, The
  European Physical Journal D-Atomic, Molecular, Optical and Plasma Physics,
  vol.~33, pp. 453--463, 2005.

\bibitem{korolkova1997quantumKerr}
Korolkova N., Pe{\v{r}}ina J., \enquote{\emph{Quantum statistics and dynamics
  of {Kerr} nonlinear couplers}}, Optics Communications, vol. 136, pp.
  135--149, 1997.

\bibitem{fiuravsek1999quantum}
Fiur{\'a}{\v{s}}ek J., K{\v{r}}epelka J., Pe{\v{r}}ina J.,
  \enquote{\emph{Quantum-phase properties of the {Kerr} couplers}}, Optics
  Communications, vol. 167, pp. 115--124, 1999.

\bibitem{ariunbold2000quantum}
Ariunbold G., Pe{\v{r}}ina J., \enquote{\emph{Quantum statistics of
  contradirectional {Kerr} nonlinear couplers}}, Optics Communications, vol.
  176, pp. 149--154, 2000.

\bibitem{korolkova1997kerr}
Korolkova N., Pe{\v{r}}ina J., \enquote{\emph{{Kerr} nonlinear coupler with
  varying linear coupling coefficient}}, Journal of Modern Optics, vol.~44, pp.
  1525--1534, 1997.

\bibitem{perina1997statistics}
Pe{\v{r}}ina~Jr. J., Pe{\v{r}}ina J., \enquote{\emph{Statistics of light in
  {Raman} and {Brillouin} nonlinear couplers}}, Quantum and Semiclassical
  Optics: Journal of the European Optical Society Part B, vol.~9, p. 443, 1997.

\bibitem{korolkova1997quantum}
Korolkova N., Pe{\v{r}}ina J., \enquote{\emph{Quantum statistics of symmetrical
  optical parametric nonlinear coupler}}, Optics Communications, vol. 137, pp.
  263--268, 1997.

\bibitem{perina1995asy}
Pe{\v{r}}ina J., Pe{\v{r}}ina~Jr. J., \enquote{\emph{Quantum statistics of a
  nonlinear asymmetric coupler with strong pumping}}, Quantum and Semiclassical
  Optics: Journal of the European Optical Society Part B, vol.~7, p. 541, 1995.

\bibitem{perina1995quantum}
Pe{\v{r}}ina J., \enquote{\emph{Quantum-statistical properties of a nonlinear
  asymmetric directional coupler}}, Journal of Modern Optics, vol.~42, pp.
  1517--1522, 1995.

\bibitem{perina1995photon}
Pe{\v{r}}ina J., Pe{\v{r}}ina~Jr. J., \enquote{\emph{Photon statistics of a
  contradirectional nonlinear coupler}}, Quantum and Semiclassical Optics:
  Journal of the European Optical Society Part B, vol.~7, p. 849, 1995.

\bibitem{perina1995non}
Pe{\v{r}}ina J., Bajer J., \enquote{\emph{Non-classical light in nonlinear
  symmetric and asymmetric couplers}}, Journal of Modern Optics, vol.~42, pp.
  2337--2346, 1995.

\bibitem{perina1996quantum}
Pe{\v{r}}ina J., Pe{\v{r}}ina~Jr. J., \enquote{\emph{Quantum statistics and
  dynamics of nonlinear couplers}}, Journal of Modern Optics, vol.~43, pp.
  1951--1971, 1996.

\bibitem{mista1997nonclassical}
Mi{\v{s}}ta~Jr. L., Pe{\v{r}}ina J., \enquote{\emph{Nonclassical light in
  symmetric nonlinear directional coupler}}, Czechoslovak Journal of Physics,
  vol.~47, pp. 629--636, 1997.

\bibitem{perina1995quantum863}
Pe{\v{r}}ina J., Pe{\v{r}}ina~Jr. J., \enquote{\emph{Quantum statistical
  properties of codirectional and contradirectional nonlinear couplers with
  phase mismatch}}, Quantum and Semiclassical Optics: Journal of the European
  Optical Society Part B, vol.~7, p. 863, 1995.

\bibitem{kowalewska2009sudden}
Kowalewska-Kud{\l}aszyk A., Leo{\'n}ski W., \enquote{\emph{Sudden death and
  birth of entanglement effects for {Kerr-nonlinear} coupler}}, Journal of the
  Optical Society of America B, vol.~26, pp. 1289--1294, 2009.

\bibitem{abbasi2013thermal}
Abbasi M.R., Golshan M.M., \enquote{\emph{Thermal entanglement of a two-level
  atom and bimodal photons in a {Kerr} nonlinear coupler}}, Physica A:
  Statistical Mechanics and its Applications, vol. 392, pp. 6161--6167, 2013.

\bibitem{leonski2004kerr}
Leo{\'n}ski W., Miranowicz A., \enquote{\emph{{Kerr} nonlinear coupler and
  entanglement}}, Journal of Optics B: Quantum and Semiclassical Optics,
  vol.~6, p. S37, 2004.

\bibitem{miranowicz2006two}
Miranowicz A., Leo{\'n}ski W., \enquote{\emph{Two-mode optical state truncation
  and generation of maximally entangled states in pumped nonlinear couplers}},
  Journal of Physics B: Atomic, Molecular and Optical Physics, vol.~39, p.
  1683, 2006.

\bibitem{kowalewska2012generalized}
Kowalewska-Kud{\l}aszyk A., Leo{\'n}ski W., Pe{\v{r}}ina~Jr. J.,
  \enquote{\emph{Generalized {Bell} states generation in a parametrically
  excited nonlinear coupler}}, Physica Scripta, vol. 2012, p. 014016, 2012.

\bibitem{el2004higher}
El-Orany F.A.A., Pe{\v{r}}ina J., \enquote{\emph{Higher-order squeezing for the
  codirectional {Kerr} nonlinear coupler}}, Physics Letters A, vol. 333, pp.
  204--211, 2004.

\bibitem{perina2017higher}
Pe{\v{r}}ina~Jr. J., Mich{\'a}lek V., Haderka O., \enquote{\emph{Higher-order
  sub-{Poissonian}-like nonclassical fields: Theoretical and experimental
  comparison}}, Physical Review A, vol.~96, p. 033852, 2017.

\bibitem{pathak2006control}
Pathak A., Garcia M.E., \enquote{\emph{Control of higher order antibunching}},
  Applied Physics B, vol.~84, pp. 479--484, 2006.

\bibitem{lee2009demonstrating}
Lee J., Kim J., Nha H., \enquote{\emph{Demonstrating higher-order nonclassical
  effects by photon-added classical states: realistic schemes}}, Journal of the
  Optical Society of America B, vol.~26, pp. 1363--1369, 2009.

\bibitem{rundquist2014nonclassical}
Rundquist A., Bajcsy M., Majumdar A., Sarmiento T., Fischer K., Lagoudakis
  K.G., Buckley S., Piggott A.Y., Vu{\v{c}}kovi{\'c} J.,
  \enquote{\emph{Nonclassical higher-order photon correlations with a quantum
  dot strongly coupled to a photonic-crystal nanocavity}}, Physical Review A,
  vol.~90, p. 023846, 2014.

\bibitem{bohmann2017higher}
Bohmann M., Sperling J., Semenov A.A., Vogel W., \enquote{\emph{Higher-order
  nonclassical effects in fluctuating-loss channels}}, Physical Review A,
  vol.~95, p. 012324, 2017.

\bibitem{loudon1987squeezed}
Loudon R., Knight P.L., \enquote{\emph{Squeezed light}}, Journal of Modern
  Optics, vol.~34, pp. 709--759, 1987.

\bibitem{miranowicz2010testing}
Miranowicz A., Bartkowiak M., Wang X., Liu Y.x., Nori F.,
  \enquote{\emph{Testing nonclassicality in multimode fields: a unified
  derivation of classical inequalities}}, Physical Review A, vol.~82, p.
  013824, 2010.

\bibitem{lee1990higher}
Lee C.T., \enquote{\emph{Higher-order criteria for nonclassical effects in
  photon statistics}}, Physical Review A, vol.~41, p. 1721, 1990.

\bibitem{gupta2006higher}
Gupta P., Pandey P.N., Pathak A., \enquote{\emph{Higher order antibunching is
  not a rare phenomenon}}, Journal of Physics B: Atomic, Molecular and Optical
  Physics, vol.~39, p. 1137, 2006.

\bibitem{an2002multimode}
An N.B., \enquote{\emph{Multimode higher-order antibunching and squeezing in
  trio coherent states}}, Journal of Optics B: Quantum and Semiclassical
  Optics, vol.~4, p. 222, 2002.

\bibitem{verma2008higher}
Verma A., Sharma N.K., Pathak A., \enquote{\emph{Higher order antibunching in
  intermediate states}}, Physics Letters A, vol. 372, pp. 5542--5551, 2008.

\bibitem{perinova2013dynamics}
Pe{\v{r}}inov{\'a} V., Luk{\v{s}} A., K{\v{r}}epelka J.,
  \enquote{\emph{Dynamics of nonclassical properties of two-and four-mode
  {Bose-Einstein} condensates}}, Journal of Physics B: Atomic, Molecular and
  Optical Physics, vol.~46, p. 195301, 2013.

\bibitem{agarwal2005inseparability}
Agarwal G.S., Biswas A., \enquote{\emph{Inseparability inequalities for higher
  order moments for bipartite systems}}, New Journal of Physics, vol.~7, p.
  211, 2005.

\bibitem{duan2000inseparability}
Duan L.M., Giedke G., Cirac J.I., Zoller P., \enquote{\emph{Inseparability
  criterion for continuous variable systems}}, Physical Review Letters,
  vol.~84, p. 2722, 2000.

\bibitem{hillery2006entanglement}
Hillery M., Zubairy M.S., \enquote{\emph{Entanglement conditions for two-mode
  states}}, Physical Review Letters, vol.~96, p. 050503, 2006.

\bibitem{hillery2006entanglementapplications}
Hillery M., Zubairy M.S., \enquote{\emph{Entanglement conditions for two-mode
  states: applications}}, Physical Review A, vol.~74, p. 032333, 2006.

\bibitem{hillery2010conditions}
Hillery M., Dung H.T., Zheng H., \enquote{\emph{Conditions for entanglement in
  multipartite systems}}, Physical Review A, vol.~81, p. 062322, 2010.

\bibitem{shchukin2005inseparability}
Shchukin E., Vogel W., \enquote{\emph{Inseparability criteria for continuous
  bipartite quantum states}}, Physical Review Letters, vol.~95, p. 230502,
  2005.

\bibitem{miranowicz2009inseparability}
Miranowicz A., Piani M., Horodecki P., Horodecki R.,
  \enquote{\emph{Inseparability criteria based on matrices of moments}},
  Physical Review A, vol.~80, p. 052303, 2009.

\bibitem{zeilinger1992higher}
Zeilinger A., Horne M.A., Greenberger D.M., \enquote{\emph{Higher-order quantum
  entanglement}}, \emph{Workshop on Squeezed States and Uncertainty Relations},
  vol. 3135, pp. 73--81, NASA Conference Publication, 1992.

\bibitem{pan2001experimental}
Pan J.W., Daniell M., Gasparoni S., Weihs G., Zeilinger A.,
  \enquote{\emph{Experimental demonstration of four-photon entanglement and
  high-fidelity teleportation}}, Physical Review Letters, vol.~86, p. 4435,
  2001.

\bibitem{mair2001entanglement}
Mair A., Vaziri A., Weihs G., Zeilinger A., \enquote{\emph{Entanglement of the
  orbital angular momentum states of photons}}, Nature, vol. 412, pp. 313--316,
  2001.

\bibitem{li2007entanglement}
Li Z.G., Fei S.M., Wang Z.X., Wu K., \enquote{\emph{Entanglement conditions for
  multimode states}}, Physical Review A, vol.~75, p. 012311, 2007.

\bibitem{jakob2001comparative}
Jakob M., Abranyos Y., Bergou J.A., \enquote{\emph{A comparative study of
  hyperbunching in the fluorescence from a bichromatically driven atom}},
  Journal of Optics B: Quantum and Semiclassical Optics, vol.~3, p. 130, 2001.

\bibitem{an1999general}
An N.B., Tinh V., \enquote{\emph{General multimode sum-squeezing}}, Physics
  Letters A, vol. 261, pp. 34--39, 1999.

\bibitem{an2000general}
An N.B., Tinh V., \enquote{\emph{General multimode difference-squeezing}},
  Physics Letters A, vol. 270, pp. 27--40, 2000.

\bibitem{hillery1989sum}
Hillery M., \enquote{\emph{Sum and difference squeezing of the electromagnetic
  field}}, Physical Review A, vol.~40, p. 3147, 1989.

\bibitem{mancini2002entangling}
Mancini S., Giovannetti V., Vitali D., Tombesi P., \enquote{\emph{Entangling
  macroscopic oscillators exploiting radiation pressure}}, Physical Review
  Letters, vol.~88, p. 120401, 2002.

\bibitem{simon2000peres}
Simon R., \enquote{\emph{{Peres-Horodecki} separability criterion for
  continuous variable systems}}, Physical Review Letters, vol.~84, p. 2726,
  2000.

\bibitem{agarwal1992nonclassical}
Agarwal G.S., Tara K., \enquote{\emph{Nonclassical character of states
  exhibiting no squeezing or sub-{Poissonian} statistics}}, Physical Review A,
  vol.~46, p. 485, 1992.

\bibitem{he2012einstein}
He Q., Drummond P., Olsen M., Reid M., \enquote{\emph{{Einstein-Podolsky-Rosen}
  entanglement and steering in two-well {Bose-Einstein}-condensate ground
  states}}, Physical Review A, vol.~86, p. 023626, 2012.

\bibitem{skrzypczyk2014quantifying}
Skrzypczyk P., Navascu{\'e}s M., Cavalcanti D., \enquote{\emph{Quantifying
  {Einstein-Podolsky-Rosen} steering}}, Physical Review Letters, vol. 112, p.
  180404, 2014.

\bibitem{shchukin2005nonclassical}
Shchukin E., Vogel W., \enquote{\emph{Nonclassical moments and their
  measurement}}, Physical Review A, vol.~72, p. 043808, 2005.

\bibitem{shchukin2006universal}
Shchukin E., Vogel W., \enquote{\emph{Universal measurement of quantum
  correlations of radiation}}, Physical Review Letters, vol.~96, p. 200403,
  2006.

\bibitem{prakash2012proposal}
Prakash R., Yadav A.K., \enquote{\emph{A proposal for experimental detection of
  amplitude nth-power squeezing}}, Optics Communications, vol. 285, pp.
  2387--2391, 2012.

\bibitem{prakash2010detection}
Prakash H., Kumar P., Mishra D.K., \enquote{\emph{Detection of
  amplitude-squared squeezing via homodyne method}}, International Journal of
  Modern Physics B, vol.~24, pp. 5547--5551, 2010.

\bibitem{prakash2006higher}
Prakash H., Mishra D.K., \enquote{\emph{Higher order sub-{Poissonian} photon
  statistics and their use in detection of hong and mandel squeezing and
  amplitude-squared squeezing}}, Journal of Physics B: Atomic, Molecular and
  Optical Physics, vol.~39, p. 2291, 2006.

\bibitem{mandel1986coherence}
Mandel L., Hong C.K., \enquote{\emph{A test for high order squeezing of a
  quantum electromagnetic field}}, F.~Hakke, L.M. Narducci, D.F. Walls
  (editors), \emph{Coherence, Cooperation and Fluctuations}, Cambridge
  University Press, New York, 1986. {p.} 254.

\bibitem{thapliyal2017nonclassicality}
Thapliyal K., Pathak A., Sen B., Pe{\v{r}}ina J.,
  \enquote{\emph{Nonclassicality in non-degenerate hyper-{Raman} processes}},
  arXiv preprint arXiv:1710.04456, 2017.

\bibitem{naikoo2017probing}
Naikoo J., Thapliyal K., Pathak A., Banerjee S., \enquote{\emph{Probing
  nonclassicality in an optically-driven cavity with two atomic ensembles}},
  arXiv preprint arXiv:1712.04154, 2017.

\bibitem{khalfin1957theory}
Khalfin L.A., \enquote{\emph{On the theory of the decay of a quasi-stationary
  state}}, Doklady Akademii Nauk SSSR, vol. 115, p. 277, 1957. [Soviet Physics
  Doklady vol. 2, p. 232 (1958)].

\bibitem{khalfin1958contribution}
Khalfin L.A., \enquote{\emph{Contribution to the decay theory of a
  quasi-stationary state}}, Soviet Physics JETP, vol.~6, pp. 1053--1063, 1958.
  [Zhurnal Eksperimental'noi i Teoreticheskoi Fiziki vol. 33, pp. 1371--1382
  (1958)].

\bibitem{khalfin1961quantum}
Khalfin L.A., \enquote{\emph{The quantum theory of unstable elementary
  particles}}, Doklady Akademii Nauk SSSR, vol. 141, p. 599, 1961. [Soviet
  Physics Doklady vol. 6, p. 1010 (1962)].

\bibitem{facchi2003three}
Facchi P., Pascazio S., \enquote{\emph{Three different manifestations of the
  quantum {Zeno} effect}}, F.~Benatti, R.~Floreanini (editors),
  \emph{Irreversible Quantum Dynamics}, pp. 141--156, Lecture Notes in Physics
  (Springer-Verlag Berlin Heidelberg), 2003.

\bibitem{abdullaev2011linear}
Abdullaev F.K., Konotop V., Shchesnovich V., \enquote{\emph{Linear and
  nonlinear {Zeno} effects in an optical coupler}}, Physical Review A, vol.~83,
  p. 043811, 2011.

\bibitem{rehacek2001quantum}
{\v{R}}eh{\'a}{\v{c}}ek J., Pe{\v{r}}ina J., Facchi P., Pascazio S.,
  Mi{\v{s}}ta~Jr. L., \enquote{\emph{Quantum {Zeno} effect in a nonlinear
  coupler}}, Optics and Spectroscopy, vol.~91, pp. 501--507, 2001.

\bibitem{thun2002quantum}
Thun K., Pe{\v{r}}ina J., K{\v{r}}epelka J., \enquote{\emph{Quantum {Zeno}
  effect in {Raman} scattering}}, Physics Letters A, vol. 299, pp. 19--30,
  2002.

\bibitem{mista2000quantum}
Mi{\v{s}}ta~Jr. L., Jel{\'\i}nek V., {\v{R}}eh{\'a}{\v{c}}ek J., Pe{\v{r}}ina
  J., \enquote{\emph{Quantum dynamics and statistics of two coupled
  down-conversion processes}}, Journal of Optics B: Quantum and Semiclassical
  Optics, vol.~2, p. 726, 2000.

\bibitem{rehacek2000quantum}
{\v{R}}eh{\'a}{\v{c}}ek J., Pe{\v{r}}ina J., Facchi P., Pascazio S.,
  Mi{\v{s}}ta L., \enquote{\emph{Quantum {Zeno} effect in a nonlinear
  coupler}}, Physical Review A, vol.~62, p. 013804, 2000.

\bibitem{luis1996zeno}
Luis A., Pe{\v{r}}ina J., \enquote{\emph{{Zeno} effect in parametric
  down-conversion}}, Physical Review Letters, vol.~76, p. 4340, 1996.

\bibitem{luis1998anti}
Luis A., S{\'a}nchez-Soto L.L., \enquote{\emph{Anti-{Zeno} effect in parametric
  down-conversion}}, Physical Review A, vol.~57, p. 781, 1998.

\bibitem{perina2004quantum}
Pe{\v{r}}ina J., \enquote{\emph{Quantum {Zeno} effect in cascaded parametric
  down-conversion with losses}}, Physics Letters A, vol. 325, pp. 16--20, 2004.

\bibitem{agarwal1994all}
Agarwal G.S., Tewari S.P., \enquote{\emph{An all-optical realization of the
  quantum {Zeno} effect}}, Physics Letters A, vol. 185, pp. 139--142, 1994.

\bibitem{kwiat1999high}
Kwiat P.G., White A.G., Mitchell J.R., Nairz O., Weihs G., Weinfurter H.,
  Zeilinger A., \enquote{\emph{High-efficiency quantum interrogation
  measurements via the quantum {Zeno} effect}}, Physical Review Letters,
  vol.~83, p. 4725, 1999.

\bibitem{itano1990quantum}
Itano W.M., Heinzen D.J., Bollinger J.J., Wineland D.J., \enquote{\emph{Quantum
  {Zeno} effect}}, Physical Review A, vol.~41, p. 2295, 1990.

\bibitem{fischer2001observation}
Fischer M.C., Guti{\'e}rrez-Medina B., Raizen M.G., \enquote{\emph{Observation
  of the quantum {Zeno} and anti-{Zeno} effects in an unstable system}},
  Physical Review Letters, vol.~87, p. 040402, 2001.

\bibitem{facchi2002quantum}
Facchi P., Hradil Z., Krenn G., Pascazio S., {\v{R}}eh{\'a}{\v{c}}ek J.,
  \enquote{\emph{Quantum {Zeno} tomography}}, Physical Review A, vol.~66, p.
  012110, 2002.

\bibitem{pascazio2001quantum}
Pascazio S., Facchi P., Hradil Z., Krenn G., {\v{R}}eh{\'a}{\v{c}}ek J.,
  \enquote{\emph{Quantum {Zeno} tomography}}, Fortschritte der Physik, vol.~49,
  pp. 1071--1076, 2001.

\bibitem{nikolic2014suppressing}
Nikoli{\'c} H., \enquote{\emph{Suppressing {Hawking} radiation by quantum
  {Zeno} effect}}, Physics Letters B, vol. 733, pp. 6--10, 2014.

\bibitem{zezyulin2012macroscopic}
Zezyulin D.A., Konotop V.V., Barontini G., Ott H., \enquote{\emph{Macroscopic
  {Zeno} effect and stationary flows in nonlinear waveguides with localized
  dissipation}}, Physical Review Letters, vol. 109, p. 020405, 2012.

\bibitem{beige2000quantum}
Beige A., Braun D., Tregenna B., Knight P.L., \enquote{\emph{Quantum computing
  using dissipation to remain in a decoherence-free subspace}}, Physical Review
  Letters, vol.~85, p. 1762, 2000.

\bibitem{tavakoli2015quantum}
Tavakoli A., Anwer H., Hameedi A., Bourennane M., \enquote{\emph{Quantum
  communication complexity using the quantum {Zeno} effect}}, Physical Review
  A, vol.~92, p. 012303, 2015.

\bibitem{supplZeno}
\enquote{\emph{See supplemental material at}},
  \url{http://link.aps.org/supplemental/10.1103/PhysRevA.93.022107}.

\bibitem{li2006continuous}
Li W.D., Liu J., \enquote{\emph{Continuous-measurement-enhanced self-trapping
  of degenerate ultracold atoms in a double well: nonlinear quantum {Zeno}
  effect}}, Physical Review A, vol.~74, p. 063613, 2006.

\bibitem{shchesnovich2010control}
Shchesnovich V., Konotop V., \enquote{\emph{Control of a {Bose-Einstein}
  condensate by dissipation: nonlinear {Zeno} effect}}, Physical Review A,
  vol.~81, p. 053611, 2010.

\bibitem{klauder1968fundamentals}
Klauder J.R., Sudarshan E.C.G., \emph{Fundamentals of Quantum Optics},
  Benjamin, New York, 1968.

\bibitem{klimov2009group}
Klimov A.B., Chumakov S.M., \emph{A Group-Theoretical Approach to Quantum
  Optics}, Wiley-VCH, Weinheim, 2009.

\bibitem{moyal1949quantum}
Moyal J.E., \enquote{\emph{Quantum mechanics as a statistical theory}},
  \emph{Mathematical Proceedings of the Cambridge Philosophical Society},
  vol.~45, pp. 99--124, Cambridge University Press, 1949.

\bibitem{hillery1984distribution}
Hillery M., O'Connell R.F., Scully M.O., Wigner E.P.,
  \enquote{\emph{Distribution functions in physics: fundamentals}}, Physics
  Reports, vol. 106, pp. 121--167, 1984.

\bibitem{miranowicz2001quantum}
Miranowicz A., Leoński W., Imoto N., \enquote{\emph{Quantum-optical states in
  finite-dimensional hilbert space. {I.} general formalism}}, \emph{Modern
  Nonlinear Optics, Part I}, vol. 119, pp. 155--193, John Wiley \& Sons, New
  York, 2001.

\bibitem{opatrny1996coherent}
Opatrn{\`y} T., Miranowicz A., Bajer J., \enquote{\emph{Coherent states in
  finite-dimensional {Hilbert} space and their {Wigner} representation}},
  Journal of Modern Optics, vol.~43, pp. 417--432, 1996.

\bibitem{mehta1965relation}
Mehta C.L., Sudarshan E.C.G., \enquote{\emph{Relation between quantum and
  semiclassical description of optical coherence}}, Physical Review, vol. 138,
  p. B274, 1965.

\bibitem{kano1964probability}
Kano Y., \enquote{\emph{Probability distribution functions relating to
  blackbody radiation}}, Journal of the Physical Society of Japan, vol.~19, pp.
  1555--1560, 1964.

\bibitem{ryu2013operational}
Ryu J., Lim J., Hong S., Lee J., \enquote{\emph{Operational quasiprobabilities
  for qudits}}, Physical Review A, vol.~88, p. 052123, 2013.

\bibitem{saffman2010quantum}
Saffman M., Walker T.G., M{\o}lmer K., \enquote{\emph{Quantum information with
  {Rydberg} atoms}}, Reviews of Modern Physics, vol.~82, p. 2313, 2010.

\bibitem{gaetan2009observation}
Ga{\"e}tan A., Miroshnychenko Y., Wilk T., Chotia A., Viteau M., Comparat D.,
  Pillet P., Browaeys A., Grangier P., et~al., \enquote{\emph{Observation of
  collective excitation of two individual atoms in the {Rydberg} blockade
  regime}}, Nature Physics, vol.~5, pp. 115--118, 2009.

\bibitem{dicke1954coherence}
Dicke R.H., \enquote{\emph{Coherence in spontaneous radiation processes}},
  Physical Review, vol.~93, p.~99, 1954.

\bibitem{dowling1994wigner}
Dowling J.P., Agarwal G.S., Schleich W.P., \enquote{\emph{{Wigner} distribution
  of a general angular-momentum state: applications to a collection of
  two-level atoms}}, Physical Review A, vol.~49, p. 4101, 1994.

\bibitem{wodkiewicz1985coherent}
Wodkiewicz K., Eberly J.H., \enquote{\emph{Coherent states, squeezed
  fluctuations, and the {SU(2) am SU(1, 1)} groups in quantum-optics
  applications}}, Journal of the Optical Society of America B, vol.~2, pp.
  458--466, 1985.

\bibitem{kitagawa1991nonlinear}
Kitagawa M., Ueda M., \enquote{\emph{Nonlinear-interferometric generation of
  number-phase-correlated fermion states}}, Physical Review Letters, vol.~67,
  p. 1852, 1991.

\bibitem{karasev1999polarization}
Karasev V.P., \enquote{\emph{Polarization coherent states in action: partial
  tomography of multimode quantum emission}}, Bulletin of the Lebedev Physics
  Institute, pp. 34--40, 1999.

\bibitem{zu2014experimental}
Zu C., Wang W.B., He L., Zhang W.G., Dai C.Y., Wang F., Duan L.M.,
  \enquote{\emph{Experimental realization of universal geometric quantum gates
  with solid-state spins}}, Nature, vol. 514, pp. 72--75, 2014.

\bibitem{monroe2014large}
Monroe C., Raussendorf R., Ruthven A., Brown K.R., Maunz P., Duan L.M., Kim J.,
  \enquote{\emph{Large-scale modular quantum-computer architecture with atomic
  memory and photonic interconnects}}, Physical Review A, vol.~89, p. 022317,
  2014.

\bibitem{jones1998implementation}
Jones J.A., Mosca M., \enquote{\emph{Implementation of a quantum algorithm on a
  nuclear magnetic resonance quantum computer}}, The Journal of Chemical
  Physics, vol. 109, pp. 1648--1653, 1998.

\bibitem{ladd2002all}
Ladd T.D., Goldman J.R., Yamaguchi F., Yamamoto Y., Abe E., Itoh K.M.,
  \enquote{\emph{All-silicon quantum computer}}, Physical Review Letters,
  vol.~89, p. 017901, 2002.

\bibitem{harneit2002fullerene}
Harneit W., \enquote{\emph{Fullerene-based electron-spin quantum computer}},
  Physical Review A, vol.~65, p. 032322, 2002.

\bibitem{gershenfeld1997bulk}
Gershenfeld N.A., Chuang I.L., \enquote{\emph{Bulk spin-resonance quantum
  computation}}, Science, vol. 275, pp. 350--356, 1997.

\bibitem{sahling2015experimental}
Sahling S., Remenyi G., Paulsen C., Monceau P., Saligrama V., Marin C.,
  Revcolevschi A., Regnault L.P., Raymond S., Lorenzo J.E.,
  \enquote{\emph{Experimental realization of long-distance entanglement between
  spins in antiferromagnetic quantum spin chains}}, Nature Physics, vol.~11,
  pp. 255--260, 2015.

\bibitem{stratonovich1957distributions}
Stratonovich R.L., \enquote{\emph{On distributions in representation space}},
  Soviet Physics JETP-USSR, vol.~4, pp. 891--898, 1957.

\bibitem{klimov2002SU}
Klimov A.B., Chumakov S.M., \enquote{\emph{On the {SU(2) Wigner} function
  dynamics}}, Revista Mexicana de Física, vol.~48, pp. 317--324, 2002.

\bibitem{chumakov2000connection}
Chumakov S.M., Klimov A.B., Wolf K.B., \enquote{\emph{Connection between two
  {Wigner} functions for spin systems}}, Physical Review A, vol.~61, p. 034101,
  2000.

\bibitem{wootters1987wigner}
Wootters W.K., \enquote{\emph{A {Wigner-function} formulation of finite-state
  quantum mechanics}}, Annals of Physics, vol. 176, pp. 1--21, 1987.

\bibitem{vourdas2003factorization}
Vourdas A., \enquote{\emph{Factorization in finite quantum systems}}, Journal
  of Physics A: Mathematical and General, vol.~36, p. 5645, 2003.

\bibitem{chaturvedi2006wigner}
Chaturvedi S., Ercolessi E., Marmo G., Morandi G., Mukunda N., Simon R.,
  \enquote{\emph{{Wigner}--{Weyl} correspondence in quantum mechanics for
  continuous and discrete systems—a {Dirac-inspired} view}}, Journal of
  Physics A: Mathematical and General, vol.~39, p. 1405, 2006.

\bibitem{leonhardt1996discrete}
Leonhardt U., \enquote{\emph{Discrete {Wigner} function and quantum-state
  tomography}}, Physical Review A, vol.~53, p. 2998, 1996.

\bibitem{paz2002discrete}
Paz J.P., \enquote{\emph{Discrete {Wigner} functions and the phase-space
  representation of quantum teleportation}}, Physical Review A, vol.~65, p.
  062311, 2002.

\bibitem{zare2013angular}
Zare R.N., \emph{Angular Momentum: Understanding Spatial Aspects in Chemistry
  and Physics}, John Wiley \& Sons, 1988.

\bibitem{agarwal1993perspective}
Agarwal G.S., \enquote{\emph{Perspective of {Einstein-Podolsky-Rosen} spin
  correlations in the phase-space formulation for arbitrary values of the
  spin}}, Physical Review A, vol.~47, p. 4608, 1993.

\bibitem{agarwal1981relation}
Agarwal G.S., \enquote{\emph{Relation between atomic coherent-state
  representation, state multipoles, and generalized phase-space
  distributions}}, Physical Review A, vol.~24, p. 2889, 1981.

\bibitem{cohen1986joint}
Cohen L., Scully M.O., \enquote{\emph{Joint {Wigner} distribution for spin-1/2
  particles}}, Foundations of Physics, vol.~16, pp. 295--310, 1986.

\bibitem{varilly1989moyal}
V{\'a}rilly J.C., Gracia-Bond{\'\i}a J.M., \enquote{\emph{The {Moyal}
  representation for spin}}, Annals of Physics, vol. 190, pp. 107--148, 1989.

\bibitem{arecchi1972atomic}
Arecchi F.T., Courtens E., Gilmore R., Thomas H., \enquote{\emph{Atomic
  coherent states in quantum optics}}, Physical Review A, vol.~6, p. 2211,
  1972.

\bibitem{margenau1961correlation}
Margenau H., Hill R.N., \enquote{\emph{Correlation between measurements in
  quantum theory}}, Progress of Theoretical Physics, vol.~26, pp. 722--738,
  1961.

\bibitem{ramachandran1996quasi}
Ramachandran G., Usha~Devi A.R., Devi P., Sirsi S.,
  \enquote{\emph{Quasi-probability distributions for arbitrary spin-j
  particles}}, Foundations of Physics, vol.~26, pp. 401--412, 1996.

\bibitem{usha2002spin}
Usha~Devi A.R., \enquote{\emph{{Spin distributions for bipartite quantum
  systems}}}, International Journal of Modern Physics A, vol.~17, pp.
  2267--2281, 2002.

\bibitem{ozdemir2001quantum}
{\"O}zdemir {\c{S}}.K., Miranowicz A., Koashi M., Imoto N.,
  \enquote{\emph{Quantum-scissors device for optical state truncation: a
  proposal for practical realization}}, Physical Review A, vol.~64, p. 063818,
  2001.

\bibitem{makhlin2001quantum}
Makhlin Y., Sch{\"o}n G., Shnirman A., \enquote{\emph{Quantum-state engineering
  with {Josephson-junction} devices}}, Reviews of Modern Physics, vol.~73, p.
  357, 2001.

\bibitem{verstraete2009quantum}
Verstraete F., Wolf M.M., Cirac J.I., \enquote{\emph{Quantum computation and
  quantum-state engineering driven by dissipation}}, Nature Physics, vol.~5,
  pp. 633--636, 2009.

\bibitem{banerjee2003general}
Banerjee S., Ghosh R., \enquote{\emph{General quantum {Brownian} motion with
  initially correlated and nonlinearly coupled environment}}, Physical Review
  E, vol.~67, p. 056120, 2003.

\bibitem{banerjee2008geometric}
Banerjee S., Srikanth R., \enquote{\emph{Geometric phase of a qubit interacting
  with a squeezed-thermal bath}}, The European Physical Journal D-Atomic,
  Molecular, Optical and Plasma Physics, vol.~46, pp. 335--344, 2008.

\bibitem{banerjee2007dynamics}
Banerjee S., Ghosh R., \enquote{\emph{Dynamics of decoherence without
  dissipation in a squeezed thermal bath}}, Journal of Physics A: Mathematical
  and Theoretical, vol.~40, p. 13735, 2007.

\bibitem{srikanth2008squeezed}
Srikanth R., Banerjee S., \enquote{\emph{Squeezed generalized amplitude damping
  channel}}, Physical Review A, vol.~77, p. 012318, 2008.

\bibitem{agarwal1998state}
Agarwal G.S., \enquote{\emph{State reconstruction for a collection of two-level
  systems}}, Physical Review A, vol.~57, p. 671, 1998.

\bibitem{shapiro1991quantum}
Shapiro J.H., Shepard S.R., \enquote{\emph{Quantum phase measurement: a
  system-theory perspective}}, Physical Review A, vol.~43, p. 3795, 1991.

\bibitem{hall1991quantum}
Hall M.J.W., \enquote{\emph{The quantum description of optical phase}}, Quantum
  Optics: Journal of the European Optical Society Part B, vol.~3, p.~7, 1991.

\bibitem{agarwal1992classical}
Agarwal G.S., Chaturvedi S., Tara K., Srinivasan V., \enquote{\emph{Classical
  phase changes in nonlinear processes and their quantum counterparts}},
  Physical Review A, vol.~45, p. 4904, 1992.

\bibitem{agarwal1996complementarity}
Agarwal G.S., Singh R.P., \enquote{\emph{Complementarity and phase
  distributions for angular momentum systems}}, Physics Letters A, vol. 217,
  pp. 215--218, 1996.

\bibitem{banerjee2007phase}
Banerjee S., Srikanth R., \enquote{\emph{Phase diffusion in quantum dissipative
  systems}}, Physical Review A, vol.~76, p. 062109, 2007.

\bibitem{banerjee2007phaseQND}
Banerjee S., Ghosh J., Ghosh R., \enquote{\emph{Phase-diffusion pattern in
  quantum-nondemolition systems}}, Physical Review A, vol.~75, p. 062106, 2007.

\bibitem{srikanth2009complementarity}
Srikanth R., Banerjee S., \enquote{\emph{Complementarity in atomic
  (finite-level quantum) systems: an information-theoretic approach}}, The
  European Physical Journal D-Atomic, Molecular, Optical and Plasma Physics,
  vol.~53, pp. 217--227, 2009.

\bibitem{banerjee2010complementarity}
Banerjee S., Srikanth R., \enquote{\emph{Complementarity in generic open
  quantum systems}}, Modern Physics Letters B, vol.~24, pp. 2485--2509, 2010.

\bibitem{srikanth2010complementarity}
Srikanth R., Banerjee S., \enquote{\emph{Complementarity in atomic and
  oscillator systems}}, Physics Letters A, vol. 374, pp. 3147--3150, 2010.

\bibitem{miranowicz2015statistical}
Miranowicz A., Bartkiewicz K., Pathak A., Pe{\v{r}}ina~Jr. J., Chen Y.N., Nori
  F., \enquote{\emph{Statistical mixtures of states can be more quantum than
  their superpositions: comparison of nonclassicality measures for single-qubit
  states}}, Physical Review A, vol.~91, p. 042309, 2015.

\bibitem{kenfack2004negativity}
Kenfack A., {\.Z}yczkowski K., \enquote{\emph{Negativity of the {Wigner}
  function as an indicator of non-classicality}}, Journal of Optics B: Quantum
  and Semiclassical Optics, vol.~6, p. 396, 2004.

\bibitem{varshalovich1988quantum}
Varshalovich D.A., Moskalev A.N., Khersonskii V.K., \emph{Quantum Theory of
  Angular Momentum}, World Scientific, Singapore, 1988.

\bibitem{banerjee2010dynamics}
Banerjee S., Ravishankar V., Srikanth R., \enquote{\emph{Dynamics of
  entanglement in two-qubit open system interacting with a squeezed thermal
  bath via dissipative interaction}}, Annals of Physics, vol. 325, pp.
  816--834, 2010.

\bibitem{einstein1935can}
Einstein A., Podolsky B., Rosen N., \enquote{\emph{Can quantum-mechanical
  description of physical reality be considered complete?}}, Physical Review,
  vol.~47, p. 777, 1935.

\bibitem{yu2010entanglement}
Yu T., Eberly J.H., \enquote{\emph{Entanglement evolution in a non-{Markovian}
  environment}}, Optics Communications, vol. 283, pp. 676--680, 2010.

\bibitem{greenberger1989going}
Greenberger D.M., Horne M.A., Zeilinger A., \enquote{\emph{Going beyond
  {Bell’s} theorem}}, M.~Kafatos (editor), \emph{{Bell’s} Theorem, Quantum
  Theory and Conceptions of the Universe}, pp. 69--72, Springer, Netherlands,
  1989.

\bibitem{bouwmeester1999observation}
Bouwmeester D., Pan J.W., Daniell M., Weinfurter H., Zeilinger A.,
  \enquote{\emph{Observation of three-photon {Greenberger-Horne-Zeilinger}
  entanglement}}, Physical Review Letters, vol.~82, p. 1345, 1999.

\bibitem{dur2000three}
D{\"u}r W., Vidal G., Cirac J.I., \enquote{\emph{Three qubits can be entangled
  in two inequivalent ways}}, Physical Review A, vol.~62, p. 062314, 2000.

\bibitem{sharma2015controlled}
Sharma V., Shukla C., Banerjee S., Pathak A., \enquote{\emph{Controlled
  bidirectional remote state preparation in noisy environment: a generalized
  view}}, Quantum Information Processing, vol.~14, pp. 3441--3464, 2015.

\bibitem{miranowicz2014phase}
Miranowicz A., Paprzycka M., Pathak A., Nori F., \enquote{\emph{Phase-space
  interference of states optically truncated by quantum scissors: generation of
  distinct superpositions of qudit coherent states by displacement of vacuum}},
  Physical Review A, vol.~89, p. 033812, 2014.

\bibitem{pathak2014wigner}
Pathak A., Banerji J., \enquote{\emph{Wigner distribution, nonclassicality and
  decoherence of generalized and reciprocal binomial states}}, Physics Letters
  A, vol. 378, pp. 117--123, 2014.

\bibitem{veitch2012negative}
Veitch V., Ferrie C., Gross D., Emerson J., \enquote{\emph{Negative
  quasi-probability as a resource for quantum computation}}, New Journal of
  Physics, vol.~14, p. 113011, 2012.

\bibitem{paris2004quantum}
Paris M.G.A., {\v{R}}eh{\'a}{\v{c}}ek J., \enquote{\emph{Quantum state
  estimation}}, Lecture Notes in Physics (Springer-Verlag Berlin Heidelberg),
  2004.

\bibitem{ibort2009introduction}
Ibort A., Man'Ko V.I., Marmo G., Simoni A., Ventriglia F., \enquote{\emph{An
  introduction to the tomographic picture of quantum mechanics}}, Physica
  Scripta, vol.~79, p. 065013, 2009.

\bibitem{filippov2011optical}
Filippov S.N., Man'ko V.I., \enquote{\emph{Optical tomography of {Fock} state
  superpositions}}, Physica Scripta, vol.~83, p. 058101, 2011.

\bibitem{man1997spin}
Man’ko V.I., Man’ko O.V., \enquote{\emph{Spin state tomography}}, Journal
  of Experimental and Theoretical Physics, vol.~85, pp. 430--434, 1997.

\bibitem{man1997damped}
Man'ko V.I., Safonov S.S., \enquote{\emph{The damped quantum oscillator and a
  classical representation of quantum mechanics}}, Theoretical and Mathematical
  Physics, vol. 112, pp. 1172--1181, 1997.

\bibitem{leonhardt1995quantum}
Leonhardt U., \enquote{\emph{Quantum-state tomography and discrete {Wigner}
  function}}, Physical Review Letters, vol.~74, p. 4101, 1995.

\bibitem{miquel2002interpretation}
Miquel C., Paz J.P., Saraceno M., Knill E., Laflamme R., Negrevergne C.,
  \enquote{\emph{Interpretation of tomography and spectroscopy as dual forms of
  quantum computation}}, Nature, vol. 418, pp. 59--62, 2002.

\bibitem{smithey1993measurement}
Smithey D.T., Beck M., Raymer M.G., Faridani A., \enquote{\emph{Measurement of
  the {Wigner} distribution and the density matrix of a light mode using
  optical homodyne tomography: application to squeezed states and the vacuum}},
  Physical Review Letters, vol.~70, p. 1244, 1993.

\bibitem{beck1993experimental}
Beck M., Smithey D.T., Raymer M.G., \enquote{\emph{Experimental determination
  of quantum-phase distributions using optical homodyne tomography}}, Physical
  Review A, vol.~48, p. R890, 1993.

\bibitem{smithey1993complete}
Smithey D.T., Beck M., Cooper J., Raymer M.G., Faridani A.,
  \enquote{\emph{Complete experimental characterization of the quantum state of
  a light mode via the {Wigner} function and the density matrix: application to
  quantum phase distributions of vacuum and squeezed-vacuum states}}, Physica
  Scripta, vol. 1993, p.~35, 1993.

\bibitem{d1994precision}
d'Ariano G.M., Macchiavello C., Paris M.G.A., \enquote{\emph{Precision of
  quantum tomographic detection of radiation}}, Physics Letters A, vol. 195,
  pp. 31--37, 1994.

\bibitem{d1994detection}
d'Ariano G.M., Macchiavello C., Paris M.G.A., \enquote{\emph{Detection of the
  density matrix through optical homodyne tomography without filtered back
  projection}}, Physical Review A, vol.~50, p. 4298, 1994.

\bibitem{d1996quantum}
d'Ariano G.M., Mancini S., Man'ko V.I., Tombesi P.,
  \enquote{\emph{Reconstructing the density operator by using generalized field
  quadratures}}, Quantum and Semiclassical Optics: Journal of the European
  Optical Society Part B, vol.~8, p. 1017, 1996.

\bibitem{mancini1996symplectic}
Mancini S., Man'ko V.I., Tombesi P., \enquote{\emph{Symplectic tomography as
  classical approach to quantum systems}}, Physics Letters A, vol. 213, pp.
  1--6, 1996.

\bibitem{mancini1995wigner}
Mancini S., Man'ko V.I., Tombesi P., \enquote{\emph{{Wigner} function and
  probability distribution for shifted and squeezed quadratures}}, Quantum and
  Semiclassical Optics: Journal of the European Optical Society Part B, vol.~7,
  p. 615, 1995.

\bibitem{lvovsky2009continuous}
Lvovsky A.I., Raymer M.G., \enquote{\emph{Continuous-variable optical
  quantum-state tomography}}, Reviews of Modern Physics, vol.~81, p. 299, 2009.

\bibitem{arkhipov2003new}
Arkhipov A.S., Lozovik Y.E., \enquote{\emph{New method of quantum dynamics
  simulation based on the quantum tomography}}, Physics Letters A, vol. 319,
  pp. 217--224, 2003.

\bibitem{arkhipov2005center}
Arkhipov A.S., Lozovik Y.E., Man'ko V.I., Sharapov V.A.,
  \enquote{\emph{Center-of-mass tomography and probability representation of
  quantum states for tunneling}}, Theoretical and Mathematical Physics, vol.
  142, pp. 311--323, 2005.

\bibitem{lozovik2004simulation}
Lozovik Y.E., Sharapov V.A., Arkhipov A.S., \enquote{\emph{Simulation of
  tunneling in the quantum tomography approach}}, Physical Review A, vol.~69,
  p. 022116, 2004.

\bibitem{arkhipov2003tomography}
Arkhipov A.S., Lozovik Y.E., Man'ko V.I., \enquote{\emph{Tomography for several
  particles with one random variable}}, Journal of Russian Laser Research,
  vol.~24, pp. 237--255, 2003.

\bibitem{man1999classical}
Man'ko O., Man'Ko V.I., \enquote{\emph{“{Classical}” propagator and path
  integral in the probability representation of quantum mechanics}}, Journal of
  Russian Laser Research, vol.~20, pp. 67--76, 1999.

\bibitem{fedorov2013feynman}
Fedorov A., \enquote{\emph{Feynman integral and perturbation theory in quantum
  tomography}}, Physics Letters A, vol. 377, pp. 2320--2323, 2013.

\bibitem{ma2012quantum}
Ma X.S., Herbst T., Scheidl T., Wang D., Kropatschek S., Naylor W., Wittmann
  B., Mech A., Kofler J., Anisimova E., et~al., \enquote{\emph{Quantum
  teleportation over 143 kilometres using active feed-forward}}, Nature, vol.
  489, pp. 269--273, 2012.

\bibitem{poyatos1997complete}
Poyatos J.F., Cirac J.I., Zoller P., \enquote{\emph{Complete characterization
  of a quantum process: the two-bit quantum gate}}, Physical Review Letters,
  vol.~78, p. 390, 1997.

\bibitem{chuang1997prescription}
Chuang I.L., Nielsen M.A., \enquote{\emph{Prescription for experimental
  determination of the dynamics of a quantum black box}}, Journal of Modern
  Optics, vol.~44, pp. 2455--2467, 1997.

\bibitem{nielsen1998complete}
Nielsen M.A., Knill E., Laflamme R., \enquote{\emph{Complete quantum
  teleportation using nuclear magnetic resonance}}, Nature, vol. 396, pp.
  52--55, 1998.

\bibitem{kuah2007state}
Kuah A.m., Modi K., Rodriguez-Rosario C.A., Sudarshan E.C.G.,
  \enquote{\emph{How state preparation can affect a quantum experiment: quantum
  process tomography for open systems}}, Physical Review A, vol.~76, p. 042113,
  2007.

\bibitem{bellomo2009reconstruction}
Bellomo B., De~Pasquale A., Gualdi G., Marzolino U.,
  \enquote{\emph{Reconstruction of {Markovian} master equation parameters
  through symplectic tomography}}, Physical Review A, vol.~80, p. 052108, 2009.

\bibitem{bellomo2010tomographic}
Bellomo B., De~Pasquale A., Gualdi G., Marzolino U., \enquote{\emph{A
  tomographic approach to non-{Markovian} master equations}}, Journal of
  Physics A: Mathematical and Theoretical, vol.~43, p. 395303, 2010.

\bibitem{bellomo2010reconstruction}
Bellomo B., De~Pasquale A., Gualdi G., Marzolino U.,
  \enquote{\emph{Reconstruction of time-dependent coefficients: a check of
  approximation schemes for non-{Markovian} convolutionless dissipative
  generators}}, Physical Review A, vol.~82, p. 062104, 2010.

\bibitem{lobino2008complete}
Lobino M., Korystov D., Kupchak C., Figueroa E., Sanders B.C., Lvovsky A.I.,
  \enquote{\emph{Complete characterization of quantum-optical processes}},
  Science, vol. 322, pp. 563--566, 2008.

\bibitem{rahimi2011quantum}
Rahimi-Keshari S., Scherer A., Mann A., Rezakhani A.T., Lvovsky A.I., Sanders
  B.C., \enquote{\emph{Quantum process tomography with coherent states}}, New
  Journal of Physics, vol.~13, p. 013006, 2011.

\bibitem{anis2012maximum}
Anis A., Lvovsky A.I., \enquote{\emph{Maximum-likelihood coherent-state quantum
  process tomography}}, New Journal of Physics, vol.~14, p. 105021, 2012.

\bibitem{fedorov2015tomography}
Fedorov I.A., Fedorov A.K., Kurochkin Y.V., Lvovsky A.I.,
  \enquote{\emph{Tomography of a multimode quantum black box}}, New Journal of
  Physics, vol.~17, p. 043063, 2015.

\bibitem{lobino2009memory}
Lobino M., Kupchak C., Figueroa E., Lvovsky A.I., \enquote{\emph{Memory for
  light as a quantum process}}, Physical Review Letters, vol. 102, p. 203601,
  2009.

\bibitem{kumar2013experimental}
Kumar R., Barrios E., Kupchak C., Lvovsky A.I., \enquote{\emph{Experimental
  characterization of bosonic creation and annihilation operators}}, Physical
  Review Letters, vol. 110, p. 130403, 2013.

\bibitem{cooper2015characterization}
Cooper M., Slade E., Karpi{\'n}ski M., Smith B.J.,
  \enquote{\emph{Characterization of conditional state-engineering quantum
  processes by coherent state quantum process tomography}}, New Journal of
  Physics, vol.~17, p. 033041, 2015.

\bibitem{liang2003tomographic}
Liang Y.C., Kaszlikowski D., Englert B.G., Kwek L.C., Oh C.H.,
  \enquote{\emph{Tomographic quantum cryptography}}, Physical Review A,
  vol.~68, p. 022324, 2003.

\bibitem{d2003spin}
d'Ariano G.M., Maccone L., Paini M., \enquote{\emph{Spin tomography}}, Journal
  of Optics B: Quantum and Semiclassical Optics, vol.~5, p.~77, 2003.

\bibitem{man1998describing}
Man'ko V.I., Man'ko O.V., Safonov S.S., \enquote{\emph{Describing spinors using
  probability distribution functions}}, Theoretical and Mathematical Physics,
  vol. 115, pp. 520--529, 1998.

\bibitem{fedorov2013quaternion}
Fedorov A.K., Kiktenko E.O., \enquote{\emph{Quaternion representation and
  symplectic spin tomography}}, Journal of Russian Laser Research, vol.~34, pp.
  477--487, 2013.

\bibitem{dodonov1997positive}
Dodonov V.V., Man'ko V.I., \enquote{\emph{Positive distribution description for
  spin states}}, Physics Letters A, vol. 229, pp. 335--339, 1997.

\bibitem{d2004quantum}
d'Ariano G.M., Paris M.G.A., Sacchi M.F., \enquote{\emph{Quantum tomographic
  methods}}, M.G.A. Paris, J.~{\v{R}}eh{\'a}{\v{c}}ek (editors), \emph{Quantum
  State Estimation}, chap.~2, pp. 7--58, Lecture Notes in Physics
  (Springer-Verlag Berlin Heidelberg), 2004.

\bibitem{liu2011experimental}
Liu B.H., Li L., Huang Y.F., Li C.F., Guo G.C., Laine E.M., Breuer H.P., Piilo
  J., \enquote{\emph{Experimental control of the transition from {Markovian} to
  non-{Markovian} dynamics of open quantum systems}}, Nature Physics, vol.~7,
  pp. 931--934, 2011.

\bibitem{harder2014tomography}
Harder G., Mogilevtsev D., Korolkova N., Silberhorn C.,
  \enquote{\emph{Tomography by noise}}, Physical Review Letters, vol. 113, p.
  070403, 2014.

\bibitem{adam2014wigner}
Adam P., Andreev V.A., Ghiu I., Isar A., Man'ko M.A., Man'ko V.I.,
  \enquote{\emph{{Wigner} functions and spin tomograms for qubit states}},
  Journal of Russian Laser Research, vol.~35, pp. 3--13, 2014.

\bibitem{adam2014finite}
Adam P., Andreev V.A., Ghiu I., Isar A., Man'ko M.A., Man'ko V.I.,
  \enquote{\emph{Finite phase space, {Wigner} functions, and tomography for
  two-qubit states}}, Journal of Russian Laser Research, vol.~35, pp. 427--436,
  2014.

\bibitem{miranowicz2014optimal}
Miranowicz A., Bartkiewicz K., Pe{\v{r}}ina~Jr. J., Koashi M., Imoto N., Nori
  F., \enquote{\emph{Optimal two-qubit tomography based on local and global
  measurements: maximal robustness against errors as described by condition
  numbers}}, Physical Review A, vol.~90, p. 062123, 2014.

\bibitem{kiktenko2014tomographic}
Kiktenko E., Fedorov A., \enquote{\emph{Tomographic causal analysis of
  two-qubit states and tomographic discord}}, Physics Letters A, vol. 378, pp.
  1704--1710, 2014.

\bibitem{fedorov2015tomographic}
Fedorov A.K., Kiktenko E.O., Man'ko O.V., Man'ko V.I.,
  \enquote{\emph{Tomographic discord for a system of two coupled nanoelectric
  circuits}}, Physica Scripta, vol.~90, p. 055101, 2015.

\bibitem{srinatha2014quantum}
Srinatha N., Omkar S., Srikanth R., Banerjee S., Pathak A., \enquote{\emph{The
  quantum cryptographic switch}}, Quantum Information Processing, vol.~13, pp.
  59--70, 2014.

\bibitem{chernega2008wave}
Chernega V.N., Man'ko V.I., \enquote{\emph{The wave function of the classical
  parametric oscillator and the tomographic probability of the oscillator's
  state}}, Journal of Russian Laser Research, vol.~29, p. 347, 2008.

\bibitem{hu2012kraus}
Hu L.Y., Wang Q., Wang Z.S., Xu X.x., \enquote{\emph{Kraus operator-sum
  representation and time evolution of distribution functions in
  phase-sensitive reservoirs}}, International Journal of Theoretical Physics,
  vol.~51, pp. 331--349, 2012.

\bibitem{bazrafkan2004tomography}
Bazrafkan M.R., Man'ko V.I., \enquote{\emph{Tomography of binomial states of
  the radiation field}}, Journal of Russian Laser Research, vol.~25, pp.
  453--467, 2004.

\bibitem{wiesner1983conjugate}
Wiesner S., \enquote{\emph{Conjugate coding}}, ACM Sigact News, vol.~15, pp.
  78--88, 1983.

\bibitem{IDQ}
\enquote{\emph{Id quantique home page}}, \url{http://www.idquantique.com/}.

\bibitem{deutsch1985quantum}
Deutsch D., \enquote{\emph{Quantum theory, the {Church-Turing} principle and
  the universal quantum computer}}, \emph{Proceedings of the Royal Society of
  London A: Mathematical, Physical and Engineering Sciences}, vol. 400, pp.
  97--117, 1985.

\bibitem{deutsch1992rapid}
Deutsch D., Jozsa R., \enquote{\emph{Rapid solution of problems by quantum
  computation}}, \emph{Proceedings of the Royal Society of London A:
  Mathematical, Physical and Engineering Sciences}, vol. 439, pp. 553--558,
  1992.

\bibitem{grover1997quantum}
Grover L.K., \enquote{\emph{Quantum mechanics helps in searching for a needle
  in a haystack}}, Physical Review Letters, vol.~79, p. 325, 1997.

\bibitem{shor1999polynomial}
Shor P.W., \enquote{\emph{Polynomial-time algorithms for prime factorization
  and discrete logarithms on a quantum computer}}, SIAM (Society for Industrial
  and Applied Mathematics) Review, vol.~41, pp. 303--332, 1999.

\bibitem{karlsson1998quantum}
Karlsson A., Bourennane M., \enquote{\emph{Quantum teleportation using
  three-particle entanglement}}, Physical Review A, vol.~58, p. 4394, 1998.

\bibitem{pathak2011efficient}
Pathak A., Banerjee A., \enquote{\emph{Efficient quantum circuits for perfect
  and controlled teleportation of n-qubit non-maximally entangled states of
  generalized {Bell}-type}}, International Journal of Quantum Information,
  vol.~9, pp. 389--403, 2011.

\bibitem{hillery1999quantum}
Hillery M., Bu{\v{z}}ek V., Berthiaume A., \enquote{\emph{Quantum secret
  sharing}}, Physical Review A, vol.~59, p. 1829, 1999.

\bibitem{wang2010hierarchical}
Wang X.W., Xia L.X., Wang Z.Y., Zhang D.Y., \enquote{\emph{Hierarchical
  quantum-information splitting}}, Optics Communications, vol. 283, pp.
  1196--1199, 2010.

\bibitem{shukla2013hierarchical}
Shukla C., Pathak A., \enquote{\emph{Hierarchical quantum communication}},
  Physics Letters A, vol. 377, pp. 1337--1344, 2013.

\bibitem{pati2000minimum}
Pati A.K., \enquote{\emph{Minimum classical bit for remote preparation and
  measurement of a qubit}}, Physical Review A, vol.~63, p. 014302, 2000.

\bibitem{nguyen2008joint}
Nguyen B.A., Kim J., \enquote{\emph{Joint remote state preparation}}, Journal
  of Physics B: Atomic, Molecular and Optical Physics, vol.~41, p. 095501,
  2008.

\bibitem{shukla2016hierarchical}
Shukla C., Thapliyal K., Pathak A., \enquote{\emph{Hierarchical joint remote
  state preparation in noisy environment}}, Quantum Information Processing,
  vol.~16, p. 205, 2017.

\bibitem{huelga2001quantum}
Huelga S.F., Vaccaro J.A., Chefles A., Plenio M.B., \enquote{\emph{Quantum
  remote control: teleportation of unitary operations}}, Physical Review A,
  vol.~63, p. 042303, 2001.

\bibitem{brassard1998teleportation}
Brassard G., Braunstein S.L., Cleve R., \enquote{\emph{Teleportation as a
  quantum computation}}, Physica D: Nonlinear Phenomena, vol. 120, pp. 43--47,
  1998.

\bibitem{yan2004scheme}
Yan F., Zhang X., \enquote{\emph{A scheme for secure direct communication using
  {EPR} pairs and teleportation}}, The European Physical Journal B-Condensed
  Matter and Complex Systems, vol.~41, pp. 75--78, 2004.

\bibitem{bouwmeester1997experimental}
Bouwmeester D., Pan J.W., Mattle K., Eibl M., Weinfurter H., Zeilinger A.,
  \enquote{\emph{Experimental quantum teleportation}}, Nature, vol. 390, pp.
  575--579, 1997.

\bibitem{sisodia2017design}
Sisodia M., Shukla A., Thapliyal K., Pathak A., \enquote{\emph{Design and
  experimental realization of an optimal scheme for teleportation of an
  $n$-qubit quantum state}}, Quantum Information Processing, vol.~16, p. 292,
  2017.

\bibitem{peters2005remote}
Peters N.A., Barreiro J.T., Goggin M.E., Wei T.C., Kwiat P.G.,
  \enquote{\emph{Remote state preparation: arbitrary remote control of photon
  polarization}}, Physical Review Letters, vol.~94, p. 150502, 2005.

\bibitem{liu2007experimental}
Liu W.T., Wu W., Ou B.Q., Chen P.X., Li C.Z., Yuan J.M.,
  \enquote{\emph{Experimental remote preparation of arbitrary photon
  polarization states}}, Physical Review A, vol.~76, p. 022308, 2007.

\bibitem{xiang2005remote}
Xiang G.Y., Li J., Yu B., Guo G.C., \enquote{\emph{Remote preparation of mixed
  states via noisy entanglement}}, Physical Review A, vol.~72, p. 012315, 2005.

\bibitem{raadmark2013experimental}
R{\aa}dmark M., Wie{\'s}niak M., {\.Z}ukowski M., Bourennane M.,
  \enquote{\emph{Experimental multilocation remote state preparation}},
  Physical Review A, vol.~88, p. 032304, 2013.

\bibitem{zhao2004experimental}
Zhao Z., Chen Y.A., Zhang A.N., Yang T., Briegel H.J., Pan J.W.,
  \enquote{\emph{Experimental demonstration of five-photon entanglement and
  open-destination teleportation}}, Nature, vol. 430, pp. 54--58, 2004.

\bibitem{riebe2004deterministic}
Riebe M., H{\"a}ffner H., Roos C.F., H{\"a}nsel W., Benhelm J., Lancaster
  G.P.T., K{\"o}rber T.W., Becher C., Schmidt-Kaler F., James D.F.V., et~al.,
  \enquote{\emph{Deterministic quantum teleportation with atoms}}, Nature, vol.
  429, pp. 734--737, 2004.

\bibitem{barrett2004deterministic}
Barrett M.D., Chiaverini J., Schaetz T., Britton J., Itano W.M., Jost J.D.,
  Knill E., Langer C., Leibfried D., Ozeri R., et~al.,
  \enquote{\emph{Deterministic quantum teleportation of atomic qubits}},
  Nature, vol. 429, pp. 737--739, 2004.

\bibitem{de2004long}
De~Riedmatten H., Marcikic I., Tittel W., Zbinden H., Collins D., Gisin N.,
  \enquote{\emph{Long distance quantum teleportation in a quantum relay
  configuration}}, Physical Review Letters, vol.~92, p. 047904, 2004.

\bibitem{lo2012measurement}
Lo H.K., Curty M., Qi B., \enquote{\emph{Measurement-device-independent quantum
  key distribution}}, Physical Review Letters, vol. 108, p. 130503, 2012.

\bibitem{kerckhoffs1883cryptographic}
Kerckhoffs A., \enquote{\emph{La cryptographic militaire}}, Journal des
  Sciences Militaires, vol.~9, pp. 5--38, 1883.

\bibitem{rivest1978method}
Rivest R.L., Shamir A., Adleman L., \enquote{\emph{A method for obtaining
  digital signatures and public-key cryptosystems}}, Communications of the ACM
  (Association for Computing Machinery), vol.~21, pp. 120--126, 1978.

\bibitem{diffie1976new}
Diffie W., Hellman M., \enquote{\emph{New directions in cryptography}}, IEEE
  transactions on Information Theory, vol.~22, pp. 644--654, 1976.

\bibitem{bennett1992quantum}
Bennett C.H., Brassard G., Mermin N.D., \enquote{\emph{Quantum cryptography
  without {Bell's} theorem}}, Physical Review Letters, vol.~68, p. 557, 1992.

\bibitem{goldenberg1995quantum}
Goldenberg L., Vaidman L., \enquote{\emph{Quantum cryptography based on
  orthogonal states}}, Physical Review Letters, vol.~75, p. 1239, 1995.

\bibitem{lo2014secure}
Lo H.K., Curty M., Tamaki K., \enquote{\emph{Secure quantum key distribution}},
  Nature Photonics, vol.~8, pp. 595--604, 2014.

\bibitem{shenoy2017quantum}
Shenoy-Hejamadi A., Pathak A., Radhakrishna S., \enquote{\emph{Quantum
  cryptography: Key distribution and beyond}}, Quanta, vol.~6, pp. 1--47, 2017.

\bibitem{brassard2000limitations}
Brassard G., L{\"u}tkenhaus N., Mor T., Sanders B.C.,
  \enquote{\emph{Limitations on practical quantum cryptography}}, Physical
  Review Letters, vol.~85, p. 1330, 2000.

\bibitem{sharma2016verification}
Sharma R.D., Thapliyal K., Pathak A., Pan A.K., De A., \enquote{\emph{Which
  verification qubits perform best for secure communication in noisy
  channel?}}, Quantum Information Processing, vol.~15, pp. 1703--1718, 2016.

\bibitem{branciard2012one}
Branciard C., Cavalcanti E.G., Walborn S.P., Scarani V., Wiseman H.M.,
  \enquote{\emph{One-sided device-independent quantum key distribution:
  security, feasibility, and the connection with steering}}, Physical Review A,
  vol.~85, p. 010301, 2012.

\bibitem{PhysRevLett.99.140501}
Boyer M., Kenigsberg D., Mor T., \enquote{\emph{Quantum key distribution with
  classical {Bob}}}, Physical Review Letters, vol.~99, p. 140501, 2007.

\bibitem{noh2009counterfactual}
Noh T.G., \enquote{\emph{Counterfactual quantum cryptography}}, Physical Review
  Letters, vol. 103, p. 230501, 2009.

\bibitem{gisin2002quantum}
Gisin N., Ribordy G., Tittel W., Zbinden H., \enquote{\emph{Quantum
  cryptography}}, Reviews of modern physics, vol.~74, p. 145, 2002.

\bibitem{bostrom2002deterministic}
Bostr{\"o}m K., Felbinger T., \enquote{\emph{Deterministic secure direct
  communication using entanglement}}, Physical Review Letters, vol.~89, p.
  187902, 2002.

\bibitem{lucamarini2005secure}
Lucamarini M., Mancini S., \enquote{\emph{Secure deterministic communication
  without entanglement}}, Physical Review Letters, vol.~94, p. 140501, 2005.

\bibitem{shukla2013improved}
Shukla C., Banerjee A., Pathak A., \enquote{\emph{Improved protocols of secure
  quantum communication using {W} states}}, International Journal of
  Theoretical Physics, vol.~52, pp. 1914--1924, 2013.

\bibitem{long2007quantum}
Long G.l., Deng F.g., Wang C., Li X.h., Wen K., Wang W.y.,
  \enquote{\emph{Quantum secure direct communication and deterministic secure
  quantum communication}}, Frontiers of Physics in China, vol.~2, pp. 251--272,
  2007.

\bibitem{banerjee2012maximally}
Banerjee A., Pathak A., \enquote{\emph{Maximally efficient protocols for direct
  secure quantum communication}}, Physics Letters A, vol. 376, pp. 2944--2950,
  2012.

\bibitem{pathak2015efficient}
Pathak A., \enquote{\emph{Efficient protocols for unidirectional and
  bidirectional controlled deterministic secure quantum communication:
  different alternative approaches}}, Quantum Information Processing, vol.~14,
  pp. 2195--2210, 2015.

\bibitem{shukla2014protocols}
Shukla C., Alam N., Pathak A., \enquote{\emph{Protocols of quantum key
  agreement solely using {Bell} states and {Bell} measurement}}, Quantum
  Information Processing, vol.~13, pp. 2391--2405, 2014.

\bibitem{jun2006revisiting}
Jun L., Yi-Min L., Hai-Jing C., Shou-Hua S., Zhan-Jun Z.,
  \enquote{\emph{Revisiting quantum secure direct communication with {W}
  state}}, Chinese Physics Letters, vol.~23, p. 2652, 2006.

\bibitem{li2006deterministic}
Li X.H., Deng F.G., Li C.Y., Liang Y.J., Zhou P., Zhou H.Y.,
  \enquote{\emph{Deterministic secure quantum communication without maximally
  entangled states}}, Journal of the Korean Physical Society, vol.~49, p. 1354,
  2006.

\bibitem{zhong2005deterministic}
Zhong-Xiao M., Zhan-Jun Z., Yong L., \enquote{\emph{Deterministic secure direct
  communication by using swapping quantum entanglement and local unitary
  operations}}, Chinese Physics Letters, vol.~22, p.~18, 2005.

\bibitem{hwang2011quantum}
Hwang T., Hwang C., Tsai C., \enquote{\emph{Quantum key distribution protocol
  using dense coding of three-qubit {W} state}}, The European Physical Journal
  D, vol.~61, pp. 785--790, 2011.

\bibitem{zhu2006secure}
Zhu A.D., Xia Y., Fan Q.B., Zhang S., \enquote{\emph{Secure direct
  communication based on secret transmitting order of particles}}, Physical
  Review A, vol.~73, p. 022338, 2006.

\bibitem{hai2006quantum}
Hai-Jing C., He-Shan S., \enquote{\emph{Quantum secure direct communication
  with {W} state}}, Chinese Physics Letters, vol.~23, p. 290, 2006.

\bibitem{yuan2011high}
Yuan H., Song J., Zhou J., Zhang G., Wei X.f., \enquote{\emph{High-capacity
  deterministic secure four-qubit {W} state protocol for quantum communication
  based on order rearrangement of particle pairs}}, International Journal of
  Theoretical Physics, vol.~50, pp. 2403--2409, 2011.

\bibitem{long2002theoretically}
Long G.L., Liu X.S., \enquote{\emph{Theoretically efficient high-capacity
  quantum-key-distribution scheme}}, Physical Review A, vol.~65, p. 032302,
  2002.

\bibitem{degiovanni2004quantum}
Degiovanni I., Berchera I.R., Castelletto S., Rastello M., Bovino F., Colla A.,
  Castagnoli G., \enquote{\emph{Quantum dense key distribution}}, Physical
  Review A, vol.~69, p. 032310, 2004.

\bibitem{nguyen2004quantum}
Nguyen B.A., \enquote{\emph{Quantum dialogue}}, Physics Letters A, vol. 328,
  pp. 6--10, 2004.

\bibitem{zhang2017quantum}
Zhang W., Ding D.S., Sheng Y.B., Zhou L., Shi B.S., Guo G.C.,
  \enquote{\emph{Quantum secure direct communication with quantum memory}},
  Physical Review Letters, vol. 118, p. 220501, 2017.

\bibitem{banerjee2017asymmetric}
Banerjee A., Shukla C., Thapliyal K., Pathak A., Panigrahi P.K.,
  \enquote{\emph{Asymmetric quantum dialogue in noisy environment}}, Quantum
  Information Processing, vol.~16, p.~49, 2017.

\bibitem{banerjee2017quantum}
Banerjee A., Thapliyal K., Shukla C., Pathak A., \enquote{\emph{Quantum
  conference}}, arXiv preprint arXiv:1702.00389, 2017.

\bibitem{yao1982protocols}
Yao A.C., \enquote{\emph{Protocols for secure computations}}, \emph{Foundations
  of Computer Science, 1982. SFCS'08. 23rd Annual Symposium on}, pp. 160--164,
  IEEE, 1982.

\bibitem{canetti2000security}
Canetti R., \enquote{\emph{Security and composition of multiparty cryptographic
  protocols}}, Journal of Cryptology, vol.~13, pp. 143--202, 2000.

\bibitem{lo1997insecurity}
Lo H.K., \enquote{\emph{Insecurity of quantum secure computations}}, Physical
  Review A, vol.~56, p. 1154, 1997.

\bibitem{thapliyal2017protocols}
Thapliyal K., Sharma R.D., Pathak A., \enquote{\emph{Protocols for quantum
  binary voting}}, International Journal of Quantum Information, vol.~15, p.
  1750007, 2017.

\bibitem{thapliyal2016orthogonal}
Thapliyal K., Sharma R.D., Pathak A., \enquote{\emph{Orthogonal-state-based and
  semi-quantum protocols for quantum private comparison in noisy environment}},
  arXiv preprint arXiv:1608.00101, 2016.

\bibitem{shukla2017semi}
Shukla C., Thapliyal K., Pathak A., \enquote{\emph{Semi-quantum communication:
  protocols for key agreement, controlled secure direct communication and
  dialogue}}, Quantum Information Processing, vol.~16, p. 295, 2017.

\bibitem{sharma2017quantumauction}
Sharma R.D., Thapliyal K., Pathak A., \enquote{\emph{Quantum sealed-bid auction
  using a modified scheme for multiparty circular quantum key agreement}},
  Quantum Information Processing, vol.~16, p. 169, 2017.

\bibitem{sisodia2017teleportation}
Sisodia M., Verma V., Thapliyal K., Pathak A., \enquote{\emph{Teleportation of
  a qubit using entangled non-orthogonal states: a comparative study}}, Quantum
  Information Processing, vol.~16, p.~76, 2017.

\bibitem{kak2006three}
Kak S., \enquote{\emph{A three-stage quantum cryptography protocol}},
  Foundations of Physics Letters, vol.~19, pp. 293--296, 2006.

\bibitem{thapliyal2018kak}
Thapliyal K., Pathak A., \enquote{\emph{Kak's three-stage protocol of secure
  quantum communication revisited: hitherto unknown strengths and weaknesses of
  the protocol}}, arXiv preprint arXiv:1803.02157, 2018.

\bibitem{huelga2002remote}
Huelga S.F., Plenio M.B., Vaccaro J.A., \enquote{\emph{Remote control of
  restricted sets of operations: teleportation of angles}}, Physical Review A,
  vol.~65, p. 042316, 2002.

\bibitem{zha2013bidirectional}
Zha X.W., Zou Z.C., Qi J.X., Song H.Y., \enquote{\emph{Bidirectional quantum
  controlled teleportation via five-qubit cluster state}}, International
  Journal of Theoretical Physics, vol.~52, pp. 1740--1744, 2013.

\bibitem{xin2010bidirectional}
Xin-Wei Z., Hai-Yang S., Gang-Long M., \enquote{\emph{Bidirectional swapping
  quantum controlled teleportation based on maximally entangled five-qubit
  state}}, arXiv preprint arXiv:1006.0052, 2010.

\bibitem{li2013bidirectional}
Li Y.h., Nie L.p., \enquote{\emph{Bidirectional controlled teleportation by
  using a five-qubit composite {GHZ-Bell} state}}, International Journal of
  Theoretical Physics, vol.~52, pp. 1630--1634, 2013.

\bibitem{shukla2013bidirectional}
Shukla C., Banerjee A., Pathak A., \enquote{\emph{Bidirectional controlled
  teleportation by using 5-qubit states: a generalized view}}, International
  Journal of Theoretical Physics, vol.~52, pp. 3790--3796, 2013.

\bibitem{li2013cqsdc}
Li Y.h., Li X.l., Sang M.h., Nie Y.y., Wang Z.s., \enquote{\emph{Bidirectional
  controlled quantum teleportation and secure direct communication using
  five-qubit entangled state}}, Quantum Information Processing, vol.~12, pp.
  3835--3844, 2013.

\bibitem{duan2014bidirectional6}
Duan Y.J., Zha X.W., Sun X.M., Xia J.F., \enquote{\emph{Bidirectional quantum
  controlled teleportation via a maximally seven-qubit entangled state}},
  International Journal of Theoretical Physics, vol.~53, pp. 2697--2707, 2014.

\bibitem{fu2014general}
Fu H.Z., Tian X.L., Hu Y., \enquote{\emph{A general method of selecting quantum
  channel for bidirectional quantum teleportation}}, International Journal of
  Theoretical Physics, vol.~53, pp. 1840--1847, 2014.

\bibitem{chen2015bidirectional}
Chen Y., \enquote{\emph{Bidirectional quantum controlled teleportation by using
  a genuine six-qubit entangled state}}, International Journal of Theoretical
  Physics, vol.~54, pp. 269--272, 2015.

\bibitem{yan2013bidirectional}
Yan A., \enquote{\emph{Bidirectional controlled teleportation via six-qubit
  cluster state}}, International Journal of Theoretical Physics, vol.~52, pp.
  3870--3873, 2013.

\bibitem{duan2014bidirectional7}
Duan Y.J., Zha X.W., \enquote{\emph{Bidirectional quantum controlled
  teleportation via a six-qubit entangled state}}, International Journal of
  Theoretical Physics, vol.~53, pp. 3780--3786, 2014.

\bibitem{dong2008controlled}
Dong L., Xiu X.M., Gao Y.J., Chi F., \enquote{\emph{A controlled quantum
  dialogue protocol in the network using entanglement swapping}}, Optics
  Communications, vol. 281, pp. 6135--6138, 2008.

\bibitem{xia2006quantum}
Xia Y., Fu C.B., Zhang S., Hong S.K., Yeon K.H., Um C.I.,
  \enquote{\emph{Quantum dialogue by using the {GHZ} state}}, Journal of the
  Korean Physical Society, vol.~48, pp. 24--27, 2006.

\bibitem{deng2003controlled}
Deng F.G., Long G.L., \enquote{\emph{Controlled order rearrangement encryption
  for quantum key distribution}}, Physical Review A, vol.~68, p. 042315, 2003.

\bibitem{shukla2012beyond}
Shukla C., Pathak A., Srikanth R., \enquote{\emph{Beyond the
  {Goldenberg--Vaidman} protocol: secure and efficient quantum communication
  using arbitrary, orthogonal, multi-particle quantum states}}, International
  Journal of Quantum Information, vol.~10, p. 1241009, 2012.

\bibitem{yadav2014two}
Yadav P., Srikanth R., Pathak A., \enquote{\emph{Two-step
  orthogonal-state-based protocol of quantum secure direct communication with
  the help of order-rearrangement technique}}, Quantum Information Processing,
  vol.~13, pp. 2731--2743, 2014.

\bibitem{shukla2013group}
Shukla C., Kothari V., Banerjee A., Pathak A., \enquote{\emph{On the
  group-theoretic structure of a class of quantum dialogue protocols}}, Physics
  Letters A, vol. 377, pp. 518--527, 2013.

\bibitem{cai2004improving}
Cai Q.Y., Li B.W., \enquote{\emph{Improving the capacity of the
  {Bostr{\"o}m-Felbinger} protocol}}, Physical Review A, vol.~69, p. 054301,
  2004.

\bibitem{deng2003two}
Deng F.G., Long G.L., Liu X.S., \enquote{\emph{Two-step quantum direct
  communication protocol using the {Einstein-Podolsky-Rosen} pair block}},
  Physical Review A, vol.~68, p. 042317, 2003.

\bibitem{an2009finite}
An N.B., Kim J., \enquote{\emph{Finite-time and infinite-time disentanglement
  of multipartite {Greenberger-Horne-Zeilinger}-type states under the
  collective action of different types of noise}}, Physical Review A, vol.~79,
  p. 022303, 2009.

\bibitem{guan2014joint}
Guan X.W., Chen X.B., Wang L.C., Yang Y.X., \enquote{\emph{Joint remote
  preparation of an arbitrary two-qubit state in noisy environments}},
  International Journal of Theoretical Physics, vol.~53, pp. 2236--2245, 2014.

\bibitem{macchiavello2002entanglement}
Macchiavello C., Palma G.M., \enquote{\emph{Entanglement-enhanced information
  transmission over a quantum channel with correlated noise}}, Physical Review
  A, vol.~65, p. 050301, 2002.

\bibitem{cao2013deterministic}
Cao T.B., Nguyen B.A., \enquote{\emph{Deterministic controlled bidirectional
  remote state preparation}}, Advances in Natural Sciences: Nanoscience and
  Nanotechnology, vol.~5, p. 015003, 2013.

\bibitem{yeo2010non}
Yeo Y., An J.H., Oh C., \enquote{\emph{Non-{Markovian} effects on
  quantum-communication protocols}}, Physical Review A, vol.~82, p. 032340,
  2010.

\bibitem{hao2012enhanced}
Hao X., Zhu S., \enquote{\emph{Enhanced quantum teleportation in
  non-{Markovian} environments}}, International Journal of Quantum Information,
  vol.~10, p. 1250051, 2012.

\bibitem{jun2012non}
Jun J.W., \enquote{\emph{Non-{Markovian} effects on entanglement swapping}},
  Journal of the Korean Physical Society, vol.~60, pp. 550--553, 2012.

\bibitem{jun2013non}
Jun J.W., \enquote{\emph{Non-{Markovian} effects on multiparticle entanglement
  swapping}}, The European Physical Journal D, vol.~67, p. 237, 2013.

\bibitem{huang2012study}
Huang P., Zhu J., He G., Zeng G., \enquote{\emph{Study on the security of
  discrete-variable quantum key distribution over non-{Markovian} channels}},
  Journal of Physics B: Atomic, Molecular and Optical Physics, vol.~45, p.
  135501, 2012.

\bibitem{bylicka2014non}
Bylicka B., Chru{\'s}ci{\'n}ski D., Maniscalco S.,
  \enquote{\emph{Non-{Markovianity} and reservoir memory of quantum channels: a
  quantum information theory perspective}}, Scientific Reports, vol.~4, p.
  5720, 2014.

\bibitem{wang2005fault}
Wang X.B., \enquote{\emph{Fault tolerant quantum key distribution protocol with
  collective random unitary noise}}, Physical Review A, vol.~72, p. 050304,
  2005.

\bibitem{wang2004quantum}
Wang X.B., \enquote{\emph{Quantum key distribution with two-qubit quantum
  codes}}, Physical Review Letters, vol.~92, p. 077902, 2004.

\bibitem{lo1999unconditional}
Lo H.K., Chau H.F., \enquote{\emph{Unconditional security of quantum key
  distribution over arbitrarily long distances}}, Science, vol. 283, pp.
  2050--2056, 1999.

\bibitem{omkar2015characterization}
Omkar S., Srikanth R., Banerjee S., \enquote{\emph{Characterization of quantum
  dynamics using quantum error correction}}, Physical Review A, vol.~91, p.
  012324, 2015.

\bibitem{omkar2015quantum}
Omkar S., Srikanth R., Banerjee S., \enquote{\emph{Quantum code for quantum
  error characterization}}, Physical Review A, vol.~91, p. 052309, 2015.

\bibitem{bellomo2007non}
Bellomo B., Franco R.L., Compagno G., \enquote{\emph{Non-{Markovian} effects on
  the dynamics of entanglement}}, Physical Review Letters, vol.~99, p. 160502,
  2007.

\bibitem{grabert1988quantum}
Grabert H., Schramm P., Ingold G.L., \enquote{\emph{Quantum {Brownian} motion:
  the functional integral approach}}, Physics Reports, vol. 168, pp. 115--207,
  1988.

\bibitem{banerjee2000quantum}
Banerjee S., Ghosh R., \enquote{\emph{Quantum theory of a {Stern-Gerlach}
  system in contact with a linearly dissipative environment}}, Physical Review
  A, vol.~62, p. 042105, 2000.

\bibitem{ban2006decoherence}
Ban M., \enquote{\emph{Decoherence of continuous variable quantum information
  in non-{Markovian} channels}}, Journal of Physics A: Mathematical and
  General, vol.~39, p. 1927, 2006.

\bibitem{maniscalco2006non}
Maniscalco S., Petruccione F., \enquote{\emph{Non-{Markovian} dynamics of a
  qubit}}, Physical Review A, vol.~73, p. 012111, 2006.

\bibitem{paz2008dynamics}
Paz J.P., Roncaglia A.J., \enquote{\emph{Dynamics of the entanglement between
  two oscillators in the same environment}}, Physical Review Letters, vol. 100,
  p. 220401, 2008.

\bibitem{piilo2008non}
Piilo J., Maniscalco S., H{\"a}rk{\"o}nen K., Suominen K.A.,
  \enquote{\emph{Non-{Markovian} quantum jumps}}, Physical Review Letters, vol.
  100, p. 180402, 2008.

\bibitem{an2007non}
An J.H., Zhang W.M., \enquote{\emph{Non-{Markovian} entanglement dynamics of
  noisy continuous-variable quantum channels}}, Physical Review A, vol.~76, p.
  042127, 2007.

\bibitem{nourmandipour2016dynamics}
Nourmandipour A., Tavassoly M.K., Rafiee M., \enquote{\emph{Dynamics and
  protection of entanglement in n-qubit systems within {Markovian} and
  non-{Markovian} environments}}, Physical Review A, vol.~93, p. 022327, 2016.

\bibitem{novais2008hamiltonian}
Novais E., Mucciolo E.R., Baranger H.U., \enquote{\emph{Hamiltonian formulation
  of quantum error correction and correlated noise: effects of syndrome
  extraction in the long-time limit}}, Physical Review A, vol.~78, p. 012314,
  2008.

\bibitem{chen2007dynamical}
Chen P., \enquote{\emph{Dynamical decoupling-induced renormalization of
  non-{Markovian} dynamics}}, Physical Review A, vol.~75, p. 062301, 2007.

\bibitem{shiokawa2007non}
Shiokawa K., Hu B.L., \enquote{\emph{Non-{Markovian} quantum error deterrence
  by dynamical decoupling in a general environment}}, Quantum Information
  Processing, vol.~6, pp. 55--79, 2007.

\bibitem{lu2010quantum}
Lu X.M., Wang X., Sun C., \enquote{\emph{Quantum {Fisher} information flow and
  non-{Markovian} processes of open systems}}, Physical Review A, vol.~82, p.
  042103, 2010.

\bibitem{xu2010experimental}
Xu J.S., Li C.F., Gong M., Zou X.B., Shi C.H., Chen G., Guo G.C.,
  \enquote{\emph{Experimental demonstration of photonic entanglement collapse
  and revival}}, Physical Review Letters, vol. 104, p. 100502, 2010.

\bibitem{orieux2014experimental}
Orieux A., d'Arrigo A., Ferranti G., Franco R.L., Benenti G., Paladino E.,
  Falci G., Sciarrino F., Mataloni P., \enquote{\emph{Experimental on-demand
  recovery of entanglement by local operations within non-{Markovian}
  dynamics}}, Scientific Reports, vol.~5, p. 8575, 2015.

\bibitem{dajka2008non}
Dajka J., Mierzejewski M., {\L}uczka J., \enquote{\emph{Non-{Markovian}
  entanglement evolution of two uncoupled qubits}}, Physical Review A, vol.~77,
  p. 042316, 2008.

\bibitem{cao2008non}
Cao X., Zheng H., \enquote{\emph{Non-{Markovian} disentanglement dynamics of a
  two-qubit system}}, Physical Review A, vol.~77, p. 022320, 2008.

\bibitem{daffer2004depolarizing}
Daffer S., W{\'o}dkiewicz K., Cresser J.D., McIver J.K.,
  \enquote{\emph{Depolarizing channel as a completely positive map with
  memory}}, Physical Review A, vol.~70, p. 010304, 2004.

\bibitem{schlosshauer2007decoherence}
Schlosshauer M.A., \emph{Decoherence: and the Quantum-to-classical Transition},
  Springer Science \& Business Media, Berlin, 2007.

\bibitem{fortes2015fighting}
Fortes R., Rigolin G., \enquote{\emph{Fighting noise with noise in realistic
  quantum teleportation}}, Physical Review A, vol.~92, p. 012338, 2015.

\end{thebibliography}

\thispagestyle{plain}


\pagebreak

\pagestyle{plain}

\phantomsection

\addcontentsline{toc}{chapter}{\textbf{LIST OF PUBLICATIONS DURING
{Ph.D.} THESIS WORK}}
\chapter*{List of publications during 
{Ph.D.} thesis work}

\noindent \textbf{\textit{Publications in International Journals}}
\begin{enumerate}
\item \textbf{Thapliyal K.}, Pathak A., Sen B., Pe{\v{r}}ina J., ``\textit{Higher-order nonclassicalities in a codirectional nonlinear optical coupler: quantum entanglement, squeezing, and antibunching}'', \href{http://link.aps.org/doi/10.1103/PhysRevA.90.013808}{Physical Review A, vol. 90, p. 013808, 2014}.
\item \textbf{Thapliyal K.}, Pathak A., Sen B., Pe{\v{r}}ina J., ``\textit{Nonclassical properties of a contradirectional nonlinear optical coupler}'', \href{http://www.sciencedirect.com/science/article/pii/S0375960114009852}{Physics Letters A, vol. 378, pp. 3431--3440,
2014}.
\item \textbf{Thapliyal K.}, Pathak A., ``\textit{Applications of quantum cryptographic switch: various tasks related to controlled quantum communication can be performed using Bell states and permutation of particles}'', \href{http://dx.doi.org/10.1007/s11128-015-0987-z}{Quantum Information Processing, vol. 14, pp. 2599--2616, 2015}.
\item \textbf{Thapliyal K.}, Banerjee S., Pathak A., Omkar S., Ravishankar V., ``\textit{Quasiprobability distributions in open quantum systems: spin-qubit systems}'', \href{http://www.sciencedirect.com/science/article/pii/S0003491615002948}{Annals of Physics, vol. 362, pp. 261--286, 2015}. 
\item \textbf{Thapliyal K.}, Verma A., Pathak A., ``\textit{A general method for selecting quantum channel for bidirectional controlled state teleportation and other schemes of controlled quantum communication}'', \href{http://dx.doi.org/10.1007/s11128-015-1124-8}{Quantum Information Processing, vol. 14, pp. 4601--4614, 2015}.
\item \textbf{Thapliyal K.}, Banerjee S., Pathak A., ``\textit{Tomograms for open quantum systems: in(finite) dimensional optical and spin systems}'', \href{http://www.sciencedirect.com/science/article/pii/S0003491616000129}{Annals of Physics, vol. 366, pp. 148--167, 2016}. 
\item \textbf{Thapliyal K.}, Pathak A., Pe{\v{r}}ina J., ``\textit{Linear and nonlinear quantum Zeno and anti-Zeno effects in a nonlinear optical coupler}'', \href{http://link.aps.org/doi/10.1103/PhysRevA.93.022107}{Physical Review A, vol. 93, p. 022107, 2016}. 
\item \textbf{Thapliyal K.}, Pathak A., Banerjee S., ``\textit{Quantum cryptography over non-Markovian channels}'',  \href{http://dx.doi.org/10.1007/s11128-017-1567-1}{Quantum Information Processing, vol. 16, p. 115, 2017}. 
\end{enumerate}



\pagebreak

\noindent \textbf{\textit{Papers presented in International/National conferences}}

\begin{enumerate}\setcounter{enumi}{8}
\item \textbf{Thapliyal K.}, Pathak A., ``\textit{Quantum Zeno and ant-Zeno effects in an asymmetric nonlinear optical coupler}'', \href{http://dx.doi.org/10.1117/12.2182653}{Proceedings SPIE 9654, International Conference on Optics and Photonics 2015, p. 96541F, 2015}.
\end{enumerate}




\noindent \textbf{\textit{Journal publications not included in the thesis}}

\begin{enumerate}\setcounter{enumi}{9}
\item Sharma R. D., \textbf{Thapliyal K.}, Pathak A., Pan A. K., De A., ``\textit{Which verification qubits perform best for secure communication in noisy channel?}'', \href{http://dx.doi.org/10.1007/s11128-015-1207-6}{Quantum Information Processing, vol. 15, pp. 1703--1718, 2016}.
\item Sharma V., \textbf{Thapliyal K.}, Pathak A., Banerjee S., ``\textit{A comparative study of protocols for secure quantum communication under noisy environment: single-qubit-based protocols versus entangled-state-based protocols}'', \href{http://dx.doi.org/10.1007/s11128-016-1396-7}{Quantum Information Processing, vol. 15, pp. 4681--4710, 2016}.
\item Giri S. K., \textbf{Thapliyal K.}, Sen B., Pathak A., ``\textit{Nonclassicality in an atom-molecule Bose-Einstein condensate:
higher-order squeezing, antibunching and entanglement}'', \href{http://www.sciencedirect.com/science/article/pii/S0378437116306197}{Physica A, vol. 466, pp. 140--152, 2017}. 
\item \textbf{Thapliyal K.}, Sharma R. D., Pathak A., ``\textit{Protocols for quantum binary voting}'', \href{http://www.worldscientific.com/doi/abs/10.1142/S0219749917500071}{International Journal of Quantum Information, vol. 15, p. 1750007, 2017}. 
\item Banerjee A., Shukla C., \textbf{Thapliyal K.}, Pathak A., Panigrahi P. K., ``\textit{Asymmetric quantum dialogue in noisy environment}'', \href{http://dx.doi.org/10.1007/s11128-016-1508-4}{Quantum Information Processing, vol. 16, p. 49, 2017}.
\item Sisodia M., Verma V., \textbf{Thapliyal K.}, Pathak A., ``\textit{Teleportation of a qubit using entangled non-orthogonal states: a comparative study}'', \href{http://link.springer.com/article/10.1007/s11128-017-1526-x}{Quantum Information Processing, vol. 16, p. 76, 2017}.
\item Sharma R. D., \textbf{Thapliyal K.}, Pathak A., ``\textit{Quantum sealed-bid auction using a modified scheme for multiparty
circular quantum key agreement}'', \href{http://dx.doi.org/10.1007/s11128-017-1620-0}{Quantum Information Processing, vol. 16, p. 169, 2017}.
\item Shukla C., \textbf{Thapliyal K.}, Pathak A., ``\textit{Hierarchical joint remote state preparation in noisy environment}'', \href{http://dx.doi.org/10.1007/s11128-017-1654-3}{ Quantum Information Processing, vol. 16, p. 205, 2017}. 
\item \textbf{Thapliyal K.}, Samantray N. L., Banerji J., Pathak A., ``\textit{Comparison of lower- and higher-order nonclassicality in photon added and subtracted squeezed coherent states}'',   \href{http://www.sciencedirect.com/science/article/pii/S0375960117307442}{Physics Letters A, vol. 381, pp. 3178--3187, 2017}. 
\item Sisodia M., Shukla A., \textbf{Thapliyal K.}, Pathak A., ``\textit{Design and experimental realization of an optimal scheme for teleportation of an n-qubit quantum state}'',  \href{https://doi.org/10.1007/s11128-017-1744-2}{Quantum Information Processing, vol. 16, p. 295, (2017)}.
\item Shukla C., \textbf{Thapliyal K.}, Pathak A., ``\textit{Semi-quantum communication: protocols for key
agreement, controlled secure direct communication and
dialogue}'', \href{https://doi.org/10.1007/s11128-017-1736-2}{Quantum Information Processing, vol. 16, p. 295, (2017)}.
\end{enumerate}





\noindent \textbf{\textit{Extended abstracts and short papers in International/National conferences}}

\begin{enumerate}\setcounter{enumi}{20}
\item \textbf{Thapliyal K.}, A.Pathak, Sen B., ``\textit{Higher order nonclassicalities in codirectional nonlinear optical coupler}'',  Proceeding of International Conference on Optics and Optoelectronics (ICOL) -2014, Dehradun, 5-8 March (2014) p. 286.
\item \textbf{Thapliyal K.}, A. Verma, Pathak A., ``\textit{A general structure of quantum channels for bidirectional controlled quantum communication schemes}'', Book of Abstracts of International Workshop and Conference on Quantum Foundations 2015, NIT Patna, India, November 28- December 4, (2015) p. 38.
\item Shukla C., \textbf{Thapliyal K.}, Pathak A., ``\textit{Hierarchical quantum communication under noisy environment}'',  Book of Abstracts of International Workshop and Conference on Quantum Foundations 2015, NIT Patna, India, November 28- December 4, (2015) p. 38.
\item \textbf{Thapliyal K.}, ``\textit{A comparative study of two different approaches of controlled quantum communication}'',  Abstract Book of Quantum Information Processing and Applications, HRI Allahabad, December 7-13, (2015).
\item \textbf{Thapliyal K.}, Pathak A., ``\textit{Nonclassical phenomena in non-linear optical couplers}'',  Book of Abstract, Student Conference on Optics and Photonics (SCOP-2016), PRL Ahmedabad, September 2-3, (2016).
\item Sharma R. D., \textbf{Thapliyal K.}, Pathak A., ``\textit{Quantum voting protocol based on controlled deterministic secure quantum communication}'',   Book of Abstract, Student Conference on Optics and Photonics (SCOP-2016), PRL Ahmedabad, September 2-3, (2016).
\item \textbf{Thapliyal K.}, Pathak A., ``\textit{Quantum communication in Himalayan region: prospects of open-air and fiber-based communication}'',  Book of Abstract of Research Papers in Physical Sciences, 86th Annual Session of the NASI, UCOST Dehradun, December 2-4, 2016 p. 4. 
\item \textbf{Thapliyal K.}, Pathak A., ``\textit{Noise against noise}'',  Abstract Book, Young Quantum-2017, HRI Allahabad, February 27-March 1, (2017). 
\item \textbf{Thapliyal K.}, Pathak A., ``\textit{Nonclassicality in hyper-Raman processes}'',   Book of Abstract, Student Conference on Optics and Photonics (SCOP-2017), PRL Ahmedabad, September 1-2, (2017) p. 23. 
\item \textbf{Thapliyal K.}, Pathak A., ``\textit{Dynamics of entanglement in various physical systems}'',  Abstract Booklet, 
International Symposium on New Frontiers in Quantum Correlations (ISNFQC18), S. N. Bose National Centre for Basic Sciences, Kolkata, January 28-February 2, (2018). 
\item Alam N., \textbf{Thapliyal K.}, Pathak A., ``\textit{Nonclassical properties of a Fabry-Perot cavity with one movable mirror}'',  Book of Abstract, International Conference on Quantum and Nonlinear Optics (QNO 2018), University of Malaya, Kuala Lumpur, Malaysia, February 2-5, (2018). 
\item \textbf{Thapliyal K.}, Pathak A., ``\textit{Effect of noise on the dynamics of quantum systems and quantum communication schemes}'', Book of Abstract, International Conference on Quantum and Nonlinear Optics (QNO 2018), University of Malaya, Kuala Lumpur, Malaysia, February 2-5, (2018). 
\item \textbf{Thapliyal K.}, Pathak A., ``\textit{Generation of nonclassical states in non-degenerate hyper-Raman processes}'',   Book of Abstract, International Conference on Quantum and Nonlinear Optics (QNO 2018), University of Malaya, Kuala Lumpur, Malaysia, February 2-5, (2018). 
\item \textbf{Thapliyal K.}, ``\textit{Optically implementable measurement device independent deterministic secure quantum communication}'',  Proceeding of the in Hundred and Fifth Session of the Indian Science Congress, Young Scientists' Award Programme, Manipur University, Imphal, March 16-20, (2018). 
\end{enumerate}




\noindent \textbf{\textit{Communicated to International Journals and not included in the thesis}}

\begin{enumerate}\setcounter{enumi}{34}
\item \textbf{Thapliyal K.}, Sharma R. D., Pathak A., ``\textit{Orthogonal-state-based and semi-quantum protocols for quantum private comparison in noisy environment}'', \href{https://arxiv.org/pdf/1608.00101.pdf}{arxiv: 1608.00101v1 (2016)}.
\item Pathak A., \textbf{Thapliyal K.}, ``\textit{A comment on the one step quantum key
distribution based on EPR entanglement}'', \href{https://arxiv.org/pdf/1609.07473.pdf}{arxiv: 1609.07473 (2016)}.
\item Naikoo J., \textbf{Thapliyal K.}, Pathak A., Banerjee S., ``\textit{Probing nonclassicality in an optically-driven cavity with two atomic ensembles}'', \href{https://arxiv.org/pdf/1712.04154.pdf}{arXiv:1712.04154 (2017)}.
\item \textbf{Thapliyal K.}, Pathak A., Sen B., Pe{\v{r}}ina J., ``\textit{Nonclassicality in non-degenerate hyper-Raman processes}'', \href{https://arxiv.org/pdf/1710.04456.pdf}{arXiv:1710.04456v1 (2017)}.
\item Alam N., \textbf{Thapliyal K.}, Pathak A., Sen B., Verma A., Mandal S., ``\textit{Lower- and higher-order nonclassicality in a Bose-condensed optomechanical-like system and a
Fabry–Perot cavity with one movable mirror: squeezing, antibunching and entanglement}'', \href{https://arxiv.org/pdf/1708.03967.pdf}{arXiv: 1708.03967v1 (2017)}.
\item \textbf{Thapliyal K.}, Pathak A., ``\textit{General structures of reversible and quantum
gates}'', \href{https://arxiv.org/pdf/1702.06272.pdf}{arxiv: 1702.06272v1 (2017)}.
\item Banerjee A., \textbf{Thapliyal K.}, Shukla C., Pathak A., ``\textit{Quantum conference}'', \href{https://arxiv.org/pdf/1702.00389.pdf}{arxiv: 1702.00389v1 (2017)}.
\item \textbf{Thapliyal K.}, Pathak A., ``\textit{Kak's three-stage protocol of secure quantum communication revisited: Hitherto unknown strengths and weaknesses of the protocol}'', \href{https://arxiv.org/pdf/1803.02157.pdf}{arXiv:1803.02157v1 (2018)}.

\end{enumerate}

%
%


\end{document}